\documentclass[10pt,oneside,openany]{book}
\usepackage{hyperref}
\hypersetup{colorlinks=true,linkcolor=blue,citecolor=blue,filecolor=blue,urlcolor=blue,}

\usepackage[utf8]{inputenc} % set input encoding (not needed with XeLaTeX)

\usepackage[a4paper, bottom=1.1in, top=1.2in, scale=0.70]{geometry} % to change the page dimensions
\usepackage{setspace}
\usepackage{Style/aas_macros}
\usepackage{multicol}
\usepackage{enumerate}
\usepackage{titlesec}

\usepackage{siunitx}
\usepackage{blindtext}
\usepackage{pdfpages}
\usepackage[font=small,labelfont=bf]{caption}
\DeclareCaptionLabelFormat{adja-page}{\hrulefill\\#1 #2 \emph{(previous page)}}

\definecolor{gray75}{gray}{0.75}
\newcommand{\hsp}{\hspace{20pt}}
\titleformat{\chapter}[hang]{\centering\huge\bfseries}{\thechapter\hsp\textcolor{gray75}{|}\hsp}{0pt}{\linespread{1.1}\selectfont\huge\bfseries}[\vspace{-1cm}]

\usepackage{relsize}

\selectfont

\usepackage{graphicx} % support the \includegraphics command and options

\usepackage{indentfirst}
\usepackage{xcolor}
\usepackage{amsmath}
\usepackage{newtxmath}
\usepackage{mathtools}
\usepackage{natbib}

\usepackage[graphicx]{realboxes}
\usepackage{anyfontsize}
\usepackage{fancyhdr}
\usepackage{epsfig}
\usepackage{graphicx}
\usepackage{amsmath}
\usepackage{theorem}
\usepackage{latexsym}
\usepackage{epic}
\long\def\/*#1*/{}

\usepackage{graphicx}
\usepackage{upgreek}
\usepackage{lipsum}
\usepackage{stfloats}
\usepackage{amsmath}	% Advanced maths commands
\usepackage[normalem]{ulem} %sout
\usepackage{zref-xr}

\usepackage{wrapfig}
\usepackage{newtxtext}
\usepackage[T1]{fontenc}
\usepackage{xcolor,colortbl}

\usepackage{graphicx}
\usepackage{siunitx}
\usepackage{hyperref}
\usepackage{lipsum}
\usepackage{bm}
\usepackage{xcolor}
\usepackage{upgreek}
\usepackage[shortlabels]{enumitem}

\usepackage{makecell}
\usepackage{anyfontsize}
\usepackage{placeins}
\usepackage{multirow}
\setlist[enumerate]{itemsep=1.5mm}
\usepackage[version=4]{mhchem}
\usepackage{tikz}

\newcommand\puteqnum{%    handy shortcut macro
  \refstepcounter{equation}\textup{(\theequation)}}

\newcommand{\rmn}[1]{\mathrm{#1}}
\newcommand{\bnabla}{\bm{\nabla}}
\newcommand{\bcdot}{\bm{\cdot}}
\newcommand{\btimes}{\bm{\times}}
\newcommand{\bfit}[1]{\textbf{\textit{#1}}}

\DeclareSymbolFont{bmisymbols}{OML}{cmm}{b}{it}
\DeclareMathSymbol{\bupsilon}{0}{bmisymbols}{"1D}

\DeclareRobustCommand{\VAN}[3]{#2}
\let\VANthebibliography\thebibliography
\def\thebibliography{\DeclareRobustCommand{\VAN}[3]{##3}\VANthebibliography}

\newcommand\T{\rule{0pt}{2.6ex}}       % Top strut
\newcommand\B{\rule[-1.2ex]{0pt}{0pt}} % Bottom strut

\newcommand{\bs}[1]{\boldsymbol{#1}}

\titlespacing\section{0pt}{4pt plus 4pt minus 2pt}{-6pt plus 2pt minus 2pt}
\titlespacing\subsection{0pt}{4pt plus 4pt minus 2pt}{-6pt plus 2pt minus 2pt}
\titlespacing\subsubsection{0pt}{4pt plus 4pt minus 2pt}{-6pt plus 2pt minus 2pt}

\newcommand\clinee{
\vspace{-0.5cm}
\begin{center}
\noindent\rule{8cm}{0.4pt}
\end{center}
\vspace{-0.5cm}
}

\def\addcontentslinex#1#2#3{% added <<<<<<<<<<<<<
    \addtocontents{#1}{\protect\contentsline{#2}{#3}{}{}\protect}}
\makeatother

\newcommand{\Chapter}[2]{%
    \chapter[#1: {\itshape#2}]{#1\\\medskip\smaller\itshape{#2}}
    \addcontentslinex{lof}{chapter}{\large Chapter \thechapter: \textit{#1}}
}

\newcommand{\ChapterX}[2]{%
    \chapter[#1 {\itshape#2}]{#1\\\medskip\smaller\itshape{#2}}
    \addcontentslinex{lof}{chapter}{\large Chapter \thechapter: \textit{#1}}
}

\newcommand\elec{\mathrm{e}}
\newcommand\pl{\mathrm{pl}}

\usepackage{cleveref}[2012/02/15]
\crefformat{footnote}{#2\footnotemark[#1]#3}
\DeclareGraphicsExtensions{.pdf}

\graphicspath{{Images/}{Images/Chapter2/}{Images/Chapter3/}{Images/Chapter4/}{Images/Chapter5/}{Images/Chapter6/}{Images/Chapter7/}}

\begin{document}

\frontmatter
\setcounter{page}{1}

\newpage

\newgeometry{left=4cm, right=4cm, top=3cm, bottom=3cm, footskip=1.15in}
\fancyfootoffset{0pt} 
\thispagestyle{empty}

\begin{center}
    \vspace*{-0.25cm}
    {\setstretch{1.2}{
    \huge \textbf{Merging galaxies and clusters:\\ Insights into the role of magnetic fields and the physics of radio relics}\par}
    }

    \vspace*{0.2cm}\par
    \rule{8cm}{0.4pt}
    \vspace*{0.2cm}\par
    
    {\LARGE Joseph Whittingham}
    \vspace*{0.4cm}
    
    {\large 
        \textbf{
        Dissertation \\
        zur Erlangung des akademischen Grades \\
        ``doctor rerum naturalium'' \\
        (Dr. rer. nat.) \\
        in der Wissenschaftsdisziplin Theoretische Astrophysik \\
        \vspace*{0.9cm}
        eingereicht an der \\
        Mathematisch-Naturwissenschaftlichen Fakult\"{a}t \\
        Institut f\"{u}r Physik und Astronomie \\
        der Universit\"{a}t Potsdam \\
        und \\
        Leibniz-Institut f\"{u}r Astrophysik Potsdam (AIP)
        }
    }
    
    \vspace*{.7cm}\par

    {\large 
        {
        Ort und Tag der Disputation:\\
        Potsdam am 27.02.2025
        }
    }
    
    \vspace*{.2cm}\par

    {\large 
        {
        Pr\"{a}dikat / Grade: \\
        \textit{Summa cum laude} (with highest honours)
        }
    }

    \vspace*{.8cm}\par

    \begin{center}
    \raisebox{-0.5\height}{\includegraphics[width=0.28\linewidth]{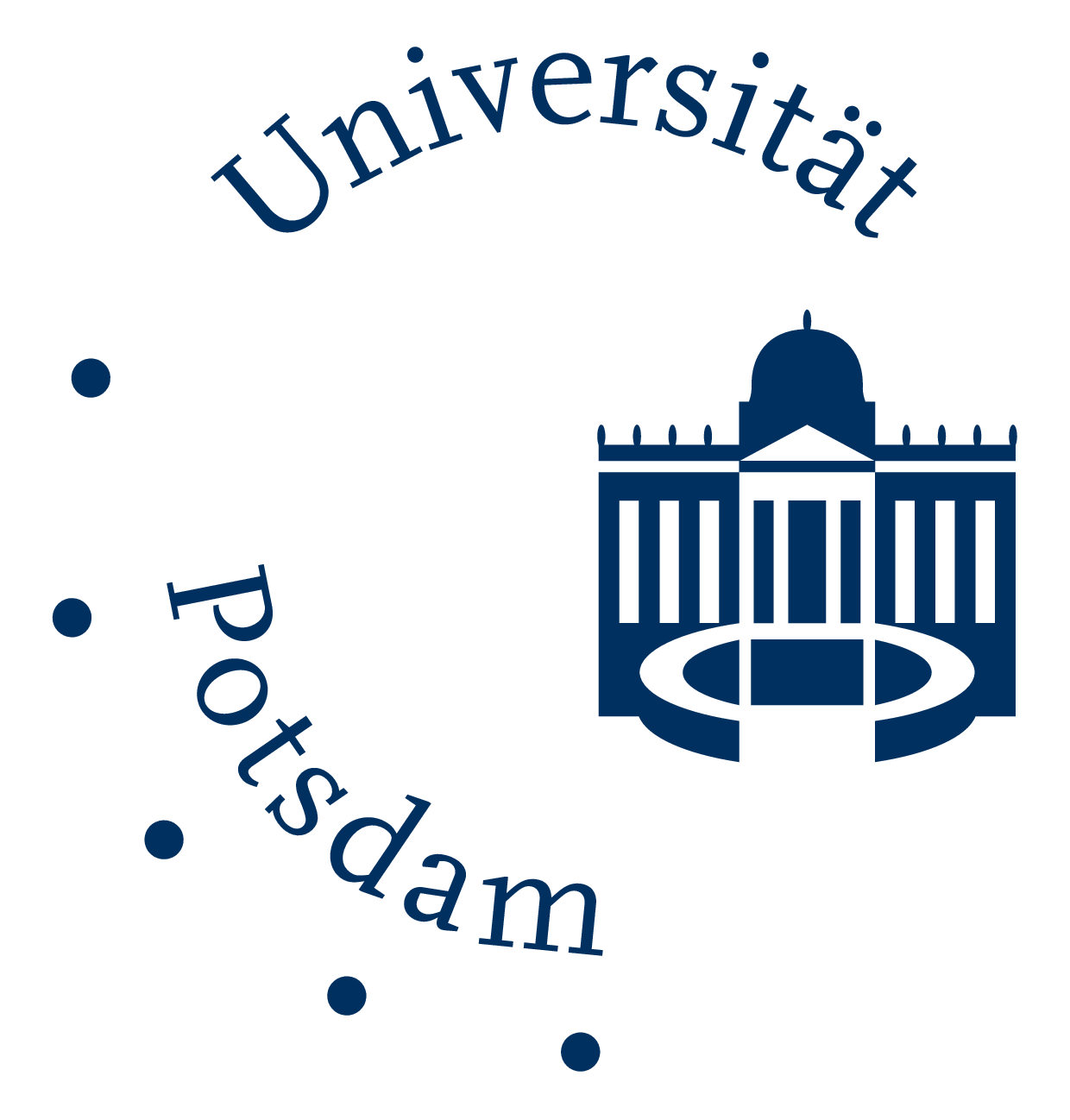}}
    \hspace{2cm}
    \raisebox{-0.5\height}{\includegraphics[width=0.25\linewidth]{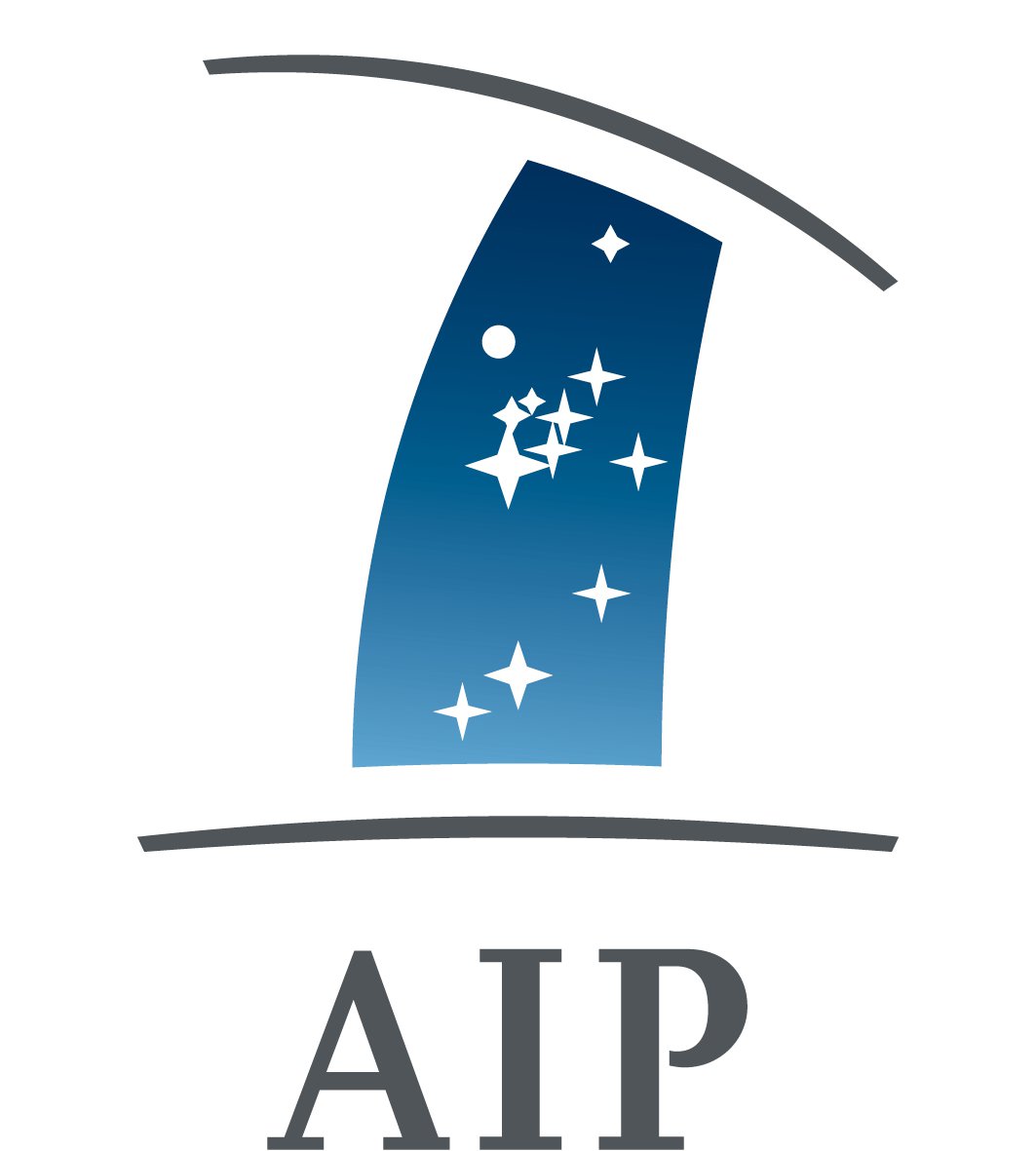}}
    \end{center}

\end{center}

\restoregeometry

{
\newpage

\newgeometry{left=2.5cm, right=2.5cm, top=3cm, bottom=2.5cm}
\thispagestyle{empty}

\begin{flushleft}
\normalsize
\hypersetup{urlcolor=black}
Unless otherwise indicated, this work is licensed under a Creative Commons License\\
Attribution 4.0 International. This does not apply to quoted content and works based on other\\
permissions.\\
To view a copy of this licence visit:\\
\href{https://creativecommons.org/licenses/by/4.0/legalcode.en}{\text{https://creativecommons.org/licenses/by/4.0/legalcode.en}}

\vfill 

\noindent
{\normalsize
\begin{tabular}{@{}ll}
Betreuer: & Prof. Dr. Christoph Pfrommer \\
Mentor: & Dr. Oliver Gressel \\
[0.5cm]
Gutachter: & Prof. Dr. Christoph Pfrommer \\
& Prof. Dr. Romain Teyssier \\
& Prof. Dr. Thomas W. Jones \\
\end{tabular}
}

\vspace{0.5cm}

Published online on the\\
Publication Server of the University of Potsdam:\\
\href{https://doi.org/10.25932/publishup-67622}{\text{https://doi.org/10.25932/publishup-67622}}\\
\href{https://nbn-resolving.org/urn:nbn:de:kobv:517-opus4-676229}{\text{https://nbn-resolving.org/urn:nbn:de:kobv:517-opus4-676229}}
\end{flushleft}

\restoregeometry

\newgeometry{left=4cm, right=4cm, top=3cm, bottom=5cm, footskip=1.25in}
\fancyfootoffset{0pt} 
\chapter*{Abstract}
\addcontentsline{toc}{chapter}{\numberline{}Abstract}

Mergers have long been understood to be a driver of galaxy and galaxy cluster evolution. They release tremendous amounts of gravitational potential energy -- $\mathcal{O}(10^{59})$ and $\mathcal{O}(10^{64})$ ergs in galaxies and clusters, respectively -- which is dissipated in powerful shock waves. In galaxies especially, the strong tidal effects can have profound effects on the remnant morphology. Although the modelling of mergers has a long history, it is only recently that it has been fully appreciated just how sensitive they are to a range of factors, including the existence of circumgalactic media (CGM), accretion along filaments, and pre-existing magnetic fields. Modelling these aspects in a cosmologically-consistent manner necessitates the use of high-resolution cosmological magnetohydrodynamic (MHD) simulations. In this work, we use such simulations to investigate two distinct merger-related phenomena: i) magnetic fields in galaxy mergers, and ii) the origin of \textit{radio relics} in galaxy clusters.

For the first topic, we isolate the impact of magnetic fields by running a series of hydrodynamic and MHD ``zoom-in'' simulations of disc galaxy mergers. We show that magnetic fields are able to have a major impact on the morphology of the remnant galaxy: remnants in MHD simulations form extended discs with flocculent spiral structures, whilst hydrodynamic simulations form compact remnants with bar-and-ring morphologies uncommon in observations. We show that a small-scale dynamo can only develop given sufficient resolution, explaining why this effect has not been seen before. Furthermore, we present a mechanism that explains \textit{how} the magnetic fields affect mergers: we find that the amplified magnetic field alters the transport of angular momentum during the merger, with subsequent resonances, stellar feedback, and accretion ultimately causing the diverging morphologies. Finally, we show that the impact of such mergers can be felt even in galaxies that have more quiescent merging histories. We conclude that magnetic fields are thus essential for the accurate simulation of disc galaxies.

For the second topic, we start by outlining the seven major problems that currently challenge our understanding of radio relics. To solve these, we use a hybrid strategy; first running zoom-in simulations of cluster mergers to identify typical shock conditions at relic distances, before using the results to inform a series of idealised shock-tube simulations. We find that, upon colliding with an accretion shock, merger shocks produce a thin, dense sheet, which propagates into upstream density fluctuations. This scenario leads to a) the formation of a Mach number distribution, b) shock corrugation, and c) a Rayleigh-Taylor instability at the contact discontinuity. These effects flatten cosmic ray electron spectra, biasing radio-derived Mach numbers high, thereby explaining the observed discrepancy with X-ray derived Mach numbers. They also lead to additional compression, which produces the observed $\upmu$G-strength magnetic fields, and turbulence behind the shock front, which invalidates laminar-flow based cooling models. Furthermore, we find that the tail of the Mach number distribution dominates the radio emission. This explains the observation of radio relics in shocks with $\mathcal{M}_\rmn{X-ray} \lesssim 2$, despite the increasing evidence for the existence of a critical Mach number of $\mathcal{M}_\rmn{crit} \approx 2.3$, below which cosmic ray electron acceleration is ineffective. Finally, we find that upstream density turbulence can explain many aspects of radio relic morphology. We hence solve four of the seven outstanding problems, with partial answers provided to a fifth.

\restoregeometry

\newgeometry{left=4cm, right=4cm, top=3cm, bottom=5cm, footskip=1.25in}
\fancyfootoffset{0pt} 

\chapter*{Zusammenfassung}
\addcontentsline{toc}{chapter}{\numberline{}Zusammenfassung}

Kollisionen gelten seit langem als treibende Kraft hinter der Entwicklung von Galaxien und Galaxienhaufen. Sie setzen enorme Mengen an Gravitationsenergie frei -- $\mathcal{O}(10^{59})$ bzw. $\mathcal{O}(10^{64})$ Erg in Galaxien bzw.\ Galaxienhaufen -- die in starken Stoßwellen dissipiert werden. Insbesondere in Galaxienkollisionen können die starken Gezeiteneffekte tiefgreifende Auswirkungen auf die Morphologie der entstehenden Galaxie haben. Obwohl die Modellierung von Galaxienfusionen eine lange Geschichte hat, ist erst vor kurzem vollständig gewürdigt worden, wie empfindlich sie auf eine Reihe von Faktoren reagieren, darunter die Existenz zirkumgalaktischer Medien (CGM), Akkretion entlang von Filamenten und bereits vorhandene Magnetfelder. Um diese Aspekte auf kosmologisch konsistente Weise zu modellieren, sind hochauflösende kosmologische magnetohydrodynamische (MHD) Simulationen erforderlich. In dieser Arbeit verwenden wir solche Simulationen, um zwei unterschiedliche Phänomene im Zusammenhang mit Kollisionen zu untersuchen: i) Magnetfelder bei Fusionen von Galaxien und ii) den Ursprung von \textit{Radiorelikten} in Galaxienhaufen.

Für das erste Thema isolieren wir die Auswirkungen von Magnetfeldern, indem wir eine Reihe von hydrodynamischen und MHD-„Zoom-In“-Simulationen von Fusionen von Scheibengalaxien durchführen. Wir zeigen, dass Magnetfelder einen großen Einfluss auf die Morphologie der entstehenden Galaxie haben können: die neufusionierten Galaxien in MHD-Simulationen bilden ausgedehnte Scheiben mit Flocken-artigen Spiralstrukturen, während hydrodynamische Simulationen kompakte Galaxien mit Balken- und Ringmorphologien bilden, die so nicht beobachtet werden. Wir zeigen, dass sich ein klein-skaliger Dynamo nur bei ausreichender Auflösung entwickeln kann, was erklärt, warum dieser Effekt bisher nicht beobachtet wurde. Darüber hinaus präsentieren wir einen Mechanismus, der erklärt, wie sich Magnetfelder auf Galaxienfusionen auswirken: Wir stellen fest, dass das verstärkte Magnetfeld den Transport des Drehimpulses während der Fusion verändert, wobei nachfolgende Resonanzen, Feedback von Sternen und Akkretion letztendlich die unterschiedlichen Morphologien verursachen. Schließlich zeigen wir, dass die Auswirkungen solcher Galaxienfusionen sogar in Galaxien vorkommt, deren Entstehungsgeschichte eher ruhig ist. Wir kommen zu dem Schluss, dass Magnetfelder daher für die genaue Simulation von Scheibengalaxien von entscheidender Bedeutung sind.

Für das zweite Thema beginnen wir mit der Darstellung der sieben Hauptprobleme, die derzeit unser Verständnis von Radiorelikten in Frage stellen. Um diese zu lösen, verwenden wir eine Hybridstrategie: Zunächst führen wir Zoom-In-Simulationen von Kollisionen von Galaxienhaufen durch, um typische Bedingungen von Stoßwellen bei den Radien zu identifizieren, wo Radiorelikte vorkommen. Danach verwenden wir die Ergebnisse, um eine Reihe idealisierter Stoßrohrsimulationen zu erstellen. Wir stellen fest, dass durch Haufenkollision verursachte Stoßwellen mit Stoßwellen interagieren, die durch Gasakkretion zustande kommen, und dabei eine dünne, dichte Schicht erzeugen, die sich in vorgelagerte Dichtefluktuationen ausbreitet. Dieses Szenario führt zu a) der Bildung einer Machzahlen-Verteilung, b) einer Faltung der Stoßwellen und c) einer Rayleigh-Taylor-Instabilität an der Kontaktdiskontinuität. Diese Effekte flachen die Elektronenspektren der kosmischen Strahlung ab und verzerren die aus Radiostrahlung abgeleiteten Mach-Zahlen zu größeren Werten, was die beobachtete Diskrepanz mit den aus Röntgenstrahlung abgeleiteten Machzahlen erklärt. Sie führen auch zu zusätzlicher Kompression, die die beobachteten Magnetfelder mit einer Stärke von $\upmu$G erzeugt, und Turbulenzen hinter der Stoßfront, die auf laminarer Strömung basierende Kühlmodelle ungültig machen. Darüber hinaus stellen wir fest, dass das Ende der Machzahlen-Verteilung die Radioemission dominiert. Dies erklärt die Beobachtung von Radiorelikten in Stoßwellen mit $\mathcal{M}_\rmn{X-ray} \lesssim 2$, trotz der zunehmenden Hinweise auf die Existenz einer kritischen Mach-Zahl von $\mathcal{M}_\rmn{crit} \approx 2,3$, unterhalb der die Beschleunigung der Elektronen der kosmischen Strahlung nicht besonders effizient ist. Schließlich stellen wir fest, dass die Dichteturbulenzen stromaufwärts viele Aspekte der Morphologie von Radiorelikten erklären können. Damit lösen wir vier der sieben offenen Probleme und liefern Teilantworten auf ein fünftes.
\restoregeometry

\let\tmp\oddsidemargin
\let\oddsidemargin\evensidemargin
\let\evensidemargin\tmp
\reversemarginpar

\addcontentsline{toc}{chapter}{\numberline{}Contents}
\tableofcontents

\mainmatter
\pagenumbering{arabic}

\chapter{Introduction}
\label{chapter:introduction}

\section{Context}

In the last half-century, numerical simulations have revolutionised our understanding of various astrophysical phenomena. This includes seminal works such as the galaxy merger simulations of \citet{toomre1972}, which showed that the distorted morphologies of many irregular galaxies can be explained by mergers, and the pioneering simulations by \citet{davis1985}, which showed that only a cold dark matter model with a cosmological constant (known as the $\Lambda$CDM model) can produce large-scale structure consistent with observations, thereby ushering in the age of the standard model of cosmology. Since these studies, the computational power available to numericists has increased exponentially, enabling simulations to become ever more detailed. This has improved them in four key ways:
\begin{enumerate}
    \item \textbf{Increased resolution:}
    
    Naturally, increased resolution increases the depth of structure that can be resolved. However, it can also improve the robustness of the simulation. It has been found, for example, that structures in galaxy simulations are only well-resolved above a threshold of ten million particles \citep{khoperskov2007}. Furthermore, some phenomena are resolution-dependent. For example, \citet{sparre2016} find that bursty star formation can only be replicated in the \citet{springel2003} interstellar medium (ISM) model above a threshold dark matter mass resolution of $m_\rmn{DM} = 5\times10^5 \rmn{M}_\odot$.\vspace{0.3cm}
    \item \textbf{Additional physics:}

    In many cases this is facilitated by the previous point as, without sufficient resolution, the effects of such physics fall under the domain of subgrid models. For example, older ISM models typically make implicit assumptions about the nature of multiphase gas, whilst models today increasingly try to model this explicitly. By modelling physics explicitly, we move away from descriptive models, which necessarily require the use of tuning parameters, and towards less arbitrary models, based upon more fundamental laws of physics\footnote{It should be noted, however, that this is only helpful as long as the model is robust to changes in resolution and random number generation.}. In galaxy and galaxy cluster simulations, non-standard physics now includes radiative transfer \citep{kannan2020}, non-equilibrium chemistry \citep{thomas2024}, neutrinos \citep{liu2018}, magnetic fields \citep{pakmor2017}, and cosmic rays \citep{Pfrommer2017}\footnote{Citations are provided as examples and by no means form an exhaustive list.}. 

    \item \textbf{Simulating the correct environment}:

    It has long been known that galaxies are ``aware'' of their environment, in that they show correlations between, for example, morphology and density \citep{dressler1980}. It is only much more recently, however, that we have realised to what extent they are affected by the circumgalactic medium (CGM) \citep{sparre2022}, accretion along filaments \citep{berlok2019}, and the impact of so-called galactic fountain flows, launched by stellar and active galactic nuclei (AGN) driven winds \citep{grand2019}. The environment also plays a major role in sculpting shock wave morphology \citep{lee2024} and, more generally, in setting the orbital parameters of mergers, as galaxy and cluster velocities are strongly affected by the tidal field. Several works have shown that the outcome of galaxy mergers is highly sensitive to these parameters \citep[see, e.g.,][]{naab2003}.
    
    \item \textbf{Simulating over cosmological time}:

    Whilst the environment plays a key role in galaxy and galaxy cluster simulations, a fully realistic simulation must also take into account the cosmology. This usually necessitates running the simulation over cosmological time, thereby allowing structure to evolve in a cosmologically-consistent manner. In this way, we can further reduce the number of free parameters available for tuning. Such simulations are then better able to reflect the physics of our own Universe.
    
\end{enumerate}

Whilst cosmological simulations have many advantages, they are computationally very expensive. This typically leads to a reduction in the maximum possible resolution. As a result, one of two techniques are typically used: firstly, cosmological ``zoom-in'' simulations are employed, which concentrate computational resources on the Lagrangian frame around an object of interest, with degraded resolution at distances beyond this zone. In doing so, the large-scale tidal field and accretion flows are accurately captured, with the object itself being resolved with enhanced resolution. The second method is to create isolated simulations using results from cosmological simulations to inform the initial conditions. This method allows for yet higher resolution, and provides an idealised sandbox, in which parameters can be easily adjusted. This, in turn, allows us to more easily understand how such parameters impact the resulting physics. Most importantly, however, basing such simulations on cosmologically-consistent results means that we make sure that we are probing \textit{relevant} physics. We have used both of these techniques in this dissertation to tackle the following topics:
\begin{enumerate}[i)]
    \item The role of magnetic fields in disc galaxy mergers
    \item The physics of radio relics in galaxy clusters
\end{enumerate}

Both of these research areas are of significant interest: mergers play a key role in galaxy evolution, but which physics is important in this process, and, consequently, what needs to be included in future simulations of galaxy evolution, is as yet unclear. On the other hand, with the advent of higher resolution observations, radio relics are being increasingly appreciated for their potential as cosmic-scale laboratories, providing insights into the acceleration of cosmic ray electrons and plasma-scale processes, as well as a window onto the intracluster medium (ICM) at the periphery of galaxy clusters. Moreover, the tools required to investigate these problems in a cosmologically-consistent manner -- namely, cosmological magnetohydrodynamic (MHD) codes \citep[e.g.][]{bryan2014, pakmor2017, katz2021} and spectral cosmic ray electron codes \citep{winner2019, boess2023} -- have finally become available in the last few years. 

To answer the role of magnetic fields in disc galaxies, we analyse the cosmological ``zoom-in'' galaxy merger simulations initially produced for my \href{https://www.aip.de/media/thesis/joseph-whittingham-master-thesis.pdf}{Masters thesis}. This comprises of a suite of eight high-resolution simulations, run with MHD and hydrodynamic variations. They are supported by four further simulations run to provide a resolution study and analysis of eight simulations of more isolated, albeit still cosmological, galaxies. For the radio relic study, meanwhile, we perform zoom-in simulations of galaxy clusters, using these to analyse the development of shocks in cluster mergers. From these results, we construct a series of idealised shock-tube simulations, where we can better resolve the underlying physics. By evolving cosmic ray electron spectra using the code \textsc{Crest} \citep{winner2019} and post-processing the result with the emission code \textsc{Crayon+} \citep{werhahn2021}, we are able to produce mock-observables \textit{ab initio}. In doing so we are able to show that magnetic fields play a major role in mergers of disc galaxies, and can propose a scenario that solves many of the problems currently challenging our understanding of radio relics.

\section{Structure of the thesis}

The thesis is structured as follows: in Chapter~\ref{chapter:theory}, I provide a summary of the theory required to understand the topics tackled in the following chapters. Specifically, I cover the basics of structure formation in the Universe, including galaxy and galaxy cluster formation (Sec.~\ref{sec:basics} and Sec.~\ref{sec:baryonic-structure}). I then provide an overview of galaxy and galaxy cluster mergers (Sec.~\ref{sec:mergers}), and the ground theory behind magnetic fields (Sec.~\ref{sec:magnetic-fields}), shocks (Sec.~\ref{sec:shocks}), and cosmic rays (Sec.~\ref{sec:cosmic-rays}).

In Chapter~\ref{chapter:paper-one}, I present the published paper \citet{whittingham2021}. Here, we build on work first presented in my Masters thesis, showing how gas-rich mergers of disc galaxies are affected by magnetic fields. We find that, given sufficient resolution, they have a substantial effect, and that this effect can be felt even in simulations of disc galaxies with more quiescent merger histories. We additionally show evidence for the existence of a small-scale dynamo in these simulations.

In Chapter~\ref{chapter:paper-two}, I present the published paper \citet{whittingham2023}. Here, we expand upon this analysis, showing \textit{how} the magnetic fields are able to affect the outcome of mergers. We find that they are able to have a significant effect on the mediation (or otherwise) of angular momentum. This affects the baryonic concentration in the merger remnants, changing the way resonances form within the disc. This, in turn, has a knock-on effect on the generation of stellar winds, which substantially affects the further accretion of gas post-merger. We discuss the conditions under which these results hold for other merger scenarios, galaxy formation models, and MHD implementations. We conclude that magnetic fields are critical for the accurate simulation of disc galaxies.

In Chapter~\ref{chapter:paper-three}, I present the submitted paper \citet{whittingham2024}. Here, we focus on radio relics, defining seven major outstanding problems regarding their origin. Using cosmological simulations of cluster mergers, we identify that merger shocks collide with accretion shocks at distances typical for radio relics. This results in the production of a dense, shock-compressed sheet. By modelling this scenario with idealised shock-tube simulations and including unresolved upstream density fluctuations, we show that this process leads to: i) the formation of a distribution of Mach numbers at the shock-front, ii) shock corrugation, and iii) the generation of a Rayleigh-Taylor instability at the trailing edge of the shock-compressed region. We show that, in turn, these effects are able to explain: i) the X-ray vs. radio Mach number discrepancy, ii) the observation of $\upmu$G-strength magnetic fields in radio relics, and iii) the inability of standard cooling models to replicate spectral index variations.

In Chapter~\ref{chapter:paper-four}, I present work which forms the basis of a paper currently in preparation. This paper expands upon the previous one by varying the relative variance, power law slope, and the injection scale of the upstream density turbulence. In doing so, we are able to show that it is primarily the relative variance that sets the Mach number distribution and impacts the radio relic morphology. We also show that, due to corrugation, the shock front is able to produce a filamentary structure in projection, which is consistent with observations. Indeed, we show that this may provide a window into the scale of turbulent injection in the ICM. Finally, we also show that, as the radio emission is predominantly produced by the tail of the Mach number distribution, a critical Mach number of $\mathcal{M}_\rmn{crit} \approx 2.3$ is compatible with observations in shocks at least as weak as $\mathcal{M} = 2$, even in upstream conditions with low relative variance.

In Chapter~\ref{chapter:additional-work}, I recap the work undertaken to make the cosmic ray spectral code \textsc{Crest} and \textsc{Arepo}'s shock-finder suitable for high-resolution cosmological simulations. These improvements have been made with a view to running simulations that show whether sufficiently energetic fossil electrons can be produced for radio relics through large-scale structure formation alone \citep[see earlier work by][]{pinzke2013}.

In Chapter~\ref{chapter:conclusions}, I summarise the major conclusions of the preceding chapters and present the outlook for the future.

Finally, in Chapter~\ref{publications_list}, I provide a list of all the publications I have been involved in during my doctoral studies, with an explicit breakdown of my contributions. This includes a breakdown of which parts of Chapters~\ref{chapter:paper-one} and~\ref{chapter:paper-two} were produced during these studies.
\ChapterX{Theory}{}
\label{chapter:theory}

\section{Basics of cosmology}
\label{sec:basics}

In this first section, we present a brief summary of the principles that inform the initial conditions of our cosmological simulations. This includes the origin of the cosmological density parameters used, and a short recap of how structure forms in a hierarchical $\Lambda$CDM Universe. Standard theory can be found in \citet{peebles1993} and \citet{peacock1999}, amongst other well-known cosmology textbooks.

\subsection{Assumptions and parameters}
\vspace{0.4cm}
Most cosmological models are based on three key assumptions:
\begin{enumerate}[i)]
    \item the \textit{cosmological} or \textit{Copernican} principle, which implies that our position in the Universe is not preferred to any other;
    \item when averaged over sufficiently large scales, the observable
properties of the Universe are isotropic, and
    \item \textit{general relativity}, which states that spacetime is a four-dimensional manifold governed by Einstein's field equations\footnote{Informally, this framework is often paraphrased as ``matter tells spacetime how to curve; curved spacetime tells matter how to move''.}.
\end{enumerate}
By the first \textit{assumption}, the second must hold for every observer in the Universe. It follows that if the Universe is isotropic around every point, it must also be homogeneous.
Isotropy is supported by observations of the cosmic microwave background (CMB, see Sec.~\ref{sec:cmb}), whose temperature fluctuations are on the order of $\Delta T / T \approx 10^{-5}$ across all lines-of-sight \citep{wmap2011, planck2014,  planck2018}. It is also supported by redshift surveys, which show that structure in the Universe is approximately isotropic and homogeneous on scales much larger than 100 Mpc \citep{blanton2017}, thus providing a successful test of the cosmological principle. General relativity, meanwhile, has produced predictions consistent with observations across a range of scales. These include, but are not limited to, the precession of Mercury's perihelion, the deflection of light by gravity \citep{dyson1920}, gravitational redshift \citep{pound1959}, gravitational time dilation (e.g. the timekeeping underlying Global Positioning Systems), the observation of gravitational waves \citep{abbott2016} and the existence and observation of the ``shadow'' of a black hole \citep{ehtc2022}. 

Assuming an isotropic and homogeneous universe, one can directly derive the Friedmann-Lema\^{i}tre-Robertson-Walker metric: 
\begin{equation}
    \mathrm{d}s^2 = -c\mathrm{d}t^2 + a(t)\mathrm{d}\bs{\sigma}^2.
\end{equation}
Here, $\mathrm{d}s^2$ is the spacetime interval, $c$ is the speed of light, $\mathrm{d}\bs{\sigma}$ is the metric for a three-dimensional maximally-symmetric space, and $a(t)$ is the scale factor, which describes the expansion or contraction of the Universe with time, $t$. Adopting this metric and applying it to Einstein's field equations, we can derive Friedmann's equations:
\begin{equation}
    \left(\frac{\dot{a}}{a}\right)^2 = \frac{8 \pi G \rho}{3} - \frac{kc^2}{a^2} + \frac{\Lambda c^2}{3},
    \label{eq:friedmann-one}
\end{equation}
and 
\begin{equation}
    \left(\frac{\ddot{a}}{a}\right) = - \frac{4 \pi G}{3} \left(\rho + \frac{3 P}{c^2}\right) + \frac{\Lambda c^2}{3},
\end{equation}
where $a = a(t)$, dots represent time derivatives, $G$ is the gravitational constant, $\rho$ is the volumetric mass density, $P$ is pressure, $\Lambda$ is the so-called cosmological constant\footnote{There have been recent claims that this ``constant'' evolves over time \citep{desi2024}.}, which can be generalised to the concept of \textit{dark energy}, and $k$ is a curvature parameter, equal to -1, 0, or 1, describing negative, zero, and positive spatial curvature, respectively. 

These equations can be solved exactly for a perfect fluid that has the equation of state:
\begin{equation}
    P = w \rho c^2,
    \label{eq:equation-of-state}
\end{equation}
where $w$ determines how the pressure scales with the rest mass energy, $\rho c^2$. Non-relativistic matter, for example, can be approximated with $w=0$, as $P \ll \rho c^2$. Meanwhile, for relativistic fermions and bosons, $w=1/3$. This implies that 
\begin{equation}
\rho_\rmn{m}(t) = \rho_\rmn{m,0} \, a^{-3}
\label{eq:matter}
\end{equation}
and
\begin{equation}
\rho_\rmn{r}(t) = \rho_\rmn{r,0} \, a^{-4},
\end{equation}
where subscripts ``m'' and ``r'' refer to non-relativistic and relativistic matter, respectively, and the subscript ``$0$'' indicates the variable at the present time. By convention, we typically set $a(t_0) = 1$.

In the following, we will use the Hubble function, defined as:
\begin{equation}
    H(a) \coloneq \frac{\dot{a}}{a}.
\end{equation}
This is sometimes parameterised by the dimensionless Hubble parameter, $h$, where $H_0 = H(a=1) = 100 h$ km s$^{-1}$ Mpc$^{-1}$, allowing for the comparison of data between different cosmological models. Inverting the Hubble constant at the current time produces the Hubble time, $t_\rmn{H} = 1 / H_0$, which is the current cosmological expansion timescale and a very rough estimate for the age of the Universe. This will be useful for comparing timescales later.

We now define the density parameter:
\begin{equation}
    \Omega_X \coloneq \frac{\rho_X} {\rho_\rmn{crit}},
\end{equation}
where $X$ represents different matter-energy components in the Universe, and $\rho_\rmn{crit}$ is the so-called critical density:
\begin{equation}
    \rho_\rmn{crit}(a) \coloneq \frac{3 H(a)^2}{8 \pi G},
\end{equation}
This parameter is important, as, if we let
\begin{equation}
    \rho_\Lambda \equiv \frac{\Lambda c^2}{8 \pi G},
\end{equation}
we may set
\begin{equation}
    \rho_\rmn{tot}(a) \equiv \rho_\rmn{m}(a) + \rho_\rmn{r}(a) + \rho_\rmn{\Lambda}
    \label{eq:rho_tot}
\end{equation}
and thus, if $\rho_\rmn{tot}(a) = \rho_\rmn{crit}(a)$, by Eq.~\eqref{eq:friedmann-one}, $k=0$, and we live in a spatially flat Universe.

Using the above definitions and scalings, we may re-write Eq.~\eqref{eq:friedmann-one} as:
\begin{equation}
    \frac{H(a)}{H_0} = E(a) = \sqrt{\Omega_\rmn{r,0}a^{-4} + \Omega_\rmn{m,0}a^{-3} + \Omega_\rmn{k,0}a^{-2} + \Omega_\rmn{\Lambda}},
    \label{eq:friedmann-with-density-parameters}
\end{equation}
where variables have their previously given definitions and we have introduced the new variable $E(a)$ to abbreviate the right-hand side\footnote{We have also assumed in this step that $P_\Lambda = -\rho_\Lambda c^2$. This is consistent with the results of \citet{planck2018}, but more complex models do exist \citep[see, in particular,][]{desi2024}.}. In order to understand the growth of the Universe we need only assign values to the cosmological density parameters, $\Omega_X$. 

\subsection{Measuring the cosmological parameters}

There are several strategies available for measuring the cosmological parameters in Eq.~\eqref{eq:friedmann-with-density-parameters}. Firstly, $H_0$ can be measured by using so-called ``standard candles'' where we have constraints on the typical luminosity of an object. In the local Universe, such standard candles are Cepheid variable stars \citep{anderson2024} and the tip of the red giant branch \citep{freedman2019}. By comparing the observed brightness with their expected brightness, we can derive the distance of the stars, $d$. This can then be used to calculate $H_0$, as
\begin{equation}
    \bupsilon_\rmn{r} = H_0 d + \bupsilon_\rmn{pec},
\end{equation}
where $\bupsilon_\rmn{r}$ is the recessional velocity along the line of sight: the speed that the object is moving away from us, and $\bupsilon_\rmn{pec}$ is the peculiar velocity: the velocity component along the line of sight unrelated to cosmological expansion. The remaining factor, $H_0 d$ is often referred to as the \textit{Hubble flow}.

Calculating $\bupsilon_\rmn{r}$ can, in turn, be done by measuring the redshift of the emitted light:
\begin{equation}
    z \coloneq \frac{\lambda_\rmn{obs} - \lambda_\rmn{emit}}{\lambda_\rmn{emit}} 
\end{equation}
where $\lambda_\rmn{emit}$ is the wavelength the light was emitted at and $\lambda_\rmn{obs}$ is the wavelength it was observed at. For local measurements and non-relativistic speeds, $z \approx \bupsilon_\rmn{r} / c$, so that, statistically,
\begin{equation}
    cz = H_0 d.
    \label{eq:hubble-approx}
\end{equation}
Note, that when the shift in wavelength is purely due to the expansion of space,
\begin{equation}
1 + z = \frac{1}{a(t)}.
\end{equation}
Hence cosmological redshift is directly linked to the size of the Universe at the time of emission and also, through Eq.~\eqref{eq:friedmann-with-density-parameters}, a measure of time elapsed since emission.

The local distance measurement with the current least uncertainty is $H_0 = 73.04 \pm 1.04$ km s$^{-1}$ Mpc$^{-1}$ \citep{riess2022}. It should be noted, however, that measurements derived from the CMB imply a significantly lower value of $H_0 = 67.66 \pm 0.42$ km s$^{-1}$ Mpc$^{-1}$ \citep{planck2018}. The origin of this discrepancy is still a cause of considerable debate \citep{hu2023}. 

Beyond the local Universe, we can use supernovae type Ia (SNIa) as a standardisable candle, as these release approximately 10$^{43}$ erg s$^{-1}$ at maximum brightness \citep{prialnik2009} and can be calibrated via the Philipps relation (a tight correlation of light-curve width and peak apparent brightness) to an absolute brightness \citep{philipps1993}. For redshifts greater than $z \gtrsim 0.1$, however, the approximation introduced in Eq.~\eqref{eq:hubble-approx} breaks down and, instead, we must use the luminosity distance:
\begin{equation}
    d_\rmn{L}(z) = (1+z)\frac{c}{H_0} \int_0^z \frac{dz'}{E(z')},
\end{equation}
where we have assumed the Universe is spatially flat. By plotting the observed distances against redshift, we find that the expansion of the Universe is actually accelerating, i.e. $\ddot{a} > 0$ \citep{riess1998}. In combination with the other parameters, this implies a non-zero $\Omega_\Lambda$ such that this component dominates Eq.~\eqref{eq:friedmann-with-density-parameters} today\footnote{More generally, it is evidence for a dark energy component with $w_\rmn{DE} < -1/3$ (see Eq.~\ref{eq:equation-of-state}).}.

The matter density parameter can be constrained by ``weighing'' galaxy clusters, which are the largest gravitationally-bound structures in our Universe\footnote{Indeed, as a result of the aforementioned expansion, they are the largest gravitationally-bound structures that will ever form.}. We can measure their mass using the weak gravitational lensing effect \citep{umestu2020}, which exploits the bending of light by massive objects, as predicted by general relativity. However, if we compare these results with measurements of the gas content, as done, for example, by using X-rays \citep{sarazin1986} or the Sunyaev–Zel'dovich effect \citep{grego2001}, we find that most of the mass in clusters is unseen. This is further supported by evidence from galactic rotation curves, which show that stars at the edges of galaxies are moving much too quickly, given the observed luminous mass \citep[see, e.g.,][]{freese2017}. These results, amongst others, are strong evidence for the existence of \textit{dark matter}. In order for the dark matter to remain ``dark'' it must not electromagnetically couple to standard baryonic matter, meaning that it does not emit photons. This has the important consequence that it cannot cool through radiative emission (see Sec.~\ref{sec:baryonic-cooling}). Moreover, dark matter appears to be collisionless. This means that, apart from gravitational forces, it interacts at most weakly with other material. The Bullet Cluster forms perhaps the most classic piece of evidence for this, with gravitational lensing studies showing that the majority of mass post-merger is in a bi-modal distribution, whilst the gas, being collisional, is predominantly in between \citep{clowe2004, markevitch2004}. 

Further methods for determining the cosmological parameters come from the theory of Big Bang Nucleosynthesis (BBN) \citep{cyburt2016} and comparing simulations with observations of large-scale structure \citep{davis1985, springel2006}. The most powerful constraints of all, however, derive from quantifying the anisotropies and power spectrum of the CMB.

\subsection{Cosmic microwave background}
\label{sec:cmb}

The CMB is space-filling radiation with a current temperature of 2.725 K \citep{fixsen2009}. This radiation dominates the photon density in the universe at the current time\footnote{Note, that locally the photon density may be dominated by stars. This is particularly the case during starbursts \citep[see, e.g., calculations in][]{owen2019}.}. Moreover, as this is tied to the neutrino density through early Universe physics \citep{kolb1990, mangano2002}, measurements of the CMB photon density allow for direct measurements of $\rho_\rmn{r, 0}$. In doing so, we find that $\Omega_\rmn{r, 0} h^2 = 4.2\times10^{-5}$ \citep{planck2014}. To understand how we can determine the remaining cosmological parameters from the CMB, we first briefly recap its origin.

It is believed that, approximately 13.8 Gyr ago, the Universe was a hot, dense plasma, in which radiation dominated and photons acted as a mediator of the temperature. This mediation was very effective, hence the near-perfect blackbody spectrum we observe today. During this radiation-dominated epoch, the mean free path of a photon was very short, with the number of free electrons effectively making the plasma opaque due to Thomson scattering. At this stage, when protons\footnote{Before $10^{-6}$ seconds, the Universe was hot enough that even protons hadn't combined yet, and quarks were still unbound \citep{kolb1990}} and electrons combined to form neutral hydrogen nuclei, photons were able to reionise the nuclei almost immediately. This can be written as:
\begin{align}
    \ce{p^+} + \ce{e^-} &\rightleftharpoons \ce{H} + \gamma.
    \label{eq:recombination}
\end{align}
As the Universe expanded, however, it cooled, thereby shifting the balance in Eq.~\eqref{eq:recombination} towards the right-hand side. When the rate of Thomson scattering approximately equalled the rate of expansion of the Universe at $z\approx 1100$ the Universe entered the epoch of \textit{recombination}, in which the above reaction froze out, via the two-photon transition. At this point, the hydrogen formed could no longer be ionised by the less energetic photons during this transition. Consequently, there were not sufficiently abundant ionised particles to scatter photons, and the Universe became effectively transparent to radiation. This left the CMB, which remains as a redshifted-imprint of the \textit{surface of last scattering}.

It is believed that the initial energy fluctuations in the Universe were purely quantum, resulting directly from Heisenberg's uncertainty principle:
\begin{equation}
    \Delta E \Delta t \geq \frac{\hbar}{2},
\end{equation}
where $E$ is energy, and $\hbar$ is the reduced Planck constant. However, at approximately $t=10^{-36}$ seconds into the Universe's history, it may have undergone a period of inflation\footnote{This is not the only model; see, e.g. the ekpyrotic model introduced in \citet{khoury2001}.}, in which space increased by 60 e-foldings\footnote{This mechanism also provides solutions to the so-called \textit{horizon}, \textit{flatness}, and \textit{magnetic monopole} problems.} The effect of this phase is seen in the CMB temperature fluctuations themselves, which are now coherent on angular scales larger than the particle horizon at recombination, implying an earlier period of causal contact. Furthermore, the fluctuations have Gaussian variance\footnote{There are occasional claims of non-Gaussian elements, which would be able to constrain models further \citep[see, e.g.][]{benoit2012}.}, which is consistent with a quantum origin \citep{white1994}.

After inflation, but before recombination, the fluctuations evolved predominantly due to two factors: i) acoustic oscillations, which result from the interplay between gravitational attraction and pressure in the photon gas, and ii) damping effects, including diffusion or Silk damping, in which photons diffused from hot, overdense regions to cold, underdense ones. The resultant anisotropies in the CMB can be measured with very high precision \citep[see, e.g.][]{wmap2011, planck2018}. Typically this is done using spherical harmonic or ``multipole'' decomposition, with temperature fluctuations shown as a function of their angular scale. An example of such a power spectrum is shown in Fig.~\ref{fig:CMB-power-spectra}. The angular peaks in the power spectrum are sensitive to the cosmological parameters in the following ways:
\begin{enumerate}[i)]
    \item \textbf{Matter:} The aforementioned oscillations result in a harmonic series, the scale of which is set by the sound horizon in the photon-baryon gas. Increased amounts of matter damp the size of the oscillations and hence reduces the height of the peaks. 
    \item \textbf{Baryonic matter:} Enhances every other peak, resulting in the second peak being suppressed in relation to the first and third. This effect is caused by the inertia provided by the baryons during oscillations\footnote{Additional baryonic effects include the movement of peaks to slightly higher multipoles due to the decrease in the oscillation frequency, and the damping of sound waves at higher multipole moments.}
    \item \textbf{Curvature:} Affects the overall spacing of the peaks due to geometric reasons, but not their height. The angular scale of the sound horizon and its harmonics is reduced or increased for lower or higher curvature, respectively.
    \item \textbf{Cosmological constant:} The effect of the cosmological constant is substantially more subtle \citep[see][for a review]{hu2008}. Its impact can be felt at lower multipoles, however, from CMB measurements it is best derived from the other three parameters following Eq.~\eqref{eq:rho_tot}.
\end{enumerate}

\begin{figure}
    \centering
    \includegraphics[width=0.8\linewidth]{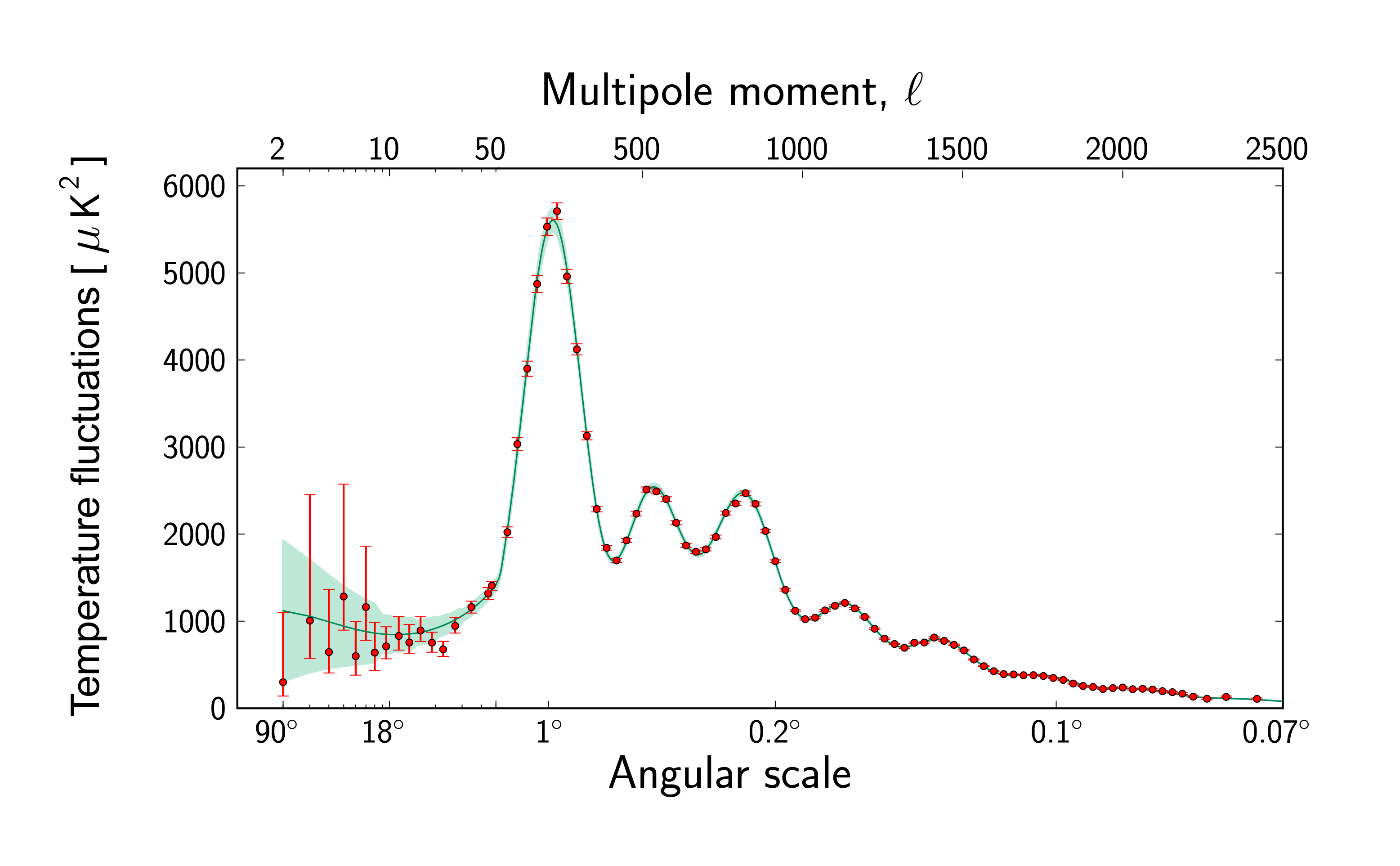}
    \caption[Angular power spectrum of the Cosmic Microwave Background]{Power spectrum showing temperature fluctuations in the CMB as a function of angular scale, as detected by the Planck spacecraft. \textit{Image credit:} \href{https://www.esa.int/ESA_Multimedia/Images/2013/03/Planck_Power_Spectrum}{ESA}.}
    \label{fig:CMB-power-spectra}
\end{figure}

In the papers presented in this thesis, we have used WMAP-9 \citep{hinshaw2013} and \citet{planck2020} measurements, respectively\footnote{Planck measurements were not available when the initial conditions for Chapters~\ref{chapter:paper-one} and~\ref{chapter:paper-two} were created.}. These agree with each other within broad margins. The cosmological density parameters for matter, baryons and a cosmological constant as calculated by the \citet{planck2020}, however, are $\Omega_\text{m} = 0.2726$, $\Omega_\text{b} = 0.0456$, $\Omega_\Lambda = 0.7274$, respectively, with $\Omega_\rmn{k,0} = 0.001 \pm 0.002$, which is consistent with our Universe having a completely flat spatial geometry. This paradigm is known as $\Lambda$CDM (i.e. cold dark matter with a cosmological constant).

\label{sec:structure}

\subsection{Growth of overdensities}
\label{sec:overdensities}

Assuming the temperature fluctuations in the CMB are adiabatic, $\frac{\partial T}{T} \sim \frac{\partial \rho}{\rho}$. Hence the temperature fluctuations directly translate to density fluctuations. The initial growth of these density fluctuations can be determined through linear theory. For this, we must first define the co-moving coordinate and velocity, where these are, respectively:
\begin{equation}
    \bs{r} \coloneq a  \bs{x} 
\end{equation}
and
\begin{equation}
    \bs{u} \coloneq \bs{\bupsilon} - H \bs{x},
\end{equation}
where $\bs{x}$ and $\bs{\bupsilon}$ are the physical coordinates and velocity, respectively. These new variables factor out the expansion of the Universe and the corresponding Hubble flow.
We now define the over-density as:
\begin{equation}
    \delta(\bs{r},t) \coloneq \frac{\rho(\bs{r}) - \rho_0}{\rho_0},
\end{equation}
where $\rho_0$ is the background density.

For an ideal fluid, in an inertial frame, density and velocity are governed by the continuity, Euler, and Poisson equations:
\begin{equation}
    \frac{\partial \rho}{\partial t} + \bs{\nabla}\bcdot\left(\rho \bs{\bupsilon} \right) = 0
    \label{eq:mass-continuity}
\end{equation}
\begin{equation}
    \frac{\partial \bs{\bupsilon}}{\partial t} + \left(\bs{\bupsilon}\bcdot\bs{\nabla}\right)\bs{\bupsilon} + \frac{\bs{\nabla} P}{\rho} = -\bs{\nabla}{\Phi}
    \label{eq:momentum-continuity}
\end{equation}
\begin{equation}
    \bs{\nabla}^2{\Phi} = 4 \pi G \rho
    \label{eq:poisson}
\end{equation}
where $\Phi$ is the gravitational potential. Equations~\eqref{eq:mass-continuity} and~\eqref{eq:momentum-continuity} represent mass and momentum conservation, respectively, whilst Eq.~\eqref{eq:poisson} is Gauss's law for gravity in differential form.

Using the transformations introduced at the start of this section, we may convert this to co-moving coordinates thusly:
\begin{equation}
    \frac{\partial \delta}{\partial t} + \frac{1}{a}\bs{\nabla}\bcdot\left[\left(1+\delta\right) \bs{u} \right] = 0
\end{equation}
\begin{equation}
    \frac{\partial \bs{u}}{\partial t} + \frac{1}{a}\left(\bs{u}\bcdot\bs{\nabla}\right)\bs{u} + \frac{\dot{a}}{a}\bs{u} = -\frac{1}{a}\bs{\nabla}{\phi} - \frac{1}{a(1+\delta)} c_\rmn{s}^2 \bs{\nabla}\delta
\end{equation}
\begin{equation}
    \bs{\nabla}^2{\phi} = 4 \pi G \rho_0 a^2 \delta,
\end{equation}
where $c_\rmn{s} = \frac{\partial P}{\partial \rho}$ is the sound speed, and $\phi = \Phi -  \langle{\Phi}\rangle = \Phi -  \frac{1}{2}a\ddot{a}|\bs{x}|^2 $ is the peculiar potential \citep[see, e.g.,][]{gnedin2011}. 

We now linearise the equations, only keeping terms of first order in $\delta$ or $\bs{u}$. This reduces our set of equations to:
\begin{equation}
    \frac{\partial \delta}{\partial t} + \frac{1}{a}\bs{\nabla}\bcdot\bs{u}= 0
\end{equation}
and
\begin{equation}
    \frac{\partial \bs{u}}{\partial t} + \frac{\dot{a}}{a}\bs{u} + \frac{1}{a}\bs{\nabla}{\phi} + \frac{1}{a} c_\rmn{s}^2 \bs{\nabla}{\delta} = 0.
\end{equation}

Through some careful algebra \citep[see, e.g.][]{peebles1993}, this reduces to a single equation that governs the growth of the overdensity:
\begin{equation}
     \frac{\partial^2 \delta}{\partial t^2} + 2 \frac{\dot{a}}{a}\frac{\partial \delta}{\partial t} - \left[4 \pi G \rho_0 + \frac{c_\rmn{s}^2}{a^2}\bs{\nabla}^2 \right]\delta = 0,
     \label{eq:overdensity-evolution}
\end{equation}
The equation has two solutions; a growing mode, $D_+$, and a decaying mode, $D_-$, where $g \equiv D_+(a)/a$ is known as the linear growth factor. To excellent approximation, for a $\Lambda$CDM universe at the current time, it can be shown that \citep{carroll1992}:
\begin{equation}
    D_+(a) = \frac{5a}{2}\Omega_\rmn{m}\left[\Omega_\rmn{m}^{4/7} - \Omega_\rmn{\Lambda} + \left( 1+ \frac{1}{2} \Omega_\rmn{m} \right) \left( 1+ \frac{1}{70} \Omega_\rmn{\Lambda} \right)  \right]^{-1}
\end{equation}

Equation~\eqref{eq:overdensity-evolution} is in the form of a damped harmonic oscillator, where the damping term is $2 \frac{\dot{a}}{a}\frac{\partial \delta}{\partial t}$. This term is called \textit{Hubble drag}, and is the slowing down of structure formation due to the expansion of the Universe.
If $\delta \ll 1$, we may further decompose Eq.~\eqref{eq:overdensity-evolution} into a set of plane waves and thereby treat it in Fourier space. For a growing solution, this produces the condition:
\begin{equation}
    4 \pi G \rho_0 >  \frac{c_\rmn{s}^2 k^2}{a^2},
\end{equation}
which states that gravitational forces must overcome pressure, as communicated by sound waves, in order for a structure to collapse. The threshold wave number, $k_\rmn{J} = \sqrt{4 \pi G \rho_0} a / c_\rmn{s}$ is known as the Jeans wave number, with the Jeans length being $\lambda_\rmn{J} \equiv 2 \pi / k_\rmn{J}$, and associated Jeans mass, $M_\rmn{J} = \frac{4 \pi}{3} \rho_0 \lambda_\rmn{J}^3$. Any density perturbation with size $\lambda > \lambda_\rmn{J}$ will grow, whilst perturbations smaller than this will oscillate.

It is instructive to see how these overdensities grow in the matter- and radiation-dominated era. For the former, we may use the Einstein-de-Sitter model as an approximation, in which $\Omega_\rmn{m} = 1$ and $H = H_0$. This results in expansion as
\begin{equation}
    a(t) = \left(\frac{3}{2}H_0 t \right)^{2/3},
\end{equation}
which implies that
\begin{equation}
    H = \frac{2}{3t}\;\;\;\; \rmn{and} \;\;\;\; \Omega = \frac{8 \pi G \rho_0}{3H^2}
\end{equation}
After making appropriate substitutions, we find that the decaying and growing solutions to Eq.~\eqref{eq:overdensity-evolution} are $\delta(t) \propto t^{-1}$ and $\delta(t) \propto t^{2/3} \propto a(t)$, respectively. Hence density fluctuations grow linearly with respect to the scale factor in the matter-dominated era. Were it not for dark matter, this would be problematic, as the perturbations would grow from the start of recombination to now as: $\delta(a_0) \sim \frac{a_0}{a_\rmn{CMB}}\delta(a_\rmn{CMB}) \approx 1100 \times 10^{-5} \approx 0.1$. Galaxies and galaxy clusters, meanwhile, require $\delta(a_0) \approx 10^5$, and so such growth is clearly insufficient.

Dark matter only couples to the radiation field through gravity, however. We may therefore use the same linearisation technique as above, but make the substitutions $\rho \rightarrow \rho + P/c^2$ in Eq.~\eqref{eq:mass-continuity} and $\rho \rightarrow \rho + 3P/c^2$ in Eq.~\eqref{eq:poisson}, as $P=\rho c^2 /3$ in the radiation-dominated era. This results in:
\begin{equation}
     \frac{3}{4}\frac{\partial^2 \delta}{\partial t^2} + \frac{3}{2} \frac{\dot{a}}{a}\frac{\partial \delta}{\partial t} - \left[8 \pi G \rho_0 + \frac{c_\rmn{s}^2}{4 a^2}\bs{\nabla}^2 \right]\delta = 0,
\end{equation}
which, if we assume scales much larger than the Jeans length again, simplifies to 
\begin{equation}
     \frac{\partial^2 \delta}{\partial t^2} + 2 \frac{\dot{a}}{a}\frac{\partial \delta}{\partial t} - \left(\frac{32 \pi G \rho_0}{3} \right)\delta = 0
\end{equation}
In the radiation-dominated era,
\begin{equation}
    H(t) = \frac{1}{2t}
\end{equation}
and hence our solutions are $\delta(t) \propto t^{-1}$ and $\delta \propto t \propto a^2$. Perturbations with $\lambda > \lambda_\rmn{J}$ are thus able to grow significantly faster in the radiation-dominated era than in the matter-dominated one. 
This is only true as long as perturbations are greater than the particle horizon, however; as baryonic perturbations enter the horizon, they interact with the photon gas, leading to the oscillations described earlier. Moreover, the growth of dark matter perturbations that enter the horizon during this era is also suppressed due to the \citet{meszaros1974} effect. The result is shown in Fig.~\ref{fig:linear_perturbations} (see caption). The overall growth of dark matter structures during the radiation-dominated era and after matter-radiation equality is, however, critical for the development of the baryonic perturbations: at the end of the recombination process when the photons do not any more prevent the baryons from collapsing, the baryonic matter falls into the gravitational potential provided by the dark matter halos. This boosts the baryonic overdensity, thereby allowing it to form structures such as galaxies in the later Universe.

\begin{figure}
    \centering
    \includegraphics[width=0.7\linewidth]{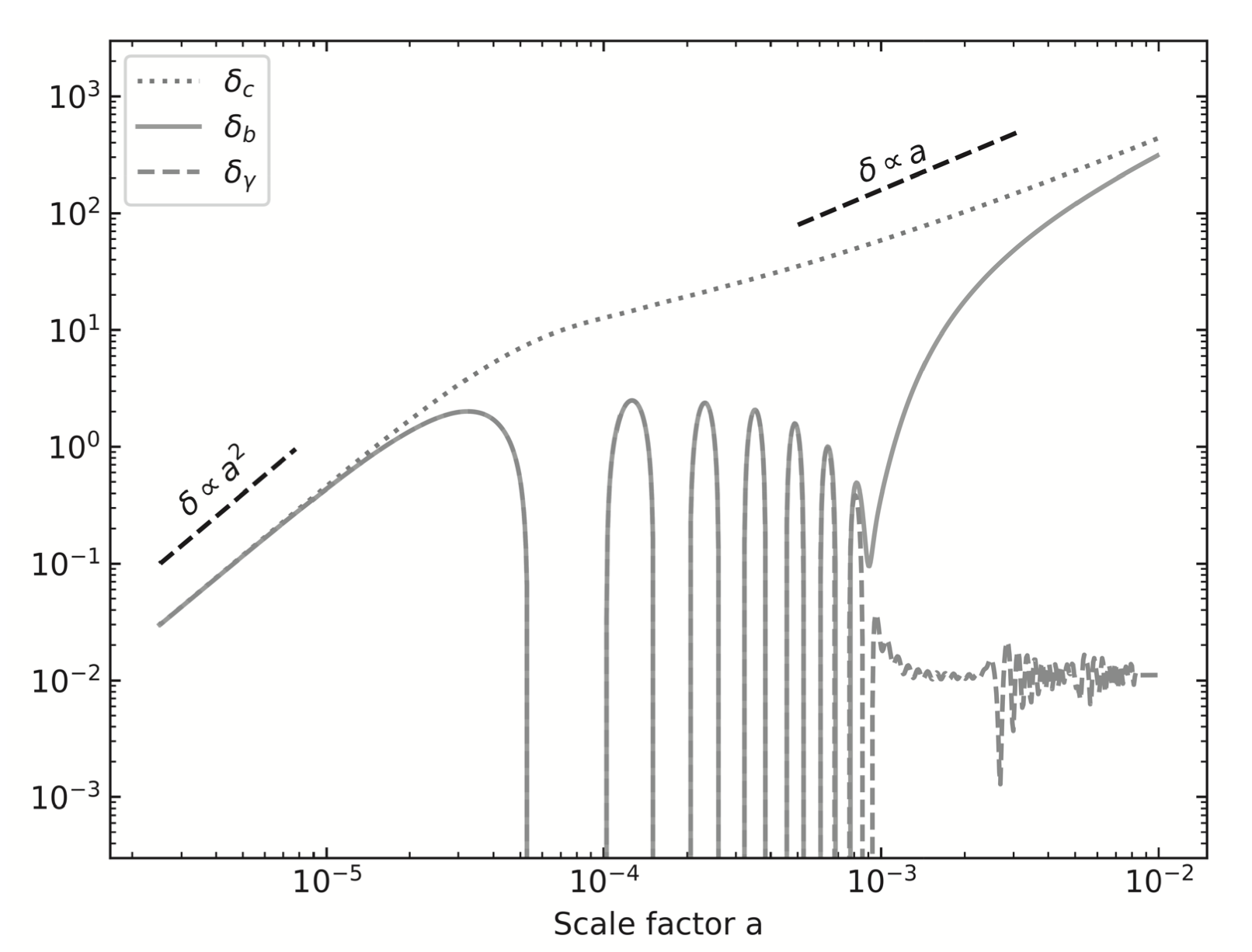}
    \caption[]{Evolution of a density perturbation following linear perturbation theory. Dotted, solid, and dashed lines show dark matter, baryonic, and radiation perturbations, respectively. All perturbations grow initially as $\propto a^2$, before being suppressed as they enter the horizon during the radiation-dominated era. During this period baryonic and radiative perturbations undergo damped oscillation. At the epoch of recombination, the photon-gas decouples, and baryonic perturbations fall into the gravitational well set by the dark matter. They then grow as $\propto a$. \textit{Image credit:} \citet{ferreras2019}.}
    \label{fig:linear_perturbations}
\end{figure}

We can describe the structures that form using the isotropic \textit{matter power spectrum}, $P({k})$, defined such that
\begin{equation}
    (2\pi)^3 P(k) \delta_\mathrm{d}(\bm{k}-\bm{k}') \equiv \langle \delta_{\bm{k}}^{} \delta_{\bm{k}'}^* \rangle,
    \label{eq:power}
\end{equation}
where $\delta_\mathrm{d}$ is the Dirac delta function, which ensures that modes with different wave vector $\bm{k}$ are uncorrelated in space.
This can be expressed in the dimensionless form:
\begin{equation}
    \Delta^2(k) \equiv \frac{k^3 P(k)}{2 \pi^2}, 
\end{equation}
which measures the density variance per logarithmic interval. Inflation generically predicts a primordial power spectrum with form $P_\rmn{i}(k) = A k^{n}$, where $A$ is a normalisation constant \citep{mo2010}. Hence for the initial power spectrum:
\begin{equation}
    \Delta_\rmn{i}^2(k) \propto k^{3+n}
    \label{eq:dimensionless-ps}
\end{equation}
For the gravitational potential, the corresponding power spectrum is:
\begin{equation}
    \Delta^2_\Phi(k) \equiv \frac{k^3 P_\Phi(k)}{2 \pi^2},
\end{equation}
By the Poisson equation, $P_\Phi(k) \equiv \langle |\delta_{\Phi,k}|^2 \rangle \propto k^{-4}$. Combining this with the result given in Eq.~\eqref{eq:dimensionless-ps}, we therefore see that:
\begin{equation}
    \Delta^2_\Phi(k) \propto k^{-4} \Delta^2_\rmn{i} \propto k^{n-1}
\end{equation}
A scale invariant spectrum, as most consistent with the cosmological principle, is consequently given by $n=1$. 
This is known as the Harrison-Zel'dovich-Peebles spectrum \citep{harrison1970, peebles1970, Zeldovich1972}. 
This generally very accurately\footnote{Note, inflationary models modify this to $n \approx 1 - 2/N$, where $N$ is the number of e-foldings \citep[see, e.g., derivations in][]{granda2019}. For 60 e-foldings, this results in a spectral slope of $n\approx0.96$. \citet{planck2020} meanwhile find $n = 0.9626 \pm 0.0057$.} describes observations by WMAP and Planck at small $k$ \citep{wmap2011, planck2020}.

As discussed above, in the radiation-dominated era, perturbations larger than the horizon grow with $\delta \propto a^2$, and hence by Eq.~\eqref{eq:power}, modes $k_1$ and $k_2$ (where $k_1 > k_2$) enter the horizon at $a_1$ and $a_2$ before matter-radiation equality, such that:
\begin{equation}
    \frac{P(k_2, a=a_2)}{P(k_1, a=a_1)} = \frac{k_2}{k_1}\left( \frac{a_2}{a_1} \right)^4 = \left( \frac{k_1}{k_2} \right)^3
\end{equation}
Perturbations entering the horizon during this era consequently fulfil the condition that $P(k) k^3 = \rmn{const.}$\footnote{This result is generally consistent with redshift and Lyman-$\alpha$ surveys, although structure formation introduces some non-linear effects at higher $k$ \citep{lee2013, oka2014}.}, and the power spectrum at the present time is hence modified to:
\begin{equation}
    P(k) \propto    
    \begin{cases}
      k & \text{for $k < k_\rmn{eq}$}\\
      k^{-3} & \text{for $k\gg k_\rmn{eq}$}
    \end{cases}
    \label{eq:matter-power-spectrum}
\end{equation}
where $k_\rmn{eq}$ is the wave number of a perturbation the size of the horizon at matter-radiation equality\footnote{More complicated models use a transfer function $T({k})$ where $P_0({k}) = A k^n T^2({k})$ is the linearly-extrapolated power spectrum at $z=0$. This deals more generically with changes to the primordial density perturbations, including the general transition between matter and radiation-dominated eras, and the dark matter free-streaming effect.}.

We can define the non-linear mass $M_*$ as the point at which the variance becomes unity for a sphere of radius $R_* = 2 \pi / k_*$, so that:
\begin{equation}
    \sigma_*^2 = \int_0^{k_*} \frac{\mathrm{d}^3 k}{(2 \pi)^3} P(k) = 1
\end{equation}
Using Eq.~\eqref{eq:matter-power-spectrum} and the fact that $\sigma^2 \propto k^3 P(k)$, it can then be shown that:
\begin{equation}
    \sigma^2 \propto    
    \begin{cases}
      \left(\frac{M}{M_*}\right)^{-4/3} & \text{for $n = 1$}\\
      1 & \text{for $n=-3$}
    \end{cases}
\end{equation}
and hence smaller objects collapse first and therefore structure in the Universe forms \textit{hierarchically}. In particular, this means that smaller objects \textit{merge} to form larger objects.

The evolution of density perturbations can be followed into the mildly non-linear regime using the \citet{Zeldovich1970} approximation. Beyond this, however, a numerical approach is required. Many cosmological simulations have now been performed, and generally these support the $\Lambda$CDM theory introduced above. In particular, dark matter simulations have shown that large-scale structure evolves to form a web-like structure consisting of voids, walls, filaments, and halos \citep{davis1985, navarro1996, springel2005c, klypin2011, ishiyama2021}. The structure produced on this scale is also in remarkable agreement with observations from spectroscopic redshift surveys \citep{york2000, colless2001, springel2006}. In order to form smaller-scale structure, however, we must introduce baryonic physics, where particles can radiate away their kinetic energy. 

\section{Physics of galaxy and galaxy cluster formation}
\label{sec:baryonic-structure}

\subsection{Cooling and heating}
\label{sec:baryonic-cooling}

We first introduce the cooling function, $\Lambda(T, Z)$, which defines the energy loss per unit time for a gas with temperature $T$ and metallicity $Z$, where this, in turn, is the mass fraction of elements heavier than helium. Here, temperature is essentially a proxy for the ionisation fraction, whilst the metallicity is a proxy for the chemical composition of the gas. The contributing processes to the cooling function are predominantly two-body interactions, and consequently it is usually normalised by $n^2$ such that $\mathcal{C} \equiv \Lambda(T, Z) n^{-2}$ is the volumetric rate of cooling, where $n$ is the gas density. 

\begin{figure}
    \centering
    \includegraphics[width=0.55\linewidth]{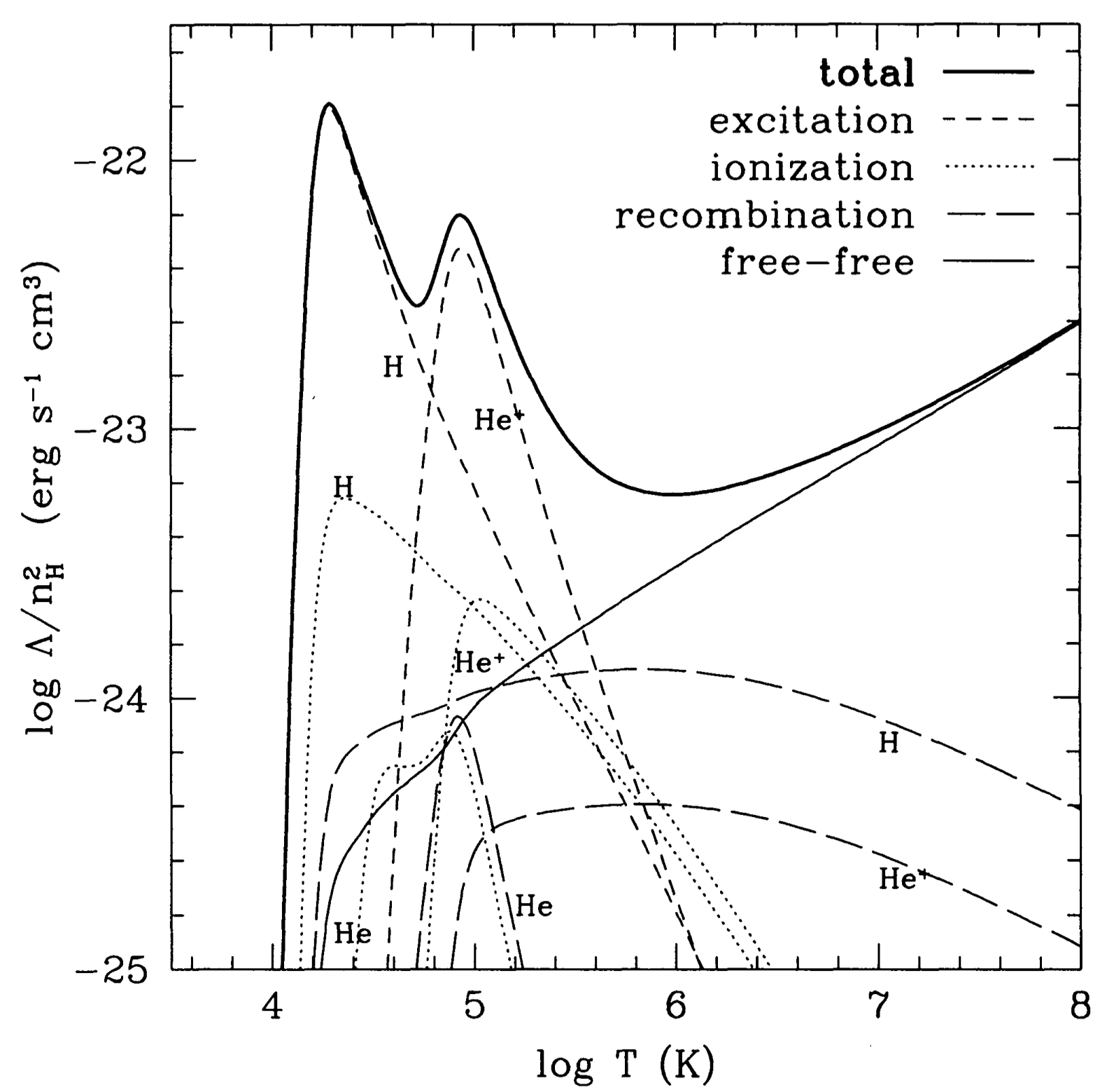}
    \caption[]{Cooling rates for a gas with primordial abundance (76\% hydrogen, 24 \% helium by mass), assuming CIE. \textit{Image credit:} \citet{thoul1995}}
    \label{fig:cooling-curve}
\end{figure}

At lower redshifts, there are four main ways that an ideal gas can cool\footnote{Above $z \gtrsim 6$, Compton scattering off of the CMB results in an additional significant contribution to the cooling rate \citep[see, e.g.][]{benson2010}.}:
\begin{enumerate}
    \item \textbf{Collisional excitation:} A free electron impacts a bound electron, thereby exciting it. As the electron decays, it emits a photon.
    \item \textbf{Collisional ionisation:} An electron is released after a sufficiently strong collision. 
    \item \textbf{Recombination:} A free electron recombines with an ion, with the excess energy being radiated away.
    \item \textbf{Free-free emission:} A free electron is accelerated by an ion; the accelerated electron emits a photon and therefore cools. This process is otherwise known as ``bremsstrahlung'' and increases such that $\Lambda \propto T^{1/2}$
\end{enumerate}
If there is a balance between recombination and ionisation, the gas is considered to be in \textit{collisional ionisation equilibrium} (CIE). Using this assumption, we can calculate the expected cooling rates from the above processes. This produces a cooling curve for a given $T$ and $Z$, as shown in Fig.~\ref{fig:cooling-curve} for a gas with primordial composition. In simulations, such curves are generally calculated ahead of time using, e.g., 
\textsc{Cloudy} \citep{ferland1998}.

The formation of baryonic structure in the Universe can be considered as a balance between the cooling timescale:
\begin{equation}
    \tau_\rmn{cool} \equiv \frac{E}{\dot{E}} = \frac{3}{2}\frac{n k_\rmn{B} T}{n^2 \Lambda},
    \label{eq:cooling-time}
\end{equation}
and the dynamical time:
\begin{equation}
    \tau_\rmn{dyn} \approx \frac{1}{\sqrt{G \rho}}
\end{equation}
Comparing these, we are left with three different regimes\footnote{See original theory given in \citet{rees1977}.}:
\begin{enumerate}[i)]
\item $H_0^{-1} < \tau_\rmn{cool}$: Cooling is ineffective, and the structure can be treated as being in hydrodynamic equilibrium.
\item $H_0^{-1} > \tau_\rmn{cool} > \tau_\rmn{dyn}$: Cooling causes the gas to contract slowly, but the system has sufficient time to respond. This leads to a quasi-hydrodynamic equilibrium state and produces a hot, pressure-supported halo, such as the circumgalactic medium (CGM) and the intracluster medium (ICM).
\item $H_0^{-1} > \tau_\rmn{dyn} > \tau_\rmn{cool}$: Cooling is efficient, and the gas cloud contracts. As the cooling time shortens with density (see Eq.~\ref{eq:cooling-time}), this leads to a feedback loop. Without additional heating, cooling loses are catastrophic, and the gas collapses on the free-fall time.
\end{enumerate}

Heating generally proceeds through the following mechanisms\footnote{These mechanisms are augmented by cosmic ray heating at low energies \citep{leite2017}, as well as kinetic and MHD effects \citep{birk1998}}:

\begin{enumerate}[i)]
\item \textbf{Photoheating:} High-energy photons from stars and AGN within the galaxy and outside it\footnote{This is the so-called \textit{UV background}, which is believed to reionise hydrogen by $z\approx 6-10$ \citep{calverley2011}.} lead to electrons being ejected from dust grains, atoms, and molecules. This heats the surrounding gas and is especially important in HI regions \citep{prialnik2009, morisset2016}.

\item \textbf{Turbulent heating:} Turbulence is injected at the scale of the largest eddy size. In incompressible flows, kinetic energy then cascades down through smaller eddies before eventually reaching the viscous scale, at which point the energy is dissipated as heat. Assuming no energy accumulates at any one scale, the eddy sizes will typically form a power law \citep{kolmogorov1941}.

\item \textbf{Shock heating:} Shocks convert kinetic energy into thermal energy (see Sec.~\ref{sec:shocks}). Of particular interest is heating by supernovae shocks and heating by the accretion shock, where cold, accreting gas encounters the hot CGM or ICM. 
\end{enumerate}

Note, that as structures form, gravitational potential energy, $U$ is converted to kinetic energy, $T$, thereby naturally producing turbulent heating. Given enough time, for a bound system, this becomes partitioned such that $\langle U \rangle + 2\langle T \rangle = 0$ according to the virial theorem. This results in larger structures becoming inherently hotter, which in turn ultimately leads to a shock forming close to the virial radius\footnote{This radius is, in turn, often used to define the boundary of a collapsed object.}.

Gas in the Universe generally has non-zero angular momentum, which is believed to be generated by tidal torques \citep{efstathiou1983}. This is important, as whilst kinetic energy is removed through cooling, angular momentum is still conserved. Given sufficient cooling, this leads to structures becoming rotationally-supported. Indeed, this, in combination with the above processes, already roughly explains the fiducial model of disc galaxies; that they have a large dispersion-supported dark matter halo, a smaller, pressurised CGM, and finally a dense disc, in which most of the star formation takes place. That the disc is not razor thin is due to the heating mechanisms discussed above \citep{wen2004, steinmetz2012}.

\subsection{Star formation}
\label{sec:star-formation}
Given sufficient densities and cooling rates, gas collapses to form \textit{giant molecular clouds} (GMCs). In these, molecular hydrogen is able to form through the increased rate of collisions between hydrogen atoms \citep{combes1999}. This allows the GMCs to cool even further, eventually resulting in run-away gravitational collapse, followed by nuclear fusion. This process lead to the formation of the first stars, so-called \textit{Population III} stars \citep{mckee2007}. Allowing for primordial gas cooling only, it is expected that the Jeans mass for these stars was between $10^2$ and $10^3$ $\rmn{M}_\odot$ \citep{abel2002}. The evolution of a star in isolation can be almost completely determined by the mass that it starts on the main sequence with \citep{prialnik2009}, and hence it is likely that these stars would have lasted only $\lesssim 100,000$ yr at most. Consequently, no Pop.\ III star has ever been observed\footnote{The observation of Pop.\ III stars forms a major part of the James Webb Space Telescope's (JWST) science goals, and candidate stars have already been found \citep{maiolino2024}.}. These stars would, however, have left a lasting impact on their environment, re-ionising the surrounding regions and increasing the metallicity of the surrounding gas. This would have strongly increased the cooling rate, thereby reducing the Jeans mass of the next population of stars\footnote{This second generation, called Pop.\ III.2 stars, are expected to have been able to form down to $40$ $\rmn{M}_\odot$ \citep{yoshida2007}.}.

Given that the initial mass of the star is so influential, there has been much work on categorising  the so-called \textit{initial mass function} (IMF) -- the probability that a star will form with a particular mass. As it turns out, this follows a (broken) power-law \citep{salpeter1955, kroupa2001, chabrier2003}. This appears to be a universal distribution \citep{chabrier2014}, although there have been some arguments for dependencies on redshift or star formation rate (SFR) \citep{vanDokkum2010, gunawardhana2011}. The SFR itself appears to be intrinsically linked to the gas density. Observationally, this is exhibited through the Kennicutt-Schmidt relation \citep{schmidt1959, kennicutt1998}, which relates the star formation and gas surface density rates: $\Sigma_\rmn{SFR} \propto \left( \Sigma_\rmn{gas} \right)^n$, where $n$ is usually set to 1.5 \citep{kennicutt2021}. Here too, however, there have been suggestions that this relation may evolve with redshift or metallicity dependence \citep{dib2011, scoville2016}. It has also been claimed that the exponent varies at the high surface density end \citep{orr2018} or between star-bursting and quenched galaxies \citep{kennicutt2021}. The normalisation of the Kennicutt-Schmidt relation is surprisingly low, with gas in MW-like galaxies forming stars with an efficiency per dynamical time of roughly $\varepsilon_\rmn{ff} = (\rmn{SFR}/M_\rmn{gas}) \tau_\rmn{dyn} \approx 1-2\%$ \citep{krumholz2007, krumholz2014}. A number of reasons have been proposed to explain this including turbulence, jets, and magnetic pressure \citep{federrath2015}.

\subsection{Feedback}
\label{sec:feedback}

\begin{figure}
    \centering
    \includegraphics[width=0.65\linewidth]{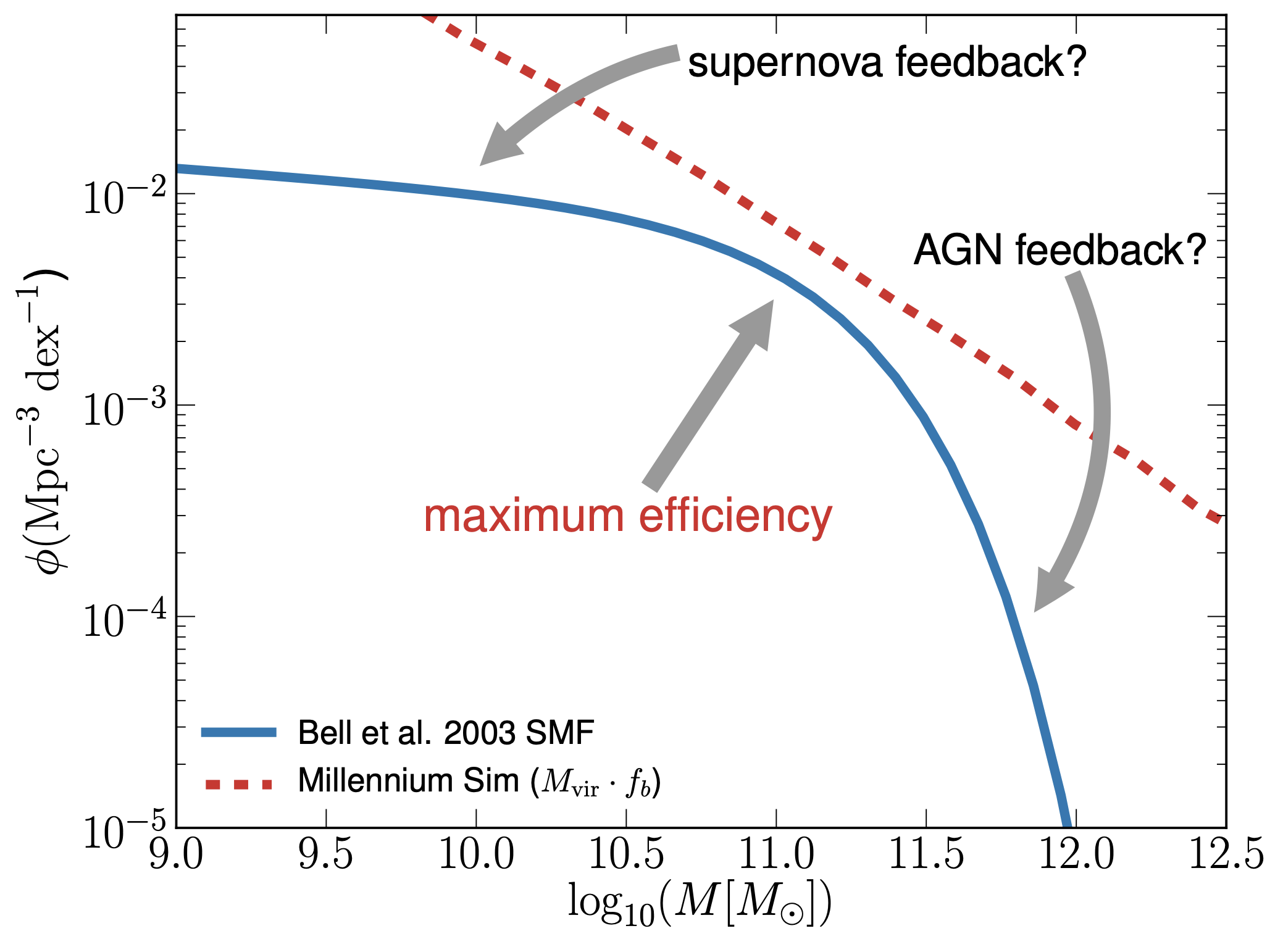}
    \caption[]{Observed galactic stellar mass function \citep {bell2003} in solid blue, and halo mass function from the Millennium simulation \citep{springel2005c} multiplied by a universal baryon fraction of $f_\rmn{b}\approx17\%$ \citep{spergel2003} in dashed red -- this marks the maximum possible stellar content as a function of halo mass. Star formation is heavily suppressed at the low and high halo-mass ends due to feedback. \textit{Image credit:} \citet{mutch2013}} 
    \label{fig:feedback}
\end{figure}

Star formation efficiency becomes even smaller in galaxies with stellar masses of $M_*\lesssim 10^{10.5} \,\rmn{M}_\odot$ and $M_*\gtrsim 10^{11.5} \,\rmn{M}_\odot$. We show this in Fig.~\ref{fig:feedback}, where we present the observed stellar mass function and the expected value if star formation was 100\% efficient. It can be seen that star formation is strongly suppressed at the low and high halo-mass end. It is generally believed that this suppression is due to feedback processes, which act in one of two ways: \textit{ejective} feedback, where gas is removed from the galaxy, and \textit{preventative} feedback, where gas is heated to the point that stars can no longer form. At the low halo-mass end, the most important feedback source is likely stellar. This can act in a variety of ways including through thermal pressure due to supernovae, radiation pressure and photoionisation, and the pressure gradient of cosmic rays that have been accelerated at supernovae remnant shocks
\citep[see, e.g.,][and references therein]{ruszkowski2023}.
These processes heat the gas, and drive stellar winds. This removes not only cold gas but also metal-rich gas from the disc, thereby reducing star formation \citep{tremonti2004}. This is believed to keep the feedback self-regulated. Stellar feedback becomes less effective at higher masses, however, owing to the increased gravitational potential \citep{efstathiou2000}; here, it is believed that AGN feedback is most effective.

Most, if not all, massive elliptical galaxies are expected to contain a super-massive black hole (SMBH) \citep{kormendy2013}. The energy output from such black holes over their lifetime is expected to be $E_\rmn{AGN} \approx 0.1 M_\rmn{BH}c^2$, where $M_\rmn{BH}$ is the mass of the black hole. This is comparable to the binding energy of the halos they reside in \citep{silk1998}. If a fraction of this energy couples with the ISM, the AGN should be able to strongly affect the evolution of the galaxy. This is supported by observations of winds launched by AGNs \citep[see][and reference therein]{proga2007, fiore2017}, and potential observations of the AGN heating the CGM \citep{heckman2014}. As before, this process is expected to cause a self-regulated feedback loop, as the impact of the AGN is directly tied to its ability to accrete gas. This accretion is usually described by an Eddington-limited Bondi-Hoyle-Lyttleton model \citep{bondi1944, bondi1952}. This model can produce fairly gentle heating at low accretion rates, but powerful quasars at high rates \citep{dimatteo2005}.

Given sufficient time in the CGM, gas can cool and rejoin the ISM. Indeed, simulations find that the majority of star formation at low-redshift takes place in gas that has been ``recycled'' like this \citep[see, e.g.,][]{oppenheimer2010}. The timescale required for recycling is inversely proportional to the halo mass \citep{oppenheimer2008}, increasing its likelihood in higher mass halos. This also shifts the balance of preventative and ejective feedback, with the former being more important in high mass halos, and the latter being more important in low mass halos\footnote{This picture is constrained, however, by the current inability of simulations to resolve the hot phase.}. 

\section{Mergers}
\label{sec:mergers}

A key process by which galaxy and galaxy clusters change is mergers. Mergers are a natural consequence of hierarchical structure formation, and form the underlying topic in this thesis. Their impact on galaxies and galaxy clusters is, however, very different. We therefore split this section accordingly.

\subsection{Galaxies}

The most famous classification sequence for galaxy morphologies is Edwin Hubble's ``tuning fork'', in which galaxies are categorised as elliptical, barred spiral, or non-barred spiral \citep{jeans1928, hubble1936}. In this diagram, ellipticals are referred to as \textit{early-type}, whilst spiral galaxies are referred to as \textit{late-type}. The temporal connotation was derived from the evolutionary sequence proposed by Sir James Jeans \citep{jeans1919}, although Hubble later distanced himself from this implication \citep{hubble1926}. We now know that elliptical galaxies are generally more massive than disc galaxies \citep{naab2017} and, moreover, that lenticular galaxies, which form the node of the ``tuning fork'', are statistically more massive than spiral galaxies \citep{freeman1970}. Additionally, observations show that, whilst the frequency of disc galaxies decreases with lower redshift, the frequency of elliptical galaxies increases \citep{couch1998, zhang1999}. This implies that the evolutionary sequence may actually be reversed; that late-type galaxies evolve into early-type galaxies.

Irregular galaxies form a final classification. These galaxies are often highly asymmetric, and hence are not consistent with any of the categories in the Hubble sequence. Even in the 1950s, several galaxies had been described as being in ``obvious interaction'' \citep{zwicky1959}. However, by the 1960s, the belief that asymmetries were due to ``tidal extensions'' had been mostly rolled back \citep{zwicky1967}, with especially strong doubts expressed in the community that gravity could produce thin features, such as the observed ``tails'' and ``bridges'' \citep[see, e.g.][]{vorontsov1964, arp1971}. This, in turn, was shown to be highly mistaken by \citet{toomre1972} and \cite{toomre1974} who showed that tidal interactions between disc galaxies were actually very likely to form such features. Moreover, they suggested that at close-range, tidal forces were likely to randomise orbits. This is particularly true of galaxies embedded in large halos \citet{toomre1977}, where particles undergo rapid angular momentum transfer due to a process known as dynamical friction \citep{chandrasekhar1942}. Taken to its most extreme, this leads to so-called ``violent relaxation'', in which stellar orbits are completely randomised and the system loses all memory of its previous configuration \citep{lynden-bell1967}. 

With the advent of hierarchical structure formation (see Sec.~\ref{sec:overdensities}), the theory that disc galaxy mergers led to the formation of ellipticals gained a lot of popularity. This was supported by the observation that elliptical galaxies tend to reside in high density regions, whilst disc galaxies are typically found in lower density regions \citep{dressler1980}. Furthermore, with the discovery that galaxies resided in dark matter halos, the impact of dynamical friction became even greater. In early simulations, computational expense restricted the number of particles allowed. This tended to result in merger remnants that rotated too quickly \citep{gerhard1981, negroponte1983}. However, with the introduction of tree codes \citep[see, e.g.][]{barnes1986}, the computational expense for modelling $N$ particles dropped from $\mathcal{O}(N^2)$ to $\mathcal{O}(N\log N)$, which allowed for higher resolution, more extended halos, which produced results in better agreement with observations \citep{barnes1986}.

The original Toomre \& Toomre simulations were run with stellar particles only. However, even at this stage, it was realised that if stars were subject to such strong tidal forces, the gas would be as well. In particular, as gas is collisional, this could result in gas densities strongly increasing in the centre of the galaxy, thereby prompting higher star formation rates. This was supported by simulations that showed that mergers were likely to trigger the formation of stellar bars \citep[see, e.g.,][]{noguchi1987}. These bars could then further extract angular momentum from the gas through tidal torques \citep{bournaud2002}.

The first accurate treatment of gas in galaxy simulations was done through the use of \textit{smoothed particle hydrodynamics}. For example, \citet{barnes1991} were able to show with this technique that a major merger could result in approximately $5\times10^9\,\rmn{M}_\odot$  being transported to the central 200 pc of the galaxy. This was expanded upon by \citet{mihos1996}, who implemented a star formation model based on the Kennicutt-Schmidt law (see Sec.~\ref{sec:star-formation}), and showed that such mergers could indeed lead to intense starbursts, where the SFR is at a rate 10 -- 100 times the level of unperturbed galaxies at the same stellar mass \citep{sanders1996}. Major mergers are now believed to be a significant contributor to the Ultra-Luminous Infra-red Galaxy (ULIRG) population, due to precisely this process \citep{draper2012}.

With the realisation that AGN feedback plays a strong role in galaxies with Milky Way masses and above (see Sec.~\ref{sec:feedback}), it was also recognised that the same merger-driven gas flows could increase black hole accretion, thereby triggering more intense AGN feedback. This was successfully modelled by \citet{springel2005a}, who showed that black hole accretion substantially increased during a major merger, which in turn suppressed the star formation rate. If the AGN grew sufficiently strong, it could remove gas from the halo, thereby quenching the galaxy. At the time, this was considered a likely method for explaining the population of \textit{red and dead} galaxies observed below the so-called \textit{SFR main sequence}, with such a ``blow-out'' scenario being codified by \citet{hopkins2008}. In recent years, however, the situation has been shown to be more nuanced, with \citet{sparre2017} showing that mergers between gas-rich disc galaxies with stellar masses of $10^{10}\,\rmn{M}_\odot$ can regrow their disc post-merger\footnote{This supports the conclusions of the earlier work by \citet{robertson2006}, who performed similar simulations without feedback.}, even when they were able to produce a starburst \citep{sparre2016} and included reasonably strong AGN feedback.

Observations match many of the predictions of simulations, including their asymmetry \citep{patton2016}, initially enhanced star formation rates \citep{li2008}, and metal poor gas content \citep{scudder2012}, resulting from dilution by accreted gas. Observation also generally support the statement that mergers typically trigger AGNs at low redshift \citep{ellison2015}. However, increasingly, observations show that post-merger galaxies still contain significant amounts of molecular gas \citep{french2015}. Indeed, statistically-speaking, this appears to be enhanced by a factor of two or more compared to non-interacting galaxies at the same stellar mass \citep{violino2018, ellison2018}. This is strong evidence that the blow-out scenario, as previously favoured, is not actually realised in our Universe. At the same time, however, there \textit{is} evidence that post-merger galaxies are quenched more frequently than non-interacting galaxies \citep{ellison2022}. The combination of these results may be evidence for preventative feedback (see Sec.~\ref{sec:feedback}).

\subsection{Galaxy clusters}

Mergers are a fundamental part of how galaxy clusters build mass \citep{peebles1993}. Unlike in galaxy mergers, however, the size of the dark matter halo in galaxy clusters prevents the stellar population from becoming too irregular in shape. A number of merger-related phenomena affect the gas content, however. These include:
\begin{enumerate}[i)]

\item \textbf{Turbulence:} Mergers on all scales increase the levels of turbulence in the cluster, with the resultant velocity power spectrum observed to be in broad agreement with Kolmogorov turbulence \citep{zhuravleva2015}. This turbulence heats the ICM \citep{zhuravleva2014} and is believed to help support a small-scale dynamo as well \citep[see][and later sections]{tevlin2024}.

\item \textbf{Shocks:} Mergers drive powerful shocks through the ICM \citep[see, e.g.][]{ryu2003, pfrommer2006, skillman2008}. These heat it (see Sec.~\ref{sec:shocks}) and accelerate particles (see Sec~\ref{sec:cosmic-rays}). Merger shocks take several forms, initially forming as a bow shock around the merging cluster \citep[see, e.g.][]{markevitch2002}, before becoming detached and forming a \textit{run-away} shock \citep{zhang2019}. Given sufficient distance, this will then collide with the accretion shock \citep{zhang2020}.

\item \textbf{Particle acceleration:} Both of the above points lead to the acceleration of particles. This is believed to result in radio halos and radio relics (see Sec.~\ref{sec:acceleration} for the respective mechanisms). Shocks should also accelerate cosmic ray protons, which should in turn produce a sizeable number of gamma-rays via inverse Compton and hadronic interactions. That these are not observed is one of the current outstanding problems regarding galaxy clusters \citep{aleksic2012, ackermann2014}.

\item \textbf{Sloshing and cold fronts:} The shifting potential during a merger can lead to the ``sloshing'' of gas. This leads to spiral-shaped ``cold fronts'', which are visible in X-ray observations \citep{markevitch2007, walker2018}. Indeed, the shift of the peak of the X-ray distribution with regard to the brightest central galaxy (BCG) is considered a good diagnostic for categorising a non-relaxed cluster \citep{hudson2010}. It has further been shown that these motions are also able to disperse AGN emission, which has been suggested as a formation mechanism for radio relics \citep{zuhone2021}. It has also been noticed that mini radio halos are often confined by such fronts \citep{giacintucci2019}.

\item \textbf{Cool-core evolution:} Clusters are categorised as ``cool core'' or ``non-cool core'' based on their central entropy profile, but it is currently unclear what sets this dichotomy\footnote{Observationally, cool cores form around $z\approx1$. However, they show remarkably little evolution after this point, even though the rest of the ICM continues to evolve self-similarly \citep{mcDonald2017}.}. There is some evidence, however, that it may be related to how mergers disrupt cooling flows \citep{burns2008, valdarnini2021}.

\item \textbf{Jellyfish galaxies:} Simulations and observations suggest that so-called ``jellyfish'' galaxies are more likely in cluster mergers \citep{owers2012, mcpartland2016, ruggiero2019}. The relative motion of such galaxies compared to the ICM leads to ram-pressure stripping, producing tails of gas that provide their namesake. Such galaxies are potentially useful for probing star formation in extreme environments \citep{sparre2022}.
\end{enumerate}

\section{Magnetic fields}

``Magnetic fields'' is the second theme that underlies the topics presented in this thesis. Here, we provide a summary that covers how they act, how we infer their strengths, and where they come from.

\label{sec:magnetic-fields}

\subsection{Dynamics}

Magnetic fields, \textbf{B}, and electric fields, \textbf{E}, are inherently interlinked, as described by Maxwell's equations:
\begin{equation}
    \bs{\nabla} \bcdot \bs{B} = 0
    \label{eq:gauss}
\end{equation}
\begin{equation}
    \bs{\nabla} \btimes \bs{B} = \frac{1}{c} \frac{\partial \bs{E}}{\partial t} + \frac{4 \pi}{c} \bs{j}
    \label{eq:ampere}
\end{equation}
\begin{equation}
    \bs{\nabla}\bcdot \bs{E} = 4 \pi \rho_\rmn{e}
\end{equation}
\begin{equation}
     \bs{\nabla} \btimes \bs{E} = - \frac{1}{c}\frac{\partial \bs{B}}{\partial t}
     \label{eq:faraday} 
\end{equation}
and Ohm's law:
\begin{equation}
     \bs{j} = \sigma \left[ \bs{E} + \frac{1}{c}\left(\bs{\bupsilon} \btimes \bs{B} \right)\right]
     \label{eq:ohm}
\end{equation}
where $\bs{j}$ is the current density, $\sigma$ is the electric conductivity, $\bupsilon$ is the velocity field, $\rho_\rmn{e}$ is the charge density, and we have given the equations in CGS units.

By combining Eqs.~\eqref{eq:ampere},~\eqref{eq:faraday}, and~\eqref{eq:ohm} we can form the induction equation for resistive MHD:
\begin{equation}
    \frac{\partial \bs{B}}{\partial t} = \bs{\nabla} \times \left(\bs{\bupsilon} \times \bs{B}\right) + \eta \bs{\nabla}^2 \bs{B},
    \label{eq:induction}
\end{equation}
where $\eta = c^2 / 4 \pi \sigma$ is the magnetic resistivity. In most astrophysical contexts, the plasma is sufficiently collisional on the scales of interest that we may treat it as fluid, and sufficiently ionised that we may treat it as a perfect electrical conductor. In this case, we may set $\eta = 0$; an approximation known as \textit{ideal MHD}. A key consequence of making such an assumption is that the magnetic field becomes flux-frozen; that is, it becomes embedded in the gas such that the field lines must move with the gas and vice versa.

Rearranging Eq.~\eqref{eq:ohm}, this assumption leads to the ideal version of Ohm's law:
\begin{equation}
\bs{E} = - \frac{\bs{\bupsilon} \times \bs{B}}{c},
\end{equation}
and hence by substitution in Eq.~\eqref{eq:ampere}, in the non-relativistic limit:
\begin{equation}
    \bs{j} = \frac{c}{4 \pi} \bs{\nabla} \times \bs{B}
\end{equation}
Thus, the Lorentz force acting per unit volume is:
\begin{equation}
    \bs{F}_\rmn{L} = \frac{1}{c}\left(\bs{j} \times \bs{B}\right) = \frac{1}{4\pi}\left( \bs{\nabla} \times \bs{B}\right)\times \bs{B}.
    \label{eq:lorentz}
\end{equation}

By using vector identities, we may further split the right-hand term into two components:
\begin{equation}
    \bs{F}_\rmn{L} = \frac{\left(\bs{B} \bs{\cdot} \bs{\nabla}\right) \bs{B}}{4\pi} - \bs{\nabla} \left( \frac{\bs{B}^2}{8 \pi} \right)
\end{equation}
Here, the first term on the right-hand side is magnetic tension -- informally, the desire of magnetic fields to straighten their field lines -- and the second term is the magnetic pressure -- their desire to push field lines apart\footnote{Pressure along field lines is exactly cancelled by the tension force, so that both effects end up only act perpendicularly to the field lines.}.

The magnetic tension effectively provides a restoring force on magnetic field line perturbations, leading to the formation of Alfv\'{e}n waves. These propagate in the direction of the magnetic field, with a group velocity of:
\begin{equation}
    \bupsilon_\rmn{A} = \frac{|\bs{B}|}{\sqrt{4 \pi \rho}},
\end{equation}
where $\rho$ is the mass density. Such waves can transfer energy; indeed, this is believed to be an important mechanism for heating the solar corona \citep{tomczyk2007}. Moreover, they play a key role in the transport of cosmic rays (see Sec.~\ref{sec:cosmic-rays}).

The strength of the magnetic field, meanwhile, is often characterised by the so-called \textit{plasma beta}, which is the ratio of thermal to magnetic pressure, i.e.:
\begin{equation}
    \beta = \frac{p_\rmn{th}}{p_\rmn{B}} = {n k_\rmn{B} T}\left(\frac{8 \pi}{\bs{B}^2}\right),
\end{equation}
where we have assumed an ideal gas with $n$ being the gas number density, $k_\rmn{B}$ being the Boltzmann constant, and $T$ being the gas temperature. When $\beta \ll 1$, the magnetic pressure dominates, and hence likely dominates the dynamics\footnote{This is not a sufficient condition, however; the magnetorotational-instability, for example, can act in high $\beta$ environments \citep{inchingolo2018}. Moreover, we are neglecting other non-thermal components.}.

If we combine the above equations with the conservation equations for mass, momentum, and energy, respectively, and include an equation for the conservation of magnetic flux, we arrive at the general equations for ideal MHD:
\begin{equation}
\frac{\partial \bs{U}}{\partial t} + \bs{\nabla}\cdot\bs{F}(\bs{U}) = 0,
\end{equation}
where
\begin{equation}
\;\;\;\;\;\;\bs{U}
=
\left(
\begin{array}{c}
\rho \\
\rho \bs{\bupsilon}\\
\varepsilon \\
\bs{B}
\end{array}
\right),\;\;\;\;\;\;
\bs{F}(\bs{U}) =
\left(
\begin{array}{c}
\rho \bs{\bupsilon} \\
\rho \bs{\bupsilon} \bs{\bupsilon}^\top + P\bs{1} - \bs{B}\bs{B}^\top\\
(\varepsilon + P)\bs{\bupsilon} - \bs{B}(\bs{\bupsilon} \cdot \bs{B}) \\
\bs{B}\bs{\bupsilon}^\top - \bs{\bupsilon}\bs{B}^\top \\
\end{array}
 \right)
\end{equation}
with the equation of state for an ideal gas:
\begin{equation}
    P_\rmn{gas} = (\gamma_\rmn{a} - 1) \varepsilon_\rmn{gas}
\end{equation}
acting as the closure relation. In turn, $P = P_\rmn{gas} + \frac{1}{2}|\bs{B}|^2$, where $P_\rmn{gas}$ is the gas pressure, and $\varepsilon = \varepsilon_\rmn{gas} + \frac{1}{2}\rho|\bupsilon|^2 + \frac{1}{2}|\bs{B}|^2$. In the above equations, we used the Heaviside-Lorentz system of units, and $\rho$, $\bupsilon$, and $\varepsilon_\rmn{gas}$ are the mass density, velocity, and energy density of the fluid, respectively, and $\gamma_\rmn{a}$ is the adiabatic index. These equations must be discretised before being implemented in simulations \citep[see, e.g.,][and references therein]{toth2000, pakmor2011}.

\subsection{Observations}

Direct measurements of magnetic fields must be done in situ, which generally limits their scope to the near-Earth environment. Within this context, the most notable in situ measurements have been performed by the \text{Parker Solar Probe}, which has recorded the magnetic field strength in the solar corona  \citep{kasper2021}, and \text{Voyager 2}, which has determined the field strength at the heliopause \citep{burlaga2019}. At distances farther than this, we must rely on indirect methods. The main methods\footnote{See \citet{beck2015} for a more exhaustive list.} are:
\begin{itemize}

    \item \textit{Zeeman splitting:} In the presence of an external magnetic field, the degeneracy between energy levels with magnetic quantum number $m_l \neq 0$ is lifted. This splits spectral lines, with the shift in frequency being directly proportional to the strength of the magnetic field. However, in order to observe this effect, the magnetic field  must be sufficiently strong and the spectral resolution sufficiently high. Consequently, this technique is usually only applied to near or compact objects. This typically means studies of our Sun, but it has also been applied to molecular clouds in external galaxies \citep{li2011} and the local diffuse ISM \citep{heiles2009}.
    
    \item \textit{Polarised starlight:} Interstellar magnetic fields cause otherwise turbulent dust grains to re-align such that their minor-axes are predominantly parallel to the field direction\footnote{This process is generally understood to result from the \citet{davis1951} effect, which is based on the principle of paramagnetic dissipation.}. As starlight passes through these grains, the dust subsequently preferentially absorbs light along its major axis, scattering this out of the line-of-sight. This results in the starlight becoming polarised parallel to the magnetic field component in the plane of the sky. Moreover, assuming the degree of polarisation of this emission is a direct result of the interplay of velocity dispersion and Alfv\'{e}n waves, the magnetic field strength can be roughly estimated using the Davis-Chandrasekhar-Fermi method \citep{davis1951b, chandrasekhar1953}.
    
    \item \textit{Polarised infrared light:} The starlight absorbed by the dust grains acts to heat them. This leads to the dust emitting in the far-infrared band, with light polarised along the major axis of the dust grains. This technique is increasingly being applied to external galaxies \citep[see, e.g.][]{lopez-rodrigues2020}, and has the advantage that the magnetic fields may continue to be analysed even when the starlight is blocked \citep{hildebrand1988}. In the Milky Way, individual stars can be probed, and hence it is possible to measure how the polarisation changes with distance. This may allow for three-dimensional tomography of the local magnetic field \citep{hoang2024}.

    \item \textit{Velocity gradient technique (VGT):} This is a relatively new technique based on the theory of MHD turbulence \citep{goldreich1995} and reconnection \citep{lazarian1999}, which state that turbulent eddies in MHD flows are increasingly anisotropic at smaller scales. The gradient of the velocity resulting from these eddies should therefore be perpendicular to the local magnetic field direction. Such gradients can be accessed in a variety of ways \citep[see][and references therein]{lazarian2024}.
    
    \item \textit{Faraday rotation:} When polarised light passes through a magnetised plasma with free electrons\footnote{This is also the main drawback of this method; analysing the magnetic field using Faraday rotation can only be done when polarised light passes through the object of interest.}, the plane of polarisation rotates by an angle, $\Delta\theta$, where:\vspace{0.4cm}\\
    \null\hfill
    $\displaystyle
    \Delta\theta = \frac{e^3 \lambda^2}{2 \pi m_\rmn{e}^2 c^4} \int B_\parallel n_\rmn{e} dl = \rmn{RM}\, \lambda^2$\hfill
    \puteqnum \vspace{0.4cm}
    \label{eq:RM}\\
    Here, $e$ is the charge of an electron, $m_\rmn{e}$ is its mass, $\lambda$ is the wavelength of the light, $B_\parallel$ is the component of the magnetic field parallel to the line of sight, and $n_\rmn{e}$ is the electron number density. The product of the integral and the constant in front of it is known as the \textit{rotation measure} (RM). Pulsar emission is often used in calculating this quantity, as emission from pulsars is typically highly polarised, and the dispersion of the signal -- i.e. the delay between different frequency components received over time -- can be used to estimate the integrated electron number density. In combination with Eq.~\eqref{eq:RM}, this produces an estimate for the electron-density weighted mean magnetic field strength along the line of sight. This quantity is particularly useful when $n_\rmn{e}$ is heavily weighted in the object of interest, such as when measuring the field strength  in external galaxies or the local ISM \citep{manchester1972}. The coherence of the Faraday rotation over larger scales also informs us about the magnetic field structure. Such analysis is particularly useful when looking for signatures of the small-scale dynamo \citep{bhat2013, tevlin2024}.
    
    \item \textit{Synchrotron radiation:} When relativistic charged particles are accelerated perpendicular to their velocity, they emit synchrotron radiation. This is naturally the case for cosmic ray electrons orbiting magnetic field lines (see Sec.~\ref{sec:cosmic-rays}). Electrons in a cosmic ray population with energy $E$ will emit most of their synchrotron emission at the critical frequency:\vspace{0.4cm}\\
    \null\hfill
    $\displaystyle
    \nu_\rmn{c} \equiv \frac{3}{4\pi} \frac{e}{m_\rmn{e} c} B_\perp \gamma^2  \approx 16 \, \rmn{MHz}\; \left( \frac{E}{\rmn{GeV}}\right)^2 \left(\frac{B_\perp}{\upmu\rmn{G}}\right),
    $\hfill
    \puteqnum \vspace{0.4cm}\\
    where $\gamma$ is the Lorentz factor (see Sec.~\ref{sec:cosmic-rays}), $B_\perp$ is the magnetic field strength in the plane of the sky \citep{beck2005}. To solve for the magnetic field, the \textit{equipartition condition} is used, which assumes that the cosmic ray proton energy density is roughly equal to the magnetic field energy density. Evidence for the validity of this assumption comes from measurements performed in the local ISM \citep{beck1996, beck2015}, however, it is unclear whether this holds in every galaxy because there is no astrophysical effect known to enforce such an equilibrium \citep{ruszkowski2023}. Synchrotron radiation is intrinsically highly polarised, and hence observations using this method can be used to infer the mean magnetic field direction in the plane of the sky. The polarisation fraction is also often used, with a higher polarisation fraction indicating greater magnetic coherence \citep{beck2009}. Such observations are, however, subject to Faraday depolarisation.
    
\end{itemize}

Using the above techniques, in combination with numerical modelling, it has been found that magnetic fields cover a wide parameter space of strengths and scales. We show a schematic that indicates some typical values for astrophysical objects in Fig.~\ref{fig:magnetic_field_vs_size}. Galaxy clusters, for example, typically have magnetic fields on the order of $\upmu$G, reducing to $\lesssim 0.1 \upmu$G in the cluster outskirts \citep{tevlin2024}. In disc galaxies in the local Universe, meanwhile, the average magnetic field strength tends to be around $10-15$ $\upmu$G, with the field roughly following the spiral arms \citep{beck2004}. This varies, however, based on the gas density and levels of turbulence. For example, in the gas-rich spiral arms and bars, the magnetic field strength is typically on the order of $20-30$ $\upmu$G, with this increasing still further to between $50$ and $100$ $\upmu$G in the central starburst regions \citep{beck2015}.

\begin{figure}
    \centering
    \includegraphics[width=0.5\linewidth]{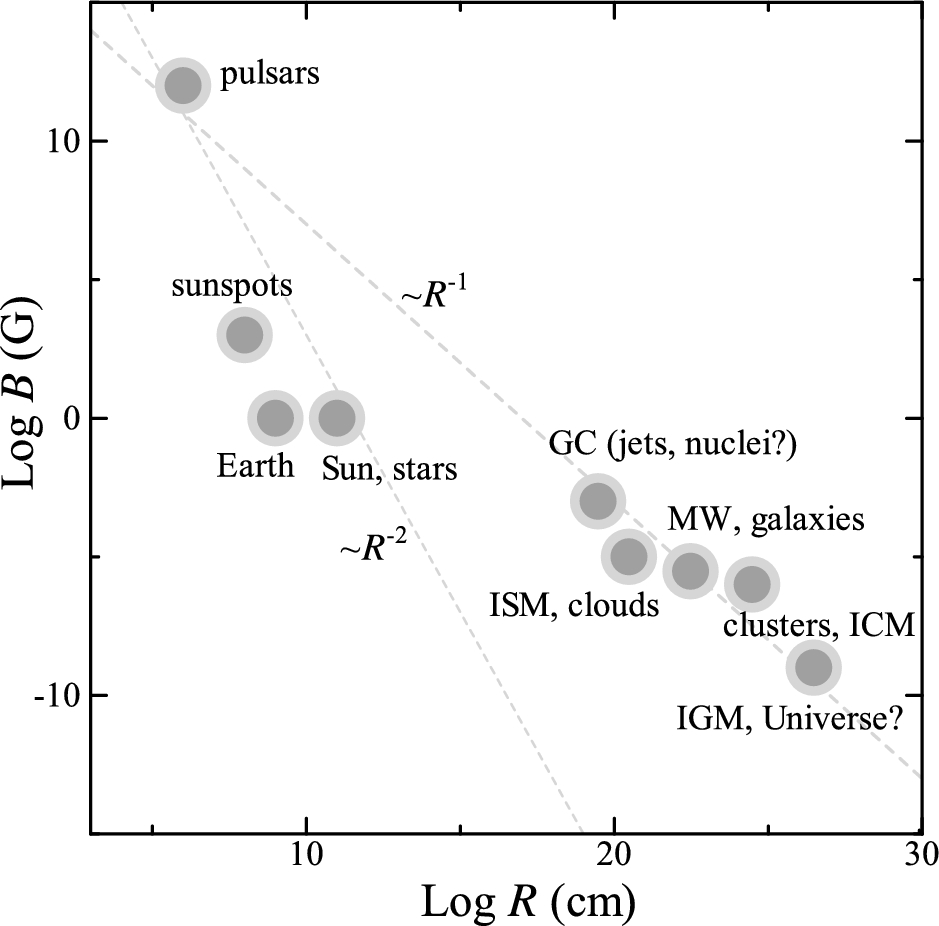}
    \caption[Magnetic field strength as a function of scale for typical astrophysical objects]{Schematic showing the typical size, $R$, and magnetic field strength, $B$, in various astrophysical objects. Dashed lines indicate scalings of the field strength with $R^{-1}$ and  $R^{-2}$, respectively. Magnetic field strengths in galaxies and clusters are typically on the order of $\upmu$G, although this varies depending on the exact conditions and spatial scale observed. \textit{Image credit:} \citet{akahori2018}.}
    \label{fig:magnetic_field_vs_size}
\end{figure}

\subsection{Seed fields}
\label{sec:seed-fields}

Equation~\eqref{eq:gauss} is also known as Gauss's law of magnetism, and states that there are no source or sinks of magnetic fields. We must thus explain how the original seed field was generated, which was subsequently amplified to the strengths we observe today. This is usually explained as being a result of one of three mechanisms:
\begin{enumerate}[i)]
    \item Generation during the very early Universe during, e.g., the electroweak phase transition \citep{vachaspati1991} or as a result of inflation \citep{turner1988}.
    \item Plasma instabilities, such as the Weibel instability, which can produce magnetic fields on kinetic scales \citep{schlickeiser2005}.
    \item Cosmic battery processes, such as the Biermann battery \citep{biermann1950, davies2000}, which can produce magnetic fields on macro-scales.
\end{enumerate}
The Biermann battery, in particular, is generally considered a likely candidate, owing to its applicability in a wide range of scenarios. This mechanism exploits the fact that, as electrons and protons have different masses, they have different inertia. Consequently, if we apply the same force to each, the electrons will accelerate more quickly, and an electric field will be generated. By Eq.~\eqref{eq:faraday}, if we can generate a curl in this field, we will produce a magnetic field. This can be done if the gradients of the electron density and the electron pressure, $P_\rmn{e}$, are misaligned. This modifies Eq.~\eqref{eq:induction} as follows:
\begin{equation}
    \frac{\partial \bs{B}}{\partial t} = \bs{\nabla} \times \left(\bs{\bupsilon} \times \bs{B}\right) - \frac{c}{e n_\rmn{e}^2}\left(\bs{\nabla} n_\rmn{e} \times \bs{\nabla} P_\rmn{e} \right),
    \label{eq:biermann}
\end{equation}
where we have assumed $\eta = 0$, and the last term on the right-hand side is the \textit{battery term}.

Assuming local charge neutrality, $n_\rmn{e} \approx n_\rmn{p} \equiv \chi \rho / m_\rmn{p}$, where $n_\rmn{p}$ is the proton number density, $\chi$ is the ionisation fraction, $\rho$ is the gas density, and $m_\rmn{p}$ is the proton mass. Furthermore, assuming the electron temperature is approximately equal to the gas temperature, $P_\rmn{e} \approx P n_\rmn{e} / (n_\rmn{e} + n_\rmn{p}) = P \chi / (1 + \chi)$, where $P$ is the gas pressure. Hence, the induction equation can be re-written:
\begin{equation}
\frac{\partial \bs{B}}{\partial t} = \bs{\nabla} \times \left(\bs{\bupsilon} \times \bs{B}\right) + \alpha \frac{\bs{\nabla} \rho \times \bs{\nabla} P}{\rho^2},
\label{eq:battery-term}
\end{equation}
where the factor $\alpha = m_\rmn{p} c / e ( 1+\chi) \approx 10^{-4}$ G s is the Biermann coupling constant \citep[see][]{kulsrud1997, davies2000}. Equation~\eqref{eq:battery-term}, however, has the same form as the vorticity equation:
\begin{equation}
    \frac{\partial \bs{\omega}}{\partial t} = \bs{\nabla} \times \left(\bs{\bupsilon} \times \bs{\omega}\right) + \frac{\bs{\nabla} \rho \times \bs{\nabla} P}{\rho^2},
\end{equation}
implying that
\begin{equation}
|\bs{B}| = \alpha |\bs{\omega}|
\label{eq:biermann-ratio}
\end{equation}
For a proto-galaxy, $|\bs{\omega}| \sim U / L \approx 30$ km s$^{-1}$ / 10 kpc $\approx 10^{-16}$ s$^{-1}$, where $U$ and $L$ are the velocity and length scale, respectively\footnote{In this case, diameter and typical rotational speed.}. Consequently, by Eq.~\eqref{eq:biermann-ratio}, the magnetic field strength generated is approximately 10$^{-20}$ G.

Such a scenario can take place in cosmological shock fronts \citep{kulsrud1997} and cosmological ionisation fronts \citep{gnedin2000}, but also within compact objects, such as stars \citep{doi2011}. This is important, as it is still unclear whether magnetic fields were created cosmologically, before being amplified in structure formation -- so-called \textit{top-down magnetogenesis} -- or first seeded in compact objects, before being injected into proto-galaxies by stellar winds, and eventually into the galactic environment -- \textit{bottom-up magnetogenesis}.

\begin{figure}
    \centering
    \includegraphics[width=0.6\linewidth]{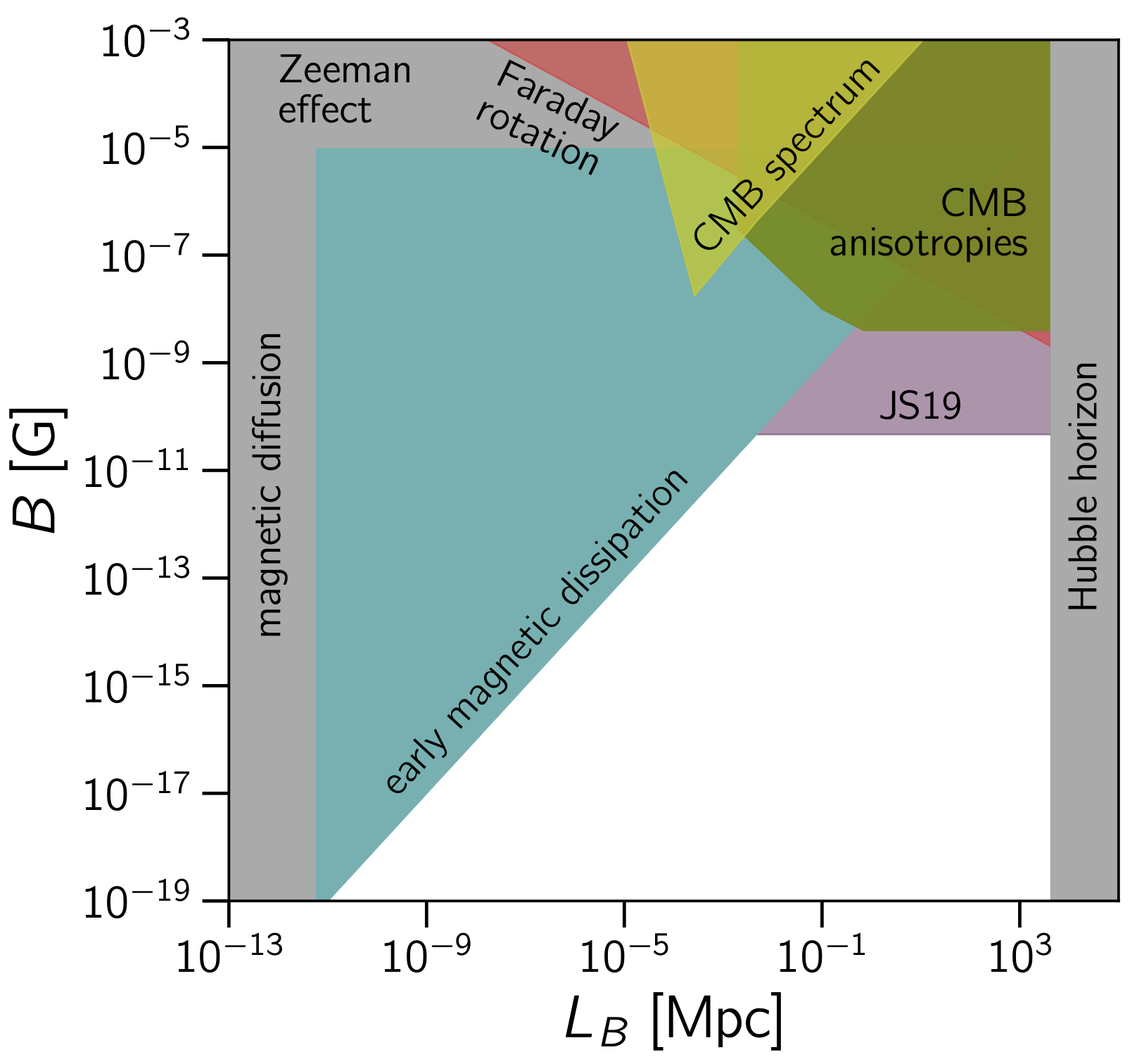}
    \caption[]{The current constraints on intergalactic magnetic fields, where $B$ is the field strength, and $L_B$ is the coherence length. \textit{Image credit:} \citet{alves2021}.}
    \label{fig:mag_parameter_space}
\end{figure}

One way of solving this problem is to measure the intergalactic magnetic field (IGMF), which is presumed to have altered very little over cosmic time. We show the current parameter space excluded and the methods by which they were excluded in Fig.~\ref{fig:mag_parameter_space}. The region labelled ``JS19'' provides the most stringent upper limit and depends on the effect that a primordial magnetic field before the epoch recombination has on small-scale baryonic fluctuations \citep{jedamzik2019}. A comprehensive list of other methods can be found in \citet{alves2021}. Blazar-based measurements are not included in Fig.~\ref{fig:mag_parameter_space}. Lower limits based on this method use the principle that stronger magnetic fields should deflect the resulting electron-positron pairs, thereby reducing the observed gamma-ray intensity. This method has been claimed to place a lower limit of roughly $7\times 10^{-16}$ G on the IGMF \citep{aharonian2023}. However, it is not clear that this is only mechanism that can reduce the gamma-ray intensity \citep{broderick2012}. Moreover, the scenario only accounting for pair deflection is predicted to produce a bow-tie feature in the gamma-ray sky, which is not observed \citep{broderick2016, broderick2018, tiede2017, tiede2020}.

\subsection{Growth and amplification }
\label{subsec:growth}

Returning to the induction equation, we see that the cross products on the right-hand side of Eq.~\eqref{eq:induction} represent the induction or advection of the magnetic field, whilst the remaining term represents its diffusion. As we will see in Sec.~\ref{subsec:growth}, in many astrophysical environments it is reasonable to assume that the advection part dominates. Indeed, this assumption is often taken to its extreme, such that we may set $\eta = 0$.

The ratio of the advection and diffusion terms is known as the \textit{magnetic Reynolds number}, $\rmn{Rm}$. This is the magnetic counterpart to the standard Reynold's numbers. Through dimensional analysis, it can be seen that
\begin{equation}
    \rmn{Rm} = \frac{U L}{\eta},
\end{equation}
where $U$ and $L$ are the typical velocity and length scales. In turbulence, this is the root-mean-square (RMS) velocity and the typical eddy scale. Assuming typical velocities in galaxies of a few 10 km/s, lengths on the order of $\sim10$~kpc, and a magnetic diffusivity of 10$^{25}$ cm$^2$ s$^{-1}$ \citep{hanasz2009b}, we arrive at $\rmn{Rm} \approx 10^4$. The values in the ICM are substantially larger, reaching $\rmn{Rm} \gtrsim 10^{27}$ and higher \footnote{See parameters estimated in table 1 of \citet{schekochihin2006}.}. The advective term would thus appear to dominate Eq.~\eqref{eq:induction}, allowing us to ignore the diffusive term. Indeed, this is generally a good approximation for an ionised gas, and thus ideal MHD applies to most of the ISM\footnote{A notable exception is in dense clouds, when ambipolar diffusion becomes important \citep{ntormousi2016}.} and is often applied to the ICM as well\footnote{Strictly speaking, the ICM today is weakly collisional, and hence Braginskii MHD is a better description \citep[see, e.g.][]{berlok2020}. With this said, \citet{tevlin2024} find that the magnetic coherence length in galaxy clusters remains consistently above the mean free path during its evolution, and hence the modelling of a small-scale dynamo in the ICM can be performed with ideal MHD.}.

A direct result of this is that magnetic fields are effectively amplified during adiabatic compression. However, if we were only to rely on this mechanism, we would be unable to explain the current $\upmu$G magnetic fields in galaxies and galaxy clusters. For example, if we approximate galaxies as resulting from spherical collapse, we should expect\footnote{Whilst collapse happens in three-dimensions, the magnetic field is only amplified in directions perpendicular to itself.} the magnetic field to scale with $B  \propto \rho^{2/3}$. However, if we treat such objects as overdensities with $\delta = 1000$, we see that the magnetic field only increases by a factor of 100 or so. By our estimate in Sec.~\ref{sec:seed-fields}, compression would produce a typical field strength of 10$^{-12}$ $\upmu$G, which is several orders of magnitude lower than required.

\begin{figure}
    \centering
    \includegraphics[width=0.8\linewidth]{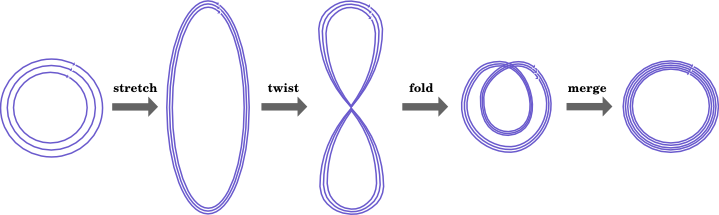}
    \caption[Schematic showing the stretch-twist-fold-merge mechanism behind the small-scale dynamo]{Schematic showing how turbulence results in magnetic field amplification through the small-scale dynamo. Turbulent motions act to stretch, twist, and fold the magnetic field lines. In the final step, the magnetic field lines re-connect, thereby increasing the flux density by a factor of two. This process continues, leading to exponential growth in the kinematic phase. \textit{Image credit:} \href{https://jennifer-schober.com/research-interests/the-turbulent-dynamo-during-the-formation-of-the-first-stars-and-galaxies/}{J. Schober}.}
    \label{fig:small-scale-dynamo-schematic}
\end{figure}

The solution, of course, is that the magnetic field is not perfectly coupled to the gas. Indeed, diffusion is critical for dynamo action\footnote{The magnetic field cannot be too diffusive, however. Indeed, numerical simulations generally find that $\rmn{Rm} \gtrsim  100$ is required for dynamo action \citep{achikanath2024}.}. One of the fastest acting dynamo mechanisms is the small-scale turbulent dynamo, otherwise known as the Zel'dovich ``stretch-twist--fold'' dynamo \citep{vauinshteuin1972}. This starts with a closed flux loop, as shown on the left-hand side of Fig.~\ref{fig:small-scale-dynamo-schematic}, which is is then stretched, twisted, and folded. In the final step, the two loops of magnetic field lines merge together through diffusion. The magnetic field now has twice as many field lines for the same volume; i.e., the field strength has increased by a factor of two. After repeating $n$ times, the strength will have grown by a factor $2^n$ and hence the dynamo increases the field strength exponentially.

Each of the steps presented in Fig.~\ref{fig:small-scale-dynamo-schematic} are important: the stretching provides the initial increase in the magnetic field strength; the twisting makes sure that the magnetic flux adds coherently in the fold step, rather than cancelling out; and the merge step makes the process irreversible. The source for these motions is turbulent kinetic energy, and the process will continue until the field strength is high enough for the magnetic field to have a significant dynamical back-reaction via the Lorentz force and magnetic tension. Indeed, one indicator for the presence of a small-scale dynamo is the anti-correlation of the magnetic field strength and the curvature, such that $|\bs{B}|| \bs{K}|^{0.5} \sim \mathrm{const.}$, where
\begin{equation}
    \bs{K} = \frac{(\bs{B}\bcdot\bs{\nabla})\bs{B}}{|\bs{B}|^{2}}
\end{equation}
is the magnetic curvature \citep{schekochihin2005}.

In the \citet{Kazantsev1968} model, the amplification timescale is initially the same as the eddy turnover time. Once the magnetic field has left the kinematic regime, however, its amplification becomes non-linear \citep{brandenburg2005}. Indeed, this provides a key problem for numerical simulations of dynamos as, if too much time is spent in the kinematic growth phase, the non-linear phase is severely truncated\footnote{We will see a realisation of this effect in Chapter~\ref{chapter:paper-one}.} \citep{donnert2018}. This can happen if the resolution in the simulation is not high enough, resulting in an eddy turnover time which is too long, or if the simulation is too diffusive; i.e., if the magnetic Reynolds number is not high enough. It has been shown analytically that the small-scale dynamo can amplify a weak seed field in a proto-galaxy by approximately 13 orders of magnitude within 300 million years \citep{arshakian2009, deSouza2010,pfrommer2022}, with minor modulations depending on whether the turbulence is compressive or incompressible \citep{federrath2010}. This then, accounts for the remaining orders of magnitude, amplifying the magnetic field strength from a seed field to the observed values.

Whilst the peak scale of the magnetic field will grow over time, eventually meeting the peak of the kinetic power, the magnetic field components generated through the small-scale dynamo will be essentially random. Disc galaxies, however, are observed to have magnetic field components that generally follow the large-scale structure \citep{beck2011}. To solve this problem the $\alpha - \omega$ dynamo is sometimes invoked \citep{steenbeck1966, parker1971}.  This dynamo works on the basis that the galaxy is both differentially rotating and stratified in the $z$-direction. In this environment, when magnetic loops rise above the galactic plane, they start to twist, eventually reconnecting such that they have converted a toroidal magnetic field component into a poloidal one. In turn, the differential rotation of the disc winds the field lines up, thereby converting the poloidal field back to a toroidal field. As the rotational energy is significantly higher than the turbulent energy, this dynamo can keep acting after the small-scale dynamo has saturated. However, this dynamo is significantly slower, and it is still unclear whether it works in a cosmological environment \citep{pakmor2017} or whether it produces magnetic field morphologies consistent with observations \citep{unger2024}.

\subsection{Impact}

The strengths measured for magnetic fields in disc galaxies at the current epoch imply that they have a dynamical role to play. Indeed, the additional pressure component is expected to help balance the disc against gravitational collapse \citep{beck2013} and affect the flow of gas in the spiral arms \citep{gomez2002}. The impact of magnetic fields can be both negative and positive for the survival times of gas clouds \citep{crutcher1999}. For example, magnetic fields may reduce mixing, through magnetic draping \citep{dursi2008,pfrommer2010} and through anisotropic transport mechanisms \citep{balbus2008}. These processes may help gas clouds to be accelerated by hot winds \citep{mccourt2015}, thereby increasing the amount of cold gas in the CGM. They may also support the growth of cold streams, which ultimately increase the star formation in the galaxy \citep{berlok2019}. Magnetic draping is also believed to stabilise the tails of jellyfish galaxies \citep{sparre2020,mueller2021}.

On the other hand, magnetic braking can help transport angular momentum outwards, which is believed to be crucial for star formation \citep{mellon2008} and may increase the metallicity pollution by massive stars \citep{meynet2011}. Meanwhile, the magneto-rotational instability (MRI) has long been known to be crucial in the dynamics of accretion discs \citep{balbus1991, balbus2008}, helping channel gas towards black holes. Finally, magnetic reconnection may be an important source of heating in the ISM and CGM \citep{birk1998}. In this way, magnetic fields are relatively unpredictable; they may either cool or heat the gas, and can help support structures or collapse them. As their dynamics are so strongly affected by their strength and the gas flows around them, it is crucial that they are modelled in an environmentally consistent manner. For disc galaxies as a whole, this means the use of cosmological simulations.

\section{Shocks}
\label{sec:shocks}

Shocks are a natural consequence of mergers, and play a pivotal role in the acceleration of cosmic ray electrons. We therefore recap the basic theory here.

Any time the bulk gas speed is higher than the sound speed, a shock will form. In this case, information about the gas properties behind the shock ``downstream'' cannot be communicated to the gas ahead of the shock ``upstream'' and a discontinuity is formed. As no gas accumulates at this discontinuity, we can use the conservation laws for mass, momentum, and energy to observe how the gas properties change across it. For a plane-parallel shock in a hydrodynamic medium\footnote{This treatment is appropriate for the shocks we discuss in this dissertation, but generally speaking a complete treatment would need to take into account magnetic and cosmic ray pressure, as well as the generation of magnetosonic waves.}, these are:
\begin{equation}
    \rho_1 \upsilon_{x,1} = \rho_2 \upsilon_{x,2},
\end{equation}
\begin{equation}
   P_1 + \rho_1 \upsilon_{x,1}^2 = P_2 + \rho_2 \upsilon_{x,2}^2,
\end{equation}
and
\begin{equation}
    \frac{\varepsilon_1 + P_1}{\rho_1} + \frac{\upsilon_{x,1}^2}{2} = \frac{\varepsilon_2 + P_2}{\rho_2} + \frac{\upsilon_{x,2}^2}{2},
\end{equation}
respectively, where subscripts ``1'' and ``2'' refer to upstream and downstream quantities, and gas properties are given in an inertial frame that co-moves with the shock-front.

Note, that the mass continuity equation allows for the trivial solution that $\upsilon_{x,1} = \upsilon_{x,2} = 0$. In this case, no mass flows across the discontinuity, and hence no shock forms. This scenario is known as a \textit{tangential discontinuity}. A \textit{contact discontinuity} is a special case of this, in which the velocity, thermal pressure, and magnetic field are all continuous; i.e. only the density and temperature change.

We may then define the Mach number, such that:
\begin{equation}
    \mathcal{M} = \frac{\upsilon_{x,1}}{c_\rmn{s,1}},
\end{equation}
which allows us to formulate the well-known Rankine-Hugoniot jump conditions for a polytropic gas:
\begin{equation}
    r = \frac{\rho_2}{\rho_1} = \frac{\upsilon_{x,1}}{\upsilon_{x,2}} = \frac{(\gamma + 1)\mathcal{M}^2}{(\gamma - 1)\mathcal{M}^2 + 2},
    \label{eq:compression-ratio}
\end{equation}
\begin{equation}
    \frac{P_2}{P_1} = \frac{2 \gamma \mathcal{M}^2 - (\gamma - 1)}{\gamma + 1},
\end{equation}
and
\begin{equation}
    \frac{T_2}{T_1} = \frac{\rho_1}{\rho_2}\frac{P_2}{P_1} = \frac{[2 + (\gamma - 1)\mathcal{M}^2][2 \gamma \mathcal{M}^2 - (\gamma - 1)]}{(\gamma + 1)^2\mathcal{M}^2},
\end{equation}
where we have assumed negligible radiative losses, and $r$ is known as the \textit{compression ratio}.

Given that $\mathcal{M} > 1$, we see that shocks lead to a deceleration of the gas in the shock-frame, as well as a pressure, density, and temperature jump. For strong shocks ($\mathcal{M} \gg 1$), we may also see from Eq.~\eqref{eq:compression-ratio} that $r \to 4$. Moreover, if we compute the difference in the up- and downstream specific entropy, we find that:
\begin{equation}
    \frac{s_2 - s_1}{ c_V} = \ln \left(\frac{P_2}{P_1}\right) - \gamma_\rmn{a} \ln \left( \frac{\rho_2}{\rho_1} \right),
\end{equation}
where $c_V$ is the specific heat capacity at constant volume. This equation shows that shocks are inherently dissipative; that is, they irreversibly convert kinetic energy into thermal energy\footnote{See, for example, that in a strong shock $P_2 / P_1 \rightarrow \frac{5}{4} \mathcal{M}^2$ , whilst $\rho_2 / \rho_1 \rightarrow 4 = \rmn{const.}$, as previously stated.}. Both of these points will also help us in Sec.~\ref{sec:DSA}.

In general, due to the low densities involved, the typical length scale for particle collisions in astrophysical environments is very large relative to the system size. Shocks in such an environment must therefore proceed in a \textit{collisionless} manner. This is usually explained through the Weibel instability \citep{weibel1959} for hydrodynamic shocks (in unmagnetised plasmas), and being a result of scattering due to magnetic inhomogeneities and additional plasma instabilities in MHD shocks. Fortunately, the Rankine-Hugoniot conditions make no assumptions on the mechanism by which the shock front itself forms, and hence the equations given remain valid regardless. Minor alterations are required, however, in the case of oblique shocks, or for gases which differ from ideal ones. The adjusted equations for a gas with cosmic rays, for example, can be found in \citet{Pfrommer2017}.

\section{Cosmic ray electrons}
\label{sec:cosmic-rays}

In the final two sections, we recap the theory necessary for our investigation into the origin of radio relics, as presented in Chapters~\ref{chapter:paper-three} and~\ref{chapter:paper-four}. We start with the basics of cosmic rays, with an emphasis on cosmic ray electrons.

\subsection{Background}

In 1912, Victor Hess conducted a series of balloon flights in order to measure radiation levels at different altitudes \citep{hess1912}, finding that levels increased as he ascended\footnote{A near simultaneous discovery was independently found by \citet{pacini1912} \citep[see additional notes in][]{ruszkowski2023}.}. This finding was later confirmed by Erich Regener and Robert Millikan, who extended the survey to higher altitudes and to deep under water \citep{millikan1926, regener1929}, thereby increasing the evidence that the radiation was of extraterrestrial origin. The recorded ionisation was initially believed to be due to gamma rays, and hence was given the name \textit{cosmic rays}. Experiments by Jacob Clay, however, showed that the cosmic ray intensity also increased towards the equator, indicating deflection by the Earth's magnetic field \citep{clay1927}. This showed that cosmic rays are actually predominantly charged, and hence are not photons. Moreover, following calculations by Bruno Rossi \citep{rossi1930}, later surveys were able to show that the intensity of cosmic rays is greater when measured towards the west \citep{rossi1934}. Extrapolating from the Lorentz force, this implies that cosmic rays measured at Earth are pre-dominantly {positively}-charged.

Today, we know through a range of ground- (e.g.\ KASCADE, Pierre Auger Observatory, Telescope array), air- (e.g.\ ATIC, CREAM, BESS), and space-based particle detectors (PAMELA, AMS-02) that \textit{primary} cosmic rays -- i.e.\ cosmic rays that originate from outside Earth's atmosphere\footnote{In contrast to secondary cosmic rays, which result from the collision of primary cosmic rays with atoms in the Earth's atmosphere, leading to a cascade of lighter particles.} -- are approximately 90\% protons, 9\% helium nuclei, and 1\% heavier nuclei. Moreover, their energy spectrum covers over 12 orders of magnitude \citep{gaisser2016}, with the spectrum dominated at low energies by those with a galactic origin and at high energies ($\gtrsim10^{17}$ eV) by cosmic rays with an extragalactic origin \citep{hillas1984, apel2012}. In the local ISM, cosmic rays are measured to be in rough equipartition with the thermal, turbulent, and magnetic energy densities \citep{boulares1990, beck2015}. This implies that they are dynamically important for the galaxy at the present time. Indeed, simulations suggest that they can launch stellar winds \citep{thomas2024} and help keep the CGM pressurised, thereby affecting gas accretion and the subsequent evolution of angular momentum in disc galaxies \citep{buck2020}.

Cosmic ray electrons also reach Earth, but these are significantly less abundant ($<1$\%) and less energetic than the cosmic ray protons, being only able to reach TeV energies \citep{archer2018}. Furthermore, cosmic ray electrons cool much faster than cosmic ray protons, and are more easily deflected by magnetic field lines. The relative difference between the two species essentially results from their large mass ratio of $m_\rmn{p} / m_\rmn{e} \approx 1836$, with electrons being correspondingly easier to accelerate. As a result of their lower energy density, cosmic ray electrons are not dynamically important. However, their short cooling times make them extremely useful for tracing cosmological shocks, and they consequently form a key part of our understanding of the formation of \textit{radio relics} (see Sec.~\ref{sec:radio-relics}).

\subsection{Dynamics}

Cosmic rays can be treated using their phase-space distribution, where this depends on position, momentum, and time: $f^\rmn{3D}(\bs{x},\bs{p}, t)$. The evolution of this phase-space distribution is described by the Fokker-Planck equation \citep{skilling1975,schlickeiser1989}:
\begin{equation}
    \frac{\partial f^\rmn{3D}}{\partial t} + (\bs{\bupsilon} + \bs{\bupsilon}_\rmn{st})\bcdot \bs{\nabla}f^\rmn{3D} = \bs{\nabla}\bcdot [\bs{D}_{xx} \bcdot \bs{\nabla}f] + \frac{p}{3}\frac{\partial f^\rmn{3D}}{\partial p} \bs{\nabla} \bcdot (\bs{\bupsilon} + \bs{\bupsilon}_\rmn{st}) + \frac{1}{p^2}\frac{\partial}{\partial p} \left[ p^2 D_{pp} \frac{\partial f^\rmn{3D}}{\partial p} \right] + S(p),
    \label{eq:fokker-planck}
\end{equation}
where $p=|\bs{p}/(m_\rmn{e} c)|$ is the absolute value of the normalised particle momentum, $\bs{\bupsilon}$ is the velocity of the background plasma that the cosmic rays are advected with, $\bs{\bupsilon}_\rmn{st}$ indicates the streaming velocity, $\bs{\bupsilon}_\rmn{A}$ is the Alfv\'{e}n velocity, $\bs{D}_{xx}$ and ${D}_{pp}$ are the respective spatial and momentum diffusion coefficients, and 
\begin{equation}
    S(p) = q(p) - \frac{1}{p^2}\frac{\partial}{\partial p} \left( p^2 \dot{p} \frac{\partial f^\rmn{3D}}{\partial p} \right) - \frac{f^\rmn{3D}}{\tau_\rmn{c}},
\end{equation}
with $q(p)$ being the source function, losses described by the cooling term $\dot{p}$, and further \text{catastrophic} losses taking place on timescale $\tau_\rmn{c}$. Adiabatic changes are described by the second term on the right-hand side of Eq.~\eqref{eq:fokker-planck}, where cosmic rays have an adiabatic index of $\gamma_{a} = 4/3$. We provide a brief overview of the other processes involved in this equation in the remaining sections of this chapter.

Being charged particles, cosmic rays are subject to the Lorentz force. Balancing this and the centripetal force produces the relativistic Larmor radius for the electron:
\begin{equation}
    r_\rmn{L} = \frac{\gamma m \bupsilon_\perp}{e|\bs{B}|},
    \label{eq:Larmor}
\end{equation}
where $\bupsilon_\perp$ is the component of the velocity perpendicular to the magnetic field, and $\gamma = \sqrt{1+|\bs{p}|^2}$ is the Lorentz factor. In typical astrophysical units, this can be reformulated as:
\begin{equation}
    r_\rmn{L} \approx 1.08 \left( \frac{E}{10^{15} \,\rmn{eV}} \right) \left( \frac{|\bs{B}|}{\upmu\rmn{G}} \right) \rmn{pc}
\end{equation}
This means that a 1 GeV cosmic ray electron has a Larmor radius of approximately 0.25 AU\footnote{At best, our simulations have resolution approximately 9 orders of magnitude greater than this, allowing us to model cosmic rays in a subgrid fashion.}.
Cosmic rays electrons gyrate at this radius with frequency
\begin{equation}
    \Omega = \frac{e |\bs{B}|}{\gamma m_\rmn{e} c}.
\end{equation}
The motion along the magnetic field is typically referred to in terms of the \textit{pitch angle cosine}, where this is:
\begin{equation}
    \cos\theta = \frac{\bs{\bupsilon}_\rmn{cr}\bcdot\bs{B}}{|\bs{\bupsilon}_\rmn{cr}||\bs{B}|},
\end{equation}
with $\bs{\bupsilon}_\rmn{cr}$ being the cosmic ray velocity.

Inhomogeneities in the magnetic field are ubiquitous. Indeed, they are frequently excited by the cosmic rays themselves \citep{kulsrud1969, bell2004, shalaby2021, shalaby2023}. We can follow the impact of such inhomogeneities by comparing their typical scale, $\lambda$, with the Larmor radius. We find that if $r_\rmn{L} \ll \lambda$, the inhomogeneities vary too slowly to scatter the cosmic ray, and it is subsequently trapped by a single field line and adiabatically follows its direction. On the other hand, if $r_\rmn{L} \gg \lambda$, the cosmic ray electron gyrates around several magnetic field lines, whose uncorrelated inhomogeneities add small-scale Lorentz forces that effectively average out. However, if  $r_\rmn{L}$ and $\lambda$ are of comparable scales, the gyration of the cosmic ray results in it becoming ``trapped'' by neighbouring inhomogeneities, and being thereby effectively scattered. This process is referred to as \textit{pitch-angle scattering} \citep{ruszkowski2023}.

Assuming that pitch-angle scattering is sufficiently effective, we can assume that the distribution function is nearly isotropised in the scattering wave frame, and we can therefore reduce our description of the phase-space distribution to a one-dimensional distribution
\begin{equation}
f(\bs{x}, p, t) = 4 \pi p^2  f^\rmn{3D}(\bs{x},\bs{p}, t)
\label{eq:1D-dist}
\end{equation}
The number and energy density of the cosmic ray electrons is then equal to:
\begin{equation}
    n_\rmn{cr} = \int f(\bs{x},{p}, t) dp,
\end{equation}
and
\begin{equation}
    \epsilon_\rmn{cr} = \int E_\rmn{kin}(p) f(\bs{x}, p, t) dp,
\end{equation}
respectively, where $E_\rmn{kin} = (\sqrt{p^2+1} - 1)m_\rmn{e} c^2$ is the kinetic energy of a particle. 

\subsection{Transport}
\label{sec:transport}
\begin{figure}
    \centering
    \includegraphics[width=0.55\linewidth]{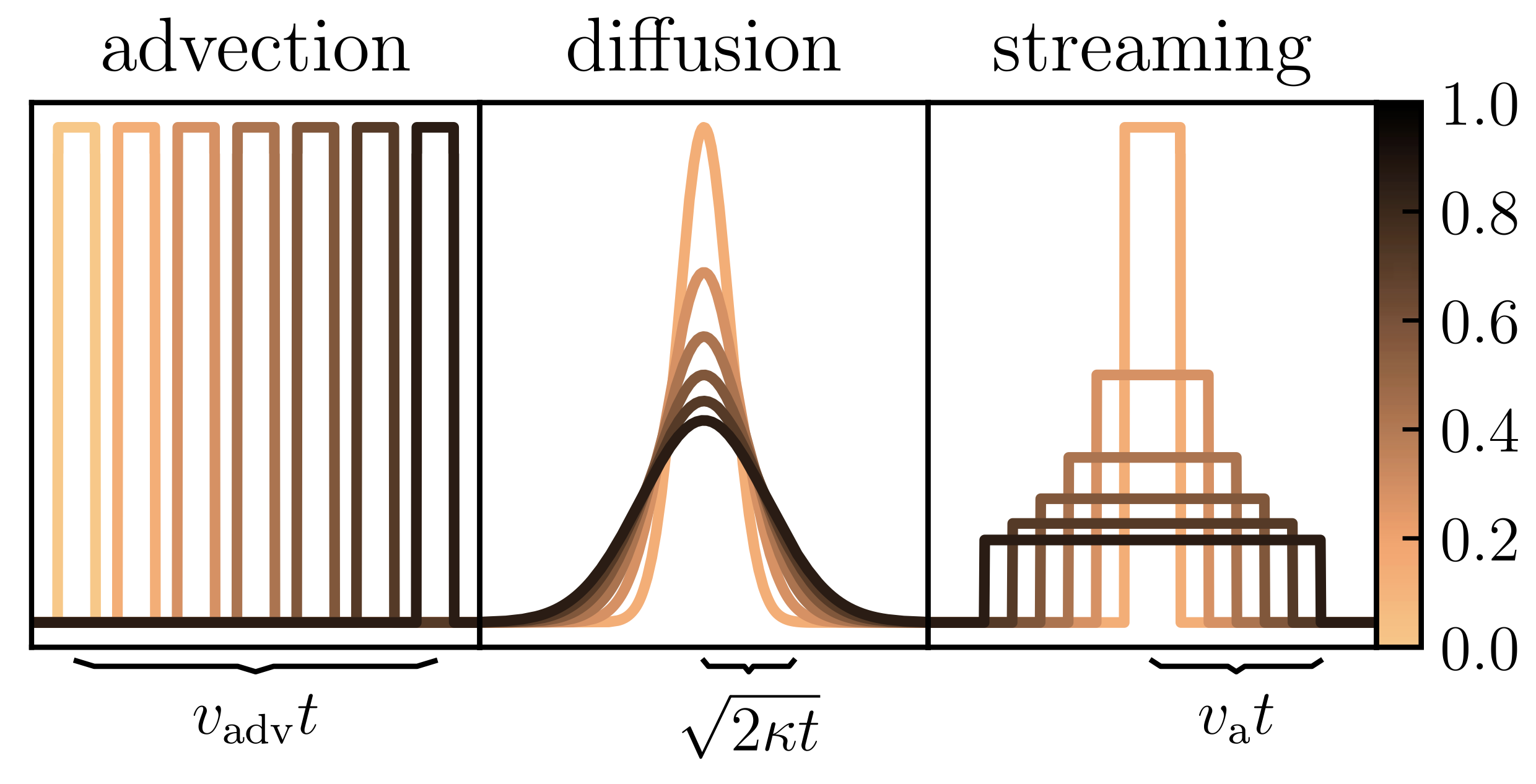}
    \caption[]{Schematic indicating how an initial distribution of cosmic rays changes as a result of advection, diffusion, and streaming, respectively. Colours indicate the same population at different times. \textit{Image credit:} \citet{thomas2020}}  
    \label{fig:cr_transport}
\end{figure}

The transport of cosmic rays follows through one of three main processes: advection, diffusion, and streaming. We describe the physics involved below, and show how these typically affect the spatial distribution in Fig.~\ref{fig:cr_transport}:

\subsubsection{Advection}

If the cosmic ray electrons are sufficiently bound to magnetic lines via the Lorentz force, they must also follow the motions of the field. In turn, as the magnetic field is essentially ``frozen in'' in most astrophysical plasmas, the field lines move with the gas. Consequently, cosmic ray electrons are advected with the gas at the fluid speed, $\bupsilon_\rmn{adv}$. 

\subsubsection{Diffusion}

The scattering of cosmic rays by Alfv\'{e}n waves causes them to diffuse. However, this diffusion is directed along magnetic field lines and hence, unlike standard thermal diffusion, is anisotropic. The one-dimensional spatial diffusion coefficient $D_{xx}$ is given in units of length$^2$ time$^{-1}$, with the general principle being expressed in Fick's second law:
\begin{equation}
    \frac{\partial N}{\partial t} = \bs{\nabla}D_{xx}\bs{\nabla} N,
\end{equation}
where $N$ is the concentration or density of cosmic rays, such that $d^2 = D_{xx} t$ expresses the typical distance, $d$, travelled by a cosmic ray in time, $t$. 

A standard value used in galaxy simulations is $3\times10^{28}$ cm$^2$ s$^{-1}$ \citep[see, e.g.][]{buck2020}, although the actual value likely varies significantly across the halo \citep{thomas2023}. In radio relics, on the other hand, the relevant diffusion coefficient is likely closer to Bohm diffusion \citep{caprioli2014}, which can be formulated as \citep{kang2016}:
\begin{equation}
    D_\rmn{Bohm}(p) = 1.7 \times 10^{19} \,\mathrm{cm}^2 \,\mathrm{s}^{-1} \left(\frac{B}{1 \upmu\mathrm{G}}\right)^{-1} p
\end{equation}

\subsubsection{Streaming}

When scatterings are sufficiently frequent, the cosmic rays distribution is isotropised in the rest frame of the Alfv\'{e}n waves. This effectively transports them at the Alfv\'{e}n speed:
\begin{equation}
\bs{\bupsilon}_\rmn{cr} = - \bs{\bupsilon}_\rmn{A} \rmn{sgn}(\bs{B} \bcdot \bs{\nabla}P_\rmn{cr}),
\end{equation}
where $P_\rmn{cr}$ is the cosmic ray pressure. This results in the adiabatic transport of cosmic ray energy.

\subsection{Acceleration processes}

In this section we will constrain ourselves to Fermi I and Fermi II acceleration; the two most important mechanisms in the production of diffusive emission in galaxy clusters. We note, however, that other mechanisms such as shock drift acceleration and magnetic reconnection are occasionally considered for these phenomena as well \citep[see, e.g.][]{kang2018, ghosh2024}. Both of the following mechanism also apply to non-thermal cosmic ray electron populations. In this case, the mechanism is referred to as \textit{re-acceleration}.

\subsubsection{Fermi I}
\label{sec:acceleration}
\label{sec:DSA}

First-order Fermi acceleration is a process that acts in collisionless, magnetised shocks\footnote{The name refers to Enrico Fermi, who proposed that particles could be accelerated by converging magnetic mirrors (see Sec.~\ref{subsec:fermi-ii}). However, this is an example of Stigler's law of eponymy, as the theory behind ``Fermi I'' was actually predominantly developed by the authors cited.} \citep{krymskii1977, axford1977, bell1978, bell1978b, blandford1978, drury1983, blandford1987}. As it can be mapped onto a diffusion process, it is also often referred to as diffusive shock acceleration (DSA). The underlying principle, is that each time electrons cross the shock front, they are scattered by the magnetic inhomogeneities, where these are generated by the shock downstream and are self-excited in the upstream. This means that the electron continually experiences an approaching magnetic mirror with speed $\Delta \bupsilon = |\bs{\bupsilon}_\rmn{u} - \bs{\bupsilon}_\mathrm{d}|$. A schematic of this scenario is shown in Fig.~\ref{fig:DSA}.

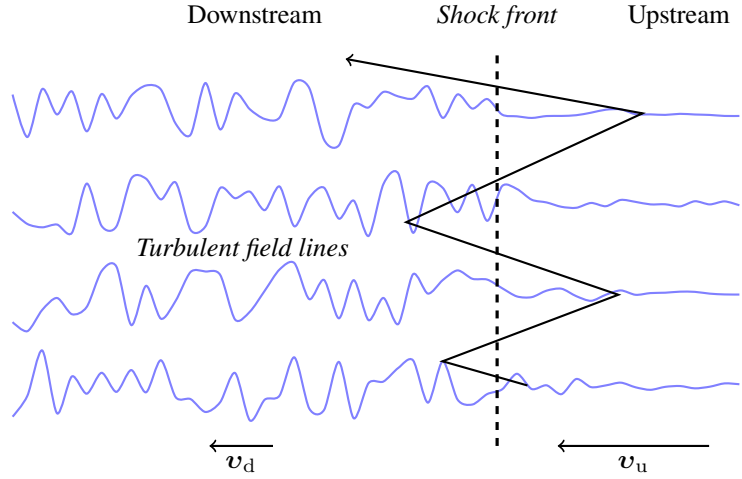
\begin{figure}
    \centering
    \begin{tikzpicture}[thick, scale=0.8]
    \pgfmathsetseed{3}
    
    \foreach \y in {0,1,2,3}
    \draw[blue!50!white,samples=50,domain=-4:8,smooth] plot 
    (-\x+2,{(atan(\x)*3.14/180.+1.57)*0.2*rand+\y*1.5});
    
    \draw[very thick, dashed] (2,-1) -- (2,5.5) node[above=0.4cm,above,anchor=base] {$\textit{Shock front}$};
    \draw (-2, 5.5) node[above=0.4cm,anchor=base] {$\text{Downstream}$};
    \draw (5, 5.5) node[above=0.4cm,above,anchor=base] {$\text{Upstream}$};
    
    \draw (-2.2, 2.25) node {$\textit{Turbulent field lines}$};
    
    \draw[->] (-1.7,-1) -- (-2.75,-1) node [midway, below] {$\bs{\bupsilon}_\mathrm{d}$};
    
    \draw[->] (5.5,-1) -- (3.,-1) node [midway, below] {$\bs{\bupsilon}_\mathrm{u}$};
    
    \draw[->] (2.5,0.0) -- +(-1.4,0.4) -- +(1.5,1.5) -- +(-2.0,2.7) -- +(1.9,4.5) -- + (-3.,5.4);
    
    \end{tikzpicture}

    \caption{Schematic representing Fermi I acceleration in the shock-frame. A cosmic ray electron diffuses across the shock front before being deflected by magnetic inhomogeneities. This increases its energy by a factor of $\mathcal{O}(\bupsilon/c)$ at each crossing. \textit{Image credit:} Adapted from work by \href{http://sprg.ssl.berkeley.edu/~pulupa/illustrations/}{M. Pulupa}}
    \label{fig:DSA}
\end{figure}
The average energy gain \textit{per crossing} is then:
\begin{equation}
    \left\langle\frac{\Delta E}{E}\right\rangle \approx \frac{2}{3}\frac{\Delta \bupsilon}{c}
\end{equation}
During each crossing and re-crossing cycle there is a finite probability that the particle will be advected downstream. Assuming a planar shock surface, it can be shown that this probability is:
\begin{equation}
     P_\rmn{esc} = \frac{4 \bupsilon_\rmn{u}}{c},
     \label{eq:p_esc}
\end{equation}
where $\bupsilon_\rmn{u}$ is the upstream speed in the shock rest-frame \citep[see, e.g.,][]{maoz2016}. The result is that if we initially inject $N_0$ particles, after $n$ cycles only $N_0 (1-P_\rmn{esc})^n$ particles remain, where these have energy $E_n$. The average energy gain \textit{per cycle} is then:
\begin{equation}
    \left\langle{\Delta E}\right\rangle \approx \frac{4}{3}\frac{\Delta \bupsilon}{c}{E} = \eta {E},
    \label{eq:dsa-gain}
\end{equation}
and for an electron to reach energy $E_n$, we require $n$ cycles so that:
\begin{equation}
    n = \ln\left(\frac{E_n}{E_0}\right) \left(\frac{1}{\ln(1+\eta)}\right)
\end{equation}
After some rearranging, this results in:
\begin{equation}
    N(\geq E) = N_0 \left(\frac{E}{E_0}\right)^\zeta,
\end{equation}
where $N(\geq E)$ is the number of particles with energy greater than $E$, and
\begin{equation}
    \zeta = \frac{\ln(1-P_\rmn{esc})}{\ln(1+\eta)}
\end{equation}
For non-relativistic shocks, by Eq.~\eqref{eq:dsa-gain}, $\eta \ll 1$, and hence $\zeta \approx -P_\rmn{esc}/{\eta}$, where we have used the fact that $\ln(1+x)\approx x$ for small $x$. 

By, recalling Eq.~\eqref{eq:p_esc} and using the jump conditions to show that $\Delta \bupsilon = \bupsilon_\rmn{u} \left(1 - \frac{1}{r}\right)$, we are left with the standard result that
\begin{equation}
    N(\geq E) = N_0 \left(\frac{E}{E_0}\right)^{-3 / (r - 1)},
\end{equation}
or, equivalently,
\begin{equation}
    N(E) \propto \left(\frac{E}{E_0}\right)^{-3 / (r - 1) \,-\,1}
\end{equation}
Hence, DSA results in a power law population with spectral slope $\alpha_\rmn{e} = - \frac{3}{r - 1} -1$, where $r$ is the compression ratio. As we showed in Sec.~\ref{sec:shocks}, in a strong shock, $r=4$, meaning that the spectral slope\footnote{If we are using a 3D distribution function, then the $4\pi  p^2$ term in Eq.~\eqref{eq:1D-dist} results in $\alpha_\rmn{e, 3D} = - 4$.} is limited to $\alpha_\rmn{e} = - 2$.

This result has generally been shown to hold in particle-in-cell (PIC) simulations, albeit with a few modifications. These include cosmic ray generated pre-cursor and post-cursor regions, which appear to limit the true maximum slope to $\alpha_\rmn{e} = -2.2$ \citep{caprioli2020}. Indeed, this is the value that is most commonly observed for strong shocks in supernovae \citep{caprioli2012}. Moreover, it has been shown that not all shocks are equal; so-called quasi-parallel shocks, where the magnetic field aligns with the shock normal, have been shown to be more efficient at accelerating cosmic ray electrons \citep{winner2020}. This is to be expected, as the cosmic ray electrons must be transported backwards and forwards across the shock front in order for DSA to be effective.  Moreover, an increasing body of evidence suggests that shocks require a minimum Mach number of $\mathcal{M}_\rmn{crit} \approx 2.3$ to result in efficient acceleration. This is believed to result from the lack of magnetic inhomogeneities produced in shocks weaker than this \citep[see, e.g.,][]{kang2019}.

There has been some discussion of how to start Fermi I acceleration for electrons, as the collisionless shock ``width'' is set by the Larmor radius of the more energetic cosmic ray protons, which is substantially bigger than the cosmic ray electron Larmor radius (see Eq.~\ref{eq:Larmor}). This would appear to prevent the cosmic ray electron from crossing the shock in the first place; an issue known as the \textit{injection problem}. Recent work, however, suggests that electrons may be injected through an intermediate-scale instability \citep{shalaby2022}.

\subsubsection{Fermi II}
\label{subsec:fermi-ii}

A related process is second-order Fermi acceleration, also known as turbulent or stochastic acceleration \citep{fermi1949}. In this case, cosmic rays are accelerated by elastic scattering off of randomly moving magnetic ``clouds''. Particles will gain or lose energy, depending on whether the collision is head-on or not, respectively. The energy gained or lost in such a collision will be: 
\begin{equation}
    \frac{\Delta E}{E} \approx 2 \,\frac{|\bs{\bupsilon}_\rmn{cl}|^2}{c^2} + \frac{2\, \bs{\bupsilon_\rmn{cr}}\bcdot\bs{\bupsilon}_\rmn{cl}}{c^2},
\end{equation}
where $\bs{\bupsilon}_\rmn{cl}$ is the cloud velocity \citep{pohl2020}. Statistically, more collisions will be head-on than not, and hence the cosmic ray gains energy. For low cloud speeds ($|\bs{\bupsilon}_\rmn{cl}|/c \ll 1$) this results in an average energy gain of:
\begin{equation}
    \left\langle \frac{\Delta E}{E} \right\rangle \approx \frac{4}{3}\left(\frac{|\bs{\bupsilon}_\rmn{cl}|}{c}\right)^2.
\end{equation}
As this change is only of second order, the resulting acceleration is relatively slow. Indeed, in galaxy and galaxy cluster environments, cooling usually acts too quickly for Fermi II acceleration to be effective from the thermal pool \citep{petrosian2001}. For non-thermal electrons, however, Fermi II is currently a strong candidate for explaining radio halos \citep{brunetti2007, fujita2015, pinzke2017}, which are associated with turbulence generated in cluster mergers. This mechanism has also been proposed to explain features in radio relics \citep{fujita2015, kang2024}, although it is currently unclear as to whether it can compete with cooling and the generally much stronger impact of Fermi I.

In the Fokker-Planck equation, the impact of Fermi II can be reduced to a diffusion equation in momentum space, where $D_{pp}$ is the diffusion tensor for this process. For isotropic diffusion, this can be reduced further to $D_{pp} = D_0 p^2$, where the details of turbulent re-acceleration are encapsulated in the constant $D_0$. As Fermi II works through stochastic collisions, $D_0$ is typically linked to measures of the levels of compressive turbulence \citep[see, e.g.][]{pinzke2017}.

\subsection{Cooling processes}
\label{sec:cooling}

Cooling of cosmic ray electrons is generally rapid and proceeds through four main processes:

\subsubsection[Coulomb cooling]{Coulomb cooling\protect\footnote{Derived from \citet{gould1975}}}

As cosmic ray electrons move through the ionised plasma, they scatter off of electrons\footnote{Scattering off of ions takes place as well, but these interactions are suppressed by the ion-to-electron mass ratio.}. When this takes place through an electrostatic potential, it is referred to as a Coulomb interaction. Such processes transfer kinetic energy to the plasma. The corresponding loss rate is:
\begin{equation}
-\frac{\mathrm{d}p_\mathrm{Coul}}{\mathrm{d}t} = \frac{3 \sigma_\mathrm{T} c n_\elec}{2 \beta_\rmn{e}^2} \Big[ \ln \left( \frac{m_\elec c^2 \beta_\rmn{e} \sqrt{\gamma - 1}}{\hbar \omega_\pl} \right)  - \left( \frac{\beta_\rmn{e}^2}{2} + \frac{1}{\gamma}\right) \ln(2) + \frac{1}{2} + \left( \frac{\gamma - 1}{4 \gamma} \right)^2 \Big],
\end{equation}
where $\sigma_\rmn{T}$ is the Thomson cross section, $\beta_\rmn{e} = \upsilon/c$ is the normalised speed, $n_\rmn{e}$ is the electron density, and $\omega_\rmn{pl}$ is the plasma frequency. The term in the brackets is the Coulomb logarithm with additional correction terms.

\subsubsection[Bremsstrahlung]{Bremsstrahlung\protect\footnote{Derived from \citet{petrosian2001}.}}

When the cosmic ray electron is decelerated or accelerated by the plasma it undergoes a process known as bremsstrahlung. The loss rate for this process is:
\begin{equation}
-\frac{\mathrm{d}p_\mathrm{brem}}{\mathrm{d}t}
= \frac{16}{3} \frac{\alpha e^4 n \gamma \chi(E)}{m_\elec^2 c^3},
\label{eq:bremsstrahlung}
\end{equation}
where $\alpha$ is the fine-structure constant, $r_0 = e^2 / m_\elec c^2$ is the classical electron radius, $n$ is the gas number density, and $\chi(E)$ is the expression given in eq. 4BN of \citet{koch1959}. In the case of thermal electrons, this process often leads to the production of X-rays.

\subsubsection[Inverse Compton]{Inverse Compton\protect\footnote{\label{rybicki-footnote}Derived from \citet{rybicki1986}.}}

When high-energy electrons collide with low-energy photons, such as those from CMB, the electron transfers energy to the photon. This increases its energy, converting it to an X-ray or a gamma ray. The loss rate for this process is:
\begin{equation}
-\frac{\mathrm{d}p_\mathrm{syn}}{\mathrm{d}t} = \frac{4 \sigma_T p^2}{3 \beta_\rmn{e} m_\elec c} U_\mathrm{ph},
\label{eq:IC}
\end{equation}
where $U_\mathrm{ph}$ is the energy density of the photon background. Note, that Eq.~\eqref{eq:IC} is, strictly speaking, only valid for $\gamma \hbar \omega \ll m_\rmn{e} c^2$, and $\sigma_T $ must be replaced with the Klein-Nishina cross section otherwise. 

\subsubsection[Synchrotron radiation]{Synchrotron radiation\protect\cref{rybicki-footnote}}

As the cosmic ray spirals around a field line, its constant acceleration results in the production of synchrotron radiation, which is generally emitted in the radio band. The loss rate associated with this process is:
\begin{equation}
-\frac{\mathrm{d}p_\mathrm{syn}}{\mathrm{d}t} = \frac{4 \sigma_T p^2}{3 \beta_\rmn{e} m_\elec c} \frac{B^2}{8\pi},
\label{eq:synchrotron}
\end{equation}
where we have assumed an isotropic distribution of pitch angles.

Assuming optically thin radiation, a power-law distribution of electrons leads to a power law in the synchrotron emission as well, with slope $\alpha_\rmn{s} = (\alpha_\rmn{e} - 1)/2$.
Furthermore, assuming continual acceleration at the same Mach number, losses from synchrotron and inverse Compton processes are expected to steepen the power-law slope to $\alpha_e = \alpha_{e,0} - 1$, where $\alpha_{e,0}$ is the initially injected slope for the electrons. For radio spectra, this equates to $\alpha_s = \alpha_{s,0} - 1/2$, where $\alpha_{s,0}$ is the initially injected slope for the synchrotron emission\footnote{Additionally useful formulae regarding synchrotron emission are given in Sec.~\ref{chapter5-subsec:crayon}.} \citep{pacholczyk1970}. Indeed, this relationship is commonly used in reverse to estimate the Mach number in radio-emitting shocks.

Note, that Eq.~\eqref{eq:synchrotron} has an identical form to Eq.~\eqref{eq:IC}. Indeed, synchrotron radiation can be considered as scattering off of virtual photons from the background magnetic field. Alternatively, the photon field can be assigned an equivalent magnetic field strength. For the CMB, this is $B_\rmn{CMB} = \sqrt{8 \pi \varepsilon_\rmn{CMB,0}}(1+z)^2 \approx 3.24 \upmu G (1+z)^2$, where $\varepsilon_\rmn{CMB,0}$ is the energy density of the CMB today, and the $(1+z)^2$ term deals with the scaling of radiation energy density (see Sec.~\ref{sec:basics}). 

We show typical cooling losses at the peripheries of the ICM resulting from Coulomb and inverse Compton processes in Fig.~\ref{fig:relic_cooling_timescales}. Note, that Coulomb cooling is more effective at lower momenta, whilst inverse Compton cooling is more effective at higher momenta. Synchrotron cooling is also more effective at higher momenta, but generally has a sub-dominant contribution at distances far from the cluster centre, due to the low magnetic field strengths here. 

\begin{figure}
    \centering
    \includegraphics[width=0.6\linewidth]{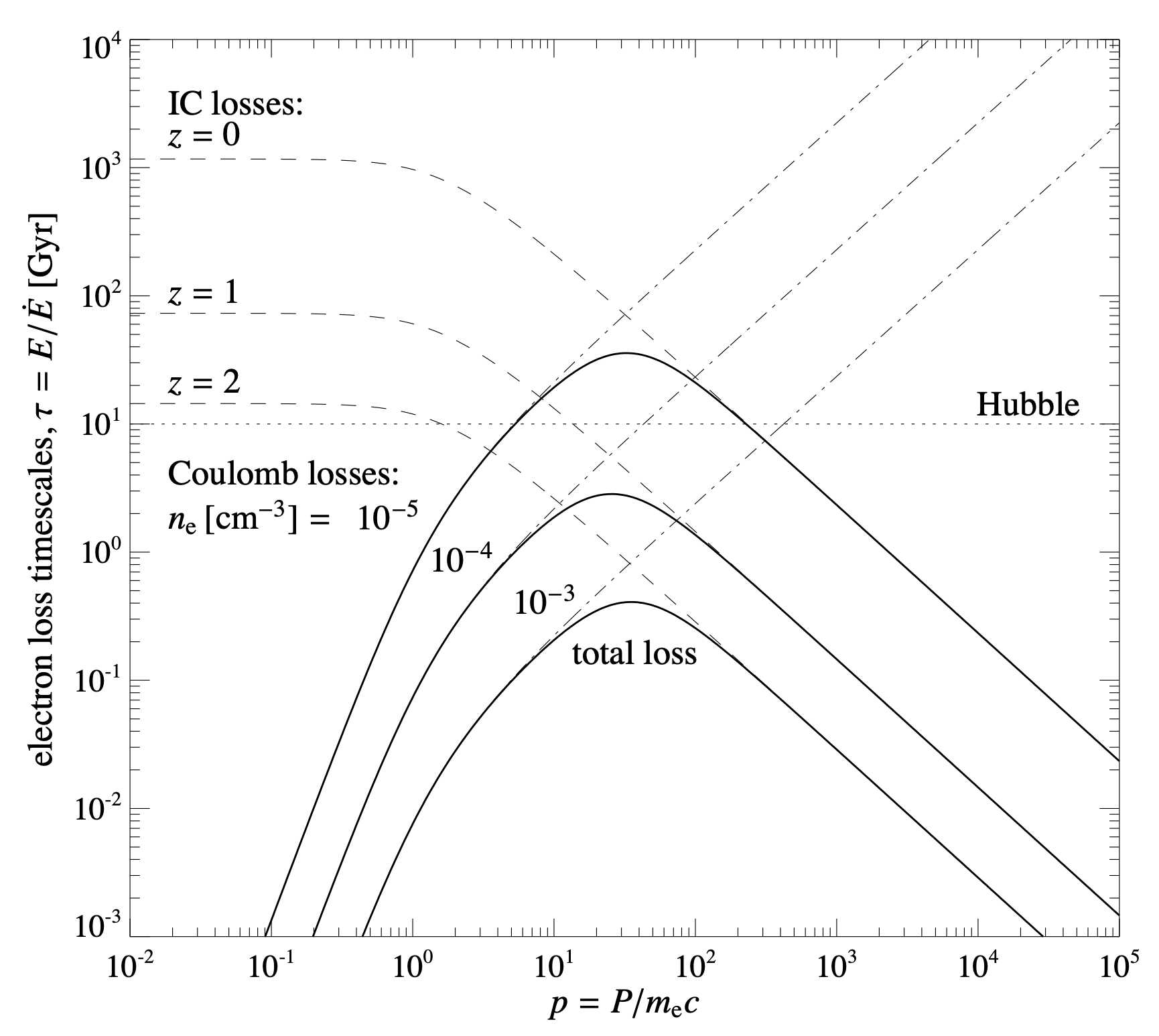}
    \caption[]{Coulomb and inverse Compton losses for varying momenta, redshift, and electron number densities. For typical ICM values, there exists a population of electrons where cooling rates are comparable to the Hubble time. These are good candidates for the so-called ``fossil'' electrons, which help boost radio relic surface brightnesses to observed values through Fermi I re-acceleration. \textit{Image credit:} \citet{pinzke2013}}
    \label{fig:relic_cooling_timescales}
\end{figure}

\section{Radio relics}

\label{sec:radio-relics}

Radio emission is common in collapsed structures, and is produced anywhere where  electrons are excited to highly relativistic energies (here: Lorentz factors in excess of $10^3$). For example, it is generated at supernovae-driven shocks, and hence the radio luminosity in galaxies is intrinsically linked to the star formation rate \citep{deJong1985, condon1992, bell2003}. Radio emission is also generated in shocks driven by AGN jets \citep[see, e.g.,][and references therein]{hardcastle2020}, and can be found filling the volume of merging clusters; a phenomenon known as a \textit{radio halo}. Most importantly for this dissertation, such mergers are also believed to produce \textit{radio relics}. 

The term ``radio relic'' has been applied to a broad class of objects. In particular, several attempts have been made to subdivide the term depending on the underlying physical scenario, with perhaps the most well-known classification attempt being given by \citet{kempner2004}. These authors provide the following three categories:
\begin{enumerate}[i)]
    \item \textbf{AGN relics:} also known as a \textit{radio ghost}, this is when an AGN lobe is still emitting, but the source AGN is no longer active. These sources have steep spectra and can be highly filamentary if they have interacted with the turbulent ICM.
    \item \textbf{Radio phoenix:} when a merger or accretion shock passes through an AGN relic, it will compress it, increasing its energy and the energy of the trapped magnetic fields such that the cosmic ray electrons emit in the radio band once more \citep{ensslin2001, ensslin2002}.
    \item \textbf{Radio gischt:} also known as a radio shock, in this mechanism, the merger shock accelerates the cosmic ray population through the Fermi I mechanism (see Sec.~\ref{sec:acceleration}). The emission therefore directly traces the shock itself \citep{ensslin1998}.
\end{enumerate}

Increasingly, the term ``radio relic'' is used to refer to the last category. Emission from this category is relatively steep (integrated $\alpha_\rmn{s} \approx -1.2$) and is often highly polarised \citep{vanweeren2010, digennaro2021}, as would be expected from synchrotron radiation from a source with a coherent magnetic field (see Sec.~\ref{sec:magnetic-fields}). This coherence is expected to be produced by the shock compression of the magnetic field \citep[see, e.g.,][]{wittor2019, dominguez-fernandez2021b}. The typical surface brightness of radio relics is on the order of $0.1$--$1$ $\upmu$Jy arcsec$^{-2}$ \citep{vanweeren2019}. However, X-ray observations show that the shocks underlying the emission are typically weak (with sonic Mach numbers $\mathcal{M} \lesssim 3$--$5$). Such shocks have a low acceleration efficiency \citep{kang2005, kang2013, mou2023}, making them incompatible with a scenario where electrons are accelerated from the thermal pool. Instead, it is generally believed that shocks re-accelerate an already semi-relativistic population of ``fossil'' electrons, which were accelerated at earlier epochs \citep{pinzke2013}. These have long cooling times at dimensionless momenta of $p\approx10$--$100$, as is shown in Fig.~\ref{fig:relic_cooling_timescales}.

\begin{figure}
    \begin{centering}
    \begin{minipage}[t][][t]{.48\textwidth}    
        \includegraphics[width=1.05\columnwidth]{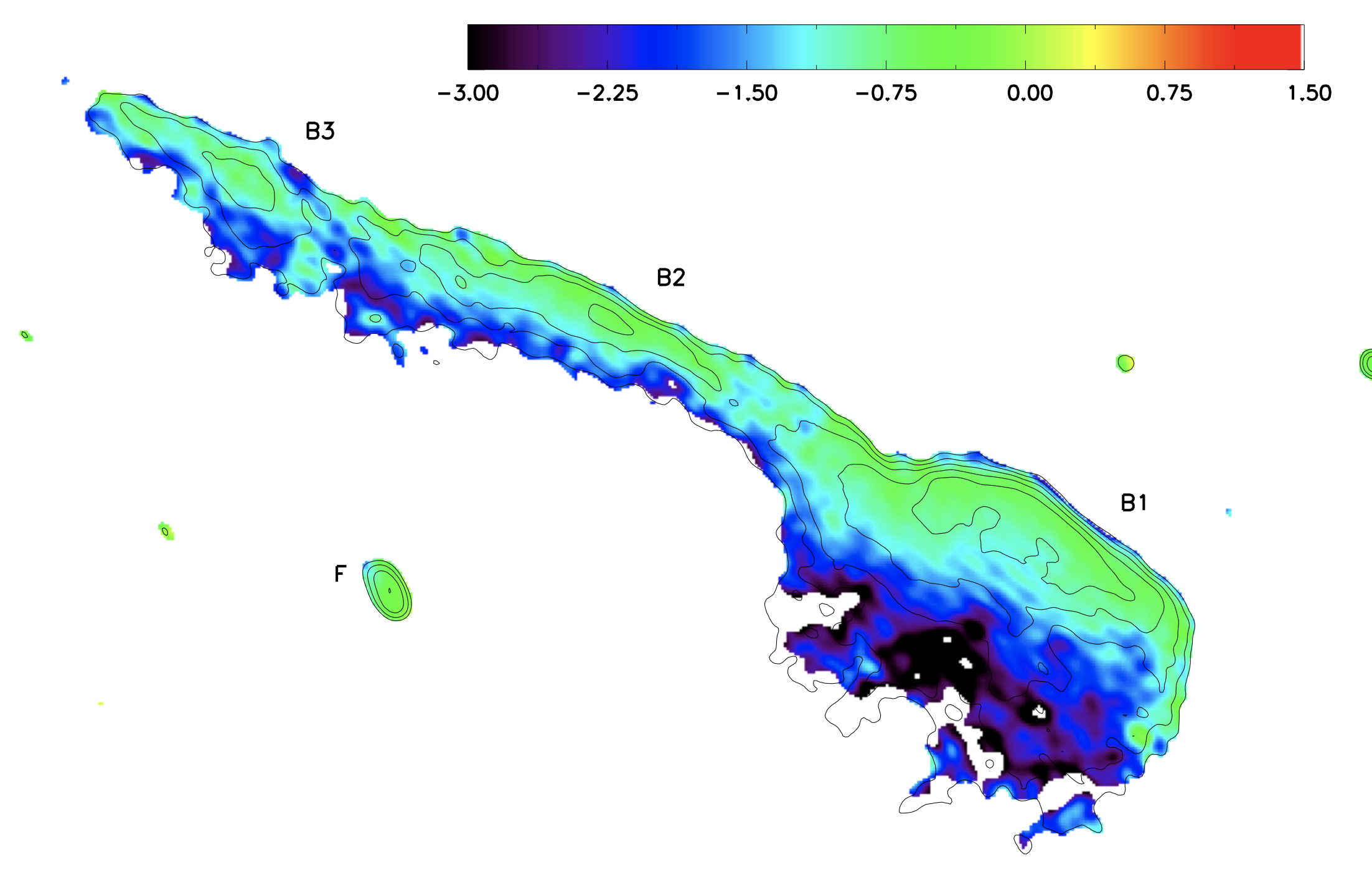}
    \end{minipage}
    \hfill
    \begin{minipage}[t][][t]{.48\textwidth}
        \centering
        \includegraphics[width=1.05\columnwidth]{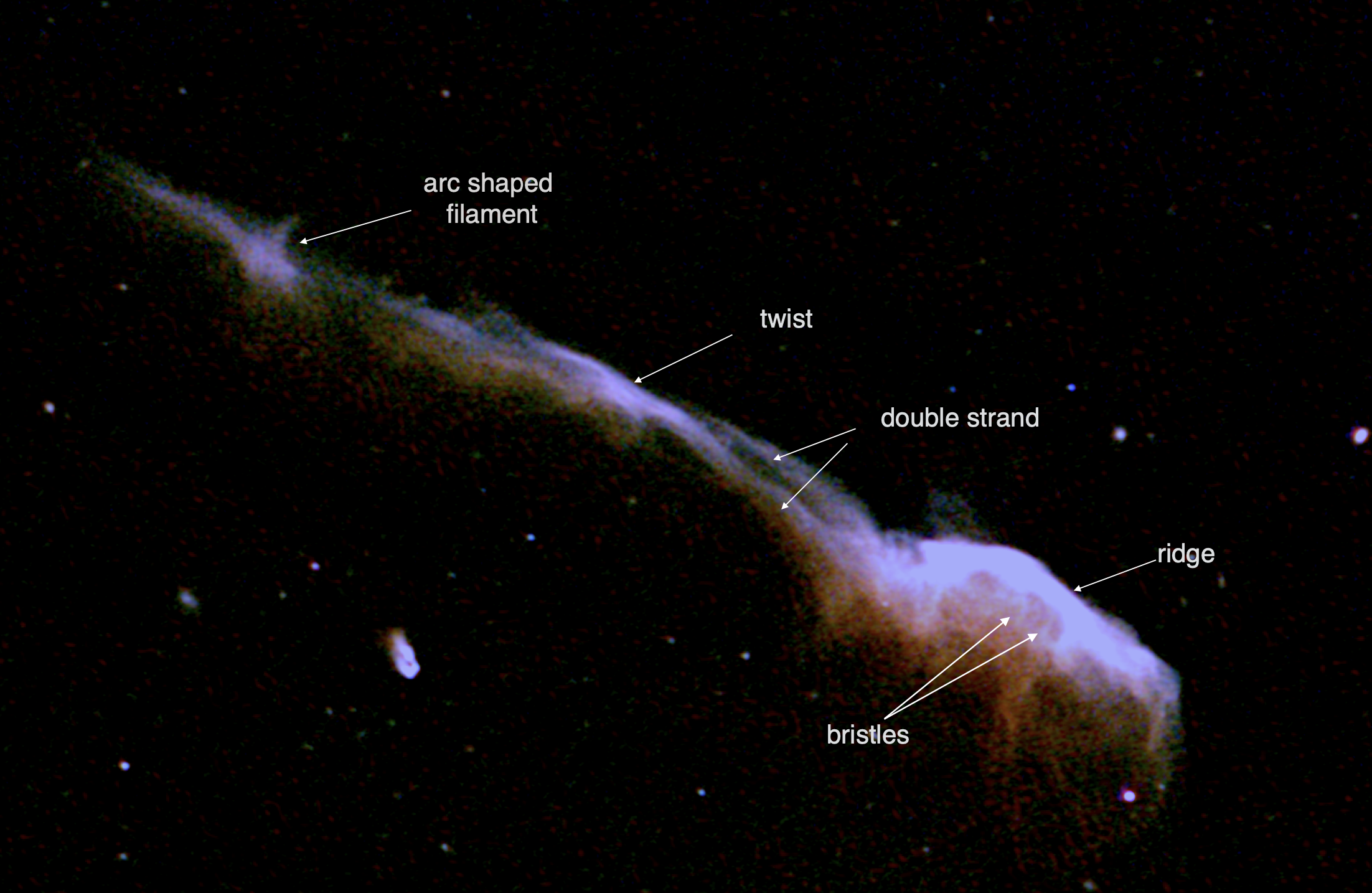}
    \end{minipage}
    \end{centering}
    \caption[]{\textit{Left:} Spectral index map for the ``Toothbrush'' radio relic, where colours give the $\alpha_\rmn{s}$ values. \textit{Right:} Intensity map for the same object. The emission is approximately 1.9 Mpc in length, and shows an array of morphological features. \textit{Image credits:} Left: \citet{vanweeren2012}. Right: \citet{rajpurohit2020}.}
    \label{fig:radio-relic}
\end{figure}

Radio relics are generally found at the edge of intracluster medium, at distances of around $1-2$ Mpc from the cluster centre. Furthermore, they exhibit varying morphologies. For example, the archetypal ``Sausage'' relic is curved \citep{kocevski2007, vanweeren2010}, whilst the ``Toothbrush'' relic is more linear in shape \citep{vanweeren2012b}. In general, relics are elongated in one direction, and are therefore often measured by their largest linear size (LSS) -- i.e.\ the largest dimension on the plane of the sky. This is typically on the order of a Mpc, but variation is strong. For example, the largest yet recorded radio relic has an LSS of approximately 3.5 Mpc \citep{hoang2021}. Radio relics can also take more irregular forms, such as Abell 2256 \citep{vanweeren2012b}, or forms where the spectral gradient is inverted, such that the fresher spectra is closer to the cluster centre \citep{riseley2022}. Moreover, with increasing resolution, relics increasingly appear filamentary and display a wide range of smaller morphological features. Examples of these features can be seen in Fig.~\ref{fig:radio-relic}. In particular, we point out the ``bristle'' and ``strand'' features, which will be revisited in Chapters~\ref{chapter:paper-three} and~\ref{chapter:paper-four}, respectively.

Whilst some scenarios have been proposed that explain specific features \citep[e.g.][]{boess2023}, what generally sets relic morphology is currently an unsolved problem. Indeed, there are several unsolved problems currently challenging our understanding of radio relics. We give a full list of the most major problems in Chapter~\ref{chapter:paper-three}.
\setcounter{equation}{0}
\thispagestyle{empty}

\graphicspath{{Images/Chapter3/}}

\Chapter{The impact of magnetic fields on cosmological galaxy mergers -- I.}{Reshaping gas and stellar discs}
\addcontentslinex{lot}{chapter}{\large Chapter \thechapter: \textit{The impact of magnetic fields on cosmological galaxy mergers -- I.}}
\label{chapter:paper-one}

\noindent This chapter is based on the published paper by Whittingham, J., Sparre, M., Pfrommer, C., and Pakmor, R. in Monthly Notices of the Royal Astronomical Society, Volume 506, Issue 1, p.229-255. The work expands upon ideas first presented in my Masters thesis. A breakdown of the exact differences is provided in Chapter \ref{publications_list}.

\textit{
Mergers play an important role in galaxy evolution. In particular, major mergers are able to have a transformative effect on galaxy morphology. In this paper, we investigate the role of magnetic fields in gas-rich major mergers. To this end, we run a series of high-resolution magnetohydrodynamic (MHD) zoom-in simulations with the moving-mesh code \textsc{arepo} and compare the outcome with hydrodynamic simulations run from the same initial conditions. This is the first time that the effect of magnetic fields in major mergers has been investigated in a cosmologically-consistent manner. In contrast to previous non-cosmological simulations, we find that the inclusion of magnetic fields has a substantial impact on the production of the merger remnant. Whilst magnetic fields do not strongly affect global properties, such as the star formation history, they are able to significantly influence structural properties. Indeed, MHD simulations consistently form remnants with extended discs and well-developed spiral structure, whilst hydrodynamic simulations form more compact remnants that display distinctive ring morphology. We support this work with a resolution study and show that whilst global properties are broadly converged across resolution and physics models, morphological differences only develop given sufficient resolution. We argue that this is due to the more efficient excitement of a small-scale dynamo in higher resolution simulations, resulting in a more strongly amplified field that is better able to influence gas dynamics.}

\section{Introduction} \label{chapter3-sec:introduction}

Radio synchrotron observations, amongst other evidence, show that spiral galaxies in the local Universe are permeated by magnetic fields \citep{beck1985}. On the galactic scale these fields are remarkably ordered and have typical strengths of a few $\upmu$G at solar radii \citep{beck2011}, rising to $50-100\;\upmu$G in nuclear starburst regions \citep{heesen2011, adebahr2013}. At these strengths, the magnetic field contributes significantly to the total pressure in the interstellar medium (ISM). Indeed, it is generally believed that the magnetic energy density in the ISM is in rough equipartition with the thermal gas, turbulent, and cosmic ray energy densities \citep{beck1996, beck2015}. Magnetic fields are therefore expected to be dynamically important for late-type galaxies today, helping to balance the disc against gravitation and directly influencing the flow of gas. On top of this, galactic magnetic fields are able to have a significant indirect impact; they can conduct collisionless particle species, such as cosmic rays, along their flux tubes \citep{Zweibel2017,thomas2020} thereby mediating the direction of momentum and energy transfer from cosmic rays to the thermal gas. This transfer enables dynamical feedback that is able to drive galactic winds, which affects galaxy formation and evolution \citep{Hanasz2013,Pakmor2016II,Ruszkowski2017,Jacob2018}.

Whilst magnetic fields may be influential at the present epoch, their direct \textit{long-term} effect on galaxy evolution is, however, still debated. It has been argued that a sufficiently strong primordial field could lead to reduced disc sizes \citep{martin-alvarez2020} and suppressed star formation rates \citep{marinacci2016}. The seed field strength required to produce these effects, though, is several magnitudes higher than that plausibly generated by battery processes in ionisation fronts or cosmological shocks \citep{kulsrad2008}. In cosmological simulations that used weaker initial seed fields, the galactic magnetic field was not able to have a significant evolutionary impact as it was amplified in a dynamo close to equipartition strengths, at which point it is quenched by magnetic tension and thus does not become strong enough to have a significant dynamical backreaction \citep{pakmor2017}. Moreover, in some simulations the field was unable to reach equipartition at all \citep{hopkins2020}, preventing it from influencing galactic development except indirectly via its impact on anisotropic transport processes. The investigations published thus far have, however, focused almost entirely on galaxies with quiescent merger histories.

In a $\Lambda$CDM cosmology -- i.e. cold dark matter with a cosmological constant, $\Lambda$ -- structure forms hierarchically and consequently few if any galaxies will remain completely untouched by interactions during their lifetime. This is especially true of more massive galaxies (with stellar mass M$_\star \gtrsim 5 \times 10^{10} \; $M$_\odot$), where mergers are suggested to be the main drivers of growth at redshifts $z\lesssim1$ \citep{bell2006, tacchella2019}. Such mergers give rise to a rapidly varying gravitational potential, which can have dramatic consequences for the galactic components and their kinematics. Being collisional, the gas component is particularly sensitive to such interactions. Indeed, it has long been recognized that mergers can draw fresh gas deep into the galaxy \citep{toomre1972, barnes1996, moreno2021}, diluting the metallicity of the existing gas \citep{scudder2012, torrey2012, bustamante2018, thorp2019, bustamante2020}, triggering starbursts \citep{farrah2003, cox2008, teyssier2010, hayward2014, luo2014} and increasing the black hole accretion rate \citep{springel2005a, sijacki2007, gabor2016}.  If the resultant feedback from these events is too strong, the gas may then be expelled from the galaxy almost entirely, quenching further star formation and transforming the galaxy into a so-called \textit{red and dead} galaxy -- an evolutionary track codified in \citet{hopkins2008}. This evolution is not necessarily pre-determined though; in fact, an increasing body of evidence shows that gas-rich mergers can instead support the growth of a stellar disc post-merger \citep{sparre2017, rodriguez-gomez2017, hani2020}. In either case, it is clear that the gas dynamics play a crucial role in the outcome.

Whilst the impact of mergers on the gas component (and vice versa) has been well-appreciated, the potential role that magnetic fields play is often neglected. However, these elements are not easily separated. Outside the densest parts of molecular clouds, the gas in galaxies is sufficiently ionised such that even the neutral component is intimately coupled to the magnetic field \citep{ferriere2001}. This has clear consequences during a merger; as gas is sheared and compressed, so too will field lines be brought closer together. Similarly, the injection of turbulence during a tidal interaction should support the development of a small-scale dynamo \citep{arshakian2009}. These processes will act to amplify the field, and help to explain why interacting galaxies show lower field regularity (the ratio of regular to random field components) and higher field strengths than non-interacting galaxies \citep{drzazga2011}. Given sufficient and rapid enough amplification, it is possible that the galactic magnetic field could reach equipartition or even become locally dominant during the merger. This would have significant ramifications for the gas dynamics and subsequent star formation.

Such a scenario has not yet been rigorously tested. Indeed, merger simulations have traditionally been performed using pure hydrodynamics only. This owes both to the well-known technical difficulties in discretising the equations of MHD whilst sufficiently maintaining the $\bs{\nabla} \bs{\cdot} \bs{B} = 0$ constraint in dynamic environments, and to the already considerable computational cost involved in sufficiently resolving the galactic interior. To increase resolution, the few MHD simulations that have focused on mergers have been run exclusively with idealised set-ups \citep{kotarba2010, kotarba2011, geng2012, moss2014, rodenbeck2016}. Whilst these can be helpful to probe the physics involved, such simulations necessarily require the somewhat arbitrary choice of a range of free parameters. This can result in cosmologically-inconsistent mass infall, tidal fields, and orbital parameters for the participating galaxies. This is problematic as such parameters have been shown to play a pivotal role in the production of the merger remnant \citep[e.g.][]{naab2003}.

The choice of free parameters may be reduced by running fully-cosmological simulations. However, to be cosmologically-consistent, the galactic environment must be resolved for several tens of Mpc. In contrast, resolving the turbulence in the ISM that drives the small-scale dynamo is expected to require close to parsec resolution \citep{dubois2010, renaud2014}. Clearly, resolving these scales at the same resolution is computationally infeasible. In this work, we attempt to reconcile the differences between these scales by running cosmological MHD `zoom-in' simulations, which focus their computational power on the object of interest and resolve the surrounding environment more coarsely. Such simulations are able to include large-scale cosmological effects, whilst resolving baryonic processes below the kpc scale. Several MHD zoom-in simulations of this kind have now been run \cite[e.g.][]{rieder2017, pakmor2017, hopkins2020, libeskind2020}. However, as already stated, until now these have all focused on relatively isolated galaxies, with any analysis on mergers being purely incidental \citep[e.g.][]{pakmor2014, martin-alvarez2018}. In this paper, we rectify this by presenting a series of high-resolution cosmological MHD zoom-in simulations of major mergers. By comparing the outcome of these simulations to hydrodynamic simulations run from the same initial conditions, we evaluate the impact of magnetic fields on galaxy mergers in a cosmologically-consistent manner for the first time.

The paper is organised as follows: in Section~\ref{chapter3-sec:methodology} we describe our methodology and present the simulations. In Section~\ref{chapter3-sec:analysis}, we present our analysis of the simulations, including: the impact of MHD on global properties (Section~\ref{chapter3-subsec:MHD-global-properties}), the impact of MHD on structural properties (Section~\ref{chapter3-subsec:MHD-disc-sizes}), and the impact of resolution on our results (Section~\ref{chapter3-subsec:resolution-study}). We support the work in this section with comparisons to simulations of galaxies with more quiescent merger histories, showing that the observed differences are particularly evident in the merger scenarios. In Section~\ref{chapter3-sec:discussion}, we suggest reasons why magnetic fields have been ineffectual in previous simulation work, discuss our results in the context of galaxy evolution as a whole, and discuss the main caveats to our results. Finally, in Section~\ref{chapter3-sec:conclusions}, we summarise our main conclusions.

\section{Methodology}
\label{chapter3-sec:methodology}

In order to observe the magnetic fields at their most effective, we have simulated a series of gas-rich major mergers. In particular, we have re-run the merger simulations first presented in \citet[]{sparre2016} with the inclusion of ideal MHD physics. These mergers were considered to be an ideal starting point as: 1) the progenitors had large, gas-rich discs, implying MHD analogues that would have strong, well-ordered magnetic fields; 2) at coalescence the gas densities were shown to reach high values, implying correspondingly strong amplification; and 3) the merger remnants in these simulations were able to rebuild their discs. The rebuilding process is naturally dependent on gas dynamics, allowing for further influence from the magnetic fields.

The set-up for these merger simulations is discussed in the following section. We discuss the set-up for the simulations of galaxies with more quiescent merger histories in Section~\ref{chapter3-subsec:methods_isolated_galaxies}. In each case, the hydrodynamic and MHD simulations use the same underlying numerical implementation, such that a hydrodynamic simulation is equivalent to an MHD simulation with seed field set to zero.

\subsection{Initial conditions and parameters}

The initial conditions for the merger simulations are the same as those presented in \citet{sparre2016, sparre2017}. These were created by selecting four galaxies from the hydrodynamic cosmological simulation Illustris \citep{vogelsberger2014, vogelsberger2014b, genel2014} that had undergone a major merger between $z=1$ and $z=0.5$, were relaxed at $z=0$, and had a final stellar and halo mass close to that of the Milky Way. Zoom initial conditions were then created with a modified version of the \textsc{N-GenIC} code \citep{springel2015} for a periodic box of side 75 co-moving Mpc~$h^{-1}$, using the WMAP-9 \citep{hinshaw2013} cosmological parameters; i.e. the density parameters for matter, baryons and a cosmological constant: $\Omega_\text{m} = 0.2726$, $\Omega_\text{b} = 0.0456$, $\Omega_\Lambda = 0.7274$, respectively, and Hubble's constant $H_0 = 100$ $h$ km s$^{-1}$ Mpc$^{-1}$ with $h = 0.704$.

Dark matter particles were given a high-resolution within a roughly spherical region around the target galaxy, with a shell of standard resolution particles following, and lower-resolution particles filling the remaining volume. The highest dark matter mass resolution in the simulation was set to:

\begin{equation}
    m_{\text{DM}} = \left( \frac{1820}{2048 \times \text{`zoom factor'}}\right)^3 \times 6.299 \times 10^6 \; \text{M}_\odot,
    \label{eq:zoomfactor}
\end{equation}

where 1820/2048 is the ratio between the number of dark matter particles per box length in Illustris relative to our standard resolution\footnote{Standard resolution is therefore equivalent to setting `zoom factor' = 1.}, and $6.299 \times 10^6 \; \text{M}_\odot$ is the finest dark matter mass resolution in Illustris. Our simulations were run with zoom factors equal to 1, 2, and 3, which corresponds to dark matter mass resolutions that are 1.4, 11.4, and 38.5 times finer than in the original Illustris run. Equivalently, simulations with a zoom factor of 3 have $\sim1.8$ times finer mass resolution than the fiducial `Level 4' Auriga simulations \citep{grand2017}.

Following \citet{springel2005b} and \citet{price2007}, the softening length, $\epsilon_\text{DM}$, was chosen to be $\sim$$1/40$ of the initial average particle spacing:

\begin{equation}
    \epsilon_\text{DM} \approx \frac{L}{40 \times 2048 \times \text{`zoom factor'}},
    \label{eq:softening}
\end{equation}
where $L$ is the box length. The softening length is a co-moving length until $z=1$, at which point it is frozen in physical units, thereby maintaining the same resolution in the simulation for $z<1$. This helps to prevent unrealistic two-body interactions at early times, whilst still allowing small-scale structure to continue to form at late-times \citep[see, e.g.][]{power2003}. As gas cells vary strongly in density, their softening length is also scaled by the mean radius of the cell. Such cells have a minimum co-moving softening length of 30~$h^{-1}$~pc (frozen at $z=1$, as before) and a maximum physical softening length of 1.1 kpc. A full list of mass resolutions and softening lengths for each zoom factor is seen in Table~\ref{tab:sim_setup}.

\begin{table}
    \centering
    \caption[Simulation resolution and softening length]{Zoom factor, finest dark matter mass resolution, finest baryon mass resolution, and softening length at $z=0$ for our highest, intermediate, and lowest-resolution runs, respectively.}
    \label{tab:sim_setup}
    \begin{tabular}{cccc}
        \hline \T
        Zoom factor & $m_{\text{DM}}$ [$\mathrm{M}_\odot$] & $m_{\text{b}}$ [$\mathrm{M}_\odot$] & $\epsilon_\text{DM}$ [kpc] \B \\
        \hline \T
        3 & $1.64 \times 10^5$ & $2.74 \times 10^4$ & 0.22 \\
        2 & $5.53 \times 10^5$ & $9.24 \times 10^4$  & 0.32 \\
        1 & $4.42 \times 10^6$ & $7.39 \times 10^5$  & 0.65 \B \\
        \hline
    \end{tabular}
\end{table}

\subsection{\textsc{Arepo}, Auriga, and MHD implementation} 
\label{chapter3-subsec:set-up}

 The simulations were evolved from $z=127$ using the Auriga galaxy formation model \citep{grand2017} and the moving-mesh code, \textsc{arepo} \citep{springel2010, Pakmor2016I, weinberger2019}. \textsc{Arepo} uses a set of mesh-generating points to define a Voronoi tessellation, on which a second-order accurate, finite-volume Godunov scheme is formulated. Mesh-generating points may be moved arbitrarily and cells can be refined and de-refined such that they maintain a target mass resolution. In this manner, cells follow the flow of mass and thereby inherit the advantages of both Lagrangian and grid-based Eulerian codes. It has been shown that \textsc{arepo} is considerably more accurate than standard smoothed-particle hydrodynamic (SPH) methods when applied to a range of computational fluid dynamic problems \citep{sijacki2012} and produces the expected Kolmogorov turbulent cascade \citep{kolmogorov1941} for subsonic turbulence, unlike standard SPH models \citep{bauer2012}. The power spectrum of turbulence has a significant impact on its ability to excite the small-scale dynamo, making this result especially germane to our investigation.

The Auriga galaxy formation model is closely based on the models of \citet{vogelsberger2013} and \citet{marinacci2014} but contains important changes with respect to stellar feedback and the inclusion of the \citet{pakmor2013} MHD implementation (both summarised in the following text). The Auriga model has been shown to be able to produce Milky Way (MW)-like galaxies with appropriate stellar masses, sizes, rotation curves, star formation rates, and metallicities \citep{grand2017}, as well as finer details such as the correct structural parameters of bars \citep{calero2019} and the existence of chemically distinct thick and thin discs \citep{grand2018}. The models for star formation, stellar feedback, and active galactic nuclei (AGN) feedback are all physically well-motivated and parameters do not require retuning between resolution levels. Earlier work has shown that this is a non-trivial result \citep{scannapieco2012}. We summarise the main features of the model here, but encourage the reader to refer to \citet{grand2017} and references therein for a more comprehensive picture.

Gas may cool in Auriga via both atomic and metal-line cooling with self-shielding corrections accounted for \citep{vogelsberger2013}. A spatially uniform UV background field is included, which fully reionises hydrogen by $z \sim 6$ \citep{faucher2009}. The ISM is described using the \citet{springel2003} model, which treats star-forming gas as a two-phase medium governed by an effective equation of state. This model is derived from the assumption that, at the onset of thermal instability, processes below the resolution limit quickly lead to a pressure equilibrium forming between the hot and cold gas phases. It is not necessary to recalibrate this model when including magnetic fields. We show this explicitly in Appendix~\ref{appendix:ISM}.

Star particles are formed stochastically above a threshold density of $n_\text{SF}= 0.13\; \text{cm}^{-3}$ with a probability that scales with the local dynamical time. Each star particle represents a single stellar population, characterised by an age and metallicity, and assuming a \citet{chabrier2003} initial mass function. Stellar evolution is treated self-consistently, with mass loss and metal yields from supernovae SNII, SNIa, and asymptotic giant branch stars calculated at each time step and distributed to nearby gas cells using a top-hat kernel. The number of SNII events is set according to the number of stars formed in the mass range $8-100\;\text{M}_\odot$. This event is modelled by converting a star-forming gas cell into a wind particle and launching it in a randomly-chosen direction with a velocity proportional to the local one-dimensional dark matter velocity dispersion \citep{okamoto2010}. Wind particles interact only gravitationally until they reach a gas cell with $n < 0.05$ $n_\text{SF}$ or exceed the maximum travel time. The particle's energy is then deposited in the gas cell with the energy being split into equal parts thermal and kinetic. This results in smooth, regular winds, which become mostly bipolar at late times, as the wind takes the path of least resistance away from the galaxy. This is opposed to the bipolar wind model of \cite{marinacci2014}, in which wind particles are explicitly assigned an initial direction pointing away from the disc \citep[see, e.g.][for further details]{pillepich2018}.

Black holes are seeded with a mass of $10^5$ M$_\odot$ $h^{-1}$ in friends-of-friends (FoF) groups \citep{davis1985} with masses greater than $5 \times 10^{10}$ M$_\odot$ $h^{-1}$ at the position of the most dense gas cell. Black hole dynamics are governed by the \citet{springel2005a} model, with accretion described by an Eddington-limited Bondi-Hoyle-Lyttleton model and an additional term modelling radio accretion based on \citet{nulsen2000}. Feedback takes place through both radio and quasar modes, with thermal energy injected isotropically into neighbouring gas cells for the quasar mode, and bubbles of gas being gently heated at locations within the halo for the radio mode. The number of black hole neighbours are doubled with each increase in zoom factor in the standard way as a compromise between maintaining the total volume of neighbours and the increasing computational expense. In both quasar and radio mode, energy is injected continuously at a rate proportional to the accretion rate.

Magnetic fields are treated in the ideal MHD approximation \citep{pakmor2011, pakmor2013}, with the divergence constraint maintained through the use of a Powell 8-wave scheme \citep{powell1999}. In theory, the divergence constraint could also be preserved at machine precision using constrained transport schemes \citep{evans1988}, such as that implemented for \textsc{arepo} in \cite{mocz2014}. In practise, however, the Powell implementation performs sufficiently well, such that it is able to accurately replicate a series of MHD phenomena. These include: the linear phase of growth of the magneto-rotational instability \citep{balbus1991, pakmor2013}; the development of a small-scale dynamo in MW-like galaxies \citep{pakmor2014, pakmor2017}; similar field strengths and radial profiles to those observed in MW-like galaxies \citep{pakmor2017}; and Faraday rotation measure strengths that are broadly consistent to those observed for MW-like galaxies, both for the disc \citep{pakmor2018} and when compared with the current upper limits available for the circumgalactic medium \citep{pakmor2020}.

For each MHD simulation, we seed a homogeneous field of $10^{-14}$ co-moving Gauss, orientated along the $z$-direction, throughout the volume at $z=127$. This choice is essentially arbitrary as, for a broad range of values, all traces of the initial field strength and configuration are erased by an exponential dynamo in collapsed haloes \citep{pakmor2014}. Our choice of initial field strength has also been shown to produce magnetic fields that are dynamically irrelevant outside of collapsed haloes \citep{marinacci2016}. We note that during a star- or wind-forming event, the magnetic energy of the associated gas cell is removed, and is assumed to be locked-up in the subsequently formed stellar macro particle. Excluding this, magnetic fields are not explicitly included in our subgrid models.

\subsection{Galaxy tracking}
\label{chapter3-subsec:galaxy_tracking}

In this work, we define haloes through the standard FoF approach and galaxies, or equivalently `subhaloes', using the \textsc{subfind} algorithm \citep{springel2001}. The distinction is useful as whilst FoF haloes may form tenuous bridges during galaxy interactions, causing them to be identified as a single structure, subhaloes are characterised by `self-boundness', meaning that structures remain essentially distinct until coalescence. The use of subhaloes hence allows us to consider the evolution of an individual galaxy until a very advanced stage of the merger. 

The primary and secondary progenitors of a merger are identified as the first and second most massive galaxies at $z=0.93$, as during this period both galaxies are relatively isolated (cf.\ \citealt{sparre2016}). In order to track the galaxies between snapshots, it is a case of identifying the galaxy that shows the most consistent trajectory to a previously identified one. In practise, this is the same as identifying the galaxy that contains the same black hole particle. Such a result is expected, as earlier work has shown that reliable merger trees can be constructed by tracking only the 10-20 most bound particles of each galaxy \citep{wetzel2009, rodriguez-gomez2015}. A short analysis of the deviations that occur is presented in Appendix~\ref{appendix:galaxy_tracking}.

\subsection{Simulations}
\label{chapter3-subsec:sims}

\begin{table*}
\caption[Merger parameters and statistics]{Table of simulation parameters and merger remnant quantities at $z=0$ ($z=0.11$ for 1605-3). The columns show: 1) simulation run name; 2) physics included; 3) stellar mass ratio of main progenitors at $z=0.93$; 4) virial mass\color{blue}\protect\footnotemark[2]\color{black}; 5) virial radius; 6) bound stellar mass; 7) inferred stellar disc mass; 8) inferred stellar bulge mass;  9) disc-to-total stellar mass ratio\color{blue}\protect\footnotemark[3]\color{black}; 10) radial scale length; 11) bulge effective radius; 12) S\'{e}rsic index; 13) optical radius\color{blue}\protect\footnotemark[4]\color{black}; 14) gas-to-total baryonic mass fraction within the optical radius. We consider columns 3-6 and 14 to be global properties, whilst 7-13 are structural properties.}
\resizebox{1.\textwidth}{!}{
\centering
{\renewcommand{\arraystretch}{1.5}
\begin{tabular}{cccccccccccccc}
\hline
\\
\textbf{Run}	&	\textbf{Physics}	&	$\dfrac{M_{*,1}}{M_{*,2}}$	&	$\dfrac{M_{200}}{10^{12}\;\mathrm{M}_\odot}$	&	$\dfrac{R_{200}}{\mathrm{kpc}}$	&	$\dfrac{M_{*}}{10^{10}\;\mathrm{M}_\odot}$	&	$\dfrac{M_\mathrm{d}}{10^{10}\;\mathrm{M}_\odot}$	&	$\dfrac{M_\mathrm{b}}{10^{10}\;\mathrm{M}_\odot}$	&	$D/T$		&	$\dfrac{R_\mathrm{d}}{\mathrm{kpc}}$	&	$\dfrac{R_\mathrm{eff}}{\mathrm{kpc}}$	&	$n$	&	$\dfrac{R_\mathrm{opt}}{\mathrm{kpc}}$	&	$f_\mathrm{gas}$	\\ 	\\ \hline

\multicolumn{14}{c}{\T \text{Highest resolution} \B} \\
1330-3M	&	MHD	&	1.92	&	1.58	&	239.41	&	10.99	&	6.99	&	3.14	&	0.69	[0.41]	&	6.51	&	1.74	&	1.05	&	27.90	&	0.35	\\	

1330-3H	&	Hydro	&	2.00	&	1.52	&	236.18	&	11.07	&	6.60	&	2.12	&	0.76	[0.43]	&	7.00	&	1.70	&	0.69	&	15.43	&	0.16	\\	
1526-3M	&	MHD	&	1.08	&	1.75	&	247.74	&	5.72	&	3.18	&	1.88	&	0.63	[0.45]	&	4.25	&	1.72	&	0.92	&	18.23	&	0.17	\\	

1526-3H	&	Hydro	&	1.10	&	1.77	&	248.39	&	5.48	&	2.77	&	1.93	&	0.59	[0.35]	&	4.00	&	1.15	&	0.74	&	12.03	&	0.19	\\	
1349-3M	&	MHD	&	1.08	&	1.46	&	233.23	&	9.92	&	4.06	&	4.50	&	0.47	[0.43]	&	4.47	&	0.90	&	0.93	&	19.35	&	0.19	\\	

1349-3H	&	Hydro	&	1.11	&	1.45	&	232.51	&	9.43	&	4.27	&	2.82	&	0.61	[0.47]	&	4.28	&	0.95	&	0.72	&	13.14	&	0.09	\\	
1605-3M	&	MHD	&	1.29	&	1.07	&	203.60	&	7.98	&	3.26	&	3.11	&	0.51	[0.24]	&	1.72	&	0.85	&	0.66	&	11.56	&	0.19	\\	

1605-3H	&	Hydro	&	1.38	&	1.04	&	201.43	&	6.86	&	1.44	&	3.56	&	0.29	[0.11]	&	1.41	&	0.74	&	0.45	&	8.36	&	0.07	\\ [2mm]
\multicolumn{14}{c}{\T \text{Intermediate resolution} \B} \\
1330-2M	&	MHD	&	2.05	&	1.56	&	238.46	&	9.39	&	5.88	&	2.22	&	0.73	[0.44]	&	7.80	&	3.50	&	1.17	&	24.72	&	0.29	\\	

1330-2H	&	Hydro	&	2.06	&	1.59	&	239.89	&	8.73	&	5.54	&	1.88	&	0.75	[0.40]	&	6.35	&	2.68	&	1.17	&	23.06	&	0.31	\\ [2mm]
\multicolumn{14}{c}{\T \text{Lowest resolution} \B} \\
1330-1M	&	MHD	&	2.17	&	1.60	&	240.23	&	7.17	&	4.34	&	1.95	&	0.69	[0.39]	&	4.19	&	2.03	&	0.74	&	21.39	&	0.41	\\	

1330-1H	&	Hydro	&	2.20	&	1.54	&	237.20	&	6.88	&	4.10	&	1.56	&	0.72	[0.42]	&	6.46	&	2.70	&	1.05	&	21.61	&	0.31	\\	\hline
\end{tabular}
}
}
\label{tab:sim_data}
\end{table*}

In total, eight high, two intermediate, and two lower-resolution merger simulations were run. Each merger simulation is given a name with the format AAAA-BC, where AAAA is the four-digit FoF group number in Illustris for the halo containing the original galaxy at $z=0$, B is the `zoom factor' of the simulation, and C is the letter `M' or `H', denoting MHD or hydrodynamic physics, respectively.  The full list of simulations run is given in Table~\ref{tab:sim_data}. For the rest of the paper, references to a part of the run name implicitly refer to all simulations that contain this part -- e.g. a reference to 1330 refers to all simulations with this prefix.

Whilst we consider only one major merger event in each simulation, the trajectories and total number of participating galaxies vary. These differences have an impact on the production of the merger remnant, and so we briefly describe the interactions here. Roughly speaking, we may separate our merger scenarios into inspiralling (1330 and 1526) and head-on (1349 and 1605) major mergers. We proxy the beginning of the merger by the time of first periapsis. For 1330, this occurs at a lookback time of $\sim$5.4 Gyr ($z\approx0.54$), for 1526 it occurs at $\sim$6.8 Gyr ($z\approx0.77$), and for the 1605 and 1349 simulations, it takes place at $\sim$6.35 Gyr ($z\approx0.69)$. Every galaxy experiences an additional mix of minor mergers and fly-bys, with most of these accompanying the major merger. With this said, some relatively significant tidal interactions take place at approximate lookback times of 4 and 1 Gyr for 1330, 2.5 Gyr for 1526, 1 Gyr for 1349, and generally at late times for 1605. It should be noted too that the merger scenario in 1349 is particularly complex, with seven galaxies of significant mass existing within 100 kpc of the main galaxy at the time of the major merger. All of these galaxies, however, either coalesce with the main galaxy or leave its neighbourhood shortly following the major merger.

\addtocounter{footnote}{+1}\footnotetext{Defined to be the mass inside a sphere in which the mean matter density is 200 times the critical density of the universe.}

\addtocounter{footnote}{+1}\footnotetext{Values are based on the circularity parameter defined in \cite{abadi2003}. The bracketed (unbracketed) values show the stellar mass ratio that kinematically belongs to the disc (doesn't belong to the bulge). See Section~\ref{chapter3-subsec:rot_support} for details.}

\addtocounter{footnote}{+1}\footnotetext{Defined, as in \citet{grand2017}, as the radius at which the $B$-band surface brightness drops below $\mu_{B} = 25$ mag arcsec$^{-2}$. This region roughly encloses the disc.}

\subsection{Comparisons to more isolated galaxies}
\label{chapter3-subsec:methods_isolated_galaxies}

In order to isolate the impact of mergers on our results, we compare our merger simulations to galaxies with more quiescent merger histories. For this purpose, we select four galaxies from the original Auriga \citep{grand2017} simulation suite (Au2, Au12, Au16, and Au23) and re-run these without magnetic fields. These galaxies have similar disc sizes and masses but all have significantly quieter merger histories than the galaxies in our simulations. This can be confirmed by observing their low accreted stellar mass fractions, $f_\text{acc}$, as given in Table 1 of \citet{grand2017}. To make sure that their growth is predominantly secular, Auriga galaxies are also selected such that they are farther than $9 \times R_\text{200}$ from any halo with a mass greater than 3 per cent of their own at $z=0$. The galaxy formation model used for these simulations is identical to that of our own. The dark matter mass resolution of these simulations is $3\times10^5\; \text{M}_\odot$, which is in between our zoom factor 2 and 3 runs. The softening lengths are therefore scaled accordingly (see Table 2 of \citealp{grand2017} for further details).

\section{Analysis}
\label{chapter3-sec:analysis}

\subsection{Impact of MHD on global properties}
\label{chapter3-subsec:MHD-global-properties}

\subsubsection{Halo and stellar mass}

In Table~\ref{tab:sim_data}, we provide a series of values that describe the merger remnant at the end of the simulation. These are given at $z=0.11$ for the 1605 simulations, as tidal disruption affects the structural properties of the remnant in the simulation at late times. For all other simulations, the data is given for $z=0$. From the table, it can be seen that the virial mass and virial radius of the merger remnant changes little, regardless of the resolution level or physics included in the simulation. Such a result is expected, as these statistics are dominated by the dark matter distribution, which feels no direct influence from magnetic fields and only a limited influence from any reorganisation of the baryonic matter. The similarity of these measures across all resolutions and physics models shows that the large-scale structure in our simulations is very well-converged. 

The small-scale baryonic structure, on the other hand, is slightly less well-converged between resolution levels. Here, the total stellar mass bound to the galaxy at the end of the simulation increases by a factor of roughly $\sim$25 per cent with each increase in zoom factor. A similar increase in stellar mass with resolution was also observed in \citet{grand2017}. In this case, it was noted that the excess mass mostly originated from stars born within the inner 5 kpc of the galaxy, where star formation is particularly susceptible to non-linear black hole accretion and related feedback loops \cite[cf.][]{marinacci2014}. Such processes will have been further affected by our treatment of black hole neighbours; as discussed in Section~\ref{chapter3-subsec:set-up}, due to computational constraints the total volume of black hole neighbours decreases slightly with increased resolution. This will increase the initial energy density of the injected thermal energy.

Whilst there are clearly some small discrepancies as function of resolution, overall we consider the stellar mass values in our simulations to be sufficiently similar for the aims of this paper. This is especially so given the non-triviality of achieving convergence in cosmological simulations and the highly dynamic nature of the systems we have modelled. In particular, the stellar mass ratio of the main progenitors is sufficiently close that the results may be robustly compared with one another.

Considering the impact of including MHD physics, we see that this too has little effect on the final stellar mass bound to the galaxy. This is notable, as theoretically magnetic fields are able to provide pressure support to the gas, reducing the fraction above the threshold density, $n_\text{SF}$, thereby suppressing star formation. Some authors have also claimed evidence of a magnetically-driven wind arising in MHD galaxy simulations \citep[e.g.][]{steinwandel2019}, which would be able to remove cold, star-forming gas from the disc with the same effect. In contrast, we see no evidence of magnetic fields suppressing star formation in our simulations. Indeed, in most cases the final stellar mass is actually slightly higher in the MHD simulation compared to the hydrodynamic analogue.

\begin{figure*}
\includegraphics[width=\textwidth]{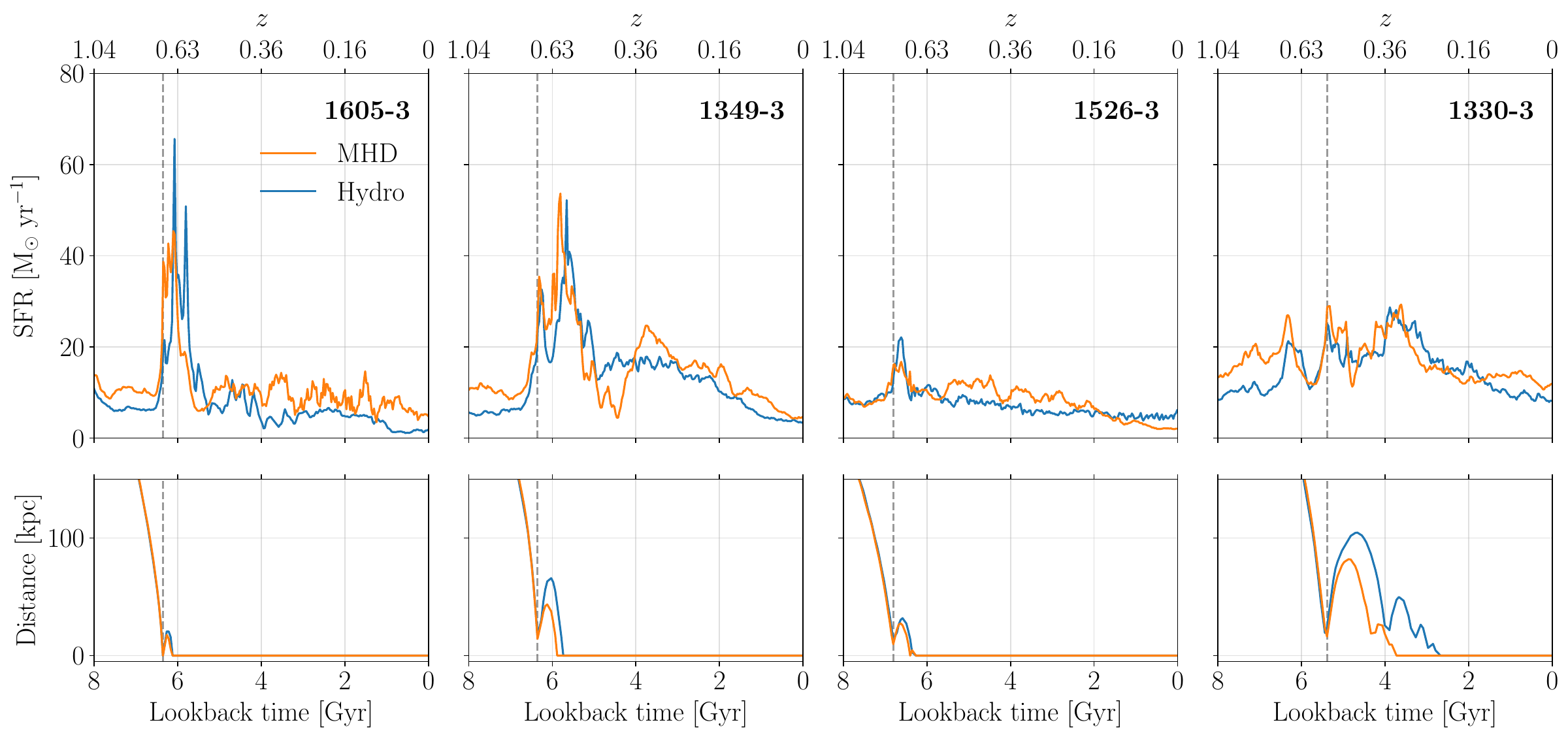}
    \caption[Star formation history and distance between merging progenitors]{\textit{Top row:} star formation history for the main galaxy in each high-resolution simulation as a function of time. The dashed vertical line marks the time of first periapsis in the MHD simulations. \textit{Bottom row:} distance between the main progenitors as a function of time for the same simulations. The star formation history of the galaxy does not change significantly with the inclusion of MHD physics.}
    \label{fig:SFR_dist}
\end{figure*}

\subsubsection{Star formation history and orbit}

This picture is reinforced in Fig.~\ref{fig:SFR_dist}, where we show the star formation history for the main galaxy in each high-resolution simulation. To produce this, all star particles that were within $R_\text{opt}$\footnote{This radius encompasses virtually all stellar material in the galaxy. We have also conducted this analysis using a fixed radius of 30 kpc for each galaxy, which produces an essentially identical result.} (see penultimate column of Table~\ref{tab:sim_data}) of the centre of the main galaxy at $z=0$ were selected. The initial masses of these particles were then binned by their formation time, with bin widths set equal to 30 Myr. This width was chosen as it provides adequate time resolution, whilst preventing the data from becoming dominated by stochastic noise resulting from our probabilistic star formation model. This method is particularly advantageous as it is independent of the galaxy tracking process. A minor disadvantage, however, is that we exclude stars that have left the galaxy after formation. In practise, though, this has a negligible impact on the final result.

In each simulation, the merger causes a sudden rise in star formation, as existing gas is compressed and cold gas is brought into the galaxy. The timing of this rise correlates closely with the merging galaxy's periapsis, with the first approach generally causing the strongest burst. The merger scenarios presented in the two left-most panels are, broadly-speaking, the most energetic, being approximately head-on. Correspondingly, they show the most enhanced star formation. In contrast, the two right-most panels show inspiralling mergers. For these mergers, even the boosted star formation rate falls well short of the starburst threshold, as defined in \citet{sparre2017}. Significant star formation continues long after coalescence in every simulation, however, with none of the galaxies being quenched. This provides yet further evidence to support the argument that gas-rich mergers are not well-described by the `traditional' merger scenario, and instead preferentially produce a star-forming remnant.

Comparing simulations that included MHD physics to those that did not, we see that the magnitude, duration, and timing of the star formation peaks change very little. In fact, any discrepancies seen between the star formation histories can be more than adequately explained by the numerically stochastic nature of our star formation model, and by variations in the merger progression, as proxied by the distance between the two main progenitors.

Apart from indicating that magnetic fields have been ineffectual in suppressing star formation, the strong similarity of the star formation histories also provides further evidence that our simulations are numerically robust. This robustness is also seen in the almost identical evolution of the distance between the two main progenitors up until the first periapsis. After this time, the trajectories deviate a little. We note that the progenitors in the MHD simulations coalesce systematically faster than in the hydrodynamic analogues, and it is possible that this is an indication of more efficient transport of angular momentum in the MHD simulations. In general, however, the differences may also be explained by the non-linear N-body dynamics at play. In particular, the apparent extended orbit of the secondary galaxy in 1330-3H may be attributed to the late merger of its black hole and the subsequent continued identification of a distinct subhalo until this time.

\subsubsection{Amplifying the magnetic field}

\begin{figure*}
    \includegraphics[width=\textwidth]{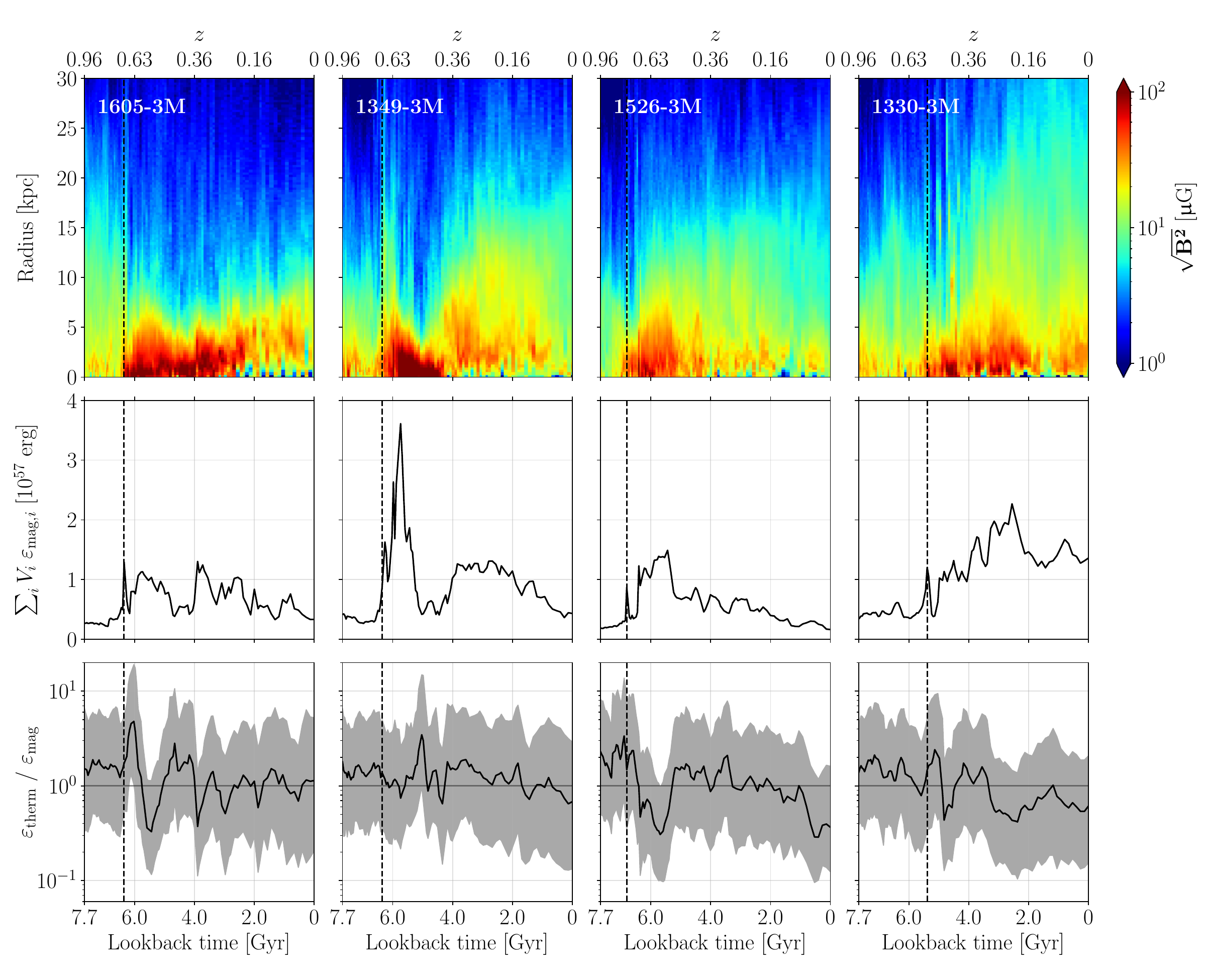}
    \caption[Radially-averaged magnetic field strength, magnetic energy, and beta values as a function of time]{\textit{Top row:} radially-binned mean magnetic field strength in the galactic disc for each high-resolution simulation as a function of time. Bins have a radial extent of 0.25 kpc and a vertical extent of $\pm$1 kpc. The dashed vertical line marks the time of first periapsis. The merger in each simulation is able to substantially amplify the magnetic field in the inner 5 kpc by up to an order of magnitude, with effects visible for several Gyr afterwards. \textit{Middle row:} total magnetic energy in a disc with radial extent $R_\text{opt}$ (as given in Table~\ref{tab:sim_data}) and vertical extent $\pm$5 kpc. \textit{Bottom row:} distribution of the ratio of thermal-to-magnetic energy density for gas cells that lie within the same disc as above. The solid black line indicates the median, whilst the grey shaded region indicates the interquartile range. A horizontal line marks the point at which the magnetic energy density is equal to the thermal energy density. The merger generally results in a temporary, but significant, increase in the fraction of gas cells where the magnetic field is dominant.}
    \label{fig:BH_mag}
\end{figure*}

The broad similarity of the star formation histories produced by each physics model could easily be explained if the galactic magnetic field was not significantly amplified during the merger. However, this is not the case. In the top row of Fig.~\ref{fig:BH_mag}, we show the evolution of the radially-binned average magnetic field strength for each high-resolution simulation. To create this, volume-weighted means of the magnetic energy density are taken, using gas cells lying in annular rings of width 0.25 kpc and vertical extent $\pm1$ kpc; a region that covers the dense gas in the disc. The mean values are then converted back into an average field strength for the corresponding radius. It may be seen that immediately after first periapsis (indicated by the dashed black line) the magnetic field strength in the inner regions of the disc ($\lesssim5$ kpc) is strongly amplified by up to an order of magnitude. During this time the radial profile of the magnetic field strength continues to be well-fit by a double exponential, as observed in \citet{pakmor2017}. As expected, the strongest amplification of the magnetic field occurs for the most energetic mergers (1605-3M, 1349-3M). Indeed, for these galaxies a number of pixels over-saturate in Fig.~\ref{fig:BH_mag}. This is particularly the case for 1349-3M, where a few pixels reach over 150 $\upmu$G with mean field strengths reaching a maximum of $310$ $\upmu$G. Field strengths this high are unusual, but are not unheard of for starburst galaxies \citep[see, e.g.][]{lacki2013}.

The remaining field strengths in our simulations are in good agreement with those expected for gas-rich merging galaxies. In particular, the early stages of amplification seen for the inspiralling galaxies (1526-3M, 1330-3M) are consistent with strengths observed for the Antennae galaxies \citep{basu2017}, whilst the evolution until coalescence is consistent with that derived from nearby interacting galaxies by \citet{drzazga2011}. Our galaxies differ in their evolution post-merger, however, as whilst \citet{drzazga2011} predicts the galactic magnetic field to weaken significantly after coalescence, in our simulations the field remains highly amplified for at least 1.5 Gyr in each instance. Furthermore, when the field strength eventually does decrease, it returns to a strength that is at least as high as that which the galaxy had pre-merger. This difference in evolution is likely to be due to the different nature of the merger scenarios that we simulate; whilst the remnants in the Drzazga scenario are quenched, those in our own simulations are not. Instead, the remnants in our simulations maintain a significant percentage of their gas content, allowing them to also maintain the strength of their magnetic field until well after the initial merger-induced starburst has passed. Assuming our simulations reflect reality, this provides a new potential observable: an observation of an unusually high magnetic field in a galaxy that otherwise has a normal or low star formation rate could be an indication that the galaxy has undergone a gas-rich major merger in its recent past.

The decrease in magnetic field strength in the inner regions after the initial period of amplification is correlated with the rebuilding of the disc. This rebuilding is seen in Fig.~\ref{fig:BH_mag} through the increase in the field strength at larger radii. In some galaxies, at late times the field strength also reduces at these larger radii, as magnetic flux is locked up in newly-formed star particles and the consumed gas is not replenished. Such a process is seen particularly for 1526-3M from ${\sim}$2 Gyr onwards. Gas is also removed periodically from the inner regions due to AGN activity, causing the magnetic field strength to `flicker' at late times. This process may be undermining the amplification of the magnetic field generally, as field strengths do not generally recover to the same level afterwards. On the other hand, there are periods when the magnetic field strength increases at late times. For example, we see an enhancement of the magnetic field strength in 1330-3M by a factor of roughly two at $z \approx 0$, relative to its value at $\sim$1.5 Gyr. The galaxy in this simulation undergoes a series of minor tidal interactions at late times. It is not clear though whether the field amplification seen is a direct result of these interactions; whilst \citet{pakmor2017} do observe that minor mergers can cause such an effect, it is difficult to distinguish this particular enhancement from similar order fluctuations seen in the other simulations.

Although not shown explicitly here, at the time of the merger the field strength in the inner regions increases by more than that expected from pure adiabatic compression. This suggests that the amplification at this time is at least in part due to a small-scale dynamo. Evidence was shown in \citet{pakmor2017} that such a dynamo was active in the Auriga galaxies. In particular, it was shown that the kinetic energy power spectrum took on a characteristic \citet{kolmogorov1941} spectrum, producing in turn a magnetic energy power spectrum of the \citet{Kazantsev1968} type, as expected from small-scale dynamo theory. Naturally, the turbulent energy available in our simulations for such amplification should be even higher, owing to the strong solenoidal and compressive forcing during the tidal interactions We will explicitly study the existence of such a small-scale dynamo in Section~\ref{chapter3-subsec:amplification_res}, showing that power spectra of the forms described here continue to be evident in our own simulations. 

In contrast to the inner regions, the magnetic field strength in the outskirts of the galaxies ($r \gtrsim$10 kpc) increases and decreases almost exactly with $\rho ^ {2/3}$. This implies that the field strength at these radii depends almost exclusively on flux conservation \citep{kulsrud2005}, and that any dynamo that may exist is already saturated. This observation is further supported by the lack of amplification and field reorganisation seen after the merger at such radii, despite the passing of several Gyr. 

In the middle row of Fig.~\ref{fig:BH_mag}, we show the total magnetic energy in a volume with radial extent $R_\text{opt}$ and vertical extent $\pm5$ kpc. This region approximately bounds the disc and its immediate neighbourhood. It can be seen that the total magnetic energy generally follows the fluctuations seen in the radial evolution above. This is expected, as the amplification of the inner regions substantially contributes to the total energy in the volume. In each case, the magnetic energy spikes at the first periapsis as the energy of the merging galaxy is included in the calculation. The energy then increases more consistently shortly afterwards, as the associated turbulence and compression works to amplify the galactic magnetic field. This amplification is substantial, and can increase the total magnetic energy by up to an order of magnitude, as can be seen for 1349-3M. This period of heightened magnetic energy generally lasts for a shorter time than the corresponding inner amplification for most simulations, as magnetic energy decreases in the surrounding volume.

The period of initial amplification is generally followed by a second, longer period of increased magnetic energy. This is a result of the rebuilding of the gas disc in the remnant, and is once again particularly clear for simulation 1349-3M. For 1605-3M, this period is also a time of high activity from the central AGN. This results in a strongly non-linear evolution of the total magnetic energy, reflecting the subsequent fluctuations of the magnetic field strength in the inner $\lesssim5$ kpc. The two periods of increased magnetic energy are not very well separated in simulation 1330-3M. Here, the phases merge as the merger takes place over a sustained duration (see Fig.~\ref{fig:SFR_dist}). This means that the merging galaxy drives turbulence, and the resulting dynamo effect, over a period of several 100 Myr.

For most simulations, the total magnetic energy decreases towards the end of the simulation, returning to a roughly pre-merger level. At this time the turbulent driving from tidal interactions has long since stopped, and there is no longer a sufficient energy budget to maintain the amplified field strength. Once again, simulation 1330-3M does not quite follow this evolution, as it continues to be harassed at late times by satellite galaxies. Indeed, a particularly close encounter takes place at around $\sim1$ Gyr, coinciding with the peak seen in the total magnetic energy here. On top of this, this galaxy retains its gas content to a greater extent (as may be seen from its $f_\text{gas}$ value in Table~\ref{tab:sim_data}), allowing it to maintain its magnetic energy as well.

\begin{figure*}
    \centering
    \includegraphics[width=\textwidth]{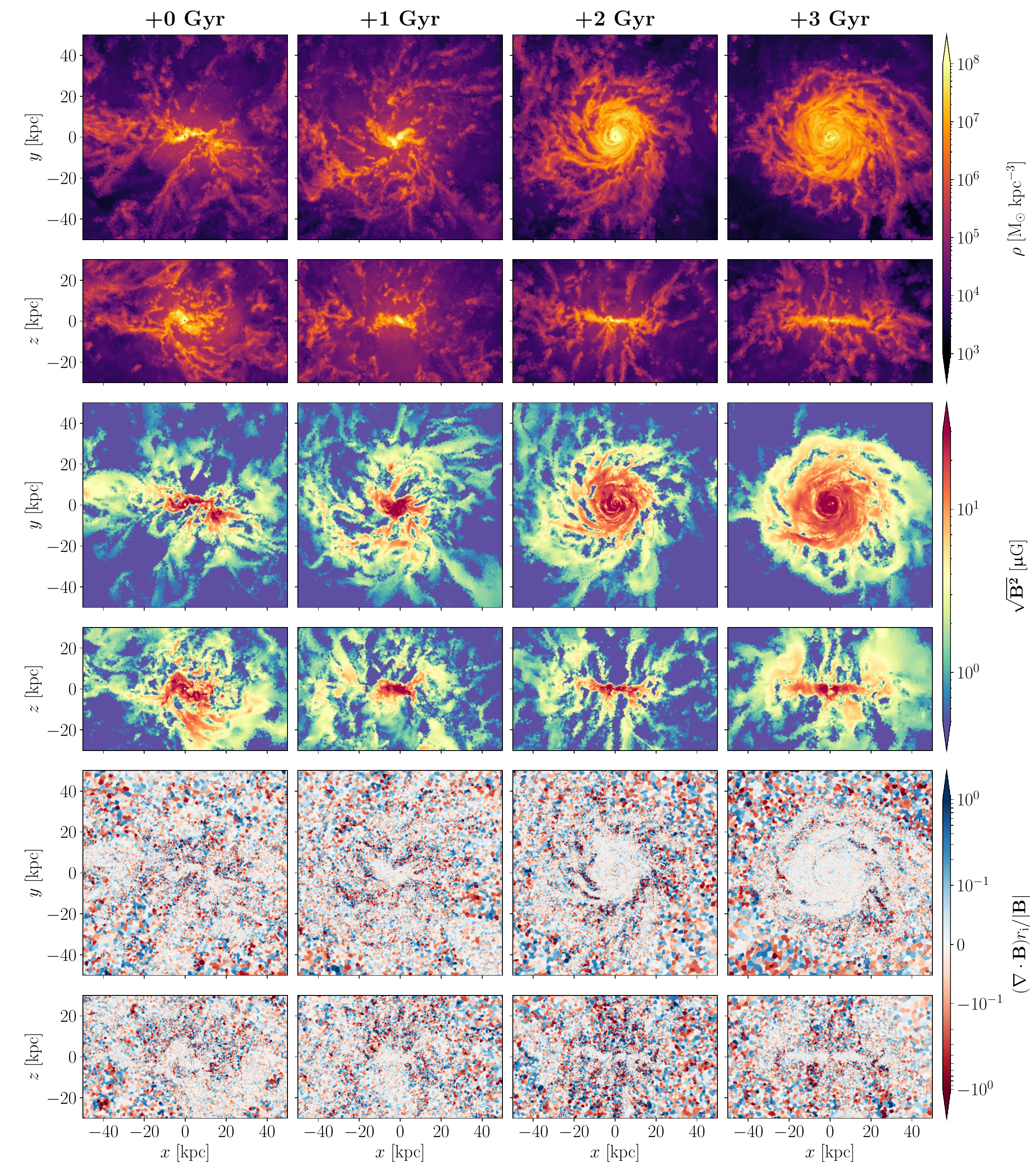}
    \caption[Time series of the galactic disc rebuilding post-merger, including magnetic field strength and relative divergence errors]{Face and edge-on slices through the main galaxy in simulation 1330-3M as it rebuilds post-merger. Timings are given from the first periapsis, showing a period when the galactic magnetic field is highly amplified. \textit{1st and 2nd row:} slices show the gas density in each cell. \textit{3rd and 4th row:} slices show the magnetic field strength in each cell. \textit{5th and 6th row:} slices show the relative divergence error in each cell. Divergence errors are highly localised and do not propagate between snapshots. Regions with higher gas density are better resolved due to our refinement scheme. These cells typically have the highest magnetic field strengths and the lowest relative divergence errors.}
    \label{fig:divergence_map}
\end{figure*}

The change in magnetic energy also changes the energetic balance of the system. In particular, it changes the ratio of thermal to magnetic pressure in the individual gas cells. This ratio varies strongly, both spatially and temporally, and is not well-captured by a radial average. Instead, in the bottom row of Fig.~\ref{fig:BH_mag}, we consider the distribution of this ratio for gas within the same volume as above. The distribution can be strongly skewed by extreme values, and so we show the median and interquartile range, rather than the mean. Larger values of this statistic imply that the gas dynamics are more affected by the thermal component, whilst smaller values imply that the magnetic fields are more influential. 

It can be seen that at all times, the gas cells cover a broad range of values, indicating that there are regions throughout the galaxy where either the thermal or magnetic pressure is dominant. In each case, however, the distribution is biased towards thermal pressure at early times, indicating that magnetic fields are, on the whole, subdominant at this time. The arrival of the secondary progenitor results in gas in the galactic neighbourhood being compressed and the production of a large amount of turbulence. This initially increases the fraction of thermal pressure, before amplification of the magnetic field swings the distribution the other way. This development takes place within a few 100 Myr, consistent with the time-scales required for a small-scale dynamo to amplify the field \citep{arshakian2009}. This evolution is less clear in 1349-3M, which may be a result of its more complex merger scenario (see Section~\ref{chapter3-subsec:sims}). After the initial period of amplification, the evolution of the distribution is highly non-linear, depending strongly on the stability of the magnetic field. As noted previously, this is seen particularly in 1605-3M, where AGN outbursts lead to oscillations in the balance of the thermal to magnetic pressure distribution. Generally, the fraction of magnetic pressure relative to thermal pressure has increased by the end of the simulation.

In Fig.~\ref{fig:divergence_map}, we show how the disc regrows after disruption. In the top and middle two rows, respectively, we show the gas density and magnetic field strength as seen in slices through the main galaxy in simulation 1330-3M. We present four times showing the development of the galaxy, starting with its state at first periapsis, with snapshots thereafter showing progressive states at increasing 1 Gyr increments. During this period, the magnetic field is at its most amplified. The angular momentum of the merging galaxy in this simulation is particularly well-aligned with the main galaxy, and consequently the disc grows quickly within the time-frame shown. The development seen is, however, qualitatively similar for all our merger simulations.

It is clear from both the gas and magnetic field strength distributions seen in the first column that the galaxy is substantially disrupted by the approach of the merging galaxy. Such disruption already generates high field strengths through compression, before a dynamo has had time to saturate. The impact of the tidal interaction scatters the gas content of the galaxy, producing filaments of high density gas well below the disc plane. In the following snapshot, this gas has begun to accrete onto the galaxy, starting the process of disc-rebuilding and increasing the magnetic field strength at the core. Already at this stage, the magnetic field has begun to take on a relatively axisymmetric profile, which only becomes smoother with time. This justifies our choice of showing radial profiles in the top row of Fig.~\ref{fig:BH_mag}. The magnetic field is generally most dominant when it is strongest, and so the dynamics at the very centre of the disc will be particularly affected. This has further ramifications, as the evolution of the galaxy as a whole is sensitive to the behaviour of the central AGN, fed by the gas in this region. The exact impact of the magnetic fields on both the central gas distribution and black hole accretion rate will be explored in an upcoming paper.

\subsubsection{Stability of MHD implementation}

Given the strong amplification of the magnetic field seen in our simulations and the ability of the field to subsequently play an important dynamical role, it is prudent to consider the evolution and impact of divergence errors. Whilst the continuum equations of MHD preserve the $\bs{\nabla} \bs{\cdot} \bs{B} = 0$ condition perfectly given an initial divergence-free field, this is not the case for the discretised versions of the equations used in our simulations. Worse still, partial differential equation solvers are generally unstable to the production of magnetic monopoles; once produced, these have a tendency to become rapidly larger in any non-trivial MHD flow, rendering simulation results unphysical \citep{pakmor2011}. As discussed in Section~\ref{chapter3-subsec:set-up}, divergence errors in our simulations are controlled using a Powell 8-wave scheme. Whilst it has been shown in previous work that this scheme is able to deal with divergence errors robustly \citep{pakmor2013}, it is sensible to analyse its performance in our simulations as well.

In the bottom third of Fig.~\ref{fig:divergence_map}, we show the distribution of the relative divergence errors as seen in a slices through the main galaxy in simulation 1330-3M. As with the panels above, this picture is qualitatively the same for all simulations. Whilst some large individual errors may be observed, they are highly localised in space. Furthermore, as observed in \citet{pakmor2013}, the sign of the divergence error is seen to alternate between neighbours when its magnitude becomes large. This alternation means that fluctuations generally cancel when considered over larger scales. Particularly large divergence errors generally stem from larger, under-resolved gradients in the local magnetic field. These, in turn, are often a due to larger cell sizes. Such cells are low density, as our refinement scheme keeps gas cells within a target mass. In contrast, the high density regions are very well resolved, and have subsequently lower relative divergence errors. This is true for a range of magnetic field strengths and means that the galactic disc, where the most substantial amplification occurs, is particularly robust to such errors. Perhaps even more importantly, the divergence errors are not seen to propagate in time. Instead, the distribution of errors in each snapshot is broadly independent of its previous distribution.

\begin{figure}
    \centering
    \includegraphics[width=0.5\columnwidth]{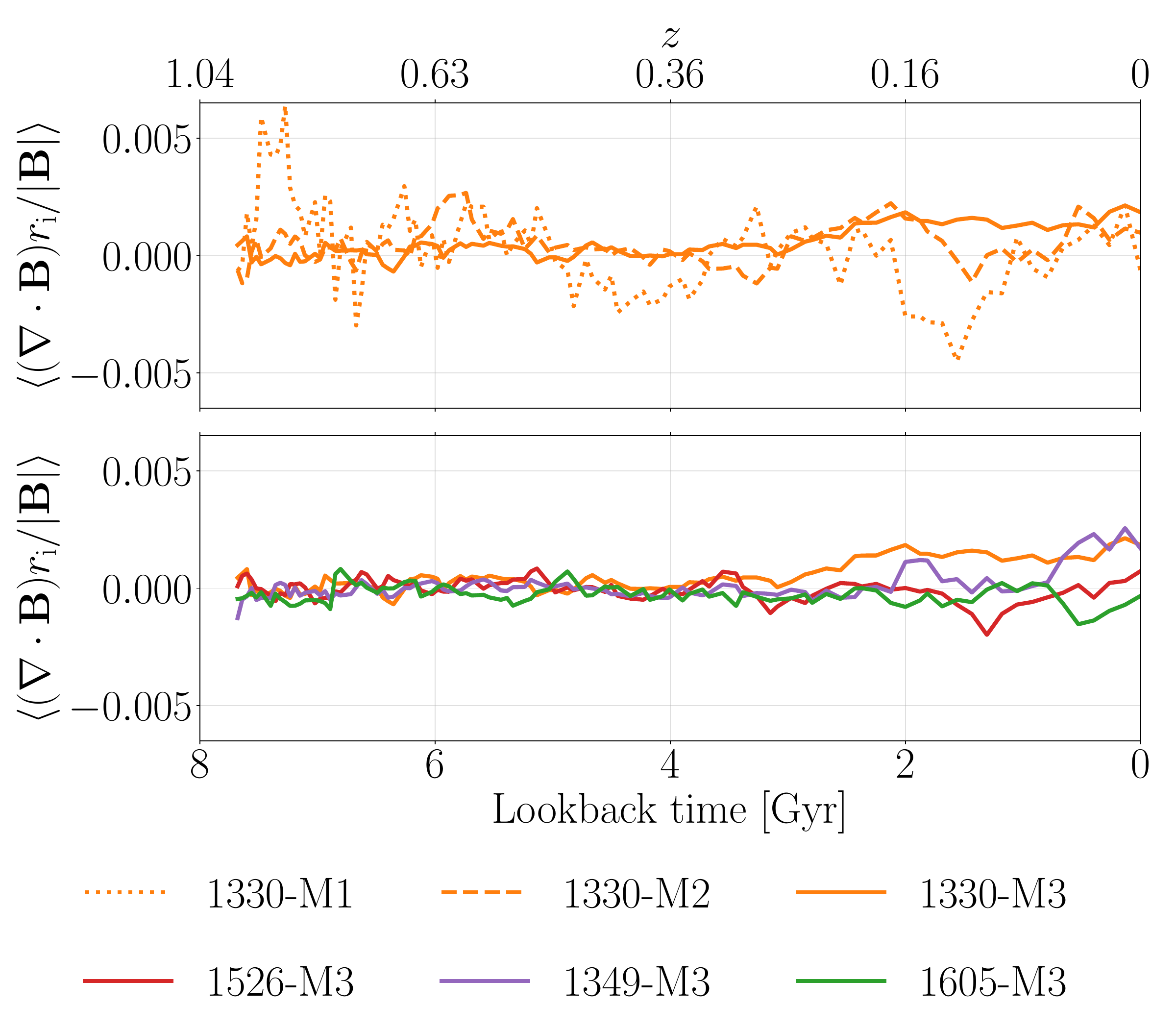}
    \caption[Mean relative divergence error as a function of time]{\textit{Top panel:} the mean of the relative divergence error, as a function of time, for simulations run with different resolution. \textit{Bottom panel:} as above, but for the highest-resolution simulations only. The mean is calculated for all gas cells that lie within 100 kpc of the main galaxy. It stays well below 1 per cent in each simulation and decreases with increased simulation resolution.}
    \label{fig:divergence}
\end{figure}

In Fig.~\ref{fig:divergence}, we show the mean of the relative divergence error in each simulation as a function of time. To calculate this, we select gas cells that lie within a radius of 100 kpc of the main galaxy at each time step, as this is well within the region of high-resolution in our simulations and also covers a volume that can affect the immediate development of the galaxy. We show the mean of the relative divergence error as we are interested in the stability of the system, rather than any particular peak values. Indeed, the Powell scheme implemented in \citet{pakmor2013} often produces higher average divergence errors than the Dedner scheme \citep{dedner2002} it replaced. The advantage of the Powell scheme, however, lies in its more effective control of such errors, which makes it more appropriate for cosmological simulations and highly dynamical systems. It can be seen in Fig.~\ref{fig:divergence} that this stability remains in our merger simulations. Indeed, the average divergence error decreases with increased resolution, as previously observed in \citet{pakmor2013}. This is in contrast to the magnetic field amplification, which increases with increased resolution (see Section~\ref{chapter3-subsec:resolution-study}). We note on top of this that there are no signs of instability developing at the time of amplification (compare to Fig.~\ref{fig:BH_mag} and Fig.~\ref{fig:mag_zoom}), indicating that these are not the source of the amplification. We conclude from this that our results are robust to our MHD implementation.

\subsubsection{Bound gas and stellar mass evolution}

Whilst the star formation history is generally very similar regardless of physics models used, the amplified magnetic fields are nevertheless able to significantly affect the gas dynamics. In Fig.~\ref{fig:bound_gas_stellar} we show the total gas and stellar mass bound to the main galaxy in each high-resolution simulation as a function of time, as well as the sum of these quantities. For each simulation, the merger results in a sharp increase in the total gas mass bound to the system, followed shortly thereafter by an increase in the bound stellar mass. The timing here is dependent on the rate at which the progenitors coalesce, as well as on the star formation history of the main galaxy. Both the gas and stellar mass evolutions exhibit a few localised peaks before full coalescence, as mass is reallocated between the merging galaxies by \textsc{subfind} \citep[cf.][]{rodriguez-gomez2015}. As the gas component is relatively diffuse, it is particularly sensitive to this reallocation process. This effect is seen for all simulations, but is most clear for the simulations where the merger took the longest. At the time of coalescence, the gas mass bound to the main galaxy is generally higher for the MHD simulations than for the hydrodynamic simulations. This supports the idea that the accelerated coalescence seen in Fig.~\ref{fig:SFR_dist} may be a result of more effective gas transport in MHD simulations at the later stages of the merger.

Assuming the galaxies exist in relative isolation post-merger, the gradient of the `stellar + gas' line provides us with information on how effective feedback is in ejecting gas from the galaxy: if gas is only being converted into stellar mass, this line will stay constant; if feedback is efficient, then this line will show a negative gradient, as gas is unbound from the system. Of course, the nature of cosmological simulations is that the galaxies do not experience complete isolation. We therefore expect to see some gas accretion after the merger; an effect that would be absent in idealised simulations. The accretion in our simulations takes place both through cosmological filaments and through further galaxy interactions, as mentioned in Section~\ref{chapter3-subsec:sims}. This effect is seen most obviously where the sum of bound stellar and gas mass continues to increase after the merger, as in 1526-3. The hydrodynamic simulation in this case shows a particularly strong increase in gas mass at late times as gas is stripped off a passing galaxy. The evolution of the stellar and gas mass is also affected by the allocation of mass by \textsc{subfind} during such interactions. In particular, this effect results in the flattening out of the bound mass evolution seen for 1349-3H around 1 Gyr and the loss of bound mass starting at 2 Gyr for 1605-3M.

\begin{figure*}
    \includegraphics[width=\textwidth]{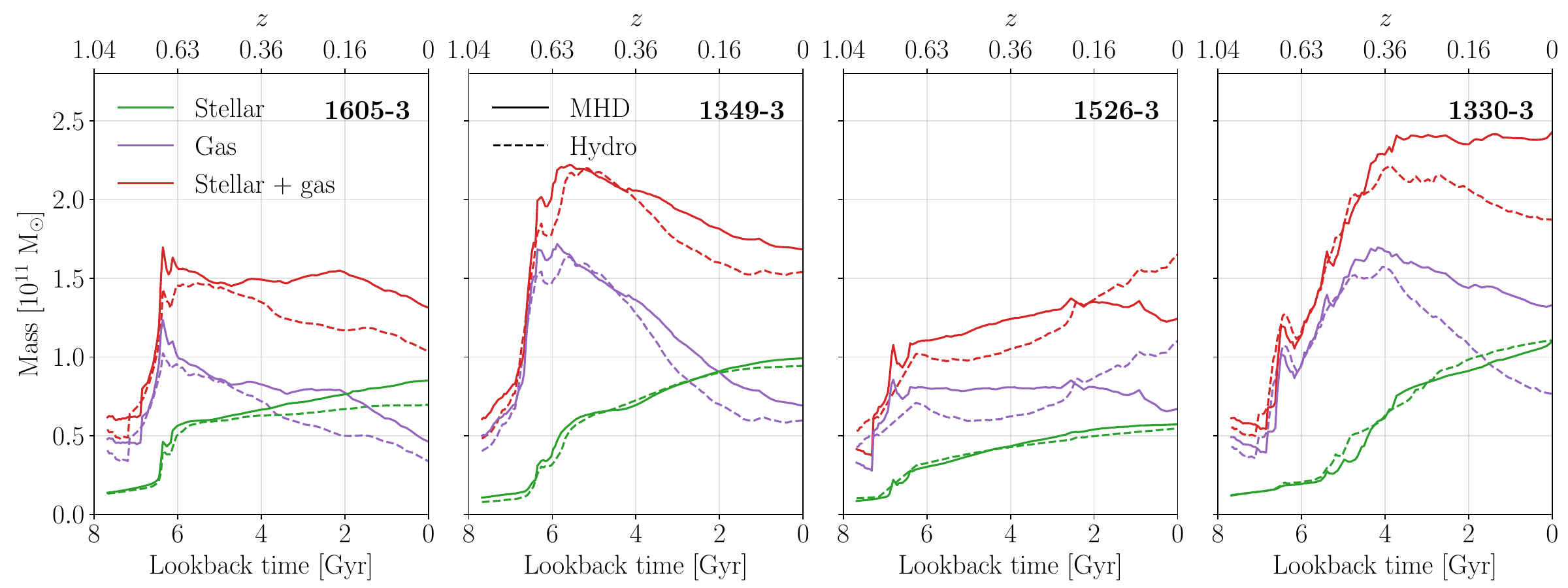}
    \caption[Total bound stellar and gas mass as a function of time]{Total stellar and gas mass bound to the main galaxy in each simulation as a function of time. Whilst the stellar mass remains similar across physics models, the bound gas mass at $z=0$ is generally higher in MHD simulations. Indeed, in some cases, almost all gas lost can be accounted for by a comparable increase in stellar mass.}
    \label{fig:bound_gas_stellar}
\end{figure*}

Despite such deviations, it is still apparent that for virtually every merger, feedback removes gas more effectively in the hydrodynamic simulation than in the MHD analogue. Indeed, for many of the MHD simulations, feedback is highly ineffective at unbinding the gas from the galaxy; frequently, the loss of gas mass may be accounted for almost entirely by the corresponding increase in stellar mass. This evolution is especially clear for simulation 1330-3M, but may also be seen in 1526-3M and 1605-3M. This provides further evidence that the winds in our simulations are not magnetically-launched. It is especially notable that we do not see magnetically-driven winds given the strength of the field reached in the disc post-merger. This result conflicts with the idealised smoothed-particle MHD simulations performed by \citet{steinwandel2019}, which did show such a wind, despite having a similar dark matter mass resolution. Whilst some differences will be due to the cosmological nature of our simulations, we suspect that the choice of numerical treatment also plays a significant role.

Finally, it is evident from Fig.~\ref{fig:bound_gas_stellar} that the differences in star formation seen in Fig.~\ref{fig:SFR_dist} do not accumulate significantly over time. Whilst the total bound gas mass diverges between physics models, the total bound stellar mass stays similar. The result is that the ratio of gas to total baryonic mass in the merger remnant (a measure of how gas-rich the remnant is) can differ substantially between the two physics models by the end of the simulations. This fraction is given in Table~\ref{tab:sim_data} for a reduced radius, showing that the loss of gas does not only affect the outer regions of the circumgalactic medium. The mechanism behind the more effective unbinding of the gas in hydrodynamic simulations can begin to be understood by considering the structural properties of the merger remnants.

\subsection{Impact of MHD on structural properties}
\label{chapter3-subsec:MHD-disc-sizes}

\subsubsection{Magnetic field and altered gas morphology}

\begin{figure*}
    \includegraphics[width=\textwidth]{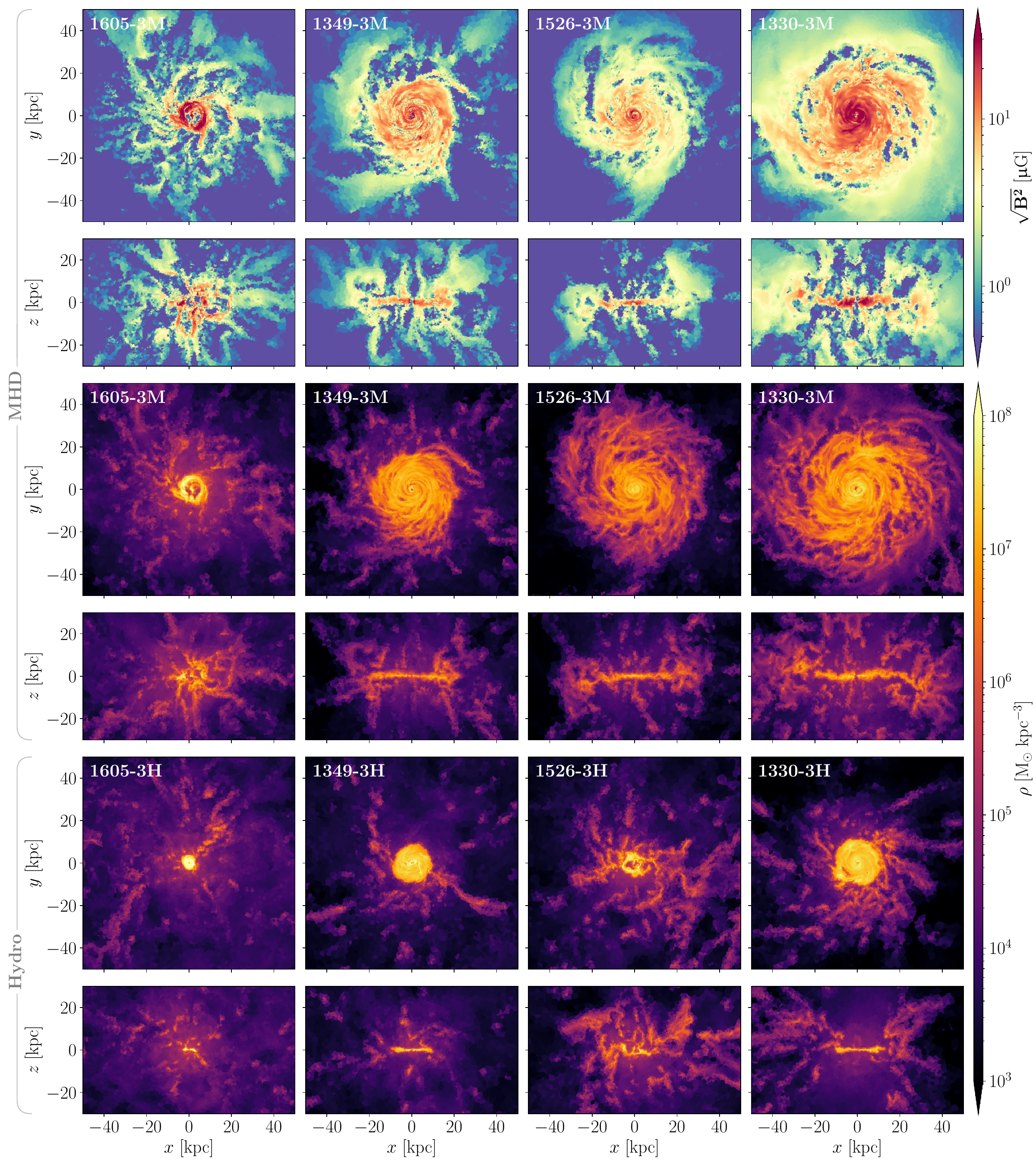}
    \caption[Slices showing magnetic field strength and gas density for each high-resolution simulation]{Face and edge-on slices through the merger remnant for each high-resolution simulation, as seen at $z=0$ ($z=0.11$ for 1605-3).  \textit{1st and 2nd row:} slices show the magnetic field strength in each gas cell. The magnetic field is broadly axisymmetric, but still shows significant amounts of small-scale structure. This roughly mirrors the corresponding gas distribution. \textit{3rd and 4th row:} slices show gas density for the MHD simulations. The gas discs have a flocculent structure, and show a shallow radial gradient. \textit{5th and 6th row:} slices show gas density for the hydrodynamic simulations. The gas discs are systematically smaller and thinner. They also exhibit a flatter density profile, which cuts off abruptly at the disc edge. The clearance of gas above the disc implies more effective stellar feedback is in action in these simulations. Simulations 1605-3M and 1526-3H have a more unusual morphology, owing to the impact of strong AGN feedback.}
    \label{fig:gas}
\end{figure*}

In Fig.~\ref{fig:gas}, we show slices taken face and edge-on through the main galaxy in each high-resolution simulation. These are shown at $z=0.11$ for the 1605 simulations (see Section~\ref{chapter3-subsec:MHD-global-properties}), and at $z=0$ for all others. In the top two rows, the slices show the magnetic field strength in each gas cell, whilst the bottom four rows show the gas density for MHD and hydrodynamic simulations, respectively. As in Fig.~\ref{fig:divergence_map}, the magnetic field profiles are observed to be generally axisymmetric on large scales. With this said, it is clear that there is a great deal of small-scale structure to be found as well. In general, this structure mirrors the flocculent nature of the underlying gas disc, with regions of dense gas correlating to regions of high magnetic field strength. This is also true for regions above and below the disc, where dense clumps are still seen to be reasonably strongly magnetised, with field strengths on the order of a few $\upmu$G.

Whilst increased density is correlated with stronger magnetic fields, the reverse effect may be seen at the centre of the galaxies. Here, quasar feedback temporarily removes gas, weakening the field. This effect is particularly clearly seen for simulation 1605-3M in Fig.~\ref{fig:gas}, where the AGN has pushed gas out of the inner $\lesssim3$ kpc, helping to produce a ring shape morphology in the face-on view, and strongly distorting the disc in the edge-on view. For the other MHD simulations, the gas discs show shallow radial and vertical gradients. The gas disc for these galaxies extends in radius well beyond the stellar disc (c.f. Fig.~\ref{fig:mocks}). At these radii, a network of filamentary gas structures with densities below $10^7\;\text{M}_\odot\; \text{kpc}^{-3}$ can be seen. These structures show gas joining the galaxy in the plane of the disc, fuelling its radial growth. The clumps of gas seen above and below the disc are typical of a fountain flow in action. Such fountain flows have also been found to be crucial for successfully growing large galactic discs in previous work \citep{grand2019}.

The merger remnants formed in the hydrodynamic simulations are systematically smaller than those formed in the MHD simulations. Furthermore, they do not display the same level of complex small-scale structure as their MHD analogues. Rather, the gas density stays high throughout the disc, with a sudden cut-off seen at the disc edge, where the density drops by some three orders of magnitude. The amount of flocculent structure seen in the face-on view beyond the stellar disc is greatly reduced too, implying that these galaxies are not receiving the high angular momentum gas they need to grow in size. This understanding will be explicitly confirmed in an upcoming work.

As well as the reduced radius, the height of the disc in the simulations without magnetic fields has also been affected, with gas discs becoming razor-thin. The lack of disruption seen in the centre of the disc implies that this is not due to AGN feedback. Indeed, the sudden cut-off in gas density in the vertical direction suggests that the disc height has been affected by stellar feedback, with wind particles coupling to the low-density gas in the region, quickly removing it. This view is supported by the reduced gas density seen above and below the disc -- taken to its extreme in 1330-3H, where gas is cleared out of a conical-shaped region -- and by the lack of disruption to the disc midplane, where the gas density is too high for our simulated wind particles to effectively couple (see Section~\ref{chapter3-subsec:set-up} for details). It is also noticeable that the gas is more effectively cleared above the disc for galaxies that have a higher stellar mass (see Table~\ref{tab:sim_data}). The increased effectiveness of the stellar winds in this case is logical, as more star formation takes place in a similar volume, leading to higher wind energy densities. Such energy densities help to explain the substantial unbinding of gas seen for hydrodynamic simulations in Fig.~\ref{fig:bound_gas_stellar}. As well as unbinding the gas, the stellar winds also help to maintain the reduced disc sizes; by clearing the gas above and below the disc, they disrupt small-scale fountain flows, further suppressing growth.

The only hydrodynamic simulation that does not fit the pattern is 1526-3H. Here, the merger remnant had a higher bound gas mass than its MHD analogue at the simulation end (see Fig.~\ref{fig:bound_gas_stellar}). However, this remnant also shows a clearly different gas morphology, displaying similar disruption to that seen for 1605-3M. This disruption indicates recent AGN feedback has taken place in this galaxy as well, and generally indicates a different evolutionary path taken compared to the other hydrodynamic simulations.

\subsubsection{Altered stellar morphology}

\begin{figure*}
    \includegraphics[width=\textwidth]{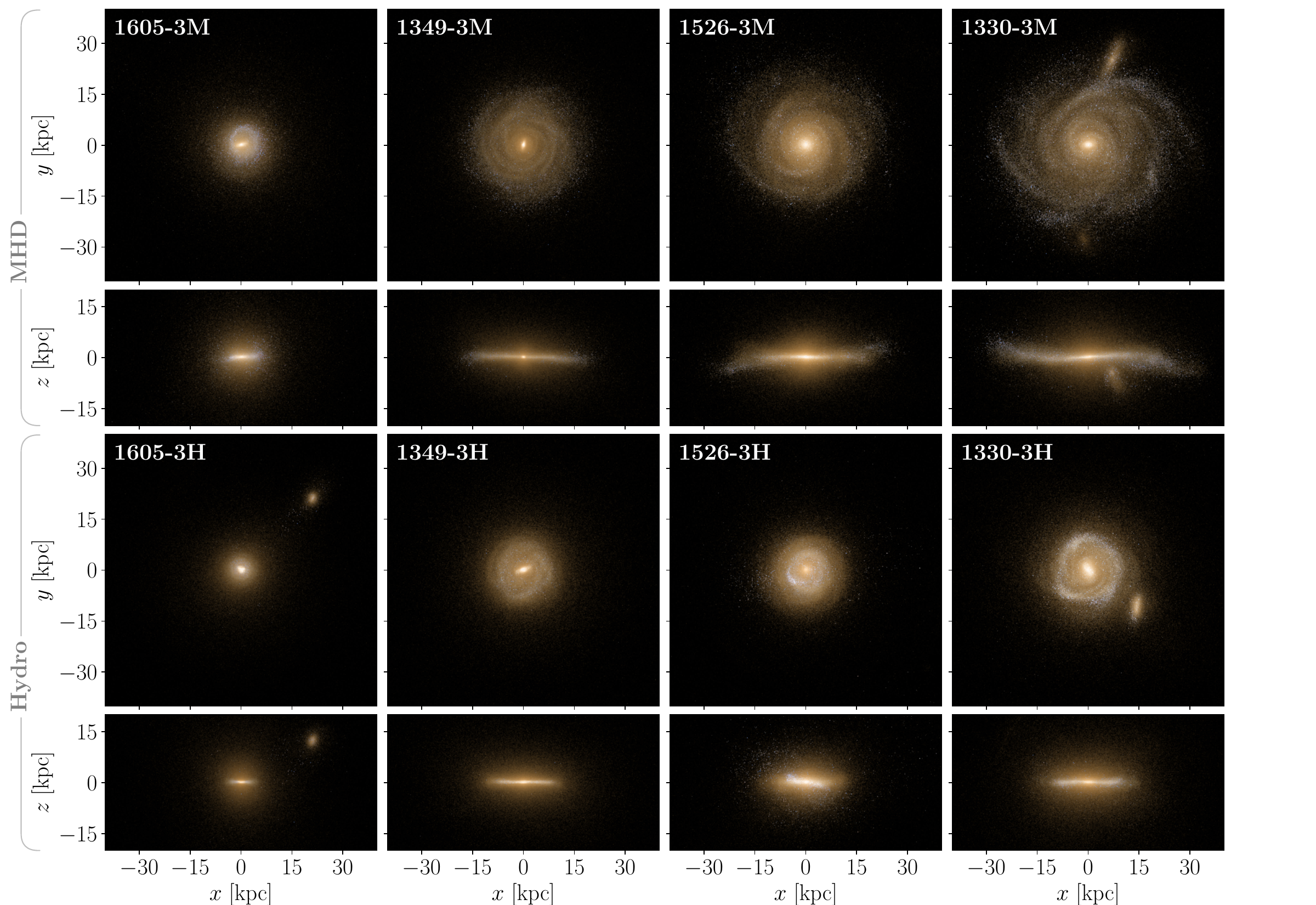}
    \caption[Mock stellar light images for each high-resolution simulation]{\textit{1st and 2nd row:} mock SDSS \textit{gri} composite images showing the merger remnants for all high-resolution MHD simulations. Remnants are seen face and edge-on at $z=0$ ($z=0.11$ for 1605-3). \textit{3rd and 4th row:} as above, but for the hydrodynamic simulations. The morphology of the merger remnant is once again systematically different between MHD and hydrodynamic runs. Whilst MHD simulations generally produce MW-like galaxies that show extended spiral-structure, hydrodynamic simulations produce a more compact disc with unusual stellar bar and ring features.}
    \label{fig:mocks}
\end{figure*}

The different gas morphologies naturally lead to different stellar morphologies. To analyse these, we produce a series of mock observational images, using the estimated photometric properties of the star particles. These properties are calculated using stellar population synthesis models based on data given in \citet{bruzual2003} and are provided in the form of mock SDSS broad band luminosities. Following \citet{vogelsberger2014}, we map the $g$, $r$, and $i$-band luminosities to the red, green, and blue channels of an RGB image, binning each channel to create a projected image. The resultant values are then scaled according to algorithms presented in \citet{lupton2004}.  A transparency factor is also set proportional to the maximum binned $g$-band luminosity. The final image does not include effects such as dust attenuation, and is therefore not a true observational mock, but it nevertheless provides much useful information. In particular, it allows us to easily identify prominent morphological features in the remnants. In Fig.~\ref{fig:mocks}, we present face and edge-on mock images of the merger remnant in all high-resolution simulations, as created in this manner. Once again the remnants are seen at $z=0$, except for the 1605 simulations, which are shown at $z=0.11$. For each image, we use the data from all star particles that exist within a depth of $\pm40$ kpc.

As expected, many of the features that were visible in the gas morphology are seen here as well. In particular, the stellar discs produced in MHD simulations are systematically larger than their hydrodynamic counterparts. Naturally, this difference is largest for the largest remnants. The discs in MHD simulations are also thicker and less sharply-defined, following the distribution of the gas. Indeed, the gas distribution is generally well-reflected in the stellar light distribution; for example, the flocculent gas structure observed for the MHD simulations in Fig.~\ref{fig:gas} is seen to support a significant amount of spiral structure here. In contrast, remnants from the hydrodynamic simulations, which had much less small-scale gas structure, show no evidence of spiral arms. Instead, they display distinctive bar and ring morphologies, more typical of barred lenticular galaxies. Morphologies of this kind are not unheard of, but are also certainly not usual for MW-size galaxies. Where they do exist, the rings are often theorised to be a result of resonant forces channelling the gas. This, in turn, is sometimes interpreted as evidence that the galaxy has undergone a mostly secular evolution \citep[e.g.][]{buta2004}. Our simulations show, however, that this must not necessarily be the case.

Once again, simulations 1605-3M and 1526-3H do not quite fit the pattern seen in the other simulations. The chaotic gas dynamics shown in Fig.~\ref{fig:gas} are here reflected by the diffuse interior stellar rings and puffed-up discs. Mergers and tidal interactions have been shown to be able to puff up stellar discs in previous work \citep[e.g.][]{welker2017}, but this is unlikely to be the case here. In particular, we note that this morphology is not seen for the other simulations, even when the remnant has experienced an interaction recently, such as in 1605 and the 1330 simulations, where interloping satellites can be seen in the mock images. Instead, this particular morphology is likely to be a result of gas being lifted above the disc midplane by AGN outbursts. Such outbursts also likely explain the diffuse nature of the ring; star formation is triggered at the edges of the outflow region where gas piles up. However, these outbursts are typically irregularly-shaped and are not consistent over time, meaning that star formation is not reinforced at the same radius as it is for the other ring galaxies. The merger remnant in 1526-3H has a particularly large scale height, which is likely to be a result of the orbit of its central black hole. This is not well-tied to the galactic centre, as in the other simulations, meaning that the AGN feedback is consequently not well-localised. We explore this issue further in Appendix~\ref{appendix:galaxy_tracking}.

\subsubsection{Rotational support and surface density profiles}
\label{chapter3-subsec:rot_support}

\begin{figure*}
    \includegraphics[width=\textwidth]{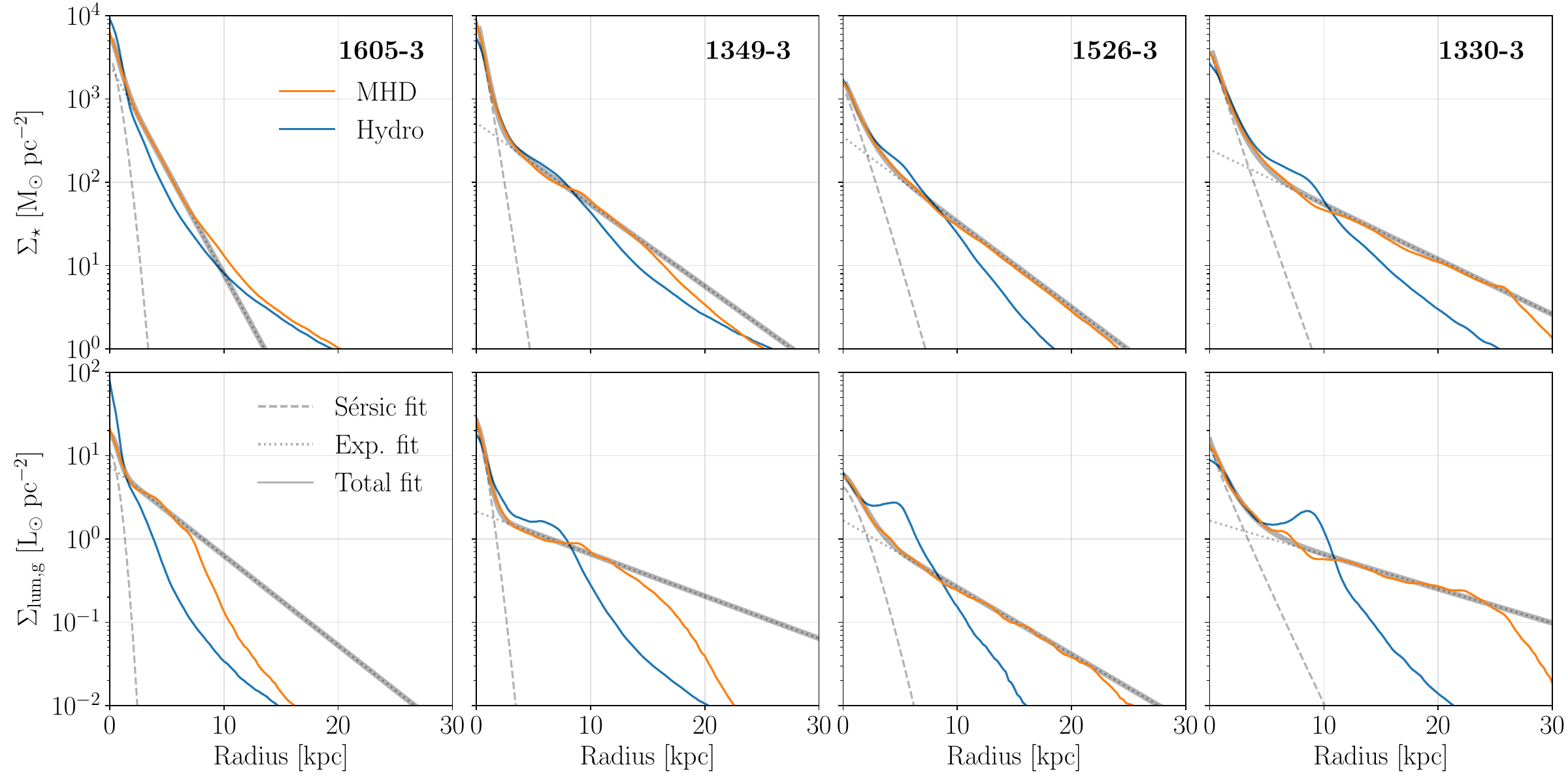}
    \caption[Stellar mass and luminosity surface density profiles for each high-resolution simulation]{\textit{Top row:} stellar mass surface density profiles for the merger remnant in each high-resolution simulation, as seen at $z=0$ ($z=0.11$ for 1605-3). Profiles are calculated over a height of $\pm5$ kpc from the midplane. \textit{Bottom row:} as above, but showing stellar luminosity surface density profiles for the mock SDSS $g$-band. Data from MHD simulations are fit simultaneously with exponential and S\'{e}rsic profiles using a non-linear least squares method. The functional form of the remnants from hydrodynamic simulations prevents a similar fit, particularly in the case of the luminosity profiles, due to the stellar ring component.}
    \label{fig:profiles}
\end{figure*}

Galaxies are often characterised by their amount of rotational support. We parameterise this here for each remnant using the orbital circularity parameter, as defined in \citet{abadi2003}. This is calculated for each star particle within $R_\text{opt}$ as $\epsilon = j_{z} / j(E)$, where $j_{z}$ is the specific angular momentum of the particle aligned with the $z$-axis, and $j(E)$ is the maximum specific angular momentum possible given the particle's specific binding energy, $E$. As in \citet{grand2017}, we then calculate the fraction of stellar mass that kinematically belongs to the disc through two different methods. For the first method, we assume that the bulge makes up twice the mass of the counter-rotating material, where counter-rotating particles are defined by having $\epsilon < 0$. Subtracting the bulge mass from the remaining stellar mass then provides the disc mass. For the second method, we infer the disc mass by summing the mass of particles with $\epsilon>0.7$.  The disc-to-total stellar mass values are shown for each remnant in Table~\ref{tab:sim_data}, with the first and second methods producing the unbracketed and bracketed numbers, respectively. We also show the inferred stellar disc and bulge masses, as given by the first method, in the same table in columns 7 and 8. For many cases, the disc-to-total mass ratio is not substantially changed between physics models. This is a result of the competing factors that produce this ratio; whilst MHD simulations produce much larger, flatter stellar discs, they also have a propensity to produce remnants with bulges. Furthermore, whilst the remnants in hydrodynamic simulations are smaller, they tend to concentrate stellar mass in a ring, thereby increasing their overall circularity fraction. The picture is further muddied, as larger galaxies are more susceptible to warping effects at the disc edge, brought on by the tidal interactions, which reduce the circularity fraction. Examples of this may be clearly seen in the edge-on mock images for 1526-3M and 1330-3M in Fig.~\ref{fig:mocks}. 

In addition to the disc-to-total values shown in Table~\ref{tab:sim_data}, we also calculate these ratios considering only stars that were born within the last 2 Gyr. By doing so, we isolate the young stellar disc in the remnant and exclude contributions from stars whose orbits were disrupted in the merger. Naturally, the resultant disc-to-total ratios tend to be marginally higher, as some of the more eccentric orbits are removed. However, the overall picture remains qualitatively the same: most disc-to-total values are very similar regardless of physics model used and any changes that do exist are not systematic.

This conclusion is somewhat in contradiction with that arrived at by \cite{vandevoort2021}, who find that disc-to-total values increase systematically with the inclusion of magnetic fields in the galaxy formation model. To an extent, this incongruity is explained by a difference in definition; instead of orbital circularity, the parameter $\kappa_\text{rot}$ is used \citep[see, e.g.][]{sales2012}, which measures the fraction of kinetic energy in ordered rotation. Calculating this parameter for our own galaxies, we find that two of the four high-resolution simulations (1330 and 1349) show a comparable increase with the addition of MHD physics. The remaining two, however, show very similar values. We note that 1330-3H and 1349-3H have particularly well-pronounced bars and we believe that it is this feature that reduces the $\kappa_\text{rot}$ parameter below that of their MHD analogues. This interpretation is also consistent with the galaxy studied in the main body of \cite{vandevoort2021}.

From a more observational perspective, the fraction of mass in the bulge and disc components may also be calculated by fitting stellar mass and luminosity surface density profiles with exponential and \citet{sersic1963} profiles. We show these profiles for the remnants from all high-resolution simulations in Fig.~\ref{fig:profiles}, with exponential and S\'ersic profiles overlaid for the MHD simulations. These fits were calculated simultaneously using a non-linear least squares method. Although not shown here, fits were also made for the hydrodynamic stellar mass surface density profiles. The radial scale length, bulge effective radius, and S\'ersic index for the stellar mass surface density profiles for all simulations are given in Table~\ref{tab:sim_data} for comparison. These quantities generally vary little between physics models, and are only systematic in the case of the S\'ersic index, which always has a lower value in the case of the hydrodynamic simulations. This is a result of the more core-like centres for these merger remnants, which most likely results from their stellar bar component.

Data is not provided for the MHD $g$-band luminosity profile fits, but it can be seen by eye that the radial scale length is much longer for the luminosity profile than for the respective mass profile. This is a result of the inside-out growth of the disc, resulting in a younger population on average at the disc edge. For the hydrodynamic simulations, the stellar ring produces an unusual `sombrero' shaped luminosity profile. This shows clearly how star formation has been concentrated in this region. This feature is to be seen to some extent in all of the broad bands available, but is particularly clear in the $g$-band, as presented here in the bottom panel of Fig.~\ref{fig:profiles}. The maxima at the ring component is followed by a dramatic drop in stellar density. This further shows how the expansion of the galaxy has been curtailed. Whilst lenticular galaxies often show some flattening of the luminosity profile at the stellar ring \citep{buta1996}, maxima and sharp drop-offs in density as seen here are highly unusual. The remnants produced in the hydrodynamic simulations are therefore not only poorly described by the standard disc profile fit, but are also atypical of observed galaxies generally.

Whilst better fit than their hydrodynamic counterparts, the luminosity profiles produced in MHD simulations are also not perfectly fit by the standard exponential and S\'ersic profiles. This is because the remnants in our simulations generally consists of a superposition of new and old stellar discs, rather than one unified disc. Furthermore, these discs are both situated on top of a stellar halo, which itself has often been extended and distorted by the merger. The result is that the remnants do not necessarily show the clearly defined up- and downwards-bending breaks needed for an exact fit, and lack a well-defined edge. This means that the integration bounds for calculating disc-to-total ratios using these fits are unclear. Consequently, we refrain from calculating disc-to-total ratios using this method.

In \citet{grand2017}, stellar mass surface density profiles were fit out to $R_\text{opt}$, defined as the radius at which the $B$-band surface brightness drops below $\mu_{B} = 25$ mag arcsec$^{-2}$. Whilst this quantity does not necessarily define the exact edge of the galaxy, as just discussed, it still provides a useful bound on the disc size. In particular, it is able to capture the difference in the disc size produced by each physics model well, and allows for a quantitative comparison between our simulations and the fiducial `Level 4' Auriga simulations. We list $R_\text{opt}$ for all our simulations in the penultimate column of Table~\ref{tab:sim_data}. By comparing the two, it can be seen that the merger remnants from the hydrodynamic simulations are significantly more compact than the fiducial Auriga galaxies. The MHD simulations, on the other hand, produce remnants that are of comparable size to the fiducial galaxies. This is not too surprising as in this case both simulations use the same physics model.

\subsubsection{Impact of MHD for more isolated galaxies}
\label{chapter3-subsec:isolated_galaxies}

\begin{figure*}
    \includegraphics[width=\textwidth]{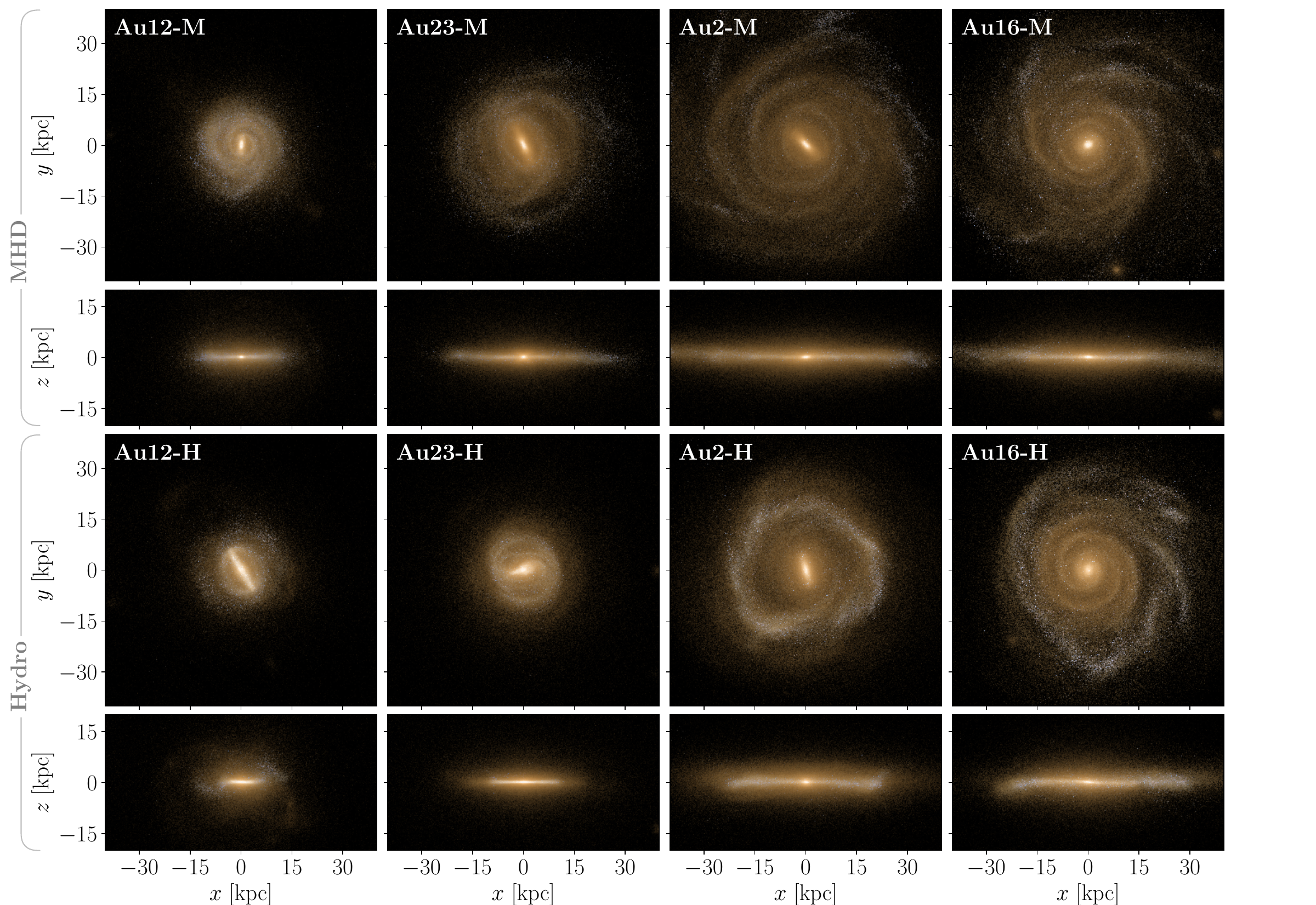}
    \caption[Mock stellar light images for the Auriga simulation variations]{As Fig.~\ref{fig:mocks}, but now showing mock SDSS \textit{gri} composite images for galaxies from the Auriga simulations \citep{grand2017}. The top row continues to show MHD simulations, whilst the bottom row shows their hydrodynamic analogues. Similar morphological features to those seen in Fig.~\ref{fig:mocks} are evident here too, but the differences are less marked. The galaxies presented have all had more quiescent merger histories than those seen in Fig.~\ref{fig:mocks}, but they have not experienced complete isolation. We interpret this as evidence that the features seen are predominantly produced by mergers and that such features are generally stable over time.}
    \label{fig:mocks-auriga}
\end{figure*}

The comparison of our simulations with the fiducial Auriga galaxies also allows us to isolate the role of mergers in producing the observed morphologies. As discussed in Section~\ref{chapter3-subsec:methods_isolated_galaxies}, four of the Auriga simulations were run using both MHD and hydrodynamic physics models. We present face and edge-on mock SDSS images for each of these in Fig.~\ref{fig:mocks-auriga}, created in the same manner as for Fig.~\ref{fig:mocks}. The simulations retain their name from the fiducial runs, albeit with the addition of an `M' or `H', indicating the inclusion of MHD or hydrodynamic physics, respectively. The simulations are also all `Level 4' in the Aquarius nomenclature \citep{marinacci2014}, with a dark matter mass resolution of $3 \times 10^5 \; \text{M}_\odot$. This mass resolution is almost exactly between that of our highest and intermediate-resolution simulations. Trivially, this lower resolution increases the minimum scale at which physical structure may form. In practice, the difference is not great enough to substantially affect the produced morphology.

At first glance, many of the features evident in our own simulations may also be observed in Fig.~\ref{fig:mocks-auriga}. For example, whilst larger stellar bars are now also seen in the MHD simulations, they are still generally more extended in the hydrodynamic simulations. This is taken to its extreme for Au12-H, where the bar transverses almost the entire length of the disc. On top of this, we continue to see stellar rings in hydrodynamic galaxies, whilst they do not appear in MHD simulations. Even the largest hydrodynamic remnant shows evidence of a stellar ring, despite this ring being fairly distorted. Coincidentally, Au16-H is the only hydrodynamic galaxy that does not display an extended stellar bar. It therefore seems likely that the bar structure is providing the resonant forces that generate and maintain the stellar ring. This follows from theoretical predictions that gas should accumulate at Lindblad resonances, under the continuous action of gravitational torques \citep{buta1996, rautiainen2000}.

Whilst there are many similarities between the Auriga galaxies and our own simulations, there are also clear differences. For example, whilst galaxies from MHD simulations are still generally larger than those in hydrodynamic simulations, this difference is nowhere near as stark as it was in our own merger simulations. We may quantify this difference by comparing the optical radii, $R_\text{opt}$, of each pair of galaxies. For our merger simulations, the MHD variant is on average 54\% larger than the hydrodynamic analogue. This relative size difference increases as the galaxies become larger. In comparison, for the Auriga galaxies the $R_\text{opt}$ values of the MHD variant are on average only 20\% larger. This difference does not increase with the size of the galaxies. Indeed, both Au2-H and Au16-H have developed quite large discs, relative to those seen in our own hydrodynamic simulations. The differences in interior morphology are also not as clear-cut. For example, Au16-H has also developed spiral arm structure in the central part of the disc, whilst no remnant in our own hydrodynamic simulations was able to form this structure. The gas distribution in a galaxy can, of course, be heavily disrupted by the existence of a stellar ring or bar component, and the weakness of these features in Au16-H has likely allowed the spiral structure to form here.

In comparing the more isolated galaxies with our own merger remnants, we conclude that the morphological differences are greatest when the merger history is most active. However, we must also bear in mind that the Auriga galaxies are not perfectly isolated. These simulations, too, are cosmological, and the galaxies are only selected to be isolated from significant tidal interactions at late times. Due to the hierarchical growth of structure in $\Lambda$CDM, these galaxies have naturally undergone mergers at earlier times in their history \citep{bustamante2018, monachesi2019}. We therefore propose that the morphological differences seen in Fig.~\ref{fig:mocks} and Fig.~\ref{fig:mocks-auriga} are primarily a result of MHD effects excited by mergers. The full mechanism for this will be explored in greater detail in an upcoming paper.

\subsection{Resolution study}
\label{chapter3-subsec:resolution-study}

\subsubsection{Divergent gas morphology}

\begin{figure*}
    \centering
    \textbf{Resolution study}\par\medskip
    \includegraphics[width=\textwidth]{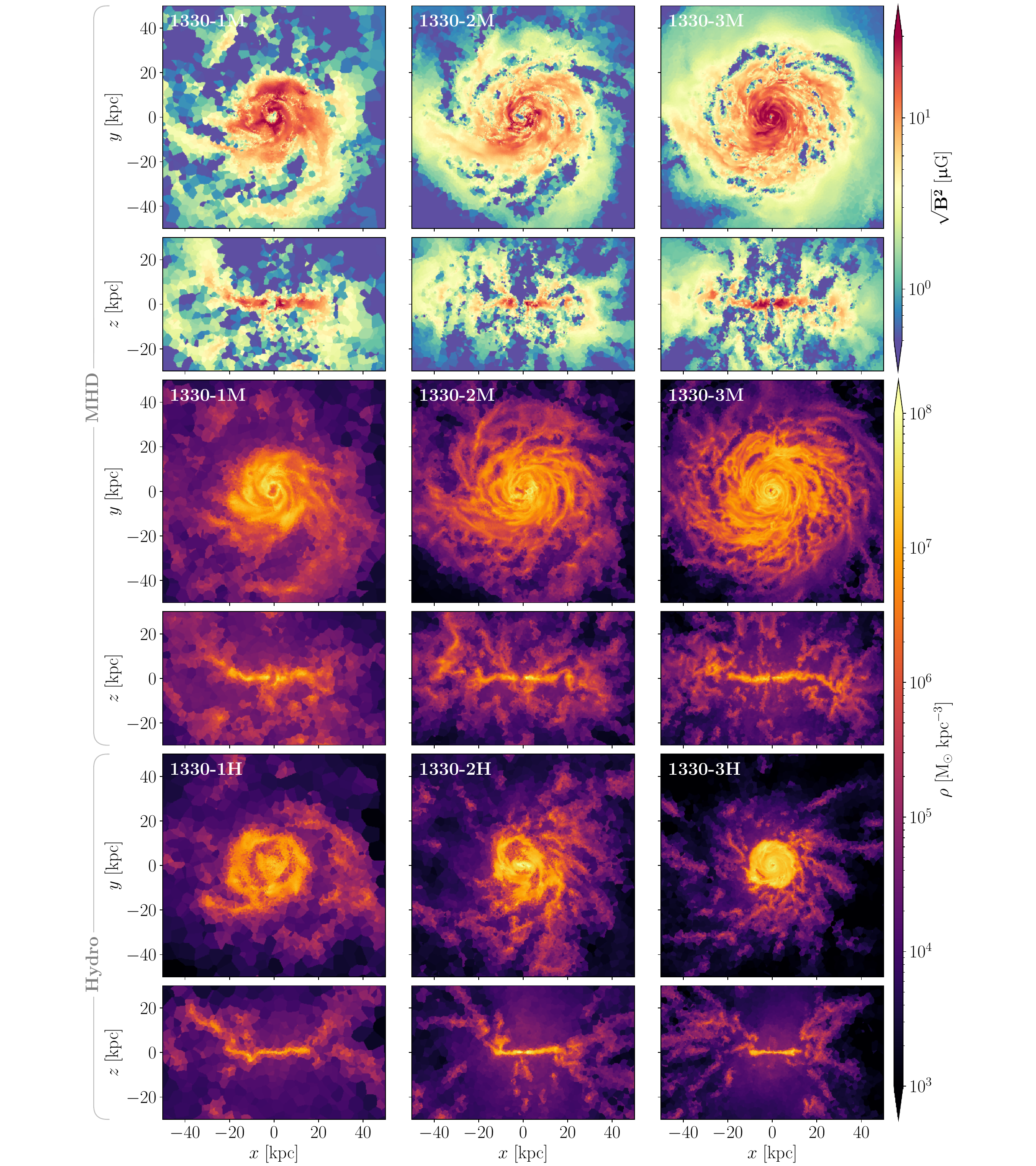}
    \caption[Slices showing magnetic field strength and gas density for simulations with varying resolution]{As Fig.~\ref{fig:gas}, but now showing face and edge-on slices through the merger remnant for simulations with increasing resolution (left to right). The magnetic field as a function of radius becomes smoother and more extended with increased resolution but is otherwise very similar. The gas morphologies, on the other hand, show divergent evolution as a function of resolution. In particular, the gas disc in the hydrodynamic simulations becomes thinner and more compact with each increase in resolution, whilst the gas disc in the MHD simulations grows slightly and becomes noticeably more flocculent.}
    \label{fig:gas_zoom}
\end{figure*}

\begin{figure}
    \centering
    \includegraphics[width=0.5\columnwidth]{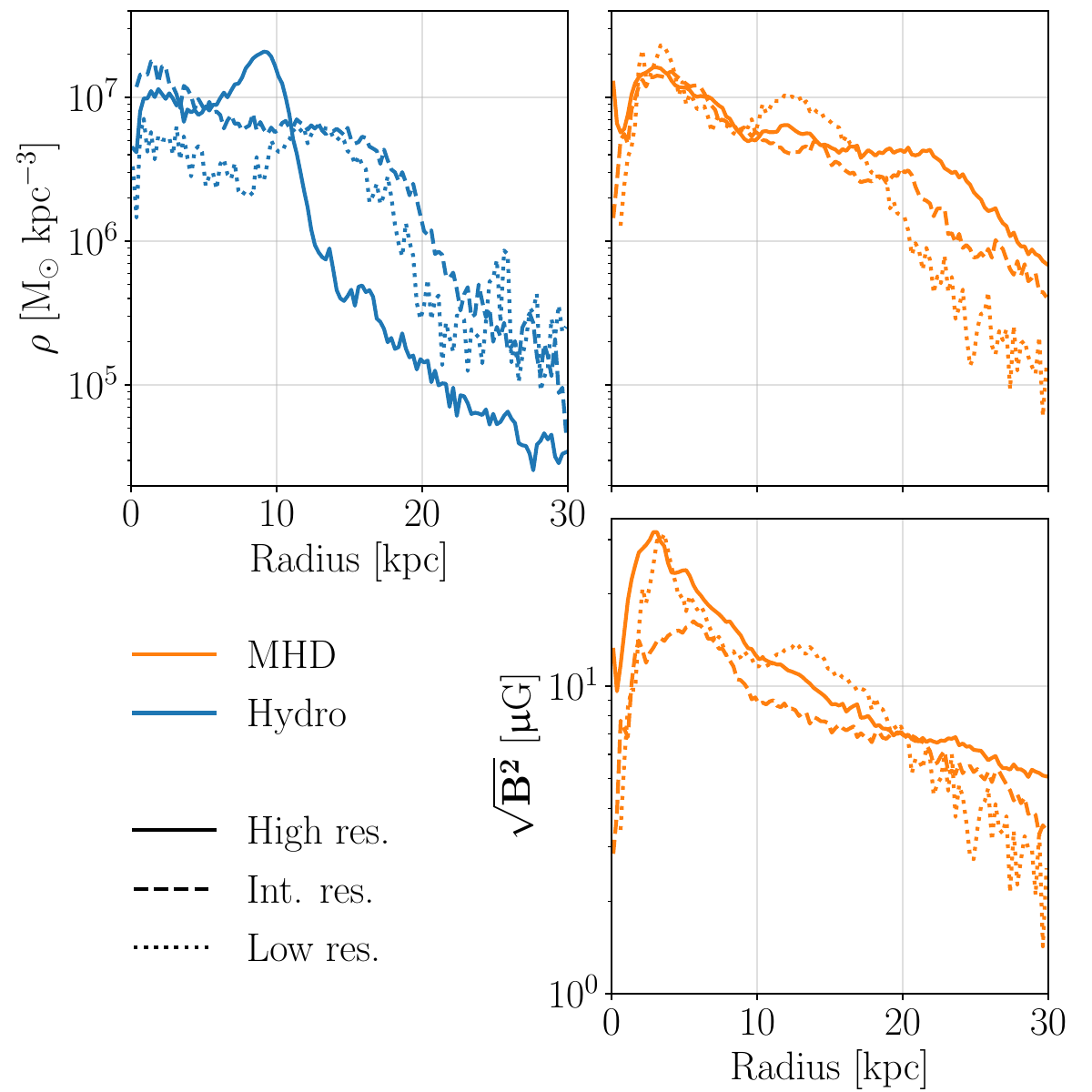}
    \caption[Radially-averaged gas density and magnetic field strength as a function of resolution]{\textit{Top row:} mean gas density as a function of radius for the merger remnants seen in Fig~\ref{fig:gas_zoom}. \textit{Bottom row:} mean magnetic field strength for the MHD simulations of the same figure. In both cases, the mean is calculated over a height of $\pm$5 kpc from the midplane. The gas disc grows in the MHD simulations whilst shrinking in the hydrodynamic simulations.}
    \label{fig:profiles-zoom-gas}
\end{figure}

In addition to our eight high-resolution simulations, we have also run two intermediate and two lower-resolution simulations. In Section~\ref{chapter3-subsec:MHD-global-properties}, we showed that the global properties of the galaxies as a function of resolution were well-converged. However, as for the high-resolution simulations, this does not quite tell the whole story. Some indication of this may already be observed in Table~\ref{tab:sim_data}, where it can be seen that the optical radius of the remnants is remarkably similar for the lower resolution runs for both MHD and hydrodynamic models. This implies that the structure of the remnants may not be converged with resolution. To investigate this effect, we repeat the analysis performed in Fig.~\ref{fig:gas} for our lower resolution simulations. In Fig.~\ref{fig:gas_zoom}, we show slices taken face and edge-on through the main galaxy in each of the 1330 simulations. Each remnant is seen at $z=0$. Once again, in the top two rows we show the magnetic field strength in each gas cell, whilst in the bottom four rows we show the gas density for MHD and hydrodynamic simulations, respectively. We quantify the results at this time by showing radial profiles of these properties in Fig.~\ref{fig:profiles-zoom-gas}.

The magnetic field profiles continue to be generally axisymmetric at lower resolution, but the radial gradient of the field strength is not as smooth. There are also some clear differences in the structure of the magnetic field outside the central 10 kpc. For example, the lowest-resolution simulation shows a rather sharp  decline in its radial profile beginning at approximately 12 kpc, reflecting the decline in gas density starting at the same radius. This effect is somewhat exaggerated by the slight rise in both the mean gas density and mean magnetic field strength shortly before this point, as seen in the right-hand panels of Fig.~\ref{fig:profiles-zoom-gas}. The strength of the magnetic field beyond 20 kpc is also lower and more erratic for the lower resolution simulations; whilst 1330-3M shows a smooth field beyond this point with an average strength of a few $\upmu$G, this is not seen in 1330-1M. The development of a smooth magnetic field at the outer edges of the disc in our highest resolution simulation likely requires the filamentary gas structure seen for this simulation in Fig.~\ref{fig:gas_zoom}. Such structure provides a sufficient average gas density to maintain the field strength, but also provides enough small-scale structure for the gas to be gravitationally bound both to itself and the disc, preventing the density cut-off seen in the lowest-resolution simulation. The development of such small-scale structure, in turn, clearly requires sufficiently high-resolution. As well as affecting structure development, the increased resolution is also likely to be the cause of the smoother, more axisymmetric field profile in the disc in 1330-3M. For this simulation, the higher resolution allows for the gradient in strength between cells to be better resolved, smoothing the resultant magnetic topology. 

Whilst an increase in resolution allows for a smoother gas density and magnetic field strength profile at the disc edge, it also helps a more heterogeneous density distribution to develop above and below the disc. This property is vital for allowing smaller cloudlets to survive the stellar wind, thereby promoting the action of a small-scale fountain flow. As noted earlier, fountain flows of this kind have been found to be crucial for the radial growth of discs in the Auriga model. The supply of gas to the outer edges in this manner may well be supporting the maintenance and growth of the filamentary gas structures at this radius as well.

Considering the disc itself, the radial density profile of the inner 10 kpc is fairly well converged for all MHD simulations as a function of resolution, as can be seen in Fig.~\ref{fig:profiles-zoom-gas}. This is not the case for the hydrodynamic simulations, which show significant variation in their profiles. With this said, there are still clear morphological similarities to be seen for all the hydrodynamic simulations in Fig.~\ref{fig:gas_zoom}. In each case, gas is seen to accumulate predominately at the centre and at the disc edge, with a sharp cut-off in density thereafter. The region above and below the gas disc is also notably lower in density for all resolutions compared to the MHD simulations. The clearance of gas from this region becomes more effective with increased resolution as the disc becomes smaller and the stellar wind becomes stronger. This also reduces the thickness of the disc.

\subsubsection{Divergent stellar morphology}

\begin{figure*}
    \centering
    \textbf{Resolution study}\par\medskip
    \includegraphics[width=\textwidth]{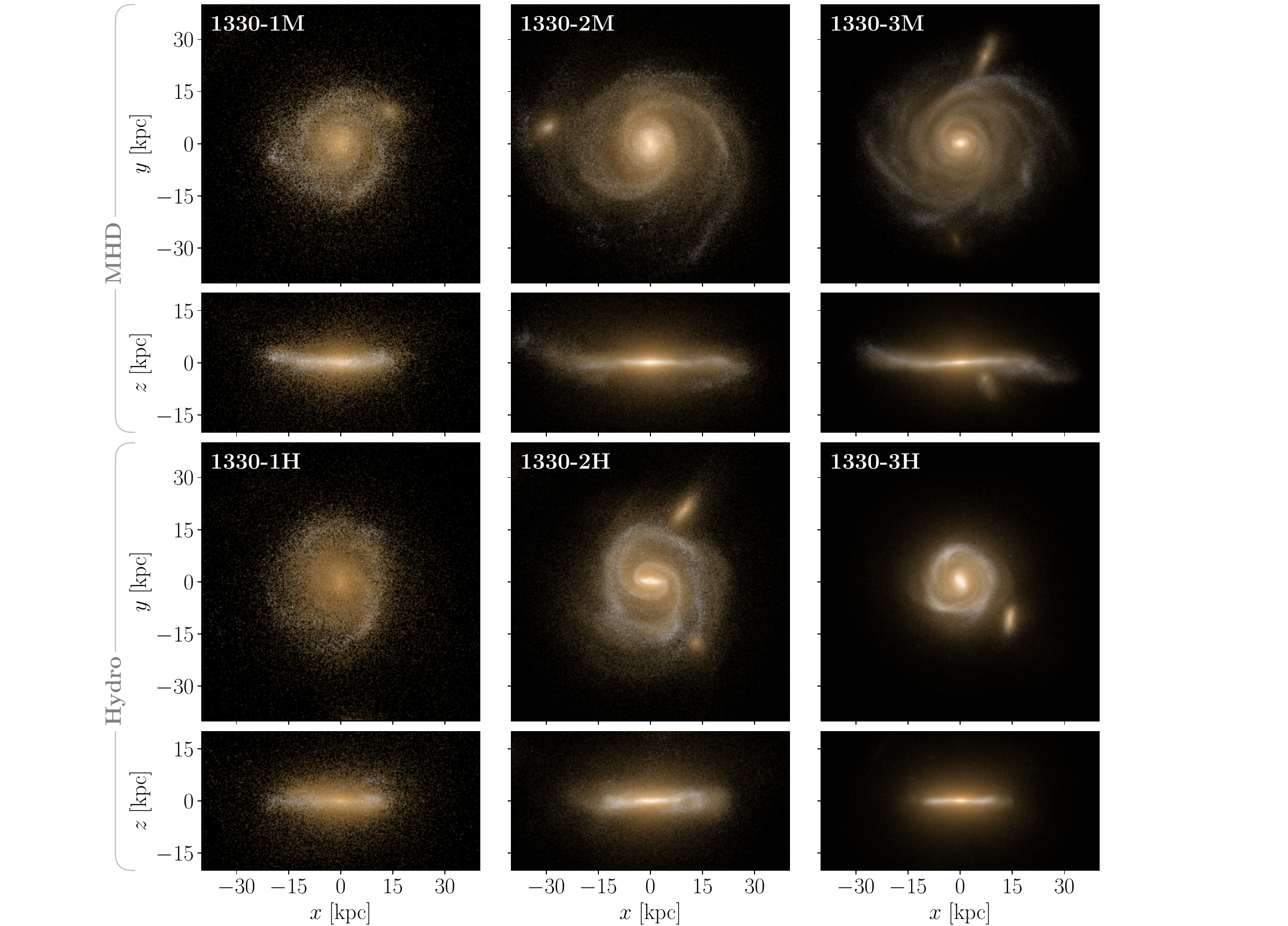}
    \caption[Mock stellar light images for simulations with varying resolution]{As Fig.~\ref{fig:mocks}, but now showing SDSS \textit{gri} composite mock images for simulations with increasing resolution (left to right). In order to conserve the average luminosity per bin, the images of the lower resolution simulations are progressively coarsened. The lowest-resolution simulations produce very similar merger remnants for both MHD and hydrodynamic models. However, by the intermediate-resolution, strong morphological differences are already apparent. This divergence continues with increasing resolution.}
    \label{fig:mocks-zoom}
\end{figure*}

The differences between the remnants as a function of resolution become even more clear when we inspect the stellar morphology. In Fig.~\ref{fig:mocks-zoom}, we show SDSS \textit{gri} composite mock images for the 1330 simulations, created in the same manner as for Fig.~\ref{fig:mocks}. The top two rows show simulations that include MHD physics, whilst the bottom two rows show simulations that include only hydrodynamic physics. In order to roughly conserve the average luminosity per bin, we have adjusted the bin size relative to the spatial resolution.

Whilst some differences could be seen in the gas structure at the lowest-resolution, the stellar morphologies at the lowest-resolution for both MHD and hydrodynamic models look very similar. Indeed, most morphological differences here can be mostly explained by the slight variations in their respective star formation histories. The radius and scale height of both lowest-resolution remnants are practically identical. The galaxies are notably more puffed up than for the higher resolution runs, but this is predominantly a result of the higher softening length used in our lower resolution simulations (see Table~\ref{tab:sim_setup}), combined with the dynamic nature of the systems. This is supported by the observation that the scale height decreases for both sets of simulations with each increase in resolution.

Whilst the lowest-resolution remnants appear very similar to one another, significant differences are already apparent at the next intermediate-resolution level. In this case, the mass resolution is eight times higher and the morphologies have diverged substantially from one another. In fact, many of the morphological differences observed in Fig.~\ref{fig:mocks} may be seen here too, but on a larger spatial scale. For example, the remnant in the MHD simulation has formed a bulge-like centre with two well-defined spiral arms connecting  to it. These fade with distance from the centre, with the galaxy showing an overall shallow radial luminosity gradient. In contrast, 1330-2H shows a bright stellar ring, with two lanes of stars connecting to a large central bar. When viewed together with Fig.~\ref{fig:gas_zoom}, it is clear that these features are strongly distorting the gas morphology. In particular, the bar coincides with the peak in the gas density, showing how it is drawing gas in from elsewhere in the disc. The accumulation of gas in this manner implies that the gas dynamics here are relatively calm. This is in contrast with the signatures of AGN outbursts observed for the remnants from the MHD simulations, as well as the particularly chaotic gas dynamics seen for 1605-3M and 1526-3H in Fig.~\ref{fig:gas}. The morphology seen for 1330-2H in Fig.~\ref{fig:mocks-zoom} is perhaps the most clear evidence seen yet that the ring morphology produced in the hydrodynamic simulations is generated from bar-driven orbital resonances. The full confirmation of this effect is, however, left to an upcoming paper. 

It may also be observed that the ring morphology is fairly robust; in the face-on panel for 1330-2H, it can be seen that the remnant is being harassed by two small satellite galaxies. Despite this, the ring morphology is still very much intact. The reinforcement of the morphology through the accumulation of gas under gravitational torques, as well as the robustness of the morphology to small gravitational perturbations, provides a further indication that the features noted are durable. This helps to explain their appearance in the more isolated galaxies analysed in Fig.~\ref{fig:mocks-auriga}. 

\begin{figure}
    \centering
    \includegraphics[width=0.5\columnwidth]{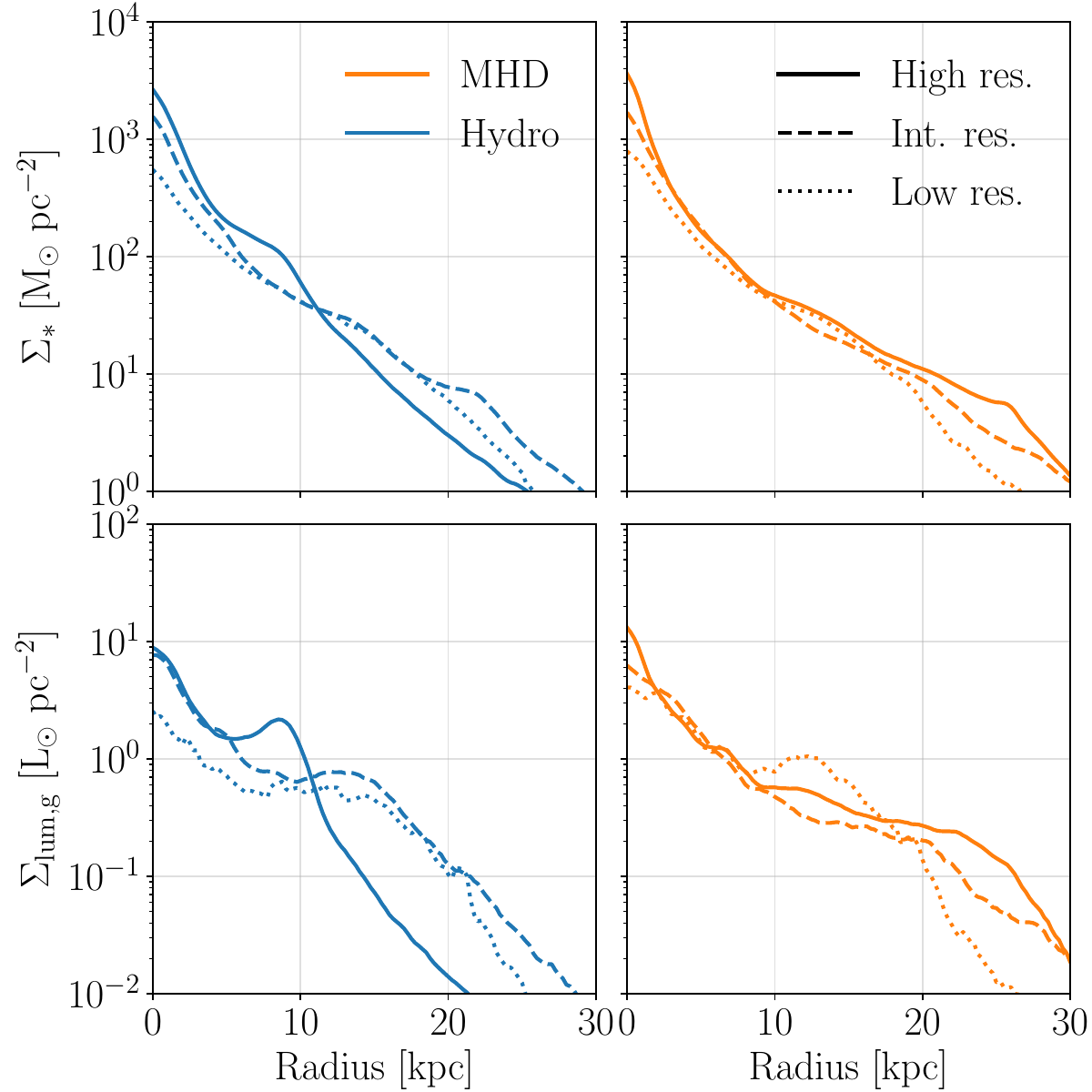}
    \caption[Stellar mass and luminosity surface density profiles for simulations with varying resolution]{As Fig.~\ref{fig:profiles}, but now showing stellar mass and luminosity surface density profiles for the 1330 simulations. Once again, the profiles produced by each physics model are very similar for the lowest-resolution, but diverge with increasing resolution.}
    \label{fig:profiles-zoom}
\end{figure}

In Fig.~\ref{fig:profiles-zoom}, we examine the differences in the structure more quantitatively. As in Fig.~\ref{fig:profiles}, we show the stellar mass surface density in the top row and stellar luminosity surface density in the mock SDSS $g$-band in the bottom row. The hydrodynamic 1330 simulations are shown on the left, whilst the MHD 1330 simulations are shown on the right. As expected, the lowest-resolution profiles are very similar. Indeed, the MHD simulation even shows evidence of a stellar ring in the luminosity profile, which was not wholly clear in the mock images. As the resolution increases, this disappears, and the remnants in the MHD simulations take on a classic disc galaxy profile. This becomes more extended with resolution, but does not change substantially in form. A slight increase in the stellar mass in the inner regions is seen with increased resolution, as was previously discussed in Section~\ref{chapter3-subsec:MHD-global-properties}. This increase is also seen for the hydrodynamic simulations. Despite the significant development in the stellar morphology seen between the low and intermediate resolution hydrodynamic simulations in Fig.~\ref{fig:mocks-zoom}, the overall radial stellar surface density profiles remain broadly similar for both physics models for the inner $\lesssim$20 kpc. The luminosity profiles, too, are not drastically different. Indeed, the full `sombrero'-style profile only develops at the highest-resolution hydrodynamical simulation. Whilst this results from a range of factors, the two key factors are likely to be the smaller softening length and slightly higher star formation rate in the highest resolution simulations. Together these factors allow for a high star-formation density, which is able to launch a strong stellar wind.

\subsubsection{Resolution study of the magnetic dynamo}
\label{chapter3-subsec:amplification_res}

\begin{figure*}
    \includegraphics[width=\textwidth]{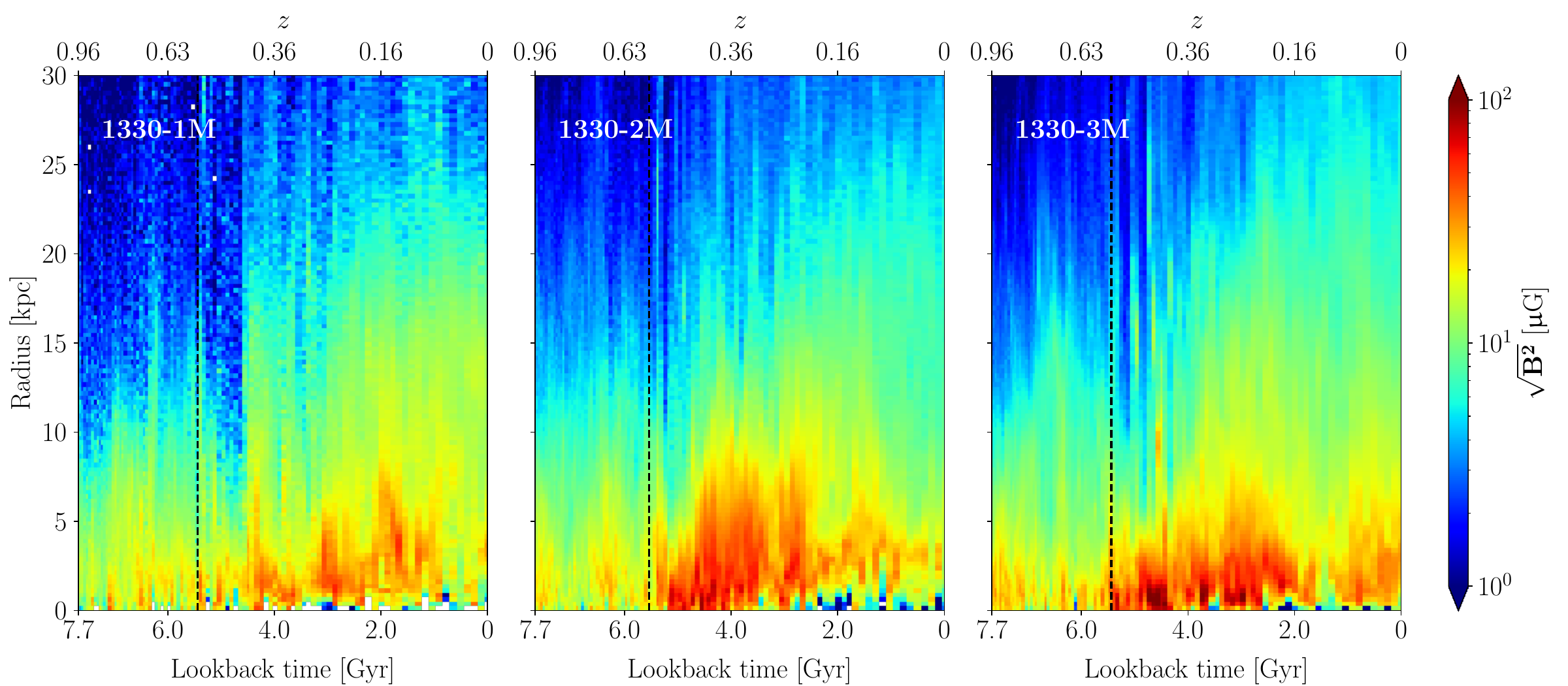}
    \caption[Radially-averaged magnetic field strength as a function of resolution]{The radially-binned mean magnetic field strength, as in Fig.~\ref{fig:BH_mag}, but for simulations with increasing resolution (left to right). The dashed vertical line marks the time of first periapsis. Whilst the outer reaches of the galactic disc show relatively well-converged evolution, the amplification in the inner regions is substantially stronger for higher resolution simulations. The resulting magnetic field strength is better able to affect the gas flow, ultimately producing a different gas and stellar morphology.}
    \label{fig:mag_zoom}
\end{figure*}

\begin{figure*}
    \includegraphics[width=\textwidth]{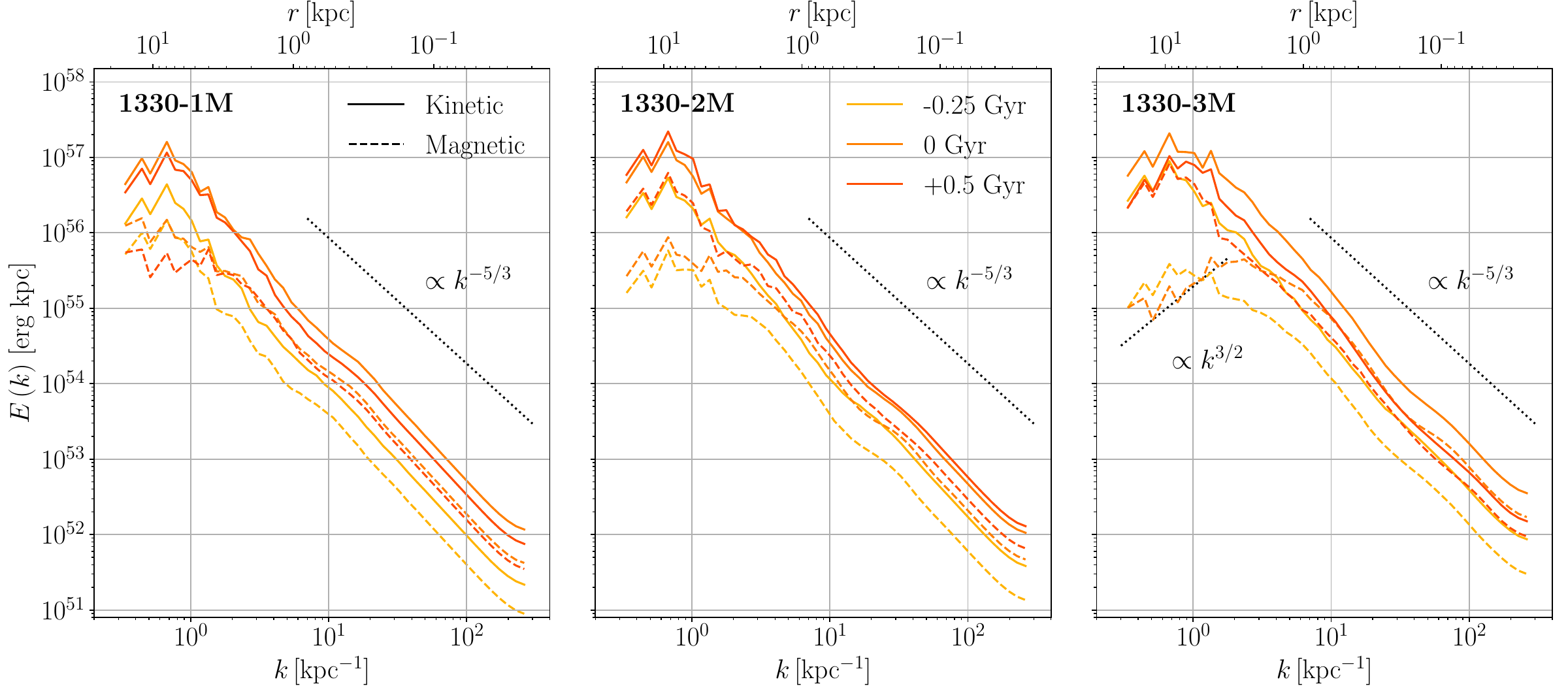}
    \caption[Kinetic and magnetic power spectra for simulations with varying resolution]{Kinetic and magnetic energy power spectra for the 1330-M simulations, calculated using all gas cells within 5 kpc of the galactic centre. Times are shown from first periapsis ($t=0$~Gyr). The black dotted lines show the slopes of a \citet{kolmogorov1941} spectrum ($\propto k^{-5/3}$)  and a \citet{Kazantsev1968} spectrum ($\propto k^{3/2}$), which are theoretically expected for a small-scale dynamo resulting from incompressible turbulence. Whilst the kinetic energy power spectrum initially evolves in a similar fashion for all simulations, the amplification of the magnetic field is more efficient with increased resolution, as seen by the increase in magnetic energy at larger scales over time.}
    \label{fig:power_spec}
\end{figure*}

As resolution increases, the galaxies transition towards a particular morphology. This transition is consistent, as seen by the regularity with which the characteristic morphological features form in the highest-resolution simulations. We therefore argue that the divergence of morphology with resolution points to the importance of including small-scale physics in the simulations, rather than towards general numerical divergence. This claim is bolstered by the global properties of the remnants, which, as previously shown, are broadly converged across all resolutions and physics models. Whilst the ability to form increased amounts of small-scale structure with higher resolution has a non-trivial impact on the remnant development, the impact of the magnetic fields themselves becomes stronger with increased resolution too. A study of this will help us understand how the morphology forms and whether we should expect even further divergence with still higher resolution.

In Fig.~\ref{fig:mag_zoom}, we show the radially-binned mean magnetic field strength as a function of time for simulations of different resolution. This was calculated in the same manner as discussed for Fig.~\ref{fig:BH_mag}. The general radial evolution of the field in the outer reaches is well-converged, showing a similar strength for all times. The inner regions, however, are clearly more strongly amplified for the higher resolution simulations; whilst the lowest-resolution simulation rarely shows mean field strengths higher than 30 $\upmu$G, strengths regularly reach between $40-50$ $\upmu$G in the intermediate-resolution simulation, and upwards of 70 $\upmu$G in the highest-resolution simulation. Such strengths massively increase the ability of the magnetic field to affect the local gas dynamics. This is particularly so shortly after periapsis, when the field strengths are highest, and the rebuilding of the stellar disc has already begun. The evolution of the merger remnant during this time is crucial to its further development, and so the impact of the increased amplification here is itself magnified.

The origin of this increased amplification lies, almost certainly, in the more efficient excitation of the small-scale dynamo. This can be seen by inspecting the evolution of the magnetic and kinetic power spectra around the time of the initial injection of turbulence. We show this for the three different resolution simulations of the 1330 galaxy model in Fig.~\ref{fig:power_spec}. Following \citet{pakmor2011, pakmor2017}, we compute these power spectra by taking the absolute square of the Fourier transforms of the components of $\sqrt{\rho}\,\bs{\bupsilon}$ and $\bs{B}/\sqrt{8\pi}$, respectively \citep[cf.][]{bauer2012}, for gas within a sphere of radius 5 kpc. This is done within a zero-padded box of size $\pm$10 kpc across, and therefore the fundamental mode has a wavelength of 20 kpc. Drops in the power spectra on scales greater than 10 kpc are an artefact of this zero-padding. By considering only gas that lies within 5 kpc of the galactic centre, we isolate the region in which the greatest amplification takes place. We show a time progression from just before the first periapsis, when most of the turbulence is injected, until shortly afterwards. These results do, however, hold for a range of times and radial cuts. In addition, we have also looked at the power spectra of \textit{specific} energies -- both kinetic and magnetic. By examining these, we confirm that only a small part of the evolution can be explained by adiabatic compression.

In each case, the approach of the merging galaxy increases the total kinetic energy in the volume by a factor of roughly 3.5; far above the usual fluctuations. This results in a shift of the kinetic power spectra upwards. The rate of injection of turbulence is sufficiently large such that for a short time afterwards the kinetic energy dominates over the magnetic energy (as was seen previously in Fig.~\ref{fig:BH_mag}). In this regime, the magnetic field is expected to grow exponentially on the corresponding eddy turnover scale, with a growth rate of $\Gamma_l \sim \bupsilon_l / l$, where $\bupsilon_l$ is the eddy velocity at scale $l$ \citep{subramanian1998}. Subsonic, incompressible turbulence, as expected in our mergers, will produce a Kolmogorov-like spectrum of $E(k) \propto k^{-5/3}$ in the inertial range, with $\bupsilon_l \propto l^{1/3}$ \citep{kolmogorov1941}. Together, this results in a growth rate of $\Gamma_l \propto l^{-2/3}$, which means that the magnetic field grows fastest on the smallest scales.

Once the magnetic field becomes strong enough to have a significant dynamical back-reaction at this scale, the exponential growth phase ends and the non-linear growth phase begins \citep{schleicher2013, schober2013}. Under the strong turbulence limit, the magnetic energy in the inertial range then follows the relations: $E(k) \propto k_\perp^{-5/3}$ and $k_\parallel \propto k_\perp^{2/3}$, where $k_\parallel$ and $k_\perp$ are the components of the wave number parallel and perpendicular to the mean magnetic field \citep{goldreich1995}. At smaller scales, such anisotropies are averaged over in our power spectra, producing both kinetic and magnetic energy spectra that follow the perpendicular scaling of $k^{-5/3}$ \citep[see, e.g.][]{beresnyak2019}. For this scaling, the total magnetic energy grows linearly in time \citep{schober2013}. As it saturates at smaller scales, the peak of the magnetic energy spectrum shifts to ever larger scales  (i.e. smaller $k$). At scales larger than this peak, the kinetic energy still dominates, resulting in the familiar Kazantsev slope of $k^{3/2}$ from a kinematic dynamo \citep{Kazantsev1968}. The most clear example of this slope in Fig.~\ref{fig:power_spec} is for 1330-3M, where the difference between the peak scales of the magnetic and the kinetic energy power spectra is greatest.

The kinetic energy itself peaks at the driving scale (or energy injection scale) of the turbulence \citep{cho2009}. This is a factor of a few larger than the greatest scales shown in Fig.~\ref{fig:power_spec}. The amplitude of the kinetic power spectra will decay with time after periapsis, and the spectra as a whole can be affected by sufficiently strong magnetic tension. This is seen once again in 1330-3M, where a strongly saturated magnetic field has shifted the kinetic energy spectrum downwards in the final time step. Indeed, it is only at this highest resolution that the magnetic field is able to saturate to this extent. Despite the initially similar evolution of the kinetic energy budget, it is clear that the magnetic energy evolution at larger scales varies strongly with resolution level. We believe this variation is a direct result of the different growth rates in each simulation; the higher resolution simulations have a lower average cell size, allowing us to resolve smaller eddies. As these eddies have a faster turnover time, the magnetic field saturates sooner at the smallest scale, allowing for the earlier onset of the non-linear growth phase. 

Such differences in growth rates did not affect the fiducial Auriga galaxies, as here the turbulent driving took place over the duration of the galaxy’s initial assembly, being likely a result of cosmic filamentary accretion and stellar feedback \citep{pakmor2017}. In this scenario, the magnetic field could saturate even in lower resolution simulations, as it was given sufficient time in which to do so. In contrast, in our simulations the turbulence is driven by the merger, and the driving time of the turbulence is therefore short compared to the growth and saturation time-scale of the magnetic field. The upshot of this is that in our lower resolution simulations, the field grows too slowly to be able to saturate in the given time frame. The consequence of this can be seen in the time progression for 1330-1M in Fig.~\ref{fig:power_spec}. Here, an increase in the kinetic energy available post-interaction leads to a decrease in the magnetic energy at $k \lesssim 2\,\textrm{kpc}^{-1}$. This happens as the dynamo is unable to saturate quickly enough at higher $k$ values given the new kinetic energy available. In contrast, 1330-2M is able to saturate at the smallest resolved scale sooner, allowing for magnetic energy to cascade to larger scales in time. The result is that the magnetic energy at $k \lesssim 2\,\textrm{kpc}^{-1}$ grows significantly. This process proceeds even more quickly for 1330-3M. In Fig.~\ref{fig:power_spec-high-res}, we show that this behaviour is true of all our high-resolution simulations.

The behaviour of the magnetic field on longer time-scales is more non-linear, as the amplified magnetic fields become better able to impact the gas dynamics and the resultant kinetic energy power spectrum. This leads to fluctuations in the strength of the magnetic field, as can be seen in Figs.~\ref{fig:BH_mag} and \ref{fig:mag_zoom}. The increased speed with which the small-scale dynamo acts in higher resolution simulations, however, helps it to respond to these fluctuations, allowing it to maintain higher magnetic field values over time. Such values are maintained for as long as the kinetic energy is available.

Due to the already high level of saturation reached in the highest-resolution simulations, we do not expect that a further increase in resolution would lead to another significant increase in the average field strength. Higher resolution would, however, result in the yet quicker completion of the exponential growth phase, which would allow the magnetic fields to start affecting star formation earlier. This will continue to be the case until the dissipation scale is resolved. It is unclear to what extent this additional influence would further alter the morphology of the remnants. We expect, though, that other resolution-dependent effects would soon become as, if not more, important, some of which we discuss in the following section.

\section{Discussion}
\label{chapter3-sec:discussion}

\subsection{Why magnetic fields have been ineffectual in previous simulations}

It is clear from the previous subsection that the influence of magnetic fields in the simulations is strongly dependent on resolution. It is also clear that the morphological differences produced by the two physics models are most distinct after a major merger. These points alone explain why previous simulations run with lower resolution, as well as simulations of isolated galaxies, have not observed a similar impact from the inclusion of MHD physics. There are, however, also other factors at play. For example, we note that our simulations included a comprehensive feedback model, including explicit AGN and stellar wind subgrid models, which may have provided a supplementary role in generating turbulence and certainly affected the accretion of gas. In contrast, the idealised MHD merger simulations that have gone before us only included implicit stellar feedback or included no feedback models at all. This is problematic, both in terms of correctly amplifying the magnetic field \citep{martin-alvarez2018, su2018} and on a more general level, as explicit inclusion of feedback has been shown over the last few years to be a crucial step to generating realistic galaxies \citep{hopkins2014, hopkins2018, marinacci2019}. 

Furthermore, we note the importance of reproducing the correct magnetic field strength as a function of radius, especially in the progenitors. As the magnetic energy density increases with $|\bs{B}|^2$, the field strength must only be lowered by a factor of a few before it becomes subdominant again, as was seen in \citet{hopkins2020}. Observations by the next generation of radio telescopes -- e.g. MeerKAT, SKA, LOFAR \citep{haverkorn2019} -- will hopefully be able to place more precise bounds on such radial profiles. A key test for simulations will be to match the Faraday rotation data from observed galaxies. Such a comparison was shown explicitly for our own MHD implementation in \citet{pakmor2018} for the Auriga galaxies.

\subsection{Relevance of this work to general galaxy evolution}

As stated in Section \ref{chapter3-sec:methodology}, our merger scenarios were specifically selected in order to produce interactions where the magnetic fields could have their greatest influence. Therefore, whilst magnetic fields have had a significant impact in these simulations, it is not expected that they should have a significant impact in \textit{every} merger scenario. For example, smaller, more gas-poor progenitors would have weaker initial field strengths, and would be less likely to be able to generate the turbulence necessary to sufficiently amplify the galactic magnetic field. This logic also applies to minor mergers, which would have a less disruptive effect on the main galaxy generally. Furthermore, our results do not apply to the `traditional' merger scenario, where gas is expelled from the galaxy post-merger. Here, there is unlikely to be sufficient time for the magnetic fields to influence the development of the merger remnant before star formation is quenched. The role of magnetic fields in these type of mergers is yet to be determined, but it likely plays a weaker part. With this said, we note that the fraction of mergers that are both major and gas-rich only increases with increasing redshift \citep{hopkins2010, man2016}. It is therefore possible that magnetic fields have a more general impact on galaxy evolution, even if they do not play a strong role in particular types of merger. Indeed, as shown in Section~\ref{chapter3-subsec:isolated_galaxies}, the impact that magnetic fields had in early mergers can be felt several Gyr later even in galaxies that have evolved relatively secularly since.

\subsection{Caveats of the work}
Whilst we believe that our results are numerically robust, there are nevertheless caveats to this work regarding physical fidelity. Firstly, we justify the use of the ideal MHD approximation in our simulations as the magnetic Reynolds number for galaxies, which characterises the relative importance of induction to diffusivity, is expected to be on the order of $\sim10^{18}$ or higher \citep{brandenburg2005}. Without magnetic diffusivity, however, the field topology would be invariant. This would be particularly problematic for simulating a mean-field dynamo, which is believed to reorder the large-scale field to become azimuthally-dominant in the disc on a time-scale of $10^8 - 10^9$ years \citep{shukurov2006}. Non-ideal MHD effects such as reconnection, which acts as a source of magnetic energy loss and potentially also as a source of heating in the galactic halo \citep{raymond1992}, would also be neglected. Whilst we do not explicitly account for magnetic diffusivity in our MHD implementation, some effects will nevertheless be modelled incidentally as a result of the inherent \textit{numerical} diffusivity in our simulations. Indeed, due to our limited resolution, the numerical diffusion in our simulations will be stronger than the physical diffusion. Getting below this scale is still well out of reach of galaxy simulations, and so the use of resistive MHD codes \citep[e.g.][]{marinacci2018} in such simulations in the near future will not be possible. Whilst the numerical diffusivity remains stronger than the physical diffusivity, non-ideal MHD effects could, however, be implemented in future work by using subgrid models, such as those employed by \citet{hanasz2009} in their simulations of isolated disc galaxies. This would increase the physical fidelity of the magnetic diffusion process, although the end result is unlikely to significantly affect the outcome of the simulated dynamos.

Some resolution-dependent MHD effects will also have been neglected in our simulations. For example, idealised MHD simulations have found that magnetic fields are able to stabilise cold streams as they pass through the circumgalactic medium \citep{berlok2019}. Such cold streams would alter the accretion history of the galaxy, and could be particularly important for star formation at high-redshift \citep{keres2005}. Magnetic fields have also been found to be able to support the growth of gas clouds above a critical size in hot winds \citep{sparre2020}. This has important ramifications for the multiphase nature of the circumgalactic medium, which further affects galaxy evolution. Both of these effects, however, once again either require either a substantial increase in resolution or the introduction of new subgrid models.

Without increasing resolution, a substantial step forward in physical fidelity could be taken by implementing cosmic ray physics in our simulation \citep{Pfrommer2017}. Cosmic rays are expected to have a comparable energy density to magnetic fields in the ISM \citep{boulares1990,Pfrommer2017b}, implying that they too can influence galactic evolution. Indeed, in \citet{buck2020}, who also used the Auriga model, cosmic rays were found to be able to significantly affect circumgalactic medium properties; altering the angular momentum distribution of the gas and the subsequent development of the stellar disc. Including cosmic ray physics in our simulations could therefore strongly affect the remnant morphology once again. Apart from increasing physical fidelity, re-running our simulations with cosmic ray physics would also allow us to directly examine whether the equipartition condition holds throughout the mergers. This could naturally have important consequences for inferences made from synchrotron emission observations in the future.

\section{Conclusions}
\label{chapter3-sec:conclusions}

In this paper, we have investigated the impact of magnetic fields on galaxy mergers. We have done this by comparing MHD and hydrodynamic simulations run from the same initial conditions. Our simulations were fully cosmologically-consistent and used a state-of-the-art zoom-in code. This allowed us to simulate the galaxy and its immediate neighbourhood to high-resolution, whilst still accounting for the influence of large-scale structure. To the best of our knowledge, this is the first time that MHD zoom-in simulations have been used in this way to study galaxy mergers. 

The impact of magnetic fields on mergers is an extremely complex problem and an accurate implementation of the physics involved is technically challenging. In order to increase the reliability of our results, we therefore built upon previously proven work. In particular, our simulations employed the Auriga galaxy formation model and were run using the \textsc{arepo} moving-mesh code. The Auriga model includes a range of physically-motivated subgrid models, with parameters that do not require retuning between resolution levels, and includes an MHD implementation that has been shown to sufficiently fulfil the divergence constraint even in dynamic environments. \textsc{Arepo}, meanwhile, has been shown to have significantly better numerical accuracy than competing codes when applied to a range of relevant physical problems, including several that have direct relevance to our investigation

In total, we ran eight high-resolution simulations, with a dark matter mass resolution $\sim38.5$ times finer than the fiducial Illustris run. We have supported these with two intermediate, and two lower-resolution runs. The global properties of each simulation are broadly converged between resolution level, indicating that the results of our simulations depend on the physics included and not on the numerical implementation. We list our major conclusions from this work below:

\begin{itemize}
    \item Structural properties of the merger remnant depend strongly on the physics model. In particular, MHD simulations produce large disc galaxies that display extensive spiral structure. This is underlain by a flocculent gas disc with a shallow radial gradient. Merger remnants from hydrodynamic simulations, on the other hand, are systematically smaller, and often display distinctive stellar bar and ring elements. The gas discs in these remnants are significantly thinner, with a flatter density profile that drops abruptly at the disc edge (Figs.~\ref{fig:gas}, \ref{fig:mocks}, \ref{fig:profiles}).
    \item Simulations of galaxies with more quiescent merger histories show similar, but less marked, morphological differences. As these galaxies are not completely isolated from tidal interactions during their lifetime, we argue that the morphological differences we observe are predominantly a result of MHD effects excited by mergers (Fig.~\ref{fig:mocks-auriga}).
    \item Global properties are not significantly affected by the inclusion of MHD physics. In particular, the star formation history of the remnant is left relatively unchanged (Fig.~\ref{fig:SFR_dist}). Indeed, rather than suppressing star formation, the total stellar mass is often marginally higher in the MHD runs at $z=0$ (Table~\ref{tab:sim_data},  Fig.~\ref{fig:bound_gas_stellar}). 
    \item Merger remnants in the hydrodynamic simulations lose significantly more gas than their MHD counterparts (Table~\ref{tab:sim_data},  Fig.~\ref{fig:bound_gas_stellar}). In fact, in some MHD simulations, the loss of gas mass can be accounted for almost entirely by the increased stellar mass. This indicates that we do not see magnetically-driven winds in our simulations. The reduced gas mass in hydrodynamic simulations is likely a result of stronger galactic winds, caused by a similar amount of star formation taking place in a more compact volume (Figs.~\ref{fig:gas} and \ref{fig:mocks}).
    \item During the merger, the mean magnetic field strength in the inner $\lesssim5$ kpc of the disc is boosted by up to an order of magnitude. At the same time, the field outside this range drops by an order of magnitude. The magnetic field can become locally dominant during this time, with the magnetic energy density able to reach several times the thermal energy density in a substantial number of gas cells. These effects are typically apparent for at least 1.5 Gyr after first periapsis and fade with the rebuilding of the disc. (Fig.~\ref{fig:BH_mag}).    
    \item The differences noted above are only observed once sufficient resolution is reached (Figs.~\ref{fig:gas_zoom}, \ref{fig:mocks-zoom}, \ref{fig:profiles-zoom}). This is explained by Fig.~\ref{fig:mag_zoom}, where it is shown that the merger amplifies the galactic magnetic field more strongly in higher resolution simulations. We interpret this as evidence that our results are dependent on the sufficient excitement of a small-scale dynamo (Fig.~\ref{fig:power_spec}).
\end{itemize}

The above results apply particularly to gas-rich major mergers, where the merger remnant is able to re-form a substantial stellar disc. The impact of such mergers on galaxy evolution taken as a whole, however, is likely to be substantial. Furthermore, the influence of magnetic fields only increases when we consider their impact on the stability of gas flows and their impact on important anisotropic processes, such as the transport of cosmic rays. Our results are therefore a clear indication that the inclusion of MHD physics is critical for reliably modelling galaxy evolution in future simulations. A full elucidation of the mechanism behind the morphological differences seen in this paper will be presented in an upcoming work.

\newpage
\section{Appendix}
\subsection[Appendix A: Impact of MHD on the ISM]{Impact of MHD on the ISM}
\label{appendix:ISM}

\begin{figure}
    \centering
    \includegraphics[width=0.5\columnwidth]{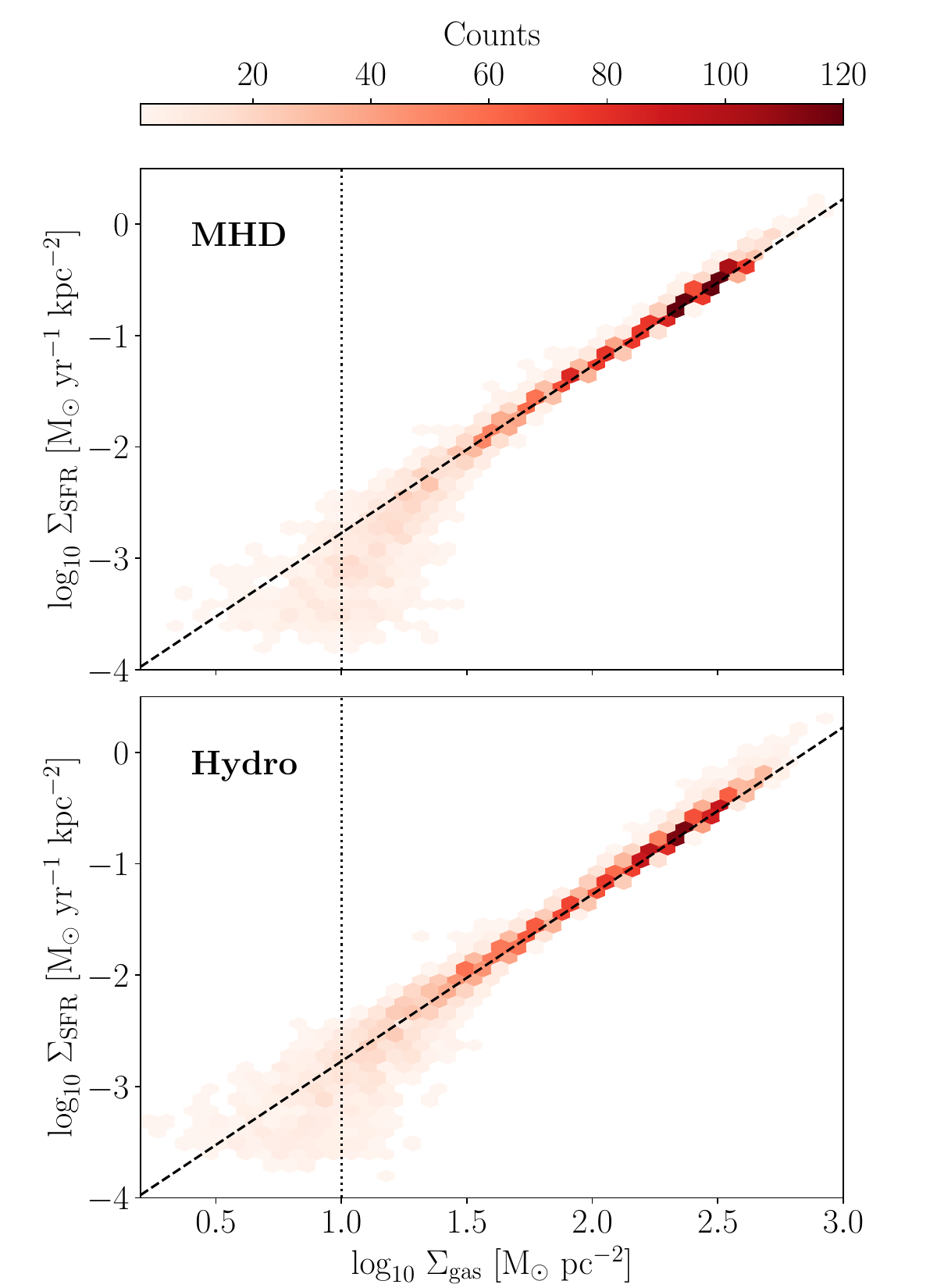}
    \caption[Kennicutt-Schmidt relation in MHD and hydrodynamic simulations]{\textit{Top panel:} star formation rate surface density as a function of gas surface density for 1349-3M, as seen at a lookback time of $\sim$4 Gyr. The dashed line shows a Kennicutt-Schmidt relation \citep{schmidt1959, kennicutt1998} with exponent 1.5. The dotted line indicates the approximate position of the cut-off in the star formation rate. \textit{Bottom panel:} as above, but for 1349-3H. Both follow the same relation, despite the strong amplification of the magnetic field in 1349-3M during this time.}
    \label{fig:kennicutt_schmidt}
\end{figure}

In Section~\ref{chapter3-subsec:set-up}, we claimed that it was not necessary to recalibrate our ISM subgrid model when including magnetic fields in the simulation. This statement is non-trivial: whilst the effective pressure in our ISM subgrid model is a function of density only in the hydrodynamic simulations \citep{springel2003}, in the MHD simulations there is an additional magnetic pressure term to consider. We can check the influence of this additional term on the ISM model by investigating its impact on the relation between the star formation rate surface density, $\Sigma_\text{SFR}$, and the gas surface density, $\Sigma_\text{gas}$. Stars form probabilistically out of our simulated ISM with a gas consumption time-scale set to match that observed by \cite{kennicutt1998} for disc galaxies in the local Universe. This results in the ISM following the well-known Kennicutt-Schmidt \citep{schmidt1959, kennicutt1998} relation of: $\Sigma_\text{SFR} \propto (\Sigma_\text{gas})^n$. If our ISM subgrid model required recalibration, the form of this relation would be dependent on the physics model used.

We show the relation between $\Sigma_\text{SFR}$ and $\Sigma_\text{gas}$ for 1349-3M and 1349-3H in Fig.~\ref{fig:kennicutt_schmidt}. The 1349 simulations were chosen due to the particularly strong amplification in 1349-3M, but naturally the results apply to all our simulations. We have chosen a time when the magnetic field is highly amplified in the MHD simulation. At later times we observe a similar relation, but with the gas density covering a much narrower range. In both cases, the surface densities are calculated by taking a face-on projection with depth $\pm$1 kpc from the midplane. We choose this depth to make sure that we are predominantly considering gas in the disc. It can be seen that for both physics models, the star formation rate surface density follows the Kennicutt-Schmidt relation with exponent $n=1.5$ \citep[cf.][]{springel2003}. This is true over a broad range of values. At lower gas surface density values, the relation becomes more scattered as the star formation threshold density is approached. The peak of the distribution is higher in the MHD simulation than the hydrodynamic analogue, which is a result of morphological differences between the two remnants. The strong similarity between both relations otherwise supports our assertion that the ISM subgrid model must not be recalibrated when introducing magnetic fields.

\subsection[Appendix B: Galaxy Tracking and Black Hole Dynamics]{Galaxy Tracking and Black Hole Dynamics}
\label{appendix:galaxy_tracking}

\begin{figure*}
    \centering
    \includegraphics[width=\textwidth]{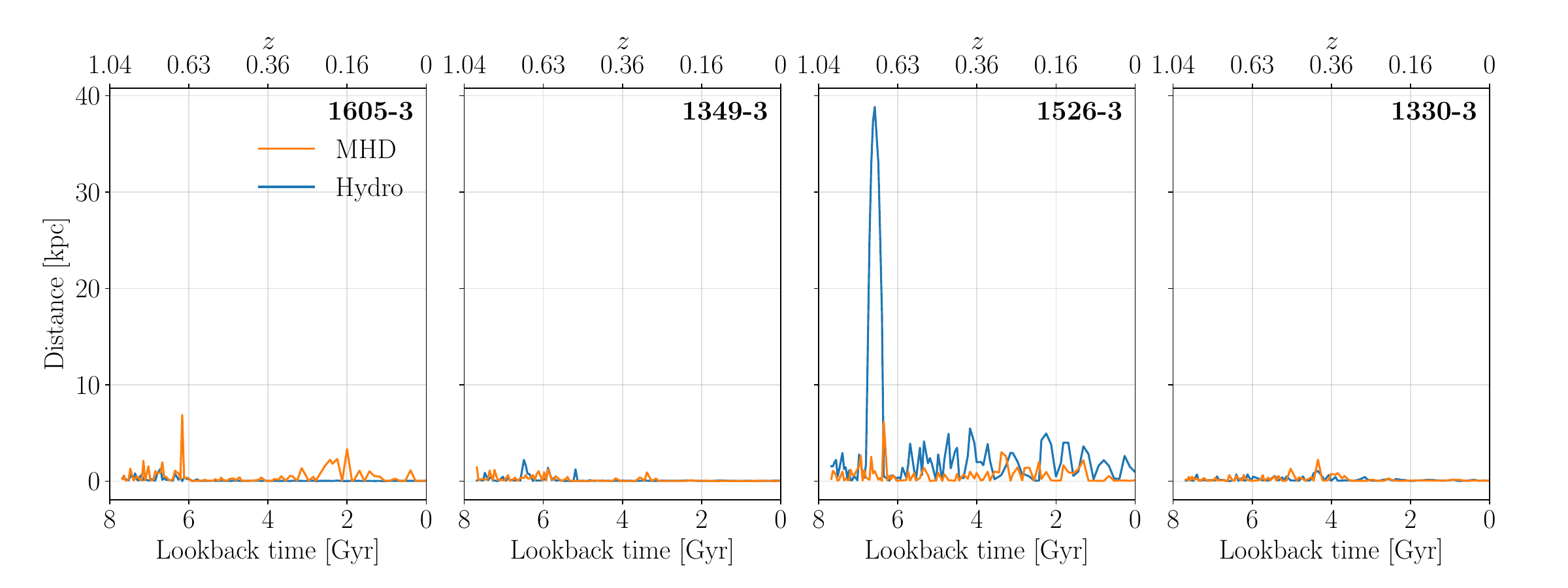}
    \caption[Distance between the galactic centre and the closest black hole as a function of time]{The distance between the galactic centre and the closest black hole for all high-resolution simulations as a function of time. In general, this distance stays well below 5 kpc, confirming the reliability of our galaxy tracking method. The `wandering' nature of the black hole in simulation 1526-3H likely contributes to the unusual morphology displayed by the corresponding merger remnant.}
    \label{fig:black_hole_dist}
\end{figure*}

In Section~\ref{chapter3-subsec:galaxy_tracking}, we claimed that tracking a galaxy between snapshots is frequently akin to tracking the black hole particle that resides in that galaxy. We support this claim in Fig.~\ref{fig:black_hole_dist}, where we show the distance between the most bound gas cell in the galaxy (i.e. the galactic centre) and the closest black hole for each high-resolution simulation. For the vast majority of snapshots, this distance is always well under 5 kpc, confirming the validity of our tracking method. An exception to this rule is simulation 1526-3H, where the black hole is not well-tied to the galactic centre. In this merger scenario, the secondary progenitor passes directly through the primary progenitor. For a short time immediately afterwards, the black hole then `hitches a ride', becoming gravitationally bound to the merging galaxy. This can be confirmed by comparing its distance from the main galaxy between 7.11 and 6.35 Gyr to the distance between the merging galaxies, as seen in Fig.~\ref{fig:SFR_dist}. 

On the black hole's return to the main galaxy, it never quite loses its newly-gained orbital energy, oscillating around the galactic centre instead. During this time, the black hole continues to accrete gas, and consequently continues to inject energy into neighbouring gas cells. The subsequent AGN outbursts disrupt the gas, producing a similar morphology to that of 1605-3M (see Fig.~\ref{fig:gas}). This similarity is unexpected as the black hole in 1526-3H grows only a quarter as large as that in 1605-3H post-merger, meaning that the AGN outbursts should be significantly less influential (see Section~\ref{chapter3-subsec:set-up}). Indeed, the black hole growth in 1526-3H is similar to that of 1349-3H and 1330-3H, both of which show no signatures of strong outbursts in their gas morphology. We therefore argue that it is the unlocalised nature of the feedback in 1526-3H that is behind the strongly disrupted morphology. This, in turn, produces the unusual stellar morphology seen in this simulation.

\subsection[Appendix C: Evidence for a Small-scale Dynamo in Action]{Evidence for a Small-scale Dynamo in Action}
\label{appendix:small-scale_dynamo}

\begin{figure*}
    \centering
    \includegraphics[width=\textwidth]{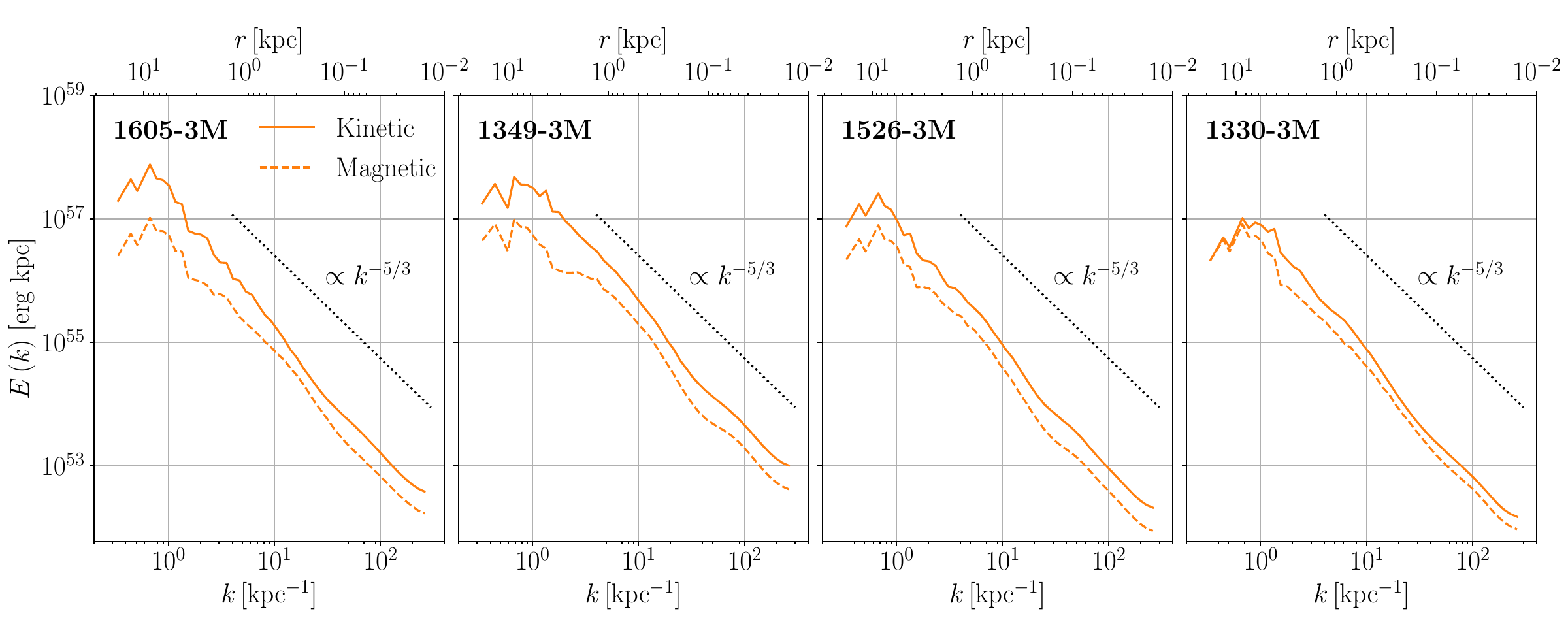}
    \caption[Kinetic and magnetic power spectra for each high-resolution simulation]{Kinetic and magnetic energy power spectra for the highest-resolution MHD simulations, calculated for gas within a sphere of 5 kpc centred on the galactic centre. The power spectra are shown at 0.5 Gyr after the time of periapsis for each simulation. The black dotted lines show the slopes of a Kolmogorov spectrum ($\propto k^{-5/3}$) \citep{kolmogorov1941}, which is theoretically expected for a small-scale dynamo resulting from incompressible turbulence. In each case, the magnetic field is strongly saturated.}
    \label{fig:power_spec-high-res}
\end{figure*}

In Section~\ref{chapter3-subsec:amplification_res}, we discussed the development of a small-scale dynamo in our simulations as a function of increasing resolution. In Fig.~\ref{fig:power_spec-high-res}, we support this with kinetic and magnetic energy power spectra for each of our high-resolution MHD simulations. These were created in the same manner as for Fig.~\ref{fig:power_spec} over the same radius of 5 kpc within a zero-padded box of 10 kpc across. The profiles are shown at 0.5 Gyr after the respective time of periapsis in each simulation (see Section~\ref{chapter3-subsec:sims}). In choosing this time, we show the power spectra at a similar point in the evolution of the magnetic field. It can be seen that for each simulation, the kinetic energy exhibits a Kolmogorov-like spectrum, which is consistent with the volume-filling phase of the gas being both turbulent and subsonic. The difference in normalisation in each plot can be mostly explained by the difference in mass evolution. At high $k$ values, the magnetic energy saturates at around 50 per cent of the kinetic energy, which is consistent with subsonic turbulent box simulations where the forcing was solenoidal \citep{federrath2016}. In each simulation, the magnetic field has grown similarly quickly. Indeed, the magnetic energy is almost fully saturated in each case, with the peak magnetic energy occurring at the driving scale of the turbulence. This means almost the entire magnetic energy spectrum is in the non-linear dynamo phase, and consequently there is little evidence of the Kazantsev-like slope at large scales. The decline of the power spectra at scales larger than 10 kpc is an artefact of the zero-padding we use. Overall, Fig.~\ref{fig:power_spec-high-res} supports our understanding of the growth of the magnetic field, as analysed in Section~\ref{chapter3-subsec:amplification_res}.
\setcounter{equation}{0}
\thispagestyle{empty}

\defcitealias{whittingham2021}{W21}

\graphicspath{{Images/Chapter4/}}

\Chapter{The impact of magnetic fields on cosmological galaxy mergers -- II.}{Modified angular momentum transport and feedback}  
\label{chapter:paper-two}

\noindent This chapter is based on the published paper by Whittingham, J., Sparre, M., Pfrommer, C., and Pakmor, R. in Monthly Notices of the Royal Astronomical Society, Volume 526, Issue 1, p.224-245. The work expands upon ideas first presented in my Masters thesis. A breakdown of the exact differences is provided in Chapter \ref{publications_list}.

\textit{The role of magnetic fields in galaxy evolution is still an unsolved question in astrophysics. We have previously shown that magnetic fields play a crucial role in major mergers between disc galaxies; in hydrodynamic simulations of such mergers, the Auriga model produces compact remnants with a distinctive bar and ring morphology. In contrast, in magnetohydrodynamic (MHD) simulations, remnants form radially-extended discs with prominent spiral arm structure. In this paper, we analyse a series of cosmological ``zoom-in'' simulations of major mergers and identify exactly \text{how} magnetic fields are able to alter the outcome of the merger. We find that magnetic fields modify the transport of angular momentum, systematically hastening the merger progress. The impact of this altered transport depends on the orientation of the field, with a predominantly non-azimuthal (azimuthal) orientation increasing the central baryonic concentration (providing support against collapse). Both effects act to suppress an otherwise existent bar-instability, which in turn leads to a fundamentally different morphology and manifestation of feedback. We note, in particular, that stellar feedback is substantially less influential in MHD simulations, which allows for the later accretion of higher angular momentum gas and the subsequent rapid radial growth of the remnant disc. A corollary of the increased baryonic concentration in MHD simulations is that black holes are able to grow twice as large, although this turns out to have little impact on the remnant's development. Our results show that galaxy evolution cannot be modelled correctly without including magnetic fields.}

\section{Introduction} 
\label{chapter4-sec:introduction}

Magnetic fields permeate the Universe at every scale yet observed. The galactic scale is, of course, no exception to this. This has been confirmed both for our own galaxy and external galaxies through a wide range of techniques, including Zeeman splitting \citep{heiles2009, li2011, mcbride2015}, stellar light polarisation \citep{heiles1996, pavel2014, berdyugin2014}, dust polarisation \citep{hildebrand1988, lopez-rodrigues2020}, Faraday rotation \citep{manchester1972, han2018}, and synchrotron radiation \citep{dumke1995, krause2009, bennett2013, planck2016}\footnote{See reviews by \citet{beck2015} and \citet{han2017} and references therein for a more comprehensive list of examples.}. The latter is particularly powerfully demonstrated through the far-infrared (FIR) -- radio correlation, which implies volume-filling magnetic fields for a vast range of galaxy sizes, masses, and morphologies \citep{lacki2010, werhahn2021, pfrommer2022}.

Disc galaxies in the local Universe are of particular interest, as observations imply that field strengths in these are on the order of ${\sim}10\,\upmu$G \citep{beck2011}. This places the energy density of the magnetic field in approximate equipartition with the turbulent, thermal, and cosmic ray energy densities in the interstellar medium (ISM) \citep{boulares1990, beck1996, beck2015}, making it a dynamically-important component at the present time. The \textit{long-term} impact of magnetic fields on galactic evolution, however, is still an open question.

To answer this question from a theoretical standpoint, we require the use of cosmological simulations, in which a full range of important environmental factors, such as accretion histories, circumgalactic media (CGM), and mergers, can be accounted for and treated self-consistently \citep{vogelsberger2020}. Many of the latest generation of cosmological simulations now include an implementation of magnetohydrodynamics (MHD), including zoom-in simulations of galaxies such as Auriga \citep{grand2017}, FIRE-2 \citep{su2017}, and those performed by \citet{rieder2017}, as well as larger box simulations such as CHRONOS++ \citep{gheller2016, vazza2017}, Illustris TNG \citep{pillepich2018}, and HESTIA \citep{libeskind2020}. However, the use of different numerical codes, seed fields, and divergence cleaning methods, amongst other factors, has led to inconclusive results. For example, in some simulations of more isolated galaxies, the magnetic field is typically subdominant for the entire runtime \citep{hopkins2020}, whilst in others the magnetic field does reach equipartition, but either only in specific density ranges \citep{ponnada2022} or only at late times \citep{pakmor2017}, thereby limiting its impact. On the other hand, in simulations that start with a sufficiently strong primordial field, magnetic fields are able to suppress star formation rates \citep{marinacci2016} and reduce disc sizes \citep{martin-alvarez2020, katz2021}. The seed strengths used in these simulations are, however, beyond the currently accepted upper limits achievable by standard battery processes \citep{gnedin2000, attia2021}.

In \citet[hereafter \citetalias{whittingham2021}]{whittingham2021}, we pointed out that mergers can raise field strengths to dynamically-important values, even in simulations that start with significantly weaker seed fields. To show this, we ran four pairs of high-resolution cosmological zoom-in simulations of major mergers between disc galaxies, using the Auriga galaxy formation model \citep{grand2017}. We showed that, under this scenario, MHD simulations produce remnants with systematically different sizes and morphologies compared to their hydrodynamic analogues. Specifically, for the merger scenarios we simulated, remnants in the MHD simulations are larger with flocculent gas discs and spiral arms, whilst those in the hydrodynamic simulations are more compact and exhibit conspicuous bar and ring elements.

In the same paper, we also revisited four pairs of high-resolution simulations, originally run by \citet{pakmor2017}. These employ the Auriga model as well, but apply it to galaxies with considerably more quiescent merger histories. Here too, however, similar, albeit more subtle, morphological differences are evident between the MHD and hydrodynamic variants. We interpreted the observation that the differences are stronger in our own simulations as evidence that this is an MHD effect excited by mergers. The observation of similar morphological differences in simulations of more isolated galaxies should not be surprising, though, as few if any galaxies will be untouched by mergers during their history. Indeed, mergers constitute a fundamental part of hierarchical structure formation – a cornerstone of $\Lambda$CDM (a cold dark matter Universe with a cosmological constant).

Significantly, for each of the eight sets of high-resolution simulations mentioned, only the MHD simulations produce galaxies consistent with observations. By analysing kinetic and magnetic energy power spectra for simulations with varying resolution, we demonstrated in \citetalias{whittingham2021} that sufficiently small-scale turbulence must be resolved in order to amplify the  magnetic fields in the necessary time frame and thereby realise this effect. We did not, however, explain \textit{how} the magnetic fields were affecting the re-growth of the disc. We answer this question in this paper.

The paper is organised as follows: in Sec.~\ref{chapter4-sec:methodology}, we summarise the merger scenarios and our numerical methods. In Sec.~\ref{chapter4-sec:analysis}, we identify how the mergers evolve differently under hydrodynamic and MHD physics models (Sec.~\ref{chapter4-subsec:how_evolution_differs}) and propose a mechanism by which magnetic fields are able to cause this effect (Sec.~\ref{chapter4-subsec:model}). We then provide evidence for this model, with particular emphasis on how magnetic fields affect angular momentum transport and subsequent orbital resonances (Sec.~\ref{chapter4-subsec:angmom} and Sec.~\ref{chapter4-subsec:resonances}) and how they alter the manifestation of stellar and black hole feedback (Sec.~\ref{chapter4-subsec:feedback} and Sec.~\ref{chapter4-subsec:AGN}). In Sec.~\ref{chapter4-sec:discussion}, we discuss the applicability of our results to different merger scenarios, numerical codes, and galaxy formation models. Finally, in Sec.~\ref{chapter4-sec:conclusions}, we summarise our conclusions.

\section{Methodology}
\label{chapter4-sec:methodology}

In this work, we analyse the four pairs of high-resolution cosmological zoom-in simulations first presented in \citetalias{whittingham2021}. These in turn, are augmentations of the original hydrodynamic simulations presented in \citet{sparre2016, sparre2017} with, in particular, the new additions of Monte-Carlo tracer particles and magnetic fields. The suite consists of four different merger scenarios, with each scenario ran twice from the same initial conditions: once with magnetic fields and once without. In each case, the same underlying numerical implementation is used, such that a hydrodynamic simulation is equivalent to an MHD simulation with the seed field set to zero. We summarise our set-up here, but direct the reader to section 2 of \citetalias{whittingham2021} and references therein for a more comprehensive description.

\subsection{Merger scenarios}
\label{chapter4-subsec:merger_scenarios}

Each simulation pair focuses on a spiral galaxy that undergoes a gas-rich major merger with another spiral galaxy between redshift $z = 0.9 - 0.5$. The merger mass ratios range between approximately 1.1 and 2 (see table 2 of \citetalias{whittingham2021} for exact details). The mergers may also be roughly separated into in-spiralling (1330, 1526) and head-on (1349, 1605), but cover a variety of impact parameters, speeds, and orbits. Post-merger, the galaxies are allowed to rebuild in relative isolation, experiencing no further events in their merger tree. We note, however, that as these are cosmological simulations, the remnants are not wholly isolated from subsequent minor tidal interactions (see section 2.4 of \citetalias{whittingham2021} for details). By $z = 0$, each remnant is able to rebuild a disc and has a final stellar mass in the range of $6-11\times 10^{10}\; \mathrm{M}_\odot$.

As discussed in section 2 of \citetalias{whittingham2021}, these mergers were specifically chosen with the intention of observing magnetic fields at their most influential; gas-rich progenitors implied strong initial magnetic fields, whilst major mergers were expected to best facilitate field amplification through compression and turbulence. Finally, the disc-rebuilding phase would elongate the time over which the magnetic fields could act. We note, however, that morphological considerations were not part of the original selection criteria \citep{sparre2016, sparre2017}.

We keep the labels for each simulation introduced in \citetalias{whittingham2021}, where a suffix of `H' or `M' represents the inclusion of hydrodynamic or MHD physics, respectively, and the first four digits represent the friends-of-friends \citep{davis1985} group number in Illustris \citep{vogelsberger2014, vogelsberger2014b, genel2014} from which the merger scenario was originally chosen. In order to be consistent with earlier analysis by \citet{sparre2016, sparre2017}, we always present data for the primary progenitor, defined here as the one with the largest stellar mass at $z = 0.93$ \citepalias[see section 2.3 of][]{whittingham2021}. We note, however, that this is a somewhat arbitrary choice, as both of the main progenitors have very similar properties pre-merger, including similar magnetic field strengths out to similar radii.

\subsection{Initial conditions and parameters}

Zoom-in initial conditions were created for each merger scenario using a modified version of the \textsc{N-GenIC} code \citep{springel2015}. In these, a volume of high resolution particles is focused on the target galaxy and its immediate surroundings, with a dark matter mass resolution equal to $1.64 \times 10^5 \; \mathrm{M}_\odot$. This is approximately 38.5 times finer than the original Illustris simulation. A buffer region envelops these particles, with yet coarser resolution particles filling the remaining volume. This volume has a side length of 75 co-moving Mpc~$h^{-1}$.

The softening length used is a co-moving length before $z = 1$, at which point it is frozen at a physical value of 0.22 kpc. For gas cells, this value is also scaled by the cell radius, with the restriction that the minimum softening length is bounded by 30~co-moving~pc~$h^{-1}$ below and 1.1 kpc above. This choice helps to prevent unrealistic two-body interactions at early times, whilst allowing small-scale structure to continue to form at late-times
\citep[see, e.g.,][]{power2003}.

The cosmological parameters were taken from WMAP-9 \citep{hinshaw2013}, with Hubble's constant $H_0 = 100 \,h~\textrm{km~s}^{-1}~\textrm{Mpc}^{-1} = 70.4$ km s$^{-1}$ Mpc$^{-1}$ and the density parameters for matter, baryons, and a cosmological constant, respectively, being $\Omega_\text{m} = 0.2726$, $\Omega_\text{b} = 0.0456$, and $\Omega_\Lambda = 0.7274$.

\subsection{\textsc{Arepo} and Monte-Carlo tracers}
\label{chapter4-subsec:arepo_mc_tracers}

The simulations were ran from $z = 127$ to $z = 0$ using the moving-mesh code \textsc{Arepo}, which employs a second-order finite-volume Godunov scheme \citep{springel2010, Pakmor2016I, weinberger2020}. Gas cells in the high-resolution region are refined and derefined so that they stay within a factor of two of the target mass, $2.74 \times 10^4 \; \mathrm{M}_\odot$. Meanwhile, mesh-generating points may be moved arbitrarily. Together, these features allow the code to behave in a quasi-Lagrangian manner, reducing the level of numerical diffusion, whilst simultaneously inheriting the robust nature of grid-based Eulerian codes. The resultant increased accuracy of this method over standard smoothed-particle hydrodynamic (SPH) methods has been well-documented \cite[see, e.g.,][]{vogelsberger2012, sijacki2012, dusan2012}. Of particular importance to this work, is the ability to replicate the Kolmogorov turbulent cascade \citep{kolmogorov1941} for subsonic turbulence, which is not achievable by standard SPH models \citep{bauer2012}. We showed in \citetalias{whittingham2021} that such a cascade was almost certainly crucial to achieving sufficient magnetic field amplification during the merger.

As \textsc{Arepo} is only a quasi-Lagrangian code, to follow the accurate flow of mass in the simulations, we employ the use of Monte-Carlo tracers \citep{genel2013}. We place five tracers per gas cell at the start of each simulation. These are then exchanged with neighbouring gas cells at a rate proportional to the mass flux across their boundaries. Tracers may also be accreted by black holes and be exchanged with star particles during star formation and stellar mass loss processes. Monte-Carlo tracers have been shown to follow the mass flux more accurately than the traditional method of Lagrangian tracers \citep{genel2013}.

\subsection{Auriga}

The galaxy formation physics in the simulations are evaluated using the Auriga model \citep{grand2017}. This model was originally built to produce Milky Way (MW)-like galaxies in zoom-in simulations, and has been able to produce appropriate stellar masses, sizes, rotation curves, star formation rates, and metallicities \citep{grand2017}, the correct structural parameters of bars \citep{calero2019}, and the existence of chemically distinct thick and thin discs \citep{grand2018}. The models for star formation, stellar feedback, and black hole feedback in Auriga are all physically well-motivated and parameters require only limited recalibration between resolution levels\footnote{Whilst parameters must not be significantly retuned, certain phenomena are nevertheless resolution-dependent; for example, the manifestation of starbursts \citep{sparre2016} and the extent of magnetic field amplification post-merger \citepalias{whittingham2021}.}. This is a non-trivial result \citep{scannapieco2012}. We summarise the model below, but encourage the reader to refer to section 2.4 of \citet{grand2017} and references therein for a more complete overview.

\subsubsection{ISM and feedback}
\label{chapter4-sec:ISM}

The ISM is described by the model of \citet{springel2003}, which assumes that hot and cold phases are in pressure equilibrium and, at the onset of thermal instability, the gas follows a stiff equation of state. In our simulations, this onset (and thus star formation) begins at a threshold gas density of $n_\text{SF}= 0.13\; \text{cm}^{-3}$. The model must not be recalibrated when magnetic fields are introduced \citepalias{whittingham2021}. To replicate Type II supernovae, wind particles are also created out of star-forming gas cells in numbers that reflect the fraction of stars formed in the mass range $8-100\;\text{M}_\odot$. These particles are launched in an isotropically random direction, with a velocity proportional to the local one-dimensional dark matter velocity dispersion \citep{okamoto2010}. Wind particles then interact only gravitationally until they reach a gas cell with $n < 0.05$ $n_\text{SF}$ or exceed a maximum travel time of approximately 25~Myr. At this point they deposit their energy in equal thermal and kinetic parts, which produces smooth, regular winds directed away from the galaxy.

Supermassive black holes are seeded with a mass of $1.4 \times 10^5$~M$_\odot$ once the mass of the corresponding friends-of-friends halo reaches $7.1 \times 10^{10}$~M$_\odot$. Seeding takes place at the position of the most bound gas cell, with dynamics set according to the \citet{springel2005a} model. Black hole accretion takes place predominantly through an Eddington-limited Bondi-Hoyle-Lyttleton model \citep{bondi1944, bondi1952}, with an additional term modelling accretion in the radio mode based on \citet{nulsen2000}. Feedback is implemented through a radio and quasar mode. For the radio mode, bubbles of gas are gently heated at random locations within the halo with a probability following an inverse square profile, whilst for the quasar mode, energy is injected isotropically into the 512 gas cells nearest the black hole. In both cases, energy is injected at a rate proportional to the black hole accretion rate.

\subsubsection{MHD implementation}

Magnetic fields are treated in the ideal MHD approximation \citep{pakmor2011, pakmor2013}, with equations solved using an HLLD Riemann solver \citep{miyoshi2005}. Divergence cleaning is handled using a Powell 8-wave scheme \citep{powell1999}. This scheme has been found to be more robust than the competing Dedner \citep{dedner2002} scheme when applied to cosmological simulations \citep{pakmor2013}. Our MHD implementation can replicate a variety of phenomena, including: the linear phase of growth of the magneto-rotational instability \citep{balbus1991, pakmor2013, zier2022}, the correct propagation of Alfv\'en and magnetosonic waves in co-moving coordinates \citep{berlok2022}, the development of a small-scale dynamo in MW-like galaxies \citetext{\citealt{pakmor2014}, \citealt{pakmor2017}, \citetalias{whittingham2021}, \citealt{pfrommer2022}},
 similar field strengths and radial profiles to those observed in MW-like galaxies \citep{pakmor2017}, and Faraday rotation measure strengths that are broadly consistent to those observed for MW-like galaxies, both for the disc \citep{pakmor2018} and when compared with the current upper limits available for the CGM \citep{pakmor2020}.

We seed magnetic fields in our initial conditions with a strength of $10^{-14}$ co-moving Gauss. This choice is essentially arbitrary, as the initial configuration and field strength is quickly erased for a broad range of values in collapsing haloes \citep{pakmor2014}. This seed strength is also dynamically irrelevant outside of collapsed haloes \citep{marinacci2016}. Magnetic energy is assumed to be locked up in wind- and star-forming events, but is otherwise not explicitly included in our subgrid models.

\section{Analysis}
\label{chapter4-sec:analysis}

\subsection{How the evolution of the merger remnant differs between physics models}
\label{chapter4-subsec:how_evolution_differs}

\begin{figure*}
    \includegraphics[width=\textwidth]{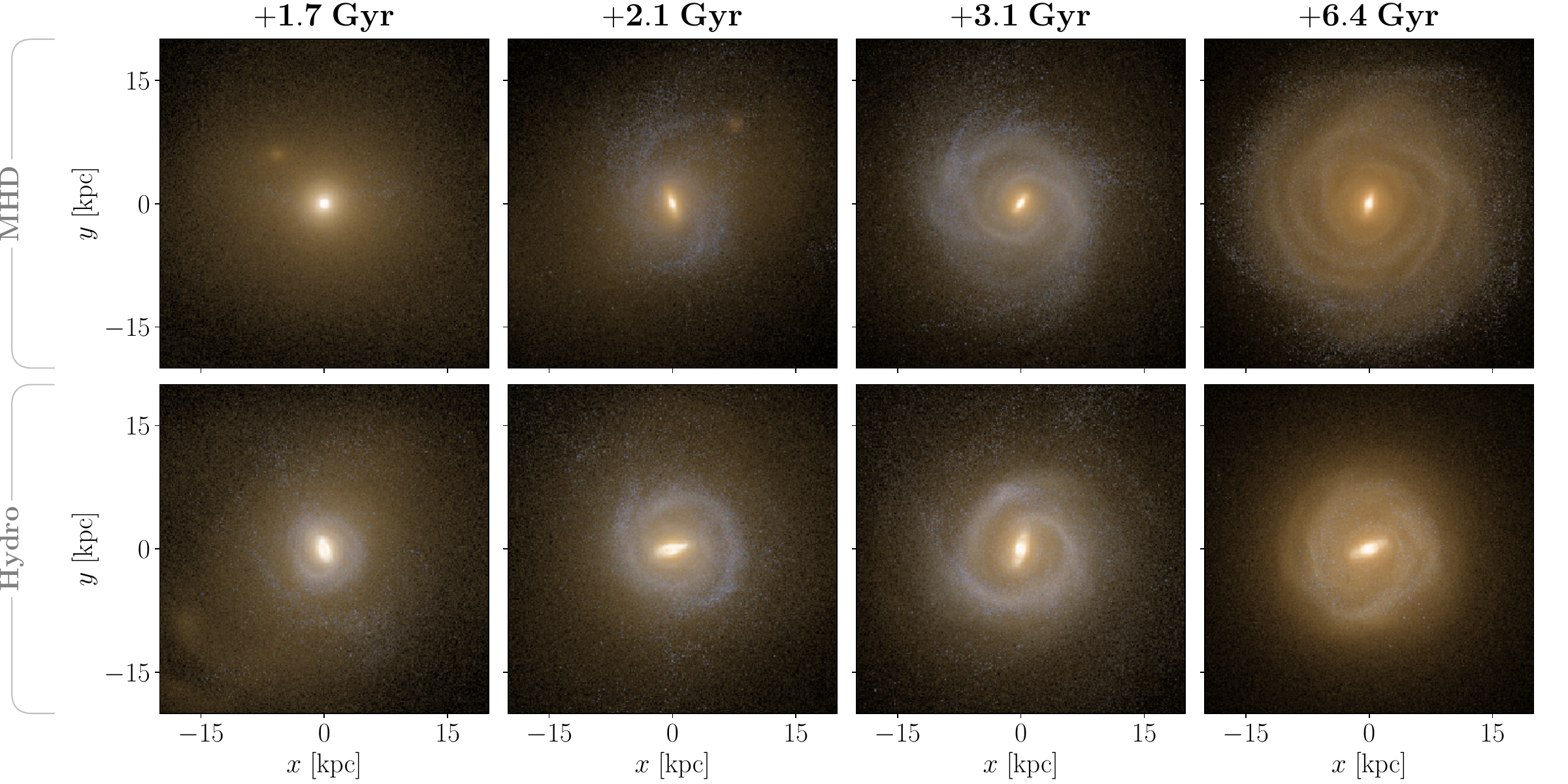}
    \caption[Mock stellar light images of the case study galaxies]{\textit{Top row:} Mock \textit{gri} composite images showing the evolution of the remnant in the 1349-3M simulation post-merger. \textit{Bottom row}: As above, but for the 1349-3H hydrodynamic simulation. Labels above each column indicate time elapsed since first closest approach. The formation of a strong bar in the hydrodynamic simulation is associated with the development of a stellar ring. The absence of a bar in the MHD simulation is associated with the formation of more varied small-scale structure. An animated version of this figure can be found \href{https://youtu.be/C9IAcu5G8Es}{here}.}
    \label{fig:mock_visual}
\end{figure*}

We start our analysis by isolating exactly how the evolution of the merger remnants differs for hydrodynamic and MHD simulations. We will use the 1349-3 simulations here as a case study, being broadly representative of the wider simulation suite. We will also focus on the stellar light morphology, which better highlights the evolution of the distinctive bar, ring, and spiral arm components. To this end, in Fig.~\ref{fig:mock_visual}, we present a series of face-on mock \textit{gri} images\footnote{The differences between the edge-on images are more subtle, and so we defer analysis of these to Appendix~\ref{appendix:edge-on-gri-images}.}. These images were created in the same manner as described in \citetalias{whittingham2021}, using the photometric properties of all star particles within $\pm30$ kpc of the midplane. For each snapshot, the time elapsed since the beginning of the merger (defined here as the time of first closest approach) is given above each column. We have chosen times such that each snapshot shows a significant step in the evolution of the remnant morphology, with the last column equivalent to $z=0$.

As stated in Sec.~\ref{chapter4-subsec:merger_scenarios}, each of our simulated remnants is able to reform a disc post-merger. However, whilst the remnant in the hydrodynamic simulation starts to rebuild a disc almost immediately, this process is initially delayed in the MHD simulation. This leads to a substantial difference in the size of the respective discs, as observed in the leftmost column of Fig.~\ref{fig:mock_visual}. Once the disc rebuilding process in the MHD simulation begins in earnest, however, progress is rapid. Indeed, the radial size of the disc in the MHD simulation ultimately outstrips that of its hydrodynamic analogue, as can be seen in the final column of the figure.

As well as the size evolution, the structural evolution of each remnant also differs; even at the earliest snapshot shown, in the hydrodynamic simulation a distinct bar and ring morphology is apparent. This ring is star-forming, as can be determined from its bluish hue, which reflects a young stellar population. The ring reaches a fairly steady form already by the second snapshot, with growth plateauing shortly thereafter. On the other hand, the bar formed in the MHD simulation is substantially weaker, and, instead of a ring, a substantial amount of small-scale structure is formed. This small-scale structure at first takes the form of distinct spiral arms before the stellar distribution eventually becomes more flocculent. In the final snapshot shown, the colours in both sets of \textit{gri} images become more yellow as the bulk of star formation has finished and the luminosity is now dominated by older stars. For the hydrodynamic simulation, this is associated with the ring structure becoming less well-defined. 

The differences observed in Fig.~\ref{fig:mock_visual} prompt three important questions, upon which we will base the analysis in this paper. These are:

\begin{enumerate}
    \item Why is the stellar population initially so much more compact in the MHD simulation?
    \item Why does a bar and ring structure form in the hydrodynamic simulation but not in the MHD simulation?
    \item Why does the remnant in the hydrodynamic simulation reach a maximum size, whilst that in the MHD simulation continues to grow?
\end{enumerate}

\subsection{Model for how magnetic fields affect mergers}
\label{chapter4-subsec:model}

\begin{figure*}
    \includegraphics[width=\textwidth]{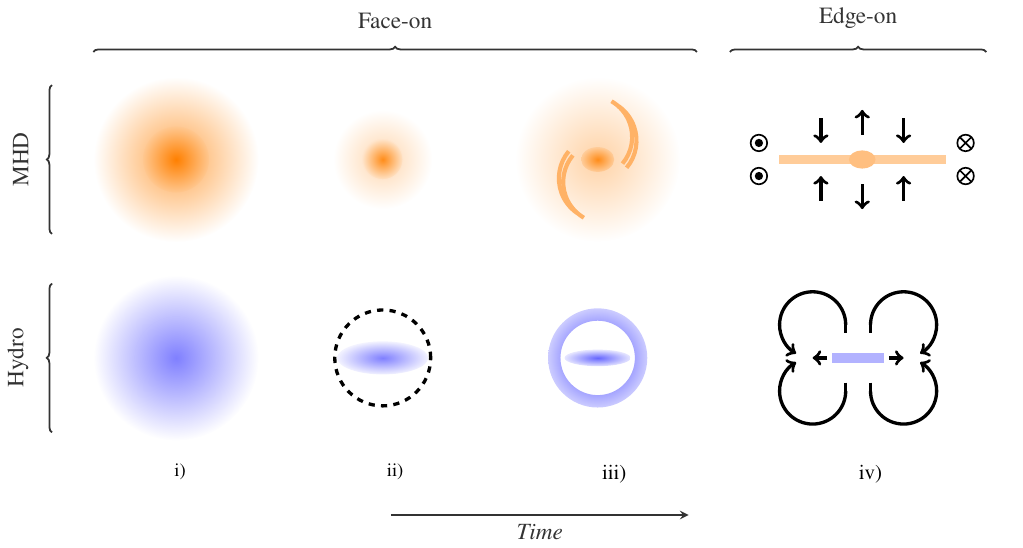}
    \caption[Schematic showing model outlined in the paper]{A schematic illustrating the key stages of development in our MHD and hydrodynamic simulations post-merger. Amplified magnetic fields are able to mediate angular momentum, which typically increases the baryonic concentration, thereby suppressing a bar instability. This leads to a fundamentally different stellar distribution and manifestation of feedback. A full description of each stage can be found in Sec.~\ref{chapter4-subsec:model}.}
    \label{fig:schematic}
\end{figure*}

Cosmological simulations are intrinsically complicated by their very nature. Accordingly, there are several factors that must be taken into account when explaining how magnetic fields are able to affect the outcome of mergers. To simplify our explanation, we first outline a streamlined model of the stages involved before presenting evidence for each of these stages in the remainder of the paper. We present a visual representation of the stages in Fig.~\ref{fig:schematic}, with descriptions given below:

\begin{enumerate}
 \item \textit{Angular momentum transfer:} The merger significantly amplifies the magnetic field, allowing it to have a strong dynamical back-reaction on the gas. This typically takes place within a few 100~Myr of the first closest approach. When the magnetic field is non-azimuthally orientated, the redistribution of angular momentum between gas cells is more effective, leading to a loss in total angular momentum in the disc and a subsequently higher central baryonic concentration.
 \item \textit{Suppression of a bar instability:} In the hydrodynamic simulations, the post-merger starburst causes the remnant to become bar-unstable. This instability is suppressed in the MHD simulations. The exact cause of this suppression depends on the magnetic field configuration; when the field is predominantly non-azimuthally orientated, the suppression originates from the generation of a strong inner Lindblad resonance caused by the increased mass concentration. In the azimuthal case, the magnetic field suppresses the instability by providing support against collapse.
 \item \textit{Resonances:} The large bar in the hydrodynamic simulation reorders the existing distribution of stars and shepherds gas towards the outer Lindblad resonance. This causes an exceptionally high star formation rate in this region. The absence of a strong bar in the MHD simulations allows the gas to remain flocculent and for spiral arm features to develop.
 \item \textit{Winds:} The high star formation rate density in the hydrodynamic simulation launches a strong stellar wind. This acts both radially away from the disc and initiates a large-scale fountain flow. Together, these winds strongly disrupt the angular momentum of accreting gas, helping to keep the disc compact. In the MHD simulations, star formation is more spread out, and stellar winds consequently have a much lower impact. Indeed, at the outskirts of the remnant, gas can be almost co-rotating, allowing it to join the disc practically in-situ. This facilitates rapid radial growth.
\end{enumerate}
The result of these steps is that the remnant in the MHD simulation forms a typical spiral galaxy with an extended radial profile, whilst in the hydrodynamic simulation, the remnant is substantially smaller and displays prominent bar and ring components. 

We present evidence for this model in the following sections. We focus on the \textit{Angular momentum transfer} stage in Sec.~\ref{chapter4-subsec:angmom}, on the \textit{Suppression} and \textit{Resonance} stages in Sec.~\ref{chapter4-subsec:resonances}, and on the \textit{Winds} stage in Sec.~\ref{chapter4-subsec:feedback}.

\subsection{How magnetic fields increase the baryonic concentration through modified angular momentum transport}
\label{chapter4-subsec:angmom}

\begin{figure*}
    \includegraphics[width=\textwidth]{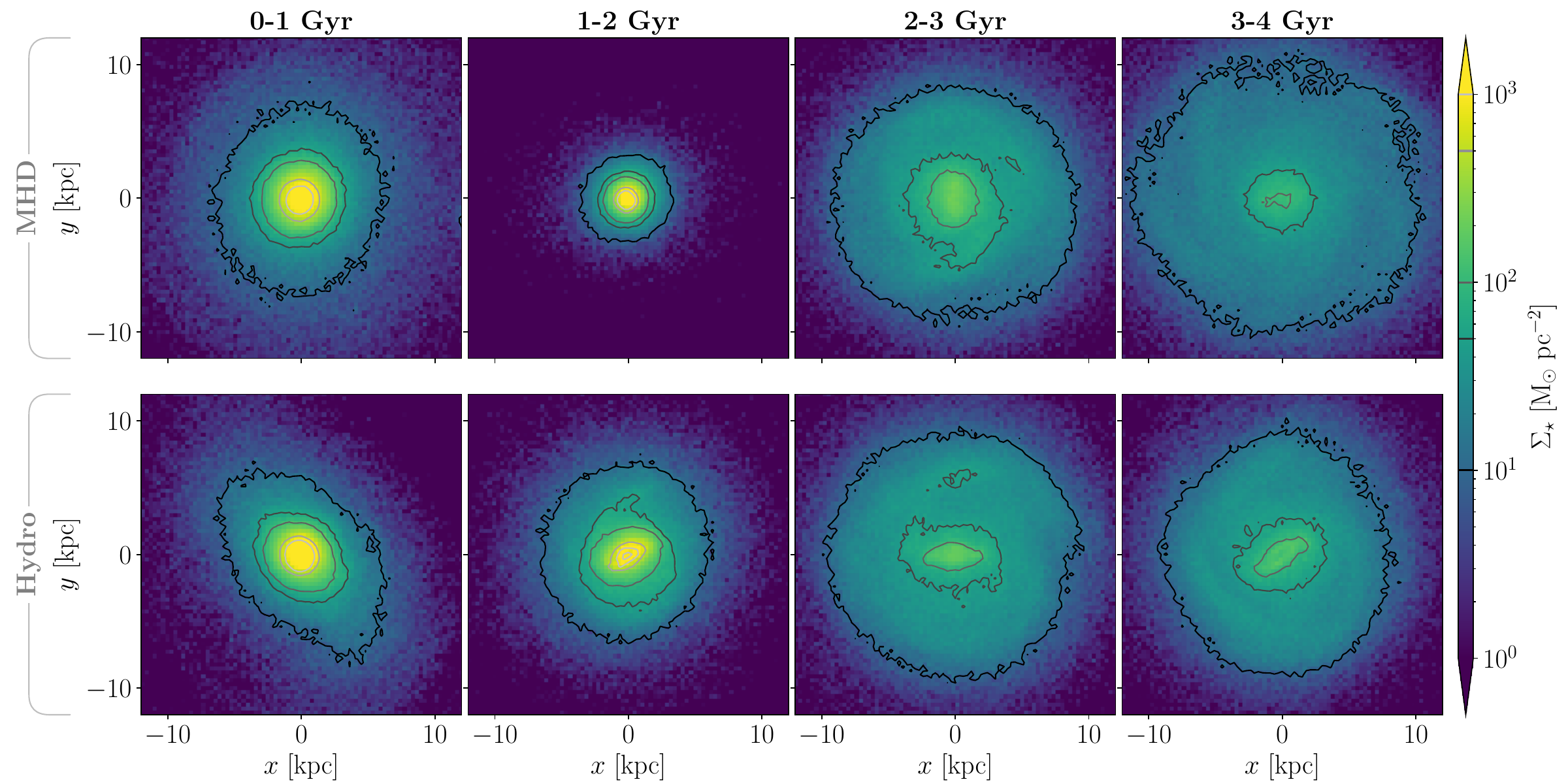}
    \caption[Stellar surface density maps as a function of time]{\textit{Top row:} Stellar surface density maps for 1349-3M, where stars have been selected such that they were formed in the previous Gyr. Maps show the distribution at +1, +2, +3, and +4 Gyr post-merger, respectively. Contours are shown at 10, 50, 100, 500, and 1000 $\text{M}_\odot\; \text{pc}^{-2}$. The projection has a vertical extent of $\pm5$ kpc from the midplane. \textit{Bottom row:} As above, but for 1349-3H. Star formation between 1 and 2 Gyr post-merger is more concentrated in the MHD simulation. This stabilises the disc against the formation of a bar. In contrast, in the hydrodynamic analogue, a large bar forms. This leads to a markedly different stellar distribution.}
    \label{fig:bar_star_ages}
\end{figure*}

To illustrate how angular momentum in the disc evolves differently between the two physics models, we start by tracing how and where stars form in the successive Gyrs post-merger. This is directly affected by how the dense gas is distributed, upon which the magnetic fields have an influence. To this end, in Fig.~\ref{fig:bar_star_ages} we show stellar surface density maps for the 1349-3 simulations, where in each panel we have selected only the stars that formed in the previous Gyr. That is to say, the first column is shown at 1~Gyr post-merger (equivalently, 1~Gyr after first closest approach) and includes stars formed between 0 and 1~Gyr post-merger, the second is shown at 2 Gyr post-merger and includes stars formed between 1 and 2~Gyr post-merger, and so forth. By binning the star formation over this time interval, we smooth over the inherent stochasticity of our underlying star formation model. This, of course, leaves up to a Gyr for the stars to move from their birth position, but in practise, we observe that migration during this time is limited.

In the first column, the distributions are approximately isotropic in both cases. This isotropy is especially strong in the case of the MHD simulation. In the hydrodynamic simulation, the distribution becomes slightly skewed as we move away from the centre. This is principally a projection effect; at the time this surface density map is made, the disc is reorientating in space as material with different orbital angular momenta is accreted. Newly-born stars at the outskirts of the disc have not yet reorientated to orbit in the plane perpendicular to the line of sight. The lack of such an effect in the MHD simulation results from angular momentum transfer facilitated by the magnetic field, which acts to keep the disc rotating coherently. We will show evidence for this in the following plot.

By the second column of Fig.~\ref{fig:bar_star_ages}, there are already noticeable differences between the two remnants. Most strikingly, the stellar population in the MHD simulation is now significantly more compact, whilst the distribution in the hydrodynamic simulation remains extended, as we saw previously in the stellar light distribution in Fig.~\ref{fig:mock_visual}. Both remnants have formed roughly the same amount of stars at this point \citepalias{whittingham2021}, implying a stellar concentration that is significantly higher in the MHD case\footnote{We will show this is true from a more quantitative standpoint in Sec.~\ref{chapter4-subsec:suppression}.}. Indeed, whilst the surface mass density increases towards the centre in the MHD case, the innermost contour in the hydrodynamic analogue actually marks the reduction of the surface density below 1000~M$_\odot\;\text{pc}^{-2}$ again. This reduction is a typical response of gas to a bar potential \citep{kormendy2004}. The existence of this bar can be seen more explicitly through the increased anisotropy of the innermost contours, as well as implicitly through the faint outline of a stellar lane traced out by the 50~M$_\odot\;\text{pc}^{-2}$ contour. As we will see later in Sec.~\ref{chapter4-subsec:stellar_ring}, the tidal impact of the bar is critical for producing the associated ring-shaped morphology in hydrodynamic remnants.

By +3 Gyr, the majority of the post-merger star formation has finished. This is reflected by the fact that stellar surface density contours in the third and fourth columns of Fig.~\ref{fig:bar_star_ages} only exist up to 100~M$_\odot\;\text{pc}^{-2}$. Nonetheless, some morphological evolution continues to take place in these panels. Firstly, it can be seen that the bar in the hydrodynamic simulation continues to develop and is supported by fresh star formation. With a keen eye, the faint traces of a star-forming ring can also be seen, close to the outermost contour\footnote{Explicit evidence of this feature will be shown in Sec.~\ref{chapter4-subsec:stellar_ring}.}. In the MHD case, on the other hand, the innermost contours become slightly more anisotropic with time, as the magnetic field dominance weakens, and the beginnings of spiral arms start to appear instead of a ring (cf. the features in Fig.~\ref{fig:mock_visual}). 

In both simulations, the evolutionary step between 1 and 2~Gyr post-merger is key to the final outcome; in the hydrodynamic simulation, a bar starts to form during this time, which goes on to have a strong tidal impact on the rest of the remnant. In contrast, in the MHD simulation, the disc appears to be stabilised against bar formation during this time through its compaction. Compaction to this extent requires a substantial reduction in the average magnitude of the gas angular momentum. This is, in turn, a direct result of the mediation of angular momentum by the magnetic field, as we show in the following figure.

\begin{figure*}
    \includegraphics[width=\textwidth]{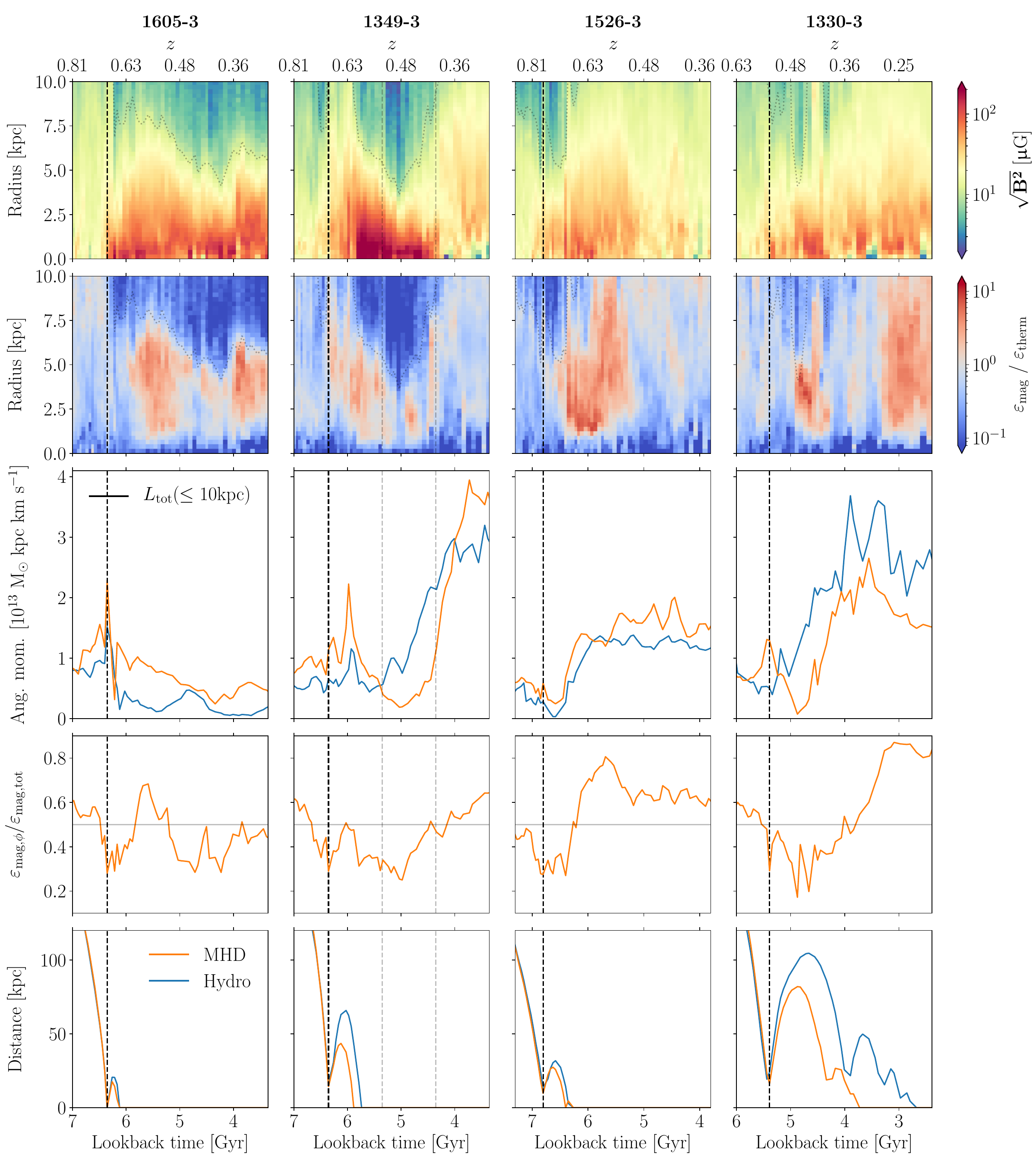}
    \caption[Evidence of modified angular momentum transport due to inclusion of magnetic fields]{\textit{Top row:} Radially-binned mean magnetic field strength of the main galaxy as a function of time, using annular rings of width 0.25~kpc  and depth $\pm 1$ kpc from the midplane. A dotted line indicates the point at which the gas density drops below $0.02\,\mathrm{M}_\odot\,\mathrm{pc}^{-3}.$ \textit{2nd row:} As above, but showing the magnetic to thermal energy ratio in each ring. \textit{3rd row:} The absolute value of the total angular momentum for all gas cells within a sphere of radius 10~kpc. \textit{4th row:} The fraction of magnetic energy density in the azimuthal component, calculated for a disc bounded by radius 10 kpc and height $\pm 1$ kpc from the midplane. \textit{Bottom row:} The distance between the two merging galaxies as a function of time. The dashed, black, vertical lines mark the time of first closest approach in each simulation. The dashed, grey, vertical lines mark +1 and +2 Gyr post-merger in 1349-3M to aid comparison with Fig~\ref{fig:bar_star_ages}. The merger-induced amplification of the magnetic field allows it to become dynamically important. When the magnetic field is predominantly non-azimuthally orientated, this leads to a more efficient redistribution of gas angular momentum between the accreting gas and that already in the disc. This, in turn, causes an initial reduction in disc size. The same mechanism hastens the progress of the merger in each example.}
    \label{fig:ang_mom}
\end{figure*}

In order to show that the magnetic fields are capable of mediating angular momentum, we first need to show that they are dynamically relevant. With this in mind, in the first row of Fig.~\ref{fig:ang_mom}, we show the mean magnetic field strength in the disc as a function of radius and time. This was created in the same manner as for figure 2 in \citetalias{whittingham2021}, using annular rings of width 0.25~kpc and a vertical extent of $\pm$1 kpc, where this volume is orientated according to the angular momentum vector of the cold gas disc. We focus on a time period for each galaxy that extends from 0.5~Gyr before the start of the merger (marked by the dashed, black, vertical lines) until 3 Gyr afterwards. The colour bar ranges from 2 to 200 $\upmu$G. This scaling covers all but the very innermost radii in 1349-3M during its period of most intense amplification. For this galaxy, the mean field strengths reach a maximum of $310$~$ \upmu$G.

As we noted in \citetalias{whittingham2021}, the first closest approach is associated with a rapid amplification of the magnetic field in all cases. An additional boost also takes place at further passages and at coalescence. This is caused by the additional compression and turbulent injection that takes place at these times. For every galaxy except 1526-3M, there are periods in which the radial extent of the amplified region reduces. This is a signature of the increase in concentration that takes place in these remnants. Such an effect can be observed, for example, for 1605-3M from $5.5-4$~Gyr, starting again at 4~Gyr; for 1349-3M from approximately $6-4$~Gyr; and for 1330-3M from $5-4.5$~Gyr. For aid of comparison between the data presented here and that in Fig.~\ref{fig:bar_star_ages}, we have added dashed, grey, vertical lines to the 1349-3M column to indicate +1 and +2~Gyr post-merger. The reduction in size of the amplified region here is clearly reflected by the reduced size of the stellar distribution in Fig.~\ref{fig:bar_star_ages}. For 1605-3M and 1349-3M, it is also reflected by the general decrease in the radius of the gas disc, as delineated by the dotted line. We use a density threshold of $0.02\,\mathrm{M}_\odot\,\mathrm{pc}^{-3}$ to measure this. All galaxies eventually experience at least a temporary regrowth of this disc as gas accretes. The concentration of the gas can, however, continue to increase during these times if the magnetic field remains strong enough. This is evident, for example, in 1349-3M at a lookback time of around $5-5.5$~Gyr.

In the second row of Fig.~\ref{fig:ang_mom}, we show the mean magnetic to thermal energy density over the same volumes as above. It can be seen that the magnetic energy density within the disc pre-merger is comparable to, if slightly lower than, the thermal energy density. However, within a short time of the merger, the magnetic energy density soon dominates. The balance between the two energy densities then fluctuates due to the back-reaction of the magnetic fields on the gas and the additional injection of turbulence by inflows\footnote{We have also analysed the magnetic to turbulent energy density, using the definition given in eq. 6 in \citet{pakmor2017}, and see similar trends, but with weaker dominance of the magnetic fields. For the magnetic to rotational energy density, magnetic fields are able to reach a similar order of magnitude when dominant in Fig.~\ref{fig:ang_mom}, but are typically a factor of a few weaker.}. The periods in which the magnetic field is dominant in each simulation are also generally reflected by a period of time in which the gas concentration increases. This, in turn, is associated with a decrease in the overall angular momentum in the disc, as we show explicitly in the next row.

In the third row of Fig.~\ref{fig:ang_mom}, we show how the magnitude of the total gas angular momentum within 10 kpc of the remnant, $L=|\bs{L}|$, evolves as a function of time for the MHD (orange) and hydrodynamic (blue) simulations, respectively. We calculate this as $\bs{L}=\Sigma_i (\bs{r}_i \bs{\times} \bs{p}_i)$, where $r_i$ is the radial distance of gas cell, $i$, from the galaxy centre and $p_i$ is its momentum. We evaluate this sum across a sphere to avoid rotating the reference frame, as was done in the upper two rows of the figure. This prevents us contaminating the sum with artificial torques.

It can be seen that in three out of four cases (i.e. all except 1526-3) the evolution of the total angular momentum differs substantially between the MHD simulation and its hydrodynamic analogue. For these, in the MHD simulation, the first closest approach (marked by the dashed, vertical lines) is always associated with a spike in the angular momentum. This is indicative of gas being brought into the 10 kpc radius by the merging galaxy. In the more head-on mergers (1605-3 and 1349-3) the total angular momentum drops shortly afterwards, as gas temporarily leaves the sphere again. This is then followed by a second spike at coalescence as the gas reaccretes. Similar temporary increases in the total angular momentum can usually be seen in the hydrodynamic simulations as well, but these are firstly not always evident and secondly, when they do exist, the spike peaks at systematically lower values. 

This behaviour can be explained by inspecting the distribution of angular momenta components\footnote{We do not show this here due to space constraints.}. For these, we observe that, in the MHD simulations, gas flows reaching the sphere are typically able to remain more coherent. This increases the ability of both matter and angular momentum to penetrate the sphere and reach the galaxy, thereby providing the larger spikes seen in total angular momentum in Fig.~\ref{fig:ang_mom}. The increased coherence of such flows is likely to be a result of them being less easily broken apart due to magnetic draping \citep{dursi2008,berlok2019}, as has been observed, for example, in simulations of jellyfish galaxies passing through the intergalactic medium \citep{sparre2020,mueller2021}. This process would also explain why the gas in 1605-3M exhibits a fairly high degree of angular momentum post-merger, whilst in 1605-3H the angular momentum reduces substantially; gas flows in this merger are strongly misaligned and therefore, in the latter, rapidly become turbulent, whilst they are shielded to a degree from this process in the MHD simulation. We note that this effect can only be realised with sufficiently high resolution.

After an initial increase in the total angular momentum in 1605-3M, 1349-3M, and 1330-3M, this quantity undergoes a sustained decline in these simulations as gas with misaligned angular momentum is accreted and is redistributed amongst the existing material. This decline is a direct measurement of the reduction in disc size of the remnants and clearly corresponds to the signatures already analysed for the upper two rows of the figure. Once again, for 1349-3M, a comparison can be made between Figs.~\ref{fig:bar_star_ages} and \ref{fig:ang_mom} with the aid of the dashed, grey, vertical lines. The reduction in magnitude of the gas angular momenta directly leads to a stellar distribution with lower angular momenta, thereby increasing the stellar concentration relative to its hydrodynamic analogue, as was observed in Fig.~\ref{fig:bar_star_ages}.

For 1349-3M and 1330-3M, sufficient angular momentum is eventually accreted such that the disc starts to grow rapidly again. For 1349-3M, this early disc-regrowth phase also provides further evidence of torquing by magnetic fields, as we show in Appendix~\ref{appendix:edge-on-gri-images}. In 1605-3M, the CGM is too disturbed by outflows (see Sec.~\ref{chapter4-subsec:feedback} and Appendix~\ref{appendix:stellar_feedback}) to provide any substantial growth, and subsequently the disc continues to mostly decrease in size, save for a brief increase at $\sim$4~Gyr. Similar outflows are likely to stop the growth of the disc in 1605-3H, which experiences a degree of accretion-driven growth post-merger between a lookback time of 5.5 and 5 Gyr before decreasing again\footnote{We note that this decrease also correlates with a period of increased AGN activity (see Sec.~\ref{chapter4-subsec:AGN}).}.

We have so far neglected 1526-3M, as it does not fit the general pattern; here, even when the magnetic field is dominant, no disc size reduction takes place. This behaviour can be understood, however, by examining the magnetic configuration in this remnant. In the fourth row of Fig.~\ref{fig:ang_mom}, we show the fraction of the magnetic energy density in the azimuthal component, where we have calculated this value using the same volumes analysed in rows 1 and 2. The grey, horizontal line marks the point at which the azimuthal component dominates over the non-azimuthal (i.e. vertical and disc-like radial) components. By comparing rows 2 and 4 of Fig.~\ref{fig:ang_mom}, it can be seen that for 1605-3M, 1349-3M and 1330-3M, the magnetic field experiences sustained periods of dynamical dominance when the field is also non-azimuthally orientated, whilst in 1526-3M, the magnetic field becomes strongly azimuthal just as it also becomes dominant. We believe that this explains why the other three MHD simulations increase their baryonic concentration, whilst 1526-3M does not.

The orientation of the magnetic field is important because of its implications for angular momentum transfer; when the magnetic field is predominantly non-azimuthally-orientated, field lines connect the disc to the CGM. In general, there will be a difference in angular velocity between these two components, which results in a magnetic tension force acting on them. When the gas in the disc is rotating faster, as is typically the case, this tension force decreases the speed of the gas in the disc whilst increasing it in the CGM. Angular momentum is thereby transported out of the disc, shrinking it. This effect will be still stronger if the infalling gas rotates oppositely to that in the disc, as then a drag force applies to both parts. Such a case will generally arise in a turbulent CGM, but will be especially influential in retrograde mergers where the majority of new material is counter-rotating relative to the existing gas disc. This is exactly the case in 1349-3M and 1330-3M, and likely the cause of the large drops seen in their total angular momentum. In 1526-3M, meanwhile, the magnetic field is predominantly azimuthally-orientated. In this case, field lines connect gas cells with similar angular momenta, which limits the impact that angular momentum redistribution can have. This results in a very similar evolution in the total angular momentum for 1526-3M and 1526-3H. However, here too, the magnetic fields have an impact, as, in connecting similar angular momenta gas, the field lines actively isolate the gas from external influences. Consequently, in 1526-3M, the magnetic field does not increase the baryonic concentration, but rather supports the disc against collapse. This encourages more isotropically distributed star formation, which also helps to stabilise the disc against bar formation \citep{sellwood2014}.

As well as affecting the baryonic concentration, the mediation of angular momentum through the magnetic fields also has a larger-scale effect. We explore this in the final row of Fig.~\ref{fig:ang_mom}, where we show the distance between the centres of the two merging galaxies as a function of time. We define the centre of each galaxy as the particle with the lowest potential in the subhalo found by $\textsc{subfind}$ \citep{springel2001}. Coalescence is then defined when $\textsc{subfind}$ can no longer identify two gravitationally ``self-bound'' subhaloes (see section 2.3 of \citetalias{whittingham2021} for further details). It can be seen that the mergers in the MHD simulations coalesce systematically faster than their hydrodynamic analogues. 

The difference in time required for coalescence is greatest in absolute terms when the merger took longest. The 1330-3 simulations are a particularly strong example of this, with the MHD simulation coalescing over a Gyr earlier than its hydrodynamic analogue. The trajectories in this case are practically identical for each pair of galaxies until first closest approach, at which point the merging galaxy in the MHD simulation loses a significant amount of angular momentum. A similar angular momentum transfer also takes place at the next two closest approaches, further quickening the rate of coalescence.

\subsection{The impact of resonances}
\label{chapter4-subsec:resonances}
\subsubsection{The suppression of a bar in MHD simulations}
\label{chapter4-subsec:suppression}

It may perhaps sound contradictory that magnetic fields act to reduce disc sizes immediately post-merger, but lead to larger sizes overall by $z = 0$. However, the compaction stage is actually critical to the remnant's future growth, having a major impact on how resonances form in the disc. To show this, in Fig.~\ref{fig:gas_concentration}, we analyse the relationship between the baryonic concentration in the disc and the subsequent formation of resonances in the 1349-3 simulations. We do this for the first 2 Gyr of the remnant's regrowth, as we identified this as a critical stage in the remnant's development in Sec.~\ref{chapter4-subsec:angmom}.

In the first row of Fig.~\ref{fig:gas_concentration}, we show radial density profiles of the gas for both MHD and hydrodynamic simulations, calculated using spherical shells of width 0.25~kpc. We previously asserted that the addition of magnetic fields leads to a higher baryonic concentration in the remnant, and we may use Fig.~\ref{fig:gas_concentration} to reevaluate this assertion from a more quantitative standpoint. For the first column, at +0.3 Gyr, we observe that the radial profiles for the inner 2.5 kpc of both physics models are very similar. At distances further out than this, it can be seen that there is a ``bump'' of higher density gas in the MHD case. The timing of this snapshot matches the angular momentum peak seen at coalescence in Fig.~\ref{fig:ang_mom}. This gas likely belongs to the inflows providing the extra angular momentum seen at this time.

In the following panel, the radial profiles begin to look more similar at radii $\gtrsim$2.5~kpc, but a strong peak in the gas density can be seen at the inner central kpc for the MHD simulation. This region is within the range affected by quasar feedback and hence is subject to a certain degree of variability as gas is expelled and then flows back following successive outbursts. The profiles we show here, however, are typical for all following times until approximately +1.7~Gyr post-merger, as shown in the third panel. That is, in the MHD simulation, the gas density outside the 2~kpc region decreases as the disc size reduces, whilst the peak gas density levels are maintained at levels typically several factors higher than in the hydrodynamic analogue.

By +2~Gyr post-merger, as seen in the final column, the peak gas densities are once again similar for both physics models. At this time, the period of most significant amplification has finished, as has the bulk of the starburst. The overall amount of stars formed during this time is approximately the same, as can be seen by inspecting the second row of Fig.~\ref{fig:gas_concentration}, where the cumulative stellar mass profiles at +2 Gyr at distances $\gtrsim$5~kpc approximately match. The distribution of stellar mass, however, is different for each physics model; in the MHD simulation, more mass is found closer to the disc centre. This divergence may appear subtle, but this change in mass concentration has a strong influence on the generation of resonances, which influence the likelihood of bar formation.

\begin{figure*}
    \includegraphics[width=\textwidth]{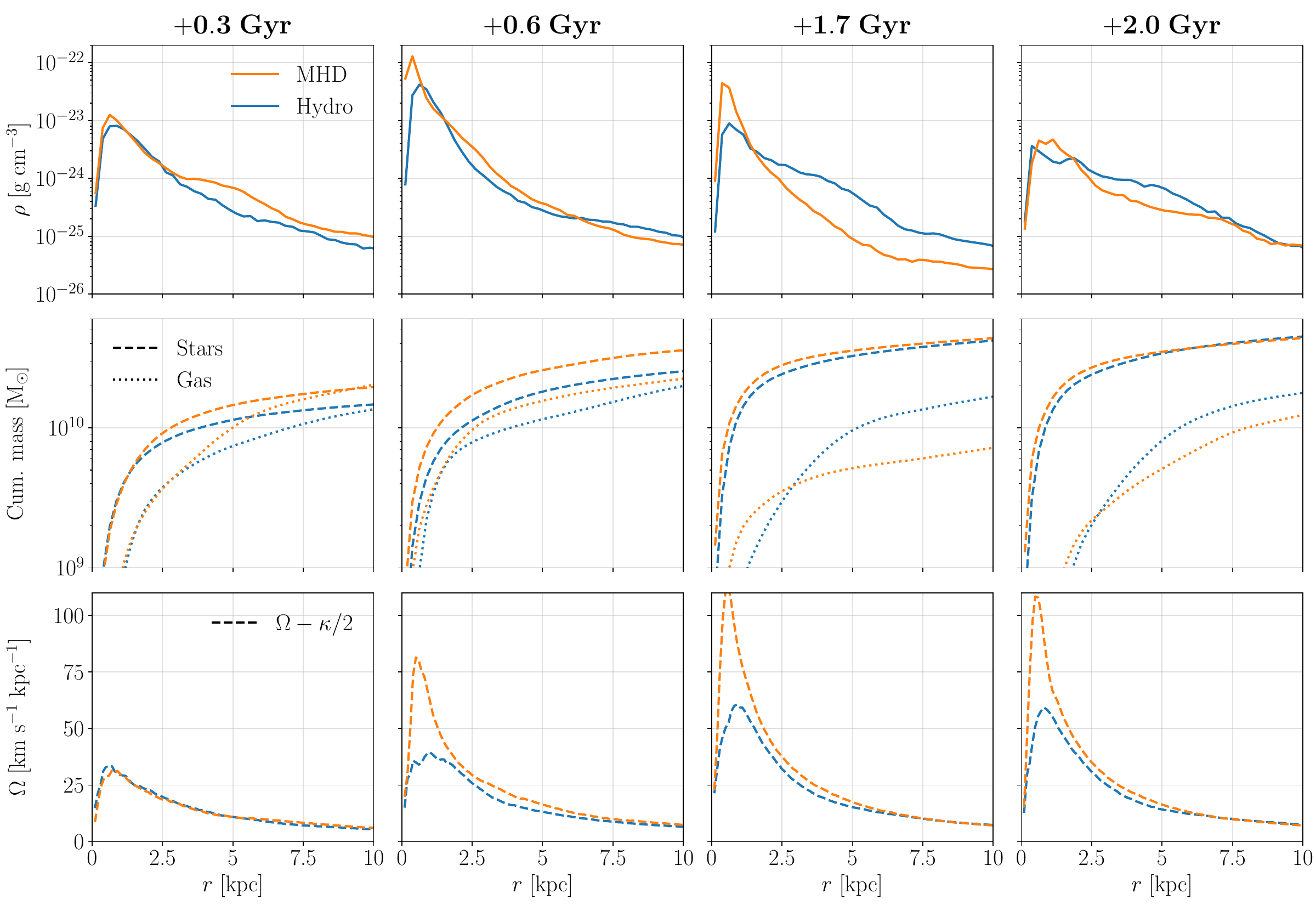}
    \caption[Radial density profiles and the subsequent generation of Lindblad resonances]{\textit{1st row:} Mean gas density as function of radius, measured in spherical shells of width 0.25 kpc for the 1349-3 simulations. \textit{2nd row:} The cumulative mass in gas and stars, calculated using the same shells as above. \textit{3rd row:} The evolution of the inner Lindblad resonance profile over time. Labels above each column indicate time elapsed since the start of the merger. The increased concentration of gas in the MHD simulation results in a more concentrated distribution of stars. This, in turn, generates a strong inner Lindblad resonance, which acts as a barrier to bar formation. In the hydrodynamic simulation, the peak of the resonant profile is low enough to be overcome and subsequently a strong bar is able to form.}
    \label{fig:gas_concentration}
\end{figure*}

In the final row of Fig.~\ref{fig:gas_concentration}, we present profiles for the inner Lindblad resonance. This resonance forms a crucial part of our current understanding of both bar formation and orbital dynamics in barred galaxies \citep[see, e.g.][]{friedl1993, weinberg2007, athanassoula2013, sellwood2014, renaud2015}. For approximately axisymmetric potentials, as inferred from Fig.~\ref{fig:bar_star_ages}, we may calculate the profile of this resonance by employing the epicyclic approximation \citep{binney2008}. Under this, orbits can be considered to be mostly circular with a small radial oscillation about a guiding centre. The frequency of this oscillation, $\kappa$, resonates if it is a multiple of the bar pattern speed, $\Omega_\text{p}$ (the angular frequency at which the bar rotates). We may write this condition as: $m (\Omega_\text{p} - \Omega) = l\kappa$, where $l$ and $m$ are integers, and $\Omega$ is the average angular frequency for an orbit at a certain radius. The inner Lindblad resonance occurs for $l=-1$ and $m=2$, implying $\Omega_\text{p}= \Omega -\kappa/2$. In this case, the star executes two radial oscillations for every rotation of the bar, meaning that it is at the same phase of its oscillation each time an end of the bar swings underneath.

To calculate $\Omega$, we use
\begin{equation}
\Omega = \bupsilon_\text{circ} / r,
\label{eq:omega}
\end{equation}
where $\bupsilon_\text{circ}$ is the circular velocity at a particular radius, $r$. Further to this, following standard theory, we make the approximation that the system is spherically symmetric, and therefore
\begin{equation}
\bupsilon_\text{circ} = \sqrt{G M(\leq r)/ r},
\end{equation}
where $G$ is the gravitational constant, and $M(\leq r)$ is the cumulative mass within radius, $r$. The validity of this approximation is weaker for non-spherically symmetric systems, but previous work has shown that the results have errors of only 5\% - 10\% in the case of more disc-like systems \citep{fragkoudi2021}. This approximation is therefore sufficient for our ends, particularly before the disc-rebuilding process has fully got underway.

Following \citet{kormendy2004}, we calculate $\kappa$ as: 
\begin{equation}
\kappa^2 = 2 \Omega \left(\Omega + \dfrac{\mathrm{d}\bupsilon_\text{circ}}{\mathrm{d}r} \right).
\label{eq:kappa}
\end{equation}

This allows us to calculate the relation $\Omega - \kappa/2$ as a function of radius, where the intersection of this profile with the bar pattern speed provides the location of the inner Lindblad resonance.

The significance of the inner Lindblad resonance lies in its influence on families of stellar orbits. There are two families, in particular, which are important for the formation of bars. In the notation of \citet{contopoulos1980}, these are the $x_1$ orbits, which are elongated parallel to the major-axis of the bar, and the $x_2$ orbits, which have lower eccentricity and are elongated orthogonally to the bar. Stable $x_1$ orbits, naturally, support the formation of a bar, whilst $x_2$ orbits act against it. The domain of each orbit swaps when passing resonant boundaries, with $x_2$ orbits able to exist between the two possible solutions for the inner Lindblad resonance \citep{combes2002}. The result of this is that the larger the range between these solutions is, the more difficult it is for a strong bar to form. This is especially so when the bar is at a nascent stage; when self-gravity is not enough to force orbits to precess at the same rate \citep{kormendy2004}.

With this in mind, the importance of the variations we identified in the mass distribution for the second row of Fig.~\ref{fig:gas_concentration} becomes clear. When we inspect the third row, it can be seen that, at first, the profiles for the inner Lindblad resonance are almost identical, as the cumulative mass profiles at small radii are also similar. However, as the mass concentration in the MHD simulation increases relative to its hydrodynamic analogue, a large divergence takes place. The result of this is that, already by +0.6~Gyr post-merger, an inner Lindblad resonance can exist in MHD simulations for pattern speeds twice as high as in the corresponding hydrodynamic simulation. This situation becomes worse as time proceeds, with the pattern speed required to avoid encountering a broad range of $x_2$ orbits quickly becoming unrealistically high\footnote{These orbits are populated, as can be seen by observing the isotropic star formation during this time evidenced in Fig.~\ref{fig:bar_star_ages} and the edge-on mock images provided in Appendix~\ref{appendix:edge-on-gri-images}.}. Furthermore, as the starburst finishes within the initial 2~Gyr post-merger \citepalias[cf. figure 1 of ][]{whittingham2021}, the inner concentration also undergoes little change after this time. The inner Lindblad resonance therefore stays strong, keeping bar formation in the MHD simulation consistently suppressed post-merger. In hydrodynamic simulations, the pattern speed required to avoid an inner Lindblad resonance is, on the other hand, much more achievable. 

\subsubsection{The formation of a stellar ring in hydrodynamic simulations}
\label{chapter4-subsec:stellar_ring}

The subsequent growth of a bar in the hydrodynamic simulation has a major impact on how gas and stellar orbits evolve in the remnant. We show evidence of this for 1349-3H in Fig.~\ref{fig:resonances}. We choose to perform this analysis at approximately 3 Gyr post-merger. At this time, the bar has been well-developed for at least a Gyr, and has had a corresponding amount of time to shape the orbits in the disc. Naturally, the pattern speed of the bar varies slightly as it evolves and couples with other modes. At the time we pick, however, the bar pattern speed has varied by no more than $\pm$1 km s$^{-1}$ kpc$^{-1}$ over the last 0.5 Gyr, meaning that the radial positions of the resonances have also stayed approximately constant over the same period. This helps us to better isolate their impact.

In panel~A of Fig.~\ref{fig:resonances}, we show the circular velocity profiles that exist at 3 Gyr post-merger, calculated under the same spherical symmetry assumptions made earlier. The solid line indicates the overall velocity profile, taking into account all matter components. The other lines, meanwhile, take into account only the contribution of stellar, dark matter, and gas components, respectively. It can be seen that stars dominate the dynamics of the central 5~kpc. This is, of course, typical of observed galaxies \citep[see, e.g.,][]{marasco2020}, but it illustrates well why subtle changes in the stellar mass concentration are able to affect the position of the inner Lindblad resonance so strongly. As the cumulative stellar mass increases away from the centre, there is a corresponding rapid increase in the total circular velocity. This increase eventually levels off, with the galaxy maintaining an overall flat rotation profile from then on, as is characteristic of disc galaxies embedded in dark matter haloes. From this point onwards, the $\mathrm{d}\bupsilon_\text{circ}/\mathrm{d}r$ term effectively becomes negligible. Inspecting Eq.~\eqref{eq:kappa}, we see that this leads to $\kappa \propto 1/r$, and therefore also the profile of the inner Lindblad resonance tends to $\Omega - \kappa/2=(1-1/\sqrt{2})\Omega\propto 1/r$. Consequently, if we require the peak of this profile to stay low, the overall velocity curve must begin to flatten later. This, in turn, is only possible if the stellar concentration is kept sufficiently low.

\begin{figure*}
    \includegraphics[width=\textwidth]{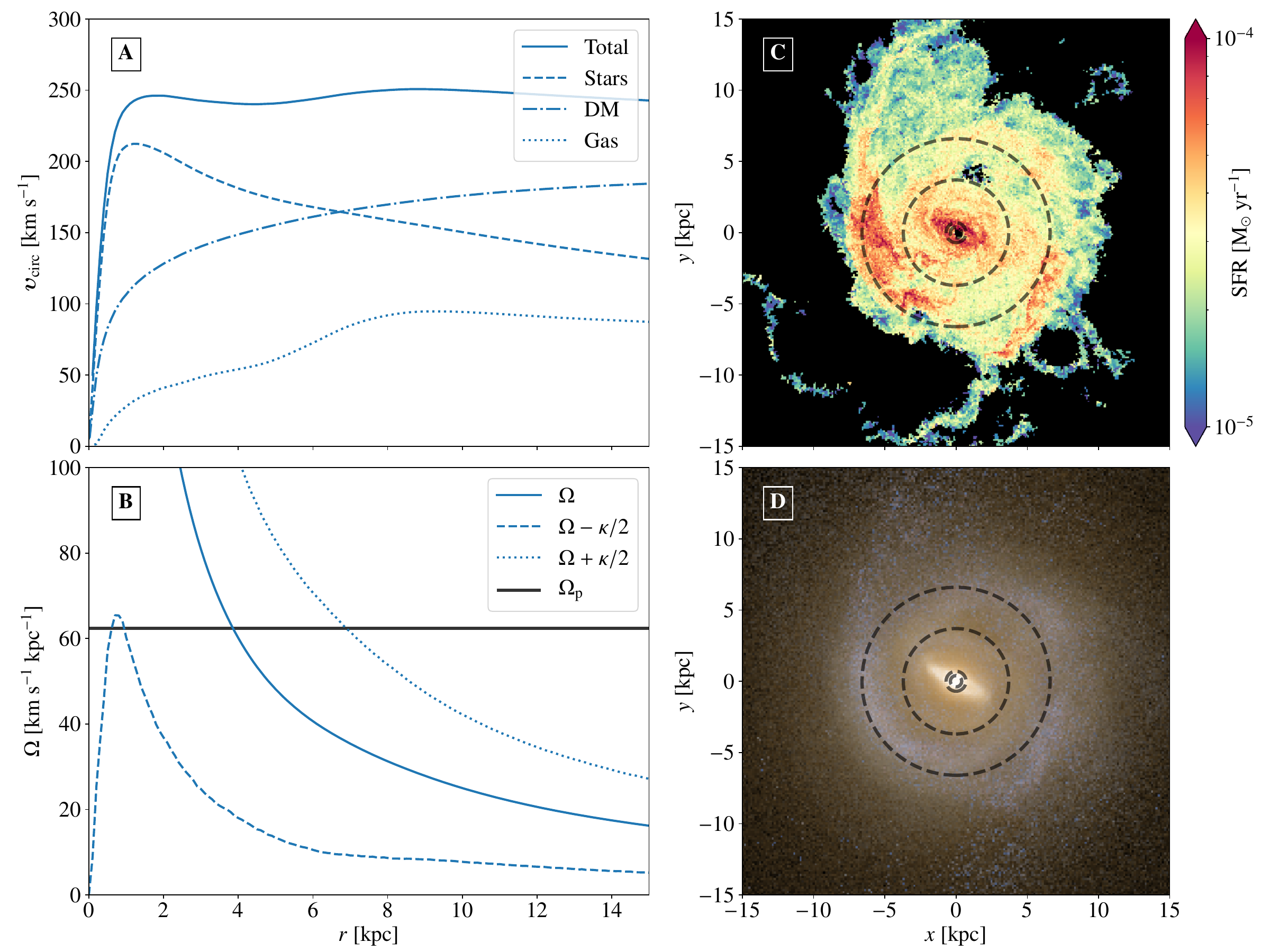}
    \caption[How Lindblad resonances generate ring-like star formation]{\textit{Top left:} The circular velocity as a function of radius in the disc for 1349-3H at 3 Gyr post-merger. \textit{Bottom left:} The corresponding inner Lindblad (dashed), co-rotation (solid), and outer Lindblad resonances (dotted) as a function of radius. The horizontal line indicates the bar pattern speed measured for this galaxy. The intersection of this line with the profiles marks the radial position of the resonances. \textit{Top right:} A slice through the midplane of the galaxy indicating the star formation rate in each cell, with the resonant positions overlain as dashed circles. \textit{Bottom right:} As above, but the background image now shows a mock \textit{gri} image. Resonances generated by the bar drive gas to the Lindblad resonances, producing a high star formation rate there. This has a pivotal role in how the hydrodynamic remnants develop.}
    \label{fig:resonances}
\end{figure*}

Using Eqs.~\eqref{eq:omega} to~\eqref{eq:kappa}, we obtain the resonant profiles observed in panel~B. In addition to the inner Lindblad resonance, we show two further resonances here: the co-rotation resonance and the outer Lindblad resonance. The condition for the former is fulfilled when an orbit's angular frequency is equal to the forcing frequency: $\Omega = \Omega_\text{p}$. For the outer Lindblad resonance, the condition is $\Omega_\text{p} = \Omega + \kappa/2$ (or $l=1$ and $m=2$, in the language introduced previously) so that the star particle once again performs two radial oscillations for each revolution of the bar, but this time lags behind in the co-rotating reference frame. The presence of these resonances is especially important for gas cells that are not on exactly circular orbits. In this case, gas between the co-rotation and outer Lindblad resonance experiences a net \textit{positive} torque from the bar, whilst that between the co-rotation and inner Lindblad resonance experiences a net \textit{negative} torque \citep{buta1996}. As the resonances are approached, the eccentricity of stellar orbits increases whilst the major axes of the dominating family of orbits rotates by 90 degrees, meaning that orbit crossing becomes inevitable \citep{sellwood1993}. However, the gas cannot interpenetrate and will therefore develop shocks at the position of these resonances, provided its sound speed is not large enough \citep{englmaier1997}, as is the case for the cold star-forming gas. The end effect is that gas is removed from the co-rotation resonance and accumulates at the two Lindblad resonances.

The position of the resonances in the disc can be determined by observing where the resonant profiles intersect with the bar pattern speed, $\Omega_\text{p}$, as indicated by the horizontal, black line in panel~B. This pattern speed was calculated using the standard method based on Fourier decomposition, as applied, for example, in \citet{fragkoudi2021}. We summarise this method in Appendix~\ref{appendix:pattern-speed}. It can be seen that solutions for the inner Lindblad resonance exist at 0.4 and 0.7 kpc, respectively, for the co-rotation resonance at 3.7 kpc, and for the outer Lindblad resonance at 6.6 kpc. As already mentioned, the pattern speed of the bar varies slightly over time. Correspondingly, the radii at which the resonances exist varies over time as well. This is particularly important for the inner Lindblad resonance, which has no solutions when the pattern speed is only a few km s$^{-1}$ kpc$^{-1}$ higher. Overall, the previously-discussed $x_2$ family of orbits is typically restrained to a radial annulus of 0.3 kpc. This is unlikely to be a significant problem for the bar, as we will see in the next two panels.

The impact of the resonances can be understood by examining panel C of Fig.~\ref{fig:resonances}. Here we show a slice through the disc midplane, with colours indicating the star formation rate in each gas cell. Regions in black show where no star formation is taking place. Overlain as dashed circles are the radial positions of the resonances. It can be seen that the regions of heightened star formation align well with the bar near the inner Lindblad resonance and at the edge of the disc near the outer Lindblad resonance. This pattern is a direct tracer of the dense gas that has accumulated at these positions under the action of continuous gravitational torques from the bar. At the outer Lindblad resonance, star formation rates are further increased by the steady accretion of gas post-merger. Indeed, it is possible to see, both above and below the disc, regions of star formation outside the disc. These are dense gas streams, which are helping to fuel star formation in the ring.

The formation of stars in this manner is extremely influential for how the remnant morphology develops. In panel D of Fig.~\ref{fig:resonances}, we show a face-on mock \textit{gri} image of the remnant, created in the same manner as in Fig.~\ref{fig:mock_visual}. Once again, the radial positions of the resonances are overlain. It can be seen the star-forming ring, as indicated by the bluish hues, lines up perfectly with the outer Lindblad resonance. Meanwhile, the co-rotation resonance is practically devoid of new stars, as dense gas has been removed from this region. The bar is also sufficiently large such that the inner Lindblad resonance lies within it. The $x_2$ orbits that would have acted against a weaker bar have therefore almost certainly been subdued by the bar's self-gravity \citep[see, e.g.,][]{kormendy2004}.

Although not presented here, we have performed similar analysis for the simulation Au2-H \citepalias[see figure~9 in][]{whittingham2021} -- a hydrodynamic analogue of one of the original Auriga galaxies. We observe that the star forming ring in this case also aligns with the outer Lindblad resonance. This galaxy is able to grow substantially larger than our merger remnants, however, with a degree of star formation even taking place beyond the outer Lindblad resonance. This is a result of the way accretion takes place in these simulations, and, in particular, the lower star formation rates that result in a more limited impact from wind particles.

\subsection{The impact of stellar feedback on accreting material}
\label{chapter4-subsec:feedback}

\begin{figure*}
    \includegraphics[width=\textwidth]{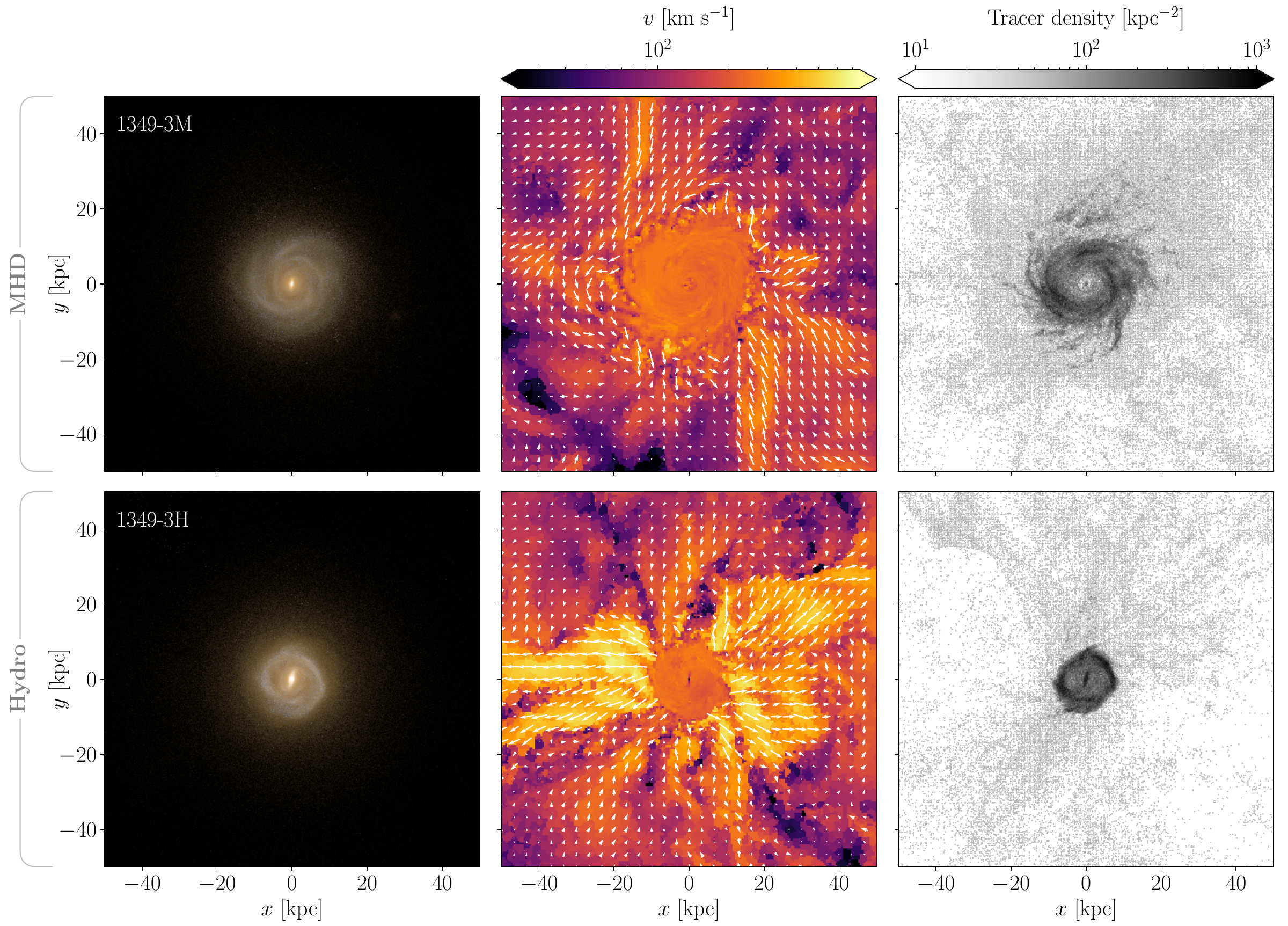}
    \caption[Gas velocity and tracer analysis showing the impact of stellar winds on accretion]{\textit{Left:} Face-on mock \textit{gri} images of the 1349-3 remnants, as seen approximately 5~Gyr post-merger (lookback time of $\sim1.4$~Gyr). \textit{Centre:} The gas velocity in the disc midplane at this time. Arrows indicate the direction, whilst colours indicate the magnitude. We have removed arrows from the approximate area of the disc. \textit{Right:} The surface density of Monte-Carlo tracers that will end up in the disc at $z=0$, where we define this as a cylinder of height $\pm1$~kpc and radius 19.35~kpc (13.14~kpc) for the MHD (hydrodynamic) simulation. In the hydrodynamic simulation, a strong stellar wind disrupts the angular momentum of gas joining the disc, keeping the disc compact. In the MHD run, however, the stellar wind is much less effective, and gas is able to join the disc almost in-situ, helping it to grow rapidly.}
    \label{fig:accretion}
\end{figure*}

The accretion of gas post-merger, particularly from the former CGM, plays a major role in the rebuilding of a galaxy's disc \citep{sparre2022}. However, as identified in Sec.~\ref{chapter4-subsec:how_evolution_differs}, the remnants in the hydrodynamic and MHD simulations grow to markedly different sizes. We will show in this section that this predominantly results from the impact of stellar winds on post-merger accretion.

As explained in Sec.~\ref{chapter4-sec:ISM}, in our simulations, stellar feedback is implemented through the use of wind particles. These are generated at star formation sites and are launched isotropically, interacting only gravitationally until they: a) reach a gas cell with a density that is 5\% of the star formation threshold density, or b) exceed the maximum travel time. At this point the particle's momentum and energy is deposited in its parent gas cell, with energy being split equally into thermal and kinetic parts. In the Auriga simulations, this leads to bipolar winds at late times \citep{grand2017}. This is emergent behaviour arising from the fact that particles encounter lower density gas more quickly when they travel away from the disc midplane; the wind thereafter takes the path of least resistance. In our own simulations, the merger-driven starburst significantly increases the overall number of wind particles formed, helping increase their influence. The result of this is, however, extremely different for the two physics models. We show this in Fig.~\ref{fig:accretion}, where we examine the impact of winds on accretion for the ``1349'' remnants. We do this specifically for a snapshot taken at approximately 5~Gyr post-merger, but our analysis may, of course, be generalised across all simulations and a broad range of times. We show this explicitly in Appendix ~\ref{appendix:stellar_feedback}. 

In the first column of Fig.~\ref{fig:accretion}, we show face-on mock \textit{gri} images, created in the same manner as in Figs.~\ref{fig:mock_visual} and~\ref{fig:resonances}. It can be seen here, that the remnant in the MHD simulation is beginning to form a disc with spiral arms, whilst that in the hydrodynamic simulation has formed the bar and ring morphology previously discussed. These different morphologies lead to a different distribution of wind particles, which alters their impact.

In the second column of the figure, we show a slice through the disc midplane, with colours indicating the magnitude of the gas velocity. Arrows indicate the plane-projected direction of this velocity, with a length scaled to the magnitude of this projection. We have removed arrows from the approximate area of the disc to highlight the dynamics of the CGM. It can be seen that the velocity distributions in each panel exhibit very different patterns; whilst the gas flows in the MHD simulation are predominantly azimuthal, in the hydrodynamic analogue, flows are preferentially radially orientated. These radial outflows are powered by wind particles resulting from the high density star formation at the disc edge, as observed previously in Fig.~\ref{fig:resonances}. As the gas density drops abruptly at the disc edge, wind particles moving in this direction may recouple almost immediately, generating strong, coherent winds, which whisk neighbouring gas away. This leads to a further drop in the gas density at this radius, as was shown quantitatively in \citetalias{whittingham2021}, helping the process to continue.

The strong outflows in the hydrodynamic simulation strongly affect the accretion of gas; because of these, inflows are restricted to areas where star formation -- and therefore the stellar wind -- is weaker, limiting the accretion rate. Moreover, the inflows that do manage to reach the disc are strongly radially-orientated, owing to the disruption of gas angular momentum in the CGM. Together, these factors limit the overall intake of high angular momentum gas in the galaxy, curtailing the growth of the disc. In contrast, star formation in the MHD simulation is spread over a much wider area, with relatively limited star formation at the disc edge. Wind particles are therefore much less effective at disrupting the gas velocity distribution in the CGM and gas that joins the disc can retain its high angular momenta.

\begin{figure}
    \centering
    \includegraphics[width=0.5\columnwidth]{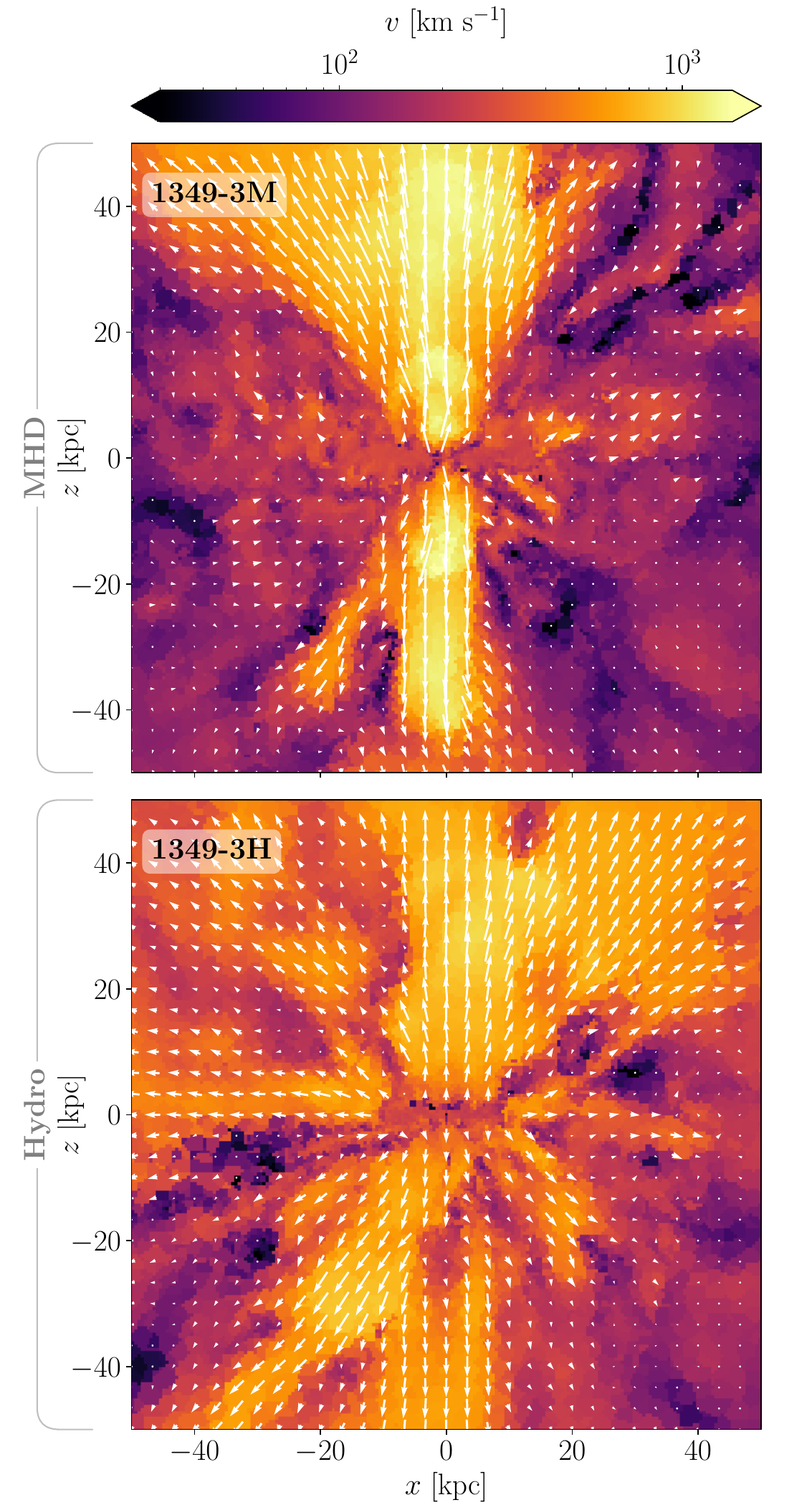}
    \caption[Edge-on slices showing gas velocity]{As the second column of Fig.~\ref{fig:accretion}, but showing the remnants edge-on. The strong stellar wind in the hydrodynamic simulation disrupts the CGM in all directions. Meanwhile, in the MHD simulation, outflows are predominantly bipolar and gas in the midplane is consequently able to keep its high angular momentum.}
    \label{fig:outflow}
\end{figure}

\begin{figure*}
    \includegraphics[width=\textwidth]{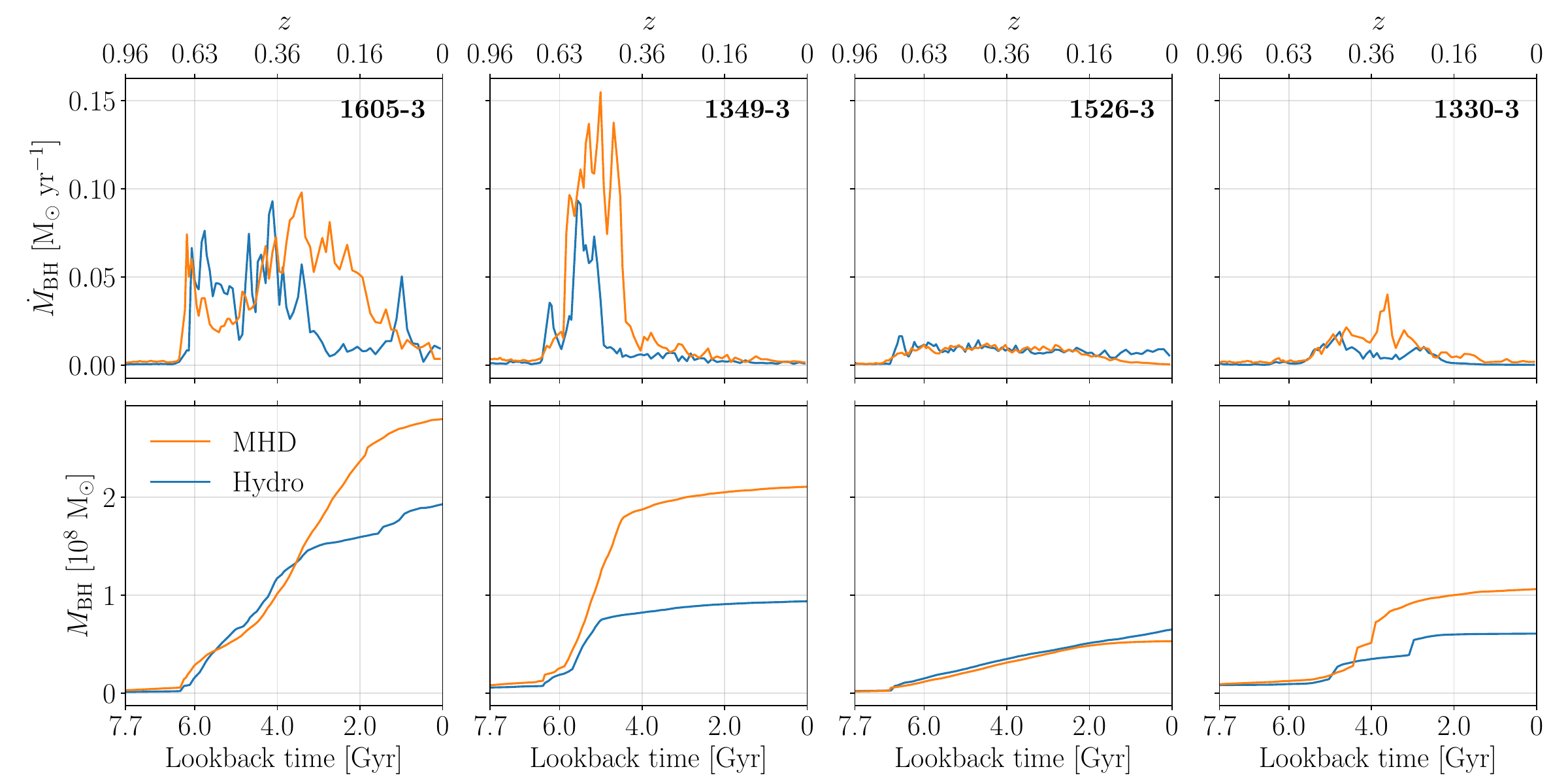}
    \caption[Black hole accretion rates and cumulative mass as a function of time]{\textit{Top row:} The black hole accretion rate in each simulation as a function of time. \textit{Bottom row:} The black hole mass in each simulation as a function of time. Black holes in MHD simulations can grow up to a factor of 2 larger than their hydrodynamic analogue, owing to the increased gas concentration in these simulations.}
    \label{fig:BH_accretion}
\end{figure*}

We illustrate the differences between how gas accretes onto each remnant in the final column of Fig.~\ref{fig:accretion}. Here we show the surface density of Monte-Carlo tracers (see Sec.~\ref{chapter4-subsec:arepo_mc_tracers}) that will end up in the disc at $z=0$. We define the disc of each remnant to be a cylinder of depth $\pm1$~kpc with a radius of 19.35~kpc and 13.14~kpc for 1349-3M and 1349-3H, respectively. These radii are the point at which the $B$-band surface brightness drops below $\mu_{B} = 25$ mag arcsec$^{-2}$ \citep[see definition of \textit{optical radius} in][]{grand2017, whittingham2021}. It can be seen that, for the hydrodynamic simulation, a large number of tracers already exist in the bar and ring regions, reflecting the high star formation density here. Outside the disc, however, the density of tracers drops strongly, with tracers only evident in thin filaments, indicating radial accretion of the like identified in the previous column. In the MHD simulation, on the other hand, there is an extensive population of tracers that exist in the immediate neighbourhood of the disc. This population provides a pool of high angular-momentum gas. This joins the disc practically in-situ, thereby enabling its rapid growth.

The full scale of the impact of wind particles can be better understood by also examining the gas dynamics above and below the disc. We show this in Fig.~\ref{fig:outflow} for the 1349-3 remnants for the same time and in the same manner as in the central column of Fig.~\ref{fig:accretion}. It can be seen that, for the hydrodynamic simulation, wind particles dominate the dynamics of practically the entire panel. This is enormously disruptive to the angular momentum of the gas. The distance at which the stellar wind is still active implies that a large-scale fountain flow is in effect, which helps to maintain the radial inflows.

In contrast to this, the velocity distribution in the MHD simulation is predominantly bipolar. This means that gas in the midplane is left mostly unaffected, as is indicated by the arrows, which show extremely small projected velocities. The outflow velocities are generally higher in the MHD simulation by a factor of a few and also originate predominantly from the centre of the disc. This is because, in these simulations, outflows are more greatly influenced by black hole feedback. We explore this in our final analysis section.

\subsection{Altered black hole feedback}
\label{chapter4-subsec:AGN}

In our simulations, as described in Sec.~\ref{chapter4-sec:ISM}, the energy released by a black hole is directly proportional to its accretion rate. This, in turn, depends on the gas density in the neighbourhood of the black hole \citep[see eq. 8 of][]{grand2017}. As gas is typically more concentrated in our MHD simulations post-merger (see Sec.~\ref{chapter4-subsec:angmom}), we should expect accretion rates to also be higher and consequently black hole feedback to be more influential. We show that the first of these statements is true in Fig.~\ref{fig:BH_accretion}.

In the first row of the figure, we show the black hole accretion rates for each simulation as a function of time, with MHD simulations shown in orange and hydrodynamic ones in blue. In each case, the arrival of the merging galaxy is associated with an uptick in the black hole accretion rate. Except for 1526-3, it is evident that accretion rates are indeed, on the whole, higher in MHD simulations. The cumulative effect of this increased accretion is that the black hole mass grows substantially larger, as we show in the second row of the figure. In this row, we show the evolution of the total black hole mass as a function of time. Whilst this evolution is dominated by accreted mass, it also includes the impact of black hole mergers. Such mergers produces the discontinuous increases seen, for example, in the 1330-3 simulations. The timing of these increases is different for different physics models, owing to the individual merger trajectories taken, as discussed in Sec.~\ref{chapter4-subsec:angmom}.

Except in 1526-3, the black hole in the MHD simulation accumulates between 1.5 -- 2 times as much gas as its hydrodynamic analogue by $z = 0$, owing to the increase in baryonic concentration in these simulations. Such density increases will clearly be at their highest in major mergers of gas-rich galaxies. However, under our model, even simulations of more isolated galaxies should exhibit mild density increases when performed with MHD (see Sec.~\ref{chapter4-subsec:angmom}). These galaxies will therefore also show heightened black hole accretion rates. This implies that increased black hole masses are a generic feature of including magnetic fields in the Auriga model. Nonetheless, even if the average black hole mass increased by a factor of two (i.e. the maximum value seen in Fig.~\ref{fig:BH_accretion}) such values would still be well within the scatter of the well-known black hole  -- halo mass relation \citep{reines2015}. The increase is also clearly only true in a statistical sense; not every remnant in Fig.~\ref{fig:BH_accretion} shows an increase.

The answer as to why the black hole in 1526-3M does not grow larger than its hydrodynamic analogue has already been identified in Sec.~\ref{chapter4-subsec:angmom}; namely, the magnetic field configuration, and therefore gas density evolution, in this galaxy is different. Here, the magnetic field becomes azimuthally-dominant just as it becomes dynamically important, unlike the non-azimuthal dominance seen in the other three MHD simulations. This means that gas is actually supported from collapse in this simulation, as seen in the angular momentum evolution provided in Fig.~\ref{fig:ang_mom}. Such support may also explain the cessation of black hole accretion in the last $\sim2$~Gyr in this simulation.

Under our black hole model, galaxies that have higher accretion rates necessarily have increased levels of quasar feedback. After the remnant has formed a disc, quasar feedback typically acts to displace gas periodically from the centre. The effect of this can be seen in Fig.~\ref{fig:accretion} through the low tracer density at the centre of the MHD remnant, and in Fig.~\ref{fig:gas_vel-sfr} through the face-on signatures of central outflows and the coincident star formation voids. However, whilst such phenomena are more frequent in the MHD simulations, their impact on the remnant evolution as a whole turns out to be limited. This, perhaps, should be expected, as whilst the black hole accretion rates in Fig.~\ref{fig:BH_accretion} are substantially higher in three out of the four pairs of simulations, morphological differences are observed between \textit{all} pairs of simulations in \citetalias{whittingham2021}; our model must also explain why the 1526-3 simulations evolve differently.

We show explicitly that quasar feedback does not explain the morphological differences in our simulations in Fig.~\ref{fig:gas_dist_evolution}. In this figure, we present a series of slices through the midplane of the 1349-3 simulations showing the gas density. In addition to the standard MHD and hydrodynamic simulations, we also include two further simulations in this figure. In these, we have switched off quasar feedback at the start of the merger (see Fig.~\ref{fig:ang_mom} for times). By doing so, we allow the galaxies to evolve normally pre-merger, and thereby isolate the impact of quasar feedback on the re-growth phase of the disc. The resulting simulation data is naturally not reflective of real galaxies, as, in particular, we remove the pressure support of quasar feedback post-merger, allowing gas to concentrate unphysically at the centre. Nonetheless, the results are instructive. We chose the 1349-3 simulations for this figure as these showed the greatest difference in accretion rates in Fig.~\ref{fig:BH_accretion}, and therefore have the greatest difference in energy output by the black hole post-merger; if quasar feedback is ineffective here, we should not expect it to be effective when the energy output is weaker.

\begin{figure*}
    \includegraphics[width=\textwidth]{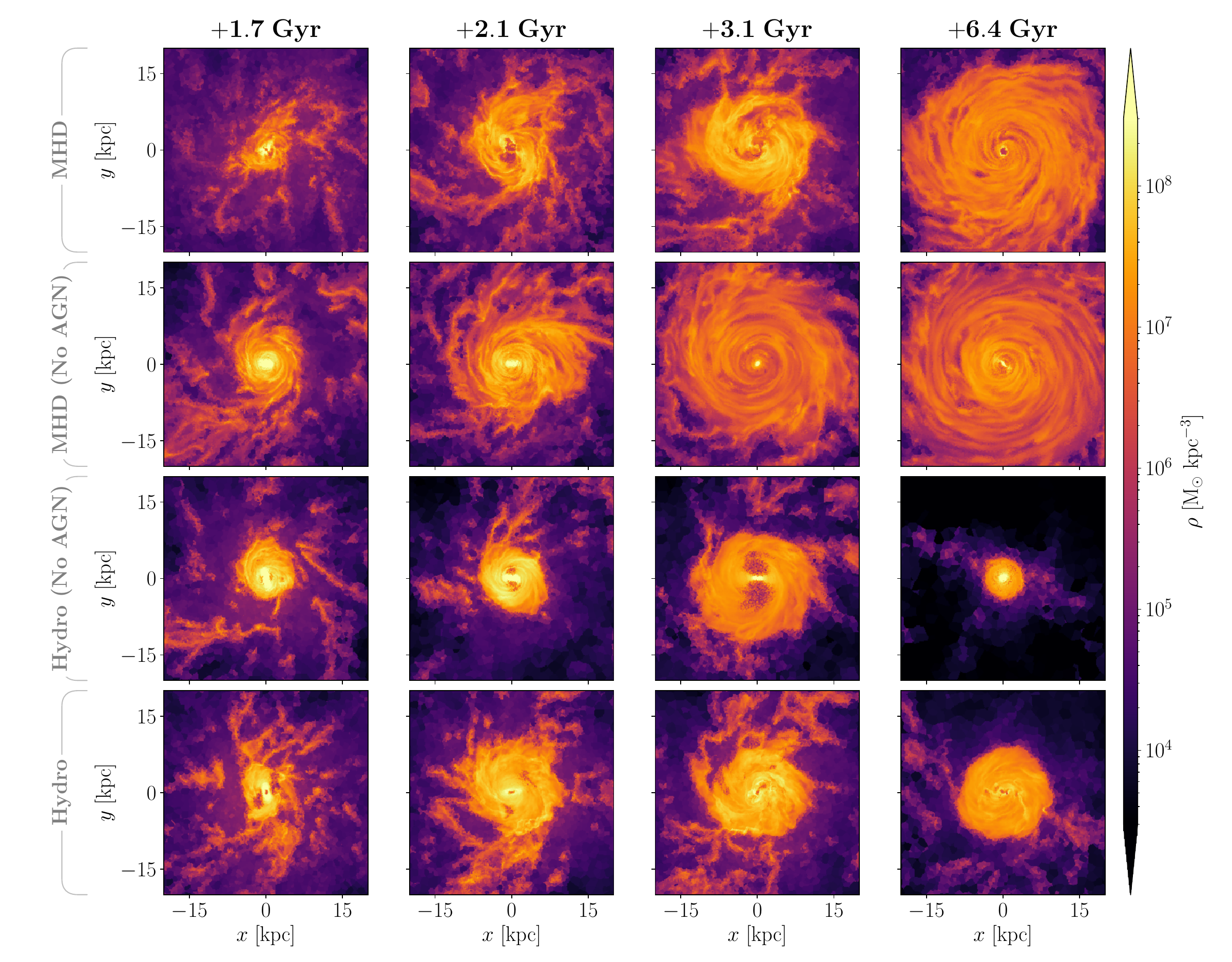}
    \caption[Face-on slices showing gas density for models with and without AGN feedback post-coalescence]{Face-on slices through the disc midplane showing the gas density in the 1349 simulations. Times are given from the start of the merger. \textit{1st row:} Standard MHD simulation. \textit{2nd row:} MHD simulation, but quasar feedback was turned off at the start of the merger. \textit{3rd row:} Hydrodynamic simulation, but quasar feedback was turned off at the start of the merger. \textit{4th row:} Standard hydrodynamic simulation. It is apparent that the morphological differences between hydrodynamic and MHD simulations only become stronger once quasar feedback is removed. Increased quasar feedback in MHD simulations can therefore not be the primary cause of the divergent evolution in the original runs.}
    \label{fig:gas_dist_evolution}
\end{figure*}

We show the four variations at different times in the process of rebuilding their disc. The physics included in each simulation is labelled on the left-hand side. The amount of time elapsed since the beginning of the merger is also given above each column, with the final column equivalent to $z=0$. In the first column of the figure, the gas discs are all of a similar size. Those simulations where quasar feedback was included appear more disrupted as their morphology has been affected by outbursts, preventing the gas from collapsing neatly into a disc. Such outbursts are particularly strong shortly after the merger, when gas reaches high densities and black hole accretion rates are correspondingly high. 

There are signatures of such outbursts in the top row of Fig.~\ref{fig:gas_dist_evolution} until past the 3 Gyr mark, as evidenced by the density irregularities in the disc until this time. Whilst the most major outbursts take place early on in the rebuilding process, they have a lasting impact. This can be seen by comparing the final disc sizes produced in the two MHD simulations; the disc in the original MHD simulation actually ends up smaller than that in the \textit{MHD (No AGN)} simulation, as outbursts post-merger disrupt the angular momentum of both accreting gas and gas already in the disc. The opposite, however, is true of the hydrodynamic simulations. Here, the final disc size in the \textit{Hydro (No AGN)} simulation is significantly smaller than in the original run. This is because in the hydrodynamic simulations, the dynamics are being more strongly affected by another component; the formation of a central bar.

Both hydrodynamic simulations form bars quickly, but this becomes particularly disruptive in the \textit{Hydro (No AGN)} variation. Here, the bar dominates the centre of the disc, sweeping up gas during its rotation, leading to strong underdensities. Such underdensities are already evident in the +2.1 Gyr snapshot, but are particularly extreme in the following snapshot, where they extend to a distance of a few kpc from the centre. The accretion of gas onto the centre of the bar, however, eventually destroys it, as can be seen in the last snapshot. At this point, support for resonant orbits is removed, and, without any black hole feedback to provide remaining pressure support, the underdensities rapidly fill in, leading to a drop in the disc size.

To summarise, even without quasar feedback, the remnants continue to evolve in ways that are distinctive to the underlying physics models. Indeed, ultimately, the removal of quasar feedback post-merger actually leads to an even larger morphological difference between the two physics models. This suggests that, rather than cause the effect, black hole feedback may actually suppress some of the morphological differences that result from including MHD physics.

\section{Discussion}
\label{chapter4-sec:discussion}

\noindent We have identified four important questions that arise from this work:
\setlist{nolistsep}
\begin{enumerate}
    \item to what extent does the discussed mechanism apply to other mergers?
    \item to what extent does it apply to other galaxy formation models? 
    \item to what extent is the numerical technique used for solving the MHD equations responsible for the results obtained?
    \item how essential are magnetic fields in our model; i.e. does the model rely intrinsically on magnetic fields, or can it be replicated through the tuning of other feedback model parameters?
\end{enumerate}
We attempt to answer these questions below.

\subsection{Applicability of the model to other merger scenarios}

The mergers analysed in this paper are all gas-rich major mergers between disc galaxies situated in MW-sized haloes. Of these properties, it is the gas-rich nature of the mergers that is the most important for our mechanism; firstly, as noted in \citetalias{whittingham2021}, sufficient gas is required in order to amplify the magnetic field through turbulence and adiabatic compression to dynamically-important levels. However, in turn, the magnetic fields in our simulations are also only able to act upon gas, and, as described in detail in Sec.~\ref{chapter4-subsec:angmom}, it is the motion of this gas in response to torques applied by the magnetic field that ultimately causes the observed morphological differences. We therefore expect magnetic fields to be less influential in gas-poor mergers, such as in the case of mergers between elliptical galaxies.

With this said, virtually all galaxies in cosmological simulations will have undergone a gas-rich merger at some point in their history. Indeed, in \citetalias{whittingham2021}, it was shown that even in the case of isolated, but still cosmological simulations, the consequences of such a merger can be felt for several Gyr after the event. Although a full application of our analysis to such simulations is outside the scope of this paper, we note that our proposed mechanism explains observed features here too, including the appearance of stellar rings at the outer Lindblad resonance (see Sec.~\ref{chapter4-subsec:stellar_ring}). It seems therefore likely that our mechanism applies more generally in a cosmological context.

\subsection{Applicability of the model to other galaxy formation models}

As described in the introduction of this paper, there are several competing galaxy formation models now available that include an implementation of MHD. However, in only a few of these have magnetic fields been able to impact the dynamics. The inability of the magnetic field to become dynamically important may be linked to a number of factors. For example, it will depend on the seed field strength chosen, the diffusivity of the numerical implementation, and the resolution of MHD phenomena such as amplification through the small-scale dynamo and magnetic draping. As described in \citetalias{whittingham2021}, the magnetic field strengths in Auriga compare favourably with observations of real disc galaxies, which bodes well for analysis of their dynamical importance. We note, too, that simulations where magnetic fields were able to become dynamically important, were able to replicate some of our results. For example, in both \citet{martin-alvarez2020} and \citet{katz2021}, which employed both different numerical methods and implementations of MHD from our own, it was identified that, given sufficiently high field strengths, magnetic fields can torque the gas, thereby reducing the size of the disc, albeit at the expense of using artificially large initial magnetic field strengths (see discussion in Sec.~\ref{chapter4-sec:numerical_technique}). Such torques form a key part of our own model.

We do not expect a different stellar feedback model to substantially affect the parts of our mechanism that relate to turbulence. For example, whilst the explicit resolution of stellar feedback could generate small-scale turbulence more quickly, thereby shortening the time taken for the magnetic dynamo to reach the non-linear amplification stage, we have already shown in \citetalias{whittingham2021} that this would be unlikely to change the saturation point of the magnetic field. We note as well that whilst stellar-driven turbulence in the CGM and on the ISM-CGM border is already captured by wind particles \citep{pakmor2020, vandevoort2021}, during the merger, turbulence in the ISM is overwhelmingly gravitationally-driven \citepalias[see, e.g., fig.\ 15 of][]{whittingham2021}. This is already fully-captured in the \citet{springel2003} ISM model.

With this said, other aspects of stellar feedback could still play a significant role. For example, more explosive stellar feedback would likely disrupt the formation of high density structures, having a particularly strong impact on the formation of stellar rings. In contrast, the wind particle implementation, as used in Auriga, allows gas to stay at high densities, as the multiphase nature of the ISM cannot be resolved and wind particles only recouple below a threshold density. Indeed, it is noticeable that in the original hydrodynamic Illustris simulation, which also used a wind particle implementation, galaxies frequently formed ring like structures \citep[see, e.g., fig.\ 1 and 13 of][]{marinacci2014}. Wind particles in this simulation were launched with a bipolar model, where particles were explicitly launched away from the disc \citep[see, e.g.][for further details]{pillepich2018}. However, as seen in Fig.~\ref{fig:outflow}, this would likely be effective enough to disrupt the angular momentum of the CGM, as required under our model. The updated Illustris TNG model \citep{nelson2019}, meanwhile, forms approximately the right frequency of barred galaxies \citep{zhao2020} and no longer forms such a large number of disc galaxies with star-forming rings (c.f.\ fig.\ 6 of \citealt{snyder2015} and fig.\ 5 of \citealt{rodriguez-gomez2019}). One of the major advances made in Illustris TNG compared to the original Illustris model was the implementation of magnetic fields. The importance of this addition may not have been fully appreciated.

Finally, we expect the cosmological nature of the simulation to play a large role. This affects many aspects of galaxy evolution. For example, as shown in \citet{sparre2022}, a substantial fraction of star formation post-merger originates from gas that was previously outside the discs of the progenitors. Without this additional gas, star formation rates would be lower, thereby reducing the impact of winds and the ability of the magnetic field to affect the disc rebuilding process. The existence of such gas also helps to maintain turbulence in the galaxy, aiding the amplification of the magnetic field, and therefore its dynamical importance, as examined in \citetalias{whittingham2021}. Furthermore, isolated simulations of galaxies are typically initialised with the magnetic field in an almost purely azimuthal or toroidal configuration. In contrast, in our own cosmological simulations, we find that the magnetic field can exhibit strong non-azimuthal components. Indeed, these are vital for producing the increased baryonic concentrations identified in Sec.~\ref{chapter4-subsec:angmom}. This points to the need to model magnetic fields self-consistently.

\subsection{Requirement on the numerical technique for resolving magnetic field growth}
\label{chapter4-sec:numerical_technique}

Cosmological magnetic fields are believed to have grown from seed fields produced in the early Universe. Typically, one of two sources are invoked for the production of such seeds: i) the Biermann battery mechanism, which is able to generate magnetic fields in proto-galaxies with typical values of $10^{-20}$ Gauss and coherence scales of several kpc, and ii) primordial magnetic fields, which could be produced with similar strengths during the epoch of cosmic inflation or during phase transitions in the post-inflation era \citep{widrow2002,kulsrud2005,brandenburg2005}. By adopting seed fields of such strength and modelling amplification in a small-scale dynamo with realistic Reynolds numbers of order $\textrm{Re}\sim10^{11}$, it is possible to theoretically explain the micro-Gauss strength of magnetic fields observed in galaxies today \citep{schober2013}. 

It turns out, however, that simulating this process explicitly is extremely computationally challenging. Indeed, current-day galaxy formation simulations are still far from resolving the necessary scales of turbulence required to amplify the magnetic field in the requisite time frame. As a result, the strength of the seed field must be artificially increased in order to make up for the missing resolution. However, at the same time, care must be taken to prevent increasing it to the point that the subsequent magnetic field unphysically modifies the process of galaxy formation e.g., by preventing gas accretion onto the forming disc through dynamically important magnetic pressure resulting from the adiabatically compressed field  \citep{marinacci2016,martin-alvarez2020,katz2021}.

Three possibilities exist to circumvent the aforementioned problems. Firstly, adaptive mesh-refinement simulations of magnetic field growth in galaxies can adopt extremely small (quasi-uniform) resolutions in the high-density regions of interest to be able to produce magnetic fields at the observed strengths \citep{martin-alvarez2022}. This method is, however, currently only appropriate for cosmological simulations of galaxies forming in isolation. Alternatively, the effective resolution can be increased by introducing a turbulent subgrid scheme, where the magnetic field growth via the small-scale dynamo is modelled below the formal grid resolution. This avoids the otherwise large numerical diffusion at the grid scale, which would preclude simulating a magnetic dynamo \citep{liu2022}. This approach, however, necessarily requires the addition of more free parameters to the overall model, which must then be tuned. The final approach is to use a moving mesh code. Because the numerical truncation error of a given numerical scheme is proportional to the sum of the absolute values of sound speed and gas velocity relative to the mesh, the numerical diffusion can be substantially reduced and the effective Reynolds number thus increased by using a Voronoi mesh that is co-moving with the flow \citep{springel2010,bauer2012}. By using this method, we substantially boost the effective resolution, enabling us to resolve the small-scale dynamo in galaxies, whilst ensuring that the magnetic fields do not artificially interfere with the collapse and formation of the galaxy \citep{pakmor2017,pfrommer2022}.

\subsection{Can the effect of our model be mimicked in hydrodynamic simulations?}

Despite the broad range of differences between galaxy formation models, each claims to be able to replicate some aspect of galaxy evolution. This implies a certain level of degeneracy in these models, given the current level of observational error attached. It is therefore natural to ask: are magnetic fields actually required for creating accurate galaxies in Auriga, as proposed here, or can their impact be replicated by another mechanism? The most likely candidate for this would be the feedback implementation, given its well-documented impact on star formation processes. We note, for example that recent work has shown that quasar feedback may help to weaken bars in Auriga \citep{irodotou2022}. This would help to reduce the likelihood of forming a star-forming ring under our model. As explained in Sec.~\ref{chapter4-subsec:AGN}, however, the overall impact is unlikely to be enough. The impact of more influential black hole feedback in a still hydrodynamic model can, furthermore, be observed in our own simulations in 1526-3H (see Appendix B of \citetalias{whittingham2021}). As can be seen in fig.~7 of \citetalias{whittingham2021}, whilst this does indeed weaken the bar, the remnant still shows a substantially different morphology compared to its MHD analogue.

Stellar feedback has also been shown to be highly influential in merger simulations for a range of models \citep[see, e.g.,][]{moreno2019, moreno2021, li2022}. However, rescaling our stellar feedback would, too, almost certainly not prevent the observed morphological divergence. This can be seen through inspection of the MHD and hydrodynamic versions of the Auriga simulations, as shown in \citetalias{whittingham2021}. For these galaxies, star formation was not as intense, and subsequently fewer wind particles were generated, reducing their effectiveness. On the one hand, this meant that the CGM was less disturbed and so the remnants could grow larger. Ultimately, however, a similar morphological divergence still takes place; hydrodynamic simulations still exhibit bar-and-ring structures whilst the remnants in the MHD simulations are predominantly MW-like.

More fundamentally, feedback and magnetic fields act in different ways; whilst feedback can transport the angular momentum of gas to large galactocentric radii, magnetic fields are able to promote inwards transport via magnetic draping \citep{lyutikov2006,dursi2008,pfrommer2010,berlok2019}, before magnetic tension forces transport and redistribute the angular momentum locally. This is inherently different and allows magnetic fields to initially reduce the size of the disc before helping to grow it substantially. In contrast, feedback through disruption can only reduce the size of the disc. We conclude from this that feedback can neither be tuned nor modified to replicate the mechanism we have presented in this paper.

\section{Conclusions}
\label{chapter4-sec:conclusions}

In this paper, we have investigated how magnetic fields are able to affect galaxy mergers in the framework of the Auriga galaxy formation model. To do this, we have analysed the simulations first presented in \citetalias{whittingham2021}. These are a series of high-resolution (dark matter resolution equal to $1.64 \times 10^5 \; \mathrm{M}_\odot$) cosmological zoom-in simulations of major mergers between disc galaxies in MW-like haloes. The mergers take place between $z=0.9-0.5$, and all remnants are subsequently able to regrow a disc. The remnant disc, however, is systematically larger in MHD simulations and also shows spiral arm features and an extended radial profile. In contrast, in hydrodynamic simulations, the remnant is compact and displays prominent bar and ring components. We have presented a mechanism in this paper that explains how magnetic fields cause this morphological divergence. Our model is provided as a schematic in Fig.~\ref{fig:schematic} and is as follows:

\begin{enumerate}
    \item Within a few 100~Myr of the first closest approach, the magnetic field becomes dynamically dominant. Non-azimuthally orientated parts of the field then effectively redistribute angular momentum between accreting gas and the gas in the disc. When the field is predominantly non-azimuthally orientated, this leads to an initial reduction in the disc size (Figs.~\ref{fig:mock_visual}, ~\ref{fig:bar_star_ages} and~\ref{fig:ang_mom}).
    \item The resultant higher baryonic concentration produces a strong inner Lindblad resonance, which suppresses the formation of a bar. When the magnetic field is predominantly azimuthally-orientated, the support it provides against collapse performs the same role. In the hydrodynamic runs, however, a large bar forms easily (Figs.~\ref{fig:mock_visual}, \ref{fig:bar_star_ages}, \ref{fig:gas_concentration}, and \ref{fig:resonances}).
    \item In the hydrodynamic simulation, the large bar shepherds gas towards the outer Lindblad resonance, resulting in a high star formation rate in this region. The absence of a strong bar in the MHD simulation, on the other hand, allows the gas to remain flocculent and for spiral arm features to develop (Figs.~\ref{fig:mock_visual},  \ref{fig:bar_star_ages}, and \ref{fig:resonances}). 
    \item The high star formation rate density in the hydrodynamic simulation launches a strong stellar wind away from the disc, disrupting the angular momentum of neighbouring gas cells, thereby keeping the remnant compact. In contrast, in the MHD simulation, winds are less effective and gas on the outskirts of the disc retains much of its angular momentum, resulting in rapid disc growth (Figs.~\ref{fig:accretion} and~\ref{fig:outflow}).
\end{enumerate}

In addition, we also find in this paper that:
\begin{itemize}
    \item Torques provided by the magnetic field are able to systematically reduce the time taken until coalescence (Fig.~\ref{fig:ang_mom}). This effect is particularly strong for inspiralling mergers, which experience several fly-bys.
    \item The increased gas concentration in MHD simulations is able to grow the central black hole up to a factor of two greater than in the hydrodynamic analogue (Fig.~\ref{fig:BH_accretion}). The subsequent increase in quasar feedback. however, does not have a significant impact on the remnant evolution (Fig.~\ref{fig:gas_dist_evolution}).
 \end{itemize}
 
Whilst the impact of magnetic fields is probably strongest under our set-up, we have shown in \citetalias{whittingham2021} that this impact is also felt in more isolated, but still cosmological galaxy simulations. Furthermore, as discussed in Sec.~\ref{chapter4-sec:discussion}, it seems highly unlikely that this impact could be replicated by a different feedback mechanism. We therefore conclude that magnetic fields are a crucial element of modelling galaxy formation in a cosmological environment, and that the modelling of disc galaxies, in particular, cannot be done correctly in purely hydrodynamic simulations.

\newpage
\section{Appendix}
\subsection[Appendix A: Edge-on mock stellar light images]{Analysis of edge-on mock stellar light images}
\label{appendix:edge-on-gri-images}
\begin{figure*}
    \includegraphics[width=\textwidth]{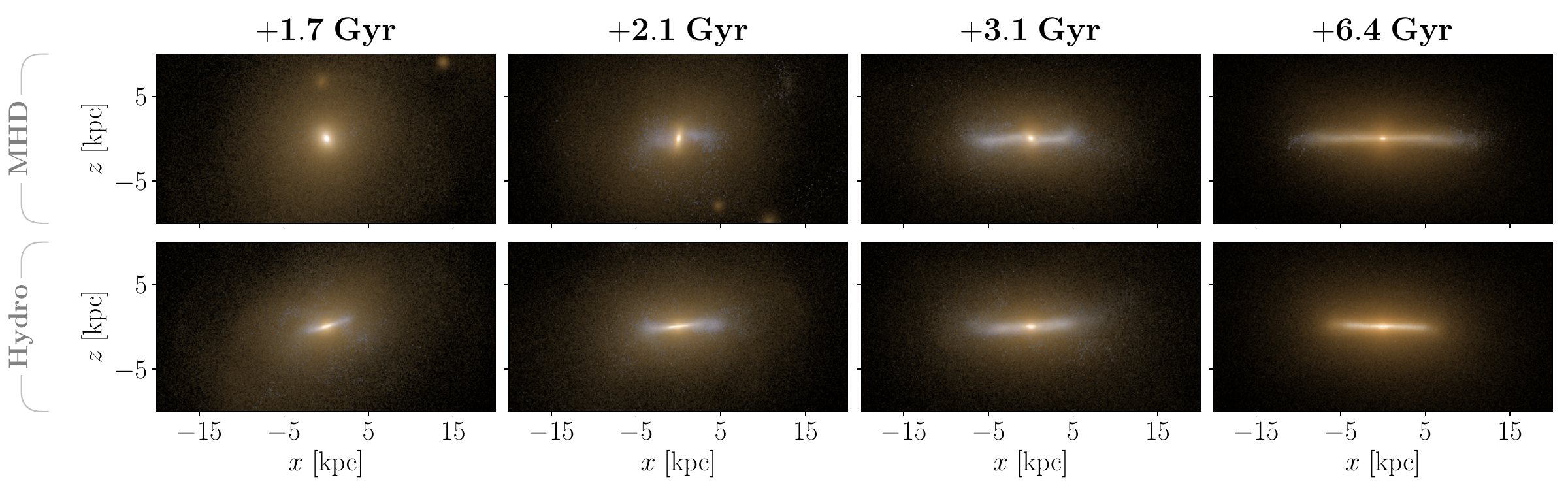}
    \caption[Edge-on mock stellar light images as a function of time]{As Fig~\ref{fig:mock_visual}, but the images are now shown edge-on. In the MHD simulation, the grow
    th of the young stellar disc takes place in misalignment with the bar. This is caused by the magnetic field torquing the gas disc.}
    \label{fig:mock_visual-edge-on}
\end{figure*}

In Fig.~\ref{fig:mock_visual-edge-on}, we show edge-on mock \textit{gri} images of the 1349-3 remnants as they evolve post-merger. These images are created in the same way as in Fig.~\ref{fig:mock_visual} and are shown at the same times. The images are orientated with the cold, dense gas disc, which, especially for the hydrodynamic simulation early on, does not always fully line up with the young stellar disc. This is due to the fact that the accreting gas has angular momentum  misaligned with said disc.

Several aspects in Fig.~\ref{fig:mock_visual-edge-on} support the analysis already provided in Sec.~\ref{chapter4-subsec:how_evolution_differs}. For example, the remnant in the MHD simulation at 1.7~Gyr post-merger is more compact than its hydrodynamic analogue. Similarly, the radial growth of the remnant in the hydrodynamic simulation stalls over time, whilst in the MHD simulation, growth is rapid following the initial compaction stage. The thickness of each disc at +6.4~Gyr ($z=0$) is similar, although in the hydrodynamic simulation the disc ends up marginally thinner. This originates from the thinner gas disc \citepalias[see fig.\ 6 of][]{whittingham2021}, which in turn arises from the stronger stellar winds in this galaxy, as analysed in Sec.~\ref{chapter4-subsec:feedback}.

Where the remnants differ in particular, however, is in the accretion of material. For example, it is clear in the second column of Fig.~\ref{fig:mock_visual-edge-on} that in the MHD simulation the bar is misaligned with the young stellar disc. This is in contrast to the bar in the hydrodynamic simulation, which sits aligned with this disc. This misalignment is a direct result of the magnetic field applying torques to the reforming gas disc, as we analysed in Sec.~\ref{chapter4-subsec:angmom}. This reduces the influence of the bar and allows stars to be born very close to the centre, thereby populating the $x_2$ orbits discussed in Sec.~\ref{chapter4-subsec:resonances}. Such orbits will ultimately suppress the bar, causing it to become noticeably smaller, as can be observed in Fig.~\ref{fig:mock_visual}.

\subsection[Appendix B: Generalising the stellar feedback analysis to all simulations]{Generalising the stellar feedback analysis to all simulations}
\label{appendix:stellar_feedback}

\begin{figure*}
    \includegraphics[width=\textwidth]{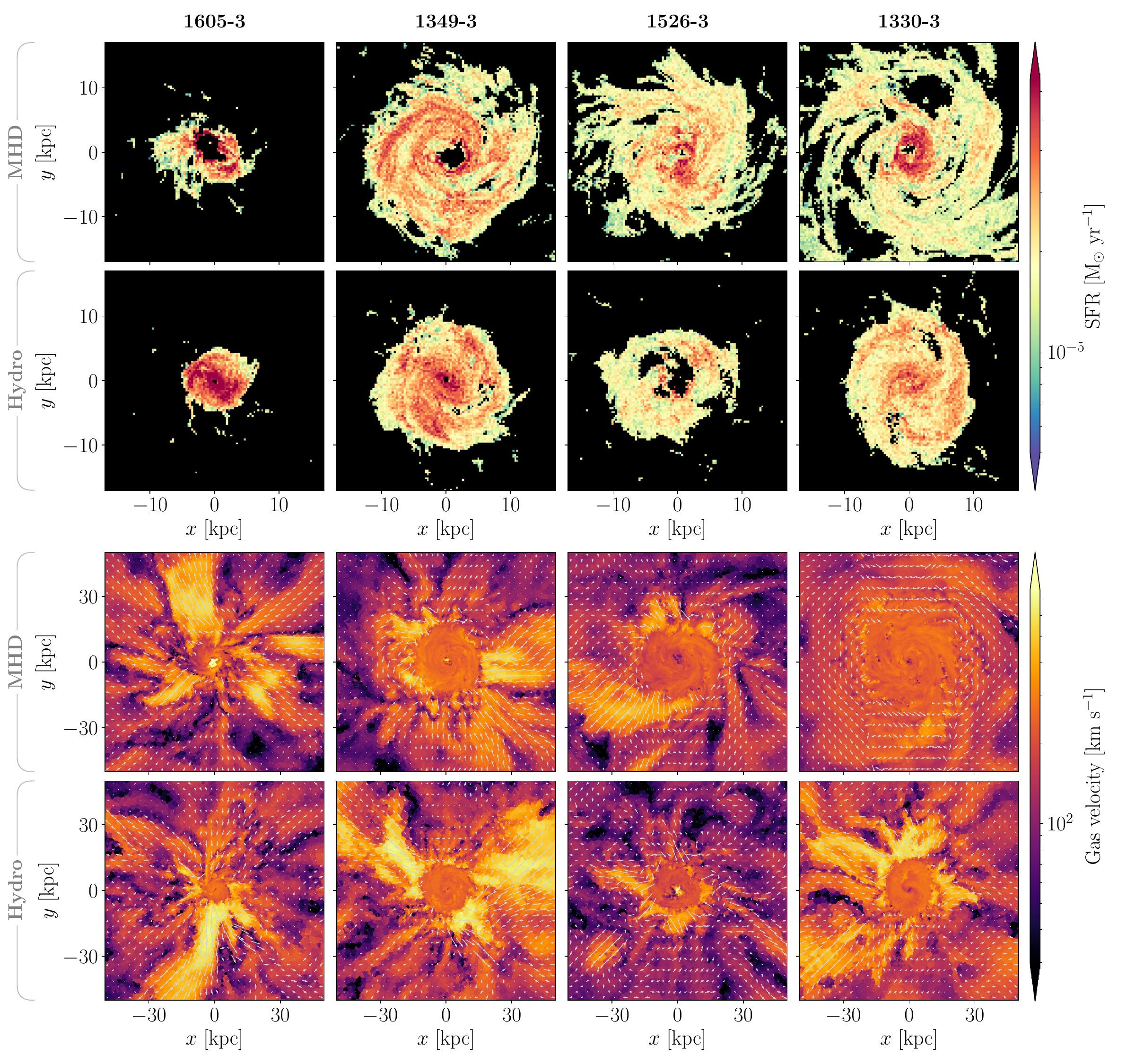}
    \caption[Slices showing star formation rates and gas velocity in each of the original simulations]{\textit{1st and 2nd row:} Face-on slices through the midplane of the disc for MHD and hydrodynamic simulations, respectively, where colours indicate the star formation rate in each cell. Remnants are seen at +4 Gyr after the beginning of the merger. \textit{3rd and 4th row:} As above, but colours indicate gas velocity, with arrows indicating the projected velocity in the CGM. The bar-and-ring structure observed for 1349-3M in earlier case studies is typical of all hydrodynamic simulations in our suite. The star forming rings strongly disrupt the dynamics of the neighbouring gas, preventing the accretion of high angular momentum gas and keeping the remnants compact. Star formation in MHD simulations, on the other hand, is more evenly distributed and the CGM is subsequently less disrupted, helping the remnant discs to grow faster radially.}
    \label{fig:gas_vel-sfr}
\end{figure*}

During this paper, we have frequently used the 1349-3 simulations as a case study. In this section, we support our claim that conclusions drawn from these studies may be generalised to the wider simulation suite. We focus, in particular, on our assertion that star formation is distributed differently in MHD simulations and that this, in turn, produces a velocity distribution in the CGM that is more conducive to the growth of the disc. We show this with the aid of Fig.~\ref{fig:gas_vel-sfr}. In the top two rows of the figure, we show the star formation rate distribution, as previously displayed in panel C of Fig.~\ref{fig:resonances}. In the bottom two rows, we show the gas velocity distribution for all simulations, displayed in the same manner as in Figs.~\ref{fig:accretion} and~\ref{fig:outflow}. Each panel shows a face-on slice through the galactic midplane, as observed at 4~Gyr post-merger.

In the upper panels, it can be seen that star formation in the MHD simulations takes on a complicated structure, reflecting the underlying flocculent gas distribution. Star formation at the edge of the disc, in particular, is strongly inhomogeneous, owing to the increased effects of stochasticity that arise with decreasing density. Star formation is, in general, distributed throughout the disc. However, some coherent structures are also evident, such as the spiral arms in 1349-3M and 1526-3M, and the bulge elements in 1526-3M and 1330-3M. Star formation in the hydrodynamic simulations, on the other hand, is typically distributed in bar and ring structures, as previously discussed in Sec.~\ref{chapter4-subsec:resonances}. Signs of this are clearly visible in three of the galaxies (1605-3H, 1349-3H, and 1330-3H). In 1526-3H, the central black hole is unusually active (see appendix B in \citetalias{whittingham2021}), resulting in a star formation void at the centre. Even here, however, there are weak signs of the bar-and-ring structure. Indeed, this structure is also visible in the optical counterpart of the galaxy \citepalias[see figure~7 of][]{whittingham2021}, albeit less clearly defined compared to the other hydrodynamic remnants.

In the lower panels of Fig.~\ref{fig:gas_vel-sfr}, it can be seen that the differences in star formation distribution generally translate to a different CGM velocity distribution. In particular, outflows are typically stronger in the hydrodynamic simulations, resulting in a more disturbed velocity distribution. The differences between the physics models are greatest when the merger scenario was most inspiralling, as is the case in the 1330 simulations. Such mergers lead to a greater retention of high angular momentum gas, helping the disc to grow more quickly. In MHD simulations, this spreads star formation over a larger area, reducing the effectiveness of wind particles. Indeed, inspection of the gas velocity distribution for 1330-3M shows that gas at the edge of the disc is rotating almost perfectly azimuthally. This in strong contrast with its hydrodynamic analogue. For mergers that were more ``head-on'', star formation rates are higher and disc growth is slower, as the CGM consists of lower angular momentum gas. Both of these factors increase the star formation rate density. This results in more effective winds and stronger disruption of the CGM. Even in this scenario, however, winds are, overall, more effective in hydrodynamic simulations, owing to the strong star-forming rings formed. These launch significantly faster outflows, which penetrate further into the CGM. Such outflows are better able to disrupt the CGM, reducing the amount of high angular momentum gas still further, thereby helping to keep the disc compact. 

Winds are stronger in the MHD simulations shown here compared to that shown in Fig.~\ref{fig:accretion} as the disc is still relatively young and star formation is still comparably high at this point in time. Choosing such a time was necessary in order to better highlight the star formation distribution in the upper two rows. It is noticeable, however, that just 1~Gyr later in Fig.~\ref{fig:accretion}, the CGM in 1349-3M has returned to a predominantly azimuthal velocity distribution, whilst the distribution in the hydrodynamic case is still highly disturbed. This shows how enduring the impact of the bar-and-ring structure is on the evolution of the hydrodynamic remnant and its environment.

\subsection[Appendix C: Calculating the bar pattern speed]{Calculating the bar pattern speed}
\label{appendix:pattern-speed}

To calculate the bar pattern speed, we take Fourier decompositions of stellar surface density projections centred on the bar potential minimum and analyse the evolution of the symmetric $m=2$ mode. In quantifiable terms, this means we calculate the components:

\begin{equation}
\label{eq:a_m}
    a_m(r) = \sum_{i=0}^N M_i \cos({m \theta_i}),
\end{equation}
\begin{equation}
\label{eq:b_m}
    b_m(r) = \sum_{i=0}^N M_i \sin({m \theta_i}),
\end{equation}
where $m$ is the Fourier mode, $r$ is the projected distance taken in cylindrical bins of 0.25 kpc from the centre, $N$ is the number of star particles in each bin, $M_i$ is the mass of an individual star particle, $i$, and $\theta_i$ is its azimuthal angle in the plane of the disc. We include all star particles within $\pm5$ kpc of the midplane for this calculation.

Using these components, we find the radial value at which the normalised bar strength, $A_2(r)$, reaches its peak, where this is calculated as:
\begin{equation}
    A_2(r) = \left( \frac{\sqrt{a_2(r)^2 + b_2(r)^2}}{a_0(r)} \right).
\end{equation}
Evaluating the components $ a_2(r)$ and $ b_2(r)$ at this peak radius, $r_\textrm{max}$, we can then calculate the angle at which the $m=2$ mode, and therefore the bar, lies in the plane of the disc:
\begin{equation}
   \theta = 0.5 \arctan{ \left(\frac{b_2(r_\text{max})}{a_2(r_\text{max})} \right)}.
\end{equation}
The instantaneous bar pattern speed is then simply:
\begin{equation}
   \Omega_\text{p} = \frac{\Delta \theta}{\Delta t},
\end{equation}
where $\Delta t$ is the time between angle calculations. For the time at which we perform our analysis, we have sufficient snapshot cadence such that pattern speeds may be recovered unambiguously. Nonetheless, we have also calculated the bar pattern speed using the Tremaine-Weinberg method \citep{tremaine1984} in order to cross-check our calculated values. This method returns similar values within an approximate error margin of 5 km$^{-1}$kpc$^{-1}$. This is within the expected accuracy bounds of this method \citep[see, e.g.,][]{fragkoudi2021}.
\setcounter{equation}{0}
\thispagestyle{empty}

\graphicspath{{Images/Chapter5/}}
	
\Chapter{Zooming-in on radio relics -- I.}{How density fluctuations explain the Mach number discrepancy, microgauss magnetic fields, and spectral index variations}
\addcontentslinex{lot}{chapter}{\large Chapter \thechapter: \textit{Zooming-in on radio relics -- I.}}
\label{chapter:paper-three}

\noindent This chapter is based on the published paper by Whittingham, J., Pfrommer, C., Werhahn, M., Jlassi, L., and Girichidis, P. in  Astronomy \& Astrophysics, Volume 706, id.A39, 29 pp.

\textit{It is generally accepted that radio relics are the result of synchrotron emission from shock-accelerated electrons. Current models, however, are still unable to explain several aspects of their formation. In this paper, we focus on three outstanding problems: i) Mach number estimates derived from radio data do not agree with those derived from X-ray data, ii) cooling length arguments imply a magnetic field that is at least an order of magnitude larger than the surrounding intracluster medium (ICM), and iii) spectral index variations do not agree with standard cooling models. We use a hybrid approach to solve these problems; first identifying typical shock conditions in cosmological simulations and then using these to inform idealised shock-tube simulations, which can be run with substantially higher resolution. We post-process our simulations with the cosmic ray electron spectra code \textsc{CREST} and the emission code \textsc{CRAYON+}, allowing us to generate mock observables \textit{ab initio}. We identify that upon running into an accretion shock, merger shocks generate a dense, shock-compressed sheet, which, in turn, runs into upstream density fluctuations in pressure equilibrium. This mechanism directly gives rise to solutions to the three aforementioned problems: density fluctuations lead to a distribution of Mach numbers forming at the shock-front. This flattens cosmic ray electron spectra, thereby biasing radio-derived Mach number estimates to higher values. We show that such estimates are particularly inaccurate in weaker shocks ($\mathcal{M} \lesssim 2$). Secondly, the density sheet becomes Rayleigh-Taylor unstable at the contact discontinuity, causing turbulence and additional compression downstream. This amplifies the magnetic field from ICM-like conditions up to $\upmu$G levels. We show that synchrotron-based measurements are strongly biased by the tail of the distribution here too. Finally, the same instability also breaks the common assumption that matter is advected at the post-shock velocity downstream, thus invalidating laminar-flow based cooling models.}

\section{Introduction}
\label{chapter5-sec:intro}

Radio relics have presented a problem for theorists ever since \citet{mills1960} reported an unusually strong spectral flux density of 57 $\upmu$Jy in the vicinity of Abell S753\footnote{This source is known today as 1401-33.}. Follow-up observations \citep[see, in particular,][]{hoskins1970, mcadam1977, wall1979, subrahmanyan2003} were able to progressively resolve this emission into two distinct regions; one roughly circular and centred on the cluster, and one thin and arc-like, sited at the periphery of the cluster. Today, these are recognised as a radio \textit{halo} and a radio \textit{relic}\footnote{The radio relics referred to in this paper have, for historical reasons, variously been called radio \textit{gischts} \citep{kempner2004}, \textit{giant radio relics} \citep{pinzke2013}, and cluster radio \textit{shocks} \citep{vanweeren2019}.}, respectively.

Even in the 1970s, it was recognised that radio relics could not be fit by traditional models \citep{wall1979}. Whilst their emission follows a power-law, as is characteristic of synchrotron radiation, the spectral index $\alpha$ (where $S \propto \nu^{\,\alpha}$, for flux density $S$ at frequency $\nu$) varies from the outer to the inner edge, typically starting at $\alpha \lesssim -0.5$ and steepening strongly thereafter \citep{feretti2012}. Inactive radio galaxies were initially favoured as sources \citep[see, e.g.,][]{goss1987, harris1993}, however, by the end of the 1990s, it was clear that not every relic could be spatially associated with one \citep{feretti1996}. Moreover, relics show no apparent spectral cut-off \citep{komissarov1994, rajpurohit2020b}, which, combined with the relatively short cooling times of MHz- and GHz-emitting electrons, implies an in-situ source \citep{ensslin2002, kang2012}.

It was eventually recognised that relics actually trace cosmological shocks \citep{ensslin1998}. In particular, they predominantly trace low Mach number shocks ($\mathcal{M} \lesssim 3-5$) driven by cluster mergers \citep[see, e.g.,][]{hoeft2007, markevitch2007}. The observation of such shocks in the X-ray band is now well-established, as is their spatial association with radio relics \citep[see, e.g.,][and references therein]{brunetti2014, vanweeren2019}. This link has been strengthened by observed correlations between the orientation of the radio relic and the merger axis \citep{vanweeren2011}, and scalings of the radio relic power with both the X-ray luminosity \citep{bonafede2012} and mass of the host cluster \citep{degasperin2014}.

Two main mechanisms have been proposed to link such shocks with the resultant highly-polarised, Mpc-sized emission. However, whilst adiabatic compression scenarios have not been wholly ruled out\footnote{Note, adiabatic compression plays a major role in the related category of radio \textit{phoenixes}, which are typically less linear in shape and can generate torus-like morphologies \citep[see, e.g.][]{ensslin2002, pfrommer2011, raja2023}.} \citep{ensslin2001, button2020}, it is questionable whether they can replicate the observed scaling relations \citep{colafrancesco2017} and spectral slopes \citep[see, e.g., evidence in][]{vanweeren2017}. Instead, the prevailing theory is that the merger shocks \mbox{(re-)accelerate} existing populations of electrons at the cluster outskirts via diffusive shock acceleration \citep[DSA; ][]{fermi1949, drury1983, blandford1987}. This theory is not without problems either, however, as such shocks are expected to have very low acceleration efficiencies ($\lesssim 1$\%) \citep{kang2005, kang2013, mou2023}. Acceleration from the thermal pool is consequently incompatible with the observed spectral intensities \citep{pinzke2013, vazza2014, botteon2020}. To solve this issue, a population of pre-existing, semi-relativistic ``fossil'' electrons are usually invoked \citep{markevitch2005, kang2011, pinzke2013, vazza2015, botteon2020b}, although some competing theories have emerged, such as turbulent re-acceleration \citep{fujita2015} and the multi-shock scenario \citep{inchingolo2022}.

Whilst first-order Fermi re-acceleration has emerged as the standard paradigm for explaining radio relics, our understanding of their origin remains far from complete. In particular, we highlight the following seven major problems: 

\begin{enumerate}
    \item \textit{What is the origin of the seed electrons needed for re-acceleration?}
    
    Four competing schools of thought currently exist: i) seed electrons are shock-heated during structure formation events / accretion shocks and therefore have an external origin \citep{pinzke2013, vazza2023}, ii) electrons are injected by radio galaxies at the extremities of a cluster \citep{degasperin2015, johnston2017, botteon2020b, vazza2023}, iii) seed electrons originate from active galactic nuclei (AGN) lobes emitted from the cluster core \citep{ensslin2001, shulevski2015, vazza2021, zuhone2021}, and iv) the electron population is common with that of radio halos, and hence turbulent re-acceleration is the explanation \citep{beduzzi2024}. Whilst these options are not necessarily mutually exclusive, they each suffer from issues. For example, respectively, i) has not yet been shown to work in a fully cosmological MHD simulation where the Fokker-Planck equation is solved explicitly, ii) radio galaxies are not always observationally evident, iii) it is unclear how AGN lobes survive thermal instabilities during buoyancy, and iv) it is unclear if radio halos can extend as far as some radio relics.

    \item \textit{What is the origin of relic morphology?}
    
    Radio relics exhibit a wide range of morphologies including irregular forms, such as in the case Abell 2256 \citep{vanweeren2012b}, more regular, arc-like forms, such as the ``Sausage'' relic \citep{kocevski2007, vanweeren2010}, and even relatively linear forms, such as in the case of the ``Toothbrush'' relic \citep{vanweeren2012}. Moreover, with increasing resolution, it is clear that most relics have a filamentary nature \citep{rajpurohit2020, rajpurohit2022}. These are often attributed to magnetic fields \citep{rudnick2022}, but shock geometry no doubt also plays a role. Numerical studies have already made some in-roads towards answering this question, with simulations of relic morphology \citep{wittor2019, wittor2021b, lee2024, nuza2024}, investigations into substructure and relic ``patchiness'' \citep{dominguez-fernandez2021, dominguez-fernandez2024}, and a demonstration of a potential formation process for so-called ``wrong way'' relics \citep{boess2023} all being published recently. We are, however, still far from being able to explain the morphology of individual relics.

    \item \textit{Mach numbers inferred from radio measurements are typically larger than those inferred from X-ray data. How do we explain this discrepancy?}

    Temperature and density jumps can be measured using X-ray observations, allowing Mach numbers to be inferred through the standard jump conditions. Meanwhile, the spectral slope of the radio emission can also be used to infer Mach numbers, assuming  DSA. However, despite, in principle, measuring the same shock, the two measurements rarely agree \citep[see][and references therein]{wittor2021, lee2024b}. Potential solutions include the fact that X-ray measurements are subject to projection effects \citep{wittor2021} and that radio emission may be biased to higher values as a result of a Mach number distribution \citep{hoeft2011, skillman2013, hong2015, roh2019, wittor2019, dominguez-fernandez2021, wittor2021}. A simulation showing a full \textit{ab initio} development of this effect, however, is missing from the literature.
    
    \item \textit{Observations appear to imply $\mu$G magnetic fields in radio relics. How is this possible given the surrounding intracluster medium (ICM) has strengths at least an order of magnitude lower?}

    Magnetic field estimates have been made in relics based on inverse Compton emission studies \citep{chen2008}, constraints provided by the width of the radio relic \citep{markevitch2005, hoeft2007, vanweeren2010}, and constraints using Faraday rotation \citep{bonafede2013}. All of these models suggest magnetic field strengths on the order of $\upmu$G, albeit with some scatter in their estimates. Observations \citep{brunetti2001, govoni2017} and simulations \citep[see, e.g.][]{skillman2013, dominguez-fernandez2019, nelson2024} of the surrounding ICM, meanwhile, typically imply magnetic fields that are only a fraction of this. This cannot be covered by standard shock compression alone \citep{donnert2016}, although there has been some evidence that radio observations bias values slightly high \citep{wittor2019}.

    \item \textit{Why are standard cooling models unable to match spectral variations?}
    
    Four cooling models are typically applied when trying to analyse radio relics \citep[see, e.g.,][]{vanweeren2012, rajpurohit2020}. The most simple of these assumes a one-time injection\footnote{In contrast to \citet{winner2019}, we will use ``injection'' and ``acceleration'' as equivalent phrases in this paper.} with subsequent cooling dependent \citep[``KP''][]{kardashev1962, pacholczyk1970} or independent \citep[``JP''][]{jaffe1973} of the CR electron pitch angle, respectively. Extensions include extended injection until the present time \citep[``CI''][]{pacholczyk1970} or until a fixed time in the past \citep[``KGJP''][]{komissarov1994}. Each of these mechanisms generates a characteristic spectral shape, the evolution of which can be probed using ``colour-colour'' diagrams \citep{katz-stone1993}. These are created by correlating spectral index maps, thereby producing a phase space diagram with $\alpha^{\nu_3}_{\nu_1}(x,y)$ as a function of $\alpha^{\nu_2}_{\nu_1}(x,y)$, where $\nu_1 < \nu_2 < \nu_3$, and $x$ and $y$ are spatial coordinates. Spectral features then translate to distinctive trajectories in this plane. However, despite fine-tuning, models are typically unable to replicate the tracks produced by observed radio relics.
    
    \item \textit{Recent studies imply that CR electron acceleration is only efficient above a critical Mach number of $\mathcal{M}_\mathrm{crit} \approx 2.3$. How do we reconcile this with reports of radio relics at shocks apparently weaker than this?}

    There are now several claims in the literature that suggest that efficient CR electron acceleration can only take place above $\mathcal{M}_\mathrm{crit} \approx 2.3$. This is typically because shocks below this value do not excite sufficiently strong plasma waves via CR electron driven instabilities. These waves are necessary in order to confine the CR electrons during their acceleration to radio-emitting energies\footnote{A notable exception is the study by \citet{vink2014}, who arrive at a similar value through an energetics argument.}. This result has been observed in particle-in-cell (PIC) simulations at the pre-acceleration phase of CR electrons in both quasi-parallel \citep{shalaby2022, gupta2024} and quasi-perpendicular shocks \citep{kang2019, ha2021, boula2024}. However, efficient acceleration may also fail at quasi-perpendicular shocks due to a lack of CR proton generated plasma waves \citep{ha2018, ryu2019} or due to conditions on the firehose instability \citep{guo2014}. Whilst shocks derived from radio data are generally consistent with the above studies, shocks derived from X-rays are highly inconsistent. Indeed, there are now a substantial number of relics reported with X-ray derived Mach numbers below $\mathcal{M} =  2.3$ \citep[see][and references therein]{wittor2021}. How can we explain this apparent inconsistency?

    \item \textit{If electrons are most efficiently accelerated at quasi-parallel shocks, why does polarization data appear to imply quasi-perpendicular acceleration?}

    Both observations \citep{masters2013, liu2019} and simulations \citep{crumley2019, winner2020, shalaby2021, shalaby2022} imply that DSA is most effective for electrons at quasi-parallel shocks\footnote{It should be noted that (stochastic) shock drift acceleration (SDA) is most effective in quasi-\textit{perpendicular} shocks \citep[see, e.g.,][]{amano2022}. However, whilst this process may aid the pre-acceleration phase, it is unclear how quasi-perpendicular shocks can facilitate electrons crossing the shock-front more than a few times. This would prevent SDA accelerating electrons to Lorentz factors of order $10^4$, as is necessary for radio synchrotron emission. DSA, therefore, appears to be the more likely candidate for acceleration in cluster shocks.} \citep[see theory introduced in][]{bykov1999}. This, it seems, is in contradiction with observations of radio relics; for example, observations of the ``Sausage'' relic appear to show the magnetic field vectors aligning almost perfectly with the shock surface \citep{vanweeren2010, digennaro2021}. This behaviour could be replicated in part by \citet{wittor2019}, and \citet{dominguez-fernandez2021b}, however, neither included magnetic obliquity dependent shock acceleration in their simulations. It remains to be seen whether these results can still be replicated in such a scenario.

\end{enumerate}

This paper marks the first in a series where we will attempt to tackle the above listed problems. In this study, we will tackle problems 3 -- 5. Our strategy is to first run cosmological simulations in order to identify typical shock conditions, and then use these to inform significantly higher-resolution shock-tube simulations. We implement, additionally, upstream density turbulence, as inferred from both observations \citep{simionescu2011, eckhert2015, ghirardini2018} and previous simulations \citep{nagai2011, zhuravleva2013, battaglia2015, angelinelli2021}. By employing this method, we combine the strengths of previous numerical approaches, resolving relevant physics that is, as yet, out of reach of current-generation cosmological simulations.

The paper is organised as follows: in Sec.~\ref{chapter5-sec:methodology}, we introduce both the codes and the simulation set-up used in this study. In Sec.~\ref{chapter5-sec:cosmological-analysis}, we present analysis of a cosmological galaxy cluster merger simulation. This is then used to inform a series of shock-tube simulations, the results of which we present in Sec.~\ref{chapter5-sec:shock-tube-analysis}. In this section, we show that upstream density turbulence results in: shock corrugation and the generation of downstream velocity turbulence (Sec.~\ref{chapter5-sec:RT}), the formation of a Mach number distribution at the shock front (Sec.~\ref{chapter5-subsec:mach-dist}), flatter CR electron spectra (Sec.~\ref{chapter5-sec:spectra}), breaking of the assumption that distance from the shock front is a reliable indicator of time cooled for individual electrons (Sec.~\ref{chapter5-sec:laminar-flow}), an amplification of the magnetic field to $\upmu$G levels (Sec.~\ref{chapter5-sec:magnetic-field}), and synchrotron emission that is better able to replicate observed colour-colour diagrams (Sec.~\ref{chapter5-sec:synchrotron-emission}). We finish the paper with a discussion of the caveats of our model (Sec~\ref{chapter5-sec:discussion}) and a summary of our main conclusions (Sec.~\ref{chapter5-sec:conclusions}).

\section{Methodology}
\label{chapter5-sec:methodology}

\subsection[Arepo]{\textsc{Arepo}}
\label{chapter5-subsec:arepo}

All simulations presented in this paper have been run with the moving-mesh code \textsc{Arepo} \citep{springel2010, Pakmor2016I, weinberger2020}. This code employs a set of unfixed mesh-generating points to construct a volume-filling Voronoi tessellation. Upon this, the equations for an ideal magnetohydrodynamic (MHD) fluid are solved using a second-order finite-volume Godunov scheme \citep{pakmor2011, pakmor2013} based on an HLLD Riemann solver \citep{miyoshi2005}. 

This method inherits the advantages of both Eulerian and Lagrangian codes. For example, cells are kept within a factor of two of a predetermined target mass (see following sections for values), being refined and merged as necessary. This means that spatial resolution increases naturally with structural complexity, allowing computational power to be concentrated where the system is most dynamic. Meanwhile, mesh-generating points follow the motion of the gas, leading to quasi-Lagrangian behaviour, which has the effect of substantially reducing numerical diffusion \citep[see, e.g.,][]{pfrommer2022}. The result is increased accuracy at the same resolution compared to competing codes; in particular, \textsc{Arepo} out-performs standard smoothed-particle hydrodynamics (SPH) \citep[see comparison studies in][]{vogelsberger2012, sijacki2012, keres2012, bauer2012}.

Especially germane to this study is the impact this has on the resolution of the turbulent cascade \citep{kolmogorov1941}. Several studies have shown that \textsc{Arepo} can reproduce this \citep[see, e.g.,][]{bauer2012, whittingham2021}, with strong evidence that it can also reproduce the closely-related growth of the magnetic coherence scale in a fluctuating small-scale dynamo, given sufficiently high resolution\footnote{Milky Way-like galaxy simulations require approximately $10^7$ particles before a sufficiently high Reynold's number is resolved \citep{pakmor2017,whittingham2021}.} \citep{pakmor2017, whittingham2021, whittingham2023, pfrommer2022, pakmor2024}. The same studies have shown that the employed Powell 8-wave divergence cleaner \citep{powell1999} is highly robust even in strongly dynamic flows.

\subsection{Cosmological simulations}
\label{chapter5-subsec:cosmological-setup}

To make sure that our shock-tubes probe the most relevant physics, we first analyse the formation of shocks in a series of cluster mergers. We use the PICO-Clusters (Plasmas In COsmological Clusters) suite (Berlok et al., in prep) for this purpose, which comprises of 25 individual zoom-in simulations. The simulated clusters vary in mass between $10^{14.9}-10^{15.5}\,\rmn{M}_\odot$ at $z=0$. Cosmological volumes are periodic with a side-length of 1~co-moving Gpc $h^{-1}$ and use a Planck-2018 cosmology \citep{planck2020}; i.e. Hubble's constant is $H_0 = 100 \,h~\rmn{km~s}^{-1}~\rmn{Mpc}^{-1} = 67.3$ km s$^{-1}$ Mpc$^{-1}$ with the density parameters for matter, baryons, and a cosmological constant being $\Omega_\text{m} = 0.316$, $\Omega_\text{b} = 0.049$, and $\Omega_\Lambda = 0.684$, respectively.

Each zoom-in simulation is evolved from $z=127$ and has a dark matter resolution of $5.9\times10^7\,\rmn{M}_\odot$ (target gas mass of  $1.1\times10^7\,\rmn{M}_\odot$) in the highest resolution region, which extends without contamination of lower resolution particles to a minimum of $3\times R_{200}$ throughout the simulation. This resolution is roughly comparable to that of the TNG300 \citep{nelson2019}, TNG-Cluster \citep{nelson2024}, and MilleniumTNG \citep{pakmor2023} simulations. Like these simulations we also use the IllustrisTNG galaxy formation model \citep{weinberger2017, pillepich2018}. This includes a \citet{springel2003} ISM, stellar and AGN  feedback \citep{springel2005a, weinberger2017}, radiative cooling and metal enrichment \citep{wiersma2009, vogelsberger2013}, magnetic fields \citep{pakmor2013, marinacci2018}, and the \citet{schaal2015} shock finder. 

CR protons and electrons are not included in this model. This is not a problem, however, as we focus at this stage purely on the study of shocks in low-redshift massive mergers; i.e. those most expected to create radio relics. Such shocks are only minimally affected by CR protons in clusters \citep{Pfrommer2017} and are negligibly affected by CR electrons, owing to their small energy content \citep{winner2019}.

The IllustrisTNG model has been shown to produce clusters that compare well with observations \citep{springel2018,pillepich2018b}. We note, in particular, that density and pressure profiles compare favourably \citep{pakmor2023}. This is important for this study, as it is these properties specifically that we will be probing in order to create higher-resolution shock-tube simulations.

\subsection{Shock-tube simulations}
\label{chapter5-subsec:shock-tube-setup}
\subsubsection{Resolution}
\label{chapter5-subsec:shock-tube-setup-resolution}

We use the up- and downstream gas properties extracted from the aforementioned cosmological simulations to create initial conditions for a series of shock-tubes. These shock-tubes are three-dimensional with a periodic volume of $1800 \times 300 \times 300$ kpc$^3$. The volume can be further separated into four distinct regions, each 450 kpc long:

$\bullet$ Region I: Low-resolution piston region

$\bullet$ Region II: Piston region with increasing resolution

$\bullet$ Region III: High-resolution upstream region

$\bullet$ Region IV: Upstream region with decreasing resolution

\noindent We illustrate these regions along with the initial shock conditions for a Mach 3 shock in dashed lines in Fig.~\ref{figure:1D-shock-tube-profiles}. Regions \text{I} and \text{IV} are necessary as our shock finder necessitates periodic boundary conditions. We must therefore allow for the reverse shock, as seen in Fig.~\ref{figure:1D-shock-tube-profiles} in region \text{IV}. To mitigate for this, we vary the resolution across the volume, so that cells in region \text{I} have side-length 75 kpc, whilst cells in region \text{III} have a side-length of 2.5 kpc. By ramping the resolution up and down in this fashion, we aim to focus our computational resources on region \text{III}, whilst preventing the accumulation of truncation errors. 

We set the target gas mass resolution in the high-resolution region of our shock-tubes to $m_\rmn{gas} \approx 1.5 \times 10^4 \,\rmn{M}_\odot$, with the exact value based on our initial density (see following section). We allow the cells here to refine during the course of the simulation, which provides sub-kpc resolution after shock-compression. We fix the resolution outside of region \text{III}, which leads to some minor variations from the expected density and pressure profiles to be evident in Fig.~\ref{figure:1D-shock-tube-profiles}. These variations do not impact the accuracy of the dynamics in region III.

\begin{figure}
\centering
\includegraphics[width=0.5\columnwidth]{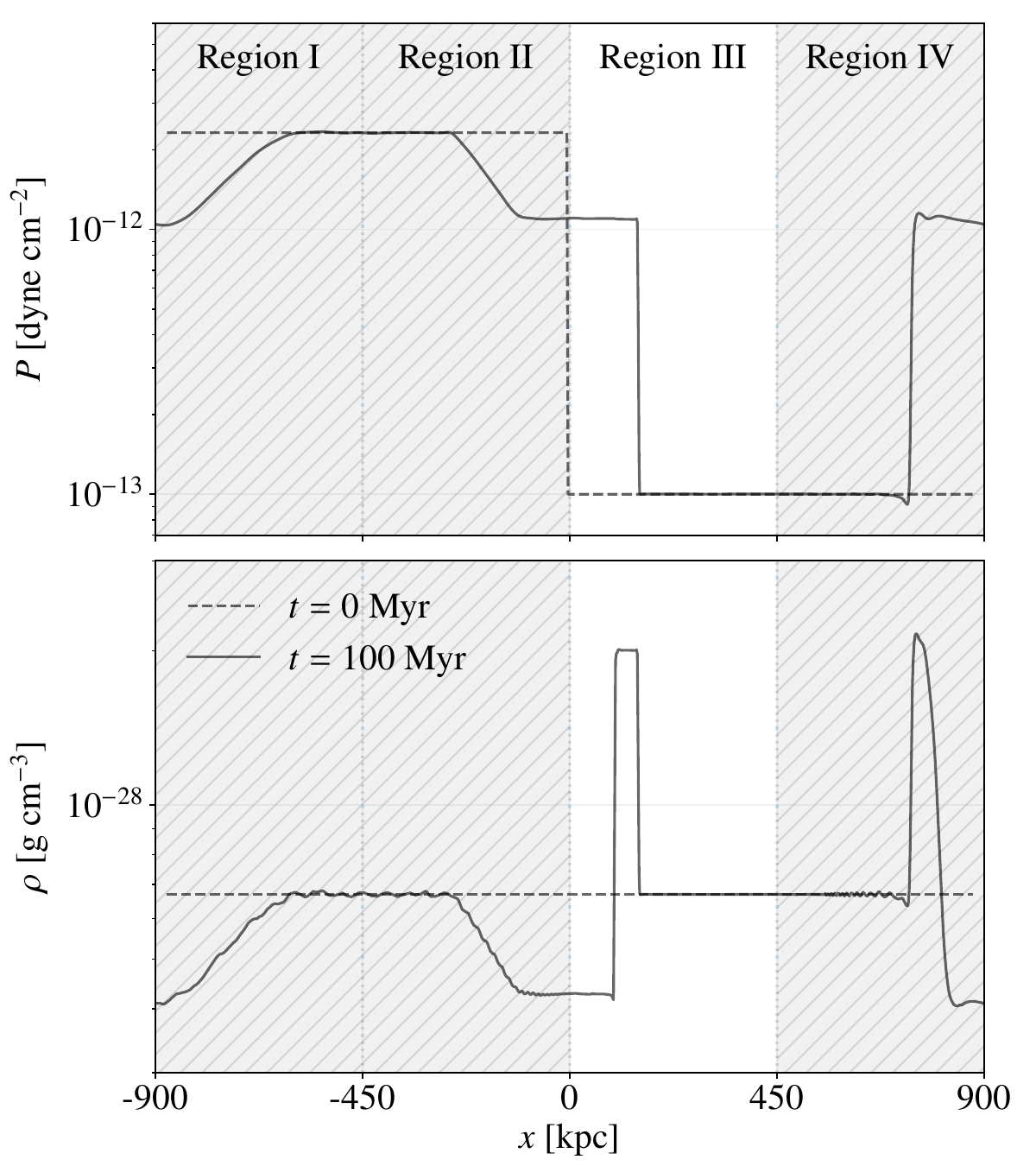}
\caption[Pressure and density profiles in the Mach 3 \textit{Flat} shock-tube simulation]{\textit{Top:} the mean pressure along the $x$-axis at $t=0$ Myr (dashed) and $t=100$ Myr (solid) in our Mach 3 \textit{Flat} simulation (see Table~\ref{tab:sims}). Dotted lines indicate the edges of various regions in the shock-tube (see Sec.~\ref{chapter5-subsec:shock-tube-setup-resolution}). We have greyed out panels which we do not analyse. \textit{Bottom:} as above, but lines show the mean density. The initial pressure discontinuity causes a shock to propagate into region III. As a result of shock compression, a dense sheet moves rightwards, bounded by the density jumps at the contact discontinuity and the shock-front. This feature is critical to the mechanism we analyse in this paper. }
\label{figure:1D-shock-tube-profiles}
\end{figure}

\subsubsection{Density and pressure values}
\label{chapter5-subsec:density-and-pressure}

As we will show in Sec.~\ref{chapter5-sec:cosmological-analysis}, at the outskirts of clusters, merger shocks are able to generate dense, shock-compressed sheets. Indeed, this will form a critical part of the model for radio relics we introduce in this paper. This scenario can be replicated by setting the mean density of all four regions to be the same and driving the shock purely by setting a jump in pressure. Indeed, this turns out to be a good approximation to the scenario observed in our cosmological simulations. Following these simulations, we set the density to $n_\mathrm{e} = 3.5 \times 10^{-5}$ cm$^{-3}$ (or, equivalently, $\rho \approx 6.7 \times 10^{-29}$ g cm$^{-3}$) and the initial upstream pressure to $P_1 = 1 \times 10^{-13}$ dyne cm$^{-2}$. Regions I and II then form the high-pressure piston, whilst regions III and IV form the low-pressure upstream region into which the shock runs. These initial conditions are shown as dashed lines in Fig.~\ref{figure:1D-shock-tube-profiles}. 

The Mach number of the shock is determined entirely in our case by the choice of pressure in the piston. Note, this differs from the classic \citet{sod1978} shock-tube set-up, in which density also plays a role. We choose to investigate the impact of $\mathcal{M}=2$ and $\mathcal{M}=3$ shocks, as these are typical values reported in radio relics \citep[see, e.g.,][]{vanweeren2017}. This fixes the initial downstream pressure to $P_2 = 1 \times 10^{-12}$ dyne cm$^{-2}$ and $P_2 = 2.32 \times 10^{-12}$ dyne cm$^{-2}$, respectively, with exact values being calculated using a Riemann solver. This set-up generates a narrow, shock-compressed region as desired (see solid lines in Fig.~\ref{figure:1D-shock-tube-profiles}). 

By initialising the shock in this way, we may test a variety of upstream conditions. Indeed, we examine a full range of variables in a companion paper (Whittingham et al., in prep.). In this current paper, however, we restrain ourselves simply to the impact of density and magnetic turbulence.

\subsubsection{Simulation variations}
\label{chapter5-subsec:sim-vars}

To unpin the impact of magnetic and density turbulence we run four separate simulations. These are listed in Table~\ref{tab:sims} and cover various permutations. Density turbulence is added only to region III, with the other regions given a constant density (see Sec.~\ref{chapter5-subsec:density-and-pressure} for values). A short buffer region without density turbulence is also provided at the start of region III to allow for the initial formation of a planar shock.

In simulations with magnetic turbulence, on the other hand, this is added to all regions in order to keep magnetic divergence to a minimum. In simulations without magnetic turbulence, the magnetic field is oriented in the $x$-direction and has a fixed strength equal to the root-mean-square (RMS) value in the turbulent runs. We do not correlate the magnetic field and density fluctuations in this initial study, leaving a study of this improvement to future work.

\renewcommand{\arraystretch}{1.2}
\begin{table}
\centering
\begin{tabular}{|l||l|l|}
\hline
\textbf{Name} & \makecell{\textbf{Turbulent density} \\ \textbf{fluctuations?}} & \makecell{\textbf{Turbulent magnetic} \\ \textbf{fluctuations?}} \\ \hline
\textit{Flat-ConstB} & No & No \\ \hline
\textit{Flat} & No & Yes \\ \hline
\textit{Turb-ConstB} & Yes & No \\ \hline
\textit{Turb} & Yes & Yes \\ \hline
\end{tabular}
\caption[Simulation variations]{Shock-tube simulations presented in this paper, ordered according to an increasing level of complexity. Our fiducial simulation is \textit{Turb}. Each simulation is run twice: once in a Mach 2 and once in a Mach 3 variation (see Sec.~\ref{chapter5-subsec:density-and-pressure}).}
\label{tab:sims}
\end{table}

\subsubsection{Generating turbulence}
\label{chapter5-subsec:generating_turb}

We generate turbulence using the method given in \citet{ruszkowski2007} and \citet{ehlert2018}. In this method, turbulence is first created in $k$-space before being converted to configuration space using a fast Fourier transform. For density turbulence, we apply \citet{kolmogorov1941} turbulence ($P(k)\propto k^{-5/3}$) below an injection scale of 150 kpc, with white noise given above this. This is similar in magnitude to the scale used in previous shock-tube studies of relics \citep[see, e.g.,][]{dominguez-fernandez2021}. We additionally implement log-normal variance, as inferred from both simulations and observations \citep[see, e.g.,][]{kawahara2008}, using a Box-Muller random variate method. Following \citet{zhuravleva2013}, we set the relative variance of the distribution to $\sigma_\rho/\mu_\rho = 0.4$, where $\sigma_\rho$ is the standard deviation and $\mu_\rho$ is the mean of the density distribution, respectively.

We expect the peak of the magnetic power spectra to lag behind that of the kinetic power spectra by a factor of a few \citep[see, e.g.][]{tevlin2024}, and so choose an injection scale here of 40 kpc. As before, we apply Kolmogorov turbulence below this scale and white noise above it. We set the RMS strength, such that in the upstream $\langle P_\mathrm{th} / P_B\rangle = 100$, where $P_\mathrm{th}$ and $P_B$ are the thermal and magnetic pressures, respectively. We find this is typical of values in the ICM and in our cosmological simulations \citep{tevlin2024}. The result is an RMS field strength of approximately 0.16~$\upmu$G. Following, \citet{ehlert2018}, we implement Gaussian variance. This means that the standard variation, $\sigma_B$, is determined by Parseval's formula:
\begin{equation}
    \sigma_B^2 = \frac{1}{N^2} \sum_\bfit{k} | \bfit{B}_{i, \bfit{k}} |^2,
\end{equation}
where $i$ covers the three spatial components, \textit{N} is the number of cells in the initial conditions, and the sum runs over all \textit{k}-values. Magnetic divergence is cleaned by projecting it out in $k$-space, i.e. by subtracting $\hat{\bfit{k}} (\hat{\bfit{k}} \bm{\cdot} \bfit{B}_\bfit{k})$.

Note, that our method of creating turbulence differs fundamentally from the method used in the shock-tubes of \citet{dominguez-fernandez2021, dominguez-fernandez2021b, dominguez-fernandez2024}, who first drive turbulence, letting it then decay over time. In the shock-tube simulations presented in this study, all cells are initially at rest; i.e. there is no initial velocity turbulence. We discuss the advantages and disadvantages of our method in Sec.~\ref{chapter5-sec:discussion}.

\subsubsection{CR physics and tracer particles}
\label{chapter5-subsec:tracers}

We use the standard \textsc{Arepo} code (see Sec~\ref{chapter5-subsec:arepo}) in our shock-tubes, augmented with the \citet{Pfrommer2017} CR proton module, in which CR protons are treated as a relativistic fluid with effective adiabatic index, $\gamma_\rmn{a}= 4/3$. We inject CR protons at shocks according to DSA theory (see Sec.~\ref{chapter5-subsec:shock-finder}) and advect them thereafter with the gas. We do not account for streaming and diffusion, as strong turbulence in the vicinity of the shock is believed to decrease the diffusion coefficient to values approaching Bohm diffusion \citep{caprioli2014}. This leaves advection the dominant transport process. CR protons are dynamically unimportant in our simulations, but are intrinsically linked to the modelling of CR electrons (see Sec.~\ref{chapter5-subsec:crest}). 

As \textsc{Arepo} is only a quasi-Lagrangian code, we also employ the use of tracer particles \citep{genel2013} for CR electron modelling. One tracer particle is placed in each upstream cell, with additional tracers placed at the same resolution at the end of region II, just behind the initial pressure discontinuity. This allows us to take advantage of the contact discontinuity that forms (see Fig.~\ref{figure:1D-shock-tube-profiles}), helping us to define the volume of cells in the injected region. In total, this results in approximately $2.9\times10^6$ tracer particles. The tracer particle functionality has been adapted to save values on-the-fly in order to be used by \textsc{Crest} \citep{winner2019}. Of particular importance is the recording of up- and downstream shock properties. For this, we use a numerical shock-finder.

\subsection{Shock finder}
\label{chapter5-subsec:shock-finder}

In both cosmological and shock-tube simulations, we use the \citet{schaal2015} shock-finder, with extensions for CRs, as presented in \citet{Pfrommer2017}. This is based on the Rankine-Hugoniot jump conditions with modifications for the case of additional CR pressure and injection. These produce the following formula:

\begin{equation}
\label{eq:mach-number}
\mathcal{M}^2 = \left(\frac{P_2}{P_1} - 1 \right) \frac{x_\rmn{s}}{\gamma_\mathrm{a, eff}(x_\rmn{s} - 1)},
\end{equation}
where $\gamma_\mathrm{a, eff}$ is the effective adiabatic index, and $x_\rmn{s}$ is the density jump at the shock.

The dissipated energy flux at the shock can be expressed as:
\begin{equation}
    \dot{E}_\rmn{diss} = \varepsilon_\rmn{diss}A_\rmn{shock}\frac{\mathcal{M}c_\mathrm{s,1}}{x_\rmn{s}},
    \label{eq:shock-dissipated-energy}
\end{equation}
where $\varepsilon_\rmn{diss}$ is the post-shock energy density minus the adiabatically compressed pre-shock energy density, $A_\rmn{shock}$ is the cell's shock surface, and $c_{\rmn{s},1}$ is the upstream sound speed \citep[see][for details]{Pfrommer2017}.

We guard against spurious shocks by further ensuring the following conditions:

\begin{enumerate}[label=(\roman*)]
    \item $\bm{\nabla} \bm{\cdot} \bm{\bupsilon} < 0$
    \item $\bm{\nabla}T \bm{\cdot} \bm{\nabla}\rho > 0$
    \item $\mathcal{M} > \mathcal{M}_\mathrm{min}$
    \item $x_\rmn{s} > x_\mathrm{s,\,min}$  
\end{enumerate}
where $\bm{\bupsilon}$, $T$, and $\rho$, are the gas velocity, temperature, and density respectively. The guards act to: i) ensure converging velocity flows, ii) filter against tangential and contact discontinuities, iii) ensure a minimum Mach number (here, $\mathcal{M}_\mathrm{min} = 1.3$), and iv) ensure a minimum density jump, respectively. This last criterion is especially important for ensuring the stability of Eq.~\eqref{eq:mach-number} at weak shocks, and is calculated using the hydrodynamic jump condition with $\mathcal{M}_\mathrm{min}$.

Shocks are numerically broadened in our simulations; typically with a thickness of 2 -- 3 cells. The cell with the minimum velocity divergence (i.e. maximum compression) is labelled the \textit{shock surface} cell, whilst the cell directly outside the shock zone is labelled the \textit{post-shock} cell. We run the shock-finder in the ``before-snapshot'' mode in the cosmological simulation and in the ``on-the-fly'' mode in our shock-tubes. 

\subsection[Crest]{\textsc{Crest}}
\label{chapter5-subsec:crest}

CR electron modelling in this study is performed using the post-processing code \textsc{Crest} \citep{winner2019}. This code uses the previously discussed tracer functionality in \textsc{Arepo} to store relevant data on the MHD timestep, allowing us to evolve the Fokker-Planck equation in the Lagrangian frame. In particular, we account for adiabatic changes; cooling via  Coulomb, bremsstrahlung, inverse Compton, and synchrotron losses; and DSA with magnetic obliquity \citep[see formulae in][]{pais2018, winner2020}. In principle, \textsc{Crest} is also capable of Fermi I re-acceleration, Fermi II momentum diffusion, and Bell amplification, but we do not use these features in the following work.

On each tracer particle, \textsc{Crest} samples the underlying CR electron spectral density field. In practice, this means that all points closest to a given tracer are represented by that tracer's spectrum. To compute a thermodynamically extensive quantity such as the CR electron energy, we therefore construct the volume associated with a given tracer by using a Voronoi tessellation of space with the tracer position as the mesh-generating point. This gives the tracer an evolving volume, $V_\rmn{cell}$, where $E_\rmn{e} = \varepsilon_\rmn{e} V_\rmn{cell}$ is the CR electron energy in that volume, and $\varepsilon_\rmn{e}$ is the CR electron energy density obtained by taking the second moment of the CR electron spectrum.

Tracers are initially assigned a purely thermal spectrum. This is acceptable, as whilst a pre-existing non-relativistic population is likely required to reach radio relic luminosities (see Sec.~\ref{chapter5-sec:intro}), the additional of this feature should only affect the normalisation of the spectra, not its shape at radio-emitting frequencies \citep[see, e.g.,][]{pinzke2013, winner2019}. modelling the impact of such a population is outside the scope of this work, and hence left to a future study.

For our cooling modules, we assume that the gas is primordial and hence that the mass fraction of hydrogen is $X_\rmn{H} = 0.76$. We also assume that the gas is fully ionised, resulting in a mean molecular weight of $\mu = 0.588$ and an ionisation fraction of $x_\rmn{e} = 1.157$. To aid comparison with observations we assume a redshift of $z=0.2$, which is the approximate redshift of the Toothbrush and Sausage radio relics \citep{vanweeren2010, vanweeren2012}. The energy density of the cosmic microwave background (CMB) is therefore set to $\epsilon_\rmn{CMB} = 8.65 \times 10^{-13}$ erg cm$^{-3}$, or equivalently a magnetic field strength of $B_\rmn{CMB} \approx 4.7$ $\upmu$G.

Throughout this work, we use dimensionless momentum, $p = \tilde{p} / (m_\rmn{e} c)$, where $\tilde{p}$ is the dimensional momentum, $m_\rmn{e}$ is the electron rest mass, and $c$ is the speed of light. We also convert the transport equation \citep[see][]{winner2019} into one-dimensional momentum space, such that $f^\rmn{1D} = 4 \pi p^2 f^\rmn{3D}$, where $f^\rmn{1D}$ and $f^\rmn{3D}$ are the one- and three-dimensional distribution functions, respectively. We let $p$ range in all simulations from $10^{-1}$ to $10^{8}$ and use 10 logarithmically-spaced bins per decade. We find this is sufficient to produce converged spectra. In the remainder of this section, we summarise the DSA implementation, having the greatest impact on our study. We direct the reader, however, to \citet{winner2019} for a more thorough discussion of all the physics represented in the code. 

CR electrons are accelerated in \textsc{Crest} in shock-surface and post-shock cells by attaching to the one-dimensional distribution function a power-law slope\footnote{Note, we define $\alpha < 0$ here, in contrast to the notation used in \citet{winner2019}.} with:
\begin{equation}
\alpha = -\frac{x_\rmn{s} + 2}{x_\rmn{s}-1}
\label{spectral-slope}
\end{equation}
This slope is limited to a maximum of -2.2, following \citet{caprioli2020}. This has a very minor impact on our results, however, given the weak shocks we simulate and the correspondingly steep spectral slopes. To avoid spurious heating before the shock due to numerical broadening, we stop updating the recorded density on encountering a shocked cell. When the tracer reaches a shock-surface or post-shock cell, the density is then updated to its post-shock value. This produces a discrete jump at the shock. 

Following \citet{webb1984} and \citet{pinzke2010}, we set the maximum momentum up to which we accelerate to:
\begin{equation}
    p_\mathrm{max} = \frac{v_\mathrm{post}}{c}\sqrt{\frac{6 \pi e}{\sigma_\mathrm{T}} \frac{B}{(B^2 + B_\mathrm{CMB}^2)} \frac{x_\rmn{s} (x_\rmn{s}-1)}{(x_\rmn{s}+1)}},
    \label{eq:p_max}
\end{equation}
where $v_\mathrm{post}$ is the post-shock velocity in the shock rest-frame, $e$ is the elementary charge, $\sigma_\mathrm{T}$ is the Thomson cross section, and $B$ is the strength of the magnetic field.

We then require that $p_\mathrm{min}$, the minimum acceleration momentum, fulfils:
\begin{equation}
\int\limits_{p_\mathrm{min}}^{p_\mathrm{max}} f_\mathrm{th}(p_\mathrm{min})p_\mathrm{min}^{-\alpha} p^{\alpha} E_\mathrm{e,\,kin}(p)\mathrm{d}p = \Delta\varepsilon_\mathrm{e},
\label{eq:p_min}
\end{equation}
where $f_\rmn{th}(p)$ is the thermal Maxwellian\footnote{See eq.~31 of \citet{winner2019}}, $E_\mathrm{e,\,kin}(p) = (\sqrt{1+p^2} - 1) m_\mathrm{e} c^2 $ is the electron kinetic energy, and $\Delta\varepsilon_\mathrm{e}$ is the increase in energy density of the CR electron population due to acceleration. We use 0.1\% of the liberated thermal energy at the shock, converting this to an energy density by dividing by the cell volume \citep[see][]{winner2019}.

The normalisation of the injected spectrum is subsequently:
\begin{equation}
C = f_\mathrm{th}(p_\mathrm{min}) p_\mathrm{min}^{-\alpha}.
\end{equation}
At the high momentum end of the spectrum, we impose a super-exponential cut-off \citep[see][]{ensslin1998, zirakashvili2007, pinzke2010}, which modifies the spectrum thusly:
\begin{equation}
\tilde{f}(p) = f(p)\left[1 + 0.66\left(\frac{p}{p_\mathrm{max}}\right)^{2.5} \right]^{1.8} \exp{\left[ -\left(\frac{p}{p_\mathrm{max}}\right)^2 \right]}
\end{equation}

As explained in Sec.~\ref{chapter5-sec:intro}, there are now several studies that suggest that CR electron acceleration can only take place above a critical Mach number of $\mathcal{M}_\mathrm{crit} = 2.3$. Subsequently, we do not accelerate electrons below this Mach number in the simulations that include upstream density turbulence\footnote{We do not apply this condition to simulations without upstream density turbulence, as the Mach number distribution in this case is insufficiently broad.}. We show in Whittingham et al. (in prep.) that this has remarkably little impact on the subsequent emission, even when the Mach number distribution peaks at $\mathcal{M} = 2$. This is due to the fact that emission is dominated by the tail of the Mach number distribution. 

\subsection[Crayon+]{\textsc{CRAYON+}}
\label{chapter5-subsec:crayon}

In the final step of our methodology, we post-process the resultant \textsc{Crest} output with the \textsc{Crayon+} code \citep{werhahn2021}, which is able to convert spectra into instantaneous emission for a range of CR electron processes. Naturally, we are most interested in the radio synchrotron emission. We summarise the salient formulae here, but encourage the interested reader to see section~2.4 and associated appendices of \citet{werhahn2021b} for a more comprehensive overview.

Following \citet{rybicki1986}, the synchrotron emissivity $j_\nu$ of a tracer at frequency $\nu$ is:
\begin{equation}
j_\nu = \frac{\sqrt{3}e^3 B_\perp}{m_\rmn{e} c^2} \int\limits_{0}^{\infty} f(p) F(\nu / \nu_c) \mathrm{d}p \;\,\propto\; B_\perp^{1-\alpha_\nu}\nu^{\alpha_\nu},
\label{eq:synchrotron-emissivity}
\end{equation}
where $B_\perp$ is the projection of the magnetic field onto the plane perpendicular to the line of sight, $\alpha_\nu$ is the radio spectral index, and $F(\nu/\nu_\rmn{c})$ is the dimensionless synchrotron kernel, with $\nu_\rmn{c}$ the critical frequency. These last two variables are, in turn, defined as:
\begin{equation}
\nu_\rmn{c} = \frac{3 e B_\perp \gamma^2}{4 \pi m_\rmn{e} c},
\label{eq:critical-frequency}
\end{equation}
where $\gamma = \sqrt{1 + p^2}$ is the Lorentz factor and:
\begin{equation}
F(x) = x \int\limits_{x}^{\infty} K_{5/3} (\xi)\mathrm{d}\xi, 
\end{equation}
where $K_{5/3}$ is the modified Bessel function of order 5/3, with $x = \nu/\nu_c$.  To get the right-hand side of Eq.~\eqref{eq:synchrotron-emissivity}, we have assumed a power-law spectrum over an appropriate range of frequencies. 

The specific radio synchrotron intensity, $I_\nu$, at frequency $\nu$ is then obtained by integrating $j_\nu$ (with units of erg s$^{-1}$ Hz$^{-1}$ cm$^{-3}$) along the line of sight, $s$:
\begin{equation}
I_\nu = \frac{1}{4\pi}\int\limits_0^{\infty} j_\nu ds
\label{eq:intensity}
\end{equation}
This gives $I_\nu$ in units\footnote{This may be converted to the more typical observational units of Jy~arcsec$^{-2}$ with the transformation 1 Jy arcsec$^{-2}$ = $2.35 \times 10^{12}$ erg cm$^{-2}$ s$^{-1}$ Hz$^{-1}$ sterrad$^{-1}$.} of erg cm$^{-2}$ s$^{-1}$ Hz$^{-1}$ sterad$^{-1}$. In the remainder of the paper, where projections are shown we have set the upper limit of the integral in Eq.~\eqref{eq:intensity} to be the depth of the box, equal to 300 kpc (or approximately $1 \times 10^{24}$ cm).

\section{Cosmological analysis}
\label{chapter5-sec:cosmological-analysis}

\begin{figure*}
    \centering
    \includegraphics[width=1.0\columnwidth]{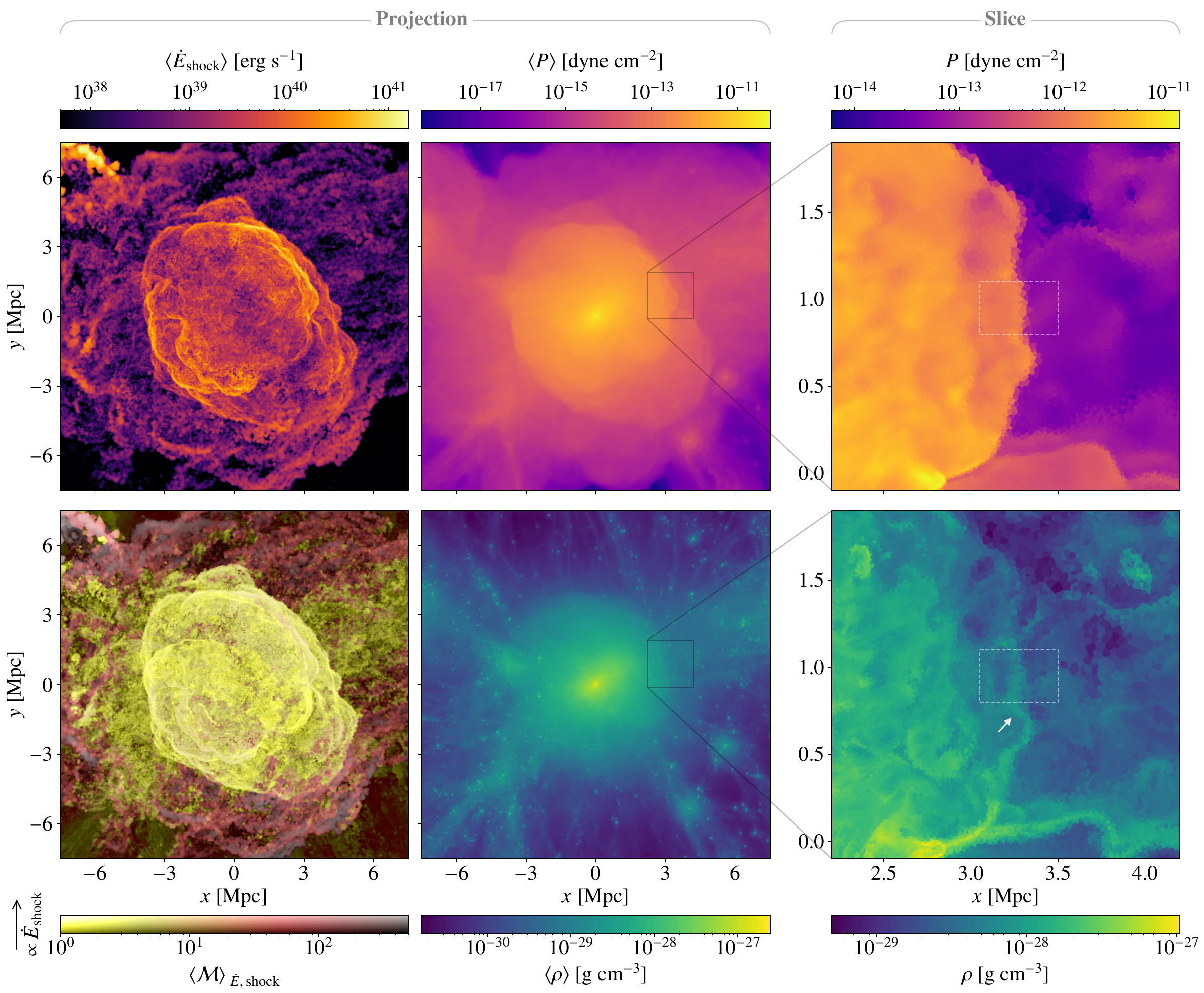}
    \caption[Slices through a cosmological simulation showing the generation of a shock-compressed density shell]{Cosmological simulation of a galaxy cluster undergoing a major merger at $z=0.14$. \textit{Four left-most panels, clockwise from top left:} projected shock-dissipated energy rate, gas pressure, gas density, and dissipation-weighted Mach number, respectively. Projections have a depth of $\pm 7.5$ Mpc from the cluster centre. \textit{Right-most panels:} enlarged cut-outs showing slices through the midplane of the projection. The white, dashed box indicates the size of the upstream region in our idealised shock-tube simulations. The merger drives a shock-wave out to the cluster outskirts. The collision of this shock-wave with an accretion shock leads to the production of a thin shell of compressed gas (see white arrow). A movie of this process can be found \href{https://youtu.be/Ka-Odrwwamo}{here}.}
    \label{figure:cosmological}
\end{figure*}

As explained in Sec.~\ref{chapter5-sec:intro} and~\ref{chapter5-sec:methodology}, the first step in our strategy is to analyse shock conditions in cosmological simulations. To this end, we present here a case study. We use a variation\footnote{The simulation presented here has been run with Arepo-1, whilst the PICO-Cluster suite is run with Arepo-2.} of \textit{Halo 0003} from the PICO-Cluster suite (Berlok et al., in prep) as it features a particularly energetic merger, making it especially likely that the generated shock wave will form a radio relic \citep[see similar shocks in, e.g.,][]{boess2023, lee2024}. However, our analysis generalises to all the cluster mergers we investigated.

The major merger in \textit{Halo 0003} takes place between $z\approx0.3$ and $z\approx0.1$. At the start, the progenitors have masses of $10^{14.8}\,\rmn{M}_\odot$ and $10^{15.2}\,\rmn{M}_\odot$, giving a mass ratio of roughly 1:2.5. We show gas conditions during the merger in Fig.~\ref{figure:cosmological}, focusing on a period when the energy dissipation in the shock is at its most intense. At this time, the shock has become detached from the merging cluster that drove it, known as the ``run-away'' phase \citep[see definitions in][]{zhang2019}. It has reached a distance of a little under $1.8 \times R_{500}$, which is within the range expected for radio relics \citep[see, e.g.][]{bagchi2011, erler2015}.

In the top left panel of Fig.~\ref{figure:cosmological}, we show the projected shock-dissipated energy, where this has a depth of 15 Mpc. The merger has led to the formation of two Mpc-size, arc-like shocks. These align with the merger axis and are relatively symmetric, as expected when the initial cluster masses are similar \citep{vanweeren2011b, lee2024b}. The shocks from the major merger are superimposed on a background of less energetic shocks. Some of these, such as those towards the lower-right are from ongoing minor mergers, whilst the majority result from accretion.

The bottom-left panel has a two-dimensional colorbar, where the colour indicates the dissipated energy-weighted Mach number, with saturation being set proportional to this energy. It can be seen that the merger shock is made up predominantly of low Mach-numbers ($\mathcal{M}\lesssim5$). Shocks at the edge of the cluster, meanwhile, are much higher ($\mathcal{M}\gtrsim100$). These shocks define the filamentary structure, within which the cluster sits. Indeed, the merging cluster has arrived from the filament evident at the top right of the panel (see \href{https://youtu.be/Ka-Odrwwamo}{movie}).

In the middle column, we show the projected pressure (top) and density (bottom) respectively. The brightest central galaxy (BCG) is located at the peak value in both panels. In-falling galaxies can also be seen as bright spots at the edge of the cluster. The merger has led to a region of increased pressure centred on the cluster. This region is well-delineated, in contrast to the more diffuse density distribution. It is also clear that the region of high pressure is bounded by the merger shocks.

To remove the effects of projection, in the right-hand column we show slices of the same quantities. We zoom in on a region of size $2\times2$~Mpc$^2$ in order to better explore the shock details. At this time and position, the outwards-moving merger shock encounters an inwards-moving accretion shock\footnote{
We leave a full investigation of the conditions under which this takes place, as well as an estimate of the frequency of such mergers, to future work. In this current study, we show only that it is a viable solution to the aforementioned radio relic problems.}. These are evident as ridges in the density and pressure slices in the corresponding \href{https://youtu.be/Ka-Odrwwamo}{movie}. The time at which the two shocks meet can be considered a Riemann problem with density, pressure and velocity specified at both sides of the discontinuity. The result of this is two-fold: firstly, an inwards travelling rarefaction wave is generated, and secondly, a thin, shock-compressed density shell forms behind the merger shock. This feature is marked by the white arrow in the bottom right panel, and is consistent with previous studies \citep{zhang2019, zhang2020}. 

It can be seen that the pressure downstream of this feature is approximately constant, whilst the density jumps. This indicates that the edge closest to the arrow is actually a contact discontinuity. The jump itself is exacerbated by the rarefaction wave, which lowers the density downstream of this discontinuity. Such waves have also been observed in previous studies of cluster mergers \citep{shi2020}. This feature will become key to several results shown in the remainder of the paper.

The subsequent evolution of the shock wave can be well-approximated as a pressure-driven shock wave on a flat density background (see Sec.~\ref{chapter5-subsec:shock-tube-setup}). This lends itself to shock-tube simulations, where we can approach the problem with significantly higher resolution. This is necessary, as whilst the overall pressure and density profiles of our simulated clusters are converged \citep[Berlok et al., in prep;][]{tevlin2024}, density turbulence is not sufficiently resolved at these radii. In both of the right-most panels, we have added a white, dashed box with dimensions 450 kpc $\times$ 300 kpc, which represents the size of the high-resolution region in our shock-tube initial conditions. This region is traversed by only $\mathcal{O}(10)$ cells, which is clearly insufficient. The shock-tube method will also allow us to precisely define the various turbulent parameters, allowing for a clearer investigation into their impact on downstream properties.

\section{Shock-tube analysis}
\label{chapter5-sec:shock-tube-analysis}

\begin{figure*}
    \centering
    \includegraphics[width=1.0\columnwidth]{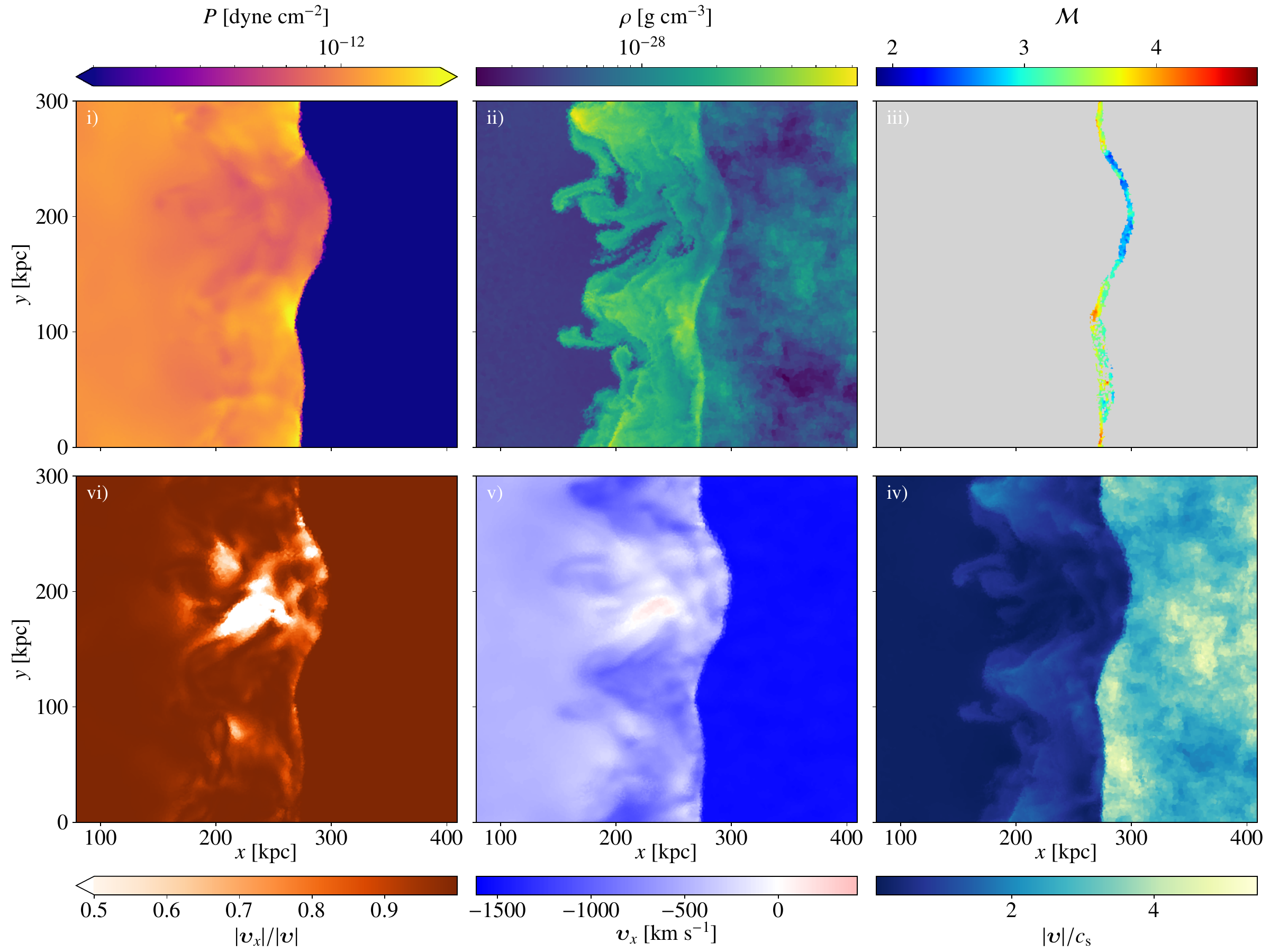}
    \caption[Slices through the fiducial shock-tube set-up showing the generation of turbulence and a Mach number distribution]{A Mach 3 shock is driven into a turbulent upstream density field. The initial pressure and density values for the shock-tube correspond to those shown in Fig.~\ref{figure:cosmological}. \textit{Clockwise from top left:} i) gas pressure, ii) gas density, iii) dissipated-energy-weighted Mach number, iv) gas speed divided by sound speed v) the $x$-component of gas velocity, vi) the fraction of the gas speed in the $x$-component. Data is shown in slices except for iii), which is a thin projection of 35 kpc. All values are shown in the shock rest frame at $t=180$ Myr. Upstream density turbulence directly leads to shock corrugation, a distribution of Mach numbers at the shock front, and downstream velocity turbulence. An animated version of this figure can be found \href{https://youtu.be/gX-urGrKNy8}{here}.}
    \label{figure:shock-tube}
\end{figure*}

Our shock-tube set-up, including initial pressure and density values, is informed by the above cosmological analysis, with values given in Sec.~\ref{chapter5-subsec:shock-tube-setup}. We note that the pressure jump dominates the white, dashed box shown in Fig.~\ref{figure:cosmological}, and so choose to make the shock exclusively pressure-driven, i.e. the downstream density is set equal to the mean upstream density. The mass resolution in the high-resolution upstream region of these simulations (i.e. region III) is approximately 730 times greater than in the cosmological simulation, allowing for the resolution of density turbulence.  

As previously mentioned, multiple studies have shown that the ICM is clumpy. We implement this in our shock-tube simulations using the method described in Sec.~\ref{chapter5-subsec:generating_turb}. The injection scale of the density turbulence is set to half the smallest box dimension, i.e. 150 kpc, and the amplitude of the fluctuations is taken from \citet{zhuravleva2013}. We remind the reader that there is no initial velocity turbulence; that is, the density fluctuations are initially at rest.

We show the state of our fiducial Mach 3 simulation at $t=180$~Myr in Fig.~\ref{figure:shock-tube}, where the shock-front is travelling from left to right. Each panel shows a 2D slice (except for panel \textit{iii)}, which is a thin projection) and $x=0$ marks the beginning of ``region III'' (see Sec.~\ref{chapter5-subsec:shock-tube-setup}). The time is chosen such that several characteristics of the shock are on display simultaneously. In particular, we draw attention to the following three consequences of including upstream density turbulence: i) the formation of downstream turbulence, ii) shock corrugation, iii) the formation of a Mach number distribution.

\subsection{Downstream turbulence and shock corrugation}
\label{chapter5-sec:RT}

Perhaps the most striking of these three is the generation of downstream turbulence\footnote{We show that our resolution of this is numerically stable in Fig.~\ref{figure:low-res-density-slice}.}. The impact of this can be seen in almost all panels, but we start with panel \textit{vi)}. Here we show the fraction of the shock-frame gas velocity in the $x$-component. To calculate the shock-frame, we take the median $x$-coordinate for all shock surface cells in each snapshot. This is then converted to an average shock velocity by dividing by the snapshot cadence, with the subsequent value being subtracted from the gas velocities. In a purely laminar flow aligned with the shock normal, there would be no deviations from $|\bupsilon_x|/|\mathbf{\bupsilon}| = 1$. This is the case in the simulations without density turbulence.

This panel is supported by panel \textit{v)}, which shows the $x$-component of the velocity. Blue shades indicate gas moving from right to left, whilst red shades indicate gas moving the opposite way. These values are, once again, both given in the frame of the shock. The reduction in speed after crossing the shock is a standard theoretical result, however, panel \textit{v)} shows that including density turbulence causes gas to move away from the shock-front at varying speeds. Indeed, in some cases, gas even moves towards the shock-front from the left. Whilst we will directly model CR electrons later in the paper, this already shows us that distance from the shock front is not a reliable indicator of time since injection.

The turbulence results predominantly from a Rayleigh-Taylor instability, which leads to the formation of density ``fingers''. These are visible, for example, in panel \textit{ii)} at roughly $y=120$ kpc and $y=290$ kpc. This spacing is no coincidence, as such fingers form at positions where the shock reaches higher density clumps, which in turn is set by the injection scale. In the non-linear regime of this instability, the resulting velocity shear causes the Kelvin-Helmholtz instability to form as a parasitic instability. The effect of this can be seen particularly clearly in the upper half of panel \textit{ii)}.

We explain the emergence of the Rayleigh-Taylor instability as follows: density fluctuations mean that the shock speed varies along its surface, with the shock accelerating into lower density regions. An initially planar shock will subsequently warp. This leads to the pressure dropping behind the most advanced part of the shock, as can be seen in panel \textit{i)}. The consequence of this is that the pressure gradient starts to point towards the far downstream. 

The density jump from the downstream behind the contact discontinuity towards the shock-compressed region, however, provides a gradient pointing the opposite way. What is more, as higher density fluctuations are less easily accelerated by the shock, the back of the compressed region also begins to warp. Indeed, as seen in Fig.~\ref{figure:shock-tube}, this happens to a greater extent than at the shock-front.

\begin{figure}
    \centering
    \includegraphics[width=0.4\columnwidth]{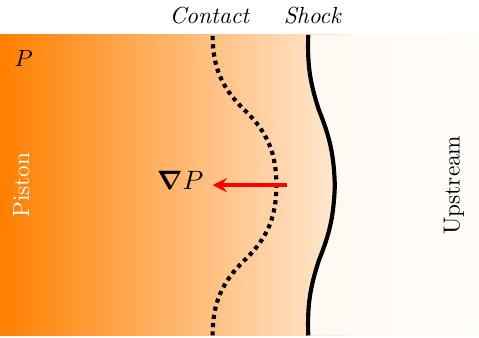}
    \\
    \includegraphics[width=0.4\columnwidth]{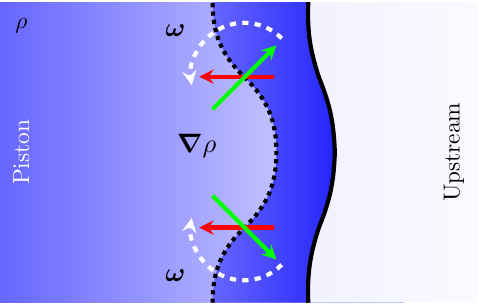}
    \caption[Schematic showing how upstream density turbulence causes velocity turbulence to be generated downstream via the Rayleigh-Taylor effect]{A schematic showing how upstream density turbulence causes velocity turbulence to be generated downstream. The corrugation of the shock front naturally leads to a misalignment of pressure and density gradients (shown here by red and green arrows, respectively). This results in a baroclinic term, which in turn induces vorticity, causing the contact discontinuity to become Rayleigh-Taylor unstable.}
    \label{figure:RTI-schematic}
\end{figure}

We show a schematic that illustrates this scenario in Fig.~\ref{figure:RTI-schematic}, with orange colours representing pressure and blue colours representing density, respectively. Pressure and density gradients have become misaligned, which, due to the inviscid vorticity equation, results in a baroclinic torque, as:
\begin{equation}
    \frac{\rmn{D}\bm{\omega}}{\rmn{D}t} = \frac{1}{\rho^2}\bm{{\nabla}}{\rho}\bm{\times}\bm{{\nabla}}{P},
\label{eq:inviscid-vorticity}
\end{equation}
where the left-hand term is the Lagrangian derivative of vorticity $\bm{\omega}$. This mechanism rapidly generates downstream velocity turbulence, and leads to a reinforcement of the finger-like structures, with eddies directed as shown by the white, dashed arrows in Fig.~\ref{figure:RTI-schematic}.

Upstream density fluctuations are critical for the formation of this scenario. Without them, there is only a very limited pressure gradient and no corrugation at the back of the shock-compressed region. The later condition in particular would mean that $\bm{{\nabla}}{\rho}$ and $\bm{{\nabla}}{P}$ were approximately anti-parallel, leading to a vanishing cross-product in Eq.~\eqref{eq:inviscid-vorticity} and hence little to no growth of the instability. We show this to be true in our simulations in App.~\ref{appendix:RTI-without-density-perturbations}.

We note that \citet{hu2022} identified the related Richtmyer-Meshkov instability as causing turbulence in their shock-tube simulations. We may discount this here as the Richtmyer-Meshkov instability takes place at the shock-front, whilst the instability we observe takes place at the contact discontinuity. Vorticity may also be generated by curved shock surfaces according to the theorem by \citet{crocco1937}. We discount this as the primary driver in this case, however, as this mechanism would produce vorticity oriented in the opposite direction. This would reduce the size of the fingers rather than increase them.

\subsection{Mach number distribution}
\label{chapter5-subsec:mach-dist}
\subsubsection{Form}
\label{chapter5-subsec:mach-form}

We now turn our attention to the third consequence of including density fluctuations: the formation of a Mach number distribution. It has long been known that shocks in clusters vary in strength \citep{ryu2003, kang2005, pfrommer2006, hoefft2008, skillman2008, vazza2009}. It is also inferred from synchrotron fluctuations that this is true within radio relics as well \citep{rajpurohit2020, rajpurohit2021}. This was supported by the shock-tube simulations of \citet{dominguez-fernandez2021}. Like them, we find that an initially well-defined Mach number decomposes into a distribution upon encountering turbulence. This can be seen in the top right panel of Fig.~\ref{figure:shock-tube}, which shows a thin projection\footnote{Numerical broadening occasionally leads to cells not being recognised as part of the shock front. Such cells are captured at the next hydro-timestep, however, meaning that no gas cells downstream remain unshocked as a result of the numerical implementation. This is shown explicitly in the appendix of Whittingham et al. (in prep.). Projecting therefore gives a more faithful indication of the coherence of the shock front.} of depth 35 kpc, where colours indicate the Mach number at the shock front.

The shock in Fig.~\ref{figure:shock-tube} is initially set up with $\mathcal{M}=3$. However, whilst many cells continue to show similar values, it is clear that there is now a distribution, with values lying approximately between $\mathcal{M}=1.9$ and $\mathcal{M}=4.8$. We quantify the development of this distribution in Fig.~\ref{figure:mach-no-pdf}. For this, we have calculated the distribution at each snapshot, where each cell has been weighted by its contribution to the overall shock surface. We show the median of these distributions in Fig.~\ref{figure:mach-no-pdf} as solid lines and provide the 25\%$-$75\% percentile range as shaded values.

\begin{figure}
    \centering
    \includegraphics[width=0.5\columnwidth]{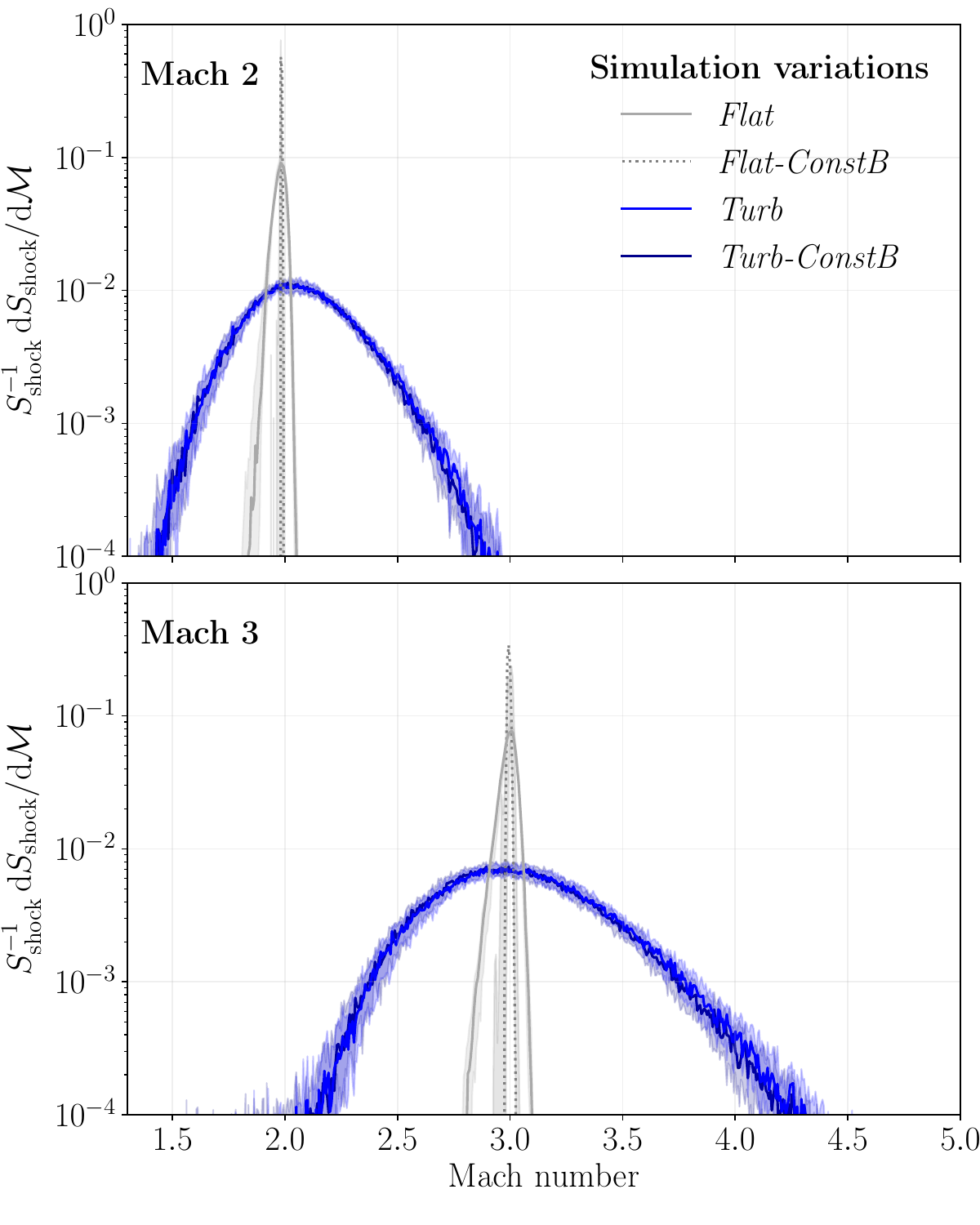}
    \caption[Shock-surface weighted Mach number distributions for each simulation]{Mach number distributions for all models, where each cell has been  weighted by its normalised contribution to the shock surface. Lines indicate the median taken over all snapshots, whilst the shaded values indicate the interquartile range. The addition of magnetic turbulence (\textit{Flat}) to a homogeneous density distribution (\textit{Flat-ConstB}) broadens the distribution only very mildly. By contrast, the addition of upstream density fluctuations in the simulation (\textit{Turb} and \textit{Turb-ConstB}) turns an extremely narrow distribution into a much broader one. The width of the distribution is proportional to the peak Mach number.}
    \label{figure:mach-no-pdf}
\end{figure}

The dotted grey line indicates the values for simulation \textit{Flat-ConstB}, which has neither density nor magnetic turbulence in the initial conditions. It can be seen that the distribution forms an approximate delta function. This indicates the base-line accuracy of our shock-finder. When we add magnetic fluctuations, as in \textit{Flat}, the distribution broadens slightly. This is due to the additional pressure fluctuations caused by the magnetic field (see Sec.~\ref{appendix:beta}). This effect is, however, totally dominated by the impact of including density fluctuations. Indeed, there is essentially no difference between the Mach number distributions in \textit{Turb-ConstB} and \textit{Turb}. The initial conditions for these simulations differ only by their inclusion of a uniform or turbulent magnetic field, respectively.

Simulations with turbulent density fields produce Mach numbers that are best fit by a skew normal distribution. This is in contrast to \citet{lee2024b}, who find that the Mach numbers in their cluster simulations fit a log-normal distribution. Naively, one might expect to see a log-normal distribution in our simulations as well, as we apply a log-normal distribution to the density fluctuations in the initial conditions, which enters Eq.~\eqref{eq:mach-number} through $x_\rmn{s}$, the density jump. The physics are, however, more complicated than this, as we will show in Sec.~\ref{chapter5-sec:mach-no-dist-origin}. 

The peak of the distribution meanwhile remains at the initial Mach number\footnote{We examine this relationship in more detail in Whittingham et al. (in prep.), however, and find this depends to an extent on the turbulent conditions.} ($\mathcal{M}=2$ in the top panel and $\mathcal{M}=3$ in the bottom). Finally, we note that the distribution remains remarkably stable over time; the difference between the 25th and 75th percentile is very low. This is in contrast to \citet{dominguez-fernandez2021} who, whilst also finding a tail towards higher Mach numbers, do not see such stability (see their fig. B2). We attribute this to our different methods of generating turbulence (see Sec.~\ref{chapter5-sec:discussion}).

\subsubsection{Origin}
\label{chapter5-sec:mach-no-dist-origin}

Returning to  Fig.~\ref{figure:shock-tube}, it can be seen that the higher Mach numbers are typically found towards the back of the shock front, whilst lower Mach numbers are found at its most advanced position. An intuitive understanding of this can be gained by inspecting panel \textit{iv)}. Here we show a slice, where colours indicate the shock-frame gas speed as a fraction of the sound speed. We expect this quantity to provide a fair approximation to the more rigorous result given by Eq.~\eqref{eq:mach-number}. We remind the reader that we have defined the shock-frame based on the median of the local shock velocities. This is reasonable as the local velocity fluctuations along the shock front are small compared to its median shock speed.

Gas is stationary in the lab-frame and so, under our approximation, approaches the shock at the same speed (see the right-hand side of panel \textit{v)}). Additionally, the gas is in pressure equilibrium (see panel \textit{i)}). Consequently, $|\mathbf{\bupsilon}|/c_\rmn{s}$ in the upstream is determined purely by the density fluctuations, as $c_\rmn{s}= \sqrt{\gamma_\rmn{a} P/\rho}$, where the adiabatic index is set to $\gamma_\rmn{a} = 5/3$. This means that denser (more rarefied) gas has a lower (higher) sound speed, respectively. The consequence of this is that we expect denser gas to produce higher Mach numbers, and more rarefied gas to produce lower Mach numbers. This is indeed the case, as can be seen by comparing panels \textit{ii)} and \textit{iv)}. As discussed in Sect.~\ref{chapter5-sec:RT}, lower density gas also causes the shock front to accelerate locally. As a result, lower Mach numbers are found further forward. We show that this is generically true across the shock front in Fig.~\ref{figure:mach-number-projection}.

\subsection{Radio vs. X-ray Mach number discrepancy}
\label{chapter5-sec:spectra}

It has been argued previously that the highest Mach numbers in a distribution are responsible for the radio emission \citep[see, e.g.,][]{hoeft2011, wittor2021, dominguez-fernandez2021}. This is currently the leading theory for relieving the tension in the so-called $\mathcal{M}_\rmn{radio} - \mathcal{M}_\rmn{X-ray}$ discrepancy, where Mach numbers derived from radio observations are found to be higher than those derived from X-ray observations. The effect has yet to be modelled end-to-end, however, using a full Fokker-Planck solver. We resolve this issue here, using the tracers discussed in Sec.~\ref{chapter5-subsec:tracers} and the CR electron post-processing code \textsc{Crest}, introduced in Sec.~\ref{chapter5-subsec:crest}.

We show the resulting volume-weighted non-thermal electron spectra at $t=250$ Myr in the upper panels of Fig.~\ref{figure:spectra-binned-by-tinj}. At this time, the shock in the Mach 3 simulations has reached the end of the high-resolution upstream region (region III). We remind the reader that we present dimensionless momenta $p= \tilde{p} /(m_\rmn{e} c)$ and that we use the 1D distribution function, $f^\rmn{1D} = 4 \pi p^2 f^\rmn{3D}$. We have further multiplied the spectra by $p^{2.2}$, which is the maximum slope of injection we allow (see Sec.~\ref{chapter5-subsec:crest}). Using this convention, such a slope would appear as a horizontal line.

\begin{figure*}
    \centering
    \includegraphics[width=1.0\columnwidth]{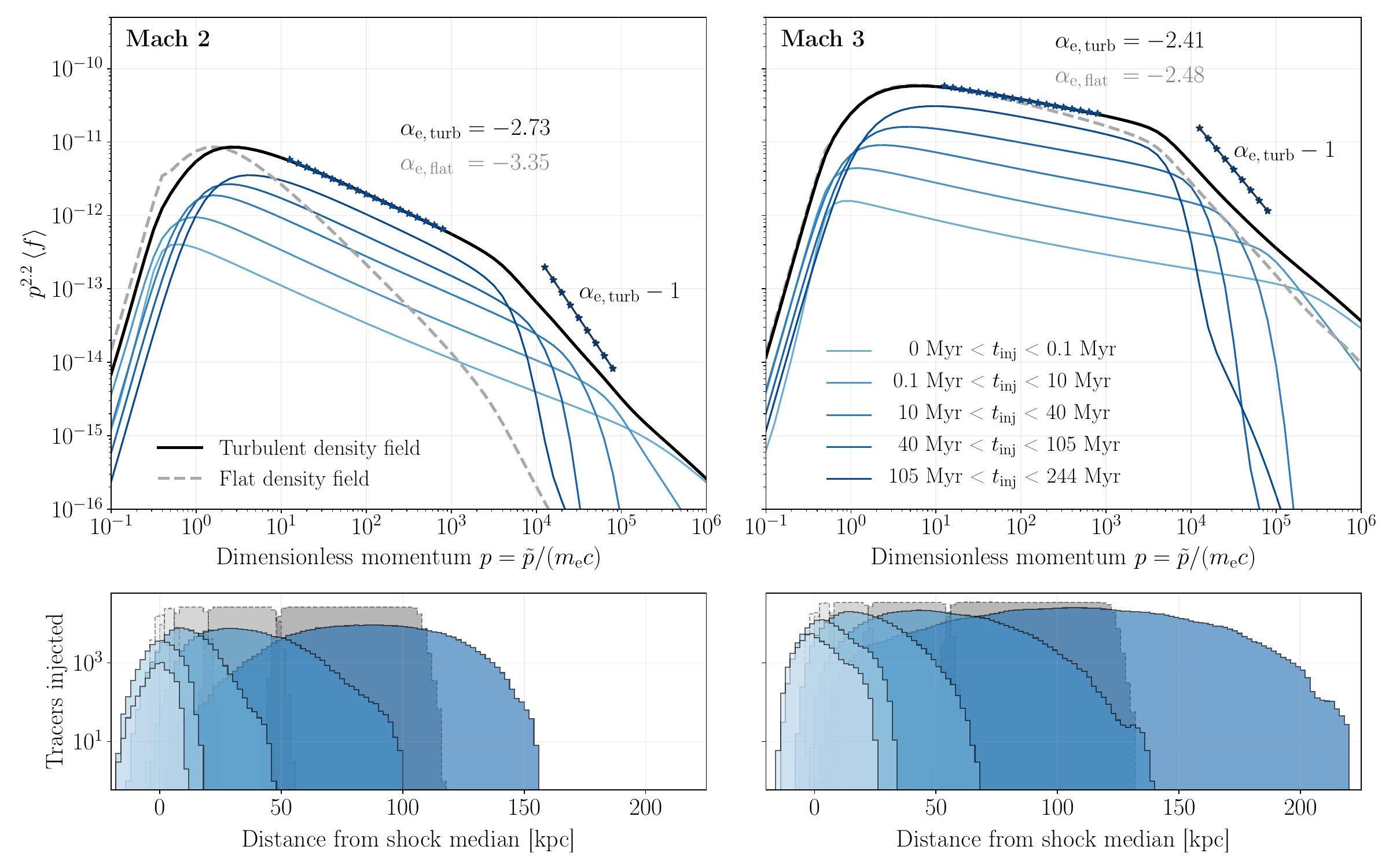}
    \caption[Non-thermal cosmic ray electron spectra for each simulation, with spectra overlaid binned by time since injection. Histograms of the positions of tracers relative to the shock-front based on these bins]{\textit{Top row:} volume-weighted non-thermal electron spectra generated from our \textit{Turb} (solid black) and \textit{Flat} (dashed grey) simulations at $t=250$ Myr. Blue lines show contributions to the \textit{Turb} spectrum, where tracers have been binned by time since injection. \textit{Bottom row:} Histograms indicating the total number of injected tracers relative to the shock front, where bins have a width of 2~kpc. Blue and grey colours represent our \textit{Turb} and \textit{Flat} simulations respectively, whilst colour saturations indicate the same time bins as above. The broadened Mach number distribution seen in Fig.~\ref{figure:mach-no-pdf} causes a shallower slope compared to the theoretical expectation (see spectral slopes, $\alpha_\rmn{e}$). This is especially noticeable for low Mach number shocks. In the case of upstream density fluctuations, substantial mixing takes place downstream, such that distance from the shock front is no longer a good indication of cooling time.}
    \label{figure:spectra-binned-by-tinj}
\end{figure*}

The grey, dashed lines show spectra from our \textit{Flat} simulations, which do not include density fluctuations. We have calculated their spectral slope, $a_\rmn{e, flat}$, in each panel by fitting a straight line between $10 < p < 10^3$ using the method of least squares. This is approximately the region unaffected by cooling. 

If we assume a single Mach number is responsible for the CR electron spectra slope, $\alpha_\rmn{e}$, we may derive the following formula from the jump conditions \citep{ensslin1998, clarke2011}:

\begin{equation}
    \mathcal{M} = \sqrt{\frac{2 (2-\alpha_\rmn{e})}{1 - 2\alpha_\rmn{e} - 3\gamma_\rmn{a}}},
    \label{eq:mach-no-slope}
\end{equation}
or equivalently
\begin{equation}
    \alpha_\rmn{e} = - \frac{2 (\mathcal{M}^2 + 1)}{\mathcal{M}^2 - 1},
    \label{eq:mach-no-slope-inverse}
\end{equation}
where we set the adiabatic index $\gamma_\rmn{a}=5/3$. These formulae predict that $\mathcal{M}=2$ and $\mathcal{M}=3$ shocks should produce slopes of $\alpha_\rmn{e}=-3.33$ and $\alpha_\rmn{e}=-2.5$, respectively. The actual values of $\alpha_\rmn{e, flat}=-3.35$ and $\alpha_\rmn{e, flat}=-2.48$, on the other hand, imply $\mathcal{M}=1.99$ and $\mathcal{M}=3.06$. Whilst some of this variation is due to the distribution of Mach numbers shown in Fig.~\ref{figure:mach-no-pdf}, at higher Mach numbers the error arises predominantly due to the fact that Eq.~\eqref{eq:mach-no-slope} has a pole at $\alpha_\rmn{e}=-2$ and consequently shallow slopes must be measured with exponentially more accuracy. With this in mind, we consider the \textit{Flat} runs to be good approximations of single Mach number shocks. That is, their slopes are what would be expected if the radio-derived Mach number matched that derived from X-ray observations, which tend to average over upstream and downstream properties.

The solid black lines in Fig.~\ref{figure:spectra-binned-by-tinj} show spectra from our \textit{Turb} runs, which include density fluctuations. It can be seen that the resultant spectral slopes, labelled $\alpha_\rmn{turb}$, are systematically shallower than in the \textit{Flat} case. Indeed, Mach numbers derived from these slopes would imply shocks with strength $\mathcal{M} = 2.56$ and $\mathcal{M} = 3.29$ for our two simulations with initial Mach numbers of $\mathcal{M} = 2$ and 3, respectively\footnote{Without the use of $\mathcal{M}_\rmn{crit} = 2.3$ (see Sec.~\ref{chapter5-subsec:crest}), the spectral slope for the Mach 2 run would be $\alpha_\rmn{e, turb}=-2.79$, implying $\mathcal{M} = 2.45$. See Whittingham et al. (in prep.) for a more in-depth discussion on the impact of using a critical Mach number.}. This is despite the fact that the shock strength continues to peak at $\mathcal{M} = 2$ and $\mathcal{M} = 3$, respectively, as seen in Fig.~\ref{figure:mach-no-pdf}.

\renewcommand{\arraystretch}{2}
\begin{table}
\centering
\begin{tabular}{|c|c||c|c|c|c|}
\multicolumn{2}{c}{}                 & \multicolumn{4}{c}{\textbf{Model}} \\ \cline{3-6}
\multicolumn{2}{c|}{}                 & \textit{Flat} & \textit{Turb} & \textit{Flat} & \textit{Turb} \\ \cline{3-6} \noalign{\vskip\doublerulesep
         \vskip-\arrayrulewidth\vskip\doublerulesep
         \vskip-\arrayrulewidth} \cline{2-6} 
\multicolumn{1}{c|}{\parbox[t]{-1mm}{\multirow{2}{*}{\rotatebox[origin=c]{90}{\textbf{Theory}}}}}
  & \textit{Initial $\it{\mathcal{M}}$} & \bf 2.00          & \bf 2.00          &  \bf3.00          & \bf 3.00          \\ \cline{2-6} 
\multicolumn{1}{c|}{} & \textit{Expected slope}             & -3.33         & -3.33         & -2.5          & -2.5         \\ \cline{2-6} \noalign{\vskip\doublerulesep
         \vskip-\arrayrulewidth\vskip\doublerulesep
         \vskip-\arrayrulewidth} \cline{2-6} 
\multicolumn{1}{c|}{\parbox[t]{8mm}{\multirow{2}{*}{\rotatebox[origin=c]{90}{\makecell{\textbf{Low} \\ \textbf{momenta}}}}}}
 & \textit{Measured slope}             & -3.35         & -2.73         & -2.48         & -2.41         \\ \cline{2-6} 
\multicolumn{1}{c|}{} & \textit{Implied $\it{\mathcal{M}}$} & \bf 1.99          & \bf 2.56          & \bf 3.06          & \bf 3.29          \\ \cline{2-6} \noalign{\vskip\doublerulesep
         \vskip-\arrayrulewidth\vskip\doublerulesep
         \vskip-\arrayrulewidth} \cline{2-6} 
\multicolumn{1}{c|}{\parbox[t]{7mm}{\multirow{2}{*}{\rotatebox[origin=c]{90}{\makecell{\textbf{High} \\ \textbf{momenta}}}}}
\hspace{3mm}} & \textit{Measured slope}               & -4.28         & -3.51         & -3.45         & -3.36       \\ \cline{2-6} 
\multicolumn{1}{c|}{\hspace{3mm}} & \textit{Implied $\it{\mathcal{M}}$} & \bf 2.03          & \bf 2.97          & \bf 3.15          & \bf 3.48          \\ \cline{2-6} 
\end{tabular}
\vspace{2mm}
\caption[Theoretical vs. measured spectral slopes and their implied Mach number]{In roman font, the spectral slopes expected for Mach 2 and Mach 3 shocks compared with the values measured from Fig.~\ref{figure:spectra-binned-by-tinj} at low ($10 < p < 10^3$) and high momenta ($10^4 < p < 10^5$), respectively. In bold, we show the Mach number that one would infer using Eq.~\eqref{eq:mach-no-slope}. The tail of the Mach number distribution causes shallower slopes and breaks the expected $\alpha_\rmn{e} - 1$ scaling at higher momenta. This causes an overestimation of the peak Mach number in the shock.}
\label{tab:inferred_mach_no}
\end{table}

It is notable that the discrepancy is greater in weaker shocks. This seems to be in contrast to our finding in Fig.~\ref{figure:mach-no-pdf}, which shows that the tail of the Mach number distribution actually extends further in \textit{stronger} shocks. This effect is outweighed, however, by the strong inverse dependence of $\alpha_\rmn{e}$ on $\mathcal{M}$, with $\alpha_\rmn{e}$ having a strong dependence on $\mathcal{M}$ at $\mathcal{M}=2$, but a substantially weaker dependence at $\mathcal{M}=3$. For example, from Eq.~\eqref{eq:mach-no-slope-inverse}, it can be seen that a shock with $\mathcal{M}=2.5$ would produce a spectral slope of $\alpha_\rmn{e}=-2.76$, which is 0.57 shallower than that of an $\mathcal{M}=2$ shock. On the other hand, an $\mathcal{M}=3.5$ shock will produce a spectral slope of $\alpha_\rmn{e}=-2.36$, which is only 0.14 shallower than an $\mathcal{M}=3$ shock. The tail of the Mach number distribution is therefore more influential in weaker shocks, as the shallower slopes from these fluctuations can more easily affect the integrated spectra at higher frequencies, thereby increasing the discrepancy between the expected and inferred Mach number.

Of course, in radio observations, only the momenta range $10^4 \lesssim p \lesssim 10^5$ is available for observation, as it is this range that produces MHz- and GHz emitting electrons. Standard theory dictates that the slope of the CR electron spectrum in this range should be $a_\rmn{e} - 1$ due to cooling\footnote{Equivalently, the spectral index of the emission will steepen by 1/2.} \citep{condon2016}. Indeed, it is this fact that is used to infer the spectral slope in radio relic observations \citep[see, e.g.][]{vanweeren2012, rajpurohit2020}. This result relies, however, upon the spectrum being produced by a single Mach number shock. We have added the theoretical slope to Fig.~\ref{figure:spectra-binned-by-tinj} for the \textit{Turb} runs to guide the eye. It can be seen when the shock is made up of a distribution of Mach numbers, the slope in this region is actually slightly shallower than $a_\rmn{e} - 1$. Once again, this is especially evident in the $\mathcal{M}=2$ run. Indeed, if we measure the slope in this range for the Mach 2 $\textit{Turb}$ run, we would infer a spectral slope of $\alpha_\rmn{e, turb}=-2.51$. This in turn produces an estimated Mach number of $\mathcal{M}=2.97$, which is almost 50\% higher compared to the true peak value. As previously, this effect originates due to the shallower slopes of higher Mach number shocks. At the higher momenta end, where $f(p)$ is lower, individual tracers can have a larger impact on the volume-weighted mean.

This effect can be more clearly seen if we plot the spectra binned by the time elapsed since their injection at the shock-front. These bins are given in the legend in the top right panel of Fig.~\ref{figure:spectra-binned-by-tinj}. We have chosen the time bins such that the number of tracers in them approximately doubles each time, which produces an approximate doubling of the amplitude of the binned spectra in regions where the cooling time exceeds the simulation run-time. The maximum time since injection is 244 Myr, as we ignore tracers injected during the initial buffer region (see Sec.~\ref{chapter5-subsec:sim-vars}). As the electrons cool rapidly at $p \gtrsim 10^4$, binning in this fashion allows us to pick out regions of momenta which are dominated by a specific age. Indeed, it is this layering that produces the overall volume-weighted slope. It can be seen that the binned spectra bend slightly, becoming slightly shallower at higher momenta. This results from the layering of different injected slopes as tracers sample the Mach number distribution. It is this bending effect that ultimately produces a slope of less than $a_\rmn{e} - 1$.

It follows from the analysis presented in this section that we expect Mach numbers derived from radio spectra to be intrinsically higher than the X-ray derived equivalents. Moreover, we expect radio observations to be particularly inappropriate for characterising the peak of the Mach number distribution when the shocks are weak ($\mathcal{M} \lesssim 2$). For ease of comparison, we collate our measurements in Table~\ref{tab:inferred_mach_no}. We investigate the relationship between the radio spectral slope and the width of the Mach number distribution in more detail in Whittingham et al. (in prep.).

\subsection{Breaking of the laminar flow assumption}
\label{chapter5-sec:laminar-flow}

\begin{figure*}
    \centering
    \includegraphics[width=1.0\columnwidth]{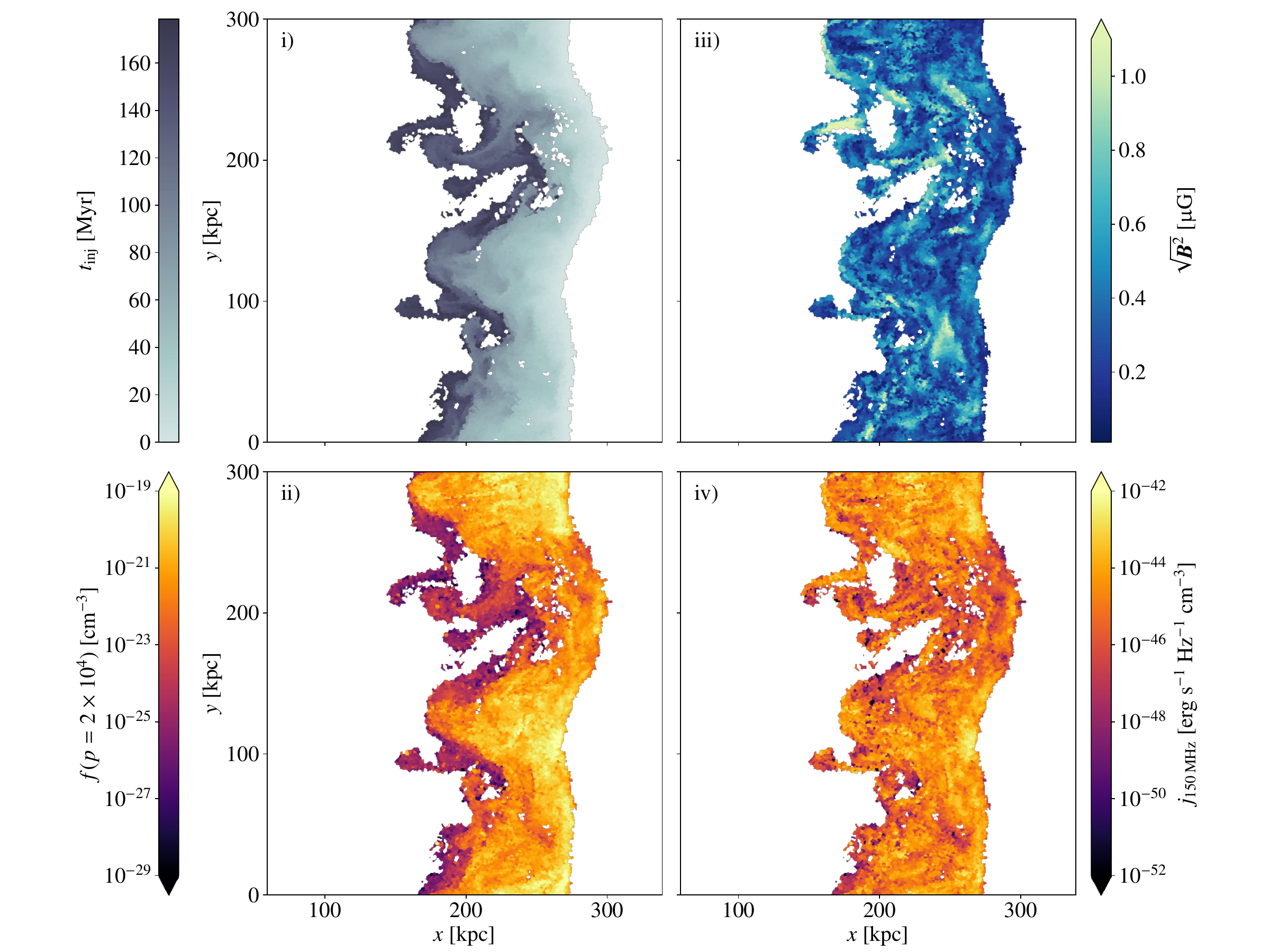}
    \caption[Slices through the fiducial shock-tube simulation showing time since injection, electron number density, magnetic fields strength, and synchrotron emissivity at 150 MHz]{Slices through our fiducial Mach 3 simulation at $t=180$ Myr showing: i) time since injection, ii) CR electron number density at $p=2\times 10^4$ (see text), iii) magnetic field strength, and iv) synchrotron emissivity at 150 MHz. Only tracers that have undergone DSA are shown. A Rayleigh-Taylor instability induces a counter-streaming plume (in the shock rest frame) that brings aged, and therefore cooled, electrons close to the shock front. These electrons are still relatively bright at 150 MHz, however, due to magnetic field amplification up to $\upmu$G levels.}
    \label{figure:slice-downstream}
\end{figure*}

It is frequently assumed that distance from the shock front is a reliable measure for time cooled \citep[see, e.g.][]{vanweeren2012, rajpurohit2020}. We now investigate whether this assumption is valid, particularly in light of the turbulence we identified in Sec.~\ref{chapter5-sec:RT}. For this purpose, we reuse the time bins just discussed, making histograms of the distance from the shock median for each binned population. These histograms are presented in the bottom row of Fig.~\ref{figure:spectra-binned-by-tinj}. The grey shaded regions indicate data from the \textit{Flat} simulations. It can be seen that whilst there is some overlap between time bins, owing to the aforementioned magnetic pressure fluctuations, the overlap is relatively minimal. This indicates that the flow is predominantly laminar. In contrast, tracers from the \textit{Turb} simulations, represented by the blue shaded regions, overlap strongly. Indeed, there are tracers in the last time bin that are still close to the shock front. 

Two further aspects further distinguish the \textit{Turb} simulations from their \textit{Flat} counterparts: firstly, there are well-populated regions of negative distance due to the corrugation of the shock front (see Sec~\ref{chapter5-sec:RT}), and secondly, the maximum distance from the shock is substantially larger in the \textit{Turb} runs. Indeed, the relative maximum distance increases proportionally with the strength of the shock: tracers reach 32\% and 67\% further in Mach 2 and Mach 3 shocks, respectively. This comes about primarily due to the inertia of the high-density fluctuations; higher density clumps experience less acceleration and can therefore better penetrate downstream. Higher Mach number shocks, meanwhile, have faster speeds, resulting in a greater distance opening up between the shock front and such clumps. This effect is further reinforced by the Rayleigh-Taylor instability as analysed in Sec.~\ref{chapter5-sec:RT}, which itself is more effective at faster shock speeds, when the downstream is more dynamic. 

The consequence of all of these effects is that, once turbulent density fluctuations are taken into account, distance from the shock front is no longer a good indicator for time since injection for individual electrons. Indeed, as we show in App.~\ref{appendix:spectra_by_dist}, it is not possible to make an analogous figure to Fig.~\ref{figure:spectra-binned-by-tinj} where the tracers have been binned by distance.

These effects can be seen particularly well in Fig.~\ref{figure:slice-downstream}, where we show slices through the injected region in our Mach 3 \textit{Turb} simulation at $t=180$ Myr. The slices presented here may be directly compared with those presented in Fig.~\ref{figure:shock-tube}; data has simply been taken from the tracers here rather than the gas cells.
In panel \textit{i)} of Fig.~\ref{figure:slice-downstream}, we show the time since injection, with lighter colours indicating more recent injection. Regions without colour indicate a lack of injection. In the shock-compressed region, this results from the shock-front falling below the critical Mach number (see discussion in Whittingham et al., in prep.). The impact of the range of post-shock gas speeds, as already shown in panel \textit{v)} of Fig.~\ref{figure:shock-tube}, can clearly be seen, with relatively freshly-injected CR electrons evident at varying distances from the shock-front. With this said, the overall distribution of CR electron ages mostly traces the shape of the Rayleigh-Taylor fingers, with extra perturbations formed by the generated turbulence. As this instability acts at the contact discontinuity, the resultant eddies are generally populated by older material. This effect is also responsible, however, for bringing the same aged CR electrons closer to the shock front.

In panel \textit{ii)} of Fig.~\ref{figure:slice-downstream}, we show the CR electron distribution function $f(p)$, where we have set $p = 2 \times 10^4$. We choose this momentum in particular as it is the one which contributes most to the 150 MHz emission\footnote{To calculate this, we invert the critical frequency formula given in Eq.~\eqref{eq:critical-frequency} to find the momentum primarily responsible for the 150 MHz emission in each tracer (assuming $\nu_\rmn{syn} = 2 \nu_\rmn{c}$, following \citealt{werhahn2021}). We then take the emission-weighted mean of the resulting distribution of momentum values.}. Older electrons have had more time to cool, and so $f(p)$ at a fixed momentum generally reflects the distribution shown in panel \textit{i)}. Some noticeable differences are evident at the shock front, however. Here it can be seen that the most advanced part of the shock has lower $f(p)$ values compared to neighbouring regions. This is due to the weaker shocks that take place here, as may be seen by comparing with panel \textit{iii)} in Fig.~\ref{figure:shock-tube}. Such shocks produce spectra with steeper slopes and lower normalisations\footnote{We do not include Mach-dependent acceleration efficiencies in our simulations, but lower Mach number shocks naturally result in less dissipated energy through Eq.~\eqref{eq:shock-dissipated-energy} and hence lower normalisations through Eq.~\eqref{eq:p_min}.}, which act to reduce $f(p)$ values. Both of these effects are also evident in Fig.~\ref{figure:spectra-binned-by-tinj}. 

Observationally, $f(p)$ is not available to us; instead, we must use synchrotron emission to infer it. However, this is modulated by the magnetic field strength of the emitting region, as can be seen by inspection of Eq.~\eqref{eq:synchrotron-emissivity}. We therefore focus now on the magnetic field downstream.

\subsection{Magnetic field amplification}
\label{chapter5-sec:magnetic-field}
\subsubsection{Strength}
\label{chapter5-sec:magnetic-field-strength}

\begin{figure}
    \centering
    \includegraphics[width=0.5\columnwidth]{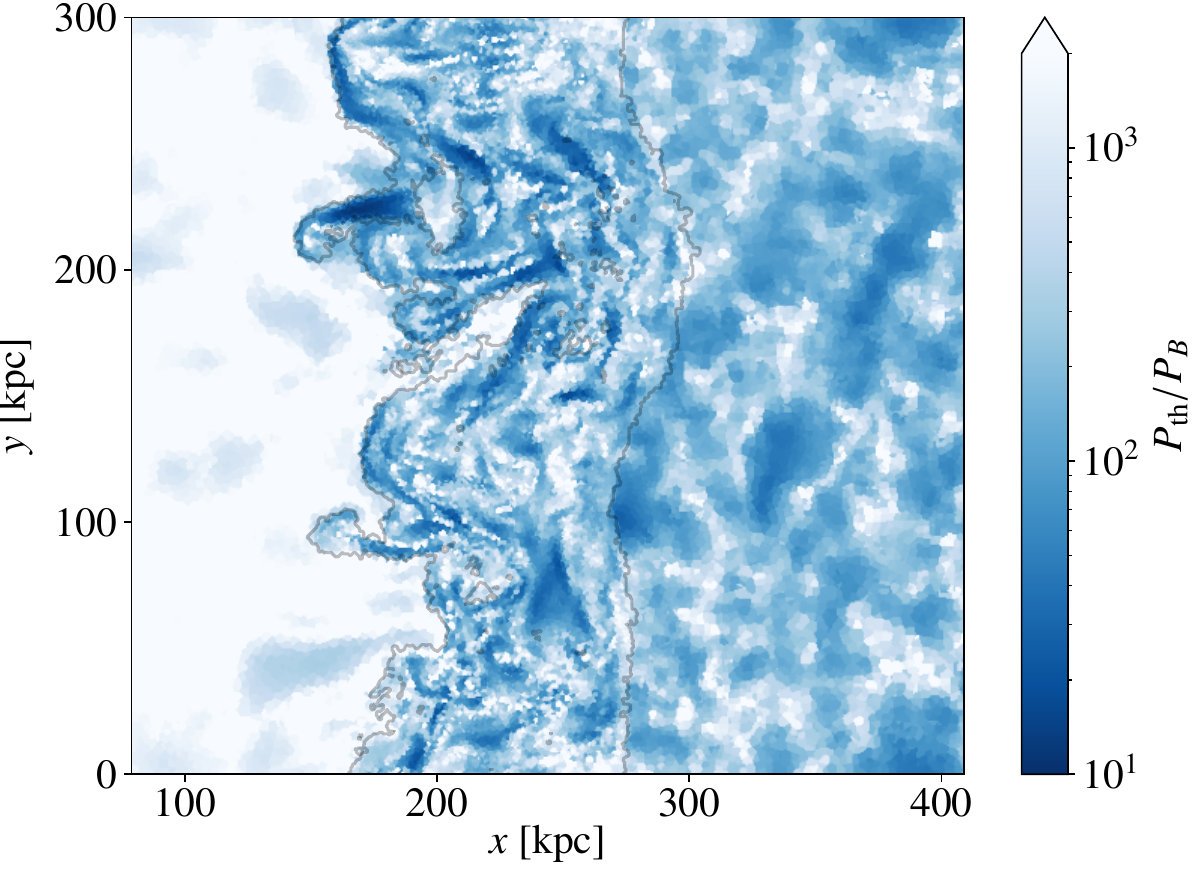}
    \caption[Slice showing plasma beta values in the fiducial simulation]{Slices through our fiducial Mach 3 simulation at $t=180$ Myr showing plasma beta values. The grey lines mark the corrugated shock surface and the contact discontinuity (i.e., the region within which tracers have been injected). Despite significant amplification, beta values are typically well above 10, limiting the ability of the magnetic field to affect dynamics.}
    \label{figure:plasma-beta-slice}
\end{figure}

As discussed in Sec.~\ref{chapter5-subsec:generating_turb}, the upstream turbulent magnetic field in our simulations has an RMS field strength of $B_\rmn{rms} \approx 0.16$~$\upmu$G. In the $\mathcal{M}=3$ case without density fluctuations we expect a shock compression ratio of $x_\rmn{s}
=3$. For an isotropic turbulent field we also expect $B \propto n^{2/3}$ and hence $B_\rmn{rms} \approx 0.31$~$\upmu$G behind the shock. This is problematic, as multiple studies now indicate that magnetic fields in radio relics are at least a factor 3$-$4 higher than this (see citations given in Sec.~\ref{chapter5-sec:intro}). In panel \textit{iii)} of Fig.~\ref{figure:slice-downstream}, we show the magnetic field strength in the injected region. It is evident that $B = 0.31$~$\upmu$G is actually at the lower end of the field strengths; indeed, the peak values exceed $\upmu$G strengths, in accordance with observed radio relics. 

Nevertheless, even at its highest strengths, the magnetic field remains dynamically subdominant, as we show in Fig.~\ref{figure:plasma-beta-slice}; maximum values in the slice shown only reach $P_\rmn{th} / P_{B} = 10$, with the majority being $P_\rmn{th} / P_{B} > 100$. We provide distributions of the plasma beta values before and after injection for the whole simulation in App.~\ref{appendix:beta}, allowing us to quantify this statement further. Moreover, the peak strengths remain below the equivalent value for the CMB magnetic field strength at redshift $z=0.2$, $B_\rmn{CMB} \approx 4.7$~$\upmu$G. This means that inverse Compton cooling dominates the CR electron losses in the downstream.

\subsubsection{Origin}
\label{chapter5-subsec:mag-field-origin}

\begin{figure*}
    \centering
    \includegraphics[width=1.0\columnwidth]{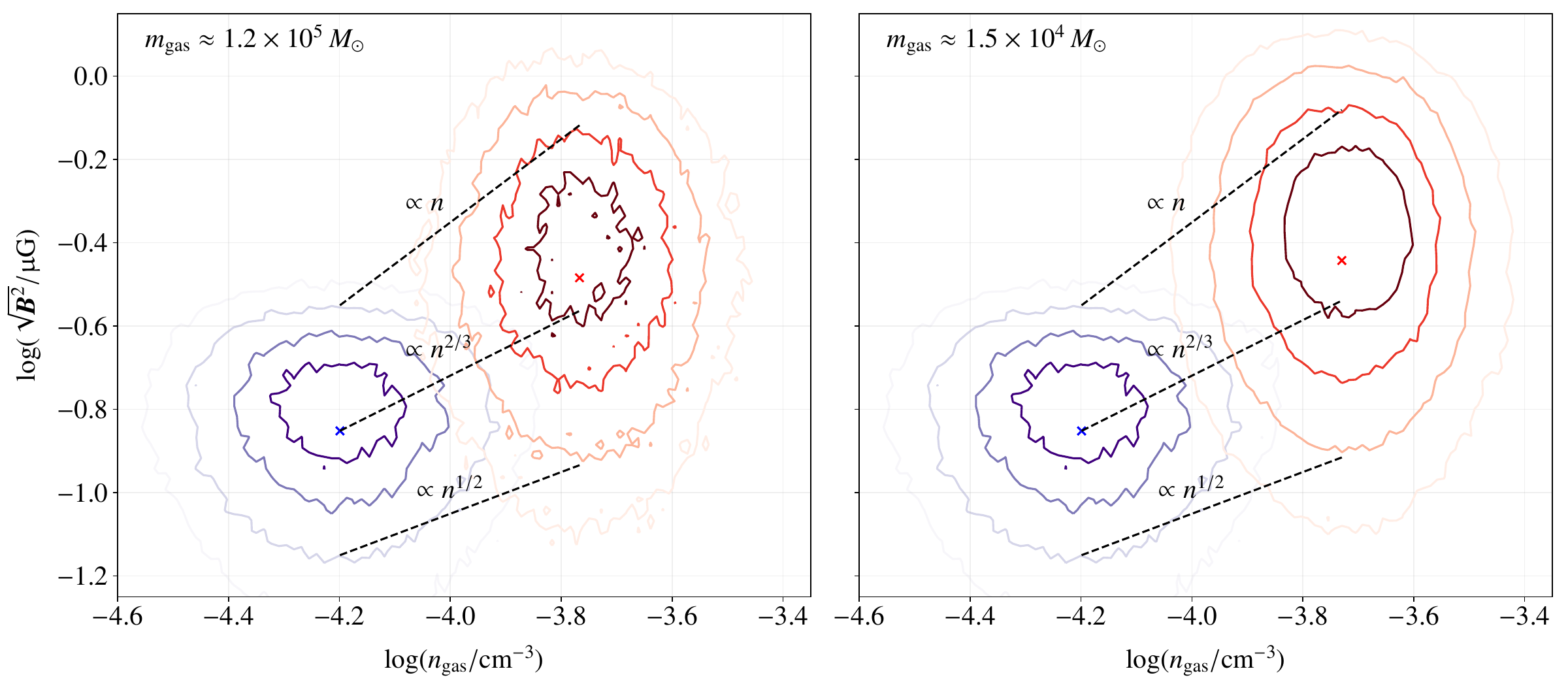}
    \caption[Phase space diagrams of magnetic field strength vs. gas number density for simulations with varying resolution]{\textit{Left:} a phase space diagram of magnetic field strength vs. gas number density for our lower resolution fiducial Mach 3 simulation. Contours cover 10\%, 25\%, 50\%, and 75\% of the population, respectively (from darker to lighter colours). Blue colours indicate the initial upstream distribution, whilst red colours show the final state in the injected region at $t = 250$~Myr. A cross marks the median of the distribution. \textit{Right:} as previous, except data is taken from the high-resolution fiducial Mach 3 simulation. Dashed lines indicate typical scaling relations due to adiabatic compression with a specific field geometry (see text). Elongation of the distribution over time can mostly be explained by such relations. However, they cannot explain the peak magnetic field values, which increase with finer mass resolution, hinting at an additional small-scale dynamo amplification.}
    \label{figure:rho-B-phase-diagram}
\end{figure*}

Both in Fig.~\ref{figure:slice-downstream} and Fig.~\ref{figure:plasma-beta-slice}, it can be seen that magnetic fields in the injected region are arranged in filament-like shapes. These are predominantly sheets seen in cross-section, which can be formed by the compression of a turbulent field. We check this statement with Fig.~\ref{figure:rho-B-phase-diagram}, where we compare the magnetic field strength vs. gas number density phase space before and after the passing of the shock surface. To do this we take all gas cells in the high-resolution upstream region (region III) at $t=0$ and all gas cells at $t=250$ Myr that are in the shock-compressed region. Contours of the distributions are shown in blue and red, respectively. On the left-hand side of Fig.~\ref{figure:rho-B-phase-diagram} we show data from the lower resolution Mach 3 re-simulation of \textit{Turb} (see Sec.~\ref{appendix:numerical-stability}), whilst the right-hand side shows data from our fiducial Mach 3 simulation. 

We have added dashed lines, which indicate the typical scaling relations of $B \propto n^\kappa$, where $\kappa$ = 1/2, 2/3, and 1, respectively. These represent compression along i) one parallel and one perpendicular axis, ii) isotropic compression, and iii) compression along perpendicular axes only, respectively. We start and end the lines at the median density in initial and final states, respectively. It can be seen that the given scaling laws are able to explain most of the increase in the height of the distribution. This is in line with work by \citet{ji2016}, who also modelled shocks propagating into upstream turbulence. We note, however, that compression cannot explain all of the evolution. In particular, we would expect compression of a turbulent magnetic field from one side to follow the scaling for isotropic compression, and therefore the median of both distributions, marked by an `x', should be joined by the $B \propto n^{2/3}$ line. Although it is close, this line undershoots the true median in the final state in both simulations. Moreover, the $B \propto n$ line takes the 25th percentile contour in the initial state only as far as the 50th percentile in the final state. This implies that the amplification is not purely compressive.

The turbulence generated downstream may be able to create a small-scale dynamo, which is highly effective at amplifying magnetic fields in the kinetic regime \citep{whittingham2021, pfrommer2022, kriel2023}. To test this, we compare the phase space distributions in both simulations; increasing the resolution allows us to resolve higher peak densities, but should not change the compressive amplification of the magnetic field. Increasing the resolution will, however, reduce the sizes of the resolved turbulent eddies, allowing more rapid amplification \citep{whittingham2021}. If we observe the two distributions in Fig.~\ref{figure:rho-B-phase-diagram}, we see that, indeed, the higher-resolution one is stretched towards higher magnetic field strengths: the peak of the red contours reaches 1.14 $\upmu$G in the lower-resolution and 1.26 $\upmu$G and higher-resolution simulation, whilst the overall maximum reaches 1.96 $\upmu$G and 2.22 $\upmu$G, respectively. This is approximately a 10\% increase. Furthermore, it can also be seen that the highest magnetic field strengths in Figs.~\ref{figure:slice-downstream} and \ref{figure:plasma-beta-slice} are associated with the turbulent eddies at the back of the shock-compression zone. We consider both of these factors to be good circumstantial evidence that a small-scale dynamo is in action.

\begin{figure*}
    \centering
    \includegraphics[width=1.0\columnwidth]{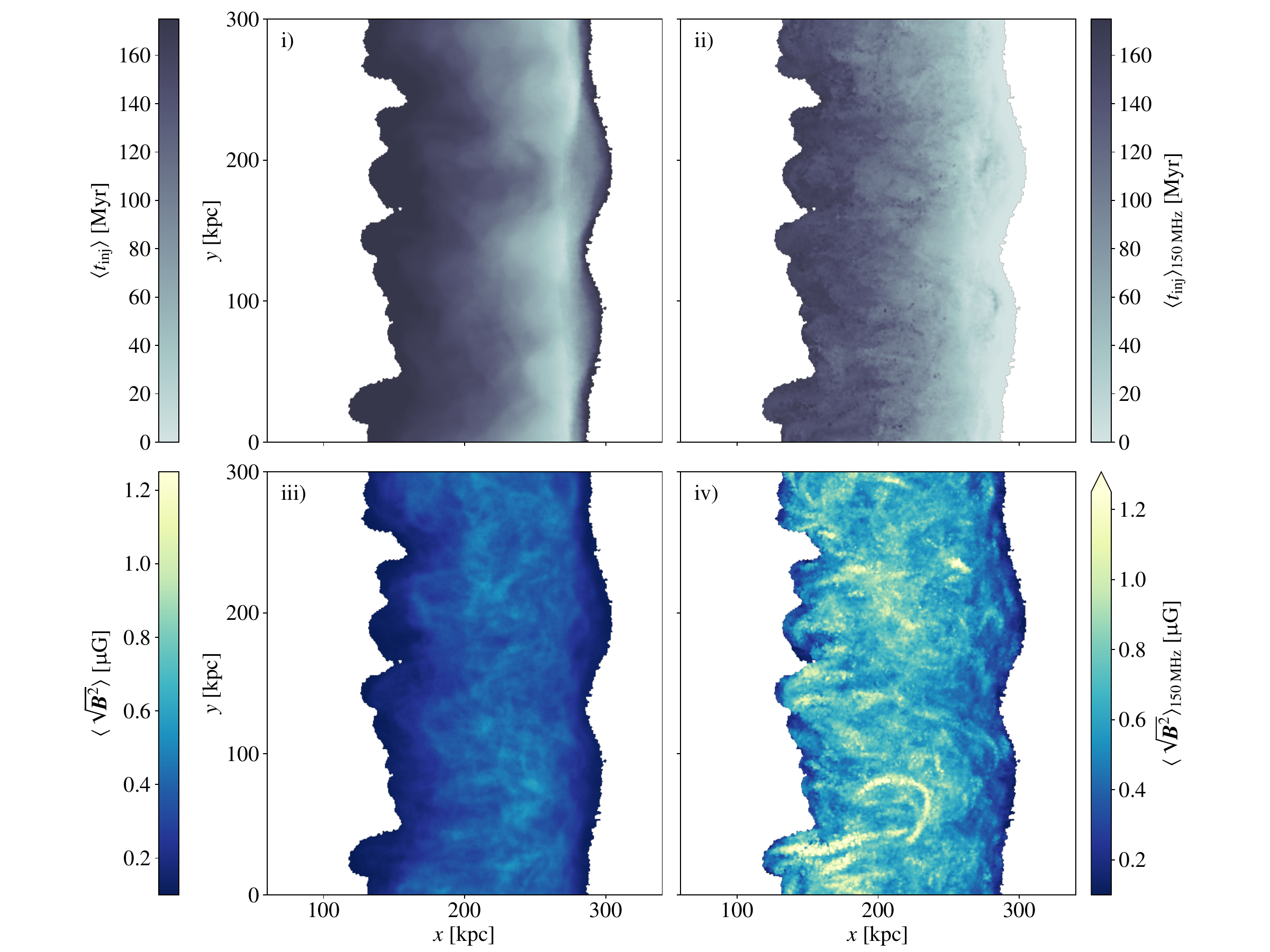}
    \caption[Projections through the fiducial simulation showing volume- and synchrotron-weighted values for the time since injection and magnetic field strength]{Projections with a depth of 300 kpc through our fiducial Mach 3 simulation at $t=180$ Myr. These show: i) volume-weighted time since injection, ii) synchrotron-weighted time since injection, iii) volume-weighted magnetic field strength, and iv) synchrotron-weighted magnetic field strength. Regions with no injected tracers have been masked (see text for further details). Synchrotron emission is calculated at $\nu = 150$~MHz and is biased towards fresher injection and stronger magnetic fields. Observations made through this channel consequently appear to show a relatively laminar flow and $\upmu$G-strength magnetic fields, even this is not typical of the downstream.}
    \label{figure:t_inj-and-B-field-projections}
\end{figure*}

\subsubsection{Impact on observational inferences}
\label{chapter5-sec:impact-on-obs}

Returning to Fig.~\ref{figure:slice-downstream}, we may now inspect how the magnetic field strength affects synchrotron emissivity, which is shown in panel \textit{iv)}. This is calculated using \textsc{Crayon+}, as explained in Sec.~\ref{chapter5-subsec:crayon}, and is shown in this figure at 150 MHz. The colorbar for this panel is given the same overall range as was given to $f(p=2\times10^4)$. It can be seen that the dynamic range, however, is substantially smaller in panel \textit{iv)}. This is due to the significant amplification that takes place towards the back of the shock-compressed region. There are two ways in which this impacts the synchrotron emissivity. Firstly, as can be seen by inspecting Eq.~\eqref{eq:synchrotron-emissivity}, there is a linear scaling of the emission with the component perpendicular to the line of sight. Secondly, through Eq.~\eqref{eq:critical-frequency}, a change in the magnetic field strength shifts the momenta at which most emission takes place. Specifically, as $\nu_\rmn{c} \propto B \gamma^2$, an increased magnetic field strength means that lower momenta, which have higher values of $f(p)$, contribute more at the same frequency of emission.  Together these effects boost emission from the rear of the shock-compressed zone, thereby evening out some of the effects of cooling. Indeed, in some regions where $f(p)$ is low, such as towards the back of the shock at $y\approx210$ kpc, the synchrotron emissivity is now higher than at the front of the shock. In general, however, the emissivity still decreases from front to back. We also note that the brightest regions are still at the front of the shock where we previously identified the highest Mach number shocks to be. The front of the shock undergoes milder magnetic field amplification and hence milder modulation, and so our analysis in Sec.~\ref{chapter5-sec:laminar-flow} regarding features close to the shock front holds for the synchrotron emission too.

We now consider how this modulation affects our inference of radio relic variables. As stated earlier, synchrotron emission is our primary window into CR electron properties in radio relics, and so it is critical to understand how observations are biased. In Fig.~\ref{figure:t_inj-and-B-field-projections}, we investigate the impact of synchrotron-weighting on the projected time since injection and projected magnetic field strength. On the left-hand side, we show the volume-weighted values, whilst on the right-hand side we show the synchrotron-weighted projections. We use the 150 MHz channel for this, as previously. We only include values where the line-of-sight crosses at least one injected particle. The colorbars in each row have the same scale, so panels may be directly compared.

In panel \textit{i)}, we project the time since injection across all tracers. We have set tracers that haven't been injected to have values equal to the simulation run-time ($t=180$ Myr). This leads to a gradual fade-out towards the left-hand side, as tracers age and as we project through non-injected tracers as well. This effect also highlights the Rayleigh-Taylor fingers, where the majority of injected tracers are. These are not as clear as in Figs.~\ref{figure:shock-tube} and \ref{figure:slice-downstream} as  we are now projecting through significant amounts of turbulence. On the right-hand side of panel \textit{i)}, the non-injected tracers reduce the projected average in the most advanced parts of the shock, leading to these regions being darker than expected. This is a useful feature, however, as it shows us that the back of the shock is predominantly planar; it is the front of the shock that corrugates, as it advances into regions of lower density. Indeed, we could see this effect in cross-section in panel \textit{iii)} of Fig.~\ref{figure:shock-tube}. We should consequently expect the brightest part of the simulated radio relic to also be relatively flat, as it is here that the shock-dissipated energy is highest (see Sec.~\ref{chapter5-sec:mach-no-dist-origin} and App.~\ref{appendix:mach-in-projection}). We will revisit this analysis in Sec.~\ref{chapter5-sec:SI-and-intensity-maps}.

\begin{figure*}
    \centering
    \includegraphics[width=1.0\columnwidth]{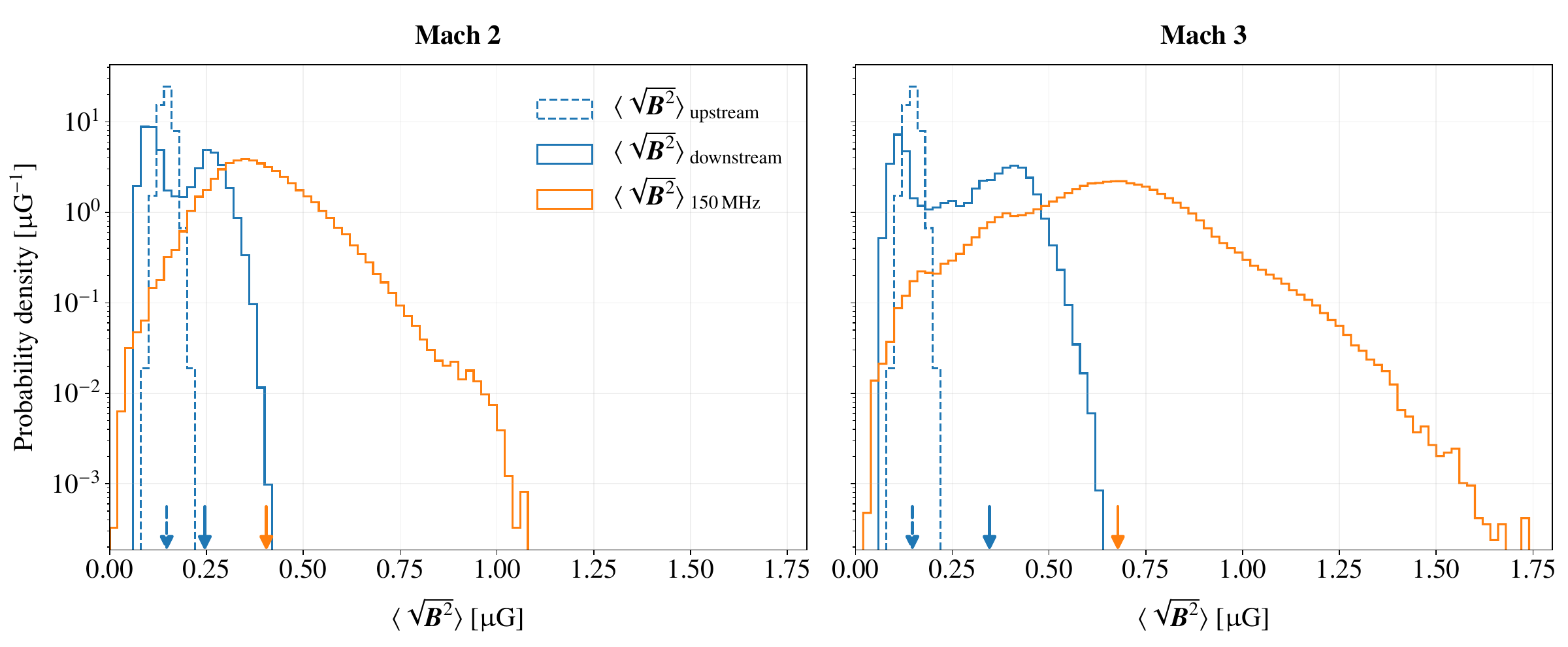}
    \caption[Probability density distributions of magnetic field strength as implied by different weightings for varying shock strength]{\textit{Left:} probability density distributions of the projected magnetic field strength in our fiducial Mach 2 simulation. Solid (dashed) lines represent projections through the shock-compressed region at $t=250$ Myr (high-resolution region at $t=0$ Myr). Blues lines represent volume-weighting, whilst orange lines represent synchrotron-weighting, where the synchrotron emission is calculated at $\nu = 150$~MHz. Arrows are placed at the RMS magnetic field strength calculated for each distribution. \textit{Right:} as previous, except data comes from the fiducial Mach 3 simulation. Synchrotron-weighting significantly overestimates the average magnetic field strength. Magnetic field values are able to reach $\upmu$G values even in weaker shocks, although amplification is more effective at higher Mach number shocks.}
    \label{figure:projected-B-histograms}
\end{figure*}

When we weight the projection by the 150 MHz channel, as in panel \textit{ii)}, we start to recover the relationship between distance from the shock and time since injection; on large enough scales, distance once again becomes a reasonable proxy for age (see Sec.~\ref{appendix:spectra_by_dist}). On kpc-scales, however, there is still substantial variance. This typically takes the form of horizontal striations. An exception to this is towards the very front of the shock, where two darker, curved regions are visible behind the shock corrugations. If we compare this panel with Fig.~\ref{figure:shock-tube}, which shows the same simulation at the same time, we can see that this is due to the turbulent transport of older material towards the shock front. As we showed in Sec.~\ref{chapter5-sec:RT}, turbulence is most effective behind the regions where the shock corrugates. 

In panel \textit{iii)}, we show the volume-weighted projected magnetic field strength. This reaches typical values of around 0.4~$\upmu$G, with peak values reaching closer to 0.65~$\upmu$G, as expected from Fig.~\ref{figure:rho-B-phase-diagram}. When we weight the projection by synchrotron emission at 150 MHz, as shown in panel \textit{iv)}, we find a significant shift towards higher field strengths. Now, a substantial amount of the projection is above 1~$\upmu$G, more closely in line with observational inferences from radio relics. Once again, the emission-weighted projection predominantly shows horizontal striations, especially towards the rear of the shock-compressed zone. Indeed, this explains the patterns seen in panel \textit{ii)}: the synchrotron emission is proportional to $B_\perp^{1-\alpha_\nu}$ (see Eq.~\ref{eq:synchrotron-emissivity}), hence CR electrons moving along these features dominate the projected quantities.

Magnetic field features stretched parallel to the $x$-axis could also be seen in Figs.~\ref{figure:slice-downstream} and \ref{figure:plasma-beta-slice}. By comparing these plots with the velocity structures evident in Fig.~\ref{figure:shock-tube}, it can be seen that the features originate from the shearing of gas. Curved filaments of higher magnetic strength are also evident, however. Through similar comparisons, it can be seen that these form due to turbulent eddies and adiabatic compression in the troughs between Rayleigh-Taylor fingers (see  Fig.~\ref{figure:slice-downstream} again). All of these mechanisms require turbulence, and this is mainly generated at the contact discontinuity. Consequently, the strongest magnetic field values in Fig.~\ref{figure:t_inj-and-B-field-projections} are seen towards the back of the shock-compressed region, with minimal extra amplification at the very front of it.

In Fig.~\ref{figure:projected-B-histograms}, we quantify just how much the synchrotron-weighted projections are biased relative to their volume-weighted counterparts. We show the initial state in the high-resolution region (region III) in dotted lines, and the final state in the shock-compressed region at $t=250$ Myr in solid lines, with blue indicating volume-weighting and orange indicating synchrotron-weighting. The distribution is initially narrow and peaks at approximately 0.16~$\upmu$G, matching the chosen RMS value (see Sec.~\ref{chapter5-subsec:generating_turb}). By construction, the initial distribution is the same in both Mach 2 and Mach 3 simulations. At the end of each simulation run, the volume-weighted distribution has broadened and now shows a double peak. This is a line-of-sight effect; with the left-hand peak representing lines-of-sight that predominantly pass through regions outside the shock-compressed region, and the right-hand peak representing the opposite scenario\footnote{Weakening of the field below the initial distribution is primarily due to tracers not respecting the contact discontinuity, leading to adiabatic expansion. This is caused by tracers not exactly following the mass flux of the gas cells \citep[see discussion in][]{genel2013}. Fortunately, this effect only has a very limited impact on our results.}. This effect biases these projections slightly low relative to the true volume-averages in the injected region (see values in Fig.~\ref{figure:rho-B-phase-diagram}). The projected volume-weighted RMS magnetic field strengths in the Mach 2 and Mach 3 simulations are 0.23~$\upmu$G and 0.33~$\upmu$G, respectively. These values are not far from the respective maximums of the distribution at approximately 0.4~$\upmu$G and 0.6~$\upmu$G. 

The peak of the synchrotron-weighted distribution is not far from the amplified peak of the volume-weighted distribution. However, unlike this distribution, the synchrotron-weighted distribution has a strong tail towards higher magnetic field strengths, reaching approximately 1.07~$\upmu$G and  1.73~$\upmu$G in the Mach 2 and Mach 3 simulations, respectively. This increases the projected RMS magnetic field strength to 0.48~$\upmu$G and 0.78~$\upmu$G in these simulations, which is more than double the volume-weighted values. Indeed, this problem gets worse at higher Mach number shocks, as the amplification increases. Figure~\ref{figure:projected-B-histograms} shows clearly that strong caution should be exercised when using CR electron observations to infer the average magnetic field strength in radio relics.

\subsection{Impact on emission}
\label{chapter5-sec:synchrotron-emission}

\subsubsection{Spectral index and intensity maps}
\label{chapter5-sec:SI-and-intensity-maps}

We now turn our attention to the projected synchrotron emission itself, and how features in intensity and spectral index maps can be explained with the knowledge we have gained over the last sections. We first focus on the intensity maps, which we show for our Mach 3 simulation at $t=180$ Myr in the upper panels of Fig.~\ref{figure:intensity-and-spectral-index}. The intensity of the emission is calculated for each pixel using Eq.~\eqref{eq:intensity}. In the left-hand panel, we focus on emission at 150 MHz. The maximum values here reach approximately $6 \times 10^{-3}$ $\upmu$Jy arcsec$^{-2}$. Typical radio relic surface brightnesses, meanwhile, are $0.1 - 1$ $\upmu$Jy arcsec$^{-2}$ \citep{vanweeren2019}. This is not problematic, however, as we are not modelling first-order Fermi re-acceleration of a fossil relativistic electron population in our simulations (see Sec.~\ref{chapter5-subsec:crest}). modelling this is expected to increase the emission at Mach 3 by a factor of 10 to 100 \citep{pinzke2013}, bringing our simulations into line with observations.

It can be seen that the peak of the emission traces an approximately straight line, with the shock corrugation being visible as promontories ahead of it. This follows from our observation in the previous section that the back of the shock is relatively flat, especially when seen in projection (see Figs.~\ref{figure:t_inj-and-B-field-projections} and~\ref{figure:mach-number-projection}). The most advanced regions of the shock in Fig.~\ref{figure:intensity-and-spectral-index} show emission that is a factor of $10-100$ weaker than the main body of the relic, indicating that they could be trickier to locate in observations; if the shock is not observed edge-on, this feature could be overpowered by the emission behind it. We will investigate the full impact of orientation on such features in future work.

\begin{figure*}
    \centering    \includegraphics[width=1.0\columnwidth]{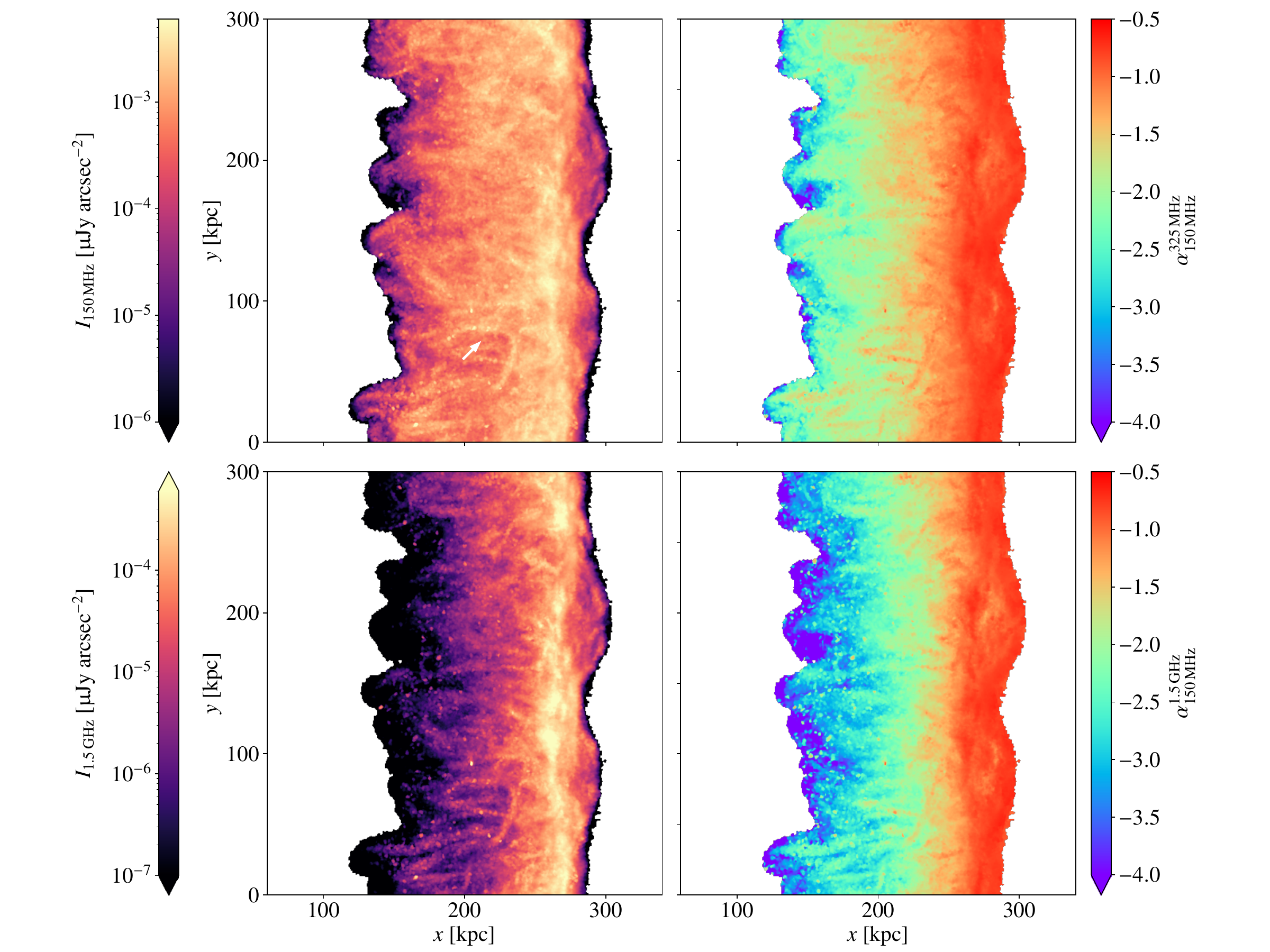}
    \caption[Synchrotron intensity and spectral index maps of the fiducial simulation]{\textit{Left column:} synchrotron intensity at 150 MHz (top) and 1.5 GHz (bottom), respectively, for our fiducial Mach~3 simulation at $t=180$ Myr. The projection depth is 300 kpc. \textit{Right column:} spectral index maps between 325 MHz and 150 MHz (top) and 1.5 GHz and 150 MHz (bottom), respectively. Intensity fluctuations in this region are a consequence of Mach number variations, whilst spectral index variations towards the shock front are predominantly a result of the corrugated shock front. Both intensity and spectral index maps show long striations predominantly orientated along the $x$-axis, resulting from the magnetic filaments observed in Fig.~\ref{figure:t_inj-and-B-field-projections}. The white arrow indicates a particularly clear example caused by adiabatic compression of the post-shock magnetic field. An animated version of this figure can be found \href{https://youtu.be/A-jQvWVClyg}{here}.}
    \label{figure:intensity-and-spectral-index}
\end{figure*}

Regions of higher intensity can be seen at the shock front at $y \approx130$ kpc and $y \approx270$ kpc. These regions are especially evident in the 1.5 GHz observations, where cooling acts faster. These regions are produced by the more intense emission created when the shock is strongest, as can be seen by comparing these features with the position of the highest projected Mach numbers in Fig.~\ref{figure:mach-number-projection}. We note that the spatial separation of these ``hotspots''\footnote{Unrelatedly, some kpc-scale hotspots can additionally be seen downstream. These are an artefact of our model, arising from the aforementioned adiabatic effects at the contact discontinuity.} is approximately equivalent to the turbulent injection scale ($l_\rmn{inj} = 150$ kpc). We postpone more detailed investigation of this correlation to the companion paper to this piece, Whittingham et al. (in prep.).

Finally, we observe that the striations towards the back of the injected region align well with the patterns seen in the synchrotron-weighted magnetic field in panel \textit{iv)} of Fig.~\ref{figure:t_inj-and-B-field-projections}. Indeed, individual features in both plots can be matched up with one another. This implies that such synchrotron ``striations'' are the result of particularly strong magnetic field lines. Evidence for the origin of these striations can be seen in the spectral index maps, which we show in the right-hand panels of Fig.~\ref{figure:intensity-and-spectral-index}. We calculate the spectral index in each pixel for these maps as:
\begin{equation}
    \alpha^{\nu_2}_{\nu_1} = \frac{\log_{10}(I_{\nu_2} / I_{\nu_1})}{\log_{10}({\nu_2 / \nu_1})},
\end{equation}
where $\nu_2>\nu_1$ so that $\alpha^{\nu_2}_{\nu_1}<0$ for a spectrum that decreases with increasing frequency.

The brightest filaments in the left-hand panels are evident as red-orange colours in the spectral index maps, indicating relatively fresh injection. Filaments with weaker emission are also evident, however, in greenish colours. It is notable that each filament shows a relatively uniform colour; the strong magnetic fields in them should lead to rapid cooling, creating a colour gradient if tracers travelled along them. Indeed, the lack of colour gradient indicates that this is not the case. We find instead that the magnetic field is amplified simultaneously along its length, temporarily brightening CR electrons of the same age. This happens in one of two main ways: i) the filament is formed close to the shock front and is then advected downstream, typically being rotated vertically due to the gas flow, or ii) compression at the contact discontinuity amplifies a region of the magnetic field simultaneously, as is the case for the horseshoe-like filament evident where the white arrow is located in panel \textit{i)}. Both formation mechanisms can be witnessed in the animated version of Fig.~\ref{figure:intensity-and-spectral-index}, linked \href{https://youtu.be/A-jQvWVClyg}{here}.

Towards the front on the injected region, it can be seen that the spectral index undergoes fluctuations. Specifically, behind the most advanced part of the shock, the spectral index temporarily reduces in value before increasing again. This feature is best seen in the $\alpha^{1.5\,\rmn{GHz}}_{150\,\rmn{MHz}}$ data, in which the effects of cooling are particularly strong. This indicates that more aged material temporarily dominates the emission, as could also be seen in the synchrotron-weighted projection of the time since injection in Fig.~\ref{figure:t_inj-and-B-field-projections}. As explained in Sec.~\ref{chapter5-sec:impact-on-obs}, this feature is produced by the transport of older material towards the shock front due to the Rayleigh-Taylor instability (see Fig.~\ref{figure:shock-tube}). As previously, the observation of this feature will be easiest when the relic is seen edge-on\footnote{An example of this feature in observations can be found in the spectral index maps presented in \citet{vanweeren2012}, where the spectral index decreases temporarily along parts of the ``handle'' of the Toothbrush.}.

The spectral index values in Fig.~\ref{figure:intensity-and-spectral-index} approximately match the values and trends observed for the Toothbrush relic \citep[see, e.g.,][]{vanweeren2012, rajpurohit2020}. To quantify this statement, however, we turn to the use of colour-colour diagrams.

\subsubsection{Colour-colour diagrams}

Colour-colour diagrams are often used in radio observations to understand the underlying acceleration and cooling processes. They are created by plotting two spectral index maps against each other. Points on this plane then measure the spectral curvature, i.e., the functional behaviour of $\rmn{d}\log_{10} j_\nu / \rmn{d}\log_{10}\nu$ in the given frequency range, which, in turn, is a measure of the spectral shape. As the CR electrons age, they trace a trajectory in the plane, indicating how the spectral shape evolves with time. This can then be compared against the trajectories produced by various models.

As discussed in Sec.~\ref{chapter5-sec:intro}, four models are typically used. These are abbreviated: KP, JP, CI, and KGJP\footnote{Examples of the characteristic spectral shapes formed by each model can be seen in figure~11 of \citet{vanweeren2012}.}. The first two of these model a one-time injection, assuming that the pitch angle distribution in each CR electron population is either fixed or continuously re-isotropised, respectively. The difference, here, is that the KP model leads to an increasingly inhomogeneously-cooled population along the line-of-sight, leading to flatter spectral curvature at later times. This generates a curved trajectory in the colour-colour plane. In contrast, the JP model produces homogeneous populations along the line of sight, generating relatively straight lines in the colour-colour plane, indicative of constant spectral curvature. The CI and KGJP models are extensions of the JP model, and assume continuous injection and injection for a limited time only, respectively. These models also produce (initially) curved trajectories in the colour-colour plane, for the same reasons as above.

In theory, such tracks are insensitive to the rate of cooling, as this only affects the rate at which the spectra develops, not its overall shape. Varying the density and magnetic field strength\footnote{The magnetic field strength is able to affect the ``break'' frequency, above which cooling dominates. As explained in Sec.~\ref{chapter5-subsec:crest}, however, we expect this frequency to be predominantly set by inverse Compton cooling.} should therefore only change the distance of the points from the shock, not their locus on the plane. The key assumption underpinning each model, however, is that lines-of-sight intersect CR electron populations that have aged for the same amount of time. This implicitly assumes an edge-on perspective and that distance from the shock front, $d$, is a reliable measure of time cooled, $t$; specifically, that $d = v_\rmn{post} t$, where $v_\rmn{post}$ is the post-shock velocity given by the jump conditions.

\begin{figure*}
    \centering
    \includegraphics[width=1.0\columnwidth]{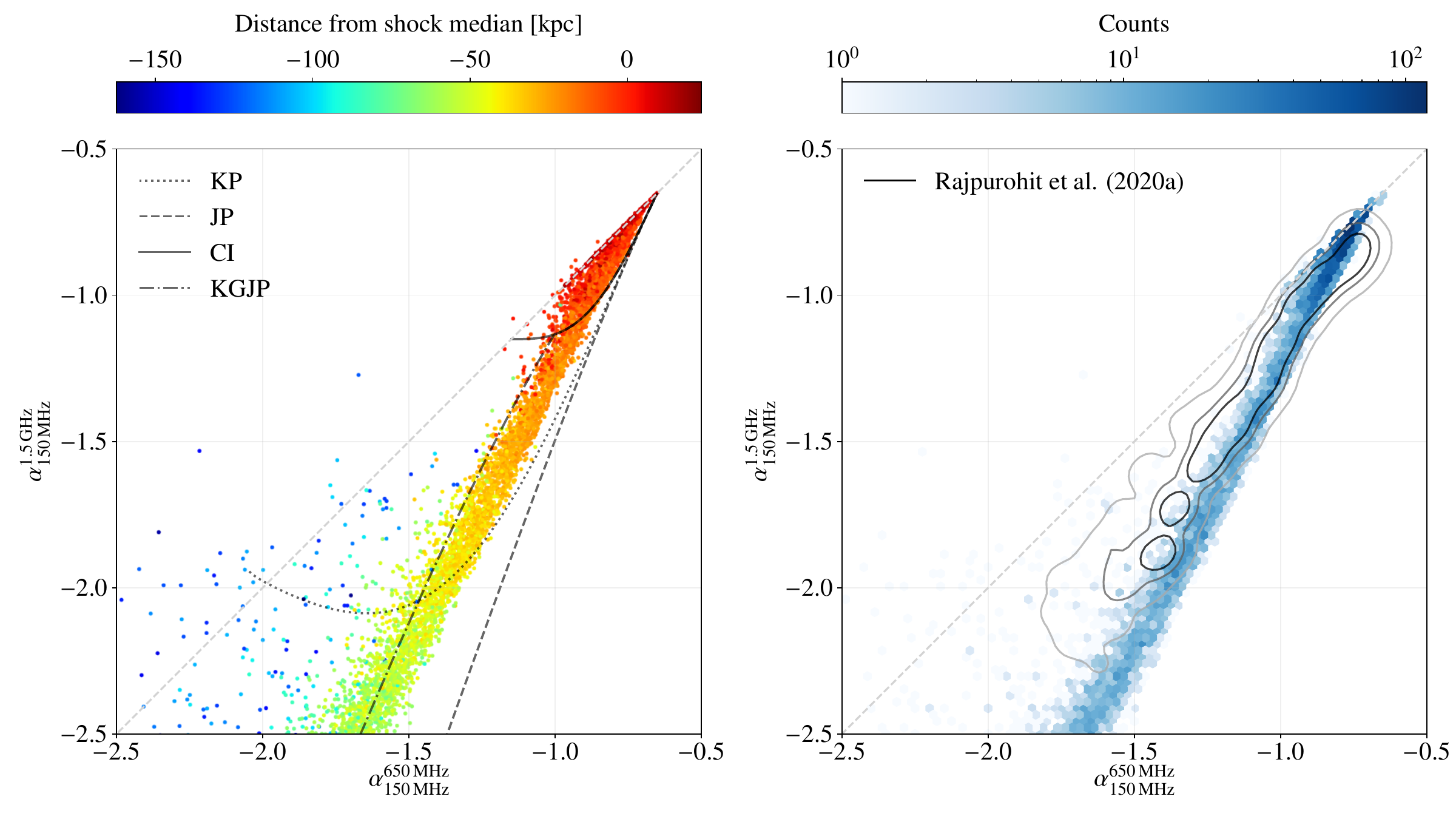}
    \caption[Colour-colour diagrams of the fiducial simulation with proposed cooling models and observed trends overlaid]{\textit{Left:} a colour-colour diagram created from our fiducial Mach 3 simulation at $t=180$ Myr using spectral indices between 1.5 GHz and 150 MHz, and 650 MHz and 150 MHz, respectively. Points are coloured based on their distance to the shock median. The diagonal grey line indicates zero spectral curvature. The remaining curves indicate spectral cooling models fitted to the B1 region of the Toothbrush radio relic by \citet[][see text for details]{rajpurohit2020}. All models begin at $\alpha^{1.5\,\rmn{GHz}}_{150\,\rmn{MHz}} = \alpha^{650\,\rmn{MHz}}_{150\,\rmn{MHz}}=-0.65$. \textit{Right:} as previous, but points have been binned over, such that colour now represents the phase space density. Contours show the analogous data observed for the B1 region by \citet{rajpurohit2020} and contain 95\%, 75\%, and 50\% of the scatter points from their figure 10 respectively. Data from the simulation contradict traditional cooling models, but are in rough agreement with the observed colour-colour diagram for a real radio relic.}
    \label{figure:colour-colour}
\end{figure*}

In Fig.~\ref{figure:colour-colour}, we show a colour-colour diagram for our fiducial Mach 3 simulation at $t=180$ Myr, created from the data presented in Fig.~\ref{figure:intensity-and-spectral-index}. In the left-hand panel, we show a scatter plot, where each point matches a pixel in the projected emission maps. We colour each point by its distance from the shock median. For reference, we include the cooling models fitted to the B1 region of the Toothbrush relic by \citet{rajpurohit2020}. We do not specifically aim to replicate this radio relic in the current work, but by coincidence our simulation approximately matches the same maximum spectral index values, with both observed and simulated values reaching $\alpha \approx -0.65$. We should therefore expect such models to also fit the simulation. It is evident, however, that none of them do.

In particular, the KP and CI models produce shapes fundamentally incompatible with the simulations. The KGJP model appears to fit within the scatter, but this, too, shows an initially curving trajectory, not evident in the simulated data, and is too steep once continuous injection is turned off. Moreover, as the models start at the data point with the highest spectral index, their trajectories should be compared with the bottom of the scatter, not its mid-point. With this in mind, it can be seen that the JP model is the closest to describing the data. This is perhaps to be expected, as our set-up is a one-time injection with cooling independent of pitch angle (see Sec.~\ref{chapter5-subsec:crest}), which matches the theoretical basis of the model. However, even the JP model has flaws; firstly the trajectory produced is too steep, and secondly, this model is unable to replicate the subtle flattening of the simulated trajectory, which begins at $\alpha^{1.5\,\rmn{GHz}}_{150\,\rmn{MHz}} \approx -1.3$. 

This ``flattening'' feature is perhaps best seen in the right-hand panel of Fig.~\ref{figure:colour-colour}, where we show the phase space density. This feature appears in the Sausage relic \citep[see figure 19 of][]{stroe2013} and in the Toothbrush relic. To show this, we overlay contours indicating the phase space density observed for the B1 region of the Toothbrush relic, as given in figure 10 of \citet{rajpurohit2020}. To produce the contours, we have binned the scatter points provided and smoothed the resulting 2D histogram with a Gaussian kernel. It can be seen that the trajectory of the observed data also flattens at approximately the same point. To help explain the origin of the feature, we present a schematic in Fig.~\ref{figure:colour-colour-schematic}.

Here, the distribution of points in the colour-colour plane is represented by the two dashed, black lines in the left-hand panel. These lines represent the approximate boundaries of the distribution of spectral indices resulting from the Mach number distribution. The trajectory is initially steep, but constant, as lines-of-sight intersect homogeneous populations of CR electrons, which have all cooled for the same amount of time. This scenario is shown in the top right panel. As time increases, the ``break'' frequency moves leftwards, as shown by the dashed grey line in the same panel. The spectrum thereby becomes increasingly steep when measured at the same frequencies. However, the spectral curvature, which is the relation of the logarithmic slope of the radio surface brightness measured between $\nu_3$ and $\nu_1$, and $\nu_2$ and $\nu_1$, stays the same. If this evolution continued, it would result in the continuation of the straight trajectory in the left-hand panel, shown by the light grey, dashed lines.

\begin{figure*}
    \centering
    \includegraphics[width=.7\columnwidth]{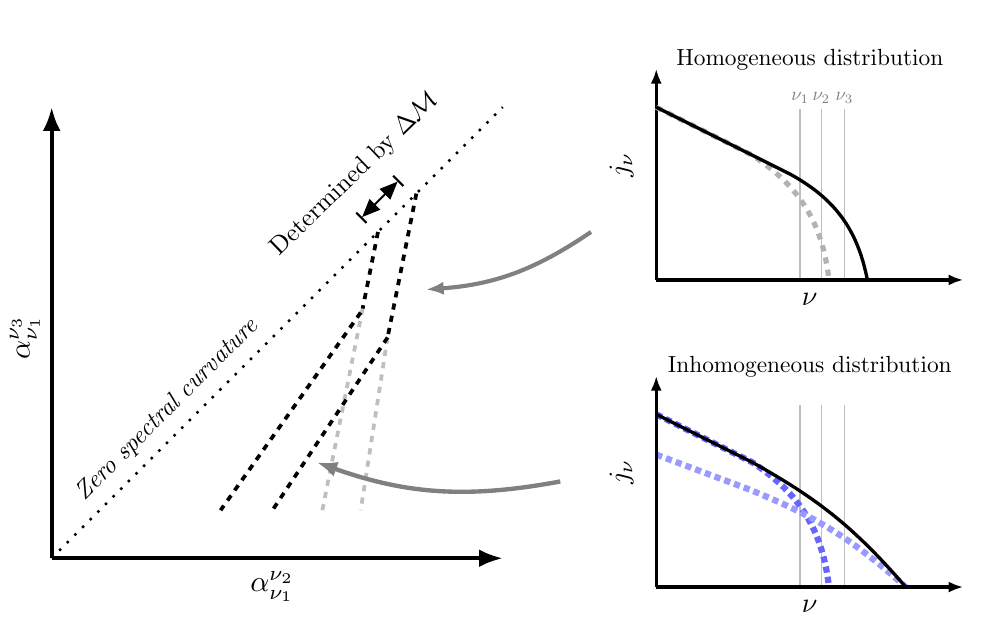}
    \caption[Schematic showing why the trajectory in the colour-colour plane flattens]{Schematic showing why the trajectory in the colour-colour plane flattens, as shown by the break in the dashed black lines in the left-hand panel, rather than following the standard JP model (dashed grey lines). Lines-of-sight initially intersect homogeneous CR electron populations (top right panel), with the same steep spectral curvature. Due to turbulence, later lines-of-sight intersect inhomogeneous populations (bottom right panel), made up of superpositions of cooled and fresher spectra (dashed purple and lilac lines, respectively). The resultant weaker spectral curvature correspondingly flattens the trajectory in the colour-colour plane. The effect is exaggerated here for explanatory purposes.}
    \label{figure:colour-colour-schematic}
\end{figure*}

Instead, however, the trajectory flattens. This is due to the turbulence in our simulations, which results in lines-of-sight intersecting inhomogeneous populations that have cooled for different amounts of time. The result of this scenario is shown in the bottom right panel. Lines-of-sight are now contributed to by both cooled spectra (purple line) and by more freshly-injected spectra (lilac line). This weakens the curvature of the integrated spectra, reducing the difference between $\alpha^{\nu_3}_{\nu_1}$ and $\alpha^{\nu_2}_{\nu_1}$ and hence flattening the trajectory in the colour-colour plane as well. This effect can be seen by comparing the binned spectra in Fig.~\ref{figure:spectra-binned-by-tinj} and Fig.~\ref{figure:spectra-binned-by-distance}, which bin by age and by distance respectively. It is clear that binning by distance produces weaker spectral curvature, as we no longer bin over homogeneously cooled populations. 

Under this interpretation, the trajectory will flatten when the intensity of the more freshly injected spectra contributes significantly to the overall intensity of the pixel. Whilst this effect begins in Fig.~\ref{figure:colour-colour} at approximately the same point for both the simulated and the observed data, we note that the overall trajectory of the observed data at $\alpha^{1.5\,\rmn{GHz}}_{150\,\rmn{MHz}} < -1.3$ is even flatter than that seen in our simulation. There are multiple possibilities to explain this, including smoothing and projection effects \citep{rajpurohit2020}, or potentially still missing physics, such as Mach number dependent electron acceleration efficiencies, insufficient mixing due to post-shock turbulence not being strong enough, pitch-angle dependent cooling in combination with an incomplete pitch-angle isotropisation process, or turbulent re-acceleration \citep{fujita2015, kang2024}. We defer a more detailed analysis of the differences to future work, where we may compare with a more like-for-like simulation.

In the right-hand panel of Fig.~\ref{figure:colour-colour}, it can be seen that both the observed and simulated data produce trajectories with similar widths. This remains fairly constant over the length of the trajectory, and will initially be set by the range of Mach numbers at the shock-front, as these produce different spectral indices (see Eq.~\ref{eq:mach-no-slope}). Points lying close to the dashed, grey line, which marks zero spectral curvature, have experienced minimal cooling, and should therefore be able to constrain the overall distribution. We will investigate this further in Whittingham et al. (in prep.).

Whilst the general width of the distribution is fairly constant, an increasingly large amount of scatter can be seen in the left-hand panel of Fig.~\ref{figure:colour-colour}, predominantly at high distances from the shock (shown in blue colours). At such distances, fewer tracers contribute to individual pixels, which leads to volume-averaged spectra that exhibit larger noise due to incomplete sampling. This occasionally leads to negative curvature, with such points lying above the dashed diagonal line. For our particular plot, this can happen when a spectrum with a steeper slope dominates the range from 150 MHz to 650 MHZ, with a shallower slope dominating at 1.5 GHz. We expect this effect to be mainly caused by the low number of tracers contributing to such pixels, and therefore do not expect such strong scatter in observations. The effect is also much reduced when we show the phase space density, as provided in the right-hand panel of Fig.~\ref{figure:colour-colour}.

\section{Discussion}
\label{chapter5-sec:discussion}

\subsection{What affects the formation of the Rayleigh-Taylor instability?}
\label{chapter5-subsection:impact-on-RTI}

The Rayleigh-Taylor instability has played a large role in the results shown in this paper. Specifically, in Sec.~\ref{chapter5-sec:magnetic-field} we showed that it plays a fundamental role in amplifying the magnetic field through compression and turbulence, whilst in Sec.~\ref{chapter5-sec:synchrotron-emission} we showed that it breaks the commonly used assumption that $d = v_\rmn{post} t$, thereby helping to explain observed trends in spectral features. Accordingly, it is important to understand under what circumstances the instability forms. 

As explained in Sec.~\ref{chapter5-sec:RT}, in order to generate a Rayleigh-Taylor instability, we require the gradients of density and pressure to be misaligned with one another. Specifically, they must be misaligned such that the induced vorticity amplifies the perturbation. In our case, this means that the density gradient must point across the contact discontinuity towards the shock-compressed region. Furthermore, the amount of vorticity generated is proportional to the size of the gradients. Indeed, the growth rate of the instability follows:
\begin{equation}
    \sigma = \sqrt{|\mathbf{a}| \left( \frac{\rho_2 - \rho_3}{\rho_2 + \rho_3} \right) k},
    \label{eq:RTI-growth-rate}
\end{equation}
where $\mathbf{a} \propto -\bnabla P$ is the acceleration of the lighter material into the denser material, the density terms in brackets constitute the Atwood number, the subscripts ``2'' and ``3'' indicate post-shock gas and gas to the left of the contact discontinuity, respectively, and $k$ is the wave number of the perturbation \citep{chandrasekhar1961}. In the scenarios presented in this paper, the most influential variable in this equation is the density, where $\rho_2$ represents the density in the shock-compressed region, and $\rho_1$ is the density of the downstream gas on the other side of the contact discontinuity. 

This difference between these variables is amplified in our simulations due to the rarefaction wave produced (see Fig.~\ref{figure:1D-shock-tube-profiles}). This wave exists in cosmological simulations as well, as can be seen in the \href{https://youtu.be/Ka-Odrwwamo}{movie} of Fig.~\ref{figure:cosmological}, but is somewhat masked by the overall radial density gradient found in the cluster. We would therefore expect a slightly slower growth of the instability in the cluster; for example, if all other variables remained the same, but $\rho_1$ increased to $1\times 10^{-28}$~g~cm$^{-3}$, we would expect the growth rate to be halved.

Slower shocks will also reduce the growth rate, as this affects $a$ in Eq.~\eqref{eq:RTI-growth-rate}. These can be generated by increasing the upstream average density or weakening the Mach number. Indeed, we have predominantly shown results from our Mach 3 simulations throughout this paper, as the growth rate here is 1.5 times faster than in the corresponding Mach 2 simulation. The turbulence is subsequently better developed in these simulations at the end of the 250 Myr runtime. Nonetheless, as we showed in Fig.~\ref{figure:projected-B-histograms}, even the Mach 2 simulations are able to produce sufficient turbulence to generate $\upmu$G strength magnetic fields. 

The instability will be totally suppressed when $\rho_1 > \rho_2$; i.e. when the density achieved due to shock compression is lower than the density downstream. As can be seen in the \href{https://youtu.be/Ka-Odrwwamo}{movie} of Fig.~\ref{figure:cosmological}, this is the case for the propagating merger shock wave during the initial stages of the cluster merger, before it reaches an accretion shock. Along with geometric and shock-dissipated energy arguments \citep{vazza2012}, as well as the density argument given above, this may help explain why radio relics are so rarely observed close to the cluster centre: the mechanism presented in this paper requires conditions only available at the cluster outskirts.

\subsection{How does our method for generating turbulence impact the results?}

Many authors have used shock-tubes to probe resolution-dependent physics. In particular, they have frequently been used to investigate how upstream turbulence affects downstream magnetic field amplification \citep[see, e.g.,][]{ji2016, hu2022, hew2023}. Few papers, however, have focused on the gas and shock properties relevant to radio relics, with the notable exception of \citet{dominguez-fernandez2021}. In contrast to our own work, the authors of this paper used the Ornstein–Uhlenbeck method to generate upstream fluctuations. This process creates density, velocity, and magnetic turbulence simultaneously \citep[see][and references therein]{federrath2010}. The turbulence then decays over time as the shock progresses through it.

The key advantage of the Ornstein–Uhlenbeck method is that turbulent properties are self-consistently modelled. Hence fluctuations are correlated, unlike in the simulations we present. Moreover, the magnetic field will display intermittency, which is not true of the Gaussian-random fields used in our initial conditions. Such intermittency is produced naturally by small-scale dynamos \citep{sur2021}, which are expected to generate the magnetic field in the ICM \citep{tevlin2024}.  Although density turbulence dominates the effects shown in this paper, magnetic fluctuations could be influential for the striations analysed in Sec.~\ref{chapter5-sec:synchrotron-emission}. We will investigate this effect in a future study.

As already stated, in our own simulations, there is no initial velocity turbulence (see Sec.~\ref{chapter5-subsec:generating_turb}). Such turbulence should exist in the ICM, however, which will also result in Mach number variations. The velocity fluctuations would need to be a substantial fraction of the shock speed, however, before they could outweigh the effect of the density fluctuations shown in this paper. We leave a precise parameter study of this effect to future work as well. 

\section{Conclusions}
\label{chapter5-sec:conclusions}

A standard theory of radio relics has emerged in the past two decades. This states that radio relics are formed by the Mpc-size shocks generated during cluster mergers and that such shocks \mbox{(re-)accelerate} electrons, enabling them to emit in the radio band. This framework has been largely successful in explaining observational data, however, several significant challenges remain. This study has focused on three of those problems:
\begin{enumerate}[label=\roman*)]
    \item What is the origin of the $\mathcal{M}_\rmn{radio} - \mathcal{M}_\rmn{X-ray}$ discrepancy if both measurement methods trace the same physical shock?
    \item How do magnetic fields reach $\upmu$G strength in radio relics (as inferred from cooling length arguments) if the surrounding ICM has field strengths an order of magnitude smaller?
    \item Why do spectral variations fail to match standard cooling models in colour-colour diagrams?
\end{enumerate}

To answer these problems, we have taken a hybrid approach, first identifying typical conditions in cosmological zoom-in simulations of cluster mergers before applying our findings to significantly higher-resolution idealised shock-tubes. In this last step, we use the CR electron spectral code \textsc{Crest} and the emission code \textsc{Crayon+}, thereby producing synchrotron data \textit{ab initio}.

From our cosmological simulations, we have identified that the merger shock typically meets an accretion shock at distances commonly recorded for radio relics. This leads to the production of a narrow, shock-compressed density sheet (Fig.~\ref{figure:cosmological}). This scenario can be modelled as a shock-tube problem (see Sec.~\ref{chapter5-subsec:shock-tube-setup}). We additionally include upstream density turbulence in our shock-tubes, which is not well-resolved at the outskirts of our zoom-in simulations, but evident in many other previous studies and observations. We choose a log-normal distribution and set the variance of the distribution according to results given in \citet{zhuravleva2013}. 

We find that density fluctuations are directly responsible for the following effects:
\begin{enumerate}[label=\roman*)]
    \item A Mach number distribution forming at the shock front (Figs.~\ref{figure:shock-tube} and~\ref{figure:mach-no-pdf}),
    \item A Rayleigh-Taylor instability forming at the contact discontinuity (Figs.~\ref{figure:shock-tube} and~\ref{figure:RTI-schematic}), and
    \item Shock corrugation (Fig.~\ref{figure:shock-tube}).
\end{enumerate}
We show how these features form in Sec.~\ref{chapter5-sec:RT}. These effects provide answers to our three main questions:

\begin{itemize}

\item \bfit{X-ray vs.\ radio discrepancy:} In Sec.~\ref{chapter5-sec:spectra}, we show that the tail of the Mach number distribution dominates the higher frequencies. In particular, it flattens the integrated injected spectra and results in the slope at radio emitting frequencies being shallower than the theoretical value of $\alpha_e - 1$ (Fig.~\ref{figure:spectra-binned-by-tinj}). Both of these effects lead to an over-estimation of the Mach number with respect to the actual peak of the distribution (Table~\ref{tab:inferred_mach_no}). The effect is particularly strong in weak shocks ($\mathcal{M} \lesssim 2$). 

\item \bfit{Magnetic field amplification:} The Rayleigh-Taylor instability causes higher levels of amplification than would be expected due to the standard theory of shock-compression alone. This boosts the magnetic field strength from ICM-like conditions up to $\upmu$G levels, although the volume-average remains significantly below this (Figs.~\ref{figure:slice-downstream} and~\ref{figure:plasma-beta-slice}). Amplification is predominantly driven by the additional adiabatic compression that results from the instability. However, there is also evidence for the existence of a small-scale dynamo in our simulations. Indeed, the peak magnetic field values cannot be reached without it (Fig.~\ref{figure:rho-B-phase-diagram}).

Additionally, we find that the projected synchrotron emission is dominated by the strongest magnetic field values encountered along the line-of-sight. This means that estimations of the magnetic field made from emission are heavily biased by these regions; values inferred from cooling length arguments are only representative of the tail of the distribution, not the volume-average (Figs.~\ref{figure:t_inj-and-B-field-projections} and~\ref{figure:projected-B-histograms}).

\item \bfit{Cooling models:} The turbulence generated by the Rayleigh-Taylor instability breaks the laminar flow assumption. This is especially true on an electron-by-electron basis, where distance from the shock front is a poor indicator of time since injection (Figs.~\ref{figure:spectra-binned-by-tinj} and~\ref{figure:slice-downstream}). The relationship is somewhat recovered when projecting the synchrotron-weighted time since injection, but even here $d = v_\rmn{post} t$ no longer holds true (Fig.~\ref{figure:t_inj-and-B-field-projections}. Turbulence leads to the mixing of spectra along the line-of-sight, which acts to flatten the spectral curvature and explains why standard cooling models do not work for radio relics (Figs.~\ref{figure:colour-colour} and~\ref{figure:colour-colour-schematic}).

\end{itemize}

In addition, we find that features in surface brightness and spectral index maps can be attributed to specific mechanisms (Fig.\ref{figure:intensity-and-spectral-index}). Specifically, the corrugation of the shock front leads to filamentary emission in projection. Mach number fluctuations, on the other hand, cause fluctuations in intensity along the shock front, which spatially correlate with the injection scale. The Rayleigh-Taylor instability can produce spectral index fluctuations towards the front of the shock. Finally, magnetic fluctuations, either advected downstream or compressed at the contact discontinuity, produce striated emission aligned with the shock normal. In the next paper in the series, we will show how these effects can be used to determine ICM conditions and address the critical Mach number question (Whittingham et al., in prep.).

\newpage
\section{Appendix}
\subsection[Appendix A: Numerical stability]{Numerical stability}
\label{appendix:numerical-stability}

\begin{figure}
    \centering
    \begin{minipage}[t][][t]{.48\textwidth}    
        \includegraphics[width=1.05\columnwidth]{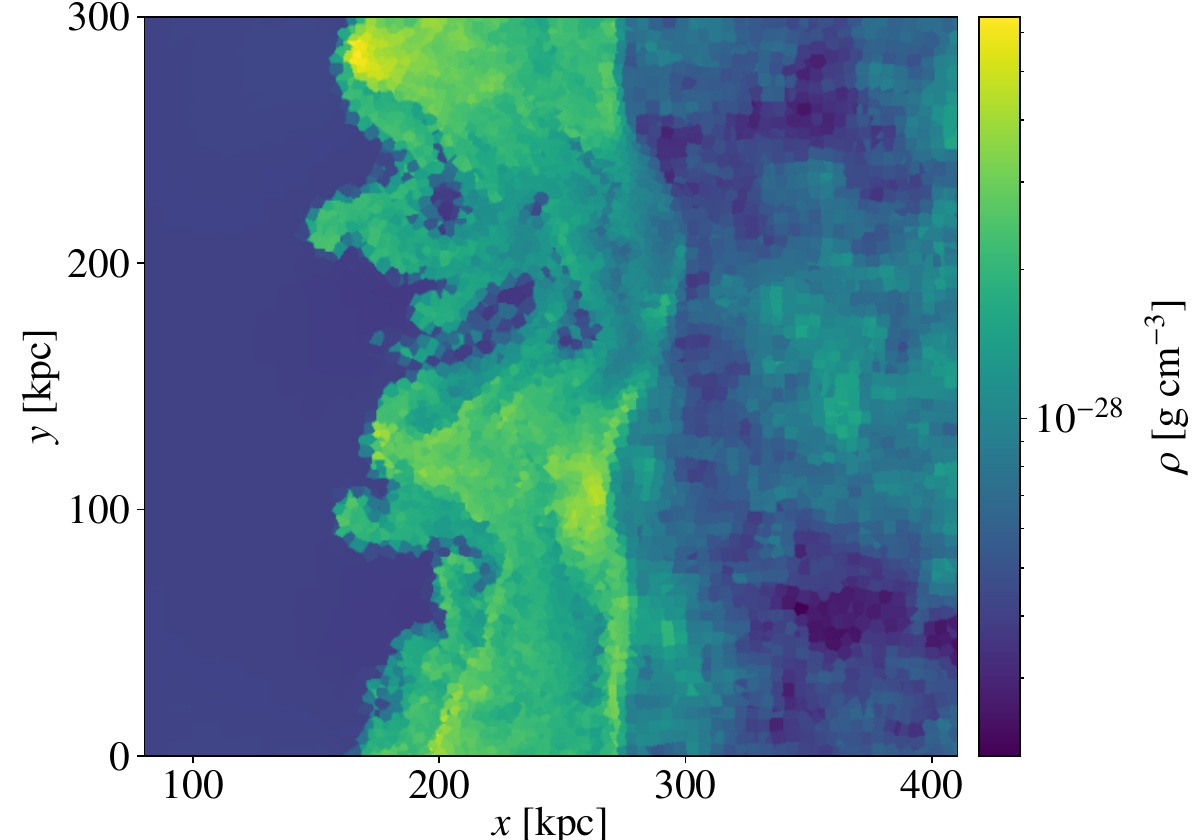}
        \caption[Slice through the low-resolution variation of the fiducial simulation showing gas density]{Version of our fiducial simulation, run with 8 times lower mass resolution. colours indicate density, with the same limits and bounds as shown in Fig.~\ref{figure:shock-tube}. Higher resolution only leads to smaller-scale structure, indicating the numerical stability of our simulations.}
        \label{figure:low-res-density-slice}
    \end{minipage}
    \hfill
    \begin{minipage}[t][][t]{.48\textwidth}
        \centering
        \includegraphics[width=1.05\columnwidth]{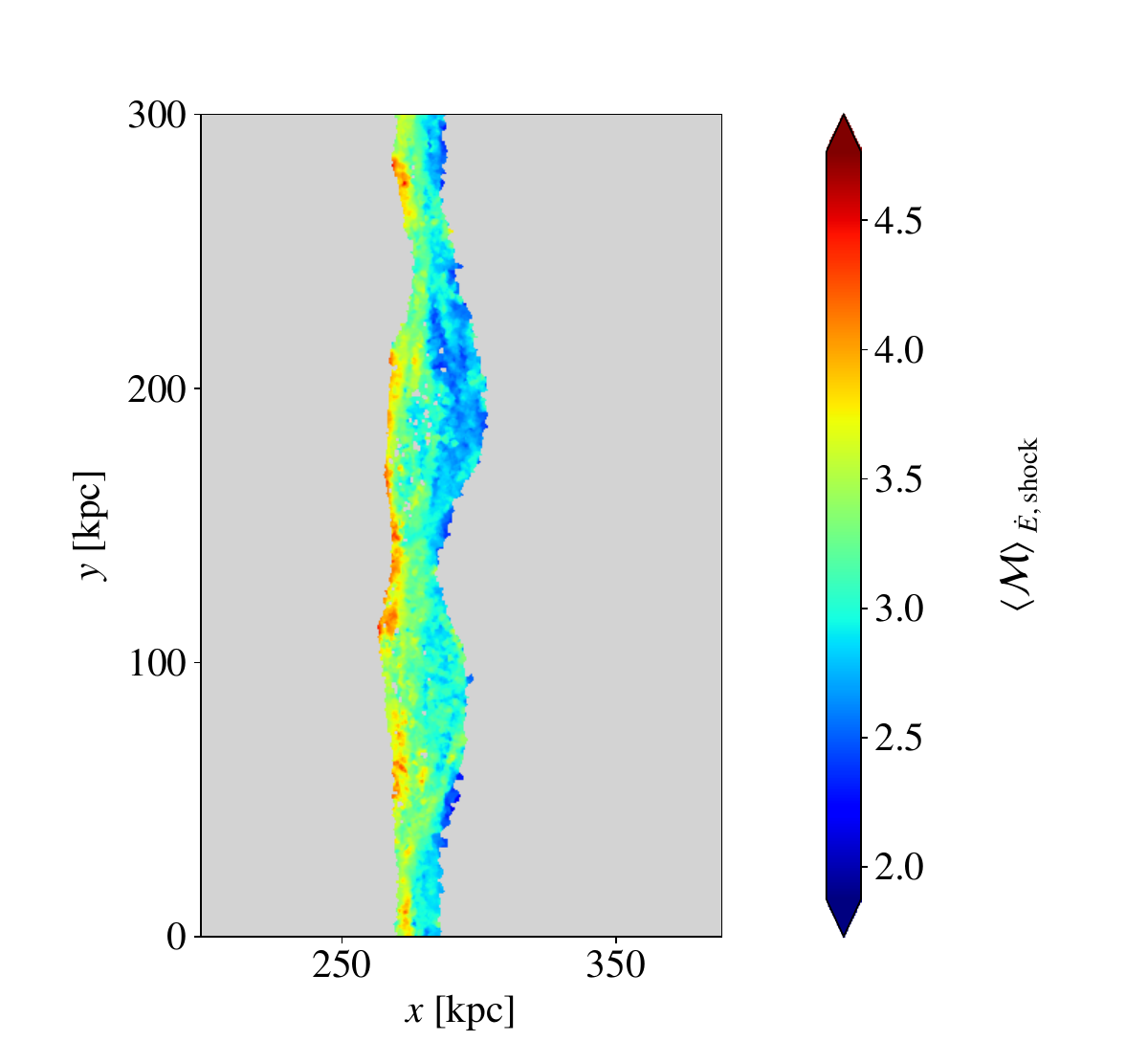}
        \caption[Projected dissipation-weighted Mach number]{Projected dissipation-weighted Mach number, shown for the same simulation and time as in Fig.~\ref{figure:shock-tube}. The projection depth is 300 kpc. Weaker Mach numbers tend to form where the shock advances ahead of the median position, whilst stronger ones form where it lags behind.}
        \label{figure:mach-number-projection}
    \end{minipage}
\end{figure}

In order to trust our results, it is critical that the simulations are numerically converged. For this purpose, we re-run our fiducial simulation, \textit{Turb}, using a target gas mass of $m_\rmn{gas} \approx 1.2 \times 10^5 \,\rmn{M}_\odot$. This is eight times lower in mass resolution than the fiducial run or, equivalently, two times lower in spatial resolution. We find that all of our results continue to hold, although the magnetic field amplification is slightly weaker, as already discussed in Sec.~\ref{chapter5-subsec:mag-field-origin}.

As evidence for convergence, we present a slice through the simulation at $t=180$ Myr in Fig.~\ref{figure:low-res-density-slice}, with colours indicating density. This may be directly compared with panel \textit{ii)} in Fig.~\ref{figure:shock-tube}. It can be seen that all structures continue to be evident, except with lower resolution. We therefore consider this to be good evidence that our simulations are indeed converged.
 
\subsection[Appendix B: Mach numbers in projection]{Mach numbers in projection}
\label{appendix:mach-in-projection}

In Sec~\ref{chapter5-subsec:mach-dist}, we argued that lower Mach numbers should be found typically in a more advanced position, and that stronger Mach numbers should be found towards the back of the shock. In panel \textit{iii)} of Fig.~\ref{figure:shock-tube}, specifically, we analysed a thin projection and found the result to be consistent. In Fig.~\ref{figure:mach-number-projection}, we show the projected dissipated-energy weighted Mach number in our Mach 3 \textit{Turb} simulation at $t=180$ Myr. This is analogous to the previous projection, except the projection depth has increased from 35 kpc to 300 kpc (i.e., the box size).

It can be seen that our argument continues to hold.
We have already shown in Sec.~\ref{chapter5-subsec:mach-dist} that the properties of the Mach distribution remain relatively stable over time. With this said, however, we must also emphasise that our argument only holds true in a statistical sense; Mach numbers do not decrease monotonically in value towards the most advanced part of the shock. This can also be seen in Fig.~\ref{figure:mach-number-projection}.

\subsection[Appendix C: The lack of a Rayleigh-Taylor instability in the Flat run]{The lack of a Rayleigh-Taylor instability in the \textit{Flat} run}
\label{appendix:RTI-without-density-perturbations}
\begin{figure*}
    \includegraphics[width=1.0\columnwidth]{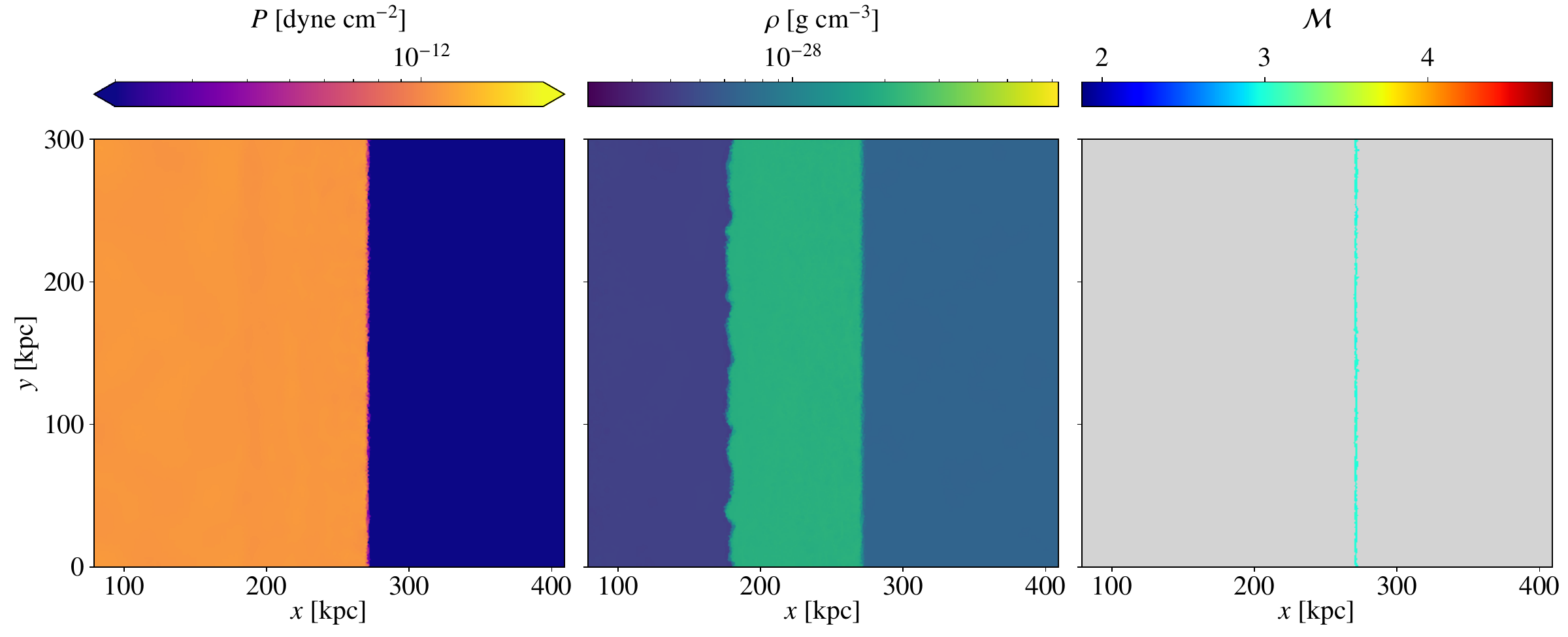}
    \caption[As the top row of Fig.~\ref{figure:shock-tube} with data from the \textit{Flat} simulation.]{As Fig.~\ref{figure:shock-tube}, but only the top row is shown, and data is taken from the \textit{Flat} simulation. Without upstream density fluctuations, the Rayleigh-Taylor instability is restricted to a very low growth rate.}
    \label{figure:shock-tube-flat}
\end{figure*}

In Sec.~\ref{chapter5-sec:RT}, we claimed that density perturbations were necessary for the formation of the Rayleigh-Taylor instability in our simulations. This instability forms a critical part of our mechanism for solving the magnetic field and cooling model problems outlined in Sec.~\ref{chapter5-sec:intro}. To show that perturbations are necessary, we show slices through our \textit{Flat} shock-tube simulation in Fig.~\ref{figure:shock-tube-flat}. This simulation has no density turbulence in the upstream initial conditions, but includes magnetic turbulence (see Sec.~\ref{chapter5-subsec:sim-vars}). The panels shown are analogous to the panels in the top row of Fig.~\ref{figure:shock-tube} and are shown at the same time (i.e. $t = 180$ Myr).

It can be seen that without upstream density turbulence, the shock front shows no curvature. Magnetic pressure fluctuations, however, lead to mild variations in density and pressure gradients at the contact discontinuity. Consequently, the Rayleigh-Taylor instability still takes place here; indeed, this forms the bumps seen in the middle panel at the trailing edge of the shock-compressed zone. The growth-rate of this instability has been severely restricted, however. This is partially due to the significantly smaller pressure gradient, and partially due to the lack of corrugation at the contact discontinuity, which helped to seed the Rayleigh-Taylor instability in Fig.~\ref{figure:shock-tube}.

\subsection[Appendix D: Plasma beta distribution]{Plasma beta distribution}
\label{appendix:beta}

In Fig.~\ref{figure:plasma-beta-hist} we quantify our earlier statement in Sec.~\ref{chapter5-sec:magnetic-field-strength} that despite significant amplification, the magnetic field does not play a major dynamical role. To do this, we show histograms of the plasma beta values in the initial upstream region (region III, in the nomenclature given in Sec.~\ref{chapter5-subsec:shock-tube-setup-resolution}) at $t=0$ Myr and in the shock-compressed region at $t=250$ Myr. As discussed in Sec.~\ref{chapter5-subsec:generating_turb}, we implement density and magnetic field turbulence in our simulations independently of one another. This leads to a wide initial distribution with a strong tail towards higher plasma beta values. The RMS magnetic field strength is picked, however, to produce a peak at $P_\rmn{th} / P_B = 100$, as can be seen in the figure.

As the region undergoes shock compression, the width of the distribution increases. This happens as regions undergo shock compression but do not necessarily increase in magnetic field strength, increasing the high-end of the distribution. On the other hand, some magnetic field amplification takes place above that expected for compression alone (see Fig.~\ref{figure:rho-B-phase-diagram}), which leads to a widening of the distribution at low plasma beta values. A minority of cells reach values $P_\rmn{th} / P_B < 10$, but no cell reaches below $P_\rmn{th} / P_B = 4$. We may therefore safely treat the downstream region as being in the kinetic regime, meaning that hydrodynamic models of the downstream remain valid.

\begin{figure}
    \centering
    \includegraphics[width=0.5\columnwidth]{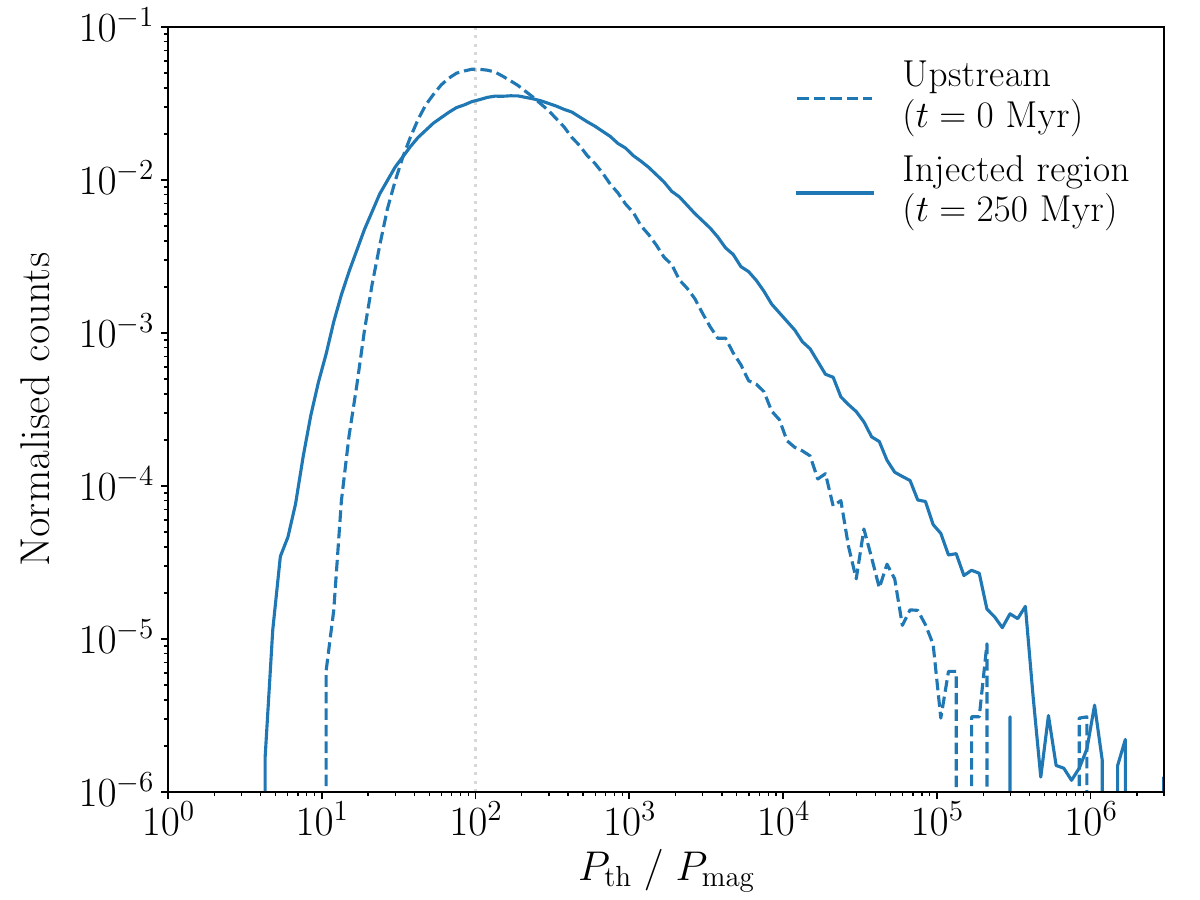}
    \caption[Histograms of the plasma beta values in the initial upstream region and the shock-injected region at the end of the simulation]{Histograms of the plasma beta values for all gas cells upstream at $t=0$ Myr (dashed) and in the shock-injected region at $t=250$ Myr (solid) for our Mach 3 fiducial simulation, weighted by the cell volume. The distribution initially peaks at 100 (marked by a dotted, grey line), but shifts towards higher values as cells experience increased gas pressure behind the shock front. Amplification downstream is, however, able to increase the magnetic field pressure up to $\sim$10\% of the gas pressure in a fraction of the cells.}
    \label{figure:plasma-beta-hist}
\end{figure}

\subsection[Appendix E: Spectra binned by distance]{Spectra binned by distance}
\label{appendix:spectra_by_dist}

\begin{figure*}
    \includegraphics[width=1.0\columnwidth]{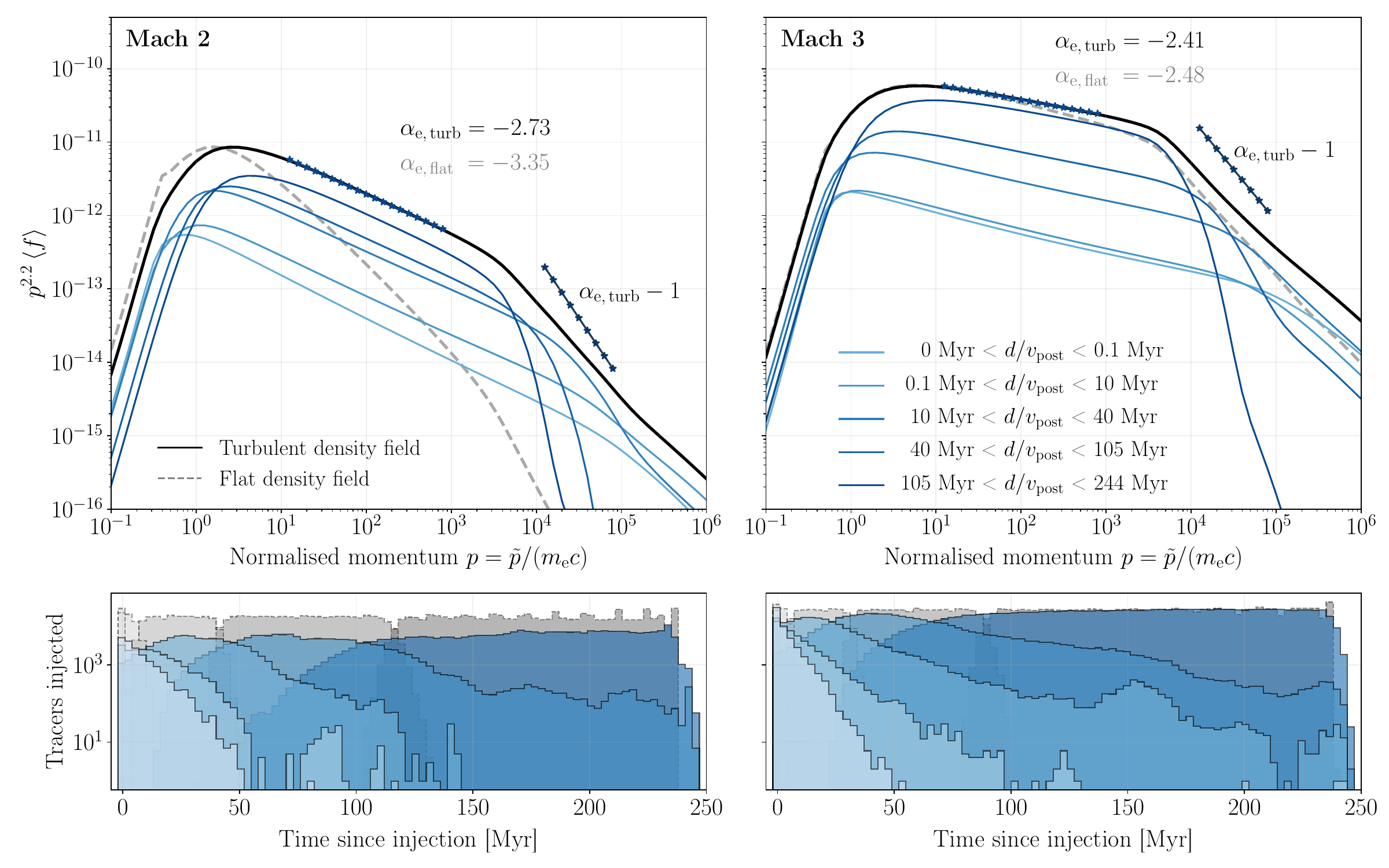}
    \caption[Non-thermal cosmic ray electron spectra for each simulation, with spectra overlaid binned by distance to the shock front. Histograms of the positions of tracers relative to the shock-front based on these bins]{As Fig.~\ref{figure:spectra-binned-by-tinj}, except tracers are now binned by distance from the median shock position, rather than time since injection. Distance bins are equal to the time bins in Fig.~\ref{figure:spectra-binned-by-tinj} multiplied by the relevant post-shock velocity (approximately 440 and 500 km/s for Mach 2 and Mach 3 shocks, respectively). Contributions binned by distance do not match those binned by time since injection due to the highly non-laminar flow. This can also be seen in the strongly overlapping distributions of time since injection within each distance bin (\textit{bottom row}).}
    \label{figure:spectra-binned-by-distance}
\end{figure*}

In Sec.~\ref{chapter5-sec:laminar-flow} we presented evidence that distance from the shock is not a reliable proxy for time since injection, at least on a tracer-by-tracer basis (see caveats given in Sec.~\ref{chapter5-sec:impact-on-obs}). We showed, in particular, that the volume-averaged spectrum is created through the layering of tracers binned by age, and that tracers in these bins overlap significantly when their spatial distributions are overplotted (see Fig.~\ref{figure:spectra-binned-by-tinj}). In Fig.~\ref{figure:spectra-binned-by-distance}, we plot the inverse; that is, we bin tracers by their distance and make histograms of the temporal distributions. To remove some degree of arbitrariness, we use the same time bins as previously, but now multiply out by the theoretical post-shock speed $\bupsilon_\rmn{post}$. We have calculated this by taking the actual shock speeds of approximately $1,000$ km~s$^{-1}$ and $1,500$ km~s$^{-1}$ in the Mach 2 and Mach 3 simulations, respectively, and applying the standard jump conditions, giving $\bupsilon_\rmn{post} \approx 440$ km~s$^{-1}$ and $\bupsilon_\rmn{post} \approx 500$ km~s$^{-1}$. This equates to distance intervals of approximately 0, 0.04, 4, 18, 46, and 107 kpc in the Mach 2 simulations, and intervals 1.14 times higher in the Mach 3 simulations. 

The layering in Fig.~\ref{figure:spectra-binned-by-distance} is clearly less well-defined than in Fig.~\ref{figure:spectra-binned-by-tinj}. Nonetheless, it still remains true that CR electrons further from the shock are, on average, more cooled. Consequently, it is still possible to pick distance intervals that dominate a particular momentum range. This is especially true at later ages, when the cooled part of the spectra is steeper. This is despite the fact that the tracers are strongly mixed, as can be seen by inspecting the bottom row of Fig.~\ref{figure:spectra-binned-by-distance}. 

The layering of spectra can be partially improved by removing the assumption that the distance bins in Fig.~\ref{figure:spectra-binned-by-distance} must be proportional to the time bins given in Fig.~\ref{figure:spectra-binned-by-tinj}. However, in order to see an improvement, the spacing of bins must be picked independently for Mach 2 and Mach 3 simulations. Moreover, we find that, regardless of binning, it is not possible to produce a figure entirely analogous to Fig.~\ref{figure:spectra-binned-by-tinj}. This implies that the standard formula of $d = v_\rmn{post} t$ cannot be adapted in a linear fashion, and that laminar flow is only a good assumption for simulations without density turbulence.
\setcounter{equation}{0}
\thispagestyle{empty}

\graphicspath{{Images/Chapter6/}}

\defcitealias{whittingham2024}{W24}
\defcitealias{dominguez-fernandez2021}{Dom\'{i}nguez-Fern\'{a}ndez et al., 2021a}
\defcitealias{wittor2021}{Wittor et al., 2021a}

\Chapter{Zooming-in on radio relics -- II.}{How relic morphology probes intracluster turbulence and plasma processes}

\label{chapter:paper-four}

\textit{Comparatively little is known about gas properties at the outer reaches of the intracluster medium (ICM) owing to the low surface brightness at these distances. Radio relics, which take place during cluster mergers, may, however, provide an effective probe. In this paper, we expand upon the shock-tube simulations first presented in Whittingham et al., 2024, varying the upstream density turbulence to understand its impact on the downstream emission. Specifically, we vary the amplitude of the fluctuations, the power law slope, and the injection scale of the turbulence. We find that radio relics are especially sensitive to the relative variance of the fluctuations. Indeed, this is the most influential factor for setting: i) the extent of the emission, ii) the Mach number distribution, ii) the level of shock corrugation, iii) the level of downstream velocity turbulence, iv) the electron spectral slope, v) the peak emission, and vi) the magnetic field strength inferred from synchrotron emission. The power law slope and the injection scale of the turbulence can impact these properties as well, however, albeit typically to a much lesser extent. Intriguingly, we find that the turbulent injection scale sets the spacing of ``threads'' and ``knots'' at the shock front. In observed radio relics, this spacing suggests an injection scale of $500-600$ kpc, which is significantly higher than the injection scale typically assumed. Finally, we use our results to investigate the impact of setting a critical Mach number, $\mathcal{M}_\text{crit}$, above which injection takes place. We find that $\mathcal{M}_\text{crit} = 2.3$, as suggested in literature, appears to be compatible with observations, even in shocks that are nominally weaker. This is because injection from the tail of the Mach number distribution dominates the projected radio emission.}

\section{Introduction}

Gas properties in the intracluster medium (ICM) are generally probed using X-ray surface brightness and spectroscopy observations \citep{mohr1999, fabian2006, churazov2012, sanders2020}, the Sunyaev-Zel'dovich effect \citep[SZE;][]{sunyaev1980, birkinshaw1999, planck2013, bender2016}, and indirectly through weak lensing measurements, which can be used to normalise X-ray observations \citep{okabe2014}. 
However, these techniques are generally most effective close to the cluster core. For example, X-ray surface brightness scales as $S_\rmn{X} \propto n_\rmn{e}^2 T_\rmn{e}^{1/2}$ \citep{gronenschild1978}, where $n_\rmn{e}$ is the electron number density and $T_\rmn{e}$ is the electron temperature, and hence this drops precipitously at the cluster outskirts. The SZE is less affected, as it is sensitive to the integrated pressure, with the Compton $y$-parameter scaling as $y \propto \int n_\rmn{e} T_\rmn{e}\mathrm{d}s$, where $s$ is the path length along the line of sight. However, the intensity of the cosmic infrared background (CIB) can cause significant spatial confusion when $y\approx10^{-6}$, as is typical at the periphery of galaxy clusters \citep[see discussion in][]{walker2019}. To maximise the signal-to-noise ratio therefore, observations using these methods are often stacked \citep[see, e.g.][]{planck2013, walker2013}, but this necessarily averages over the data. Galaxy clusters are host to a range of radio phenomena, however, which may also provide a window into the ICM conditions. In particular, radio relics may be able to probe the ICM conditions at distances of approximately $1-2$ Mpc from the cluster centre with substantially higher signal-to-noise than the above techniques and angular resolution on the order of arcseconds\footnote{\citet{deGasperin2022}, for example, are able to resolve structures down to 3 kpc at $z=0.05$.} at frequencies $\nu \approx 1.4$~GHz \citep[see, e.g.][]{rajpurohit2020, deGasperin2022}.

Radio relics are extended ($\sim1$ Mpc) sources of emission often found at the outskirts of merging clusters. They have typical surface brightnesses on the order of $\upmu$Jy arcsec$^{-2}$ at 1.4 GHz, and steep spectra with spectral indices of $\alpha < -1$ (where $S_\nu \propto \nu^\alpha$, with flux density $S_\nu$). Emission is also usually highly fractionally polarised, with values over 20\% being common \citep[see examples in][]{wittor2019}. As a result of this polarisation and the power law emission in the radio band, radio relics are generally believed to be a result of synchrotron emission from (re-)accelerated electrons. In particular, they are believed to result from diffusive shock acceleration \citep[DSA; ][]{krymskii1977, axford1977, bell1978, bell1978b, blandford1978} taking place during galaxy cluster merger shocks (see summary of developments given in \citealt{whittingham2024} -- hereafter, \citetalias{whittingham2024} -- and the review by \citealt{brunetti2014}).
Surveys suggest that approximately 10\% of all clusters host at least one radio relic \citep{jones2023} and over 60 have been observed in total \citep{wittor2021}. These come in a wide range of shapes and sizes, with arc-shaped relics being relatively common, observed both as single \citep[see, e.g.][]{bonafede2018} and double relics \citep{koribalski2024}. The variations in relic morphology are perhaps best illustrated by Abell 2256 \citep{vanweeren2012b,owen2014}, the ``Sausage'' relic \citep{kocevski2007, vanweeren2010}, and the ``Toothbrush'' relic \citep{vanweeren2012}, which have irregular filamentary, arc-like, and linear forms, respectively. 

Simulations have shown that, on the Mpc scale, shock morphology can produce many of these forms\footnote{There have also been claims that turbulence due to filamentary accretion plays a role in some examples \citep{vanweeren2013}.} \citep{hoefft2008, hoeft2011, skillman2011, nuza2017, lee2024, lee2024b}. It is likely, however, that the spatial and energy distribution of cosmic ray (CR) electrons also plays a role. For example, sloshing motions may be able to displace or shred AGN bubbles, dispersing their electron content throughout the cluster, thereby producing arc-like forms \citep{zuhone2021, botteon2024}. Moreover, mechanisms have been proposed that explain specific scenarios, such as the origin of so-called ``wrong-way round'' relics, in which the spectral gradient points towards the cluster centre, rather than away from it \citep{riseley2022, boess2023}, and the origin of the surprisingly straight shock-front evident in the Toothbrush relic \citep{brueggen2012}. 

Radio relics also exhibit a range of morphological features on smaller scales as well. In particular, we highlight the following two:
\begin{itemize}
    \item \textbf{Downstream emission length:} Radio relics are significantly wider than expected from standard shock theory \citep{jones2023}. Additionally, the discrepancy appears to increase with decreasing frequency. For example, \citet{kang2017} find that the brush of the Toothbrush relic is $\sim2\times$ wider than the standard theory result at 650 MHz, and \citet{deGasperin2020} find it is $\sim4\times$ wider at 58 MHz. Similar results were also found for the Sausage relic \citep{kang2016b} and Abell 3667 \citep{deGasperin2022}. Furthermore, the width of the downstream emission often varies across the shock front. This is particularly noticeable in the Toothbrush relic, which lends it its name. This variation has been replicated in part in simulations \citep[e.g.][]{wittor2019}, but an understanding as to what causes this effect is still lacking.

    \item \textbf{Filamentary nature:} As resolution increases, it has become apparent that radio emission in galaxy clusters is generally filamentary in nature \citep[see, e.g.][]{botteon2020b, brienza2021, knowles2022, giacintucci2022}. Radio relics are no exception to this, with several reports of filamentary structure \citep{owen2014, vanweeren2017, diGennaro2018, rajpurohit2020, deGasperin2022}. Filamentary emission in radio relics is observed in the downstream, with orientations roughly parallel to the shock direction \citep[see, e.g., the ``brush'' section of the Toothbrush relic][]{rajpurohit2020}, but it is also observed in edge-on relics at the shock-front, where there have been multiple reports of ``double'' strand features, \citep{rajpurohit2018, raja2024}. There is still considerable debate as to what causes these structure, but it is generally attributed either to magnetic filaments \citep[see, e.g.,][]{rudnick2022} or to the shock morphology. Recently, \citet{wittor2023} attempted to decompose the projected emission using Minkowski functionals and concluded that both the magnetic field and shock front are likely inherently filamentary in nature.
    
\end{itemize}
The ultimate origin of these morphological features is still unclear, but it appears to be linked to both plasma processes and the gas properties through which the shock propagates. Simulations have been run to test the impact of upstream magnetic turbulence\footnote{The authors included density turbulence as well, but focussed their analysis on the magnetic turbulence.} \citep{dominguez-fernandez2021}, Fermi I re-acceleration \citep{dominguez-fernandez2024}, adiabatic compression \citep{wittor2021b}, and the viewing angle \citep{rajpurohit2021}.

To compound matters, recent particle-in-cell (PIC) simulations, have suggested that a critical Mach number exists with $\mathcal{M}_\rmn{crit} \approx 2.3$, below which CR electron acceleration is inefficient \citepalias[see arguments summarised in][]{whittingham2024}. This implies that shocks below this value should not produce radio relics \textit{at all}. This is in apparent contradiction with observations; for example, of the 21 radio relic examples collated by \citet{wittor2021}, at least 70\% have X-ray derived Mach numbers below this threshold\footnote{This number varies slightly depending on whether temperature or pressure calculations are used, and whether error bars are accounted for.}, with approximately two thirds having $\mathcal{M}_\rmn{X-ray} < 2$. The X-ray derived Mach number can be underestimated when the viewing angle is oblique, as the projected temperature and pressure jumps decrease \citep{wittor2021}, but it should generally trace the peak of the Mach number distribution \citepalias{dominguez-fernandez2021, wittor2021, whittingham2024}. This discrepancy must therefore also be explained.

In \citetalias{whittingham2024}, it was shown that the addition of upstream density fluctuations had a major impact on radio relic properties. In particular, they led to filamentary emission being produced at the shock front and in the downstream. It was also shown that the downstream became more extended as a result of turbulence and velocity motions induced by a Rayleigh-Taylor (RT) instability at the contact discontinuity in the downstream regime. Finally, the density turbulence led to the formation of a Mach number distribution, which substantially impacted subsequent radio spectra and hence properties inferred from it as well. In the current paper, we extend the simulations first presented in \citetalias{whittingham2024}, modifying them to investigate how exactly the relic morphology is sensitive to the turbulent parameters and whether it is sensitive to the use of a critical Mach number. The paper is organised as follows: in Sec.~\ref{chapter6-sec:methodology} we recap our methodology and show which turbulent properties we change, in Sec.~\ref{chapter6-sec:analysis} we analyse the impact, including purely hydrodynamic effects (Sec.~\ref{chapter6-subsec:hydro}), impact on spectra (Sec.~\ref{chapter6-subsec:spectra}), and impact on emission (Sec.~\ref{chapter6-subsec:emission}). Finally in Sec.~\ref{chapter6-sec:conclusions}, we summarise our conclusions.

\section{Methodology}
\label{chapter6-sec:methodology}

In \citetalias{whittingham2024}, we used cosmological zoom-in simulations of major mergers in clusters to investigate how shocks evolved in the cluster outreaches. We found that merger shocks typically collide with accretion shocks in the distance range observed for radio relics \citep[see also][]{zhang2020}. This leads to the formation of a thin, shock-compressed region, which is RT unstable at the contact discontinuity. The point that the shocks collide can be modelled as a Riemann problem, in which density is kept constant, but pressure varies between the up- and downstream. The scenario can hence be modelled using an idealised shock-tube set-up, which can achieve spatial resolution significantly higher ($\sim100\times$) than that possible in cosmological simulations. We developed the initial conditions for these shock-tubes in \citetalias{whittingham2024} and extend them here. We first briefly recap the set-up, but encourage the interested reader to see section 2 of \citetalias{whittingham2024} for a more comprehensive overview.

\subsection[Arepo]{\textsc{Arepo}}

The simulations are modelled with the moving-mesh code \textsc{Arepo} \citep{springel2010, Pakmor2016I, weinberger2020}. This solves the equation for ideal magnetohydrodynamics (MHD) using a second-order finite-volume Godunov scheme \citep{pakmor2011, pakmor2013} and an HLLD Riemann solver \citep{miyoshi2005}. Mesh-generating points move approximately with the flow\footnote{Mass flux in and out of the cell prevents the code from being truly Lagrangian. This is desired, however, to, e.g., regularise the shapes of the Voronoi cells and prevent them from becoming too elongated. To recover Lagrangian trajectories, we employ the use of tracer particles (see Sec.~\ref{chapter6-subsec:crayon_and_crest}).}, whilst cells are allowed to re-fine and de-refine, so that their gas mass stays within a factor of two of a target mass. In our simulations, the target gas mass is set to $m_\rmn{gas} \approx 1.5 \times 10^4 \,\rmn{M}_\odot$, which results in sub-kpc resolution in the downstream. In this manner, resolution naturally tracks the more dynamic parts of the simulation. This reduces the overall numerical diffusion and thereby increases the numerical Reynold's number \citep[see discussion in][]{pfrommer2022}. This, in turn, increases the ability of the code to accurately model turbulent flows; indeed several studies \citep[see, e.g.][]{bauer2012, pakmor2017, whittingham2021} have shown that it can replicate the analytical relations for a turbulent cascade \citep{kolmogorov1941}.

The $\mathbf{\nabla}\bcdot\bs{B}=0$ condition is enforced with a Powell 8-wave divergence cleaner \citep{powell1999}. This 
has proven to be highly robust in dynamic flows. Indeed, \citet{pakmor2013} find that the Powell scheme is more robust than the competing scheme proposed by \citet{dedner2002}, and \citet{whittingham2021} find that it does not produce spatially- or temporally-correlated divergence errors in this regime. Moreover, the numerical diffusion produced is within the range necessary for simulating dynamo action \citep{pakmor2017, pakmor2024, whittingham2021}. In particular, it was shown in \citet{pfrommer2022} that idealised galaxy simulations using \textsc{Arepo} were able to produce magnetic curvature statistics consistent with analytical relations for a small-scale dynamo \citep{schekochihin2005}. This is important as an accurate modelling of the magnetic field is critical for our radio synchrotron results (see Sec.~\ref{chapter6-subsec:crayon_and_crest}).

We run our \textsc{Arepo} simulations with the \citet{schaal2015} shock-finder in on-the-fly mode and the CR physics module implemented by \citet{Pfrommer2017}, where CRs are treated as a fluid with adiabatic index $\gamma_\rmn{a}=4/3$. This enables us to simulate CR acceleration via DSA as well as their post-shock advection. The Mach number at a shock is calculated using the Rankine-Hugoniot jump conditions, with appropriate modifications for additional CR pressure and acceleration:
\begin{equation}
    \mathcal{M}^2 = \left(\frac{P_2}{P_1} - 1 \right) \frac{x_\rmn{s}}{\gamma_\mathrm{a, eff}(x_\rmn{s} - 1)},
    \label{eq:mach-no-chapter6}
\end{equation}
where $\gamma_\mathrm{a, eff}$ is the effective adiabatic index, and $x_\rmn{s}$ is the density jump at the shock. The dissipated energy flux is then:
\begin{equation}
    \dot{E}_\rmn{diss} = \varepsilon_\rmn{diss}A_\rmn{shock}\frac{\mathcal{M}c_\mathrm{s,1}}{x_\rmn{s}},
    \label{eq:shock-dissipated-energy2}
\end{equation}
where $\varepsilon_\rmn{diss}$ is the post-shock energy density minus the adiabatically compressed pre-shock energy density, $A_\rmn{shock}$ is the cell's shock surface, and $c_{\rmn{s},1}$ is the upstream sound speed \citep[see][for further details]{Pfrommer2017}. Shock finding is limited to $\mathcal{M} \geq 1.3$, and 10\% of the dissipated energy is used to accelerate the CR protons. As the diffusion coefficient in the vicinity of shocks is believed to approach Bohm diffusion \citep{caprioli2014, kang2016}, we do not account for streaming or diffusion. 
CRs are dynamically unimportant in our simulations, but are coupled to our CR electron implementation.

\subsection[Crest and Crayon+]{\textsc{Crest} and \textsc{Crayon+}}
\label{chapter6-subsec:crayon_and_crest}

CR electrons are modelled spectrally using \textsc{Crest} \citep{winner2019}. This is a post-processing code that uses the tracer functionality in \textsc{Arepo} \citep{genel2013} to store necessary variables on the MHD timestep. In total, we use $2.9\times10^6$ Lagrangian tracer particles, with an initial upstream spatial resolution of 2.5 kpc. The Fokker-Planck equation is then solved in the Lagrangian frame, with a spectrum being assigned to each tracer. In this way, the tracers represent a sampling of the underlying CR electron spectral density field. To assign values to the simulation volume as a whole, we treat the tracers as mesh-generating points and apply a Voronoi tessellation.

In the following work, we account for adiabatic changes; cooling\footnote{\textsc{Crest} stitches the analytic result to the spectrum when one cooling process dominates and calculates the result numerically otherwise. This reduces the need to subcycle when electron cooling timescales become shorter than the MHD time step.} via  Coulomb, bremsstrahlung, inverse Compton, and synchrotron losses; and DSA with magnetic obliquity \citep[see formulae in][]{pais2018, winner2020}. We assume an isotropised pitch-angle distribution, which allows us to work in one-dimensional momentum space, where the one- and three-dimensional distribution functions are related by $f^\rmn{1D} = 4 \pi p^2 f^\rmn{3D}$. Here, we have used the dimensionless momenta $p = \tilde{p} / (m_\rmn{e} c)$, where $\tilde{p}$ is the dimensional momentum, $m_\rmn{e}$ is the electron rest mass, and $c$ is the speed of light. Momenta are solved between $p=10^{-1}$ and $p=10^{8}$, with 10 logarithmically-spaced bins being used per decade. We find that this is sufficient for spectral convergence.

Tracers are initially assigned a thermal spectrum. To model DSA at shocks, we inject a CR electron momentum power-law spectrum at the shock-surface and post-shock cells\footnote{Definitions of these are given in \citet{schaal2015} and section 2.4 of \citet{whittingham2024}.} by attaching a power-law slope with
\begin{equation}
\alpha = -\frac{x_\rmn{s} + 2}{x_\rmn{s}-1}
\end{equation}
to the one-dimensional distribution function, where $x_\rmn{s}$ denotes the shock compression ratio. This is limited to a maximum slope of $\alpha=-2.2$, following results by \citet{caprioli2020}. Unless otherwise stated, we also do not accelerate CR electrons below $\mathcal{M}_\rmn{crit} = 2.3$, following results from recent PIC simulations (see references given in the introduction of \citetalias{whittingham2024}). We dynamically choose the minimum and maximum momenta for acceleration and model a super-exponential cut-off due to catastrophic cooling losses. The formula for these are presented in section 2.5 of \citet{whittingham2024}. The normalisation is subsequently set by the energy injected, where we use 0.1\% of the liberated thermal energy (see Eq.~\ref{eq:shock-dissipated-energy2}). Note, that we do not model Fermi I re-acceleration, which is expected to be required to produce surface brightnesses in line with observations \citep{pinzke2013, vazza2014, botteon2020}. This, however, should generally only affect the normalisation of the spectrum, and not the spectral index at radio-emitting frequencies\footnote{We note, though, that this may have some mild effects when taken in projection. For example, \citet{dominguez-fernandez2024} found that radio relics become less ``patchy'' when re-acceleration is modelled.} \citep{pinzke2013, winner2019}.

For cooling, we assume that the gas is fully ionised and of primordial composition. This results in a hydrogen mass fraction of $X_\rmn{H} = 0.76$, a mean molecular weight of $\mu = 0.588$, and an ionisation fraction of $x_\rmn{e} = 1.157$. To aid comparison with observations we assume a redshift of $z=0.2$, which is the approximate redshift of the Toothbrush and Sausage radio relics \citep{vanweeren2010, vanweeren2012}. The energy density of the cosmic microwave background (CMB) is therefore set to $\epsilon_\rmn{CMB} = 8.65 \times 10^{-13}$ erg cm$^{-3}$, or equivalently a magnetic field strength of $B_\rmn{CMB} \approx 4.7$ $\upmu$G.

To generate mock emission, we use the code \textsc{Crayon+} \citep{werhahn2021}. This is capable of converting CR electron spectra into instantaneous emission for a variety of processes. Here, however, we focus solely on radio synchrotron emission. Full details are presented in section~2.4 and associated appendices of \citet{werhahn2021b}. However, the main result ultimately follows \citet{rybicki1986}, such that the synchrotron emissivity $j_\nu$ of a tracer at frequency $\nu$ is:
\begin{equation}
j_\nu = \frac{\sqrt{3}e^3 B_\perp}{m_\rmn{e} c^2} \int\limits_{0}^{\infty} f(p) F(\nu / \nu_c) \mathrm{d}p\propto B_\perp^{1-\alpha_\nu}\nu^{\alpha_\nu},
\end{equation}
where $B_\perp$ is the projection of the magnetic field onto the plane perpendicular to the line of sight, $\alpha_\nu$ is the radio spectral index, and $F(\nu/\nu_\rmn{c})$ is the dimensionless synchrotron kernel, with $\nu_\rmn{c}$ being the critical frequency. The exact definitions of these last two variables can be found in \citet{werhahn2021b} and \citet{whittingham2024}. The specific radio synchrotron intensity, $I_\nu$, at frequency $\nu$ is then obtained by integrating $j_\nu$ along the line of sight, accounting for the aforementioned Voronoi tessellation. Spectral indices, in turn, are calculated as:
\begin{equation}
    \alpha^{\nu_2}_{\nu_1} = \frac{\log_{10}(I_{\nu_2} / I_{\nu_1})}{\log_{10}({\nu_2 / \nu_1})},
\end{equation}
where $\nu_2>\nu_1$ so that $\alpha^{\nu_2}_{\nu_1}<0$ for a spectrum that decreases with increasing frequency.

\subsection{Shock-tube}

The shock-tube used in the following simulations has a periodic volume of $1800\times300\times300$~kpc$^3$. Our analysis, however, is confined to a high-resolution region of length 450 kpc (see discussion in section 2.3 of \citetalias{whittingham2024}). We define the start of this region to be $x=0$ kpc, such that cells with $x < 0$~kpc are the initial downstream, and those with $x > 0$~kpc are the initial upstream.
For cells with $x < 0$ kpc, the pressure is set to be relatively high (see below). This means that it acts as a piston, driving a shock across the high-resolution region. The density, however, is set to the same mean value throughout the box. Following analysis of cosmological simulations in \citetalias{whittingham2024}, we set the density to $\rho \approx 6.7 \times 10^{-29}$ g cm$^{-3}$ (equivalently a thermal electron number density of $n_\mathrm{e} = 3.5 \times 10^{-5}$ cm$^{-3}$) and choose the initial upstream pressure to be $P_1 = 1 \times 10^{-13}$ dyne cm$^{-2}$. Shocks in observed radio relics cover a range of strengths, although they are generally weak \citep[see, e.g.,][]{vanweeren2017, wittor2021b}. We thus choose to investigate both $\mathcal{M}=2$ and $\mathcal{M}=3$ shocks. This requires the initial downstream pressure to be $P_2 = 1 \times 10^{-12}$ dyne cm$^{-2}$ and $P_2 = 2.32 \times 10^{-12}$ dyne cm$^{-2}$, respectively, where exact values are calculated using a Riemann solver. 

This set-up aims to replicate the narrow, shock-compressed region observed in our cosmological simulations (see section 3 of \citetalias{whittingham2024}). However, we also implement upstream density turbulence in our shock-tubes. Indeed, this turbulence was key to the mechanism we presented in \citetalias{whittingham2024}, which provided solutions to the \textit{Mach number discrepancy}, \textit{Magnetic field}, and \textit{Spectral variation} problems. The existence of such turbulence is motivated by both observations \citep{simionescu2011, eckhert2015, ghirardini2018} and simulations \citep{nagai2011, zhuravleva2013, battaglia2015, angelinelli2021}. To generate this turbulence, we use the method of \citet{ruszkowski2007} and \citet{ehlert2018}. The advantage of this method is that the turbulence is first created in $k$-space, thereby allowing precise control over its properties. 

\begin{figure}
    \centering
    \hspace*{-0.6cm}
    \includegraphics[width=0.7\columnwidth]{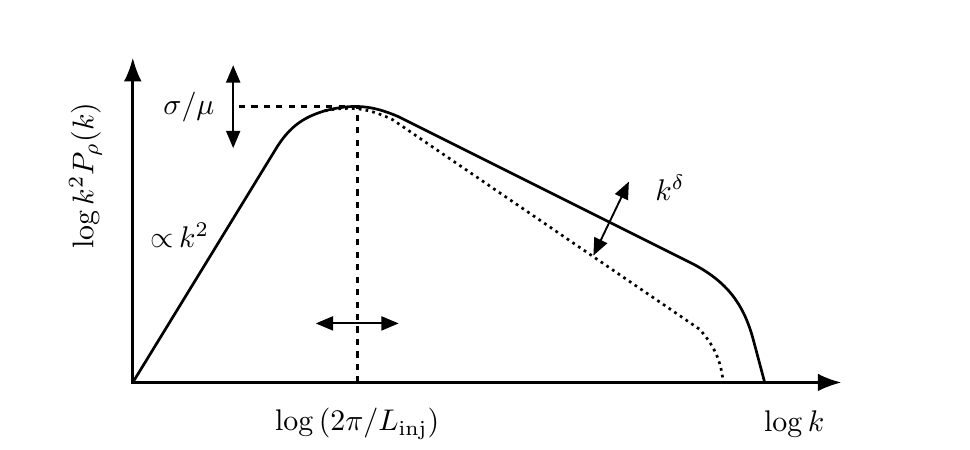}
    \caption[Power spectra schematic, illustrating how we vary the upstream density turbulence]{Schematic showing how we vary the 3D power spectra, $P_\rho$, of the upstream density turbulence in our simulations. We probe the impact of relative variance ($\sigma/\mu$), the injection scale ($L_\text{inj}$), and the power law slope ($\delta$). Power on spatial scales larger than the injection scale is set to white noise.}
    \label{figure:power-spectra}
\end{figure}

\begin{figure}
    \centering
    \includegraphics[width=0.5\columnwidth]{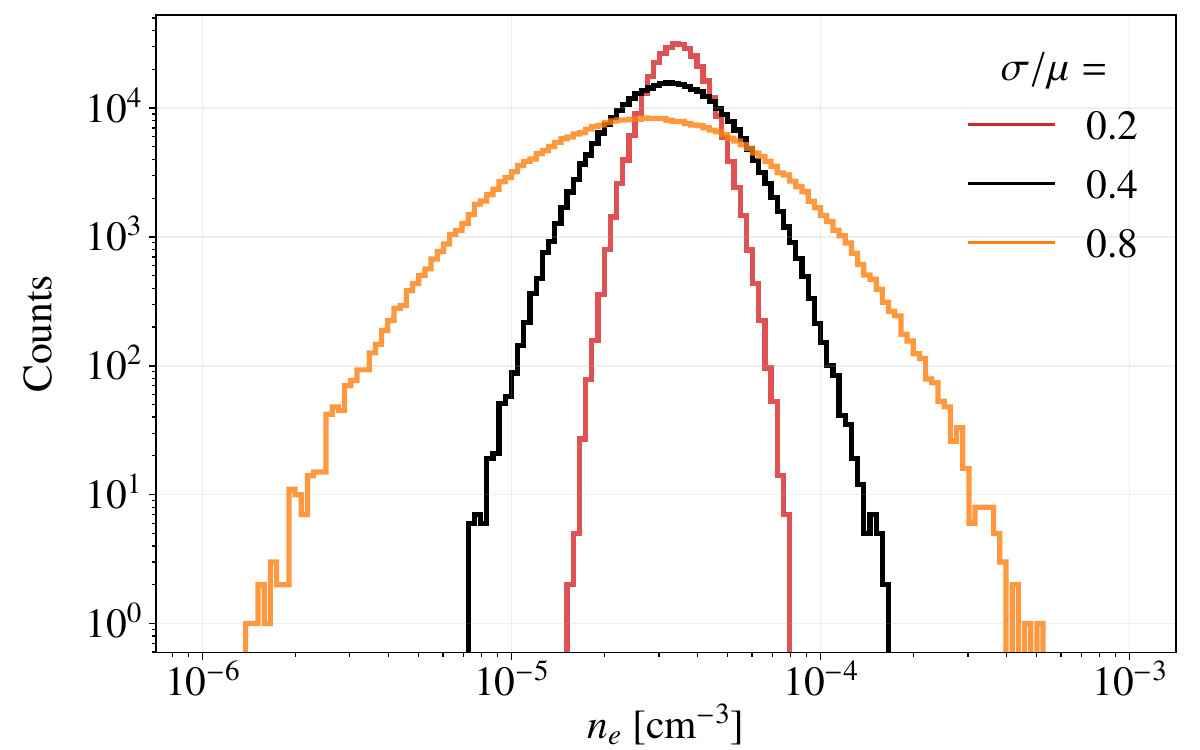}
    \caption[Histograms indicating how we vary the upstream density variance]{Histograms of the electron number density, calculated using all gas cells in the initial upstream conditions. Our fiducial value is $\sigma/\mu=0.4$ (see text). Each distribution is log-normal and has a mean density set to $3.5 \times 10^5$~cm$^{-3}$.}
    \label{figure:electron-density-pdf}
\end{figure}

The power spectra of the density turbulence $P_\rho(k)$ can be altered in three main ways. We illustrate these in Fig.~\ref{figure:power-spectra}. Firstly, the turbulent injection scale, $L_\rmn{inj}$ can be changed, where increasing this increases the typical length scale of the perturbations. For $k < 2 \pi / L_\rmn{inj}$, we apply white noise (i.e.\ the power spectra of the perturbations become scale independent). Secondly, we can vary the spectral slope of the turbulence, here parametrised by $\delta$, where a lower value results in a steeper slope, thereby enhancing the power of density fluctuations on the injection scale relative to those on smaller scales. Finally, the relative variance, $\sigma / \mu$, can be altered. This sets the typical amplitude of the perturbations, with increased variance increasing the maximum and minimum density peaks in the box\footnote{Note, when $\sigma / \mu \rightarrow 0$, we recover a ``flat'' upstream, with no density turbulence at all.}. This can be seen in Fig.~\ref{figure:electron-density-pdf}, where we show histograms for the electron number density in the upstream, including how this varies between simulations. The variance is set to have a log-normal distribution, as inferred from both simulations and observations \citep[see, e.g.,][]{kawahara2008}. We implement this using a Box-Muller random variate method. The temperature of the fluctuations is adjusted accordingly, so that the volume is in pressure equilibrium.

In the current work, we explore the parameter space available, varying each parameter to investigate its impact on the downstream turbulence and emission. Our fiducial simulation uses $\sigma / \mu = 0.4$ following figures 2 and 4 of \citet{zhuravleva2013}, who investigated turbulent gas properties in simulations of merging and relaxed clusters; $\delta=-5/3$ (i.e.\ \citealt{kolmogorov1941} turbulence); and $L_\rmn{inj} = 150$~kpc, which is half the smallest box dimension and the typical scale used in previous shock-tube studies of relics \citep[see, e.g.,][]{dominguez-fernandez2021}. We present a full list of the simulation variations in Table~\ref{tab:simulation_vars}. 

Each of our simulations also include magnetic turbulence. Here, the injection scale is set to 40 kpc. This is consistent with simulations, which suggest that the magnetic power peaks on scales a factor of a few lower than the kinetic power \citep{tevlin2024}. The root-mean-square (RMS) strength is set to $\approx0.16$~$\upmu$G, which results in a plasma beta of $\beta = \langle P_\mathrm{th} / P_B\rangle = 100$ for the upstream of the fiducial simulation, where $P_\mathrm{th}$ and $P_B$ are the thermal and magnetic pressures, respectively. This is consistent with ICM simulations \citep{skillman2013, dominguez-fernandez2019, nelson2024, tevlin2024} and observations \citep{brunetti2001, govoni2017}. Following, \citet{ehlert2018}, Gaussian variance is implemented, with the standard variation being fixed as a result of Parseval's theorem. Magnetic divergence is removed from the initial conditions by projecting it out in $k$-space (see \citetalias{whittingham2024} for further details).

\renewcommand{\arraystretch}{1.2}
\begin{table}
    \centering
    \begin{tabular}{l || c | c | c}
    \hline 
\makecell{\textbf{Simulation} \\ \textbf{name}}  & \makecell{\textbf{Relative} \\ \textbf{variance}} & \makecell{\textbf{Power} \\ \textbf{law slope}} & \makecell{\textbf{Turbulent} \\ \textbf{injection scale [kpc]}} \\ \hline 
Fiducial                   & 0.4               & -5/3            & 150                        \\[5pt] 
$\sigma/\mu = 0.2$         & 0.2               & -5/3            & 150                                 \\
$\sigma/\mu = 0.8$         & 0.8               & -5/3            & 150                                  \\[5pt]
$\delta=-2.5/3$            & 0.4               & -2.5/3          & 150                                 \\
$\delta=-10/3$             & 0.4               & -10/3           & 150                              \\[5pt]
$L_\rmn{inj}=75$ kpc  & 0.4               & -5/3            & 75                                  \\
$L_\rmn{inj}=200$ kpc & 0.4               & -5/3            & 200 
    \end{tabular}
    \caption{The parameter space explored by our simulations, where each row indicates an upstream model. Note, that each simulation is run both with a $\mathcal{M}=2$ and a $\mathcal{M}=3$ shock (see text). The critical Mach number is set to $\mathcal{M}_\rmn{crit}=2.3$, unless otherwise stated.}
    \label{tab:simulation_vars}
\end{table}

\section{Analysis}
\label{chapter6-sec:analysis}

\subsection{Hydrodynamic effects}
\label{chapter6-subsec:hydro}

\subsubsection{Downstream turbulence}
\label{chapter6-subsec:downstream-turbulence}

We start our analysis by focussing on the impact of the different upstream models on the development of downstream density turbulence. To this end, in Fig.~\ref{figure:density-slices} we show slices through the $\mathcal{M}=3$ simulations\footnote{We exclude the \textit{Fiducial} simulation for space reasons, but the analogous data can be found in the movie \href{https://youtu.be/gX-urGrKNy8}{here}.} at $t=250$~Myr, where colours indicate gas density. At this point, the shock has almost left the high-resolution region, but some indication of the upstream initial conditions can be seen towards the right-hand side of each panel. From left to right, we vary the relative variance, the power law slope, and the turbulent injection scale, respectively. 

Starting with the left-hand column, it can be seen that, although a RT instability is evident at the contact discontinuity in the $\sigma/\mu =0.2$ run, it is substantially more developed in the $\sigma/\mu =0.8$ run. In \citetalias{whittingham2024}, we showed that this instability develops for two reasons:
\begin{enumerate}[i)]
\item Firstly, higher density fluctuations are less easily accelerated in the downstream. This means they penetrate further, and cause the contact discontinuity to corrugate. This helps to seed the RT instability.
\item Secondly, the shock-front accelerates into lower density regions, causing it to corrugate as well. This results in a pressure gradient across the contact discontinuity, which does not align with the density gradient. This drives the RT instability.
\end{enumerate}
Knowing this, the difference between the two simulations is to be expected; the $\sigma/\mu =0.8$ has significantly lower density regions, increasing the shock-front corrugation\footnote{This can be better seen in Fig.~\ref{figure:turbulence-slices}.}, and thereby better triggering the instability. Furthermore, it has significantly higher density peaks, which are better able to penetrate the downstream, thereby seeding the instability more effectively. These two factors lead to a larger downstream extent, as well as substantially greater mixing, with low density regions that were originally behind the shock-compressed region able to reach, in places, right up to the shock-front.

\begin{figure*}
    \centering
    \includegraphics[width=1.0\columnwidth]{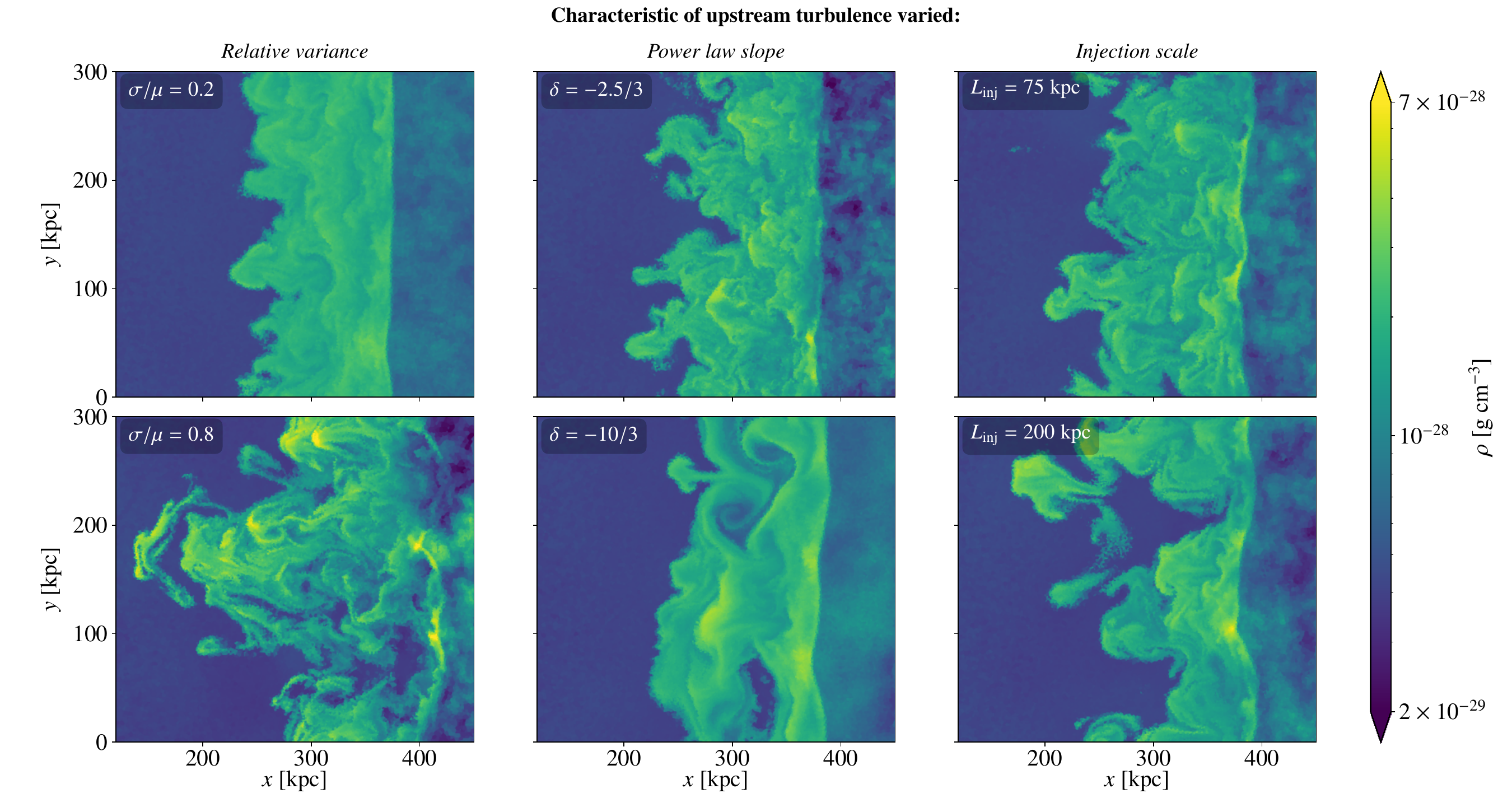}
    \caption[Slices through our $\mathcal{M}=3$ simulations at $t=250$~Myr, showing how varying the upstream density turbulence affects the generation of downstream substructure]{Slices through our $\mathcal{M}=3$ simulations at $t=250$~Myr, where colours indicate gas density and the shock travels from left to right. In each column we vary only one characteristic of the upstream initial conditions. Increasing the relative variance extends the maximum length of the downstream, but can also bring material behind the contact discontinuity closer to the shock front. Steepening the power spectra, meanwhile, suppresses small-scale power so that the typical eddy size and the scale of the substructure downstream are increased. Finally, increasing the length scale of injection increases the width of the RT ``fingers''.}
    \label{figure:density-slices}
\end{figure*}

In the middle column, we investigate the impact of the power law slope. The key differences here are the scale over which the downstream substructure varies and the subsequent scale of the eddies generated. In particular, the $\delta=-10/3$ simulation shows significantly smoother variation compared to the $\delta=-2.5/3$ simulation. This is especially evident when we observe the contact discontinuity; whilst the $\delta=-2.5/3$ simulation generates many RT ``fingers'', the contact discontinuity in the $\delta=-10/3$ simulation is significantly smoother and flatter. As discussed at the end of \citetalias{whittingham2024}, the growth rate of the RT instability is:
\begin{equation}
    \sigma = \sqrt{|\mathbf{a}| \left( \frac{\rho_2 - \rho_3}{\rho_2 + \rho_3} \right) k},
\end{equation}
where $\mathbf{a} \propto -\bnabla P$ is the acceleration of the lighter material into the denser material, the density terms in brackets constitute the Atwood number, the subscripts ``2'' and ``3'' indicate post-shock gas and gas to the left of the contact discontinuity, respectively, and $k$ is the wave number of the perturbation \citep{chandrasekhar1961}. The instability therefore grows fastest on small scales (large $k$). These have significantly lower amplitudes in the $\delta=-10/3$ simulation, and hence the instability is likely being suppressed here\footnote{An alternative interpretation is that the RT instability still exists, but is acting only at smaller $k$ values.}. 

\begin{figure*}
    \centering
    \includegraphics[width=1.0\columnwidth]{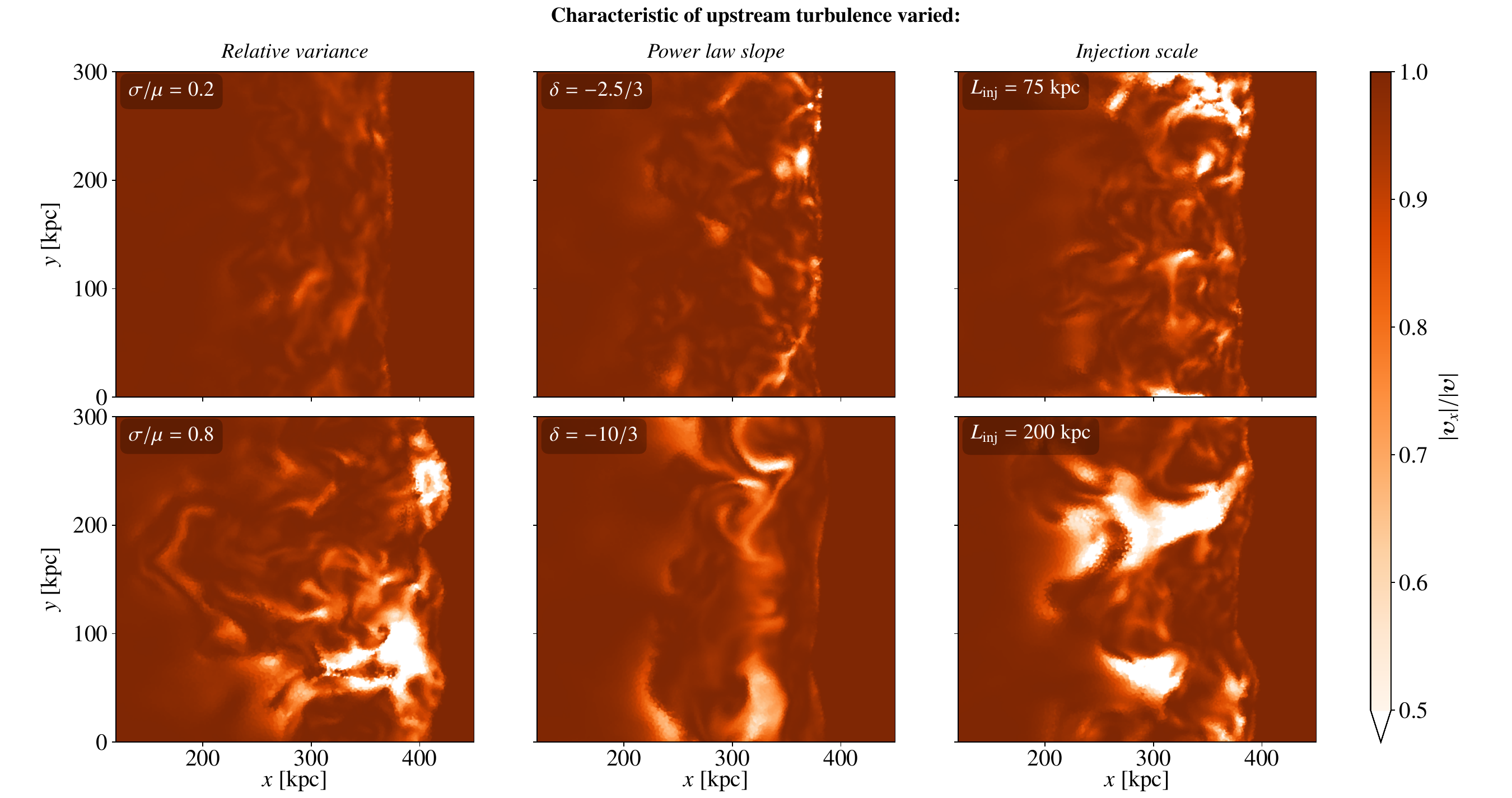}
    \caption[Slices through our simulation suite at $t=250$~Myr, showing how varying the upstream density turbulence affects the generation of downstream velocity turbulence]{As Fig.~\ref{figure:density-slices} but colours show the fraction of the gas speed in the $x$-component, with velocities calculated in the shock-frame. Increasing the relative variance, the steepness of the power law, or the scale of injection all substantially increase the amount of turbulence downstream.}
    \label{figure:turbulence-slices}
\end{figure*}

This is not to say that there is no downstream turbulence formed in the $\delta=-10/3$ simulation; indeed, it is clear that large vortices are being formed. These result from the density fluctuations experiencing different levels of acceleration based on their mass. The resulting velocity shear gives rise to the Kelvin-Helmholtz effect. Furthermore, because the power law slope is relatively steep, the scale of the density fluctuations is generally well defined. This produces a well-defined scale for the velocity shear, which, in turn, gives rise to large, well-formed eddies. Indeed, the largest eddy scale is approximately 75 kpc, which is half the injection scale. This is expected when there is a velocity shear on both sides of a fluctuation.

In the final column, we present simulations where only the injection scale of the turbulence has been varied. This varies the average length-scale between upstream perturbations. This affects shock corrugation; as the shock advances into the underdense regions, the shock-front becomes corrugated on a similar length-scale. In a similar, but subtly different manner, increasing the injection scale also increases the typical distance between the density peaks, which broadens the length-scale that the contact discontinuity corrugates on. Note, that this is different from the previous case, where the length-scale stayed the same, but the small scale power was suppressed. In this case, the RT instability continues to act but, for a higher $L_\rmn{inj}$, the induced vorticity is now more spread out, and hence the width of the RT ``fingers'' is increased. This allows for a more consistent velocity field, which produces more pronounced peaks and troughs along the shock-compressed region.

We now turn our attention to the induced downstream {velocity} turbulence. In Fig.~\ref{figure:turbulence-slices}, we also show slices at $t=250$ Myr, but colours now indicate the fraction of gas speed in the $x$-component, where all velocities are calculated in the shock-frame. Lower values of this parameter indicate increased turbulence, with more motion in the $y$- and $z$-directions. On the other hand, $|\bupsilon_x| / |\bupsilon| = 1$ indicates perfectly laminar flow in the $x$-direction. It can be seen that, in all cases, upstream density perturbations result in the formation of at least some degree of velocity turbulence. The greatest impact, however, is seen between the $\sigma / \mu$ variations, with $\sigma / \mu = 0.2$ showing the lowest levels, and $\sigma / \mu = 0.8$ exhibiting the highest. This makes sense as, in the most extreme case, $\sigma / \mu = 0$ would result in no upstream density turbulence, and hence a laminar flow. 

The figure also shows the difference between RT-driven turbulence, where $|\bupsilon_x| / |\bupsilon|$ exhibits considerable variance over short length-scales, and the Kelvin-Helmholtz driven turbulence evident in $\delta=-10/3$, where $|\bupsilon_x| / |\bupsilon|$ varies over longer length-scales. Moreover, in $\delta=-10/3$, the low $|\bupsilon_x| / |\bupsilon|$ values clearly correlate with the position of the vortices analysed in Fig.~\ref{figure:density-slices}. The impact of the RT effect, meanwhile, can be seen particularly well for $\sigma/\mu=0.8$ and $L_\rmn{inj}=200$~kpc simulations, where low $|\bupsilon_x| / |\bupsilon|$ values correlate strongly with the acceleration of low density gas towards the shock-front. This results from the velocity fields created by the RT-induced vorticity\footnote{See schematic in \citetalias{whittingham2024}.}. 

\begin{figure}
    \centering
    \includegraphics[width=0.5\columnwidth]{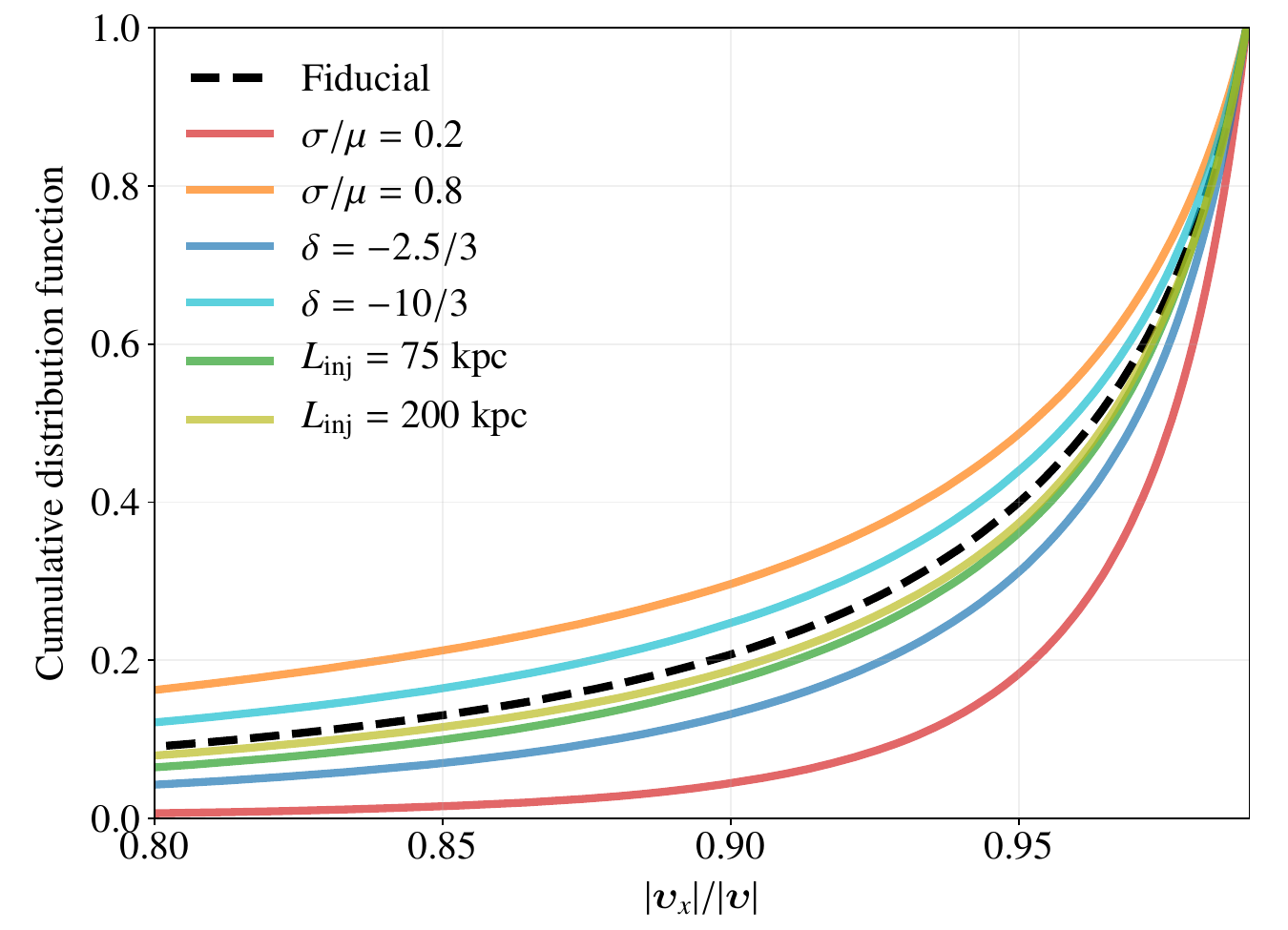}
    \caption[Cumulative distribution functions showing the relative amount of downstream velocity turbulence at $t=250$ Myr]{Cumulative distribution functions showing the relative amount of downstream velocity turbulence at $t=250$ Myr in the injected region of each of our Mach 3 simulations. Velocity turbulence is measured here by the fraction of the shock-frame gas speed in the $x$-component, where decreasing values of this fraction indicate higher levels of turbulence. Increasing the relative variance has the greatest impact on the production downstream turbulence, followed by the gradient of the power spectra.}
    \label{figure:turbulence-pdf}
\end{figure}

Figure~\ref{figure:turbulence-slices} provides only spatial information for a slice at $z=150$ kpc. To gain a more general overview, in Fig.~\ref{figure:turbulence-pdf} we show the cumulative distribution function of $|\bupsilon_x| / |\bupsilon|$ for all $\mathcal{M}=3$ simulations at $t=250$ Myr, where we have selected all cells in the shock-compressed region, and the contribution of each cell has been weighted by its mass. It can be seen that, once again, the $\sigma/\mu$ simulations show the greatest difference. For example, whilst 30\% of the gas mass in the $\sigma/\mu = 0.8$ simulation has $|\bupsilon_x| / |\bupsilon| < 0.9$, this is true of only 5\% of cells in the $\sigma/\mu = 0.2$ simulation. The remaining models are substantially closer to the fiducial model; in particular, it appears that changing the injection scale has little to no impact on the overall amount of turbulence. Changing the power spectra, meanwhile, makes a small difference. We note that the $\delta=-2.5/3$ simulation shows lower turbulence levels than the fiducial simulation. We attribute this to the increased amount of small-scale density fluctuations in this simulation, which prevent the Raylor-Taylor instability from generating velocity fields on larger scales. This, in turn, limits its ability to impact the bulk flow.

\subsubsection{Shock corrugation}
\label{subsec:shock-corrugation}

\begin{figure*}
    \centering
    \includegraphics[width=1.0\columnwidth]{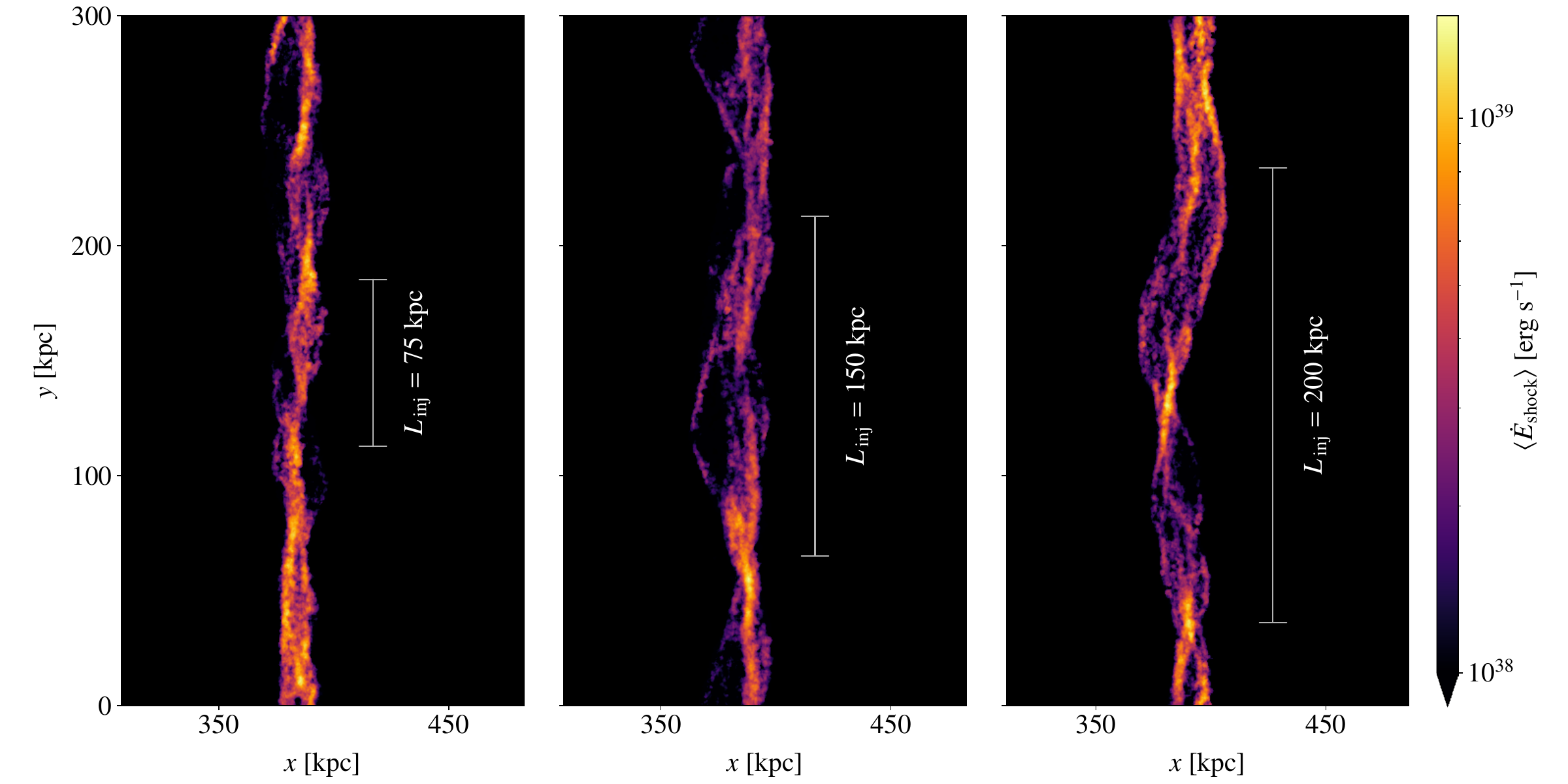}
    \caption[Projected shock fronts in simulations where we vary the scale of injection]{Projected shock-dissipated energy for simulations where we vary only the scale of injection. Each panel has a projection depth of 300 kpc and shows a Mach 3 variation at $t=250$~Myr. White brackets indicate the to-scale injection length used. The projected shock surface shows ``thread'' and ``knot'' features, similar to those observed in real radio relics. The injection scale sets the size of these features.}
    \label{figure:shock-filaments}
\end{figure*}

In the previous section, we showed that the shock-front corrugates on length scales set by the upstream turbulent injection scale. This leads to the production of morphological features when seen in projection. To show this, in Fig.~\ref{figure:shock-filaments} we present the projected shock-dissipated energy for simulations where we vary only the scale of injection. Once again, the simulations are shown at $t=250$ Myr and are the $\mathcal{M}=3$ variations. The morphologies are intricate and appear highly filamentary. Indeed, many of the features that exist in observed radio relics are evident here too; there are noticeable ``double strand'' features, where a gap in emission forms between two arcing shock fronts; there are bright ``knots'' or ``twists'', formed predominantly at the intersection of shock fronts as seen in projection; and there are ``bay'' features, where the emission becomes weak enough that the shock front fades from view (see, for example the top of the left-hand panel).

Moreover, because these features ultimately originate from the corrugation of the shock-front, their spacing is also approximately set by the turbulent injection scale. To show this we have added annotations in white, with each bracket beginning at a ``knot''. These are 75, 150, and 200 kpc long, respectively, and illustrate the turbulent injection scale in each panel. It can be seen that the morphological features generally have the same scale, with the bracket usually ending at a ``knot'' as well. The effect is clearest for the simulations with larger scales of injection, where there are fewer fluctuations along the line of sight. It is also clearer when the fluctuations consistently have the same size; in the right-hand panel, for example, the turbulent injection scale is $L_\rmn{inj} = 200$~kpc, whilst the depth and height of the box is only 300~kpc. Consequently, some features over 100~kpc scales. Even, here, however, there are clearly structures that vary over twice this length.

The Toothbrush relic is a good example of a relic which exhibits the features described above. This has a largest linear size (LLS) of approximately 1.9~Mpc \citep{rajpurohit2020}, which implies that the double strand features are between $500-600$ kpc long. The Sausage relic also exhibits filamentary structure and has an LLS of $\approx2$ Mpc \citep{diGennaro2018}. Here, the distinction between features is weaker, but by mapping between the brightest points, we find a typical length scale of $\sim500$~kpc. We note that this is significantly higher than the value generally used in idealised simulations of radio relics \citep{dominguez-fernandez2021} and at least $2-3$ times higher than that commonly assumed by the community (van Weeren, private comm.). This value is, however, consistent with measurements of eddies in simulations of merging clusters, which find eddies with length scales up to 1 Mpc in the centre \citep{minitai2014, vazza2017b, tevlin2024}. Moreover, the well-defined nature of the ``strands'' in the Toothbrush relic implies that we are not projecting along many fluctuations; indeed a structural depth of no more than $1$~Mpc seems likely based on the implied injection scale. This fits the structure within the virial size of the cluster.

\begin{figure*}
    \centering
    \includegraphics[width=1.0\columnwidth]{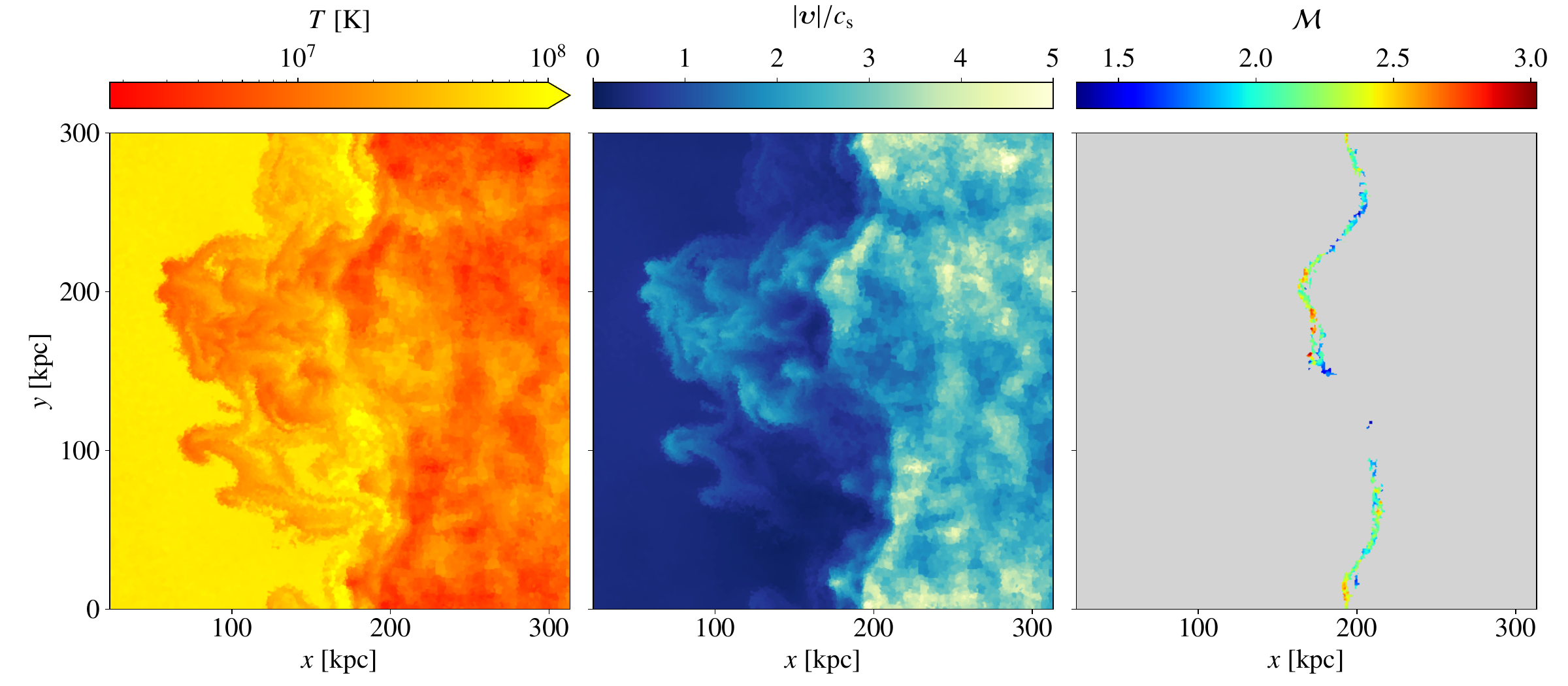}
    \caption[Figure showing how holes develop in the shock front]{Slice through the Mach 2 high relative-variance simulation ($\sigma/\mu = 0.8$) at $t=170$~Myr. The two left-most panels show slices through the midplane of the box, with colours indicating gas temperature and gas velocity as a function of sound speed (in the shock-frame), respectively. The right-most panel is a thin projection with depth 35 kpc, where colours indicate the emission-weighted Mach number. Higher temperatures result in higher sound speeds, which reduces the strength of the shock. At its most extreme, this effect causes gaps to from in the shock-front, as it temporarily transforms into a pure sound wave.}
    \label{figure:shock-gaps}
\end{figure*}

As previously mentioned, the shock corrugates in the first place due to the varying upstream density. In our simulations, the upstream is in pressure equilibrium, which means that the density is inversely proportional to the temperature. We show the resultant temperature fluctuations in the left-hand panel of Fig.~\ref{figure:shock-gaps}. However, this also means that the sound speed increases in lower density material. In our simulations, we are able to reach a regime where the shock speed becomes comparable to the upstream sound speed. This situation is shown in the middle panel, where we present the combination with the slowest shock speed (the $\mathcal{M}=2$ simulation) and the highest upstream sound speeds (i.e.\ the highest relative variance). At this point, the shock wave temporarily transforms into a pure sound wave, thereby fracturing the shock-front (see the final panel). Some radio relics, as well as their underlying shocks, do show signs of being broken apart like this \citep[see, e.g.][]{urdampilleta2018, zhang2020b} but most are typically coherent along their length. As we will see in Sec.~\ref{chapter6-subsec:emission}, however, a missing region of acceleration does not necessarily translate to an incoherent emission structure. 

We quantify the level of shock corrugation for all simulations in the right-hand side of Fig.~\ref{figure:mach-no-pdf--all-sims}. Here, we show the fraction of the shock surface within a given distance of the median position. The solid lines indicate the median over all snapshots, whilst the shaded regions represent the interquartile range. We have additionally added the fiducial simulation as a dotted line, and arrows which indicate the 90th percentiles. There is some mild evolution between the Mach 2 (top panel) and Mach 3 (bottom panel) simulations. This is particularly strong for the $\delta=-10/3$ simulation, and results from the aforementioned effect, where higher upstream sounds speeds cause more fragmentation. This effect typically takes place close to the shock median, leading to more cells being counted at further distances. Despite this, the arrows marking the 90th percentile change little overall, and are well within the error of margin set by the initial upstream cell size (2.5 kpc). Indeed, in general, the panel shows statistically, what we have already analysed above. In particular, it confirms that the relative variance has the largest impact on the shock corrugation. This is followed by the change of spectral slope. Steepening this results in smoother perturbations, which produce ``V''-shaped features in the shock-front, as they slow the shock-front down.

\subsubsection{Mach number distribution}

\begin{figure*}
    \centering
    \includegraphics[width=1.0\columnwidth]{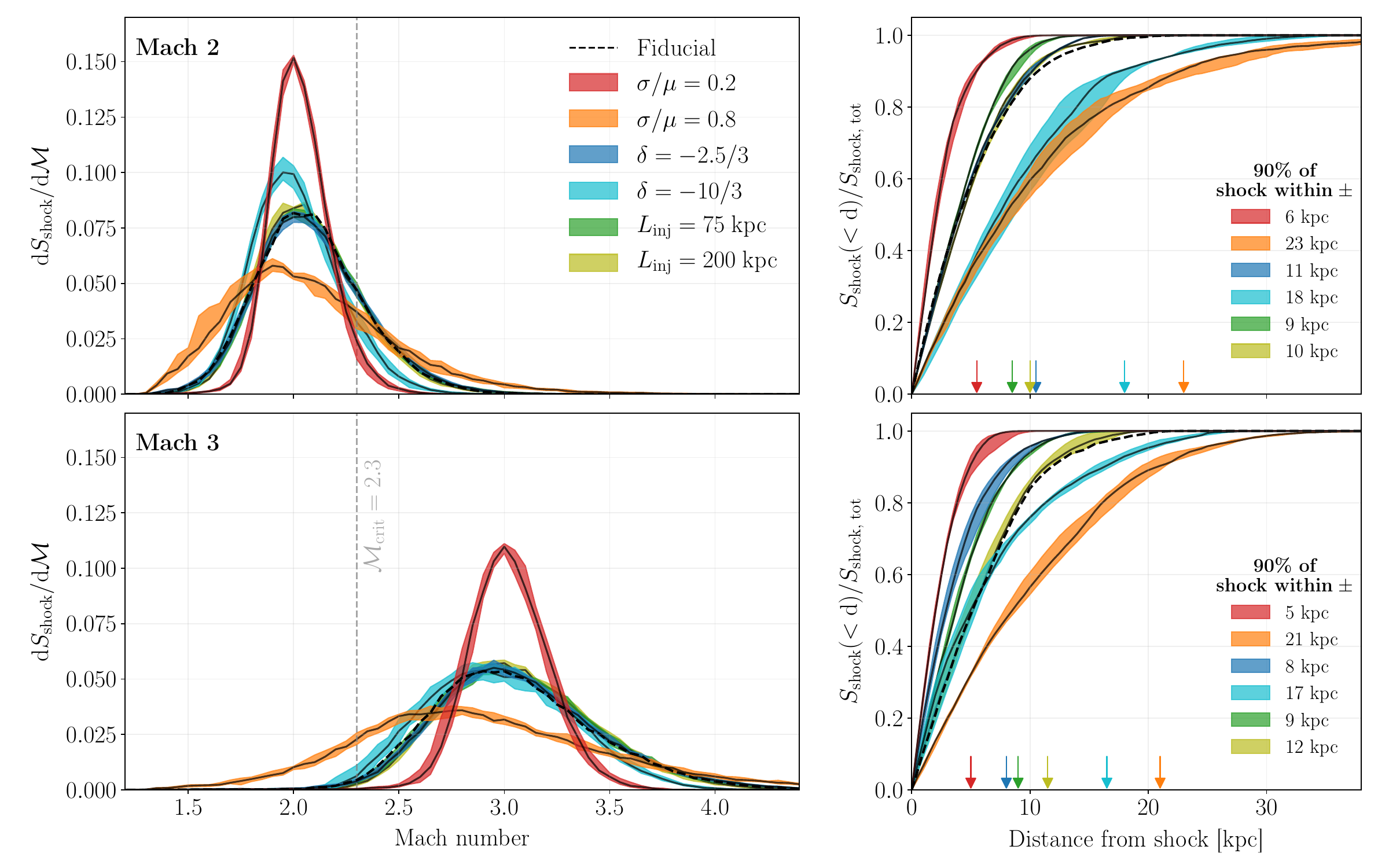}
    \caption[Mach number distributions and shock corrguation histograms for all variations]{\textit{Left:} Mach number distributions weighted by each gas cell's contribution to the shock surface. Black lines indicate the median taken over all snapshots, whilst the shaded values indicate the interquartile range. The grey, vertical, dashed line marks the critical Mach number above which we inject CR electrons. Colours indicate the simulation variation, with data in the top and bottom panels coming from our Mach 2 and Mach 3 simulations, respectively. \textit{Right:} here, the lines indicate the fraction of the shock surface within a given distance of the shock median so that a flatter distribution indicates a higher level of shock corrugation. The width of the Mach number distribution is predominantly set by the relative variance, although increasing the power law steepness also has a small effect. A similar statement is true for the level of shock corrugation as well, as these features have the same underlying cause.}
    \label{figure:mach-no-pdf--all-sims}
\end{figure*}

As we showed in \citetalias{whittingham2024}, the shock corrugation is inherently linked to the formation of a Mach number distribution. This is due to the aforementioned effect, where the upstream sound speed fluctuations in our simulations are a function of density only. The local variations in velocity across the shock front can generally be treated as perturbations as they are small relative to the median shock speed. Subsequently, to first order, the shock strength is $|\bupsilon|/c_\rmn{s}$, where $\bupsilon$ is the gas speed in the median shock frame, and $c_\rmn{s}$ is the sound speed. This is shown in the middle panel of Fig.~\ref{figure:shock-gaps}. It can be seen that hotter, less dense gas generally produces lower Mach numbers, whilst colder, denser gas generally produces higher Mach numbers. At the same time, however, underdense regions cause the shock-front to advance, whilst overdense regions pin it back. This means that we expect more corrugated shock-fronts to produce larger Mach number distributions.

The result of this correlation can clearly be seen in the left-hand side of Fig.~\ref{figure:mach-no-pdf--all-sims}. Here, we show the median Mach number across all snapshots, where each cell is weighted by its contribution to the overall shock surface. Once again, the shaded regions represent the interquartile range. It can be seen that by far the largest impact is had by the relative variance, with the $\delta=-10/3$ simulation resulting in a significantly smaller modulation\footnote{This is slightly biased to lower Mach numbers, which results from the ``V''-effect, just described; cells towards the front of the shock tend to produce lower Mach numbers \citepalias[see][]{whittingham2024}.}. Each distribution can be well fit with a skew normal function. The increasing narrowness of this function with decreasing variance makes sense, as when $\sigma / \mu \rightarrow 0$ we will have no upstream density turbulence and, analytically, a single Mach number as a result \citepalias[see][]{whittingham2024}.

\subsection{Spectra}
\label{chapter6-subsec:spectra}

\subsubsection{Impact on spectra}

In \citetalias{whittingham2024}, we showed that the generation of a Mach number distribution resulted in significantly shallower integrated spectra compared to the non-turbulent control case. This effect was stronger for weaker Mach numbers, as the spectral slope produced by an individual tracer follows the relation:
\begin{equation}
    \alpha_\rmn{e} = - \frac{2 (\mathcal{M}^2 + 1)}{\mathcal{M}^2 - 1}.
\end{equation}
This means that as $\mathcal{M}\rightarrow1$, $\alpha_\rmn{e}\rightarrow -\infty$. Consequently the difference between spectral slopes at weak shocks is that much greater, allowing the tail of the Mach number distribution to more strongly influence the overall spectral slope. This effect also resulted in the cooled part of the spectrum being shallower than the ``$\alpha_\rmn{e}-1$'' slope expected from theory, as higher Mach numbers with shallower slopes increasingly dominated the normalisation. In the top row of Fig.~\ref{figure:spectra-all-models}, we show the volume-weighted non-thermal CR electron spectra for all variations. The Mach 2 simulations are shown on the left and the Mach 3 are on the right. The spectra are multiplied by $p^{2.2}$, so that the maximum slope allowed would appear as a flat horizontal line. We also include the ``Flat'' simulation, originally presented in \citetalias{whittingham2024}, which has no upstream density turbulence, for reference. We have applied no critical Mach number to this model, whilst the others have $\mathcal{M}_\rmn{crit}=2.3$ as standard.

\begin{figure*}
    \centering
    \includegraphics[width=1.0\columnwidth]{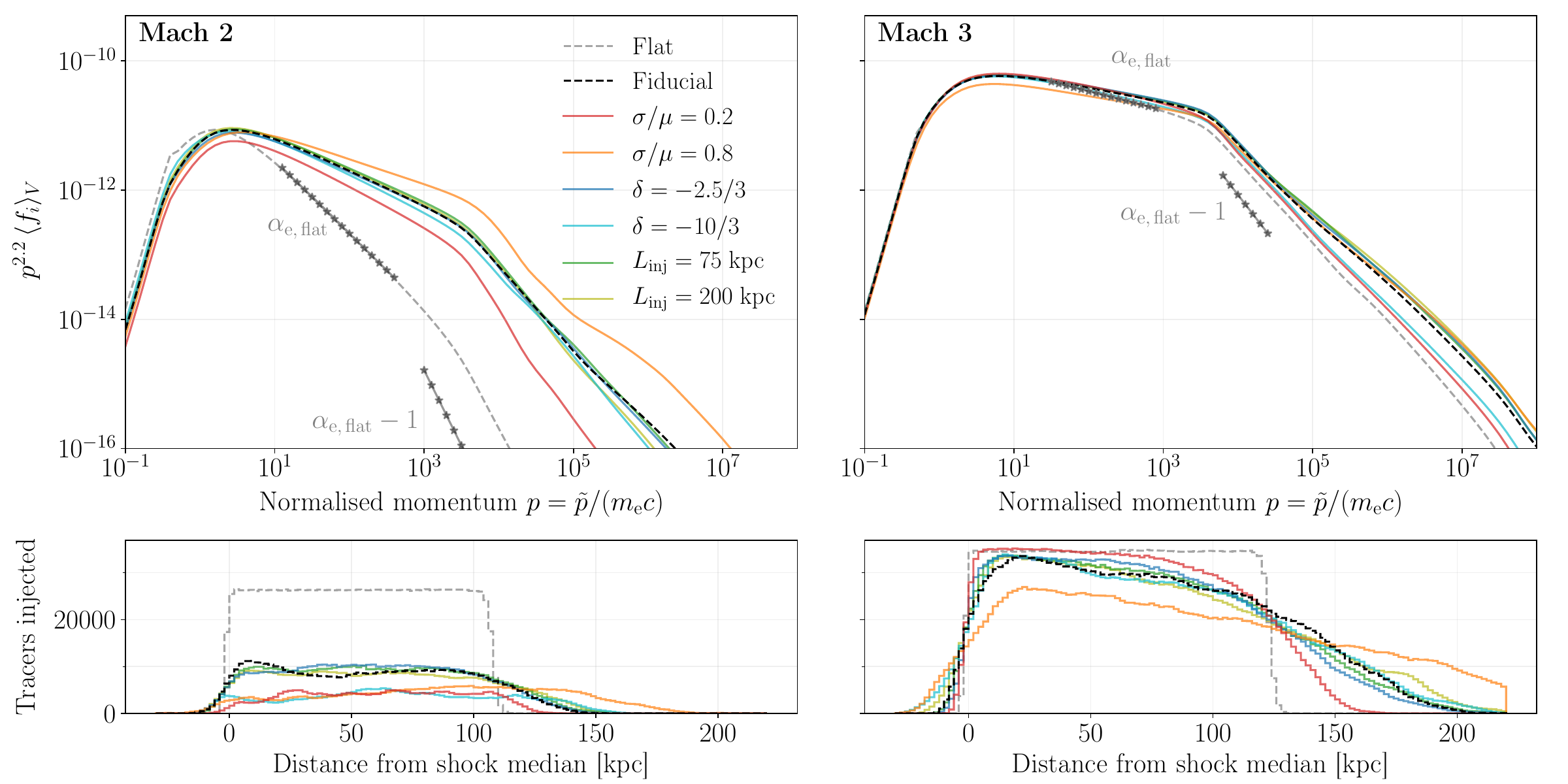}
    \caption[Volume-weighted non-thermal CR electron spectra and histograms of tracers as a function of distance for all simulation variation]{\textit{Top row}: volume-weighted non-thermal CR electron spectra for all simulation variations. \textit{Bottom row:} histograms indicating the number of injected tracers binned by their distance to the median shock position. Mach 2 and Mach 3 initial shock variations are shown on the left and right, respectively. Adding a turbulent upstream density field results in flatter spectra (relative to $\alpha_\text{e,\,flat}$). This is especially important in weaker shocks, where the difference in CR electron number density at radio-emitting frequencies can be substantial. The relative variance, once again, is the most influential factor in determining both spectra and the distribution of tracers relative to the shock.}
    \label{figure:spectra-all-models}
\end{figure*}

It can be seen that our results presented in \citetalias{whittingham2024} also translate here. Namely, adding upstream density turbulence significantly flattens the spectral slope, but this process is more effective in weaker shocks. It can also be seen that, again, the relative variance has the greatest impact. This is to be expected, as this was the most influential factor for setting the Mach number distribution. In particular, the larger the relative variance, the longer the tail of this distribution, and consequently the flatter the resulting spectra. The long tail also causes more fluctuations, as individual tracers are better able to affect the integrated spectrum\footnote{This can affect the cooled part of the slope, $\alpha_\rmn{e}-1$. We will investigate the exact impact and how it changes between the models in future work.}. It is likely that this effect is at least is partially numerical, resulting from the low number of tracers at this part of the Mach number distribution.

Most of the Mach 3 simulations produce a similar result, with the exception of the ``Flat'', ``$\sigma/\mu=0.2$'', and ``$\delta=-10/3$'' cases. This follows both from the explanation given above -- that the slope is more weakly affected by fluctuations in the Mach number in stronger shocks -- but also by the similarity of the Mach number tail, as seen in Fig.~\ref{figure:mach-no-pdf--all-sims}. It can be seen in that figure that whilst the width of the distribution increases for higher relative variance, the overall distribution shifts to the left\footnote{It is possible that this results from aligning the mean density rather than the peak of the distribution, as seen in Fig.~\ref{figure:electron-density-pdf}. We will check this in future work.}. This means that, for the Mach 3 simulations, the tail of the $\sigma/\mu=0.8$ distribution is closer to the other models. This results in similar peak Mach numbers and hence similar slopes.

In the bottom panels of Fig.~\ref{figure:spectra-all-models}, we show histograms of the number of injected tracers vs.\ their distance from the median shock position. The grey, dashed line indicates the distribution in the ``Flat'' simulation, and represents our control simulation; without minor pressure fluctuations due magnetic turbulence \citepalias[see][]{whittingham2024}, this would form an almost perfect rectangle. It can be seen that all of our turbulent models extend to the left of this box (i.e. show negative distance values) and ramp up to their peak value afterwards. This is indicative of the shock corrugation. Furthermore, in each case, the distribution extends beyond the right of the ``Flat'' distribution, with this extension most sensitive to the relative variance. This follows from the extension we saw in Fig.~\ref{figure:density-slices}. Finally, each of these simulations shows fluctuations, which results from the velocity turbulence we previously analysed. These distributions therefore represent a general statistical confirmation of the features we analysed in Sec.~\ref{chapter6-subsec:hydro}. 

Finally, we note that the number of tracers injected in the Mach 2 simulations is significantly smaller than those injected into the ``Flat'' run. This is predominantly a result of the critical Mach number we enforce, where tracers encountering a shock with $\mathcal{M} < \mathcal{M}_\rmn{crit} = 2.3$ are not injected\footnote{The $\sigma/\mu=0.8$ run, particularly, is additionally affected by the high sound speed phenomena described earlier.}. As can be seen in Fig.~\ref{figure:mach-no-pdf--all-sims}, this affects a substantial fraction of tracers, with roughly 75\% of tracers in the $\sigma/\mu=0.8$ simulation falling foul of this criteria, and approximately 95\% of tracers in the $\sigma/\mu=0.2$ simulation. The obvious question arising from this is \textit{how does it affect the production of radio emission?}

\subsubsection{Impact of a critical Mach number}

\begin{figure*}
    \centering
    \includegraphics[width=1.0\columnwidth]{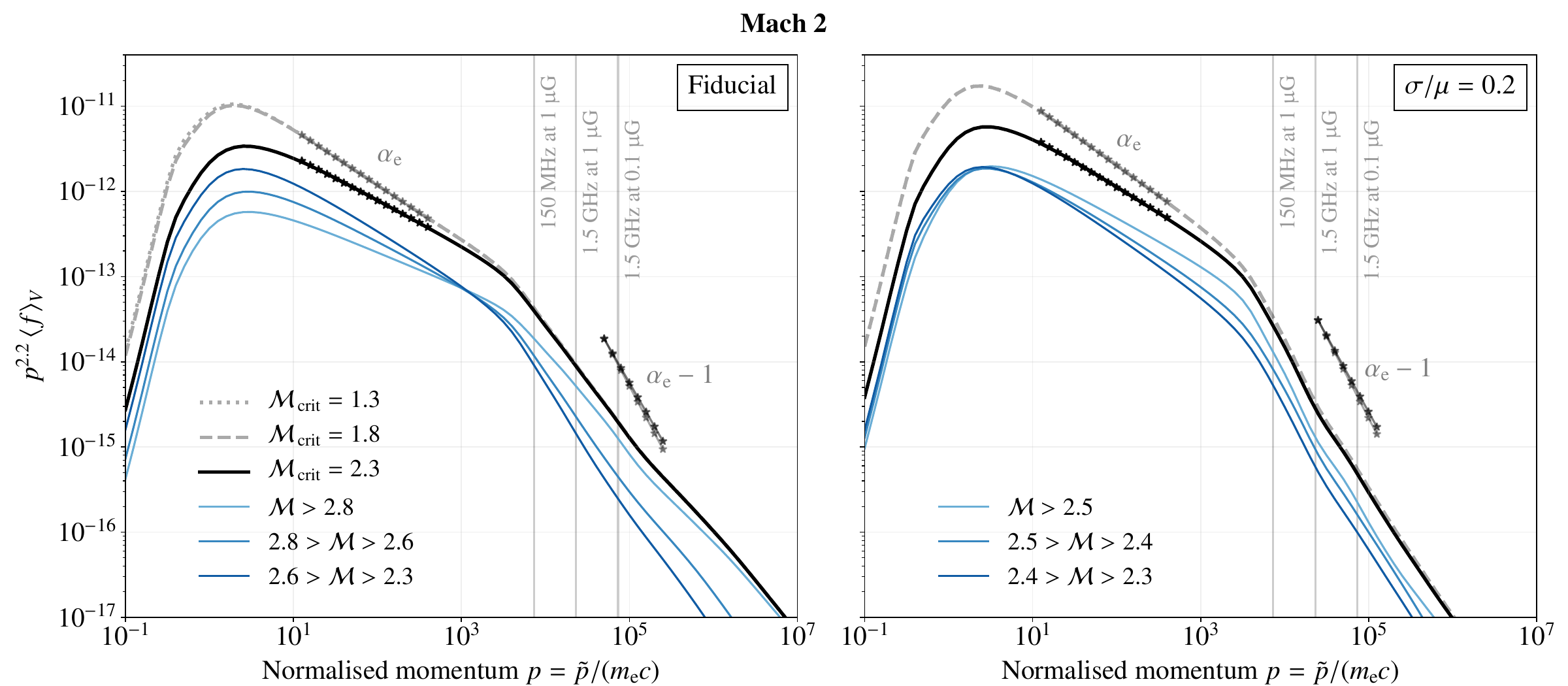}
    \caption[Non-thermal CR electron spectra for each simulation, with spectra overlaid binned by maximum Mach number encountered.]{\textit{Left:} volume-weighted non-thermal CR electron spectra from our fiducial Mach 2 simulations. Dotted, dashed, and solid lines indicate spectra generated with critical Mach numbers of 1.3, 1.8, and 2.3, respectively. Blue lines show contributions to the $\mathcal{M}_\mathrm{crit} = 2.3$ spectrum, where tracers have been binned by the highest Mach number they encountered. Vertical, grey lines indicate momenta that contribute most to a given synchrotron frequency at a given magnetic field strength. \textit{Right:} as previous, except spectra are from the lower relative-variance ($\sigma/\mu=0.2$) Mach 2 simulations, which have a narrower Mach number distribution (see Fig.~\ref{figure:mach-no-pdf--all-sims}). Radio synchrotron frequencies are dominated by the tail-end of the Mach number distribution, to the extent that there is little to no difference between the radio emission generated assuming $\mathcal{M}_\mathrm{crit} = 1.3$ and that generated assuming $\mathcal{M}_\mathrm{crit} = 2.3$.}
    \label{figure:spectra-sorted-by-mach-no}
\end{figure*}

To answer this question, we re-run the Mach 2 fiducial and $\sigma/\mu=0.2$ simulations but now vary the critical Mach number. We choose these simulations as they represent, respectively: i) our best guess for the true ICM parameters at radio relic distances, being informed by cosmological simulations, and ii) the simulation with the greatest number of tracers affected by the $\mathcal{M}_\rmn{crit} = 2.3$ cut-off. We set $\mathcal{M}_\rmn{crit}$ = 1.3\footnote{$\mathcal{M}_\rmn{crit} = 1.3$ corresponds to the minimum threshold of our numerical shock finder.}, 1.8, and 2.3. These are shown by the dotted grey, dashed grey, and black lines in Fig.~\ref{figure:spectra-sorted-by-mach-no}, respectively. It can be seen that, in both simulations, whilst there is a mild reduction in the normalisation at low momenta\footnote{This leads to a mild change in the implied cooling slope $\alpha_\rmn{e} - 1$, as discussed in \citetalias{whittingham2024}.}, there is barely any difference between the lines when cooling starts to dominate ($p \gtrsim 10^4$). This is important for radio relics, as it is this region that radio observations are sensitive to. To illustrate this, we have added a series of vertical grey lines, where these represent the characteristic synchrotron frequency at a given momenta and magnetic field strength:
\begin{equation}
    \nu_\rmn{syn} = \left(\frac{e B}{2\pi m_\rmn{e} c}\right) \gamma^2
    \label{eq:critical-freq}
\end{equation}
where $e$ is the elementary charge, $\gamma = \sqrt{ 1 + p^2}$ is the Lorentz factor, and the term in brackets is the electron gyro-frequency. This represents where most of the synchrotron emission is produced for a given frequency. We have chosen approximate frequencies that radio relics are observed at, and have based the magnetic field strengths on those found in \citet{whittingham2024}. Note, that $\nu_\rmn{syn} \propto B$, and hence an order of magnitude change in either shifts the lines by the same increment. It is unlikely that observations will be able to probe significantly above the cooling-dominated part of the spectrum, as magnetic field strengths with $B > 2 \upmu$G are rare (see Sec.~\ref{chapter6-subsec:emission}) and observations in the ultra-low frequency range ($\lesssim 100$ MHz) are technically challenging\footnote{Observations in this range also have comparably poorer spatial resolution, meaning that they are less able to focus on an area dominated by a strong magnetic field.}. Consequently, we see that the relevant part of the spectra is essentially unaffected by setting $\mathcal{M}_\rmn{crit} = 2.3$.

To understand why this is, we plot the contribution of tracers, where we have binned these by the highest Mach number they encountered\footnote{As discussed in Sec.~\ref{chapter6-subsec:crayon_and_crest}, injection takes place in shock-surface and post-shock cells. Tracers can consequently encounter mild fluctuations in the Mach number as they pass through these whilst the system evolves.}. This binning differs between the two panels. In the left-hand panel we choose bins with $2.3< \mathcal{M} < 2.6$, $2.6< \mathcal{M} < 2.8$, and $\mathcal{M}>2.8$. It can be seen that the lowest Mach numbers contribute more to the lowest normalised momenta, whilst the situation is reversed for the highest Mach numbers. Indeed, most of the spectra at radio emitting frequencies can be explained by Mach numbers with $\mathcal{M} > 2.8$. This bin is almost $\Delta\mathcal{M} = 1$ larger than the true peak Mach number, and represents $<10$\% of the Mach number distribution. For the $\sigma/\mu=0.2$ simulation, the Mach number distribution is narrower, and so we choose more finely-spaced bins. Here, it can be seen that tracers that encountered shocks with $\mathcal{M}>2.5$ produce most of the radio emission. This represents \textit{only $\sim$1.5\% of the Mach number distribution}.

\subsection{Emission}
\label{chapter6-subsec:emission}

\subsubsection{Impact of a critical Mach number, cont.}

These results translate directly to emission maps. We show this in Fig.~\ref{figure:spectra-and-intensity-map--m_crit}, where we present synchrotron intensity maps at 150 MHz and spectral index maps between 325 MHz and 150 MHz\footnote{These are typical values used in radio relic observations, but are also close to the edge of the cooling-dominated part of the spectra. They are therefore most likely to be affected by a change of Mach number (see Fig.~\ref{figure:spectra-sorted-by-mach-no}).}. Here, we have applied $\mathcal{M}_\rmn{crit} = 1.3$ and $\mathcal{M}_\rmn{crit} = 2.3$ in the top and bottom rows, respectively, with the Mach 2 fiducial simulation presented on the left-hand side and the Mach 2 $\sigma/\mu = 0.2$ simulation presented on the right. It can be seen that changing the critical Mach number has little to no impact on the bulk of the emission. In the fiducial simulation, it is evident that most of the emission that is lost originates from the most advanced parts of the shock. This is consistent with the analysis presented in \citetalias{whittingham2024}, where we showed that the shock is weakest at this point. However, the emission levels in this region are approximately $\mathcal{O}(10^3)$ weaker than the peak emission, which is well below the sensitivity of current radio telescopes\footnote{The dynamical range resolved between the brightest and weakest regions in radio relic observations is typically no more than a factor of 50 \citep[see, e.g.][]{deGasperin2022}.}. Moreover, in Fig.~\ref{figure:spectra-and-intensity-map--m_crit}, the simulated radio relic is observed perfectly edge-on; if the relic was observed at an even slightly inclined angle, this region would be dominated by the higher emission behind it. Nonetheless, for an edge-on observation, this figure implies that radio relics can only exhibit ``double strand'' features when the underlying shock is at least as strong as $\mathcal{M}_\rmn{crit}$ \citepalias[see figure B.1.\ of][]{whittingham2024}. This feature should also produce a steepening of the spectral index towards the shock-front, as visible in the top right panel for the fiducial Mach 2 simulation \citetalias[see analysis in][]{whittingham2024}.

\begin{figure}
    \begin{centering}
    \begin{minipage}[t][][t]{.48\textwidth}    
        \centering
        \medskip \par \medskip \par 
        \medskip\small{$\textbf{\;\;Fiducial\;\;} \textbf{(Mach 2)}$}\medskip \par 
        \includegraphics[width=1.05\columnwidth]{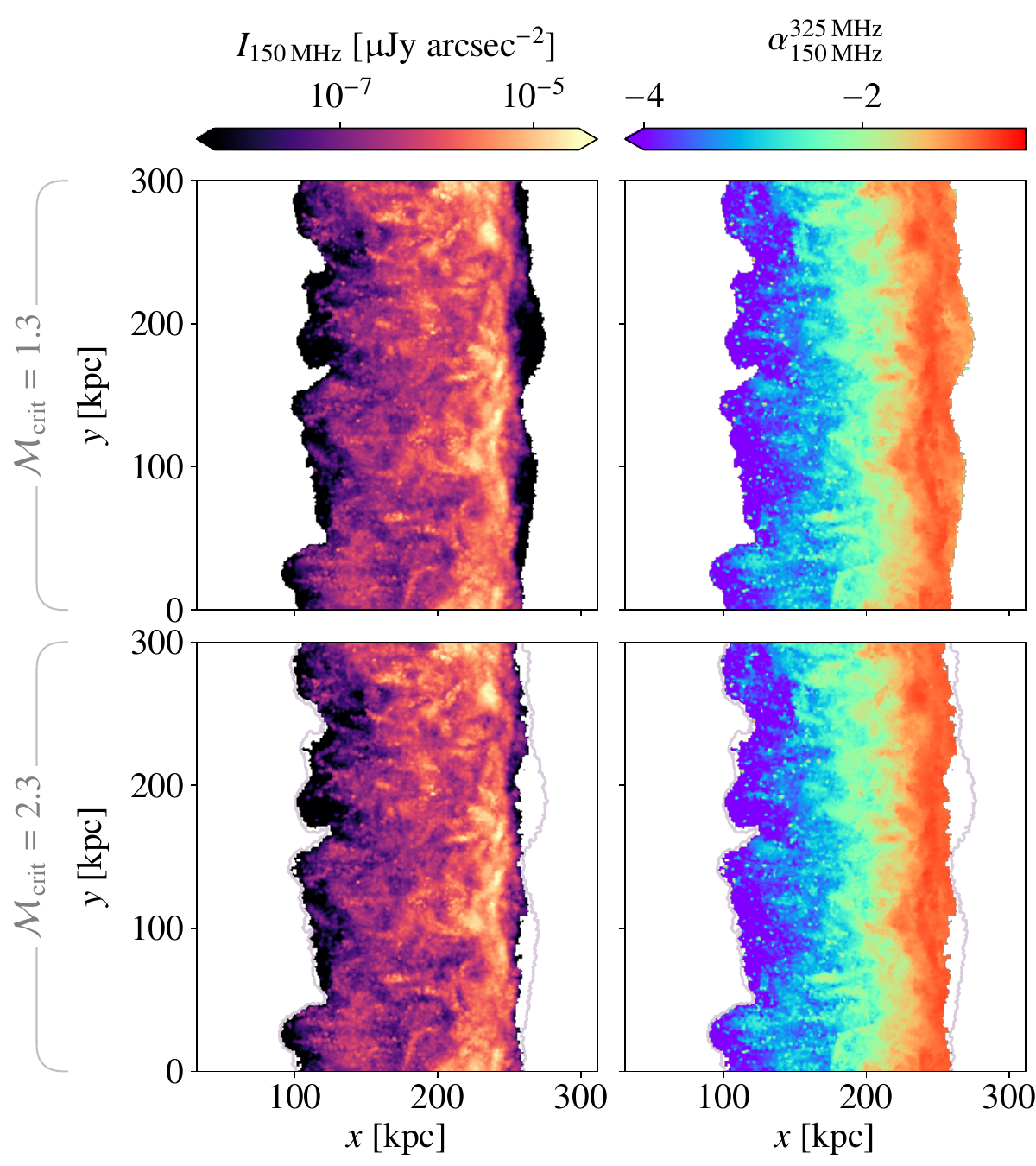}
    \end{minipage}
    \hfill
    \begin{minipage}[t][][t]{.48\textwidth}
        \centering
        \medskip \par \medskip \par 
        \medskip\small{\;\;$\mathbf{\bm{\sigma/\mu =} 0.2\;\;} \textbf{(Mach 2)}$}\medskip \par 
        \includegraphics[width=1.05\columnwidth]{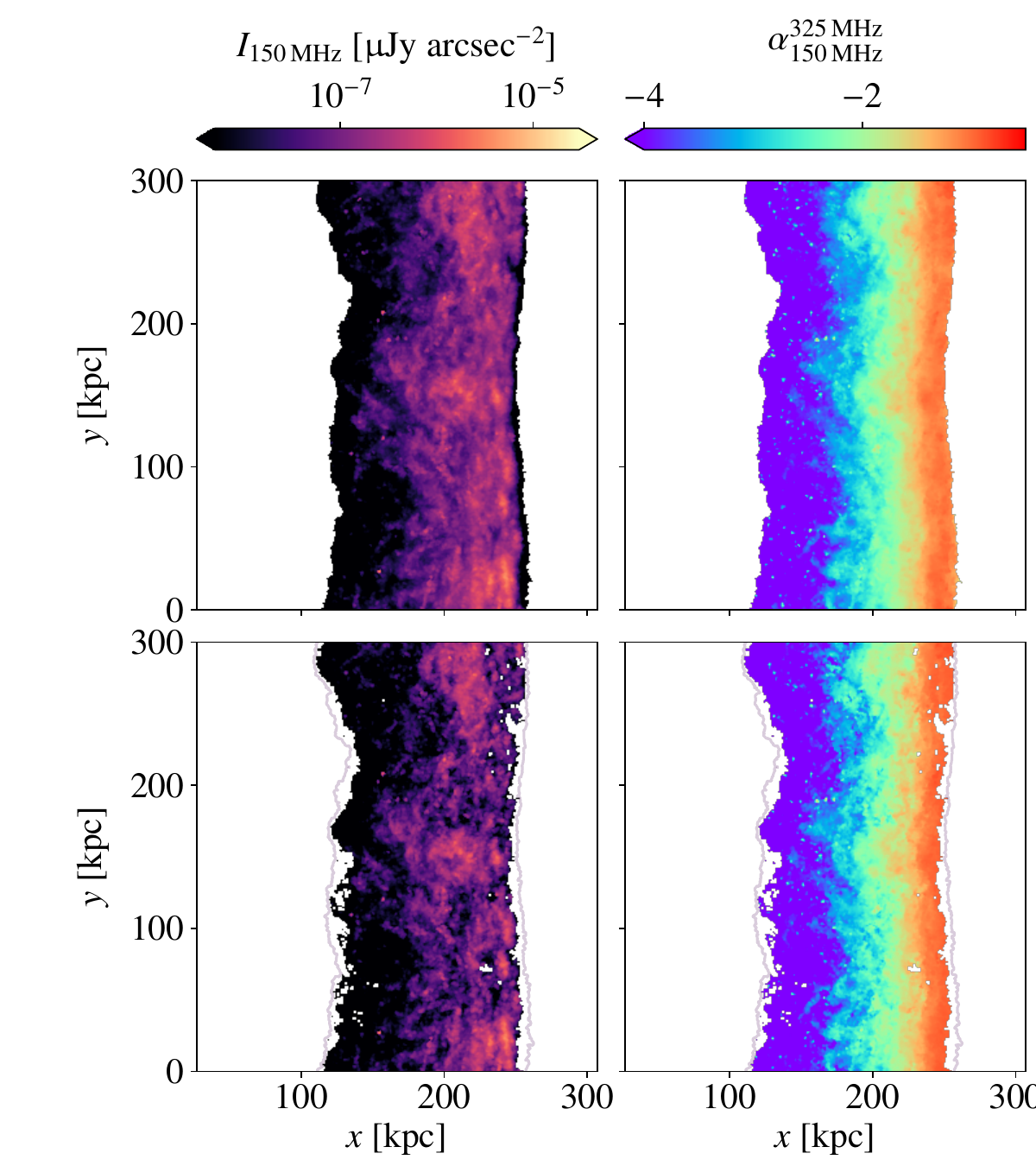}
    \end{minipage}
    \end{centering}
    \caption[Synchrotron intensity and spectral index maps showing that a critical Mach number of $\mathcal{M}_\rmn{crit} = 2.3$ is compatible with observations of radio relics at shocks nominally weaker than this]{\textit{Left:} Synchrotron intensity at 150 MHz ({left column}) and spectral index maps between 325 MHz and 150 MHz (right column) for our fiducial Mach 2 simulation at $t=250$ Myr. The top row has been generated using $\mathcal{M}_\rmn{crit}=1.3$, whilst the bottom row uses $\mathcal{M}_\rmn{crit}=2.3$. Contours in the bottom row indicate the previous extent of the emission. \textit{Right:} as previous, but data is from our lowest relative-variance ($\sigma/\mu = 0.2$) Mach 2 simulation. Although only the very tail of the Mach number distribution contributes to the $\mathcal{M}_\mathrm{crit} = 2.3$ variation, differences between each row are minimal.}
    \label{figure:spectra-and-intensity-map--m_crit}
\end{figure}

In the right-hand panels, we show the data for the $\sigma/\mu = 0.2$ simulation. It can be seen that similar conclusions apply here as well. In particular, increasing the critical Mach number to $\mathcal{M}_\rmn{crit}=2.3$ excludes some emission, but this generally takes place where the emission levels were already very low. Once again, there is a negligible impact on the spectral index values (excluding the lack of emission). For this simulation, the emission does become slightly more patchy, and the total intensity is also slightly reduced. This is consistent with the more noticeable gap between the solid black and dashed grey lines in Fig.~\ref{figure:spectra-sorted-by-mach-no} for this simulation. However, once again, to current telescopes, the differences would be negligible. Moreover, the picture is likely further complicated in reality by pre-existing spatially-varying CR electrons, which may also modulate the downstream emission levels. We therefore conclude that a critical Mach number of $\mathcal{M}_\rmn{crit}=2.3$ is entirely consistent with observations of radio relics at X-ray derived Mach numbers below this -- at least to $\mathcal{M}=2$ and likely significantly further given the breadth of the Mach number distributions analysed in Fig.~\ref{figure:mach-no-pdf--all-sims} and the impact of obliquity-dependent effects. 

\subsubsection{Mock observations}

We now extend our analysis to all the simulations in Table~\ref{tab:simulation_vars}, with the critical Mach number once again being set to $\mathcal{M}_\rmn{crit}=2.3$. In Fig.~\ref{figure:mach-2-variations} and Fig.~\ref{figure:mach-3-variations}, we show synchrotron intensity and spectral index maps for our Mach 2 and Mach 3 simulations, respectively. Intensity maps are shown at 150 MHz, and the spectral index maps are calculated between 1.5 GHz and 150 MHz. Each simulation is shown at $t=250$ Myr. In each row we vary one characteristic of the upstream turbulence, with lower magnitudes of the relevant parameter being found on the left-hand side, and higher magnitudes on the right-hand side. We start with the Mach 2 simulations in Fig.~\ref{figure:mach-2-variations}. 

\begin{figure*}
    \centering
    \small\textbf{Mach 2}\par\medskip 
    \includegraphics[width=1.0\columnwidth]{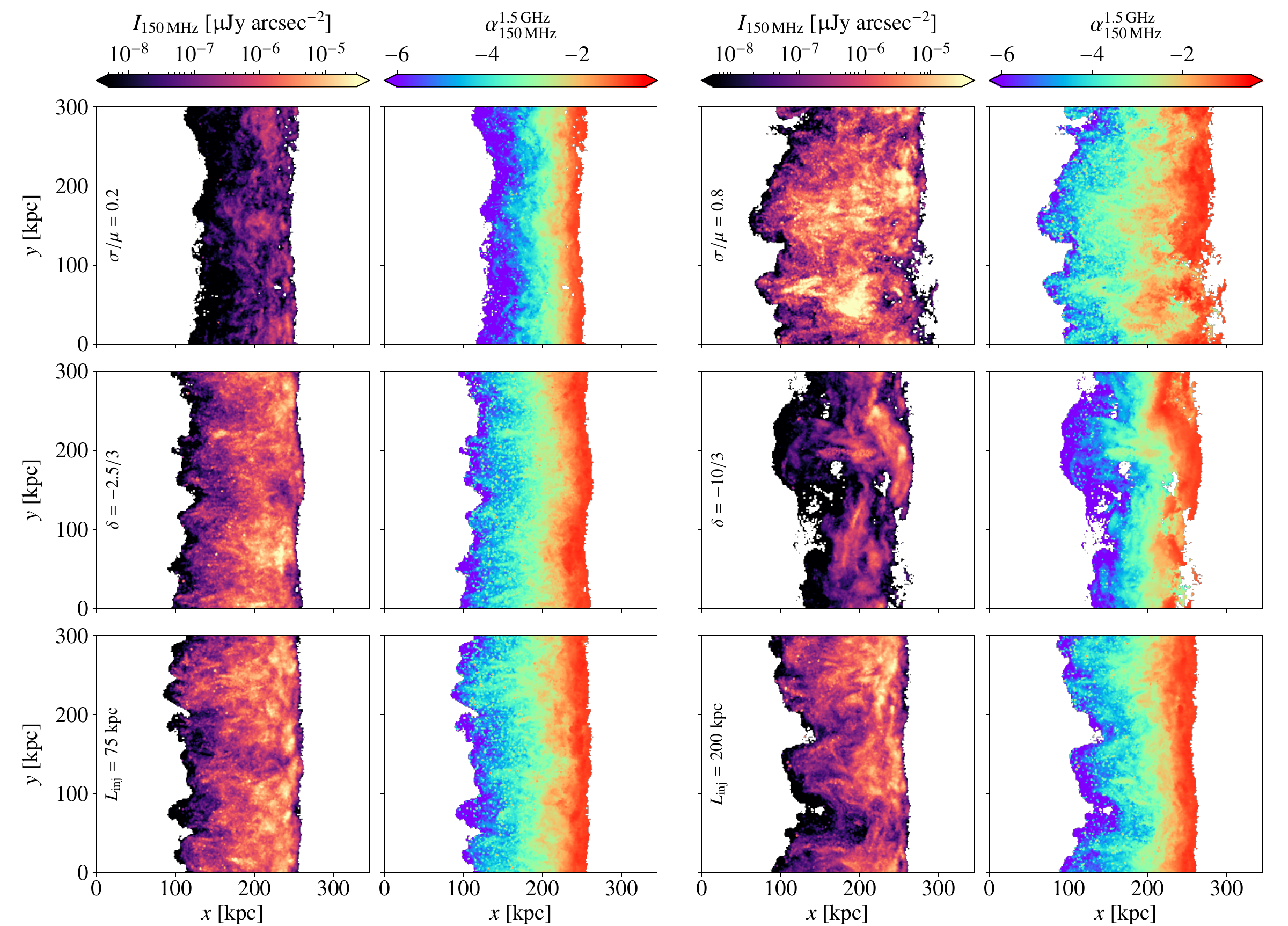}
    \caption[Synchrotron intensity and spectral index maps for all the Mach 2 simulation variations]{Synchrotron intensity maps at 150 MHz and spectral index maps taken between 1.5 GHz and 150 MHz for each of our Mach 2 variations shown at $t=250$ Myr. We vary one characteristic of the upstream density field in each row. From top to bottom, we vary the relative variance, the power law slope, and the injection scale, respectively. Increasing the relative variance both extends the emission and increases its average intensity. Steepening the power law slope meanwhile increases patchiness, which is a result of the large eddy sizes generated downstream. Finally, increasing the injection scale increases the width of the downstream RT ``fingers''.} 
    \label{figure:mach-2-variations}
\end{figure*}

In the first row, we alter the relative variance of the upstream density fluctuations. This has three main effects:
\begin{enumerate}[i)]

\item \textbf{Peak emission:} It can be seen that increasing the relative variance significantly increases the peak emission, by a factor of roughly 100. It also increases the average intensity, with substantially higher emission seen further downstream. This is consistent with Fig.~\ref{figure:spectra-all-models}, where we showed that increasing the relative variance leads to flatter spectra, with correspondingly higher normalisation at radio-emitting frequencies. 

\item \textbf{Extent of emission:} Increasing the relative variance also significantly extends the downstream emission, which is consistent with our hydrodynamical analysis in Fig.~\ref{figure:density-slices}. Here, the emission in the  $\sigma/\mu = 0.8$ simulation extends twice as far as in the $\sigma/\mu = 0.2$ analogue. This ratio only increases if we apply a lower limit to the intensity; for example, if we measure only emission with $I_\rmn{150\,MHz} > 10^{-7}$~$\upmu$J~arcsec$^{-2}$, we find that the $\sigma /\mu = 0.8$ simulation is barely affected, whilst the extent of the emission in the $\sigma/\mu = 0.2$ simulation is roughly halved. This is consistent with observations by \citet{kang2017} and \citet{deGasperin2022}, who find that the ``brush'' section of the Toothbrush radio relic is $\sim2\times$ wider than expected by standard cooling theory at 650 MHz\footnote{The shortened cooling timescale at higher frequencies should significantly reduce the extent of the emission in the $\sigma /\mu = 0.8$ simulation, as the back of the shock-compressed region is generally populated with older electrons \citepalias[see][]{whittingham2024}.}, and $\sim4\times$ wider at 58 MHz, respectively. Moreover, changing the relative variance may explain the varying downstream width in some radio relics. For example, it would naturally explain the ``handle'' and ``brush'' of the Toothbrush relic. This interpretation is also consistent with polarisation data, which shows that the brush is substantially less polarised compared to other parts of the relic\footnote{This is usually interpreted as being a result of Faraday depolarisation as a result of the ``brush'' section of the relic being behind the cluster \citep[see, e.g.][]{rajpurohit2020b, rajpurohit2022, hoeft2022}. In our interpretation, however, at least some of the lower polarisation is intrinsic, resulting from the increased downstream velocity turbulence (see Fig.~\ref{figure:turbulence-pdf}).} \citep{kierdorf2017}.

\item \textbf{Mixing:} Finally, as velocity turbulence is substantially more developed in the $\sigma/\mu = 0.8$ simulation (see Fig.~\ref{figure:turbulence-slices}), the mixing of CR electrons of different ages is also stronger. This is evident in the spectral index maps, where the gradient is significantly less coherent in the higher relative variance run. At its strongest, this turbulence disrupts the observation of fresh spectra close to the shock-front, as can be seen at $y\approx80$~kpc. It also leads to a situation where the highest intensity emission is not necessarily where the freshest spectra is, as would be expected in a simulation without upstream density turbulence.

\end{enumerate}

In the second row, we vary the power law slope. On the left-hand side, the slope is shallower, and hence there are more fluctuations on small-scales. Whilst on the right-hand side, the slope is steeper, and hence the density fluctuations have less small-scale power. In Fig.~\ref{figure:density-slices} and~\ref{figure:turbulence-slices}, we showed that this changed the scale of the typical eddies produced downstream. 
This effect can be seen particularly clearly in the $\delta=-10/3$ simulation in Fig.~\ref{figure:mach-2-variations}. Here, the emission is significantly patchier, and less intense on average. 
This is, at least partially, a result of the size of the eddies in the $\delta=-10/3$ simulation. As shown in Fig.~\ref{figure:density-slices}, these have outer scales of $L_\rmn{inj} / 2 = 75$ kpc and spiral inwards, as is typical of Kelvin-Helmholtz generated vortices. The large eddies result in patchier emission, as they surround regions without any injected tracers. Moreover, it causes curved emission features, as seen at $y\approx 180$ kpc, and curved boundaries and ``notches'' in the spectral index maps, as seen close to the shock front. When these eddies form too close to the shock front, they also act to disrupt it, as can be seen in both the intensity and spectral index maps at $y \lesssim 100$ kpc.

Finally, in the third row, we vary the scale of turbulent injection. Once again, the features here reflect our analysis in Fig.~\ref{figure:density-slices} and~\ref{figure:turbulence-slices}. In particular, the width of the downstream emission varies according to the size of the RT ``fingers'' formed, where this width is approximately equal to $L_\rmn{inj}$, as long as $L_\rmn{inj}$ is greater than the typical eddy scale. This effect is, however, only really noticeable for relatively steep ($\alpha < -2$) spectral indices; above this range, it is significantly harder to determine the differences. This puts the effect beyond the typical limits for current observations \citep[see, e.g.,][]{rajpurohit2020}. Moreover, the formation of such ``fingers'' in the emission maps requires a depth of projection through the radio relic such that $l_\rmn{depth} \lesssim L_\rmn{inj}$, so that lines of sight will generally only pass through a single RT ``finger''. When observational capabilities increase, this may help place constraints on $l_\rmn{depth}$, and thereby help constrain the typical dimensions of radio relics.

\begin{figure*}
    \centering
    \small\textbf{Mach 3}\par\medskip 
    \includegraphics[width=1.0\columnwidth]{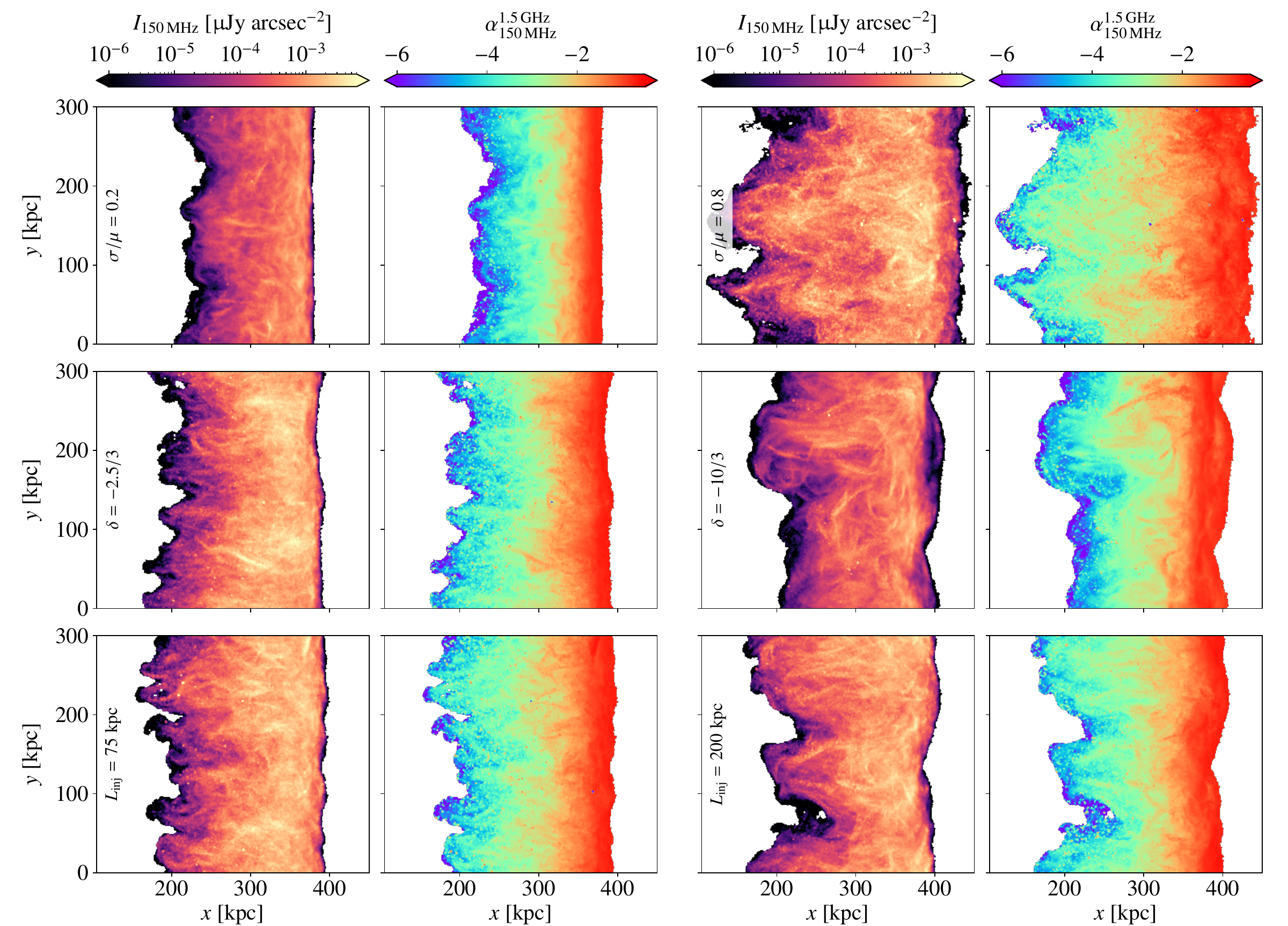}
    \caption[Synchrotron intensity and spectral index maps for all the Mach 3 simulation variations]{As Fig.~\ref{figure:mach-2-variations} but for our Mach 3 simulations. In general, the same conclusions hold, although differences between each variation are now more subtle. Emission is more extended in each case relative to the corresponding Mach 2 simulation. Filamentary emission due to magnetic field amplification is also more evident.}
    \label{figure:mach-3-variations}
\end{figure*}

In Fig.~\ref{figure:mach-3-variations}, we show the analogous plot for our Mach 3 simulations\footnote{Note, we have re-scaled the colorbar here by the average increase in emission, but have kept the same dynamic range.}. Generally, our conclusions from Fig.~\ref{figure:mach-2-variations} apply here as well. In particular, higher relative variance increases the length of the downstream emission, a steeper power law slope produces more curved emission features, and a larger turbulent injection scale leads to large RT ``fingers''. There are some developments from Fig.~\ref{figure:mach-2-variations} as well, however. In particular, we highlight the following changes:
\begin{enumerate}[i)]

\item \textbf{More extended and less patchy emission:} The emission is more extended and less patchy in every case. Both of these properties result mainly from the increased shock speed, where this is approximately 1.5$\times$ the Mach 2 case. This increases the speed of the tracers downstream, resulting in more effective mixing and fresher spectra at farther distances downstream. At the same time, the shock advances farther, extending the distance between the shock front, and also increasing the overall number of tracers injected. This makes it less likely that a line-of-sight will only pass through low energy-density CR electrons.

\item \textbf{Curved shock front:} As the peak of the Mach number distribution is higher than $\mathcal{M}_\rmn{crit} = 2.3$ in these simulations, the vast majority of tracers are injected in all models. Indeed, by comparing with Fig.~\ref{figure:mach-no-pdf--all-sims}, we can see that it is really only $\sigma/\mu=0.8$ that is affected in any significant way. Consequently, it is only in this case that the shock-front is not particularly smooth, resulting from the low Mach numbers in this region. In the remaining simulations, the shock-front is fully represented by the injected tracers. This means that the projected emission reflects the shock corrugation, as analysed in Sec.~\ref{subsec:shock-corrugation} (in contrast to the patchier shock-fronts seen in Fig.~\ref{figure:mach-2-variations}). This is particularly evident at the shock-front in the $L_\rmn{inj}=75$~kpc and $L_\rmn{inj}=200$~kpc simulation and can be seen to a slightly lesser extent in the intensity map of the $\delta=-10/3$ simulation.

\item \textbf{Filamentary nature:} Filaments, as analysed in \citetalias{whittingham2024}, are evident in both the intensity and spectral index maps for all simulations. This is in contrast to the Mach 2 simulations, which generally produced relatively patchy emission instead. This is likely a result of the aforementioned increased mixing, which reduces emission patchiness, coupled with the increased magnetic field strengths (see Sec.~\ref{subsec:inferred-magnetic-field}) resulting in brighter filaments. It can be seen that, except in the case of the $\delta=-10/3$ simulation, the filaments are generally aligned with the shock propagation direction. For, $\delta=-10/3$ the filaments tends to be more curved, with a substantial number laying parallel to the shock surface. This is a result of the large eddies, as analysed in Fig.~\ref{figure:density-slices}. 

\end{enumerate}

\subsubsection{Inferred magnetic field}
\label{subsec:inferred-magnetic-field}
 
\begin{figure*}
    \centering
    \includegraphics[width=1.0\columnwidth]{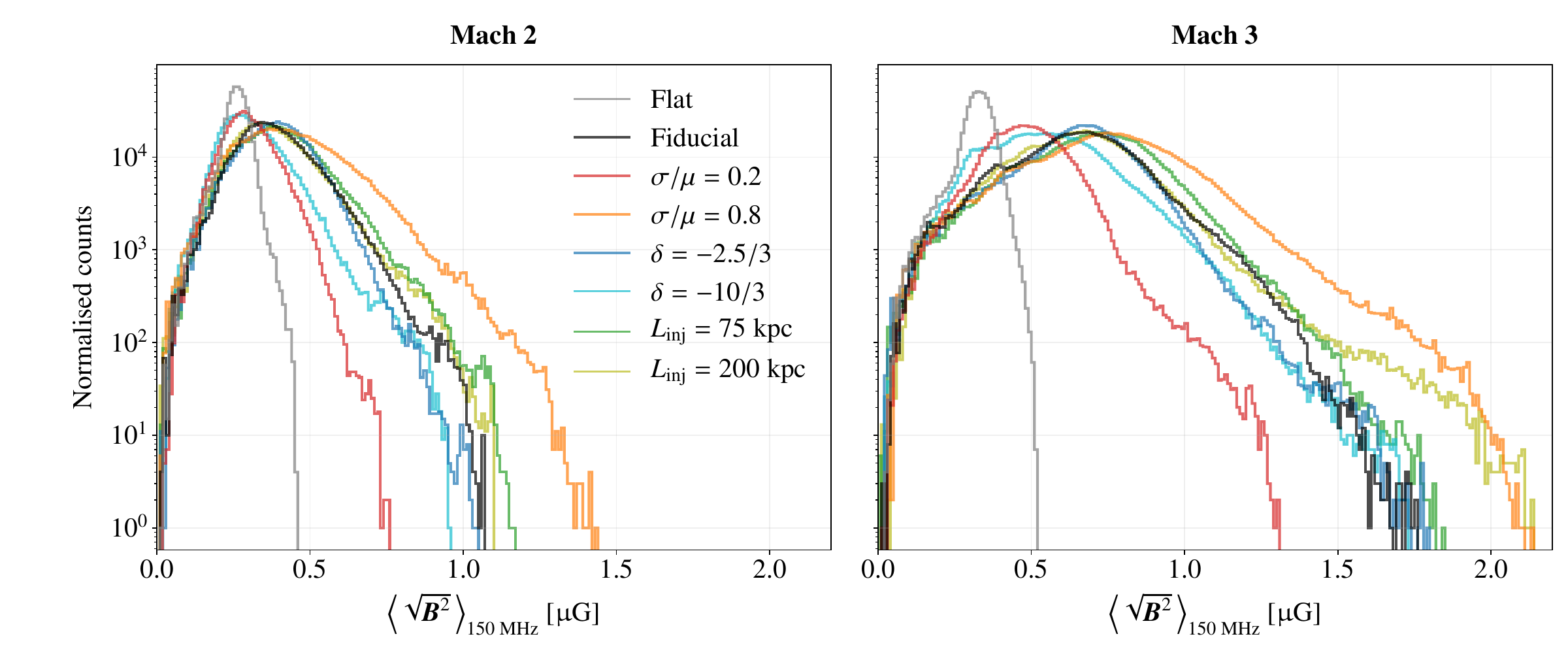}
    \caption[Probability density distributions of the projected magnetic field strength for all simulation variations, where projections are weighted by the synchrotron emission at $\nu = 150$ MHz ]{\textit{Left:} Probability density distributions of the projected magnetic field strength, where projections are weighted by the synchrotron emission at $\nu = 150$ MHz. Each line represents a Mach 2 simulation at $t = 250$ Myr. \textit{Right:} as previous, but for the Mach 3 simulations. Relative variance has the greatest impact on the distribution. In many cases, the tail end of the distribution easily reaches $\upmu$G-strength.}
    \label{figure:projected-B-field-histogram}
\end{figure*}

In \citetalias{whittingham2024}, we showed that the projected synchrotron emission is dominated by the strongest magnetic field values along the line-of-sight. This likely causes the filamentary emission as shown above. The effect is particularly well seen by plotting probability density functions of the synchrotron-weighted magnetic field strength. We do this in Fig.~\ref{figure:projected-B-field-histogram} for all models using the 150 MHz channel. We show the Mach 2 simulations of the left-hand side, and the Mach 3 simulations on the right-hand side. We also include data from the ``Flat'' simulations, which have no upstream density fluctuations \citepalias[see][]{whittingham2024}. It can be seen that simulations with upstream density turbulence produce higher maximum magnetic field values than the ``Flat'' simulation, reaching $\upmu$G strengths for many different parameters. Moreover, the distribution of values peaks at higher strengths too. Once again, it can be seen that the relative variance is the most influential factor, with the highest relative variance producing the highest average field strengths. This likely originates from the compression produced in this simulation (see Fig.~\ref{figure:density-slices}), as we showed in \citetalias{whittingham2024} that amplification of the field from the initial conditions was predominantly produced by adiabatic compression. 

We also showed in \citetalias{whittingham2024}, however, that the peak magnetic field values could \textit{not} be explained by adiabatic compression alone, thereby hinting at the action of a small-scale dynamo. We can see the impact of this in the right-hand panel, where magnetic field strengths increases with Mach number. This is partially caused by the extra adiabatic compression, generated by the higher density peaks, and partially by a small-scale dynamo, which is more efficient in more turbulent environments. Indeed, this likely produces the bulging of the probability distribution functions to the right at normalised counts $\lesssim 10^2$. Of particular interest here are the peak values of the $L_\rmn{inj} = 200$~kpc simulation, which match those of the $\sigma/\mu = 0.8$ simulation in the right-hand panel, but not in the left. This can be understood by comparing the turbulence generated by both simulations in Fig.~\ref{figure:turbulence-slices}; RT-generated flows in the $L_\rmn{inj} = 200$~kpc simulation produce substantial turbulence between and at the base of the RT ``fingers''. This is not as effective at slower shock speeds and unable to take place in the $L_\rmn{inj} = 75$~kpc simulation, where RT ``fingers'' cannot properly form. Additional simulations will be required to see whether this turbulence increases at $L_\rmn{inj} = 500$~kpc and above, which appears to be a more likely value for real radio relics (see Sec.~\ref{subsec:shock-corrugation}).

Finally, we note that, as previously, the $\delta=-10/3$ simulation shows a slightly different trend. Here, magnetic field strengths peak at lower values compared to the other simulations with a relative variance of $\sigma/\mu \geq 0.4$. We attribute this to the 
reduction of small-scale turbulence (see Sec.~\ref{chapter6-subsec:downstream-turbulence}).

\subsubsection{Spectral flux density}

\begin{figure}   
    \centering
    \includegraphics[width=0.5\columnwidth]{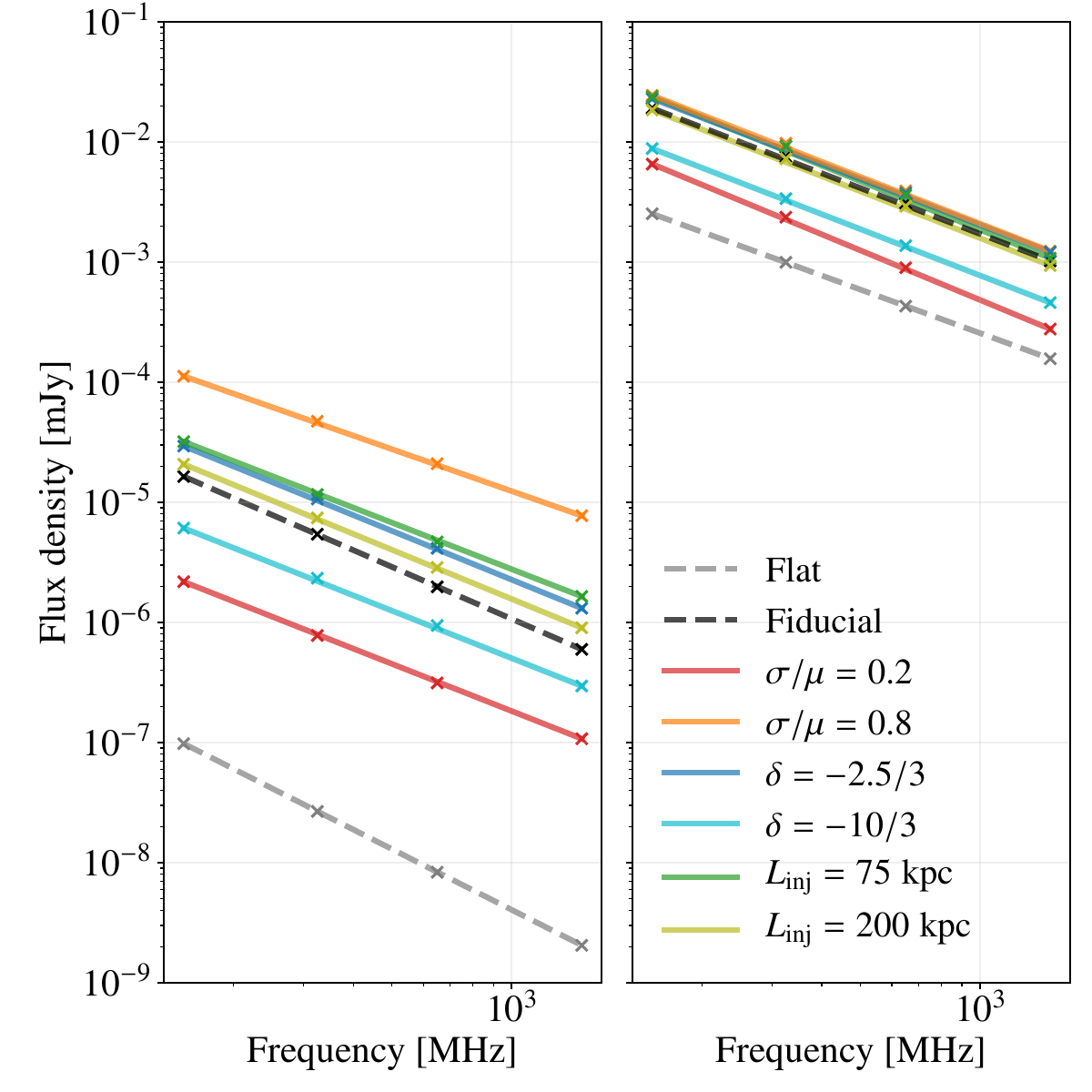}
    \caption{\textit{Left:} the flux density at 150 MHz, 325 MHz, 650 MHz, and 1.5 GHz for our Mach 2 models assuming an LLS equivalent to the Toothbrush relic (see text). 
    \textit{Right:} as previous, but for our Mach 3 variations. Adding upstream turbulence can substantially increase the spectral flux density; this is especially true of low-Mach number shocks. Nonetheless, Fermi-I re-acceleration of a fossil CR electron population is still required in order to match the observed spectral brightness of real relics.}
    \label{figure:flux-density}
\end{figure} 

In the final section of our analysis, we investigate the impact of our models on the spectral flux density. To this end, in Fig.~\ref{figure:flux-density}, we show the flux density at 150 MHz, 325 MHz, 650 MHz, and 1.5 GHz as crosses, with the Mach 2 models shown on the left-hand side and the Mach 3 models shown on the right. This is calculated as:
\begin{equation}
    S_\nu = \int_{\Omega_S} I_\nu (\Omega) \mathrm{d}\Omega \equiv \langle I_\nu \rangle \Omega_S,
\end{equation}
where is $I_\nu$ is the specific intensity at frequency $\nu$ (see Sec.~\ref{chapter6-subsec:crayon_and_crest}), and $\Omega_S$ is the solid angle that the relic extends on the sky, where we have assumed that this is sufficiently small so that we can use the flat-sky approximation. To aid comparison with observations, we have assumed that our simulated radio relics are at the location of the Toothbrush radio relic, such that $\ang{;;1}$ corresponds to a physical scale of 3.64 kpc \citep{rajpurohit2020}, and have assumed that the emission covers the same extent as this relic, i.e. 1860~$\times$~422 kpc \citep{rajpurohit2020}. This ignores the relative downstream extent of each simulation, but provides a good first approximation of the differences. Finally, we have joined the 150 MHz and 1.5 GHz data points with a straight line.

It can be seen that, first of all, virtually all of the models produce power-law emission, as observed for real radio relics \citep{rajpurohit2020b, rajpurohit2022}. Only the Mach 2 ``$\delta = -10/3$'' model deviates from this to any significant degree, with some indication of negative curvature, as data points for 325 MHz and 650 MHz fall slightly above the straight line. This is, however, far from conclusive, and so does not necessarily rule this model out. It can also be seen that, as in previous examples, the relative variance of the turbulence has the greatest impact. Indeed, the highest relative variance we have modelled, $\sigma/\mu=0.8$, has a flux density at 150 MHz a factor of $10^3$ higher than the ``Flat'' simulation. This rises to a factor of almost $10^4$ by 1.5 GHz. This is predominantly a result of the shallower spectral slope of the CR electron spectrum, as shown in Fig.~\ref{figure:spectra-all-models}. At Mach 3, the trend continues, albeit the difference between the $\sigma/\mu=0.8$ ad the ``Flat'' simulation is now little more than a factor of 10. This is also consistent with our analysis of Fig.~\ref{figure:spectra-all-models}. Indeed, in the Mach 3 simulations, there is very little difference between the simulations with $\sigma/\mu=0.4$ and $\sigma/\mu=0.8$ (see Table~\ref{tab:simulation_vars}). This implies that the ability of turbulence to boost the spectral flux beyond this point is probably limited. We conclude therefore, that turbulence does not replace the need for Fermi-I re-acceleration from a fossil CR electron population \citep[see, e.g.][]{pinzke2013}. Indeed, observations of the Toothbrush relic find that the spectral flux density at 150 MHz is $S_\rmn{150\,MHz} \approx 4\times10^3$~mJy, which is a further five orders of magnitude above\footnote{We note, however, that this could be reduced by implementing a minimum intensity cut, such that pixels below this value are removed from the average. This would likely also produce a fairer comparison with observations.}.

\begin{figure}  
   \centering
    \begin{tabular}{c||cc}
    \hline
    \textbf{Simulation name}   & \textbf{Mach 2} & \textbf{Mach 3} \\ \hline
    Flat                       & -1.68  & -1.21  \\
    Fiducial                   & -1.44  & -1.27  \\
    $\sigma/\mu = 0.2$         & -1.31  & -1.38  \\
    $\sigma/\mu = 0.8$         & -1.16  & -1.30  \\
    $\delta=-2.5/3$            & -1.35  & -1.28  \\
    $\delta=-10/3$             & -1.21  & -1.28  \\
    $L_\rmn{inj}=75$ kpc       & -1.29  & -1.35  \\
    $L_\rmn{inj}=200$ kpc      & -1.36  & -1.30  \\
    \end{tabular}
    \caption{Table indicting the spectral indices measured for the lines in Fig.~\ref{figure:flux-density}.}
    \label{tab:spectral-slopes}
\end{figure}

In Table~\ref{tab:spectral-slopes}, we have recorded the spectral indices of the lines presented in Fig.~\ref{figure:flux-density}. There is a general trend in these indices to get shallower with increasing Mach number, but this is inconclusive, owing to the large scatter. More important, however, is the fact that all the simulations with upstream density fluctuations produce slopes within the range observed for real radio relics \citep[see, e.g.][]{vanweeren2011, vanweeren2011b, kierdorf2017}. In particular, the Mach 2 ``$\sigma/\mu = 0.8$'' simulation exactly matches the slope recorded by \citep{rajpurohit2020b} for the Toothbrush radio relic. Radio observations imply $\mathcal{M}=3.7$ \citep{rajpurohit2020b}, whilst X-ray observations imply a significantly weaker shock of $\mathcal{M}\approx1.5$ \citep{ogrean2013, vanweeren2016, itahana2017}. This radio relic therefore appears to be an extremely likely case of a low Mach number shock with a broad Mach number distribution.

\section{Conclusions}
\label{chapter6-sec:conclusions}

Radio relics represent a potentially underutilised probe of ICM conditions in the cluster outskirts ($\gtrsim1$~Mpc). They have $\upmu$Jy~arcsec$^{-2}$ surface brightnesses at 1.4 GHz and exhibit a wide range of morphological features, which develop as a result of interplay with their environment. Whilst it seems likely that the large-scale ($\sim$Mpc) morphology is set by the underlying shock, it is still unclear what generates features on scales smaller that this. Moreover, recent results from PIC simulations have led to the suggestion that radio relics should be unable to form at all below in shocks with $\mathcal{M} < \mathcal{M}_\rmn{crit} = 2.3$, which is in apparent contradiction with X-ray observations, which suggest lower Mach numbers for the majority of theses systems. In this paper, we show that such ``small'-scale features are intrinsically linked to the upstream density turbulence. To show this, we have expanded on the idealised shock-tube simulations first presented in \citet{whittingham2024}, which, in turn, are constructed based on the analysis of shocks in cosmological simulations. In our expansion, we vary the upstream density field by precisely controlling the parameters of its power spectra. Specifically, we vary the:
\begin{enumerate}[i)]
\item relative variance
\item power law slope, and
\item injection scale
\end{enumerate}
of the density turbulence (see Fig.~\ref{figure:power-spectra} and~\ref{figure:electron-density-pdf}). We explore this parameter space by testing each parameter individually. This results in six simulations, with an additional ``Fiducial'' simulation probing the midpoint, where this represents our best guess of the true values in observed radio relics, and is informed by cosmological simulations and observations. We have further tested the parameter space by running each simulation twice; once with a Mach 2 shock and once with a Mach 3 shock, where these values represent the expected shock strength in the absence of density fluctuations.

Each of these simulations is then post-processed with the CR electron spectral solver \textsc{Crest} \citep{winner2019}. In our setup, we use this to model advection, DSA with magnetic-obliquity dependence, and cooling via Coulomb, bremsstrahlung, inverse Compton, and synchrotron losses. Emission is then generated from each spectrum using the \textsc{Crayon+} code \citep{werhahn2021}. We thereby produce synchrotron emission maps with a minimum freedom of input parameters.

We find that the relative variance is the most influential factor, with a higher relative variance producing:
\begin{enumerate}[i)]
\item more extended emission in the downstream (Figs.~\ref{figure:density-slices}, ~\ref{figure:mach-2-variations}, and~\ref{figure:mach-3-variations}),
\item a broader Mach number distribution (Fig.~\ref{figure:mach-no-pdf--all-sims}),
\item greater shock corrugation (Figs.~\ref{figure:turbulence-slices} and~\ref{figure:mach-no-pdf--all-sims}),
\item higher downstream velocity turbulence (Figs.~\ref{figure:density-slices} and~\ref{figure:turbulence-slices}),
\item a shallower electron spectral slope (Fig.~\ref{figure:spectra-all-models}),
\item increased peak emission and total surface brightness (Figs.~\ref{figure:shock-gaps},~\ref{figure:mach-2-variations},~\ref{figure:mach-3-variations}, and~\ref{figure:flux-density}), and
\item a more amplified magnetic field (Fig.~\ref{figure:projected-B-field-histogram}).
\end{enumerate}
In particular, we point out that i) solves a well-known problem regarding the downstream emission length, which is found to be incompatible with standard theory \citep[see, e.g.][]{kang2016, kang2017, deGasperin2020, deGasperin2022}.

We find that the remaining variables are less influential, however, a mild effect is produced by steepening the slope of the density power spectrum. In particular, this leads to larger eddies, which produced curved filaments in emission maps (Figs.~\ref{figure:density-slices},~\ref{figure:mach-2-variations}, and~\ref{figure:mach-3-variations}). Moreover, we find that the scale of turbulent injection sets the corrugation of the shock along its front (Figs.~\ref{figure:shock-filaments}, and~\ref{figure:mach-3-variations}). This corrugation is observable in real edge-on radio relics through the length scale of ``double-strand'' features and the spacing of ``knots'' in emission. By comparing observations with our simulations, we find that the turbulent injection scales of between 500--600~kpc are common. This is significantly higher than the value commonly assumed.

Finally, we have used our simulations to probe the \textit{critical Mach number} problem. We find that electron spectra at radio-emitting frequencies are dominated by the tail of the Mach number distribution (Fig.~\ref{figure:spectra-sorted-by-mach-no}). Indeed, 99\% of the total emission is produced by the top 10\% of the distribution. The result is that, for a Mach 2 shock, radio relics produced using a critical Mach number of $\mathcal{M}_\rmn{crit} = 2.3$ are observationally indistinguishable from those produced when $\mathcal{M}_\rmn{crit} = 1$ (Fig.~\ref{figure:spectra-and-intensity-map--m_crit}). Taken together, with the well-known obliquity-dependence of X-ray derived Mach numbers \citep[see, e.g.][]{wittor2021}, and the typical width of the Mach number distribution as shown in this paper, we find that there is no contradiction between $\mathcal{M}_\rmn{crit} = 2.3$ and current observations.

Our results show that radio relics are highly sensitive to ICM conditions. In future work, we will apply them to the modelling of observed radio relics, thereby further improving our understanding of their environment and formation process.
\ChapterX{Additional work}{}
\label{chapter:additional-work}

\section{\textsc{Crest}}

The cosmic ray electron spectra code, $\textsc{Crest}$, has been crucial to the results given in the preceding two chapters. This was originally published in \citet{winner2019} and was employed in \citet{winner2020} to show that supernovae observations are most consistent with quasi-parallel (rather than quasi-perpendicular) electron acceleration. These two papers acted as proof of concept for the code. However, it was not yet ready for high-resolution production-runs or for cosmological simulations. During my doctoral studies I have extended and updated the code to allow just this. In particular, \textsc{Crest} is now at the stage that it can now be used to investigate the origin of fossil electrons in radio relics. This will allow to revisit the results presented in \citet{pinzke2013} using a fully spectral solver.

During my doctoral studies, I have made the following updates to \textsc{Crest}\footnote{The development of \textsc{Crest} has very much been a team effort. Debugging and code development of \textsc{Crest} has been predominantly led by myself, Maria Werhahn, and L\'{e}na Jlassi. Whilst implementation and conceptual developments have also involved Christoph Pfrommer. Finally, Philipp Girichidis wrote the original framework for the test pipeline (see Sec.~\ref{sec:test-pipeline}.), and was involved in conceptual development during the original code development by Georg Winner. Here, I provide a list of changes that are either wholly or predominantly a result of my own work.}:

\subsection{Re-structuring the input data}

\textsc{Crest} is a post-processing code, meaning that data is recorded from the simulation, and then spectra are evolved afterwards. This is possible, as the cosmic ray electrons are not dynamically important in the simulation (see Sec.~\ref{sec:cosmic-rays}). This method has the advantage that parameters and models can be altered without the need to re-run the whole simulation. Due to the short cooling times for cosmic ray electrons at low and high momenta, however, data must be saved on the (magneto)hydro-timestep. This can lead to large amounts of recorded data over time, particularly in high-resolution, dynamic, or long simulations. This problem was compounded in the original version by the data structure used.

In the previous set-up, a relatively simple data structure was implemented, in which data were saved in blocks. Each block represented a single timestep and contained arrays, where each array stored a single variable type. The arrays, in turn, were ordered by the tracer ID, and were always $N_\rmn{tp}$ entries long, where $N_\rmn{tp}$ is the number of tracers used in the simulation. As the whole file was saved in binary format, the data recorded during the trajectory of an individual tracer could then be read by ``striding'' over blocks, as a variable for a given tracer was always in the same place within a block. This structure allowed for relative simple MPI parallelisation. However, it meant that \textsc{Arepo} simulations needed to be run with a uniform timestep, so that tracers were always active \citep[see][for details]{genel2013} and always recorded data -- else the ``striding'' method would fail. This naturally led to much superfluous data being recorded.

In the updated version of \textsc{Crest}, \textsc{Arepo} simulations can be run with tree-based timesteps, where particles are active only if the dynamical time requires it. Data is hence recorded in irregular sizes, depending on the number of tracers active. To make the navigation of this data more user-friendly, the recorded data are now saved using the \href{https://github.com/HDFGroup/hdf5}{HDF5 format} \citep{HDF5}. As well as allowing for the attachment of metadata, including variable names, this format also includes built in libraries to maximise data compression and deal with read- and write functionality. To minimise the size of recorded files, data is now stored in arrays for each variable, with timesteps added end-to-end\footnote{Earlier experiments with so-called ``ragged arrays'' were unable to provide enough data compression.}. A schematic providing an example of this format is shown in Fig.~\ref{fig:crest-array}. To allow \textsc{Crest} to read the file, the starting index of each new timestep is stored in a separate array (here referred to as ``Index''). This mapping is represented by the arrows in Fig.~\ref{fig:crest-array}. The data for the $n$th timestep can then be found between \textit{Index}[$n-1$] and \textit{Index}[$n$] -1, where we assume zero-based indexing, and where \textit{Index}[-1] is implicitly assumed to be 0.

Data for a given timestep is stored corresponding to the indexes just described. In Fig.~\ref{fig:crest-array}, this means that data is arranged in columns, where each column represents an individual tracer particle. Variables are only stored when \textsc{Arepo} has been run with the appropriate configuration flag. For example, shock data necessary for \textsc{Crest}'s DSA routine is only stored when \textsc{Arepo} is compiled with \mbox{\textit{COSMIC\_RAYS\_SHOCK\_ACCELERATION}}. Data for this routine is further kept to a minimum, by only storing related variables when the corresponding ``shock flag'' indicates a shock-surface of post-shock cell, respectively; for example, variables such as energy injected, and pre- and post-shock densities are only stored when the corresponding shock flag is triggered. The mapping is represented by the  bottom set of arrows in Fig.~\ref{fig:crest-array}.

\begin{figure}
    \centering
    \includegraphics[width=0.6\linewidth]{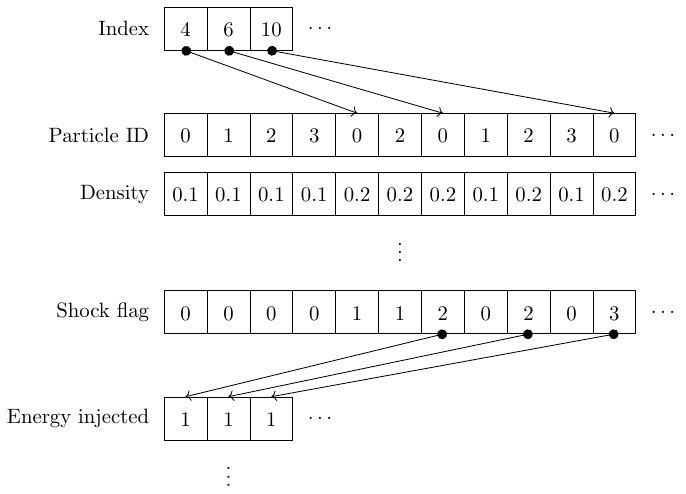}
    \caption[]{Schematic showing how tracer data is stored in \textsc{Crest}. Values are for reference purposes only. An index provides the start of the next timestep ``block'', with all data in a block ordered by particle ID. Shock data is only saved when the shock flag has a value of 2 or 3 (indicating shock-surface and post-shock cells, respectively.). In this way, the data saved is kept to a minimum. Ellipses represent array dimensions extending to further timesteps and variables.}
    \label{fig:crest-array}
\end{figure}

Finally, certain data, such as the tracer position, the individual magnetic field components, and shock directional components\footnote{The angle between the magnetic field and shock direction is stored separately for use in the obliquity-dependent shock acceleration module.}, are not used within \textsc{Crest}. However, these are useful to record for use in plotting scripts or post-processing with the emission code \textsc{Crayon+}. As they will only be used in combination with \textsc{Crest} output files, they only need to be stored at the same frequency. By default, such data is now only stored at \textsc{Arepo} snapshot times, under the rationalisation that \textsc{Crest} output will be predominantly compared with \textsc{Arepo} results\footnote{It is, however, still possible to save this data at every timestep, if the user requires.}. Additional code has been added to \textsc{Crest} to make sure that its output synchronises appropriately\footnote{Similarly, the old methods, in which \textsc{Crest} outputs at every timestep or after a fixed time period are also still possible. Such choices are available through compilation flags.}.  

In total, these changes have led to a significant reduction in the data saved per simulation. For example, in initial tests for a cluster simulation with dark matter mass resolution $m_\rmn{DM} = 10^{10} \, \rmn{M}_\odot$ and 2 million tracers, the data stored was reduced by a factor of 12, producing around 140 GB over approximately 1700 timesteps. This makes cosmological simulations with \textsc{Crest} feasible. Moreover, the savings relative to the original version are only expected to become larger for higher resolution simulations, which typically have a deeper timestep hierarchy, owing to the stronger density peaks and faster dynamics that can be resolved. 

\subsection{Re-structuring the code}

In order to deal with the change of data structure, the read-in and distribution processes in \textsc{Crest} were completely overhauled. This is one of many changes to the code structure that have taken place:

\begin{itemize}
    \item \textbf{MPI processes:} The code now loops through all tracer files and is expected to be run using multiple processes. The root node is, by default, set only to read and distribute the data, whilst the remaining nodes evolve the spectra\footnote{Testing has shown that this is the most efficient way to run the code.}. 
    
    \item \textbf{Load balancing:} \textsc{Crest} uses a semi-analytic method to evolve spectra, ``stitching'' the analytic solution on when a single cooling process dominates, and calculating the numerical solution otherwise. As numerical computation dominates the runtime, the code now keeps track of how many momentum bins were updated numerically in the previous timestep. Tracers are then ordered accordingly before being  distributed amongst the processors. This helps balance the processor load, increasing the efficiency of the parallelisation.

    \item \textbf{Unification:} The initial implementation of \textsc{Crest} had several submodules \citep[see functionality given in][]{winner2019}, but it was not necessarily possible to use these simultaneously; instead, the code contained various if-statements, with different code run depending on the version number used. The code has now been updated so that there is only one version, which allows for all (non-conflicting) compilation options to be run together, as desired.

    \item \textbf{Modularisation:} The code has been heavily modularised, so that functions are grouped by their purpose, and, where reasonable, point towards existing implementations. This is done instead of simply copying the necessary code, as was previously common throughout. This will make it easier to implement new modules in the future, and reduces the likelihood of new bugs being introduced. It also increases the ease with which functionality can be tested. Finally, it allows for individual parts of the code to be switched on or off with ease; for example, the various cooling processes (see Sec.~\ref{sec:cooling}) implemented in \textsc{Crest} can all be made active independently from one another.

    \item \textbf{Rationalisation:} Previously, \textsc{Crest} contained many iterations of code, including functionality that had been superseded by better methods. This cluttered the code, making it difficult to read. Where functionality was no longer needed, it has been taken out. 

\end{itemize}

\subsection{Additional changes}

A range of other changes have also been made to the code. My own total commits number roughly 300, and so I include only a summary of the changes, grouping them into the following distinct categories:

\subsubsection{Conceptual change}

Previously, tracers were assigned a volume and a mass during the initial conditions, where this was fixed over time. This was done as the source term in the Fokker-Planck equation, in particular, is given in units of inverse volume. During acceleration, the energy deposited in the tracer was then multiplied by the ratio of the tracer volume, $V_\rmn{tp}$, to the cell volume, $V_\rmn{cell}$ (see section 3.2 of \href{https://www.aip.de/media/thesis/georg-winner-phd-thesis.pdf}{Winner et al., 2020}). This was problematic as the overall energy injected into the cosmic ray electrons could easily be over- or under-estimated, depending on the number of tracers in a cell and their respective volumes. Moreover, creating projected emission maps from the tracers would only work as long as they were relatively uniformly-spaced, as $V_\rmn{tp}$ was not actually mapped to spatial coordinates.

To solve this problem, we now treat tracers as a sampling of the underlying cosmic ray electron spectral energy density field. The tracer's volume can then be defined at every snapshot time by constructing a Voronoi tessellation, with itself as the mesh-generating point; in other words, all points closest to a tracer are mapped to its spectrum. This creates a volume-filling field, such that integrated and line-of-sight quantities are well-defined. It also means that we can simply use the energy density on the right-hand side of Eq.~\eqref{eq:p_min}, solving the issue described in the previous paragraph. 

\subsubsection{New functionality}

The code has been updated with much new functionality. This includes:

\begin{itemize}
    \item \textbf{Cosmological conversions:} The code now has the ability to convert from co-moving \textsc{Arepo} quantities (see typical scalings \href{https://www.tng-project.org/data/docs/specifications/}{here}) to physical CGS quantities, as required by the formulae implemented in \textsc{Crest}. This adapts automatically to the chosen units in \textsc{Arepo}. This update is critical for applying \textsc{Crest} to cosmological simulations.
    
    \item \textbf{Ability to deal with weak shocks:} The previous iteration of \textsc{Crest} used the analytic approximation for solving Eq.~\eqref{eq:p_min}. This relied on the incomplete Beta function \citep[see, e.g.][]{pinzke2013}, which is only valid for spectral slopes between $- 2 > \alpha_e > -3$, and becomes increasingly inaccurate close to these limits. The spectral slopes probed in Chapters~\ref{chapter:paper-three} and~\ref{chapter:paper-four} are often substantial weaker than this. To overcome this, a numerical integration has been implemented. This is given a maximum relative integration error to achieve, which can be altered through input parameters.
    
    \item \textbf{Maximum spectral slope:} Recent PIC simulations suggest that cosmic ray generated pre-cursor and post-cursor regions limit the maximum spectral slope to $\alpha_\rmn{e} = -2.2$ \citep{caprioli2020}. This value is also supported by observations of supernovae shocks \citep{caprioli2012}. Consequently, we have implemented this result into \textsc{Crest}.

    \item \textbf{Minimum Mach number:} The minimum Mach number required for DSA can now be altered, based on a density jump condition, or reading in from stored data\footnote{The ability to record this value is also new functionality, and is set through \textsc{Arepo} config flags.}, if available. This functionality was used, in particular, to show the impact of $\mathcal{M}_\rmn{crit}$ in Chapter~\ref{chapter:paper-four}. 
    
    \item \textbf{New output options:} \textsc{Crest} now outputs the maximum recorded Mach number encountered and the elapsed time since last injection. This functionality was also heavily used in the last two chapters.

    \item \textbf{Allow higher tracer counts:} Edits have been made to the way memory is allocated in \textsc{Arepo} to allow for over 2 million tracers\footnote{Previously the system was limited due to the number of variables saved on tracer particles.}. This was required for the high-resolution simulations shown in the last two chapters. 

    \item \textbf{File splitting:} As the total recorded data can now be easily in excess of 10 GB, file splitting has been implemented, so that tracer data is written to a new file once above a user-defined threshold. This, in particular, makes files more easily transferable.

    \item \textbf{Restart functionality\footnote{Ultimately fixed by Maria Werhahn.}:} \textsc{Crest} writes restart files at the end of its run, so that it can be run over several sessions. This functionality has been tested with the test pipeline (see Sec.~\ref{sec:test-pipeline}).

    \item \textbf{Tuneable performance parameters:} \textsc{Crest} now allows for some manual setting of parameters related to performance; for example, the number of times the code is allowed to subcycle during the evolution step.

    \item \textbf{Manually set tracer IDs:} Tracer IDs can now be set through loading of the initial conditions, rather than being allocated automatically by \textsc{Arepo}. This will allow iteration of tracer initial conditions in the future, so that tracers can be placed, for example, to cover only a region of interest.

    \item \textbf{Extensions started for the future:} The code is now set up to record the photon and electron number density live from \textsc{Arepo} simulations. This will be particularly useful when using \textsc{Crest} to study cosmic ray electrons in galaxies with high star formation rates (Werhahn, in prep.).

\end{itemize}

\subsubsection{Debugging features}

\textsc{Crest} now includes several features to help with debugging:

\begin{itemize}
    \item \textbf{Error-checking during run-time:} The input data are now checked to make sure that values are physical (e.g.\ positive densities). Data is also checked during runtime to catch errors early (e.g.\ invalid timestep values). This reduces the risk that \textsc{Crest} produces unknowingly unphysical results.
    \item \textbf{Save during crash:} When errors are caught, \textsc{Crest} saves the current arrays to a dump file before crashing. This helps the user to reproduce the error for debugging. 
    \item \textbf{CPU time logged in \textsc{Arepo}:} The time spent in \textsc{Arepo} on \textsc{Crest} functionality is logged and saved to the standard file. This helps show whether \textsc{Crest} is using an unexpectedly large amount of resources.
    \item \textbf{New variables:} New variables can be recorded such as the parent gas cell ID\footnote{The logging of these values must be turned on with compile flags.}. This helps to confirm that the \textsc{Crest} input data matches with the \textsc{Arepo} simulation.
\end{itemize}

\subsubsection{Bug fixes}

The code has been heavily tested during development, and has been subjected to a wide range of input options during the projects presented in Chapter~\ref{chapter:paper-three} and Chapter~\ref{chapter:paper-four}. The results of this has been that several corrections have been made to the code. These include:

\begin{itemize}
    \item \textbf{Mach number not recorded in post-shock cells:} This was stored in a different variable, and consequently not being properly recorded by the code. This led to a reduction in the overall energy injected into the spectra by roughly a factor of two.
    
    \item \textbf{Variables stored on other tasks:} Similarly, the shock direction, pre-, and post-shock velocities were not being recorded from the \textsc{Arepo} output when they were stored on tasks other than the root node.
    
    \item \textbf{Correction to Bremsstrahlung formula:} This had been implemented but not tested. The corrected formula is shown in Eq.~\eqref{eq:bremsstrahlung}.
   
    \item \textbf{Injected momentum value:} The minimum momentum to start injection at, $p_\rmn{inj}$, is limited to 3.5$\times$ the peak thermal momentum $p_\rmn{th}$ \citep{pinzke2013, winner2019}. However, this criteria was not being properly enforced; indeed, the code allowed values down to $p_\rmn{inj} = p_\rmn{th}$. The maximum momentum is also not capped at $p_\rmn{inj} = 10 p_\rmn{th}$, as beyond here we are in the regime of relativistic shocks, which \textsc{Crest} is not equipped for.
    
    \item \textbf{Exponential cut-off value:} Catastrophic cooling losses are implemented using a super exponential cut-off above a given momentum, $p_\rmn{max}$, where this is calculated according to Eq.~\eqref{eq:p_max}. However, this was not what had been coded up; the code attempted to account for oblique shocks, which is in contradiction with the plane-parallel shock assumed in the DSA implementation, and would produce negative values under a variety of circumstances. The code has now been corrected.
    
    \item \textbf{Missing break conditions:} Some loops did not have appropriate termination conditions leading to the code ``hanging''. Conditions have generally been corrected in \textsc{Crest} to prevent this, with ``breaks'' added to prevent infinite loops in the rare case that the code cannot converge to a solution. 
    
    \item \textbf{Memory leaks:} Array handling, in particular, has been fixed to prevent a loss of memory during runtime.
    
    \item \textbf{Density freezing:} Shocks in \textsc{Arepo} are numerically broadened. To prevent spurious adiabatic effects during Fermi I (re-)acceleration and to correct the cooling rates, \textsc{Crest} ``freezes'' the density on encountering a shock zone cell and sets values to those of the post-shock cell after encountering a shock surface cell. An implementation of this had previously been attempted in \textsc{Crest}, but this had not properly been tested. Moreover, the exit conditions for the implementation were wrong. This meant that in converging flows, where tracers rapidly move across shock fronts, the implementation could actually \textit{lead} to spurious heating. The code has been re-written and tested accordingly (see Fig.~\ref{fig:density-freezing}).
\end{itemize}

\begin{figure}
    \centering
    \includegraphics[width=0.5\linewidth]{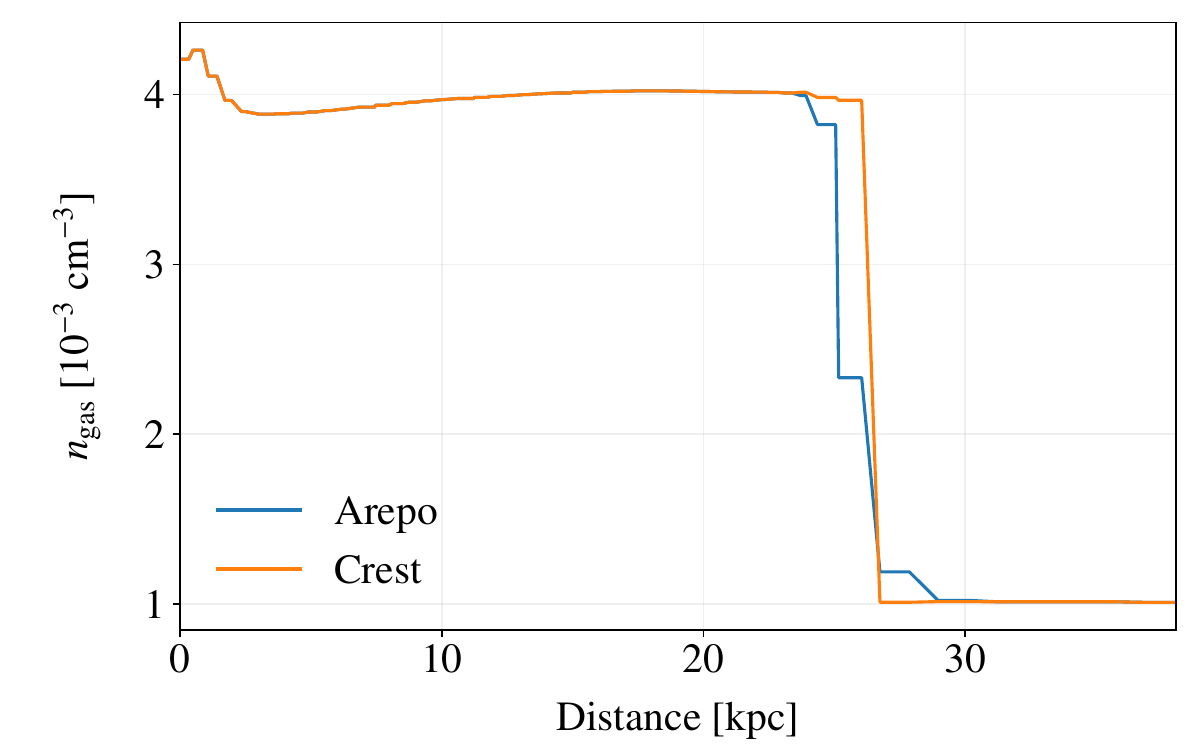}
    \caption[]{Gas number density vs. distance, for a shock-tube test. The blue curve shows the data stored from the \textsc{Arepo} simulation, whilst the orange curve shows how \textsc{Crest} interprets the data. Densities are frozen if in a shock-zone cell, and take their post-shock value if in a shock-surface or post-shock cell. This helps prevents spurious adiabatic effects during the shock and produces more accurate cooling.}
    \label{fig:density-freezing}
\end{figure}

\subsubsection{User friendliness}

The ultimate aim of the \textsc{Crest} project is to provide it as a tool for the community. Consequently, it is important that the code is user-friendly and easy to edit. To this end, where necessary, comments have been added to explain the input, method, and output of functions, and citations have been added to explain where physical formulae implemented originate from. This extends to the parameter and configuration files, which now have commented templates. In making these changes, the code has become more intuitive for first-time users. Additionally, the following user-orientated changes have been made:

\begin{itemize}
    \item \textbf{``Default'' mode:} The code has been given a well-defined ``default'' mode, in which it assumes an initially thermal spectrum with all cooling modules on. These can then be individually switched off, and extra modules activated as required.
    
    \item \textbf{Validation of parameter and compilation options:} Similarly to \textsc{Arepo}, \textsc{Crest} now only allows parameters to be given for activated compilation options. This makes it clearer which modules  \textsc{Crest} is actually being run with. Configuration options are also validated at compile time using pre-processor directives. This prevents conflicting options being loaded (e.g.\ different output modes) or options being loaded without the necessary parent option (e.g.\ obliquity-dependent shock acceleration without the Fermi I module).
    
    \item \textbf{Parameter and compilation options shown at start-up:} \textsc{Crest} now prints the input parameter values and compilation flags to screen on start-up. This makes it more obvious how \textsc{Crest} is being run.

    \item \textbf{Header flags:} Flags are stored in the header of the tracer file so that \textsc{Crest} can verify if the data exists for the given compilation option (e.g.\ whether positions have been saved every timestep). The code then exits if the option is not possible, or provides a warning if an option is possible but has not been chosen (e.g.\ if data for shock acceleration is saved but not being used). Finally, this method allows the code to automatically convert from cosmological data, without the need to be recompiled.

    \item \textbf{Warnings:} Extra warnings have been added, including if the user is potentially using their computational resources inefficiently; e.g.\ assigning too many processors to the given number of tracers.
    
    \item \textbf{Statistics:} Statistics can now be output to screen at intervals more regular than the snapshot time. This is especially useful when snapshot times are far apart relative to the number of timesteps. The statistics have also been extended to show the fraction that are active, the total number of tracers evolved so far, the fraction that have undergone various processes since the last statistics output, the file and timestep currently being read, and various statistics regarding the computational balance across processors. Additional, total wallclock times for various parts of the code are output at the end of the runtime.
        
    \item \textbf{Arepo file formats and termination:} The code now looks for \textsc{Arepo} files in standard formats, including where these have been split into multiple files. Moreover, the code now automatically terminates when no more \textsc{Arepo} input files are found.
    
\end{itemize}

\subsection{Test pipeline}
\label{sec:test-pipeline}

In order to make sure that \textsc{Crest} stays accurate we have developed a test pipeline. The framework for this was initially written by Philipp Girichidis, and has since been extended and updated by the other members of the \textsc{Crest} team (myself, Maria Werhahn, and L\'{e}na Jlassi). This tests a large amount of \textsc{Crest}'s functionality, with an emphasis on remaking the plots in \citet{winner2019}, which are generally tied against analytic results. The pipeline currently consist of 16 tests, which cover a range of modules, including subgrid injection (i.e.\ from unresolved supernovae), Fermi I and Fermi II acceleration, and adiabatic expansion, as well as free-cooling (via Coulomb, inverse Compton, and synchrotron processes) and steady-state regimes. The pipeline is constructed so that it can be set up with relative ease for any new user, and requires the minimum of expertise with the code. Indeed, with only a few bash commands, the pipeline will set up, compile, and execute \textsc{Crest} and additionally make analysis plots based on the output. This makes the pipeline an ideal tool for teaching new users how to use \textsc{Crest} and interpret its output. In the future, the pipeline will be combined with further automation and unit testing of individual functions.

\section{Shock finder}

The shock finder used in \textsc{Arepo} was introduced in \citet{schaal2015} with alterations made to allow for cosmic ray acceleration in \citet{Pfrommer2017}. This functionality is heavily used within \textsc{Crest} and was also key to the results presented in the preceding two chapters. Here too, however, the code was initially in proof-of-concept form, as cosmic ray shock acceleration had never been run in a cosmological simulation, and the DSA routine had not been so intensely applied to weak shocks. As a result, the following correction and updates were required:

\subsection{Numerical stability}

\begin{figure}
    \centering
    \includegraphics[width=0.5\linewidth]{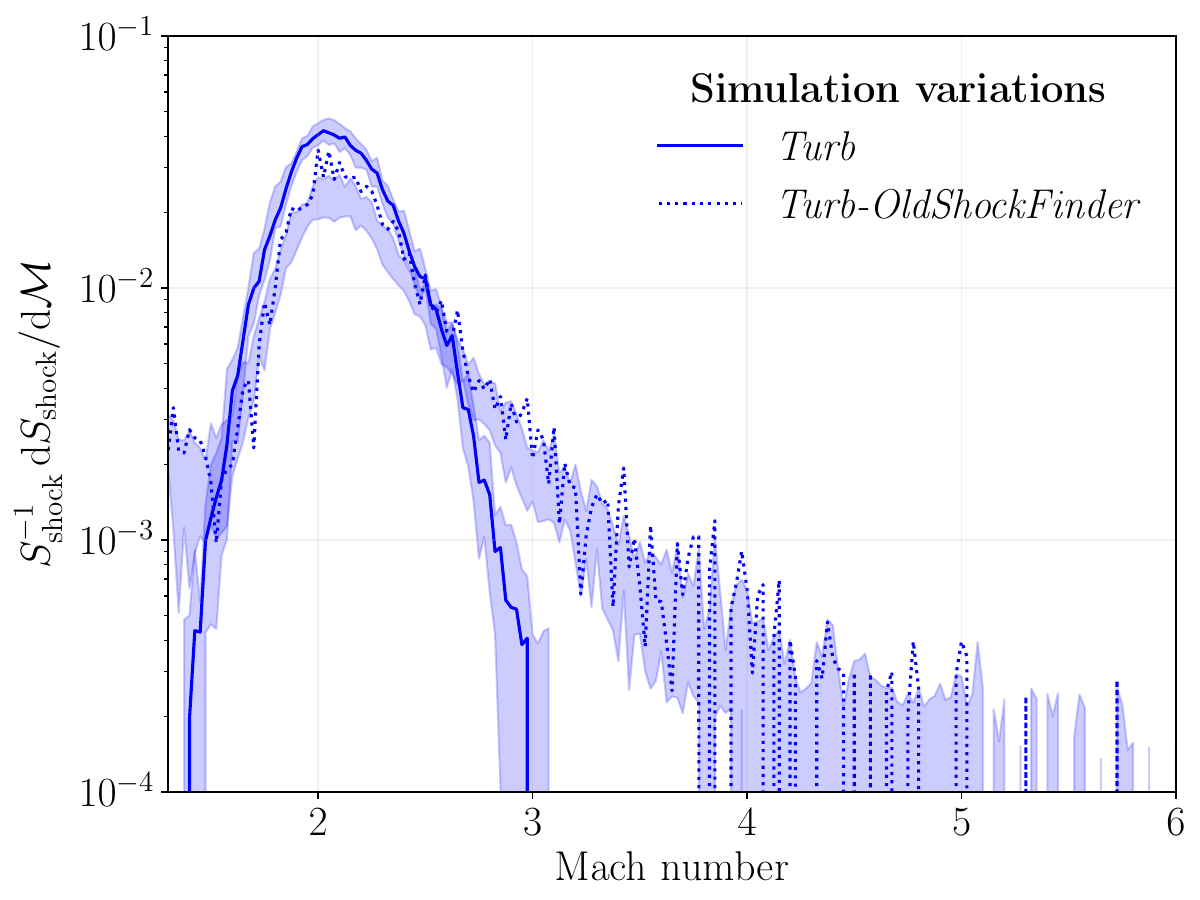}
    \caption[]{Mach number distributions generated from the Mach 2 \textit{Turb} simulation introduced in Chapter~\ref{chapter:paper-three}, with values weighted by their normalised contribution to the shock surface. Lines indicate the median taken over all snapshots, whilst the shaded values indicate the interquartile range. The dotted (solid) line shows the values produced by the original (updated) shock finder. The original algorithm was unstable at low density contrasts, leading to artificially high Mach numbers being produced.}
    \label{fig:shock-finder-mach-pdf}
\end{figure}

    The formula for the Mach number in a gas with a cosmic ray component is given in Eq.~\eqref{eq:mach-number}. After testing, it became apparent that it is numerically unstable at low compression ratios due to the last term in the equation:
    \begin{equation}
        \frac{x_\rmn{s}}{\gamma_\mathrm{a, eff}(x_\rmn{s} - 1)},
    \end{equation}
    i.e., when $x_\rmn{s}$ is close to 1. This regime had not been thoroughly tested before the shock-tube project presented in Chapters~\ref{chapter:paper-three} and~\ref{chapter:paper-four}. To fix this, we now require a minimum density jump, which is calculated by using the hydrodynamic jump conditions; specifically Eq.~\eqref{eq:compression-ratio}, where $\mathcal{M}$ is set to the minimum Mach number allowed\footnote{By default this is $\mathcal{M} = 1.3$.}. 
    
    In our case, this has a major impact on the resulting Mach number distribution. We show this in Fig.~\ref{fig:shock-finder-mach-pdf}, where the dotted line indicates the original data, before implementing the consistency check, and the solid line indicates the distribution afterwards. The lines here indicate the median value taken over all snapshot, where each cell is normalised by its contribution to the overall shock surface. The shaded values indicate the associated interquartile range. It can be seen that before the update, the Mach number distribution has a large tail towards high Mach numbers\footnote{Preliminary results show that a similar, albeit weaker, impact is had in cosmological simulations.}. These are artificial and result from the aforementioned instability. Although this problem affected relatively few tracers, it unduly biased the integrated spectra due to the correspondingly flatter spectra and increased normalisation through Eq.~\eqref{eq:shock-dissipated-energy}. After the fix, the Mach number distribution shows a skew-normal distribution, as analysed in Chapters~\ref{chapter:paper-three} and~\ref{chapter:paper-four}. 

\subsection{Computational efficiency}
    
    The shock finder works by determining candidate shock zones using the criteria given in Sec.~\ref{chapter5-subsec:shock-finder}. ``Rays'' are then sent out in the direction of the temperature gradient. Once a ray leaves the shock zone, it records the post-shock values and reverses its direction. This continues until the pre-shock is reached. The original algorithm implemented to do this assumes that both post- and pre-shock cells are in the currently active mesh. This is, however, not necessarily the case in standard simulations. The quick fix applied was for the shock finder to reconstruct the mesh so that it included all cells, before returning the mesh to its previous state afterwards. This allowed the code to continue without segmentation faults. 
    
    However, this method rapidly became problematic in cosmological simulations with cosmic ray shock acceleration activated; shock-acceleration requires the shock finder to be in ``on-the-fly'' mode, in which it is activated at every timestep. As simulation resolution increased, this led to unsustainable computational times. An example of this is in shown in Fig.~\ref{fig:shock-finder-walltime}, where the blue lines represent a simulation run with this configuration of the shock-finder. The solid line represents the total time taken by the simulation, and the dotted line represents the time taken purely by the mesh-making routine.

    To fix this, the rays now check whether the cell they are in is active or not. If a cell steps outside the active mesh, as determined by its assigned local timestep, it is discarded\footnote{An alternative config flag allows for the shock values to the pre- and post-shock values to be assigned to the first cell encountered outside the active mesh.}. This is conceptually more rigorous; the active mesh effectively contains all cells that can communicate with each other on the given dynamical time. By extending this mesh to include inactive cells, the ``quick'' fix allowed cells beyond this to impact dynamics. Careful analysis has shown that discarding the cell also has a negligible effect on the overall statistical distribution and total injected energy in cosmological simulations. Importantly, however, doing so significantly shortens the runtime of the simulation, making high-resolution cosmological simulations with cosmic ray physics feasible. 

\begin{figure}
    \centering
    \includegraphics[width=0.5\linewidth]{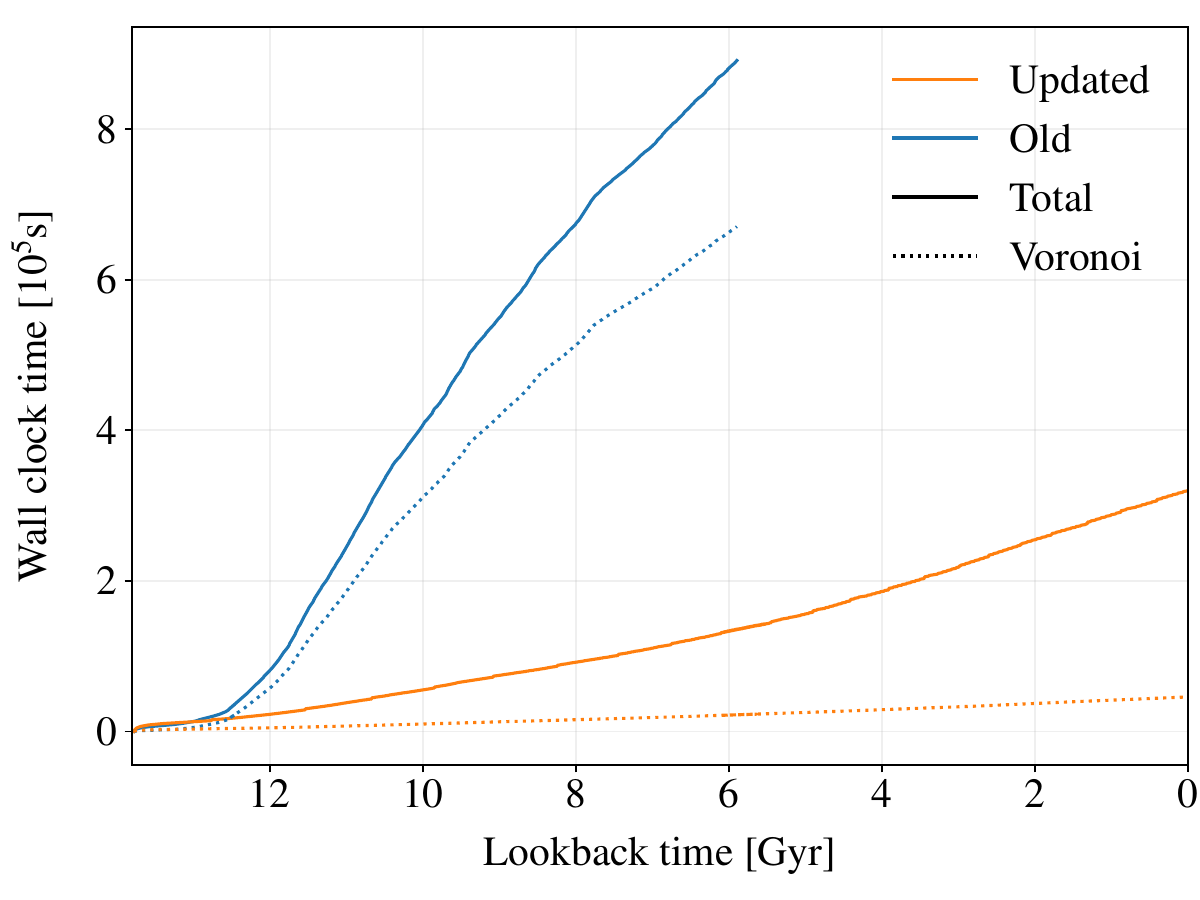}
    \caption[]{Wall clock time vs.\ lookback time for two high-resolution cosmological simulations with $m_\rmn{DM} = 2\times10^7 \,\rmn{M}_\odot$ run with on-the-fly shock acceleration. Solid lines show the total wall clock time, whilst dotted lines show just the contribution due to Voronoi-mesh construction. The simulation run before the shock-finder update (blue) needed to be stopped at $t_\rmn{lt} \approx 6$ Gyr due to the increasingly unaffordable computational cost. After the update, mesh-calls were substantially reduced, and the simulation could run to completion.}
    \label{fig:shock-finder-walltime}
\end{figure}

\subsection{Additional changes}
    The following additional changes were also implemented:

\begin{itemize}
    \item \textbf{Shock finder compatibility:} The shock finder was not previously able to work with both cosmic rays and radiative cooling of the thermal gas (i.e.\ star formation) activated. This was primarily a mixture of incompatible variables being used and variables not being correctly mapped between functions. Now that this has been corrected, the shock finder can be run in radiative simulations with cosmic ray physics\footnote{Note that the Mach finder in its current formulation assumes an adiabatic fluid, which is an assumption that breaks down when radiative cooling is strong. In the future, the Mach finder will be extended to take this case into account as well.}. The \citet{springel2003} model, with its stiff equation of state in star-forming cells, will of course prevent this being applied to cells in the ISM. However, this feature is important for modelling shocks beyond this, where the gas properties can still be affected by stellar feedback processes (e.g. the modelling of accretion and merging shocks in galaxy and galaxy cluster simulations run with radiative physics).

    \item \textbf{Minimum Mach number:} Whilst a minimum Mach number check had already been implemented, it did not cover every eventuality, and was additionally performed after \textsc{Crest} recorded the value. Consequently, Mach numbers were occasionally output both in \textsc{Arepo} snapshots and the tracer files with values below $\mathcal{M} = 1.3$. This has now been corrected.

    \item \textbf{Variables in post-processing:} The shock finder in post-processing mode has been updated to output cosmic-ray related variables, such as the pre- and post-shock cosmic ray pressure. This has aided with debugging.

    \item \textbf{Manually choose shock parameters:} Parameters such as memory, the maximum number of ray steps, and the minimum Mach number can all be chosen manually now, if required. This was previously only possible whilst using the shock finder in post-processing mode, but was also helpful during debugging.

\end{itemize}
The above changes have made it possible both to run accurate cosmic ray shock tubes with low Mach-number shocks, and to run radiative cosmological simulations with cosmic ray shock acceleration. This represents a major step towards updating the \citet{pinzke2013} paper using current state-of-the-art codes. 
\chapter{Conclusions and Outlook}
\label{chapter:conclusions}

\vspace{-0.3cm}

\noindent This work has been primarily focussed on two topics:
\begin{enumerate}
    \item Magnetic fields in galaxy mergers
    \item The physics of radio relics in galaxy clusters
\end{enumerate}
Both of these have been the subject of a great deal of research over the previous decades (see the introductions to the thesis chapters). However, it is only recently that it has been possible to investigate these areas using cosmological-consistent simulations, as employed during this thesis. 

Using such simulations has two key advantages. Firstly, we can be sure that we are simulating scenarios consistent with $\Lambda$CDM. This is important for making sure that the simulations are representative of the real Universe, rather than providing answers that are potentially only of theoretical interest. Secondly, cosmological simulations allow for a full consideration of the environmental context in a self-consistent manner. This second point has turned out to be key to our results in both fields. For example, we have shown that for magnetic fields to affect mergers they require sufficient amplification, which, in turn, requires levels of turbulence and accretion unlikely to be realised in idealised simulations. On the other hand, in our simulations of merging clusters, their cosmological nature meant that we were able to observe the interaction of shocks with material accreting at the cluster outskirts. It is this interaction that forms the basis for the mechanism presented in this thesis, with the mechanism, in turn, solving many of the major problems currently challenging our understanding of radio relics.

We summarise below the key findings presented in this thesis. Additionally, we provide a brief overview of how the work presented here can be extended in the future, and how it fits within current observational developments.

\vspace{-0.3cm}

\section{Magnetic fields in mergers}
\subsection{Results}

With only a couple of notable exceptions \citep{martin-alvarez2020, katz2021}, it has generally been assumed that magnetic fields are dynamically unimportant in galaxies over the majority of their lifetime\footnote{Note, however, that in the exceptions given, the magnetic fields strength was artificially amplified, with seed strengths set above the currently accepted upper limits achievable by standard battery processes \citep{gnedin2000, attia2021}.}. Indeed, a common argument is that magnetic fields only reach equipartition at late times ($z \lesssim 0.5$), particularly in the outskirts of the disc \citep{pakmor2017}. This is well after so-called ``cosmic noon'' at $2<z<3$, when star formation reaches its peak. Hence, the argument goes, magnetic fields cannot affect star formation, and are relegated to only having an indirect impact via their influence on anisotropic transport processes (see Sec.~\ref{sec:magnetic-fields}). In the work presented in this thesis, we have shown that this argument is flawed because it neglects the impact of mergers.

In Chapter~\ref{chapter:paper-one}, we have shown that major mergers are able to amplify magnetic fields in gas-rich disc galaxies to dynamically important strengths within a few 100 Myr of the first pericentric passage. This amplification is generated partially by adiabatic compression, but requires the impact of a small-scale dynamo to reach its full potential. By calculating kinetic and magnetic power spectra taken at 250 Myr before, during, and 500 Myr after the first pericentric passage, we have shown that the kinetic turbulence is consistent with a \citet{kolmogorov1941} cascade, whilst the magnetic power spectra responds in a manner consistent with a small-scale dynamo, including the production of a \citet{Kazantsev1968} slope at large $k$. Moreover, we have shown that the speed of amplification is dependent on the resolution, which is a direct result of modelling smaller eddies at finer resolution, with correspondingly quicker eddy turnover timescales. Consequently, in simulations with higher resolution, the magnetic field can better take advantage of the turbulence before it decays, meaning that it is amplified to higher, more dynamically important levels. In our simulations, we find that the dark matter mass resolution must be $m_\rmn{DM} \gtrsim 5\times10^5 \rmn{M}_\odot$ for the magnetic field to have a significant impact the outcome of the merger.

We investigate how the morphology of the remnant changes in Chapter~\ref{chapter:paper-one} and show what the magnetic field does to cause these changes in Chapter~\ref{chapter:paper-two}. We find that the magnetic field acts in various ways, but that its largest impact is on the alteration of transport of angular momentum between gas cells. We find that, once sufficiently amplified, and if it is predominantly non-azimuthally orientated, the magnetic field typically acts to transport angular momentum outwards. This increases the baryonic concentration, creating a strong inner Lindbland resonance, which disrupts the formation of a bar. Without a strong bar, the disc is then able to form typical small-scale baryonic structure such as spiral arms. The remnant thereby eventually forms a MW-like morphology. In contrast, in hydrodynamic simulations, the inner Lindblad resonance is weak, allowing a strong bar to form. This goes on to dominate the orbital dynamics in the galaxy, leading to a strong stellar ring being formed. In Auriga model, this results in a powerful stellar wind that disrupts the angular momentum of the gas in the CGM, thereby preventing the further growth of the remnant. This results in a stunted, compact disc being formed, with morphology not representative of observations.

Additionally, we find that the increase in concentration in MHD simulations results in an increased black hole accretion rate, with black holes able to grow twice as large relative to hydrodynamic simulations\footnote{This still places black hole mass values well within the scatter of the well-known black hole  -- halo mass relation, see, for example, \citet{reines2015}.}. This plays only a very minor role in setting the ultimate morphology of the remnant, however. Indeed, by editing simulations to stop quasar feedback at the onset of the merger, we find that the AGN may actually mask some of the effects of magnetic fields.

Finally, by studying hydrodynamic and MHD variations of the original Auriga galaxies, we find that the effect observed in merger remnants can be seen in more isolated, but still cosmological, simulations as well, albeit typically in more subtle ways. Specifically, hydrodynamic simulations show extended bars and rings and are systematically smaller than their MHD equivalents. This makes sense, as virtually all galaxies will experience interactions at high redshifts, and we have shown that the effect of such mergers can persist for several Gyr afterwards. We therefore conclude that magnetic fields are an essential part of disc galaxy modelling.

\subsection{Extensions and outlook}

There are a number of ways in which this work could be extended:

\begin{itemize}
    \item \textbf{High red-shift simulations:} As yet, we have limited our research to $z<1$. We could, however, validate our claim that even galaxies with relatively quiescent merger histories are affected by the mechanism outlined in Chapters~\ref{chapter:paper-one} and~\ref{chapter:paper-two} by studying the Auriga variations at higher redshift ($1<z<2$). Moreover, we could run simulations focussed on disc-galaxy mergers at very high redshift ($z>2$). This would be particularly timely given the recent discovery of an unexpectedly large number of star-forming disc galaxies at such redshift by the James Webb Space Telescope (JWST) \citep{kuhn2024, yan2024}.
    
    \item \textbf{Additional physics:} Simulations at very high redshift would likely require a multiphase ISM and radiative physics, to account for the more complex gas dynamics. Alternatively, we could extend our merger simulations to include cosmic ray physics. This would be analogous to the study by \citet{buck2020}, but in the context of mergers. 
    
    \item \textbf{Alternative galaxy formation models:}
    More generically, to prove that our results are not confined to the Auriga model, we could re-run our simulations using different galaxy formation models and numerical codes. In particular, it is possible that the formation of a stellar ring and the subsequent stellar winds may not hold for models that include a multiphase ISM or more explosive stellar feedback\footnote{We expect other parts of the mechanism will hold, however; see the discussion section in Chapter~\ref{chapter:paper-two}.}. Any alternative models would, however, need to produce magnetic field strengths consistent with observations. This is especially important as the magnetic energy density increases with as $\sim|\mathbf{B}|^2$.
    
    \item \textbf{Apply analysis to TNG50:} Our investigation was based on a relatively limited number of case studies\footnote{Specifically, $4\times2$ merger studies, $2\times2$ extra resolution studies, and $4\times2$ more isolated galaxy studies.}. This is mostly due to the increased computational expense of zoom-in simulations, as required to reach sufficient resolution whilst including cosmologically-consistent accretion and turbulence. However, since we published, the TNG50 simulations have been released \citep{nelson2019b}. These are MHD cosmological box simulations, which should have sufficient resolution to reproduce the mechanism we have described. Moreover, the TNG model for disc galaxies is very similar to the Auriga model. We could therefore compare our case studies with disc galaxies in the TNG50 volume. This would help to increase the statistical confidence of our result. Curiously, despite theoretically having resolution too low to replicate our results, analysis by \citet{pillepich2018} showed that trends between hydrodynamic and MHD TNG100 variations and trends between TNG100 and the original Illustris simulations are consistent with our results; that for MW-like galaxies at $z=0$ the MHD versions are: i) larger, ii) have a higher gas fraction, and iii) have a larger black hole mass (see their figures 4 and 8).

    \item \textbf{Cosmic ray electron modelling:} By using \textsc{Crest}, we can produce mock radio and spectral intensity maps, that will help better tie simulations to observations. In particular, we could test the equipartition theory; although this assumption is often used to infer magnetic field strengths, it is unclear whether it holds in galaxies, let alone in mergers, where the thermal, turbulent, and magnetic energy densities vary rapidly.
    
    \item \textbf{Extension of the dynamo analysis:} it is of particular interest in the dynamo theory community as to whether magnetic helicity is conserved. Whilst, in theory, it should could be strictly conserved under the equations of ideal MHD, this is not necessarily true for reconnection events. In a similar vein, we could extend the analysis in \citet{pfrommer2022} to include analysis of magnetic curvature statistics, thereby better supporting our argument that a small-scale dynamo is in action in our simulations. Finally, work by \citet{kriel2023} suggests that the peak-scale of magnetic turbulence is set by shocks, and that the small-scale dynamo in our simulation is acting in the compressive, super-sonic turbulence. We could directly apply their analysis tools to our simulations to verify this theory or otherwise.
    
    \item \textbf{Comparison to the Milky Way:} Recent statistical fitting has shown that the local magnetic field is just as well fit by a coherent magnetic spur as by a grand design spiral \citep{unger2024}. The simulations presented in this thesis provide an ideal candidate to compare this result with mock data, as they are high resolution simulations of MW-like galaxies that evolve over cosmological time. We could therefore apply their analysis tools to our simulations to produce mock observations here as well.
    
    \item \textbf{Study of magnetic field morphology:} More generally, the morphology of the  magnetic field presents several questions that could be answered with the simulations presented. For example, it was already noted in the Masters thesis that the magnetic field evolves over time, rather than producing the static morphologies typically assumed in analytic dynamo theory. Our simulations could be used to investigate the origin of these morphologies. Moreover, we showed in Chapter~\ref{chapter:paper-two} that the magnetic field can be either azimuthally- or non-azimuthally dominant. This appears to be linked to the merger orbits, and the resulting accretion patterns. Magnetic field morphology could therefore be indicative of merger history. Finally, many galaxies exhibit X-shaped field structures when observed edge-on \citep[see, e.g.][]{stein2020}. Our simulations could be used to probe the origins of this feature as well, seeing whether it is produced by a stellar wind \citep{chiu2024} or otherwise.
\end{itemize}

At current, there are large uncertainties surrounding the magnetic field strength in both isolated and merging galaxies over cosmic time. This will hopefully be clarified by the next generation of radio telescopes, including MeerKAT, SKA, and LOFAR \citep{haverkorn2019}, thereby putting better constraints on simulations. In particular, we re-iterate that an important future test for simulations will be to match observed Faraday rotation data. This requires better data, but also better modelling.

\section{Radio relics}
\subsection{Results}

In the second half of this thesis, we have turned our attention to the origin of radio relics. Radio relics are large, Mpc-sized regions of radio emission found at the periphery of merging galaxy clusters. They are highly polarised and are believed to tracer merger shocks, with the shocks re-accelerating populations of ``fossil'' seed electrons. However, whilst the general paradigm for producing radio relics is accepted, our understanding of their origins is far from complete. To this end, we collated a list of the major outstanding problems in the field. These are:
\begin{enumerate}
    \item What is the origin of the seed electrons needed for re-acceleration?
    \item What is the origin of relic morphology?
    \item How do we explain the observed discrepancy between X-ray derived Mach numbers and radio-derived Mach numbers?
    \item How do we explain the implied $\upmu$G-strength magnetic fields in radio relics, given that the surrounding ICM has strengths at least an order of magnitude lower?
    \item Why are standard cooling models unable to match spectral variations?
    \item How do we reconcile recent PIC results, which imply a critical Mach number of $\mathcal{M}_\rmn{crit} \approx 2.3$, when radio relics have been observed with X-ray derived Mach numbers weaker than this?
    \item If electrons are most efficiently accelerated at quasi-parallel shocks, why does polarization data appear to imply quasi-perpendicular acceleration?
\end{enumerate}

In this thesis we have dealt with problems 2 -- 6. Our strategy has been to investigate how merger shocks develop in cosmological simulations of clusters. We have then used this information to build a series of idealised shock-tube simulations, which can be run with significantly higher resolution. In particular, we have included upstream density fluctuations in our shock-tube simulations, which cannot be resolved at sufficiently fine scales in current cosmological simulations. We thus resolve physics in our shock-tube simulations that remains unresolved in the cosmological simulations.

We have found that, during major mergers, the merging cluster drives a bow shock, which eventually evolves into a so-called ``run-away'' shock. This is consistent with the idealised cluster simulations by \citep{zhang2019}. We further observe that, at typical distances reported for radio relics, this runaway shock can encounter an accretion shock, produced by infalling material. This leads to the production of a thin, outwards-travelling, shock-compressed density sheet. The point where these shocks collide can be modelled as a Riemann problem, where we only need to set a pressure discontinuity, with the mean upstream and downstream densities being set to the same value. 

We generate density turbulence for our shock-tubes using methods based on \citet{ruszkowski2007} and \citet{ehlert2018}. This allows us to precisely define the relative variance, power-law slope, and the scale of turbulent injection, without adding velocity turbulence. We can thereby better isolate the impact of density turbulence and its characteristics on our results. 

By using the tracer functionality in \textsc{Arepo}, we collect all gas variables in the shock-tube simulations necessary for evaluating the Fokker-Planck equation (see Sec.~\ref{sec:cosmic-rays}) in the Lagrangian frame. We then evaluate cosmic ray electron spectra with \textsc{Crest}, before post-processing the results with \textsc{Crayon+}, an emission code, thereby producing non-thermal emission data from our simulations \textit{ab initio}. We initially assign all tracers with a thermal spectra and model acceleration of the electrons at shocks using the DSA formalism (see methods section of Chapter~\ref{chapter:paper-three}). We also model the effect of adiabatic compression and expansion, as well as Coulomb, bremsstrahlung, inverse Compton, and synchrotron cooling. These are all independently tested through our pipeline (see Chapter~\ref{chapter:additional-work}). We do not include the effect of Fermi I re-acceleration, although \textsc{Crest} is technically capable of this. This should only impact the normalisation of the spectra, however, not its shape at radio-emitting frequencies \citep[see, e.g.,][]{pinzke2013, winner2019}, which we predominantly study in Chapters~\ref{chapter:paper-three} and~\ref{chapter:paper-four}.

We find that the addition of upstream density perturbations directly leads to: i) the formation of a Mach number distribution at the shock front, and ii) shock corrugation. Additionally, as the shock-wave is not density driven, the resulting density gradient at the contact discontinuity points towards the shock-compressed region. This leads to a misalignment between density and pressure gradients, which results in a Rayleigh-Taylor instability. This instability is enhanced by the corrugation of the contact discontinuity, which again follows from the density perturbations. The result is the generation of velocity turbulence behind the shock-front.

These three effects result in several solutions for the aforementioned problems:
\begin{itemize}
    \item \textbf{$\mathcal{M}_\rmn{X-ray} - \mathcal{M}_\rmn{radio}$ discrepancy:} The tail of the Mach number distribution dominates the spectral features. In particular, this part of the distribution is more influential at higher momenta, leading to a flattening of the integrated cosmic ray electron spectra. However, the standard $\alpha_\rmn{e} - 1$ slope is also not reproduced in the cooled part of the spectrum, being flatter than expected. This leads to radio-derived Mach numbers being biased high. This effect is particularly strong for weak shocks ($\mathcal{M} \lesssim 2$).
    \item \textbf{$\bm\upmu$G-strength magnetic fields:} The Rayleigh-Taylor instability leads to additional compression, which results in magnetic field amplification up to $\upmu$G levels. We note, however, that the peak magnetic field strengths cannot be reproduced by compression alone, and present a resolution-study which implies the existence of a small-scale dynamo in our simulations, as generated by the velocity turbulence. Furthermore, we show that the synchrotron emission is strongly weighted by the tail of the distribution of magnetic field values. This means that synchrotron emission weighted field strengths significantly over-estimate the volume-average strength.
    \item \textbf{Failure of cooling models:} Most spectral cooling models are based upon the assumption that lines-of-sight intersect homogeneous populations of cosmic ray electrons. This implicitly assumes that distance from the shock front is a good indication of time cooled. We show that velocity turbulence in our simulations breaks this assumption. Such turbulence leads to superpositions of fresh and older populations along the line-of-sight, thereby reducing the spectral curvature, and flattening the observed trajectory in colour-colour diagrams.
    \item \textbf{Critical Mach number:} We show that the spectra at radio-emitting frequencies is dominated by the tail of the Mach number distribution. This means that observed radio relics are entirely consistent with a critical Mach number, as long as a sufficient amount of the Mach number distribution lies above the critical value. We produce radio emission maps that show that this is true even for very narrow distributions where the peak of the Mach number distribution is at $\mathcal{M} = 2$. 
    \item \textbf{Origin of relic morphology:} By varying the power spectra of the upstream density turbulence, we show that relative variance plays a key factor in setting radio relic morphology. Specifically, this is the most significant factor for setting : i) the Mach number distribution, ii) the level of shock corrugation, iii) the electron spectral slope, iv) the extent of the emission, v) the patchiness of the emission, and vi) the inferred magnetic field strength from synchrotron emission. We show that relic morphology is significantly less sensitive to the power law slope and the injection scale. However, with this said, we find that the turbulent injection scale is directly correlated with the spacing of “threads” and “knots” at the shock front. This implies that this value may be directly inferred from observations. Indeed, by applying this method to observations of the Toothbrush and Sausage radio relics, we find that injection scales of between 500 and 600 kpc are common. This is $3 - 4$ times higher than the injection scale usually assumed. We note, however, that the above factors only provide half the answer, as other details such as shock and fossil electron morphology will no-doubt also play a role in setting relic morphology.
\end{itemize}

In total, this represents four solutions to the initial seven problems, with partial answers provided towards a fifth.

\subsection{Extensions and outlook}

As previously, there are several ways in which our work could be extended:
\begin{itemize}
    \item \textbf{Answering the remaining questions:} Initial extensions of this work will naturally focus on answering the remaining problems. In particular, we note that our shock-tubes are ideal for investigating how polarisation relates to magnetic-obliquity dependent shock acceleration. The two remaining problems (``the origin of seed electrons'' and ``radio relic morphology'') will require cosmological simulations, however. Fortunately, through the work presented in Chapter~\ref{chapter:additional-work}, \textsc{Crest} is now suitable for such simulations.
    
    \item \textbf{Investigation into the merger shock vs.\ accretion shock scenario:} Whilst we have shown that this scenario provides solutions to at least four of the questions we identified, and we observed the collision of merger and accretion shocks at distances observed for real radio relics, a full investigation into the scenario in a cosmological context remains to be performed. This investigation would need to address questions such as: how common is this scenario? And does the shock morphology and length match that of real radio relics? We could also improve our shock-tube simulations by modelling the collision of two shocks, rather than running the simulation from the point of collision, as we currently do.
    
    \item \textbf{Inclusion of Fermi I re-acceleration:} Whilst not included in our initial paper, \textsc{Crest} is technically capable of modelling Fermi I re-acceleration. Our shock-tube simulations can therefore be used to find the normalisation required for fossil electron spectra in order to match the surface brightness of real radio relics. Estimates already exist in the literature \citep[see, e.g.][]{pinzke2013}, but do not exist for our specific scenario. To reduce the number of assumptions, this extension could be combined with results from the ``origin of seed electrons'' project, with spectra produced from cosmological simulations (pre-merger shock) being used as the initial spectra in the shock-tubes. Such a simulation could help provide better constraints on the acceleration efficiency necessary for radio relics.

    \item \textbf{Inclusion of Fermi-II re-acceleration:} Similarly, \textsc{Crest} is capable of Fermi II re-acceleration, although currently only with a fixed diffusion coefficient. The downstream velocity turbulence seen in our simulations would certainly be sufficient for turbulent re-acceleration, but it is unclear whether this acts fast enough to result in a substantial contribution to the overall spectra and emission. Nonetheless, something is required to further flatten the trajectory of our simulated radio relics in the colour-colour plane, and  models in the literature do occasionally point towards turbulent re-acceleration \citep[see, e.g.][]{fujita2015, kang2024}. Their contribution therefore deserves investigation.
    
    \item \textbf{Inclusion of pitch-angle dependent cooling:} Another potential solution for the colour-colour plane issue is the inclusion of pitch-angle dependent cooling, in combination with an incomplete pitch-angle isotropisation process. At current, \textsc{Crest} assumes an isotropic distribution of pitch angles.
    
    \item \textbf{Inclusion of non-Gaussian magnetic fields:} It is generally believed that the magnetic field at the outskirts of the cluster results from the small-scale dynamo process within it \citep[see, e.g., simulations in][]{tevlin2024}. Such magnetic fields should be intermittent \citep{sur2021}, in contrast to the magnetic field in our initial conditions, which are Gaussian. Increasing the intermittency of the magnetic field could, in particular, increase the likelihood of magnetic filaments being formed in the resulting relic. This is particularly worth checking given our result that relics can have filamentary-looking morphologies purely due to the corrugation of the shock-front alone.

    \item \textbf{Inclusion of velocity turbulence:} The initial conditions in our shock-tube simulations have no velocity turbulence. This was to our advantage as, by removing another variable, we were better able to isolate the cause of the underlying physics. In particular, we could show that a Mach number distribution is produced purely by the density fluctuations alone. Because the shock velocities are larger than the expected turbulent velocities, we do not expect the inclusion of velocity turbulence to substantially change our conclusions. Nonetheless, we expect a degree of velocity turbulence at the cluster outskirts, and the impact of this on our results should be investigated.

    \item \textbf{Modelling of a full-size radio relic:} Finally, we have, so far, only simulated shock-tubes with a maximum depth and height of 300 kpc, and with uniform turbulent parameters in the upstream conditions. We showed, however, in Chapter~\ref{chapter:paper-four} that the turbulent injection scale is likely larger than 300 kpc in real radio relics and that this could have a minor effect on our results. Real relics are also frequently inhomogeneous along their length \citep[see, e.g., the Toothbrush relic][]{rajpurohit2020}, and this could well be produced by variations in the upstream conditions, as we showed in our study. Modelling specific relics would help us understand whether we truly understand what is setting their morphology, as well as helping us interpret observations.
\end{itemize}

Observations of radio relics have improved significantly in the last twenty years with resolution increasing at both ends of the radio spectrum. For example, at the high-frequency end the Karl G. Jansky Very Large Array (VLA), the Australia Telescope Compact Array (ATCA) and the European VLBI Network (EVN) are all in operation\footnote{The Australian Square Kilometre Array Pathfinder (ASKAP) is also in operation, but more typically used for broader surveys, owing to its poorer resolution.}, providing arcsecond or better resolution at 1 GHz and above. Meanwhile, at the low-frequency end, Low Frequency Array (LOFAR) and the LOFAR Two-Meter Sky Survey (LoTSS) can provide better than 10 arcsecond resolution at 150 MHz\footnote{The Murchison Widefield Array (MWA) also covers this range, but, like ASKAP, is typically only used for larger structures due to its arcminute resolution.}. The upgraded Giant Metrewave Radio Telescope (uGMRT) and the MeerKAT telescope cover these frequency ranges and have produced enormously detailed emission maps in recent years. Indeed, observations appear to show increasing substructure in radio relics with increasing resolution. This is unlikely to cease with the Square Kilometer Array (SKA), which is due to come online in 2027, and will provide observations across a range from 50 MHz to 15 GHz at resolution an order of magnitude higher than the current generation of radio telescopes with significantly increased sensitivity \citep{braun2019}.

As discussed in Chapter~\ref{chapter:theory}, X-ray observations are vital for understanding radio relics, as they extend out to large cluster-centric radii and can be used to measure density and temperature. This can, in turn, be used to measure the Mach number of a shock. At current, the X-ray Multi-mirror Mission (XMM) Newton telescope, Chandra X-ray Observatory (CXO), and the X-ray Imaging and Spectroscopy Mission (XRISM) are the three highest resolution-resolution telescopes available, with the first two having provided the majority of X-ray derived Mach numbers for radio relics in the last two decades. The increased spectral resolution of XRISM, which was launched in September of last year, should allow it to more finely measure density and temperature fluctuations in the denser cluster regions. However, the spatial resolution and sensitivity of this telescope is likely too low for observations of the cluster outskirts. Nonetheless, XRISM will provide a key proof-of-concept survey ahead of the launch of the New Advanced Telescope for High-Energy Astrophysics (NewAthena) and the Advanced X-ray Imaging Satellite (AXIS), which are due to launch in the 2030s. The results of these will be key for constraining the initial conditions of our own simulations. 

Taken together, constraints on ICM conditions and on radio relic structure should increase significantly in the next ten years, providing new challenges for simulations. It is, however, also up to simulators to provide predictions ahead of such observations. Such predictions will be helpful in providing targets for observation and are typically more powerful in constraining models.

\section{Development of \textsc{Crest}}

\textsc{Crest} is a spectral cosmic ray electron code originally written by Georg Winner. Its numerical techniques were first verified in a proof-of-concept paper \citep{winner2019}. It was then applied to 3D Sedov simulations, where it was used to show that supernovae results are consistent with electrons being more efficiently accelerated in quasi-parallel shocks \citep{winner2020}. Whilst the code results were generally accurate, the code was not yet fit for high-resolution simulations, nor was it suitable for cosmological simulations. During this thesis, I have significantly restructured \textsc{Crest} and the input files, as saved by \textsc{Arepo}. This has reduced the size of the typical file produced in low-resolution cosmological simulations by an order of magnitude, with savings being even greater than this for time bins with deeper hierarchies. Additionally, I have corrected formulae, made code segments more robust, and increased the code's capabilities. A list of the major improvements are given in Chapter~\ref{chapter:additional-work}.

Development work on \textsc{Crest} has been supported by L\'ena Jlassi and Maria Werhahn. Based on code initially provided by Philipp Girichidis, together we have developed a test pipeline, which provides integration testing for many of \textsc{Crest}'s functionalities. Finally, I have made a few minor but significant updates to the shock-finder in \textsc{Arepo} \citep[see][]{schaal2015}. This includes significantly reducing the computational cost of the code, as well as making it more logically consistent. This opens the way to running high-resolution cosmological simulations with \textsc{Crest}.

\newpage
\chapter{Commented publications list}
\label{publications_list}

In this chapter, I outline my and my co-authors' contributions to the three first-author papers that make up Chapters~\ref{chapter:paper-one},~\ref{chapter:paper-two}, and~\ref{chapter:paper-three} of this thesis. I also define my contributions to further publications not included in this thesis, but written during my doctoral studies. Chapters~\ref{chapter:paper-one} and~\ref{chapter:paper-two} expand upon concepts first introduced in my \href{https://www.aip.de/media/thesis/joseph-whittingham-master-thesis.pdf}{Masters thesis}. I therefore give an explicit breakdown of the updated material here.

\textit{Submitted dates provided where publication took place after thesis defence.}

\vspace{1cm}
\textit{Papers presented in Chapters~\ref{chapter:paper-one},~\ref{chapter:paper-two}, and~\ref{chapter:paper-three}:}
\vspace{-0.2cm}
\clinee
\textbf{J. Whittingham}, M. Sparre, C. Pfrommer, and R. Pakmor. \textit{The impact of magnetic fields on cosmological galaxy mergers - I. Reshaping gas and stellar discs}. In: MNRAS, Volume 506, Issue 1, pp.229-255 (published Sept. 2021). doi: \href{https://doi.org/10.1093/mnras/stab1425}{10.1093/mnras/stab1425}. arXiv: \href{https://arxiv.org/abs/2011.13947}{2011.13947 [astro-ph.GA]}.

CP and MS conceptualised the project. MS provided the initial conditions and guidance running the simulations. I oversaw the development of the project, ran the simulations and post-processing (power spectra) code, created the figures, and prepared the manuscript. All contributed to the discussion and interpretation of the results, and were involved in the editing process.

As stated above, this paper expands upon on work initially started during my Masters. Additional analysis was performed to quantify the growth of the magnetic field and to show evidence for the existence of a small-scale dynamo in the simulations, with an emphasis on the study of power spectra. The evolution of the gas and stellar distributions was also quantified, which proved a substantial step towards the development of a model, as presented in the following paper. Further work was carried out to show the numerical stability of the simulations, as well as the impact (or lack thereof) of magnetic fields on the Springel \& Hernquist ISM's ability to replicate the Kennicutt-Schmidt relation. On the basis of the firmer understanding resulting from the above work, the text was fully re-written.

\noindent\textit{New material:} Figures 3, 4, 8, 11, 13, 14, 15, 16, 18, as well as the bottom two rows of Figure 2, and the upper two rows of 6 and 10 were fully developed during my doctoral studies. Tables 1 and 2, as well as the calculations of the statistics shown here, are also new developments. All text is original.

\newpage
\clinee
\textbf{J. Whittingham}, M. Sparre, C. Pfrommer, and R. Pakmor. \textit{The impact of magnetic fields on cosmological galaxy mergers - II. Modified angular momentum transport and feedback}. In: MNRAS, Volume 526, Issue 1, pp.224-245 (published Nov. 2023). doi: \href{https://doi.org/10.1093/mnras/stad2680}{10.1093/mnras/stad2680}. arXiv: \href{https://arxiv.org/abs/2301.13208}{2301.13208 [astro-ph.GA]}.

I proposed the general idea of this paper, ran the additional simulations required, created the figures, and produced the manuscript. Simulations used the initial conditions provided by MS, as previously. CP assisted, in particular, with developing the structure of the paper. All contributed to the discussion and interpretation of the results, and were involved in the editing process.

This paper also includes some work originally produced during my Masters. Even more substantial changes took place here, however, with further investigation leading to the whole model being re-written; in the Masters thesis, I suggest that additional fuelling of the AGN is behind the difference in morphologies. In this paper, however, I show that this is explicitly not the case, and that the ultimate cause is the redistribution of gas due to the magnetic fields, leading to the growth (or otherwise) of Lindblad resonances. To show this, amongst varied analysis, I ran two additional simulations with quasar feedback stopped artificially at the coalescence of the galaxies. 

\noindent\textit{New material:} Figures 1 -- 6, 8, 10, 11, and 12 were fully developed during my doctoral studies. All text is original.

\vspace{2cm}

\clinee
\textbf{J. Whittingham}, C. Pfrommer, M. Werhahn, L. Jlassi, and P. Girichidis. \textit{Zooming-in on radio relics -- I. How density fluctuations explain the Mach number discrepancy, microgauss magnetic fields, and spectral index variations}. In: A\&A, Volume 706, id.A39, 29 pp. (published Feb. 2026, submitted Nov. 2024). doi: \href{https://doi.org/10.1051/0004-6361/202453002}{10.1051/0004-6361/202453002}. arXiv: \href{https://arxiv.org/abs/2411.11947}{2411.11947 [astro-ph.HE]}.

I developed the concept for this paper with CP. I created the initial conditions, with turbulence generated using a modified version of code initially written by Kristian Ehlert. I ran the simulations and the post-processing codes. Cosmic ray electron spectra were created using \textsc{Crest}, which is developed by a team made up of myself, MW, LJ, and PG. Synchrotron emission was generated using \textsc{Crayon+}, a code created by MW. Discussion and interpretation of the results was predominantly performed by myself and CP. I created all figures (with Kamlesh Rajpurohit providing data for comparison in Figure 13), and prepared the manuscript. All authors contributed to the editing of the script.
 
\newpage
\vspace{1cm}
\textit{Second-author publications during my doctoral thesis:}
\vspace{-0.2cm}
\clinee
M. Sparre, \textbf{J. Whittingham}, M. Damle, M. H. Hani, P. Richter, S. L. Ellison, C. Pfrommer, and M. Vogelsberger. \textit{Gas flows in galaxy mergers: supersonic turbulence in bridges, accretion from the circumgalactic medium, and metallicity dilution}. In: MNRAS Volume 509, Issue 2, pp.2720-2735 (published Jan. 2022), doi: \href{https://doi.org/10.1093/mnras/stab3171}{10.1093/mnras/stab3171}. arXiv:
\href{https://arxiv.org/abs/2110.03702}{2110.03702 [astro-ph.GA]}.

In this paper we use the simulations presented in Chapters~\ref{chapter:paper-one} and~\ref{chapter:paper-two} to investigate the dynamics of gas during a merger, with particular emphasis on: i) the properties of the bridge during the merger, ii) the impact on the starburst, and iii) the impact on the metallicity in the galaxy post-merger. We find that gas in the bridge is super-sonic and turbulent, but also star-forming. This means that bridges formed during mergers are excellent places to test theories of turbulence-dependent star-formation efficiency. Using tracer analysis, we also find that, whilst the ISM contributes the majority of gas to the bridge, between one third and one half of the gas originates from the CGM instead. Gas from the CGM also plays a major role in igniting the starburst in mergers, with up to half of the total SFR fuelled by gas that was originally outside the disc. Idealised simulations of mergers will necessarily miss such effects. Finally, the accretion of large amounts of (near-)pristine gas dilutes the metallicity in the disc. This explains why merging galaxies are not well-captured by the so-called ``fundamental metallicity relation''.

I ran the simulations analysed in the paper, and provided data for tracking the main subhalos. I also discussed the overall results with MS, and edited the paper.

\vspace{1cm}
\clinee
Y. Hu, \textbf{J. Whittingham}, A. Lazarian, C. Pfrommer, S. Xu, and T. Berlok. \textit{Anisotropic Velocity Fluctuations in Galaxy Mergers: A Probe of the Magnetic Field}. In: APJ, Volume 983, Issue 1, id.32, 17 pp. (published April 2025, submitted Oct. 2024), doi: \href{https://doi.org/10.48550/arXiv.2410.08157}{10.48550/arXiv.2410.08157}. arXiv: \href{https://arxiv.org/abs/2410.08157}{2410.08157 [astro-ph.GA]}.

The Velocity Gradient Technique (VGT), developed by Yue Hu and Alex Lazarian amongst others, is a potentially powerful method for inferring the magnetic field morphology in external galaxies and galaxy clusters. It works on the basis that turbulent eddies become increasingly anisotropic as the magnetic field grows stronger. This implies, for sub-Alfv\'{e}nic conditions, a statistical offset of 90 degrees between the magnetic field direction and the velocity gradient, whcih can, in turn, be inferred using spectroscopic observations and so-called ``velocity caustics''. Whilst the technique has been shown to work in idealised simulations of turbulence, it lacked independent verification with galaxy simulations, where dynamics are significantly more complicated. This paper attempts to fill that gap. We find that the turbulence is indeed anisotropic in our simulations. By producing synthetic observations with the VGT, we also find that it generally captures the projected magnetic field direction.

I provided the simulation data, as well as a framework code for analysing them. This includes code that was used to produce face- and edge-on projections for figures 1, 7, and 8. On top of this, I contributed to discussion and interpretation of the results, and suggested figure 2 and the top of figure 6. I also wrote section 3.1, which discusses the simulation set-up. I was heavily involved in editing the paper, which included reformulating its structure.

\vspace{3cm}
\textit{Further publications during my doctoral thesis:}

\clinee
M. Werhahn, P. Girichidis, C. Pfrommer, and \textbf{J. Whittingham}. \textit{Gamma-ray emission from spectrally resolved cosmic rays in galaxies}. In: MNRAS Volume 525, Issue 3, pp.4437-4455 (published Nov. 2023). doi: \href{https://doi.org/10.1093/mnras/stad2105} {10.1093/mnras/stad2105}. arXiv: \href{https://arxiv.org/abs/2301.04163}{2301.04163 [astro-ph.HE]}.

In this paper, the \textsc{Crayon+} code was applied to isolated \textsc{Arepo} galaxy simulations, with the aim of understanding the impact of spectrally-resolved CR transport on synthetic gamma-ray emission maps. It was found that spectrally-resolved transport led to more extended emission at high energies ($\sim$100 GeV), due to the faster diffusion of the associated CR protons. However, integrated CR proton spectra showed little difference between this model and a non-spectral steady-state model. Both models were able to recover the observed far-infrared (FIR) -- gamma-ray luminosity correlation, with the spectral model having marginally more success.

I was involved in meetings that led to the conceptualisation of this paper. I also contributed to the editing
process after preparation of the manuscript.

\vspace{1cm}
\clinee
L. Tevlin, T. Berlok, C. Pfrommer, R. Y. Talbot, \textbf{J. Whittingham}, E. Puchwein,
R. Pakmor, R. Weinberger, and V. Springel. \textit{Magnetic dynamos in galaxy clusters: the crucial role of galaxy formation physics at high redshifts}. In: A\&A, Volume 701, id.A114, 28 pp. (published Sept. 2025, submitted Oct. 2024), doi: \href{https://doi.org/10.48550/arXiv.2411.00103}{10.48550/arXiv.2411.00103}. arXiv: \href{https://arxiv.org/abs/2411.00103}{2411.00103 [astro-ph.GA]}.

In this paper, we show that galaxy formation physics are crucial to the development of the magnetic field in clusters. Specifically, it is shown that the ICM magnetic field is amplified initially in the ISM of the proto-cluster through a supersonic small-scale dynamo, before winds and ram-pressure interactions of galaxies with the ICM transport the field to the surrounding environment. At lower redshifts, mergers and accretion inject turbulence, thereby leading to the generation of a subsonic small-scale dynamo, which helps to maintain the magnetic field strength. It is additionally shown that, over the evolution of the cluster, the particle mean free path in the cluster is below the typical magnetic coherence scale. This means that MHD is a valid formalism for the study of magnetic fields in clusters.

I worked on the initial conditions for the simulations featured in this paper, running initial variations and validating their output. I also suggested the investigation into whether seeding of the ICM magnetic field was done primarily through stellar- or AGN-driven winds, which became the basis for fig. 7 and the attached analysis. I had discussions with the lead author, which helped, in particular, to shape the analysis based on fig. 6. Finally, I was involved in the editing of the paper after preparation of the manuscript.
\newpage
{
\def\bibfont{\footnotesize}
\addcontentsline{toc}{chapter}{\numberline{}Bibliography}
\begin{multicols}{2}{
\bibliographystyle{Bibliography/apa_url}
\bibliography{Bibliography/bibliography,Bibliography/chapter1,Bibliography/chapter2,Bibliography/chapter6,Bibliography/chapter7}
}
\end{multicols}}
\clearpage

\backmatter
\chapter*{Acknowledgements}
\addcontentsline{toc}{chapter}{\numberline{}Acknowledgements}

There is a long list of people who have supported me during my PhD, and have consequently shaped this thesis in one way or another. First of all, I would like to thank my supervisor Christoph Pfrommer. Christoph has put countless hours into scientific discussion, proof-reading, meetings, and organisation -- more so than ever during the write-up of this thesis. He has pushed each of the papers to be their best possible version, and has applied the same standard to this thesis. I have become a better physicist because of it and am genuinely proud of the work we have done together. On a different, but no less important note, I am very grateful for your understanding and support when I needed to return to be with my family in England during a couple of difficult times.

I would also like to thank Martin Sparre, who has been a good friend throughout, but was an especially good mentor during my Masters thesis and at the start of the PhD. He showed me the ropes and was the one who recommended me to be taken on as a doctoral student in the first place. Thank you -- I hope we stay in contact for a long time.

I would like to thank Oliver Gressel, Philipp Richter, Silke Kuba, and the astrophysics office at Uni Potsdam, who have all helped me through the necessary admin (including right up to the last minute!)

I would like to thank my colleagues at AIP, who have provided a lively and enjoyable working environment. There have been many, some who are no longer here, but I valued our interactions a great deal. In particular, though, I would like to thank Léna Jlassi and Maria Werhahn, who were crucial to the development of \textsc{Crest}, and I would like to thank Philipp Girichidis for constructing the initial test pipeline that we have used so heavily. I would also like to thank Thomas Berlok, whose code \textsc{Paicos} was used heavily in this thesis and who was instrumental in setting up the cluster simulations. Additionally, I would like to thank Georg Winner, who wrote the initial code, gave tutoring sessions about it in the early days, and even gave up his time for an online Q\&A session after he had left science.

I owe a great thanks to my family and my friends both here and abroad. You bring so much joy to my life and it would be totally unbalanced without you. I'm really very grateful that you have all been so patient in the last year, when my workload started to accelerate, and I became less and less available -- I hope you haven't felt too neglected. If I start listing people I will surely miss someone out, but you know who you are. I have missed you all, and greatly look forwards to spending more time together again.  

Finally, I would like to thank Ana and her wonderful daughter Sunna. I must have seemed surgically-attached to my laptop at times over the last year. You have been patient, supportive, and have believed in me, even when I was less sure. You mean the world to me. Thank you.

\chapter*{Statement of originality}

\addcontentsline{toc}{chapter}{\numberline{}Statement of Originality}
This thesis is based on research carried out at the Leibniz-Institut f\"{u}r Astrophysik Potsdam (AIP). No part of this thesis has been submitted elsewhere for any other degree or qualification. All work presented is my own unless otherwise stated.

\end{document}